# 符号积分系统概论

# 目 录

# 前言

计算机代数系统(Computer Algebra System, CAS)

  A computer algebra system(CAS) is a software program that facilitates symbolic mathematics. The core functionality of a CAS is manipulation of mathematical expressions in symbolic form.

  计算机代数系统（英文：*Computer Algebra System*，简称 *CAS*）是进行符号运算的软件。这种系统的核心功能是对符号表示的数学表达式进行实现。

计算机代数系统处理的表达式类型（Types of expressions）

  多变元多项式
  标准函数（三角函数、指数函数等等）
  特殊函数（Γ 函数、Bessel 函数等等）
  由各种表示式合成的函数
  表示式的导函数、积分、和与积
  以表示式为系数的级数
  表示式构成的矩阵

计算机代数系统能够进行的符号运算（Symbolic manipulations）

  表示式的简化
  对表示式求值
  表示式的变形：展开、积、幂次、部份分式表法、将三角函数表为指数函数等等。
  对单变元或多变元的微分。
  带条件或不带条件的整体最佳化。
  部份或完整的因式分解。
  求解线性方程组或一些非线性方程式。
  某类微分方程或差分方程的符号解。
  求某些函数的极限值。
  一些函数的定积分或不定基分，包括多变元的情形。
  泰勒展开式、罗朗展开式与 Puiseux 展开式
  某些函数的无穷级数展开式。
  对某些级数求和。
  矩阵运算。
  数学式的显示，通常借着 TeX 之类的系统达成。

计算机代数系统的其它功能（Additional capabilities）

  通常计算机代数系统还能进行一些数值运算：
  函数的确切求值。
  高精度求值，例如计算 $2^{1/3}$ 到小数点后 10000 位。
  线性代数的数值运算。
  描绘二维或三维的函数图形。

  在数值运算方面，计算机代数系统的速度通常较 Matlab、GNU Octave 或 C 语言中以同等方式实作的程式慢。这是因为计算机系统几乎总是对符号表示式运算，故不能充分利用 CPU 的既有指令。

  许多计算机代数系统内建高阶编程语言，以供使用者扩充功能，或设置个人的操作

模式。

常见的计算机代数系统

| Computer algebra systems | |
|---|---|
| Retail(商业) | Algebrator<br>ClassPad Manager<br>LiveMath<br>Magma<br>Maple<br>Mathcad<br>Mathematica<br>MuPAD (MATLAB symbolic math toolbox)<br>TI InterActive!<br>WIRIS |
| Open source(开源) | Axiom<br>Cadabra<br>CoCoA<br>DoCon<br>Eigenmath<br>FriCAS<br>GAP<br>GiNaC<br>Macaulay2<br>Mathomatic<br>Maxima<br>OpenAxiom<br>PARI/GP<br>Reduce<br>Sage<br>SINGULAR<br>SymPy<br>Xcas<br>Yacas |
| Free/shareware(免费) | Fermat<br>KANT |
| Discontinued | Derive<br>DCAS<br>Macsyma<br>muMATH |

# 第一章

这一章中我们主要针对计算机代数系统中的符号积分系统的现状进行介绍,并介绍主流计算机代数系统内部的符号积分系统的实现。符号积分系统是计算机代数系统中能够对不定积分,定积分,路径积分等积分表达式进行处理的模块。符号积分系统的基本功能为不定积分功能,在文献中 $Symbolic\ Integration$ 在绝大多数情况下代表的也是不定积分的概念,同时不定积分也是我们研究的重点。所以在之后的叙述中,符号积分代表对象为符号积分系统的不定积分部分,定积分等其它部分在之后章节进行介绍。

## 1.1 符号积分的发展历史

**1)数学方面(Mathematica Side)**

1800s Abel, Liouville 为基本定理的陈述与证明做出了奠基性的工作,刘维尔第一个用分析的语言严格陈述并证明了核心定理—刘维尔定理。

1910 and 1913 Mordukhai Boltovskoi (two books on this subject)
提出了一个解决问题的新思路,因为只在苏联发表,影响范围很小,但不能否认其工作的开创性。

1948 Ritt 延续了前人的工作,之后的工作由 Kolchin,和 Rosenlict 完成。

1970 Rosenlict's PhD students: Risch, Singer, Bronstein 为刘维尔定理的代数化和使用话做出了突出的贡献,Rosenlict 和 Risch 分别用代数的语言证明了刘维尔定理和强刘维尔定理,Bronstein 完善了理论,并在 IBM T.J. Watson Research Center 实现了几乎完整的初等函数的不定积分方法,这在当时是不定积分最完整的实现。之后 Bronstein 完成了专著"Symbolic Integration—Transcendental Functions"这是对于超越函数不定积分介绍最为系统的一本专著,第一版和第二版于 1997 和 2004 年出版,但是 2005 年 6 月 6 日,由于心脏病突发,Bronstein 离开了认识,计划中关于代数函数不定积分的总结的专著因此搁浅。

**2)计算机方面(Computational Side)**

对一个函数进行符号微分运算是较为简单的事情,早在 20 世纪 50 年代计算机诞生之际,Kahrimanian 和 Nolan 就各自独立地写出了符号微分程序。

1953 two early programs were written

    Kahrimanian at Temple University(天普大学)

    Nolan at MIT

1961 Slagle's thesis in Lisp->SAINT

这是第一个符号积分程序,其特点是采用人工模拟求积分的启发性算法,可以成功求解大部分大学一年级微积分考试试题。

1960's Moses's(MIT) in Lisp-> SIN

SIN 和 SAINT 相比,除了人工智能算法以外,还使用了一般性的 Risch 算法的求解过程。

    Manove's implementation(algebraic manipulation)

在实际实现方面值得一提的是 SIN 系统,主要是因为其提出的符号积分系统的设计理念。SIN 提供的系统框架理念是对于被积函数进行分层次的处理,可以通过拆解被积函数,凑微分,模式匹配等方法对被积函数进行细致的处理,分层次多次处理的实现理念使各种算法相互独立,同时又相互补充,充分发挥各种算法的优势。虽然实现的效果有一定的局限性,但是系统的框架思想成为了经典。会在下一章中较为详细地介绍 SIN 系统的设计理念。

# 1.2 主流计算机代数系统的符号积分模块实现

这一部分主要介绍 Mathematica，Maple 和 Sage 这三款主流代数软件中符号积分模块的具体实现的方法。其中 Mathematica 和 Maple 是最为常见也是最为成功的两款商业计算机代数系统，Sage 则是开源计算机代数系统中最为成功的典范，其设计理念是整合各个开源计算机代数系统的模块，从而组合而成一个完整的软件。之后将会介绍国内计算机代数系统的一些发展情况作为补充和对比。并在下一小节以 Mathematica 和 Maple 为例指出现有计算机代数系统中符号积分模块中存在的问题和可能的改进手段和一些改进的方向。

### 1.2.1 Mathematica（最新版本：8.0.4，October 24, 2011）
Mathematica 中不定积分的计算过程为：

对于不定积分,只要被积函数和积分能被用初等函数,指数积分函数,多对数及其相关函数表示,则使用推广 Risch 算法。

对于其它不定积分,则使用带模式匹配的启发式进行化简。

Mathematica 中的算法囊括了诸如 Gradshteyn Ryzhik 之类标准参考书中的所有不定积分。

不包含奇点的定积分主要通过求不定积分的界来实现.

许多其它的定积分使用了 Marichev-Adamchik Mellin 变换方法来实现.其结果最初常常表示由 Meijer G 函数表出,从而可以使用 Slater 定理转换为超几何函数然后进行化简.

Integrate 使用了约 500 页 Mathematica 代码和 600 页 C 代码.

### 1.2.2 Maple（最新版本：16.00 / March 28, 2012）
Maple 中不定积分的计算过程为：
($i$)首先，Maple 用传统方法处理一些特殊的形式，如多项式，有理式、形如$(\sqrt{a+bx+cx^2})^n$和$Q(x)(\sqrt{a+bx+cx^2})^n$的根式，以及形如$P_n(x) \cdot lnx$或$\frac{P_1(x)}{Q_1(x)}ln\frac{P_2(x)}{Q_2(x)}$的表达式；

($ii$)如果传统方法难以奏效，Maple 将应用 Risch—Norman 算法，以避免在包含三角函数和双曲函数的积分中引入复指数和对数；

($iii$)如果仍然无法得到答案，Maple 将应用 Risch 算法，这将无法避免地在结果表达式中引入有关积分变量的$Root\ of$的表达式；

($iv$)如果最终还是没有找到解析表达式，Maple 会把积分式作为结果返回。

### 1.2.3 Sage
Sage 是目前 Maple 和 Mathematica 最大的潜在竞争对手。Sage's notebook is awesome. Sage 用 Maxima 作为自己的一部分。使用 Maxima 进行微积分运算，因为 Maxima 经过足够的测试，大多数时间运行稳定可靠，运行速度快，但是 Maxima 自身也存在着问题，Maxima 使用 Lisp 语言编写的，是及其困难（不可能）使用 Python 语言进行扩展。所以 SymPy 则是

项目长远发展的一个途径。但是在我们实现和 Maxima 一样的基本功能之前（not that far away），并且达到可比的速度（still quite a lot of work to do）并且使 SymPy 成为 Maxima 的替换物之前，SymPy 不会成为 Sage 的默认选项。

为什么不像 Maple/Mathematica 那样使用自己的语言呢？

我们想要使用一个用"古典"语言编写出的计算机代数系统，例如 C++，C，Python（也许还有 Java，Ruby，C#）。有很多实现的符号系统，但是目前为止，只有三个库（libraries），GiNac，Giac（both for C++）和 Sage（Python）满足了这个要求。甚至 GiNac 和 Giac 也是无比复杂（成千上万行代码），并且很难进行扩展。不幸的是，SymPy 尽力地使自己简单，同时保持为一个具备完整特征的计算机代数系统。这就意味着，我们优先选择那些能够以最少的行数实现的代码，同时，我们避免多次实现同一个功能。不会像使用 C++那样保持一个较高的运算速度，但是可以很容易地用自己的心酸大或者函数进行扩展。同时编写的代码只是一个普通的 Python 程序段，只是将 SymPy 作为一个通常的模式，所以不会纠结于 CAS 自身的语言，通常计算机代数系统的语言并没有很好地适应编程的需要。

之后，我们也许会针对一些特殊的情况，实现一些特殊的算法，将它们用 C++进行重写。一个整体的 CAS 是否能和一个特定用途的 CAS 模块运行的一样快。目标就是解决 SymPy 中的问题，在算法中实现，之后如果成功运行，但是运行速度不够快，就用一种更适合境况的语言进行实现。

总结如下：SymPy 是一个 Maple 和 Mathematica 的开源替代品，运行速度足够快（以 C++语言编写的情况下），可以将你的想法变成程序新的扩展，可以用 Python 调用，用于解决实际生活中存在的问题（as Maple/Mathematica can）。Sage 也在努力达到这个目标，但是借助一条不同的途径，SymPy 致力于一个成为小型的，普通的 Python 模块，而 Sage 则是致力于将每一个有用的开源数学软件包组合在一起，同时提供一个一致，清晰的截面。因此 Sage 的大型化，导致了另外的一个问题，目前 Sage 不能便捷地使用 Debian 进行打包。同时我们认为 SymPy 类目前在微积分功能方面是比 Sage 简单多的（因为 Sage 使用 Maxima 作为后端程序，所以 Sage 的速度更快）

SymPy 的另一个优势在于是用纯的 SymPy 语言编写的，所以 SymPy 具有完美的多平台特性，SymPy 轻巧，并且容易安装，使用。

SymPy 可以作为 Sage 的备选模模式使用（SymPy 包含在 Sage 2.7 中，之后作为默认的模式选项）。Sage 备受赞扬因为其最终可以使用所有的 python 语言编写的 CAS，SymPy 的目标则小有不同—SymPy 依然保持一个简单但是功能强大的 Python 模式，重点不在于速度，而在于简化和多功能。

代码依然在改进，但是现在功能已经足够强大—例如，我们相信 SymPy 有计算符号极限的最短的开源代码（大约 300 行）

Sage 的未来发展方向：
1.完善积分算法，使 SymPy 能够积出所有可积函数
2.完善技术展开
3.实现渐进展开
3.指数对象（张量）
4.用 C++，C 或 Cython 重写内核，将这个作为一个可选的模式，如果用户想要加快运行速度（Python 内核会一直作为默认值出现）

Sage 作为一个开源软件，通过大量开源数学软件包的整合实现一个完整的计算机代数系统的功能，其中主要数学功能的实现方式如下：

| | |
|---|---|
| 代数 | GAP，Maxima，Singular |
| 代数几何 | Singular |
| 任意精度计算 | MPIR，MPFR，MPFI，NTL |
| 算术几何 | PARI/GP, NTL，mwrank，ecm |
| 微积分 | Maxima，SymPy，GiNaC |
| 组合数学 | Symmetrica，Sage-Combinat |
| 线性代数 | ATLAS，BLAS， LAPACK，NumPy，LinBox，IML，GSL |
| 图论 | NetworkX |
| 群论 | GAP |
| 数值计算 | GSL，SciPy，NumPy，ATLAS |
| 数论 | PARI/GP，FLINT，NTL |
| 统计计算 | R, SciPy |

Sage 中包含的其它包

| | |
|---|---|
| 命令行 | IPython |
| 数据库 | ZODB， Python Pickles，SQLite |
| 图形界面 | Sage Notebook， jsmath |
| 图像 | Matplotlib，Tachyon3d，[[GD, Jmol |
| 交互式编程语言 | Python |
| 网络 | Twisted |

我们可以看到 Sage 的微积分功能主要通过 Maxima，SymPy，GiNaC 三个软件包进行实现，下面分别进行介绍，以便了解 Sage 的具体实现。

**Maima（最新版本 December 19, 2011: Maxima 5.26.0.）**

Maxima 是一个为了实现符号和数值表示实现的系统,能够实现的功能包括微分,积分，Taylor 级数，Laplace 变换，常微分方程，线性方程组，多项式，集合，数表，向量，矩阵和张量。Maxima 利用准确的分解，任意精度的整数和浮点数输出非常精确的结果。Maxima 可以在二维或三维情况下 plot 函数和数据。

Maxima 的源代码可以在许多系统上进行编译，包括 Windows，Linux 和 MacOSX。源代码和为 Windows 和 Linux 系统预编译好的二进制文件在 SourceForge file manager 目录下可以找到。

Macsyma 是在二十世纪六十年代末期在 MIT 产生的具有传奇色彩的计算机代数系统，Maxima 则是 Macsyma 的改良版本，这是 Macsyma 唯一一个仍公开发布并且有着活跃用户交流的版本,这要归功于其开源的特性。Macsyma 在当时极富革命性，许多之后的系统，如 Maple 和 Mathematica 都受到了它的影响。

Macsyma 的 Maxima 分支由 William Schelter 保持下来，从 1982 年直到他于 2001 年去世。在 1998 年，他得到了在 GNU 下公开源代码的许可。Maxima 的幸村要归功于他的努力和他的技术。我们十分感谢他奉献出自己的时间和专业知识使得原始的 Macsyma 代码

仍然发挥着作用。也是他带动了一批使用者和开发者将 Maxima 带到了更广大受众的面前。
我们持续不断的更新 Maxima，修补漏洞，改进代码和文档。我们欢迎来自 Maxima 用户群体的建议和贡献。大部分的意见会列在 Maxima mailing list

**SymPy（最新版本 29 Jul 2011 Version 0.7.1s）**
SymPy 是一个符号数学的开源 python 代码库。其致力于在成为一个完整的计算机代数系统的同时尽可能地保持代码的简洁，从而使得代码简明易懂同时具有很好的扩展性。SymPy 是完全用 Python 语言编写的，不需要其它的扩展库。

Sympy 语言特点
目前，SymPy 内核有 13000 左右行代码（包括扩展内容和 docstrings），基本功能包括：
基本的代数运算 *,/,+,-，**
基本的化简操作
展开
基本函数
复数
微分
Taylor（Laurent）级数
替换
任意精度的整数，有理数和浮点数
无交换符号
模式匹配
SymPy 模式（73000 行代码包括文档）
更多的函数
极限
使用 extened Risch-Norman heuristic 的积分
多项式（多项式除法，最大公约数，无平方因子的分解，groebner 基，分解）
解方程（代数方程，微分方程，方程系统）
符号矩阵（行列式，LU 分解）
Pauli，Dirac 代数
几何模式
绘图 2D 或 3D 模式
扩展的测试（15000 行代码，142 个文件），可以测试 SymPy 的每一个单一功能

Sympy 相关的项目
Sage：包含了 SymPy
mpmath：包含在 SymPy 中
piglet：包含在 SymPy 中
SymPycore：另一个 Python 的计算机代数系统
symbide：包含在 SymPy 中
sfepy
symfe
scipy.org

**Ginac**

GiNaC 是一个 C++库，是基于 GNU 环境。GiNaC 是 GiNaC is Not a CAS 的递归迭代的缩写。GiNaC 是作为 xloops 的替代引擎，xloops 现在是由 Maple 支持。但是，GiNaC 并不受限于高能物理应用。GiNaC 的设计与其它 CAS 相比是革命性的，GiNaC 并不尝试提供扩展的代数功能和一种简单的编程语言。取而代之 GiNaC 接受了一种已有的语言—C++并且用一系列的代数功能对 C++进行了扩展。

### 1.2.4. MMP（最新版本：3.0，2006 年 4 月 1 日）

MMP 是国内较为成型的一个计算机代数系统，由于设计理念的差异，在功能上和传统意义下的计算机代数系统差别较大，下面进行详细的介绍。

MMP 又称数学机械化自动推理平台，以符号计算为支撑，以数学机械化算法为核心，以我国在数学机械化方面的最新研究成果为依托，希望能够为科学研究、工程应用、教学提供一个应用数学机械化的理论与方法的强有力的软件工作平台和开发平台.

目前 MMP 主要功能如下：
    支撑系统, 包括内存管理、图形界面、编程环境.
    符号计算系统, 包括任意长度的数系统, 多项式运算,符号线性代数.
    核心模块, 包括多项式、常微、偏微分方程系统的吴特征列方法与投影定理.

应用模块, 包括
    MMP/Geometer: 几何定理自动证明与发现, 几何自动作图
    MMP/DiffEquation: 微分方程求解
    MMP/Identity: 组合恒等式自动证明
    MMP/Blending: 过渡曲面自动生成
    MMP/6R-Robots: 6R 机器人模拟

作为"973"项目，"数学机械化与自动推理平台" 经过五年执行于 2003 年 12 月 9 日进行了结题验收。通过本项目的执行，数学机械化在理论研究、应用研究与软件平台开发都取得了实质性突破。并在此基础上，申请了新的"973"项目"数学机械化及其在信息技术中的应用"，得到科技部批准立项。这一项目计划研究数学机械化理论、高效算法及其在信息安全、生物特征识别、几何建模等信息领域中的应用，并以此为基础开发智能型数学机械化网络软件。

MMP 具体的系统特征为：
    系统函数列表：
    数，表达式，多项式，链表，矩阵，方程求解
    吴特征列方法：
    代数情形的特征列方法，微分情形的特征列方法，代数和微分情形的映射运算
    几何自动推理：
    几何命题的输入与转换，几何定理自动证明与发现，微分几何定理证明
    实根分离与不等式自动发现
    组合恒等式证明
    遗传算法
    其他一般函数

## 1.3 现有计算机系统中存在的问题

这一部分主要介绍现有计算机代数系统中出现的问题，分别以 Maple 和 Mathematica 为例进行说明，并介绍一些尝试性的改进方案。

**1.3.1. Maple**

下面我们讨论在积分学当中的一个微妙的漏洞,在大多数计算机代数系统中都会出现这个问题，甚至于在许多教科书和积分表中这种情况也是长期存在。

>f:=1/2(2+sin(x));

$$f := \frac{1}{2+sin(x)}$$

>F:=int(f,x);

$$F := \frac{2}{3}\sqrt{3}arctan(\frac{1}{3}(2tan(\frac{1}{2}x)+1)\sqrt{(3)}$$

>limit(F,x=Pi,right),limit(F,x=Pi,left);

$$-\frac{1}{3}\pi\sqrt{(3)}, \frac{1}{3}\pi\sqrt{(3)}$$

关于函数$f(x)$的积分仅在一些区间上是正确的，因为$F$是不连续的，虽然由微积分的基本定理可知当$f$连续时$F$应该是连续的。进一步的讨论$F$的不连续点：

>discount(F,x);

$$\{2\pi\_Z1+\pi\}$$

因此，$F$在$x=(2n+1)\pi$处有跳跃间断点。这显然是可怕的，因为$f$的原函数$F$居然不连续。通过数学分析方法可知，Maple 在给出该问题的结果时使用了"万能代换"公式，即$tan(\frac{x}{2})=t$，尽管这一公式常常在将三角函数有理化时起到至关重要的作用，但使用该公式的代价是明显的：$tan(\frac{x}{2})$在$x=(2n+1)\pi$ $n\in Z$处的不连续性会"传染"给整个过程。这便是出现前面所述问题的原因。事实上，函数$\frac{1}{2+sin(x)}$的原函数不是初等函数。

另外一个有趣的现象是 Maple 对符号积分的缺陷：

>diff(f(x)*g(x),x);

$$(\frac{d}{dx}f(x))g(x)+f(x)(\frac{d}{dx}g(x))$$

>int(%,x);

$$\int(\frac{d}{dx}f(x))g(x)+f(x)(\frac{d}{dx}g(x))dx$$

显然，Maple 对此无能为力。

我们利用 Maple 中的基础函数如$op, nops, type, has, int$以及数值分析理论和方法,提出了新的符号积分算法，并编写了相应的 Maple 程序，成功解决了 Maple 下符号积分及三角函数积分缺陷，并将研究结果应用于有关定理自动推证中。

1.在 Maple 下关于符号积分依然存在缺陷，例如：

>P:=diff(f(x)*g(x),x);

$$\frac{d}{dx}f(x)g(x)+f(x)\frac{d}{dx}g(x)$$

但对上式再积分，Maple 就无能为力了：

>int(P,x);

$$\int \frac{d}{dx}f(x)g(x) + f(x)\frac{d}{dx}g(x)dx$$

应用更为广泛的计算机代数系统 Matlab 的符号计算的核心是 Maple 的，因此同样存在上述问题。基于此，我们利用 Maple 系统中的基础函数如 $op, nops, type, has, int$ 以及数值分析理论和方法，提出了新的符号积分算法，并编写了相应的 Maple 程序，成功解决了 Maple 中符号积分缺陷。

### 1.1 原函数为乘积表达式

定理 1：函数组 $\{f_1(x), f_2(x), \ldots, f_n(x)\}$ 中任取 $m(m \leq n)$ 个函数，不妨设这 $m$ 个函数为 $L = \{f_1(x), f_2(x), \ldots, f_m(x)\}$，对于 $\forall f_i(x) \in L$ 的 $p_i$ 次方的乘积表达式：

$$F(x) = \prod_{i=1}^{m} f_i(x)^{p_i} (p_i \in R)$$

的导函数为：

$$F'(x) = F'_1(x) + F'_2(x) + \ldots + F'_i(x) + \ldots + F'_m(x),$$

则 $F'(x)$ 中至少存在一个操作数 $F'_i(x)(i \in \{1, 2, \ldots, m\})$，将其导数降低 1 阶后与 $F(x)$ 至多相差一个常数因子。

证明：由函数乘积的求导公式 $(f(x)g(x))' = f'(x)g(x) + f(x)g'(x)$ 即可得证。

为了更清楚地理解定理 1 的内容，下面给出两个例子。

例：

$$F_1(x) = f_1^3(x)f_2^9(x)x^3$$

对函数 $F_1(x)$ 求导得：

$$F'_1(x) = 3f_1^2(x)f'_1(x)f_2^9(x)x^3 + 9f_1^3(x)f_2^8(x)f'_2(x)x^3 + 3f_1^3(x)f_2^9(x)x^2$$

其操作数分别为：

$$3f_1^2(x)f'_1(x)f_2^9(x)x^3, 9f_1^3(x)f_2^8(x)f'_2(x)x^3, 3f_1^3(x)f_2^9(x)x^2$$

上式中前两个操作数去掉导数符号分别为：

$$3f_1^2(x)f_2^9(x)x^3, 9f_1^3(x)f_2^8(x)x^3$$

与 $F_1(x)$ 只相差常数分别为 3,9。

另外，对于有理式或有理函数 Maple 通常会自动化简，即消去分子，分母中的公约数或公因式。但对于更复杂的含有无理数或无理式的多项式却不能进行这种化简，如下例：

例：求下面两函数的比值：

$$c := \sqrt{3}(\frac{d}{dx}f(x)) + g(x)$$

$$d := 3(\frac{d}{dx}f(x)) + \sqrt{3}g(x)$$

>c/d；

$$\frac{\sqrt{3}(\frac{d}{dx}f(x)) + g(x)}{3(\frac{d}{dx}f(x)) + \sqrt{3}g(x)}$$

利用强行化简命令 $simplyfy$ 或者 $normal$ 也得到同样的结果，均未得到我们想象的 $\frac{\sqrt{3}}{3}$。

为了解决该问题，我们给出下述定理，定理中所谓的函数多项式是指+项至少为 2 项的

函数表达式，而后文中的函数单项式（简称单项式）指没有+项的函数表达式

定理 2：任给两个同类函数多项式$F_1(x), F_2(x)$，设$L_1, L_2$是$F_1(x), F_2(x)$分别展开后所有操作数组成的列表，若$F_1(x) = cF_2(X)$，则对$\forall F_{1i}(x) \in L_1, \exists F_{2j}(x) \in L_2$，使得$F_{1i}(x) = cF_{2j}(x)$（其中$c$为常数）。

证明：利用$op$获取两个函数多项式的操作数列表$L_1, L_2$，再将$L_1, L_2$中的操作数两两比较即可得证。

由定理 2 可知，要判定两个函数多项式$F_1(x), F_2(x)$是否相差一个常数因子，可以通过寻找函数多项式$F_1(x)$在$F_2(x)$中的对应项获取。例如$F_{11}(x)$为$F_1(x)$中的任一项，$F_2(x)$中与之对应的项为$F_{21}(x)$。当两者相差一个常数因子，即存在常数$c = \frac{F_{11}(x)}{F_{21}(x)}$时，只需判定函数$F_1(x), cF_2(x)$是否相等即可（在 Maple 中判断两个表达式是否相等很容易）。

## 1.2 原函数为任意函数多项式

任意函数多项式

$$F(x) = \sum_{i=1}^{n} F_i(x) = \sum_{i=1}^{n} \prod_{j=1}^{m} f_{ij}^{p_{ij}}(x) \ p_{ij} \in R)$$

求导后可得:

$$F'(x) = \sum_{i=1}^{n} F_i'(x) = \sum_{i=1}^{n} \sum_{k=1}^{n_i} \prod_{j=1, j \neq k}^{n_i} p_{ik} f_{ij}^{p_{ij}}(x) f_{ik}^{p_{ik}-1}(x) f_{ik}' \ (n_i \in Z; p_{ij}, p_{ik} \in R)$$

若$F_i(x)$是$F_i'(x)$的一个原函数，显然对$F'(x)$的积分问题可以类似 7.2.1 中对$F'(x)$中的项$F_i'(x)$求积分方法解决。但是，$F'(x)$中的项$F_i'(x)$与$F_j'(x)$可能相互交叉，从而使问题复杂化。解决该问题的关键是找到$F_i'(x)$的所有操作数，在此，我们采用贪婪法和回溯法设立了两个列表$intop, outop$分别存储可积分项和不可积分项的所有操作数。为了更清楚地表达我们的算法设计思路，下面给出一个具体的算例。

## 2. 三角函数积分及其机械化

三角函数积分是一类重要的积分问题。常用方法是利用诱导公式将被积函数转换为可积形式，其中利用万能公式$tan\frac{x}{2} = t$将被积三角函数有理化是一种重要的方法，因为有理函数总是可积的。

例：求积分

$$\int \frac{1}{2 + sinx} dx$$

解：令$tan\frac{x}{2} = t$，则：$x = 2arctant, sinx = \frac{2t}{1+t^2}, dx = \frac{2}{1+t^2} dt$

从而：

$\int \frac{dx}{2+sinx} = \int \frac{dt}{1+t+t^2} = \frac{2\sqrt{3}}{3} arctan \frac{\sqrt{3}}{3}(2t+1) + C$

$\qquad = \frac{2\sqrt{3}}{3} arctan \frac{sqrt3}{3}(2tan\frac{x}{2} + 1) + C$

$$\stackrel{\mathrm{d}}{=} F(x) + C$$

由这一方法得到的结果是否正确呢？为了讨论问题方便，先给出不定积分（原函数）的定义。

定义：设函数$f(x)$与$F(x)$在区间$I$上有定义，若$F'(x) = f(x)$，则称函数$F(x)$为$f(x)$在区间$I$上的一个原函数，原函数全体称为$f(x)$的不定积分。

显然，函数$f(x)$的原函数$F(x)$必可导，但例中按照万能代换所求得的积分结果是否满足这一基本要求呢？有

$$\lim_{x \to \pi^+} = -\frac{\sqrt{3}}{3}\pi, \lim_{x \to \pi^-} F(x) = \frac{\sqrt{3}}{3}\pi$$

可知：(7.10)中的$F(x)$在$\pi$处不连续，且$F(x)$在$x = \pi$处无定义。进一步推导可知，$F(x)$在$(2k+1)\pi(k \in Z)$处均不连续。从而，$F(x)$不是$f(x)$的原函数。产生这一错误结果的原因是因为求解过程额度错误还是万能代换方法导致的必然结果呢？

为了说明问题，下面给出有关文献中的几个例子（为了讨论的方便，此处将积分常数$C$略去）：

例：

$$\int \frac{1}{5 - 3cosx}dx = \frac{1}{2}arctan(2tan\frac{x}{2}) \stackrel{\mathrm{d}}{=} F(x)$$

$$\lim_{x \to \pi^+} F(x) = -\frac{\pi}{4}, \quad \lim_{x \to \pi^-} F(x) = \frac{\pi}{4}$$

由上述例子可见，这些被积三角函数为连续函数（也是初等函数）的积分计算结果均不连续，当然也不可导。

另一方面，由求解过程中的万能公式$tan\frac{x}{2} = t$可知，此式成立的前提是$x \neq (2k+1)\pi$。因此，万能公式虽然可将三角函数有理化，在一些情况下（如被积函数仅在$x \neq (2k+1)\pi$处连续等）可能会得到正确结果，但对于在$x = (2k+1)\pi$处连续的三角函数积分后必然会得到错误的结果。

2.1 主要定理及应用

对于含有三角函数的积分，由于$secx, cscx, tanx, cotx$都可以化为$sinx$和$cosx$的函数。所以，只要讨论$\int R(sinx, cosx)dx$型积分就够了，根据数学分析有关理论与方法，可得如下定理：

定理1：设$\int R(sinx, cosx)dx$是三角函数有理式的不定积分，且$R$连续，则通过万能公式$tan\frac{x}{2} = t$求得的积分结果在$x = (2k+1)\pi(k \in Z)$处必存在第一类间断点。

定理2：设$G(x) = \int R(sinx, cosx)dx$是三角函数有理式的不定积分，且$R$连续，设$tan\frac{x}{2} = t$，若$F(tan\frac{x}{2})$是万能变换后积分计算结果，记$F(-)$，$F(+)$分别为其在$x = (2k+1)\pi(k \in Z)$处的左右极限，则：

$$G(x) = \int R(sinx, cosx)dx = \begin{cases} F(tan\frac{x}{2}) - F(-), & x \in [2k\pi, 2k\pi + \pi) \\ 0, & x = (2k+1)\pi \\ F(tan\frac{x}{2}) + F(+), & x \in (2k\pi + \pi, 2k\pi + 2\pi] \end{cases}$$

应用上述研究，易得前文所述例子的积分结果。

$$\int \frac{dx}{2+sinx} = \begin{cases} \frac{2\sqrt{3}}{3}arctan\frac{\sqrt{3}}{3}(2tan\frac{x}{2}+1) - \frac{\sqrt{3}}{3}\pi, & x \in [2k\pi, 2k\pi+\pi) \\ 0, & x = (2k+1)\pi \\ \frac{2\sqrt{3}}{3}arctan\frac{\sqrt{3}}{3}(2tan\frac{x}{2})+1) + \frac{\sqrt{3}}{3}\pi, & x \in (2k\pi+\pi, 2k\pi+2\pi] \end{cases}$$

3.结论与讨论

数学基础永远是科学计算的支撑，任何对于数学基础的轻视都可能带来错误。我们在关于函数积分的处理增加了"$F'(x) = f(x)$"的判断语句，以保证$F(x) = \int f(x)dx$的正确性。而这一点恰是 Maple 没有注意到的。

由于 Maple 符号计算中引入积分常数相当于引入一个变量，对于计算极为不便，尤其是对结果作进一步处理时。因此,我们的三角函数积分机械化计算结果中没有包含积分常数。如果需要积分常数$C$，按照"结果$+C$"的形式输入即可。

当然，由于三角函数积分本身的困难以及计算机代数系统的复杂性，机械化计算结果并非总是最简单的形式，特别是 Maple 下的不等式输出不符合人们的习惯，但这并不影响结果的正确性。

**1.3.2. Mathematica**

这部分内容主要是针对 Mathematica 调试过程中出现的问题进行说明和总结，并指出下一步发展的方向。其中包含对于整个符号积分发展的思考以及少部分定积分模块的内容，非常有借鉴参考的价值。

1.介绍

算法集成的概念非常好理解，有不同的算法针对不同类型的积分和积分约束。麻烦的情况有以下几种：

确定（也许条件）收敛

识别并处理附加奇点（对奇点进行消除）

实现复杂的算法，例如，一些可能需要深层的递归或者依靠非平凡的变换

对于参数的处理

对于遗留代码的处理

在这份报告中，我将会描述并且阐明在 Mathematica 积分代码基础上开展工作所遇到的问题。这份报告主要讨论在 Mathematica 5 到 Mathematica 6 的发展过程中的情况。

2.Mathematica 积分代码的基本结构

2.1 不定积分

不定积分代码主要由$Risch$算法[2,5]的部分实现组成，另外还有大量的查表的方法。前者处理初等积分情况时并没有考虑代数扩张和一些简单低次数扩张的情况。后者处理含有指数，三角函数，双曲，椭圆积分和含有特殊函数的积分，特别是在$Risch$方法无效的情况下。真实的情况可能更加复杂一些，$Risch$的算法可能会对部分的积分进行处理，将剩余部分交给查表的方法。查表的部分进行变换，并且递归地调用积分代码。Mathematica 的实现刚开始依靠$Risch$代码，之后通过扩展的查表的方法处理特殊函数和一些$Risch$不能处理初等函数的积分。

除了上面提到的内容，一些积分变换是在最前面的。理念就是一些积分可以被转换成更容易处理的形式，但是等效为模掉一个乘法常量（例如将$\sqrt{x^2}$变换为$x(\frac{\sqrt{x^2}}{x})$），将第二个因

子视为一个微分常量。

同样有效的是数学等效变换，例如因式分解和部分分式展开。值得注意的是这样的变换可以互相转化，并且很难决定对于给定的被积函数来说哪一种变换是真正有效的。所以必须避免盲目的尝试，如果盲目尝试的话，会尝到递归之神无尽愤怒之下的恶果。在 Mathematica 代码中，先进行尝试再进行决定，从而在各种各样的情况下，判定上述变换是否有效，如果有效的话就使用这种变换。需要强调的是，最好的情况是进行一个递归的过程，最坏的情况是一个十足的递归过程。

最后，没有根本性的依据将被积函数分成 $Risch$ 处理部分和非 $Risch$ 处理部分。另一个方法就是采用查表的方法（例如通过模式匹配）作为第一步，接下来通过 $Risch$ 实现解决问题，接着通过更深的模式匹配。为是这种想法变得有效，第一步的查找需要快捷（换句话说，需要在最终必将失败的情况下，尽早跳出。）

## 2.2 定积分

定积分通过一系列以下列举的方法进行实现

特殊情况下的等值算法

牛顿—莱布尼兹方法（简而言之:计算一个不定积分，接下来带入具体的值，也许需要通过极限运算避免路径奇点的存在）。Mathematica 在很多地方都使用了这种方法。第一个地方就是针对有理函数×三角函数或者有理函数×指数函数这种形式。

牛顿—莱布尼兹方法针对被积函数中含有对数函数或多重对数函数。这种情况下的代码比针对前面那种情况设计的代码更加复杂。

一种一般化的牛顿—莱布尼兹积分算法的实现。仍然是更加复杂的代码。

一个定积分算法的实现通过 $MeijerG$ 函数的卷积实现。这需要我们我们从 0 到 ∞ 积分。同时也需要被积函数可以表示一个积分变量的幂，同时乘以一个或两个 $MeijerG$ 函数的形式。因为一个 $UnitStep$ 函数可以被一个 $MeijerG$ 函数表示出来，我们也许放弃半无限的边界要求无论被积函数是怎样的，否则除了一个 $MeijerG$ 函数外需要这个限制。这种方法尤其使用了一些变换策略以便处理代数，三角，对数，指数函数等等。因此它有自己的特殊的树状代码子集。

描述以上全部实现的最好的方法就是称之为一个 $polyalgorithm$，可以针对不同的情况调用以上任一或全部的算法，具体的调用依赖于启发式算法的判断。

## 2.3 定积分的两个基本方法的简单介绍和具体例子

## 3.积分代码的软件工程问题

以下是我在修复 Mathematica 符号积分代码时遇到的问题。我猜想在任意通用性符号积分代码中都能发现这些问题。

不同的模块儿由不同的人完成，所以并不总是通过标准化的语言进行表达。因此，各种各样的具有相同的潜在通用功能的代码段被重复发明。在一些情况下，这种情况造成的严重后果远远超过了代码膨胀这一种结果，比如不一致的 Bug，不同的未声明的潜在假设。

代码的很大一部分都是遗留代码，完成时间在 15 年左右，并且是由不同的人完成。这些代码仍然在几十个源代码文件中传播（一定意义下着也是一个好的方面，因为它们提供了一个模板）。没有一个人能够理解这部分代码的全部，并且其中一些部分在今天是没有任何一个人可以理解的—这也是遗留代码的一个极大地缺陷。我发现在这种情况下，代码的主体部分被研发它的团队完全理解是不太可能的。

一些步骤是十分不稳定的。尤其当一些看上去毫无关联的函数比如 Together（给出有

理函数的典范表示）进行小改动时，这些步骤也会随之发生改变。同样的例子还有 Apart（部分因式分解，由于 Integrate 的需要，其中有一些不明确的运行结果），Factor，Solve，Simplify 和一些其他函数。在很多情况下我都遇到了这个问题，我得出结论为这是符号积分特有的一种问题，而不是一种简单的实现细节的块效应问题。从整体上讲，被积函数的变换，伴随着从数学角度，不可能给出所有可能的表达式的典范表示，为了一定目的进行强制转换的困难使得这个问题是一个令人苦恼的问题。事实上，这种类型的策略例如通过使用 $L'Hospital$ 法来进行积分的极限求根，如果没有足够小心的话，能够导致无限递归。

  实际情况下存在功能和速度的权衡问题。如果想要尝试特定的变换，例如为了处理特定类型的问题。但是没有一般情况下有效的方法来界定在转换过程中可能出现的情况。因此一些界定的类型是需要的，例如，根据时间与操作数。

  许多积分问题可以用到关于被积函数在无穷处的假设，遗留代码的一些部分没有认识到假设的使用，并且得到与之相反的结果（不出意外的话，这也是很多 Bug 产生的原因）。甚至当代码要进行恰当的判定时，例如通过给定的假设判定无穷远处的函数行为，会出现上面这种情况，如果没有进行适当的限制，极限求根的过程会被迫终止。但是当及时对时间进行限制的情况下，内存或者操作数会被更多的调用，怎么能处理一个异常的中间结果？使用警告信息？放弃该结果？假设是正常情况并继续？例如假设在无穷远点收敛？

  符号积分是算法数学中最复杂的部分之一。依赖其它函数，同时探究被积函数可能的变换形式出现的错综复杂的情况确实说明了符号积分的实现是非平凡的。因此将代码部分充分地记录下来是非常重要的。这是一条在整个软件开发领域都适用的规则，但是其重要性不能在积分理论的情况下被夸大（这对于积分代码来说也是重要的）。

  当我开始在代码主体上进行工作时，在 Mathematica 内核（大约 2000 个函数）上超过四分之一的 Bug 是定积分方面的（Integrate 函数的一部分）。这一模块在模块内和模块外都产生了一些问题。首先，函数得到了广泛的应用，因此从头开始是一件困难的事情。另外，函数结构使得很难对问题进行分类（换言之，很难一叶障目不见森林）。此外，需要对代码进行的巨大的修补，强烈暗示在代码中存在着极大的问题，至少在较短的一段时间内。最后需要注意的是问题的规模可能超过任何一个开发者的能力—我能够证明确实超过了我的能力范畴。

## 4.实现不定积分中出现的问题

  我们将会讨论一些在不定积分出现的特定问题。

### 4.1 递归的诅咒

  经常一个积分可能会被被分割为两个部分，开玩笑地讲第一部分是"完成"部分，第二部分是"未完成"部分。第二部分可能会被不断地改写，并且返回原始的函数或者原始函数多种多样的形式，因此使程序的运行进入分裂式的递归运行。一个积分可能会引出这样的运行结果：$\int \frac{sin(x)}{(x+1)\sqrt{a-x}}dx$ 在 Mathematica 的一些旧版本上会出现无限递归的问题。我们解决这个问题依靠哈希函数标识出之前已经在递归中尝试过的形式，避免重复递归，最终导致死循环。

  另一个原因是成对逆变换的应用。例如，一段程序，将三角函数转化为指数函数，之后的处理可能将指数函数转化回三角函数。我们通过使用 Mathematica 中的 $Block$ 豁域结构以避免这个陷阱，当我们进行这样的转换时，我们会通过 $Block$ 锁住双重的处理操作。

  一个令人高兴的一面是，在实质性的错误被修正之后，尤其是在积分代码中模式匹配部分中的错误被修正后，Mathematica 可以比过去解决更多的问题。这有一个运行良好的例子，

这要归功于对于被积函数进行的一些变换。

$$\int \frac{\text{Tan}[x]^2}{\sqrt{\left(1-\frac{\text{Sin}[x]^2}{2}\right)\left(1-\frac{2\text{Sin}[x]^2}{5}\right)}} dx$$

$$-\left(4i\sqrt{\frac{5}{3}}\text{Cos}[x]^2 \left(\text{EllipticE}\left[i\text{ArcSinh}\left[\sqrt{\frac{3}{5}}\text{Tan}[x]\right], \frac{5}{6}\right] - \text{EllipticF}\left[i\text{ArcSinh}\left[\sqrt{\frac{3}{5}}\text{Tan}[x]\right], \frac{5}{6}\right]\right)$$
$$\sqrt{(4+\text{Cos}[2x])\text{Sec}[x]^2}\sqrt{2+\text{Tan}[x]^2}\right) / \left(\sqrt{25+14\text{Cos}[2x]+\text{Cos}[4x]}\right)$$

### 4.2 孤立无援减少函数分支

很多情况下我们都需要进行变换，变换之后就会将被积函数变为多值函数。结果导致我们得到一个积分仅仅对于多值函数的一个分支成立（例如：变换导致在一些地方为 1，在另一些地方为-1，例如 $\sqrt{x^2} \to x$）因此，我们尝试恢复正确的因子。这并不总是直接的，常常导致结果形式的明显增加。而且在所有情况下正确的进行回复变换造成的影响并不是简单的。

### 4.3 变换

在许多情况下，如何求出一个特殊被积函数类的不定积分是非常明确的，例如，有理三角函数。但是需要的变换必须小心地进行，这是为了避免积分复杂度的激增或者最终结果表示形式的激增。一个明显的例子是 $\int \frac{x}{(sin(2x))^2+cos(x)} dx$。Mathematica 4.0 给出的结果长度与较新的版本相比答案长度约为五倍。

### 4.4 扩展还是不扩展，不确定

这是一个变换重要而特殊的一种情况。因为问题是在定积分中相关问题的一个子集，所以我们将在定积分部分进行叙述。

## 5. 总则

这里有一些需要在定积分设计和实现的过程中考虑的问题。这些问题的基础是我们的研究和对已有代码进行的修改，但是我相信这个总则适用的范围将会更广，例如，数值求和，和相关的 *polyalgorithm* 微积分计算。

做出方法选择时，先尝试哪一个？这对于速度可能有严重的影响。例如，一个快速的 **MeijerG** 卷积计算，如果选择通过先求不定积分，再求极限的话，那么可能会变得很慢。当然，反之也有可能发生。或者方法在速度上具有优势，但是结果形式可能会极其复杂。

什么时候或者怎样进行输入的预化简？

什么时候或者怎样化简结果？

特殊方法可能对于速度和结果的简化有很大的帮助。但是它们可能使出现 Bug 的机会激增。因此，需要考虑好用这种方法能够处理的重要的函数类是什么。

一些方法的"慢"是固有的。例如，精细的条件（这在一些情况下是必需的为了防止它们失效）需要一定程度的CAD支持。甚至求极限值的牛顿—莱布尼兹方法可能也会比较慢。我们就需要解决这样的一个问题—什么时候采用这样的技术，怎样避免其导致许多输入的中止。

为了避免上述技术出现的问题，我们发现对于特定的代码设立时间限制是很有必要的。（积极的一方面：它们成功率很高，如果失败，那么就放弃这种方法，尝试其它方法）。这引起了一些新的问题。其中之一就是 **TimeConstrained** 需要的异步处理会中断正常的处理，这是不完善的，并且偶尔会导致内核崩溃。另一个问题是答案存在平台依赖的问题，并且是严重不可控的。未来的一个可能方向就是减少这种现象的发生：可能会因为对于操作数的数量限制而中止运行速度较慢的代码而不是被异步问题中止。

第五个版本的不定积分的功能在总体上讲比过去的版本更加强大。这是由速度的降低为代价换来的：它尝试的变换种类更多，这导致所有的牛顿—莱布尼兹代码都有运行变慢的风险。现在我们预置一些不定积分的结果，但这至多只是对于问题的一个部分的修正。

临时的标准，可能是一个问题：进行越多的改进，就会带来紧密相关的事情出现问题的可能性增大。

# 第二章

这一章主要进行 SIN，Albi，Rubi 三套符号积分系统的介绍，介绍 SIN 系统主要是提供一种很具有启发性的计算机代数系统符号积分模块的设计理念。SIN 和 SAINT 系统虽然提出了一个开创性的分阶段解决问题的框架，但是能够解决的积分问题的范围是非常有限的。SIN 和 SAINT 的模式匹配的形式只有进行更加细致的模式分割和提高模式匹配的效率才能使其功能更加完善。目前符号积分系统的最新发展方向为 Rubi(Rule Based Integrator)和 Albi (Algorithm Based Integrator)。其中 Rubi 系统采取的是一种递归式的算法设计，将之前符号积分系统的多层次，多形式匹配的方式融合为一张积分规则表，通过积分规则的细化和反复递归的过程，提高了匹配的效率。相对 Rubi 系统，Albi 系统则是从计算机代数的角度出发，利用刘维尔定理对积分结构进行猜测，借助代数知识求解符号积分问题。与 Rubi 系统相比较，Albi 系统实现起来较为困难，但是可以解决 Rubi 系统解决起来较为困难的难点—超越函数部分。从理论上弥补了 Rubi 系统的不足。Albi 和 Rubi 的理论出发点虽然截然不同，但能够互相补充。不仅使 SIN 和 SAINT 的理念得到了很好的发展，同时也利用计算机代数的知识寻找到了一条解决符号积分问题的崭新途径。

## 2.1 SIN 系统

与同时代的 SAINT 一样，SIN 也是一种典型的启发式程序的设计思路。虽然 SAINT 强调启发式模式匹配，但在很大程度上，SIN 优于 SAINT，因为其更节省时间，算法理论更为强大。但是不能否认 SAINT 仍是一种很具有启发性的计算机代数系统框架。

下面主要介绍 SIN 系统，SIN 系统解决符号积分问题的三个阶段为

Stage 1: Simpke Problem

  (1)被积分式的拆解，将其分成单独的分式积分，然后求和

  (2)如果被积分式为（$\sum u_i(x)$）$^n$，n 较小，进行展开，使用第一个思路进行解决

  (3)凑微分

Stage 2: Special Methods

  11 种特殊模式的解决方案，将经过 Stage 1 处理过的被积分式进行匹配，将问题转化为 Simple Problem ，再回到阶段一解决

Stage 3:

  (1)分部积分法

  (2)Edge（Educated GuEss）heuristic 算法(启发式猜想方法)

The EDGE heuristic generates a guess for the form of the integral based on the form of the integrand. The guess is differentiated and undetermined coefficients in it are obtained by matching the derivative with the integrand. We devised the EDGE heuristic independently of RIsch's work on the Liouville theory. RIsch's algorithm id clearly superior to the EDGE heuristic and later versions od SIN have used subses of the Risch algorithm in the third stage.

流程框图：

### Stage 1

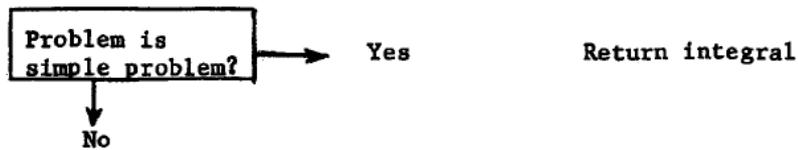

### Stage 2

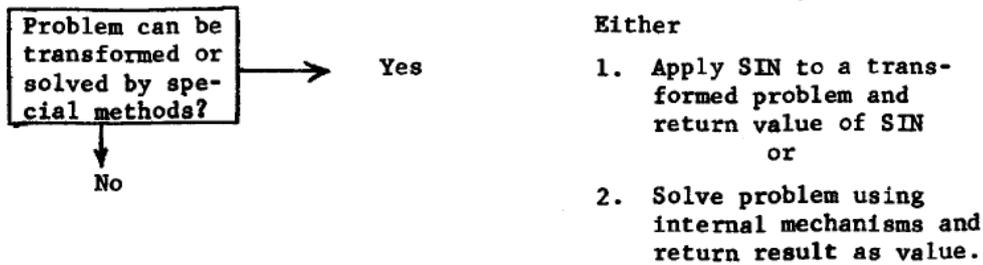

### Stage 3

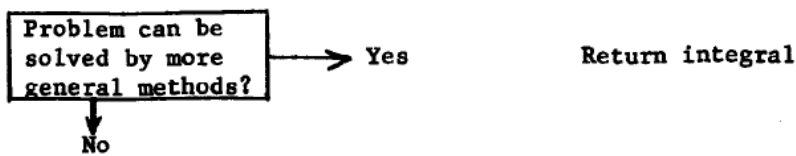

Return notice of failure

Figure 1 - The 3 Stages of SIN

SIN 提供了一个使用的符号积分模块的模型，同时 SIN 有良好的扩容性，一些较好的方法都可以被吸纳进来，成为 Stage1,2 or 3 的一个补充分支。与此同时有另外的两种思路被提出。

    From algebraic manipulation led to Manove's implementation and to Horowitz' and Tobey's reexamination of the Hermite algorithm for integrating rational functions.

    From mathematics, led to Richardson's proof of the unsolvability of the problem for a class of functions and for Risch's decision procedure for the elementary functions. Generalizations of Risch's algorithm to a class of special functions and programs for solving differential equations and for finding the definite integral are also described.

    这两种的方法的区别在于第一种方法的重点在于 integrate rational functions，第二种方法的重点在于 Risch's mathod（域扩张的理论）。第二种方法是包含 rational functions 这一个方面的。

    介绍 SIN 系统的目的是为了更好的理解 Rubi 和 Albi 两种算法之间的关系。Rubi（Rule Based Integrator）使用规则匹配的方式，通过检索积分表进行积分的化简和求值。Albi(Algorithm Based Integrator)则使用专门的算法对复杂有理函数、超越函数等 Rubi 无法完全求解的被积表达式进行积分。

    在介绍本章之后部分之前给出关键性概念—超越函数和代数函数的定义：代数函数，代数函数是包括加、减、乘、除和开方等基本算符的数学函数。非代数函数则被称为超越函数。超越函数与代数函数相反，是指那些不能满足任何多项式方程的函数，也就是说函数不能表

示为有限次的加、减、乘、除和开方的运算。对代数函数进行不定积分可能产生超越函数。

下面给出 SIN 系统在实现过程中的两种具体思路：

  1. 首先是层次式的，第一层是积分表，第二层是 DDM（凑微分法），第三层是 SIN 的转化方法，第四层是有理函数，即用最快的方法解决简单的问题，之后逐次考虑复杂的问题。关于 SIN 的转化问题，按一定次序检测，转化之后先用前两层处理，不行的话，按一定规律进入下一种算法的匹配，如果转化之后的是有理函数，则直接进入下一步。

  2. 首先一个主 SIN 控制台，分为 StageI 和 StageII，StageI 含有积分表和一些基本规则，关于 DDM 是基础，是否优先于积分表处理有待进一步考虑。
SIN 控制台先考虑 StageI，无法得到结果，或者问题分解成几个部分，部分无法解决，则进入 StageII，一旦匹配，给出转化后的被积分式。重新调用主控制台解决子问题。
  下面是具体的流程框图：

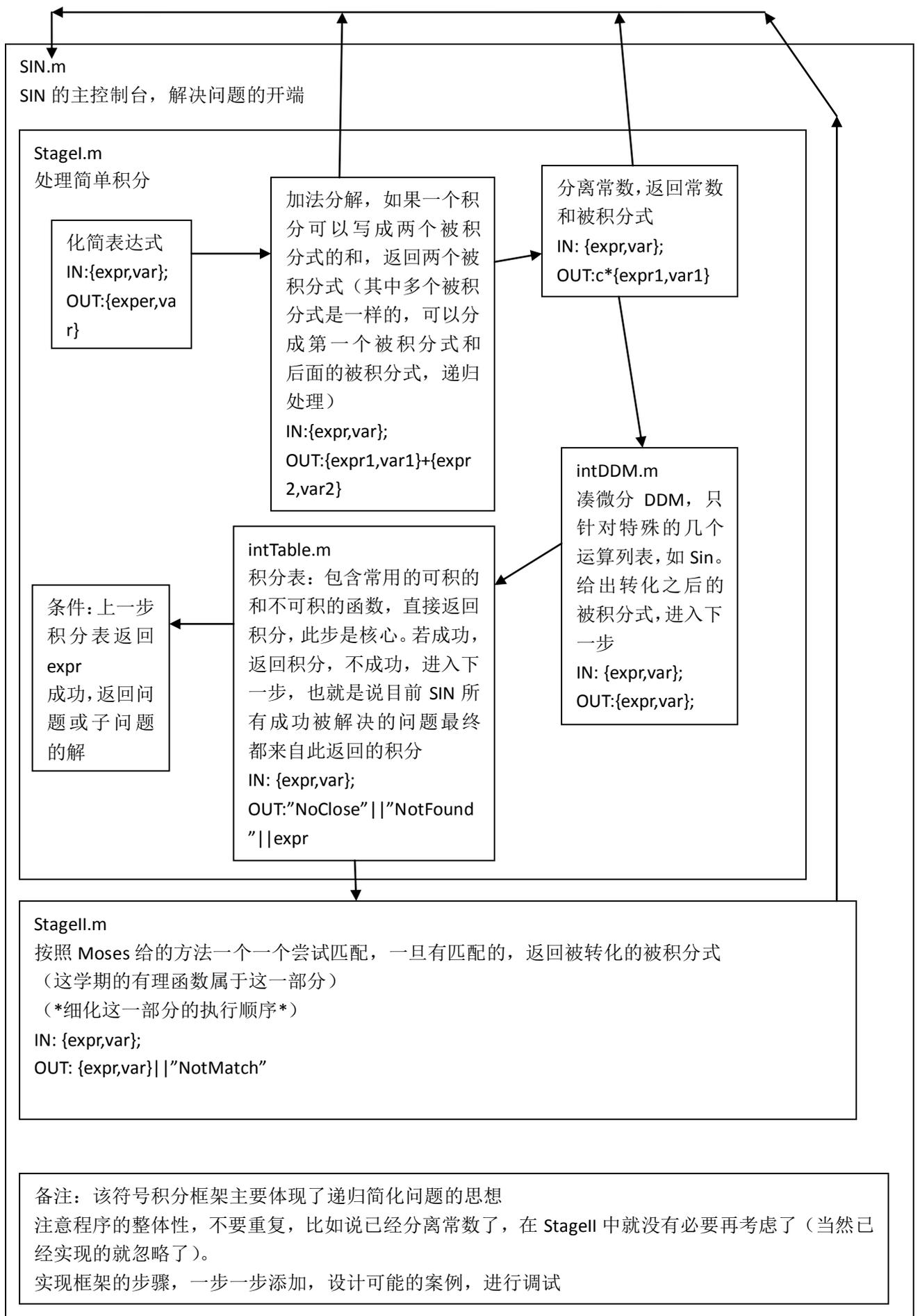

Albi 系统理论部分的核心定理—刘维尔定理由刘维尔在 19 世纪首次进行了陈述并进行了严格的证明。但是刘维尔的证明是基于分析的语言,在实际操作过程中只有通过在相应的代数结构上引入微分映射,将定理的陈述转化为代数条件下的陈述,才能产生具有使用价值的算法。刘维尔定理的第一个完整的代数证明于上世纪七十年代由 Rosenlicht 发表,之后刘维尔定理的强化形式的首次证明由 Rosenlicht 的学生 Risch 完成,Risch 将两个证明进行了整理,得到了一个针对纯超越初等函数的算法。之后,该领域完整而系统的归纳总结及补充由 Bronstein 完成,Bronstein 于 2004 年出版了"Symbolic Integration —Transcendental Functions" 一书。这是迄今为止对于 Albi 系统最为完整的梳理与总结。之后由于理论难度的提升和符号积分领域整体发展相对缓慢的客观因素,符号积分领域一直没有新的重大的理论突破。其后的发展主要出现在 Albi 理论所衍生出一些分支上,研究工作基本上是对 Albi 理论进行局部的补充和完善。

Albi 系统的理论部分虽然完善,但是实际的实现过程仍不成熟。Albi 系统的完整实现是目前符号积分领域的前沿课题,主流的计算机代数系统如 Mathematica,Maple 都以 Albi 系统作为自身符号积分功能的核心,但是由于 Mathematica,Maple 都为商业软件,无法得到具体的实现过程。所以研究只有从理论着手,选择一个合适的框架,自己摸索出一套可行的实践方案。

现阶段,Albi 针对超越函数的积分理论和算法相对系统成熟,而代数函数方面理论难度大,在算法理论方面还没有进行系统的阐述,形成成熟的算法框架。所以还停留在一个理论的层面,以计算机代数的思想解决代数函数的符号积分问题暂时是不成熟的,所以自然引入了 Rubi 这套实现效果优良的积分系统,处理代数函数的不定积分问题。但是这不意味这 Rubi 系统对于超越函数是无计可施的,在实际运行过程中,Rubi 的运行效果也是非常优异的。只不过 Rubi 在理论方面的局限性决定这套积分系统的局限性,Albi 在理论上的成功与完备决定了其必将成为符号积分领域的主流,换言之 Rubi 系统只是 Albi 系统的近似实现,但是其近似实现的效果也是非常不错的。

下面给出 Albi 系统针对超越函数情形的完整理论过程,在之后的章节会略微地介绍针对代数函数不定积分的一些理论与算法实现。

# 第 1 章  代数基础

在这一章我们回顾一些贯穿本书的基本代数结构与算法。这一章不被预期能代替一门基本的抽象代数教程，并且我们假定你已经了解群、环、多项式的定义和基本性质。我们在这里仅仅回顾那些在抽象代数课程上不被经常提起但在这里用处很广的一些概念的定义和一些多项式的算法。因为那些都是在计算机代数里有名的算法，在这里我们就不再详细证明，仅仅涉及一点。为了对代数与代数方法有更好的了解，包括那些计算多项式最大公因子的算法，我们推荐几本在文献中提及的文章。如果读者对代数有一些较好的理解认识，你可以跳过这一章，以后需要也可以查询。

## 1.1 群、环、域

一个代数结构通常被赋予一个或多个运算，运算所遵循的被叫做运算法则。为了不重复列举一些代数结构所满足的运算法则，我们定义了一些常用的代数结构，也就是给其一个简洁的名字。群、环、域就是这样的一种结构，我们这一节回顾它们的定义。

**定义 1.1.1.** 一个群 $(G, \circ)$ 就是一个非空集合 $G$，在上面定义了一种运算：$G \times G \to G$，并且运算满足如下规则：

**(i)（结合律）** $\forall a, b, c \in G, a \circ (b \circ c) = (a \circ b) \circ c.$

**(ii)（单位元）** $\exists e \in G, \forall a \in G, e \circ a = a \circ e = a.$

**(iii)（逆元）** $\forall a \in G, a \circ a^{-1} = a^{-1} \circ a = e.$

另外，$\circ$ 被称为交换的，如果 $a \circ b = b \circ a$；$(G, \circ)$ 是一个交换群（Abel 群），如果 $G$ 是一个群，且 $\circ$ 是交换的。

**例 1.1.1.** 设 $G = GL(\mathbb{Q}, 2)$ 是所有有理系数行列式值非零的 $2 \times 2$ 矩阵，假设 $\circ$ 就是通常的矩阵乘法。如此，$(G, \circ)$ 就是一个群：结合律明显成立，单位元是单位矩阵，$G$ 中矩阵的逆由下式给出：

$$\begin{pmatrix} a & b \\ c & d \end{pmatrix} = \frac{1}{ad - bc} \begin{pmatrix} d & -b \\ -c & a \end{pmatrix}$$

矩阵的逆行列式值非零, 亦在 $G$ 中。注意 $(G, \circ)$ 不是一个交换群，因为

$$\begin{pmatrix} 1 & 1 \\ 0 & 1 \end{pmatrix} \circ \begin{pmatrix} 0 & 1 \\ 1 & 0 \end{pmatrix} = \begin{pmatrix} 1 & 1 \\ 1 & 0 \end{pmatrix}$$

但是

$$\begin{pmatrix} 0 & 1 \\ 1 & 0 \end{pmatrix} \circ \begin{pmatrix} 1 & 1 \\ 0 & 1 \end{pmatrix} = \begin{pmatrix} 0 & 1 \\ 1 & 1 \end{pmatrix}$$

**例 1.1.2.** 假设 $G = \mathcal{M}_{2,2}(\mathbb{Q})$ 表示所有有理系数的 $2 \times 2$ 矩阵的集合，$\circ$ 表示通常的矩阵加法。显然 $(G, \circ)$ 是一个交换群，且单位元为零矩阵。

**定义 1.1.2.** 环 $(R, +, \cdot)$ 是一个集合，在其上定义了 $+ : R \times R \to R$ 和 $\cdot : R \times R \to R$ 两种运算：

**(i)** $(R, +)$ 是一个交换群.

**(ii)** （结合律） $\forall a, b, c \in G,\ a \cdot (b \cdot c) = (a \cdot b) \cdot c.$

**(iii)** （乘法单位元） $\exists i \in G, \forall a \in G, i \cdot a = a \cdot i = a.$

**(iv)** （分配律） $\forall a, b, c \in G, a \cdot (b + c) = (a \cdot b) + (a \cdot c)$, 且 $(a + b) \cdot c = (a \cdot c) + (b \cdot c).$

$(R, +, \cdot)$ 是一个交换环，如果它是一个环且 $\cdot$ 是可交换。另外，我们定义 $R$ 的特征是 0，如果对任意整数 $ni \neq 0$，或为 $m$，如果是 $m$ 是使 $mi = e$ 成立的最小正整数。假设 $R$ 和 $S$ 是环，它们是同态的，如果存在映射 $\phi : R \to S$, $\phi(e_R) = e_S$, $\phi(i_R) = i_S$，并且 $\phi(a + b) = \phi(a) + \phi(b)$ 和 $\phi(ab) = \phi(a) \cdot \phi(b)$ 对于任意 $a$, $b \in R$。$R$ 和 $S$ 是同构的，如果 $\phi$ 是一个双射。

在这之后，只要 $(R, +, \cdot)$ 是一个环，我们用 0 表示关于 $+$ 的单位元，用 1 表示关于 $\cdot$ 的单位元，并且对于任意 $a$, $b \in G$，用 $ab$ 代替 $a \cdot b$。

**例 1.1.3.** 假设 $G = \mathcal{M}_{2,2}(\mathbb{Q})$ 表示所有有理系数的 $2 \times 2$ 矩阵的集合，$\circ$ 表示通常的矩阵加法，$\cdot$ 表示通常的矩阵乘法。$(R, +, \cdot)$ 是一个环，但并不是一个交换环（见例 1.1.1）。因为，

$$ni = n \begin{pmatrix} 1 & 0 \\ 0 & 1 \end{pmatrix} = \begin{pmatrix} n & 0 \\ 0 & n \end{pmatrix}$$

对于任一个正数 $n$ 均为非零矩阵，$R$ 的特征为 0.

**例 1.1.4.** 设 $R = \mathbb{Z}_6$(整数对模6 的 6 个同余类构成的集合)，在其上定义 $+$ 和 $\cdot$ 表示模 6 的同余类的加法和乘法。$(R, +, \cdot)$ 是一个交换环，并且映射 $\phi : \mathbb{Z} \to \mathbb{Z}_6 : \phi(n) = n(\text{模 } 6)$，是一个环同态映射。因为在 $\mathbb{Z}_6$ 中 $1+1+1+1+1+1 = 0$，并且对于任意 $n, 1 < n < 6$，$n1 \neq 0$，$\mathbb{Z}_6$ 特征为 6。注意到 $2 \cdot 3 = 0$，然而 $2 \neq 0$，$3 \neq 0$，因此，我们不能从等式 $ab = 0$ 简单地推出 $a = 0$ 或 $b = 0$。交换环如果能如此简化，将是非常实用，普遍的性质，于是我们给那些具有这性质的交换环一个特殊的名字。

**定义 1.1.3.** 整环 $(R, +, \cdot)$ 是一个交换环，$0 \neq 1$，并且 $\forall a, b \in R$, $a \cdot b = 0 \implies a = 0$ or $b = 0$。

**例 1.1.5.** 设 $R = \mathbb{Z}[\sqrt{-5}] = \{a + b\sqrt{-5};\ a, b \in \mathbb{Z}\}$，$+$ 和 $\cdot$ 表示复数的加法和乘法。$(R, +, \cdot)$ 是一个整环。

下面我们开始考虑因式分解的问题，比如说，把整环中的一个元素用整环中其它元素的乘积表示。

**定义 1.1.4.** 设 $(R, +, \cdot)$ 是一个整环，$x, y \in R$。我们说 $x$ 整除 $y$，并写成 $x|y$，如果 $y = xt, t \in R$。$R$ 中的元素被称为单位，如果 $x|1$。$R$ 中所有单位的集合写成 $R^*$。我们说 $z \in R$ 是 $x_1, x_2, \ldots x_n$ 的最大公因子，记为 $z = gcd(x_1, x_2, \ldots x_n)$，如果 $z$ 满足如下条件：

**(i)** $z|x_i, 1 \leq i \leq n$,

**(ii)** $\forall t \in R, t|x_i, 1 \leq i \leq n \Longrightarrow t|z.$

另外，我们说 $x, y$ 是互质的，如果它们的最大公因子是 $R^*$ 中的一个单位。

**例 1.1.6.** 设 $R = \mathbb{Z}[\sqrt{-5}]$（如例 1.1.5），$x = 6$，$y = 2 + 2\sqrt{-5}$。通过规范论证我们可以知道 $x, y$ 在 $R$ 范围内无最大公因子。定义映射 $N : R \to \mathbb{Z}$，$N(a + b\sqrt{-5}) = a^2 + 5b^2$，$a, b \in \mathbb{Z}$。我们可以得到对任意 $u, v \in R$，$N(uv) = N(u)N(v)$，在 $R$ 中 $u|v$ 意味着在 $\mathbb{Z}$ 中 $N(u)|N(v)$。假设 $z \in R$ 是 $x$，$y$ 的最大公因子，$n = N(z) > 0$。然后，由 $n|N(x) = 36$ 和 $n|N(y) = 24$ 可知，在 $\mathbb{Z}$ 中 $n|12$。如果，在 $R$ 中有，$2|x$，$2|y$，那么在 $\mathbb{Z}$ 中 $4 = N(2)$。另外，$R$ 中 $1 + \sqrt{-5}|y$，且

$$6 = 2 \cdot 3 = (1 + \sqrt{-5})(1 - \sqrt{-5}) \tag{1.1}$$

因此在 $R$ 中 $1 + \sqrt{-5}|x$，在 $\mathbb{Z}$ 中 $6 = N(1 + \sqrt{-5})|n$。在 $\mathbb{Z}$ 中 $12|n$，可得 $n = 12$。因 $z = a + b\sqrt{-5}$，$a, b \in \mathbb{Z}$，$N(z) = a^2 + 5b^2 = 12$，所以 $a^2 \equiv 2(mod 5)$。但是 $\mathbb{Z}_5$ 中的平方数只有 $0$，$1$ 和 $4$，所以这个方程无解，即 $x$，$y$ 在 $R$ 中无最大公因子。

尽管最大公因子并不总是存在，但它们一旦存在，均可唯一分解为单位的乘积。

**定理 1.1.1.** 设 $(R, +, \cdot)$ 是一个整环，$x, y \in R$。如果 $z$ 和 $t$ 均是 $x$，$y$ 的最大公因子，则 $z = ut$，$t = vz$，$u, v \in R^*$。

*定理 1.1.1 的证明.* 假设 $z$ 和 $t$ 都是 $x$ 和 $y$ 的最大公因子。那么，$t|z$，因为 $t|x$，$t|y$，并且 $z = gcd(x, y)$。因此，$z = ut$，$u \in R$。相似地，$z|t$，因此 $t = vz$，$v \in R$。因此，$z = ut = uvz$，即 $(1 - uv)z = 0$。如果 $z \neq 0$，则 $1 = uv$，$uv \in R^*$。如果 $z = 0$，那么 $t = vz = 0$，于是 $z = 1t$ 且 $t = 1z$。 □

**定义 1.1.5.** 设 $R$ 是一个整环。一个非零元素 $p \in R \setminus R^*$ 被称为素数，如果对任意 $a, b \in R$，$p|ab \longrightarrow p|a$ or $p|b$。一个非零元素 $p \in R \setminus R^*$ 被称作不可约因子，如果对任意 $a, b \in R$，$p|ab \longrightarrow a \in R^*$ or $b \in R^*$。

**例 1.1.7.** 设 $R = \mathbb{Z}[\sqrt{-5}]$（如例 1.1.5），检查 $R$ 中元素 $2$，$3$，$1 + \sqrt{-5}$ 和 $1 - \sqrt{-5}$ 均是不可分解的。等式 (1.1) 表明相同的元素可能有不同的分解形式。因此，对于那些具有唯一因式分解形式的整环，被赋予一个特殊的名字。

**定义 1.1.6.** 唯一分解整环 (UFD) $(R, +, \cdot)$ 是一个整环，对于任一个非零元素 $x \in R \setminus R^*$，均有 $u \in R^*$，互质的不可约因子 $p_1, \ldots, p_n \in R$ 和正整数 $e_1, \ldots, x_n$ 使 $x = u p_1^{e_1} \cdots p_n^{e_n}$。而且，这种分解对于 $u$，单位 $p_i$ 和指数的排列是唯一的。

**例 1.1.8.** 设 $R = \mathbb{Q}[X,Y]$ 是以 $X$，$Y$ 为变量的所有有理系数的多项式的集合。对于通常的多项式加法 + 和乘法 ·，$(R,+,\cdot)$ 是一个典型的唯一分解整环。

对于任意一个整环，素数一定是不可约的。反过来，却不一定正确，但在唯一分解整环中还是成立的。于是，当我们考虑的对象是一个唯一分解整环时，$x$ 的质因式分解和最简因式分解具有相同的意义。

**定理 1.1.2.** 设 $(R,+,\cdot)$ 是一个整环，对于 $R$ 中的每一个素数 $p$ 均是不可约的。如果 $R$ 是一个唯一分解整环，则对于 $R$ 中的每一个不可约元素 $p$ 均是素数。

另外，最大公因子在唯一分解整环中总是存在，可由最简因式分解得到。

**定理 1.1.3.** 如果 $R$ 是一个 $UFD$，那么任意 $x,y \in R$，均有最大公因子。

*定理 1.1.3 的证明.* 设 $x,y \in R$，并首先假定 $x = 0$。$y|y$，$y|0$，并且在 $R$ 中整除 $x$ 和 $y$ 的必定整除 $y$，因此 $y$ 是 $x$ 和 $y$ 的最大公因子。相似地，如果假定 $y = 0$，则 $x$ 是 $x$ 和 $y$ 的最大公因子。现在假定 $x \neq 0$，$y \neq 0$，并且 $x = u\prod_{p \in \mathcal{X}} p^{n_p}$ 和 $y = v\prod_{p \in \mathcal{Y}} p^{m_p}$ 是 $x$ 和 $y$ 的最简因式分解，其中 $\mathcal{X}$ 和 $\mathcal{Y}$ 是不可约因子的有限集。我们选择单位 $u$ 和 $v$，以便使能整除 $x$ 和 $y$ 的任何不可约因子都在 $\mathcal{X} \cap \mathcal{Y}$ 中。然后使

$$z = \prod_{p \in \mathcal{X} \cap \mathcal{Y}} p^{min(n_p, m_p)} \in R \tag{1.2}$$

我们有

$$x = zu \prod_{p \in \mathcal{X} \cap \mathcal{Y}} p^{n_p - min(n_p, m_p)} \prod_{p \in \mathcal{X} \setminus \mathcal{Y}} p^{n_p}$$

因此 $z|x$。一个相似地公式即可表明 $z|y$。假设对于某一 $t \in R$，$t|x$，$t|y$，并使 $t = w\prod_{p \in \mathcal{T}} p^{e_p}$ 是它的最简因式分解，其中 $\mathcal{T}$ 是一个不可约因子的有限集。对于 $p \in \mathcal{T}$，我们有 $x = tb = p^{e_p}ab$，$a,b \in R$，于是有 $s \in R^*$，使得 $sp \in \mathcal{X}$。用 $ws^{-e_p}$ 代替 $w$，我们可以由最简因式分解的唯一性猜测到 $p \in \mathcal{X}$ 和 $e_p \leq n_p$。相似地，我们由 $t|y$ 可以得到 $p \in \mathcal{Y}$ 和 $e_p \leq m_p$。于是，对于任意 $p \in \mathcal{T}$，有 $\mathcal{T} \subseteq \mathcal{X} \cap \mathcal{Y}$，$e_p \leq min(n_p, m_p)$。从而，

$$z = tw^{-1} \prod_{p \in \mathcal{T}} p^{min(n_p, m_p) - e_p} \prod_{p \in (\mathcal{X} \cap \mathcal{Y}) \setminus \mathcal{T}} p^{min(n_p, m_p)}$$

这意味着 $t|z$，于是 $z = gcd(x,y)$。 □

下面是 Guass 发现的一个经典结论，多项式可以被唯一地分解成不可约因式之积。

**定理 1.1.4.** 如果 $R$ 是一个 $UFD$，那么多项式环 $R[X_1,\ldots,X_n]$ 是一个 $UFD$。

**定义 1.1.7.** 设 $(G,\circ)$ 是一个单位元为 $e$ 的群。我们说 $H \subseteq G$ 是 $(G,\circ)$ 的一个子群，如果：

**(i)** $e \in H$,

**(ii)** $\forall a,b \in H, a \circ b \in H$,

**(iii)** $\forall a \in H, a^{-1} \in H$.

实际上，给定 $G$ 的一个子集 $H$，检查它是不是 $G$ 的一个子群，下面两种方法等效：$H$ 是否具有上面列出的三条性质，或 $H$ 是否满足是非空集合，且对任意 $a,b \in H$，$a \circ b^{-1} \in G$。

**例 1.1.9.** $G = GL(\mathbb{Q}, 2)$ (如例 1.1.1)，$\circ$ 表示矩阵乘法，设 $H = SL(\mathbb{Q}, 2)$ 是 $G$ 的一个子集，其中包含所有行列式值为 1 的矩阵。$H$ 有单位矩阵，即非空，且对于任意 $a, b \in H$，$a \circ b^{-1}$ 的行列式值是 $a$ 与 $b$ 的行列式值之商，即为 1，故 $H$ 是 $G$ 的一个子群。

**定义 1.1.8.** 设 $(R, +, \cdot)$ 是一个交换环。$R$ 的子集 $I$ 被称为理想，如果 $(I, +)$ 是 $(R, +)$ 的子群，且 $a \in I$，对任意 $R$ 中 $x$ 均有 $xa \in I$。设 $x_1, \ldots, x_n \in R$，一个以 $x_1, \ldots, x_n$ 为基的理想是包含 $x_1, \ldots, x_n$ 的最小理想，记做 $(x_1, \ldots, x_n)$。一个理想 $I \subseteq R$ 被称作主理想，如果 $I = (x)$，$x \in R$。

实际上，以 $x_1, \ldots, x_n$ 为基的理想即是 $x_1, \ldots, x_n$ 在 $R$ 上所有线性组合的集合。

**定理 1.1.5.** 设 $(R, +, \cdot)$ 是一个交换环，$x_1, \ldots, x_n \in R$。则有，

$$(x_1, \ldots, x_n) = \{a_1 x_1 + \cdots + a_n x_n; a_1, \ldots, a_n \in R\}$$

*定理 1.1.5 的证明.* 设 $I = a_1 x_1 + \cdots + a_n x_n, a_1, \ldots, a_n \in R$。对任意 $i$，有 $x_i \in I$。让 $a = \sum_{i=1}^n a_i x_i \in I$，$b = \sum_{i=1}^n \in I$。我们有 $a - b = \sum_{i=1}^n (a_i - b_i) x_i \in I$，所以 $(I, +)$ 是 $(R, +)$ 的一个子群。对于任意 $x \in R$，我们有 $xa = \sum_{i=1}^n (xa_i) x_i \in I$，所以 $I$ 是 $R$ 中包含 $\{x_1, \ldots, x_n\}$ 的一个理想。现在设 $J$ 是 $R$ 中包含 $\{x_1, \ldots, x_n\}$ 的任一个理想，并设 $a = \sum_{i=1}^n a_i x_i \in I$。对于每一个 $i, x_i \in J$，由 $RJ \subseteq J$ 可知 $a_i x_i \in J$，由 $(J, +)$ 是一个群可知 $a \in J$。到此即证，$I \subseteq J$，有 $I = (x_1, \ldots, x_n)$。 □

**例 1.1.10.** $R = \mathbb{Q}[X, Y]$ (如例 1.1.8)，设 $I = (X, Y)$。经检验，我们可以得知 $I$ 并不是主理想，所以说 $R$ 中的理想并不都是主理想。自然地，那些任一理想均是主理想的整环得到了一个特殊的名字。

**定义 1.1.9.** 主理想整环 (PID) 是一个整环，它的任一理想均是主理想。

**例 1.1.11.** 用 $R = \mathbb{Q}[X]$ 表示所有以 $X$ 为单变量的实多项式。$(R, +, \cdot)$ 是一个主理想整环，$+$ 和 $\cdot$ 依次表示多项式的加法和乘法。

最后，我们来说在这本书中用处最大的整环，在其上可以进行欧几里得带余除法。

**定义 1.1.10.** 欧几里得整环 $(R, +, \cdot)$ 是指一个整环，在其上定义了一种映射 $\nu : R \setminus \{0\} \to \mathbb{N}$ 满足：

**(i)** $\forall a, b \in R \setminus \{0\}, \nu(ab) \geq \nu(a)$.

**(ii)** (欧几里得带余除法)$\forall a, b \in R, b \neq 0, \exists q, r \in R$，使 $a = bq + r$，并且 $r = 0$ 或 $\nu(r) < \nu(b)$.

映射 $\nu$ 被称为欧几里得整环的大小函数。

**例 1.1.12.** 环 $(\mathbb{Z}, +, \cdot)$，在其上定义整数通常的加法与乘法，显然它是一个欧几里得整环，其欧几里得映射为 $\nu(a) = |a|$，事实上，这也是我们叫它欧几里得整环的原因。

尽管主理想整环和欧几里得整环是定义在任一个整环上的，但其实它们之间有线形层次关系。

**定理 1.1.6.** 每一个欧几里得整环均是主理想整环 (PID)。

**定理 1.1.7.** 每一个主理想整环 (PID) 均是唯一分解整环 (UFD)。

既然每一个主理想整环 (PID) 均是唯一分解整环 (UFD)，而由定理 1.1.3 可知在 UFD 中最大公因子总是存在，故在 PID 中也总是存在最大公因子。下面我们说明，两个元素的最大公因子与它们产生同样的理想。

**定理 1.1.8.** 如果 $R$ 是一个 PID，那么对任意 $x, y \in R$，有 $(x, y) = gcd(x, y)$。

*定理 1.1.8 的证明.* 设 $x, y \in R$，且 $z$ 是 $R$ 中理想 $(x, y)$ 的生成元，换句话说，$(z) = (x, y)$。那么，$x \in (z)$，存在 $u \in R$ 使得 $x = zu$，这即是说 $z|x$。相似地，$y \in (x)$，$z|y$。除此之外，$z \in (x, y)$，所以存在 $a, b \in R$ 使得 $z = ax + by$。设 $t \in R$，且 $t|x$，$t|y$。那么存在 $c, d \in R$，使得 $x = ct$ 和 $y = dt$。因此，$z = act + bdt = (ac + bd)t$，$t|z$，这即是说，$z = gcd(u, v)$。 □

最后，让我们来回忆一下有关域的重要定理与结论。

**定义 1.1.11.** 域 $(F, +, \cdot)$ 是一个交换环，并且 $(F \setminus \{0\}, \cdot)$ 是一个群，换言之，$F$ 中的每一个元素均是 $F$ 的单位 ($F^* = F \setminus \{0\}$)。

**例 1.1.13.** 设 $R = \mathbb{Z}_5$(整数对模 5 的 5 个同余类构成的集合)，在其上定义 $+$ 和 $\cdot$ 表示模 6 的同余类的加法和乘法。$(F, +, \cdot)$ 是一个域。

**例 1.1.14.** 设 $R$ 是一个整环，定义在 $R \times R \setminus \{0\}$ 的关系 $\sim$ ($(a, b) \sim (c, d)$) 表示 $ad = bc$。易知 $\sim$ 是定义在 $R \times R \setminus \{0\}$ 上的一个等价关系，定义在等价类集合的运算即为通常的

$$\frac{a}{b} + \frac{c}{d} = \frac{ad + bc}{bd} \text{ 和 } \frac{a}{b} \times \frac{c}{d} = \frac{ac}{bd}$$

$a/b$ 表示 $(a, b)$ 的同余类。这个域被称为 $R$ 上的商域。例如，$\mathbb{Z}$ 的商域是 $\mathbb{Q}$，多项式环 $D[x]$ 的商域是有理函数域 $D(x)$，$D$ 是一个整环。

**定义 1.1.12.** 设 $F \subseteq E$ 是一个域。元素 $\alpha \in E$ 在 $F$ 上是代数的，如果对于一些非零多项式 $p \in F[X]$，$p(\alpha) = 0$，否则元素就被称为在 $F$ 上是超越的。$E$ 被称为 $F$ 的一个代数扩展，如果 $E$ 中的每一个元素在 $F$ 上都是代数的。

**定义 1.1.13.** 域 $F$ 被称为代数封闭的，如果对于每一个多项式 $p \in F[X] \setminus F$，都存在 $\alpha \in F$ 使得 $p(\alpha) = 0$。域 $E$ 被称为 $F$ 上的一个代数封闭，如果 $E$ 是在 $F$ 上代数封闭的 $F$ 的代数扩展。

注意到如果 $F$ 是代数封闭的，那么对于像 $p = c \prod_{i=1}^{n}(X - \alpha_i)^{e_i}$ 的在 $F$ 上的任何线性因子 $p \in K[X] \setminus K$ 有：由定义知 $p$ 在 $F$ 上有一个根 $\alpha$，并且由归纳法知 $p/(X - \alpha)$ 是 $F$ 上的一个线性因子。这些有关代数封闭的基本结论都是 $E$ 可得的一些结论。Steinitz 曾说过：它们存在，并且唯一。

**定理 1.1.9.** *([54]Chap.VII § 2,[92] § 10.1) 每一个域 $F$ 都有一个代数封闭,并且 $F$ 的任意两个代数封闭都是同构的。*

鉴于上述定理,我们便可以谈域 $F$ 上的代数封闭,我们把它记为 $\overline{F}$。在这节我们最后提及的结论是 Hilbert's Nullstellensatz,这在算法中是不需要的,但是在排除出现在不定积分中的超越常量是需要的。我们在这写出它的两种经典表示形式。

**定理 1.1.10.** *(Weak Nullstellensatz,[92] § 16.5) 设 $F$ 是一个代数封闭域,多项式环 $F[X_1,\ldots,X_n]$ 的一个理想 $I$ 和 $F^n$ 的一个子集 $V(I)$ 由下式给出:*

$$V(I) = \{(x_1,\ldots,x_n) \in F^n \ s.t. p(x_1,\ldots,x_n) = 0, \forall p \in I\}. \tag{1.3}$$

那么, $V(I) = \emptyset \iff 1 \in I$。

**定理 1.1.11.** *(Nullstellensatz,[54] Chap.X § 2,[92] § 16.5).(Weak Nullstellensatz,[92] § 16.5) 设 $F$ 是一个代数封闭域,多项式环 $F[X_1,\ldots,X_n]$ 的一个理想 $I$ 和 $F^n$ 的一个子集 $V(I)$ 由 (1.3) 给出。对于任意的 $p \in F[X_1,\ldots,X_n]$,如果对于每一个 $(x_1,\ldots,x_n) \in V(I)$ 均有 $p(x_1,\ldots,x_n) = 0$,则存在某一整数 $m > 0$ 使 $p^m \in I$。*

## 1.2 欧几里得除法及伪除法

设域 $K$ 中的未知变量为 $x$。首先,我们描述一种经典的多项式分解方法:已知 $A, B \in K[x]$, $B \neq 0$,存在唯一的 $Q, R \in K[x]$ 使得 $A = BQ + R$,同时 $R = 0$ 或 $\deg(R) < \deg(B)$。这说明当 $K$ 是域时,多项式环 $K[x]$ 是欧几里得整环,其次数是其大小函数的次数。我们分别称 $Q$ 和 $R$ 为 $A$ 除以 $B$ 的商、$A$ 模 $B$ 的余数。

---

**PolyDivide**$(A, B)$ (* 欧几里得多项式分解 *)
(* 给定域 $K$ 和 $A, B \in K[x]$ 且 $B \neq 0$,返回 $Q, R \in K[x]$ 使得 $A = BQ + R$,同时 $R = 0$ 或 $\deg(R) < \deg(B)$。*)
 $Q \leftarrow 0, R \leftarrow A$
 **while** $R \neq 0$ 且 $\delta \leftarrow \deg(R) - \deg(B) \geq 0$ **do**
  $T \leftarrow \frac{lc(R)}{lc(B)} x^\delta, Q \leftarrow Q + T, R \leftarrow R - BT$
 **end while**
 **return** $(Q, R)$

---

**例 1.2.1.** 下面是 $A = 3x^3 + x^2 + x + 5$ 关于 $B = 5x^2 - 3x + 1$ 在 $\mathbb{Q}[x]$ 上的欧几里得分解:

| $Q$ | $R$ | $\delta$ | $T$ |
|---|---|---|---|
| $0$ | $3x^3 + x^2 + x + 5$ | $1$ | $\frac{3}{5}x$ |
| $\frac{3}{5}x$ | $\frac{14}{5}x^2 + \frac{2}{5}x + 5$ | $0$ | $\frac{14}{25}$ |
| $\frac{3}{5}x + \frac{14}{25}$ | $\frac{52}{25}x + \frac{111}{25}$ | $-1$ | |

所以,
$$A = B\left(\frac{3}{5}x + \frac{14}{25}\right) + \left(\frac{52}{25}x + \frac{111}{25}\right)$$

这个算法要求系数是来自于域的，这是因为在 $K$ 中两个首项系数进行了除法运算。如果 $K$ 是一个整环，那么 $A$ 的首项系数并不总是被 $B$ 的首项系数整除，因而欧几里得分解并非总存在。例如，对于上面的例子来说，在 $\mathbb{Z}[x]$ 上 $A$ 对于 $B$ 的欧几里得除法就是无意义的。但是在 $\mathbb{Z}[x]$ 上 PolyDivide 算法是适用于 $25A$ 和 $B$ 的，因为所有的求商运算在 $\mathbb{Z}$ 上是整除的。总之，给定整环 D 以及 $A, B \in D[x]$，令 $b = \text{lc}(B)$，$\delta = \max(-1, \deg(A) - \deg(B))$，那么 $b^{\delta+1}A$ 对于 $B$ 的 PolyDivide 运算在 $D$ 上是可以整除的，返回的 $Q$ 和 $R$ 分别称为 $A$ 除以 $B$ 的伪商、$A$ 模 $B$ 的伪余数。它们满足 $b^{\delta+1}A = BQ + R$，同时 $R = 0$ 或 $\deg(R) < \deg(B)$。$A, B$ 产生的伪商、伪余数分别记为 $\text{pquo}(A, B), \text{prem}(A, B)$。实际上，正如下面算法所做的，$A$ 反复乘以 $b$ 比 $A$ 一次性乘以 $b^{\delta+1}$ 效率更高。

---

**PolyPseudoDivide**$(A, B)$　（* 欧几里得多项式伪分解 *）
(* 给定一个整环 D 以及 $A, B \in D[x]$ 且 $B \neq 0$，返回 $\text{pquo}(A, B)$ 和 $\text{prem}(A, B)$。*)
　$b \leftarrow lc(B), N \leftarrow \deg(A) - \deg(B) + 1, Q \leftarrow 0, R \leftarrow A$
　**while** $R \neq 0$ 且 $\delta \leftarrow \deg(R) - \deg(B) >= 0$ **do**
　　$T \leftarrow lc(R)x^\delta, N \leftarrow N - 1, Q \leftarrow bQ + T, R \leftarrow bR - TB$
　**end while**
　**return** $(b^N Q, b^N R)$

---

**例 1.2.2.** 对于例 1.2.1中的 $A$ 和 $B$，我们有 $b = 5, N = 2$，并且：

| $Q$ | $R$ | $\delta$ | $T$ | N |
|---|---|---|---|---|
| 0 | $3x^3 + x^2 + x + 5$ | 1 | $3x$ | 1 |
| $3x$ | $14x^2 + 2x + 25$ | 0 | 14 | 0 |
| $15x + 14$ | $52x + 111$ | $-1$ | | |

所以
$$25A = B(15x + 14) + (52x + 111)$$

## 1.3 欧几里德算法

设 $D$ 是一个欧几里得整环且 $\nu : D \setminus \{0\} \to N$ 为其大小函数。$D$ 上的欧几里得除法可用来计算 $D$ 的任两个元素的最大公约数。基本的想法可以回溯到欧几里得，他用下面的方法计算两个整数的最大公因子：若 $a = bq + r$，则 $gcd(a, b) = gcd(b, r)$。如下定义序列 $(a_i)(i \geq 0)$:

$$a_0 = a, a_1 = b, 且对 i \geq 2 (q_i, a_i) = \textbf{EuclideanDivision}(a_{i-2}, a_{i-1})$$

由于对任何 $x \in D$ 有 $gcd(x, 0) = x$，故上述序列的最后一个非零元素是 $a$ 和 $b$ 的最大公因子。又对于 $a_i \neq 0 (i \geq 1)$，有 $a_{i+1} = 0$ 或 $\nu(a_{i+1} < \nu(a_i))$，故序列只能有有限个非零元素。这便引出了通过重复欧几里得除法来计算 $gcd(a, b)$ 的算法。

```
Euclidean(a,b)    (* 欧几里德算法 *)
(* 给出欧几里德整环 D 和 a,b ∈ D, 返回 gcd(a,b) *)
  while b ≠ 0 do
    (q,r) ← EuclideanDivision(a,b)            (* a = bq + r *)
    a ← b
    b ← r
  end while
  return a
```

**例 1.3.1.** 对 $D = \mathbb{Q}[x]$ 中的

$$a = x^4 - 2x^3 - 6x^2 + 12x + 15 \text{和} b = x^3 + x^2 - 4x - 4$$

使用欧几里德算法:

| $a$ | $b$ | $q$ | $r$ |
|---|---|---|---|
| $x^4 - 2x^3 - 6x^2 + 12x + 15$ | $x^3 + x^2 - 4x - 4$ | $x - 3$ | $x^2 + 4x + 3$ |
| $x^3 + x^2 - 4x - 4$ | $x^2 + 4x + 3$ | $x - 3$ | $5x + 5$ |
| $x^2 + 4x + 3$ | $5x + 5$ | $\frac{1}{5}x + \frac{3}{5}$ | $0$ |
| $5x + 5$ | $0$ | | |

故 $5x+5$ 是 $a$ 和 $b$ 在 $\mathbb{Q}[x]$ 中的最大公因子。

欧几里德除法可以不仅仅返回 $a$ 和 $b$ 的最大公因子，通过简单的拓展还可以找出 $D$ 中使得 $sa + tb = \gcd(a,b)$ 的元素 s 和 t。由定理 1.1.8, $\gcd(a,b)$ 属于 $a$ 和 $b$ 生成的理想，故这样的元素始终存在。

```
ExtendedEuclidean(a,b)    (* 拓展欧几里德算法 *)
(* 给出欧几里德整环 D 以及 a,b ∈ D, 返回 s,t,g ∈ D 使得 g = gcd(a,b) 并
且 sa + tb = g *)
  a_1 ← 1, a_2 ← 0, b_1 ← 0, b_2 ← 1
  while b ≠ 0 do
    (q,r) ← EuclideanDivision(a,b)            (* a = bq + r *)
    a ← b, b ← r
    r_1 ← a_1 - qb_1, r_2 ← a_2 - qb_2
    a_1 ← b_1, a_2 ← b_2, b_1 ← r_1, b_2 ← r_2
  end while
  return (a_1, a_2, a)
```

**例 1.3.2.** 考虑和例 1.3.1中相同的 $a$ 和 $b$

| $a$ | $b$ | $q$ | $r$ |
|---|---|---|---|
| $x^4 - 2x^3 - 6x^2 + 12x + 15$ | $x^3 + x^2 - 4x - 4$ | $x - 3$ | $x^2 + 4x + 3$ |
| $x^3 + x^2 - 4x - 4$ | $x^2 + 4x + 3$ | $x - 3$ | $5x + 5$ |
| $x^2 + 4x + 3$ | $5x + 5$ | $\frac{1}{5}x + \frac{3}{5}$ | $0$ |
| $5x + 5$ | $0$ | | |

| $a_1$ | $a_2$ | $b_1$ | $b_2$ |
|---|---|---|---|
| 1 | 0 | 1 | $-x+3$ |
| 1 | $-x+3$ | $-x+3$ | $x^2-6x+10$ |
| $-x+3$ | $x^2-6x+10$ | $\frac{1}{5}x^2-\frac{4}{5}$ | $-\frac{1}{5}x^3+\frac{3}{5}x^2+\frac{3}{5}x-3$ |

因此 $5x+5$ 是 $a$ 和 $b$ 在 $\mathbb{Q}[x]$ 中的最大公因子，并且

$$(-x+3)a + (x^2-6x+10)b = 5x+5 \tag{1.4}$$

若只需要系数 s 和 t 中的一个，那么我们不妨使用计算该系数的变形的拓展欧几里德算法：

---

**HalfExtendedEuclidean**$(a,b)$    (* 半拓展欧几里德算法 *)
(* 给出欧几里德整环 $D$ 以及 $a,b \in D$，返回 $s,g \in D$ 使得 $g = \gcd(a,b)$ 且 $sa \equiv g(mod\, b)$*)
  $a_1 \leftarrow 1, b_1 \leftarrow 0$
  **while** $b \neq 0$ **do**
    $(q,r) \leftarrow$ **EuclideanDivision**$(a,b)$              (* $a = bq+r$ *)
    $a \leftarrow b, b \leftarrow r$
    $r_1 \leftarrow a_1 - qb_1, a_1 \leftarrow b_1, b_1 \leftarrow r_1$
  **end while**
  $mathbf return\ (a_1, a)$

---

这个"半"算法可用作拓展欧几里德算法的一个高效替代品，因为第二个系数可以通过下式得到：

$$t = \frac{g-sa}{b}$$

其中除法总是整除的。

---

**ExtendedEuclidean**$(a,b)$    (* 拓展欧几里德算法——"半/全"版本 *)
(* 给出欧几里德整环 $D$ 以及 $a,b \in D$，返回 $s,t,g \in D$ 使得 $g = \gcd(a,b)$ 并且 $sa+tb=g$ *)
  $(s,g) \leftarrow$ **HalfExtendedEuclidean**$(a,b)$         (* $sa \equiv g(mod\ b)$ *)
  $(t,r) \leftarrow$ **EuclideanDivision**$(g-sa, b)$            (* $r$ 必为 $0$ *)
  **return** $(s,t,g)$

---

**例 1.3.3.** 再一次计算例 1.3.1中 $a$ 和 $b$ 的拓展 gcd，我们有：

1. $(s,g) =$ **HalfExtendedEuclidean**$(a,b) = (-x+3, 5x+5)$

2. $g - sa = x^5 - 5x^4 + 30x^3 - 16x$

3. $(t,r) =$ **PolyDivide**$(g-sa, b) = (x^2 - 6x + 10, 0)$

从而我们再一次得到了(1.4)。

拓展欧几里德算法也可以用来解丢番图方程

$$sa + tb = c \tag{1.5}$$

其中已知 $a, b, c \in D$ 求 $s, t \in D$。(1.5)有解当且仅当 $c$ 在 $a$ 和 $b$ 生成的理想中，也即 $c$ 是 $\gcd(a,b)$ 在 $D$ 中的倍数。拓展欧几里德算法先求解方程 $sa + tb = \gcd(a,b)$，剩下只需要对各解乘上 $c/\gcd(a,b)$ 来得到(1.5)的一个解。值得一提的是，当 $c$ 在 $a$ 和 $b$ 生成的理想中时，(1.5)解的个数和 $D$ 中元素个数相同（当 $a$ 和 $b$ 非零时），这是因为对任何 $d \in D$ 有 $sa+tb = (s+bd)a+(t-ad)b$。为了保证前一个只有两个参数的拓展欧几里德算法没有不清楚的地方，我们也将这个算法称作"拓展欧几里德算法"。正如之前提及的，当只需要一个系数时，可以用半拓展版本的算法。在此展示描述和往后拓展使用中，我们认为各拓展版本的欧几里德算法返回的解 $s$ 或者 $(s,t)$ 都满足 $s=0$ 或者 $\nu(s) < \nu(b)$。在多项式环中（有 $\nu(p) = \deg(p)$），一个重要的推论是如果 $\deg(c) < \deg(a) + \deg(b)$，那么我们有 $t=0$ 或 $\deg(t) < \deg(a)$。更进一步，若有 $\deg(s) < \deg(b)$ 且 $\deg(t) \geq \deg(a)$，那么 $\deg(c) = \deg(sa+tb) = \deg(tb) = \deg(t) + \deg(b) \geq \deg(a) + \deg(b)$。

---

**ExtendedEuclidean**$(a,b,c)$　　(* 拓展欧几里德算法——丢翻图版本 *)
(* 给出欧几里德整环 $D$ 以及 $a,b,c \in D$,　$c \in (a,b)$，返回 $s,t \in D$ 使得 $sa+tb = c$ 且或者 $s = 0$ 或者 $\nu(s) < \nu(b)$ 成立 *)
　　$(s,t,g) \leftarrow$ **ExtendedEuclidean**$(a,b)$　　　　　　　　　　(* g=sa+tb *)
　　$(q,r) \leftarrow$ **EuclideanDivision**$(c,g)$　　　　　　　　　　(* c=gq+r *)
　　**if** $r \neq 0$ **then**
　　　　**error** "$c$ 不在 $a$ 和 $b$ 生成的理想中"
　　**end if**
　　$s \leftarrow qs, t \leftarrow qt$
　　**if** $s \neq 0$ **and** $\nu(s) > \nu(b)$ **then**
　　　　$(q,r) \leftarrow$ **EuclideanDivision**$(s,b)$　　　　　　　　(* s=bq+r *)
　　　　$s \leftarrow r, t \leftarrow t + qa$
　　**end if**
　　**return** $(s,t)$

---

**例 1.3.4.** $a$ 和 $b$ 仍同例 1.3.1，我们在 $\mathbb{Q}[x]$ 中解 $sa+tb = x^2-1$。使用 **ExtendedEuclidean**：

1. $(s,t,g) =$ **ExtendedEuclidean**$(a,b) = (-x+3, x^2-6x+10, 5x+5)$

2. **PolyDivide**$(x^2-1, 5x+5) = ((x-1)/5, 0)$

3. $s \leftarrow qs = (-x^2+4x-3)/5$

4. $t \leftarrow qt = (x^3-7x^2+16x = 10)/5$

从而得到如下的解：

$$(\frac{-x^2+4x-3}{5})a + (\frac{x^3-7x^2+16x-10}{5})b = x^2-1 \tag{1.6}$$

---

**HalfExtendedEuclidean**$(a,b,c)$    (* 半拓展欧几里德算法——丢翻图版本 *)
(* 给出欧几里德整环 $D$ 以及 $a,b,c \in D$, $c \in (a,b)$, 返回 $s \in D$ 使得 $sa \equiv c (\bmod\ b)$ 且或 $s = 0$ 或 $\nu(s) < \nu(b)$ *)
  $(s,g) \leftarrow$ **HalfExtendedEuclidean**$(a,b)$             (* $sa \equiv g (\bmod\ b)$ *)
  $(q,r) \leftarrow$ **EuclideanDivision**$(c,g)$                 (* $c = gq + r$ *)
  **if** $r \neq 0$ **then**
    **error** "$c$ 不在 $a$ 和 $b$ 生成的理想中"
  **end if**
  $s \leftarrow qs$
  **if** $s \neq 0$ **then**
    $(q,r) \leftarrow$ **EuclideanDivison**$(s,b)$                (* $s = bq + r$ *)
    $s \leftarrow r$
  **end if**
  **return** $s$

---

如前所述,"半"版本能给出比拓展丢翻图版本更高效的替代品,因为第二个系数可以通过

$$t = \frac{c-sa}{b}$$

得到,并且除法总能整除。

---

**ExtendedEuclidean**$(a,b,c)$    (* 拓展欧几里德算法——"半/全"丢翻图版本 *)
(* 给出欧几里德整环 $D$ 以及 $a,b,c \in D$, $c \in (a,b)$, 返回 $s,t \in D$ 使得 $sa + tb = c$ 且或者 $s = 0$ 或者 $\nu(s) < \nu(b)$ 成立 *)
  $s \leftarrow$ **HalfExtendedEuclidean**$(a,b,c)$            (* $sa \equiv c (\bmod\ b)$ *)
  $(t,r) \leftarrow$ **EuclideanDivision**$(c-sa,b)$           (* $r$ 必为 0 *)
  **return** $(s,t)$

---

**例 1.3.5.** 用例 1.3.1的 $a$ 和 $b$ 在 $\mathbb{Q}[x]$ 中解 $sa + tb = x^2 - 1$,有

1. $s =$ **HalfExtendedEuclidean**$(a,b,x^2-1) = (-x^2 + 4x - 3)/5$

2. $c - sa = x^2 - 1 - sa = (x^6 - 6x^5 + 5x^4 + 30x^3 - 46x^2 - 24x + 40)/5$

3. $(t,r) =$ **PolyDivide**$(c-sa,b) = ((x^3 - 7x^2 + 16x - 10)/5, 0)$

从而我们回到了(1.6)。

    鉴于拓展欧几里得算法可用来求解丢番图方程,我们也可用其来计算部分分式分解。令 $d \in D \setminus \{0\}$ 且令 $d = d_1 \ldots d_n$ 为 $d$ 的任意满足 $\gcd(d_i, d_j) = 1$对$i \neq j$ 分解(不

一定化为不可约的)。从而对任何 $a \in D \setminus \{0\}$ 有 $D$ 中的 $a_0, a_1, \ldots, a_n$ 使得对 $i \geq 1$ 有 $a_i = 0$ 或者 $\nu(a_i) < \nu(d_i)$，并且

$$\frac{a}{d} = \frac{a}{\prod_{i=1}^n d_i} = a_0 + \sum_{i=1}^n \frac{a_i}{d_i}。$$

这样的分解称为遵从分解 $d = \prod_{i=1}^n d_i$ 的部分分式分解，并且其计算可归结到求解形如(1.5)的方程，进而归结到欧几里德算法。更进一步地，首先由欧几里德除法写下 $a = da_0 + r$，其中或者 $r = 0$ 或者 $\nu(r) < \nu(d)$。若 $n = 1$，则 $a/d = a_0 + r/d$ 已经是所需形式；否则由于对 $i \neq j$ 有 $\gcd(d_i, d_j) = 1$，我们有 $\gcd(d_1, d_2 \cdots d_n) = 1$，固由拓展欧几里德算法，我们可以找到 $D$ 中的 $a_1$ 和 $b$ 使得

$$r = a_1(d_2 \cdots d_n) + bd_1 \tag{1.7}$$

并且有 $a_1 = 0$ 或者 $\nu(a_1) < \nu(d_1)$。我们可以递归地寻找 $b_0, a_2, \ldots, a_n \in D$ 使得或者 $a_i = 0$ 或者 $\nu(a_i) < \nu(d_i)$，并且

$$\frac{b}{d_2 \cdots d_n} = b_0 + \sum_{i=2}^n \frac{a_i}{d_i}.$$

用 $d$ 除(1.7)然后加上 $a_0$，我们有

$$\frac{a}{d} = a_0 + \frac{r}{d} = a_0 + \frac{a_1}{d_1} + \frac{b}{d_2 \cdots} = (a_0 + b_0) + \sum_{i=1}^n \frac{a_i}{d_i}$$

值得注意的是在多项式环的情况下，若在(1.7)中 $\deg(r) < \deg(d) = \deg d_1 + \deg(d_2 \cdots d_n)$ 且 $\deg(a_1) < \deg(d_1)$，那么 $\deg(b) < \deg(d_2 \cdots d_n)$，从而 $b_0 = 0$。

---

**PartialFraction**$(a, d_1, \ldots, d_n)$   (* 部分分式分解 *)
(* 给出欧几里德整环 $D$，正整数 $n$ 和 $a, d_1, \ldots, d_n \in D \setminus \{0\}$ 对 $i \neq j$，返回 $a_0, a_1, \ldots, a_n \in D$ 使得
$$\frac{a}{d_1 \cdots d_n} = a_0 + \sum_{i=1}^n \frac{a_i}{d_i}$$
并且或者 $a_i = 0$ 或者 $\nu(a_i) < \nu(d_i)$ 对 $i \geq 1$。*)
  $(a_0, r) \leftarrow$ **EuclideanDivision**$(a, d_1 \cdots d_n)$          (* $a = (d_1 \cdots d_n)a_0 + r$ *)
  **if** $n = 1$ **then**
    **return** $(a_0, r)$
  **end if**
  $(a_1, t) \leftarrow$ **PartialFraction**$(t, d_2, \ldots, d_n)$          (* $\nu(a_1) < \nu(d_1)$ *)
  **return**$(a_0 + b_0, a_1, a_2, \ldots, a_n)$

---

**例 1.3.6.** 我们来计算

$$f = \frac{a}{d} = \frac{x^2 + 3x}{x^3 - x^2 - x + 1} \in \mathbb{Q}(x)$$

的遵从分解 $d = (x+1)(x^2 - 2x + 1) = d_1 d_2$ 的部分分式分解。我们有：

1. $(a_0, r) = \mathbf{PolyDivide}(a, d) = (0, x^2 + 3x)$
2. $(a_1, t) = \mathbf{ExtendedEuclidean}(x^2 - 2x + 1, x + 1, x^2 + 3x) = (-\frac{1}{2}, \frac{3x+1}{2})$
3. $(b_0, a_2) = \mathbf{PartialFraction}((3x+1)/2, x^2 - 2x + 1) = (0, (3x+1)/2)$

故 $f$ 的部分分式分解是

$$\frac{x^2 + 3x}{x^3 - x^2 - x + 1} = \frac{-1/2}{x+1} + \frac{(3x+1)/2}{x^2 - 2x + 1}$$

通过结合欧几里德除法，我们可以得到部分分式分解的改进：令 $m \geq 1$ 且 $d \in D \setminus \{0\}$，则对任何 $a \in D \setminus \{0\}$，存在 $a_0, a_1, \ldots, a_m \in D$ 使得对 $j \geq 1$ 或者 $a_j = 0$ 或者 $\nu(a_j) < \nu(d)$，并且

$$\frac{a}{d^m} = a_0 + \sum_{j=1}^{m} \frac{a_j}{d^j}$$

这样的分解叫做$a/d^m$ 的 $d$ 进制展式。用欧几里德除法写出 $a = dq + a_m$，其中或者 $a_m = 0$ 或者 $\nu(a_m) < \nu(d)$，从而

$$\frac{a}{d^m} = \frac{dq + a_m}{d^m} = \frac{q}{d^{m-1}} + \frac{a_m}{d_m}$$

若 $m = 1$，那么上式就是满足 $a_0 = q$ 的所需表达式。否则我们递归的寻找 $a_0, a_1, \ldots, a_{m-1} \in D$ 使得对 $j \geq 1$ 或者 $a_j = 0$ 或者 $\nu(a_j) < \nu(d)$，并且

$$\frac{q}{d^{m-1}} = a_0 + \sum_{j=1}^{m-1} \frac{a_j}{d_j}$$

因此，

$$\frac{a}{d^m} = \frac{q}{d^{m-1}} + \frac{a_m}{d^m} = a_0 + \sum_{j=1}^{m} \frac{a_j}{d^j}$$

现令 $d \in D \setminus \{0\}$，并令 $d = d_1^{e_1} \cdots d_n^{e_n}$ 为 $d$ 的任何分解（不一定要不可约），其中对 $i \neq j$ 有 $\gcd(d_i, d_j) = 1$，并且各 $e_i$ 为正整数。下面，对任何 $a \in D \setminus \{0\}$，我们可首先计算 $a/d$ 的遵从 $d = b_1 \cdots b_n$ 的部分分式分解，其中 $b_i = d_i^{e_i}$：

$$\frac{a}{d} = a_0 + \sum_{i=1}^{n} \frac{a_i}{b_i} = a_0 + \sum_{i=1}^{n} \frac{a_i}{d_i^{e_i}}$$

然后计算每个加数的 $d_i$ 进制展式得到

$$\frac{a}{d} = \frac{a}{\prod_{i=1}^{n} d_i^{e_i}} = \tilde{a} + \sum_{i=1}^{n} \sum_{j=1}^{e_i} \frac{a_{ij}}{d_i^j}$$

其中 $\tilde{a} \in D$ 并且对每个 $i$ 和 $j$ 或者 $a_{ij} = 0$ 或者 $\nu(a_{ij}) < \nu(d_i)$。这个分解称作$a/d$ 的遵从因子分解 $d = \prod_{i=1}^{n} d_i^{e_i}$ 的完全部分分式分解，或在用 $d$ 的不可约因数分解（注…）时简称为 $a/d$ 的完全部分分式分解。

```
PartialFraction(a, d_1, ..., d_n, e_1, ..., e_n)    (* 完全部分分式分解 *)
(* 给出欧几里得整环 D, 正整数 n, e_1, ..., e_n 和 a, d_1, ..., d_n ∈ D \ {0}对i ≠ j,
返回 a_0, a_{1,1}, ..., a_{1,e_1}, ..., a_{n,1}, ..., a_{n,e_n} ∈ D 使得
```

$$\frac{a}{d_1^{e_1}\cdots d_n^{e_n}} = a_0 + \sum_{i=1}^{n}\sum_{j=1}^{e_i}\frac{a_{ij}}{d_i^j}$$

并且或者 $a_{ij}=0$ 或者 $\nu(a_{ij}) < \nu(d_i)$。 *)
  $(a_0, a_1, \ldots, a_n) \leftarrow$ **PartialFraction**$(a, d_1^{e_1}, \ldots, d_n^{e_n})$
  **for** $i \leftarrow 1$ **to** $n$ **do**
    **for** $j \leftarrow e_i$ **downto** $1$ **do**
      $(q, a_{ij}) \leftarrow$ **EuclideanDivision**$(a_i, d_i)$        (* $a_i = d_i q + q_{ij}$ *)
      $a_i \leftarrow q$
    **end for**
    $a_0 \leftarrow a_0 + a_i$
  **end for**
  **return**$(a_0, a_{1,1}m, \ldots, a_{n,1}, \ldots, a_{n,e_n})$

**例 1.3.7.** 我们来计算遵从因子分解 $d = (x+1)(x-1)^2 = d_1 d_2^2$ 的

$$f = \frac{a}{d} = \frac{x^2+3x}{x^3-x^2-x+1} \in \mathbb{Q}(x)$$

的完全部分分式分解。对 $a, d_1, d_2$ 和指数 1 和 2 使用 **PartialFraction**, 我们有:

$$(a_0, a_1, \ldots, a_n) = \textbf{PartialFraction}(x^2+3x, x+1, (x-1)^2) = (0, -\frac{1}{2}, \frac{3x+1}{2})$$

然后:

| $i$ | $j$ | $a_i$ | $d_i$ | $q$ | $a_{ij}$ | $a_0$ |
|---|---|---|---|---|---|---|
| 1 | 1 | $-1/2$ | $x+1$ | 0 | $-1/2$ | 0 |
| 2 | 2 | $(3x+1)/2$ | $x-1$ | $3/2$ | 2 | 0 |
| 2 | 1 | $3/2$ | $x-1$ | 0 | $3/2$ | 0 |

从而 $f$ 的完全部分分式分解是

$$\frac{x^2+3x}{x^3-x^2-x+1} = \frac{-1/2}{x+1} + \frac{2}{(x-1)^2} + \frac{3/2}{x-1}$$

    这里我们展示的计算部分分式分解的算法可以回溯到 19 世纪的 Hermite。这里我们就不对有理函数的替代算法和更好的实现展开了, 其它实现和它们的复杂度请参见 [1,45]。

## 1.4 结式和子结式

    在这节中, 我们讨论两个多项式的结式。尽管它们起源于十九世纪人们对非线性等式系统的认识, 但是结式在积分中扮演了一个重要的角色。在整节中, $R$ 是一个交换

1 代数基础

环，$x$ 是 $R$ 上一个不定元。

**定义 1.4.1.** 设 $A, B \in R[x] \backslash \{0\}$。$A = a_n x^n + \cdots + a_1 x + a_0$，$B = b_m x^m + \cdots + b_1 x + b_0$，其中 $a_n \neq 0$，$b_m \neq 0$，$n$ 和 $m$ 不全为 0。$A$ 和 $B$ 的 Sylvester 矩阵是按如下定义的一个 $n + m \times n + m$ 矩阵：

$$S(A,B) = \begin{pmatrix} a_n & \cdots & \cdots & \cdots & a_1 & a_0 & & & \\ & \ddots & & & & & & & \\ & & a_n & \cdots & \cdots & \cdots & a_1 & a_0 \\ b_m & \cdots & b_1 & b_0 & & & & & \\ & \ddots & & & & & & & \\ & & \ddots & & & & & & \\ & & & \ddots & & & & & \\ & & & & b_m & \cdots & b_1 & b_0 \end{pmatrix} \begin{matrix} \left.\vphantom{\begin{matrix}1\\1\\1\end{matrix}}\right\}\text{m rows} \\ \\ \left.\vphantom{\begin{matrix}1\\1\\1\\1\\1\end{matrix}}\right\}\text{n rows} \end{matrix}$$

其中，$A$ 式出现了 $m$ 行，$B$ 式出现了 $n$ 行。

$A$ 和 $B$ 的结式是 $S(A, B)$ 的行列式。

**例 1.4.1.** 设 $R = \mathbb{Z}[t]$，$A = 3tx^2 - t^3 - 4 \in R[x]$，$B = x^2 + t^3 x - 9 \in R[x]$。$A$ 和 $B$ 的 Sylvester 矩阵为

$$S(A,B) = \begin{pmatrix} 3t & 0 & -t^3 - 4 & 0 \\ 0 & 3t & 0 & -t^3 - 4 \\ 1 & t_3 & -9 & 0 \\ 0 & 1 & t^3 & -9 \end{pmatrix}$$

$A$ 和 $B$ 的结式为

$$det(S(A,B)) = -3t^{10} - 12t^7 + t^6 - 54t^4 + 8t^3 + 729t^2 - 216t + 16$$

两个多项式的结式的第一个有用的性质是与它们的根有关。

**定理 1.4.1.** *([54] Chap.V § 10,[92] §5.9).* 设 $\alpha_1, \ldots, \alpha_n$，$\beta_1, \ldots, \beta_m$，$a$ 和 $b$ 都是 $R$ 中的元素，且 $a \neq 0$，$b \neq 0$，$A = a(x - \alpha_1) \cdots (x - \alpha_n)$，$B = b(x - \beta_1) \cdots (x - \beta_m)$。有下式成立：

$$resultant(A,B) = a^m b^n \prod_{i=1}^{n} \prod_{j=1}^{m} (\alpha_i - \beta_j) = a^m \prod_{i=1}^{n} B(\alpha_i)$$

$$= (-1)^{nm} b^n \prod_{j=1}^{m} A(\beta_j) = (-1)^{mn} resultant(B,A)$$

由此可知，在整环 $R$ 上两个行列式的结式为 0 当且仅当它们在 $R$ 的商域上的代数封闭中有共同的零。

**推论 1.4.1.** *([54] Chap.V § 10,[92] §5.8). 设 $R$ 是一个整环，$K$ 是 $R$ 的商域，$K$ 的代数封闭为 $\overline{K}$。则，对于任意 $A, B \in R[x] \setminus \{0\}$,*

$$resultant(A, B) = 0 \iff \exists \gamma \in \overline{K}, 使得 A(\gamma) = B(\gamma) = 0$$

*推论 1.4.1 的证明.* 设 $A, B \in R[x] \setminus \{0\}$，且

$$A = a \prod_{i=1}^{n}(x - \alpha)^{e_i} 和 B = b \prod_{j=1}^{m}(x - \beta)^{f_j}$$

分别是 $A$ 和 $B$ 在 $\overline{K}[x]$ 上的质因式分解。由 $A \neq 0 \neq B$ 可知 $a \neq 0 \neq b$，因此从定理 1.4.1 我们可知

$$C = resultant(A, B) = a^M b^N \prod_{i=1}^{n}\prod_{j=1}^{m}(\alpha_i - \beta_j)^{e_i f_j}$$

其中，$M = \sum_{j=1}^{m} f_j$，$N = \sum_{i=1}^{n} e_i$。$\overline{K}$ 是一个域，如果 $C = 0$，那么对于某一 $i_0$ 和 $j_0$ 满足 $a_{i_0} - b_{j_0} = 0$。但若 $\gamma = \alpha_{i_0} = \beta_{j_0}$，有 $A(\gamma) = B(\gamma) = 0$。相反地，如果对某一 $\gamma \in \overline{K}$，$A(\gamma) = B(\gamma) = 0$，由 $\overline{K}$ 是一个域和存在 $i_0$ 和 $j_0$ 使得 $a_{i_0} - b_{j_0} = 0$，那么 $a_{i_0} - b_{j_0} = 0$，$C = 0$。 □

两个多项式的结式的另一个性质是有关它们产生的理想。

**定理 1.4.2.** *([54] Chap.V § 10,[92] §5.8). 对于任意 $A, B \in R[x] \setminus \{0\}$，均有 $S, T \in R[x]$ 使得 $resultant(A, B) = SA + TB$。*

由此可得，唯一分解整环上的多项式的结式为 0 当且仅当它们有一个非平凡的公因子。

**推论 1.4.2.** *([92] §5.8). 设 $R$ 是一个唯一分解整环。则，对于任意 $A, B \in R[x] \setminus \{0\}$,*

$$resultant(A, B) = 0 \iff deg(gcd(A, B)) > 0$$

子结式是源自 Sylvester 矩阵的子矩阵的多项式。

**定义 1.4.2.** 设 $A, B \in R[x] \setminus \{0\}$，$n = deg(A)$，$m = deg(B)$，$S$ 是 $A$ 和 $B$ 的 Sylvester 矩阵，$j$ 是一个满足 $0 \leq j < min(n, m)$ 整数。令 $_jS$ 为一个按如下删除规则从 $S$ 得到的一个 $n + m - 2j \times n + m$ 矩阵：

(i) 删除从 $m - j + 1$ 到 $m$ 行 (即 $A$ 的最后 $j$ 行)

(ii) 删除 $m + n - j + 1$ 到 $m + n$ 行 (即 $B$ 的最后 $j$ 行)

另外，对于 $0 \leq i \leq j$，设 $_jS_i$ 是删除 $_jS$ $m + n - 2j$ 到 $m + n$ 列（除了 $m + n - j - i$ 列）一个 $m + n - 2j \times m + n - 2j$ 的方阵。那么 $A$ 和 $B$ 的 $j^{th}$ 子结式为

$$S_j(A, B) = \sum_{i=0}^{j} det(_jS_i) x^i \in R[x]$$

从定义我们可以看出，对于每一个 $j$，$deg(S_j(A,B)) \leq j$。依据标准术语称法 [60]，如果 $(S_j(A,B) < j)$，那么称其为异常的，$(S_j(A,B) \geq j)$，称其为规范的。另外，如果 $_0S_0 = S$，那么 $S_0(A,B) = resultant(A,B)$。

**例 1.4.2.** 设在 $\mathbb{Z}[x]$ 中 $A = x^2 + 1$，$B = x^2 - 1$。$A$ 和 $B$ 的 Sylvester 矩阵为

$$S(A,B) = \begin{pmatrix} 1 & 0 & 1 & 0 \\ 0 & 1 & 0 & 1 \\ 1 & 0 & -1 & 0 \\ 0 & 1 & 0 & -1 \end{pmatrix}$$

由定义 1.4.2 可知子矩阵：$_0S =\,_0S_0 = S(A,B)$，$_1S = \begin{pmatrix} 1 & 0 & 1 & 0 \\ 1 & 0 & -1 & 0 \end{pmatrix}$，$_1S_0 = \begin{pmatrix} 1 & 1 \\ 1 & -1 \end{pmatrix}$ 和 $_1S_1 = \begin{pmatrix} 1 & 0 \\ 1 & 0 \end{pmatrix}$ 那么 $A$ 和 $B$ 的子结式为 $S_0 = det(_0S_0) = 4 = resultant(A,B)$，且异常的 $S_1 = det(_1S_0) + det(_1S_1)x = -2$。

子结式的另一个有用的性质是当次数不减少时，它们与环同态映射是可以交换的；当 $A$ 和 $B$ 中只有一个次数减少时，与环同态映射也有一个可知的对应关系：由任一个环同态映射：$\sigma : R \to S$ 可以得到一个多项式环的同态映射 $\overline{\sigma} : R[x] \to S[x]$，由下式表示：

$$\overline{\sigma}(\sum a_j x^j) = \sum \sigma(a_j) x^j \tag{1.8}$$

下面的定理描述了在 $A$ 和 $B$ 的最高次系数不会都被 $\sigma$ 映射为 0 的时候，如何由 $S_j(A,B)$ 计算出 $S_j(\overline{\sigma}(A), \overline{\sigma}(B))$。

**定理 1.4.3.** *([64] § 7.8) 设 $\sigma : R \to S$ 是一个环同态映射，$\overline{\sigma} : R[x] \to S[x]$ 由等式 (1.8) 给出，且 $A, B \in R[x] \setminus \{0\}$。如果 $deg(\overline{\sigma}(A)) = deg(A)$，那么*

$$\overline{\sigma}(S_j(A,B)) = \sigma(lc(A))^{deg(B)-deg(\overline{\sigma}(B))} S_j(\overline{\sigma}(A), \overline{\sigma}(B))$$

*其中，$0 \leq j < min(deg(A), deg(\overline{\sigma}(B)))$。*

特别提醒，当 $A$ 或 $B$ 是首一的，或者 $deg(A) = deg(\overline{\sigma}(A))$ 且 $deg(B) = deg(\overline{\sigma}(B))$ 的时候，$\overline{\sigma}(S_j(A,B)) = S_j(\overline{\sigma}(A), \overline{\sigma}(B))$。定理 1.4.3 可被用于特化环同态，如果 $R$ 是 $R = D[t_1, \ldots, t_n]$ 的形式，其中 $t_i$ 都是独立的不定元；$S$ 是包含 $D$ 的环，$\alpha_1, \ldots, \alpha_n$ 是 $S$ 中给定的元素，并且 $\sigma : R \to S$ 是一个环同态映射，它在 $D$ 上是恒等映射，且将 $t_i$ 映射为 $\alpha_i$。在这种情况下，定理 1.4.3 说明，要计算给定参数 $t_i$ 的子结式的值，可以转化为计算相应的两个最初多项式的子结式代入特定值的结果。

**例 1.4.3.** 设在 $\mathbb{Z}[t][x]$ 中，$A = 3tx^2 - t^3 - 4 \in R[x]$，$B = x^2 + t^3 x - 9$。$A$ 和 $B$ 的 Sylvester 矩阵为

$$S(A,B) = \begin{pmatrix} 3t & 0 & -t^3-4 & 0 \\ 0 & 3t & 0 & -t^3-4 \\ 1 & t_3 & -9 & 0 \\ 0 & 1 & t^3 & -9 \end{pmatrix}$$

并且由定义 1.4.2 可知子结式：${}_0S ={}_0 S_0 = S(A,B)$，${}_1S = \begin{pmatrix} 3t & 0 & -t^3-4 & 0 \\ 1 & t^3 & -9 & 0 \end{pmatrix}$，${}_1S_0 = \begin{pmatrix} 3t & -t^3-4 \\ 1 & -9 \end{pmatrix}$，和 ${}_1S_1 = \begin{pmatrix} 3t & 0 \\ 1 & t^3 \end{pmatrix}$。所以，$A$ 和 $B$ 的子结式是

$$S_0(A,B) = resultant_x(A,B) = det({}_0S_0) = -3t^{10}-12t^7+t^6-54t^4+8t^3+729t^2-216t+16$$

$$S_1(A,B) = det({}_1S_1)x + det({}_1S_0) = 3t^4x + t^3 - 27t + 4$$

现在考虑赋值映射，$t \to 1$，即同态映射 $\sigma : \mathbb{Z}[t] \to \mathbb{Z}$，给定 $\sigma(t) = 1$ 和 $\sigma(n) = n$，对于 $n \in \mathbb{Z}$。我们有 $\overline{\sigma}(A) = 3x^2 - 5$，和 $\overline{\sigma}(B) = x^2 + x - 9$，则由定理 1.4.3 可知 $S_0(\overline{\sigma}(A), \overline{\sigma}(B)) = resultant_x(3x^2-5, x^2+x-9) = \overline{\sigma}(S_0(A,B)) = 469$，$S_1(\overline{\sigma}(A), \overline{\sigma}(B)) = \overline{\sigma}(3t^4x + t^3 - 27t + 4) = 3x - 22$。

## 1.5 多项式剩余序列

现在我们引入多项式剩余序列这个概念，它是计算最大公因子和结式的欧几里得算法的概括。在这一节中，设 $D$ 是一个整环且 $x$ 是其中的变量。

**定义 1.5.1.** 设 $A, B \in D[x]$，$B \neq 0$ 且 $\deg(A) \geq \deg(B)$。$A$ 与 $B$ 的多项式剩余序列是 $D[x]$ 中的一个序列 $(R_i)_{i \geq 0}$ 满足

**(i)** $R_0 = A, R_1 = B$

**(ii)** 对 $i \geq 1$，

$$\beta_i R_{i+1} = \begin{cases} 0 & \text{当} R_i = 0 \\ prem(R_{i-1}, R_i) & \text{当} R_i \neq 0 \end{cases}$$

其中 $(\beta_i)_{i \geq 0}$ 是 $D$ 中的非零序列。

从定义中可以清楚地看到对于任意 $i \geq 1$，或者 $R_{i+1} = 0$ 或者 $\deg(R_{i+1}) < \deg(R_i)$，因此

**(i)** 一个多项式剩余序列只有有限个非零元素。

**(ii)** 如果 $R_i \neq 0, R_j \neq 0, \deg(R_i) = \deg(R_j)$ 且 $i, j \geq 1$，那么 $i = j$（即只有 $R_0$ 和 $R_1$ 才可能次数相同）。

**定义 1.5.2.** 设 $A, B \in D[x]$。称 $A$ 相似于 $B$，如果存在 $a, b \in D \setminus \{0\}$ 使得 $aA = bB$。

从多项式剩余序列的定义中我们可以看到，不同的 $\beta_i$ 序列会产生不同类型的多项式剩余序列。例如，令 $\beta_i = 1$ 得到的多项式剩余序列就是 $A$ 与 $B$ 的连续伪余项序列，它被称为 $A$ 与 $B$ 的欧几里得多项式剩余序列。令 $\beta_i$ 为 $D$ 中 $prem(R_{i-1}, R_i)$ 系数的最大公因子，所得到的多项式剩余序列称为 $A$ 与 $B$ 的原始多项式剩余序列。有一个重要结论是，如果 $D$ 是唯一分解整环，那么多项式剩余序列中最后一个非零元素相似于 $A$ 与 $B$ 的最大公因子。

**定理 1.5.1.** 设 $D$ 为唯一分解整环，$A, B \in D[x]$，$B \neq 0$ 且 $deg(A) \geq deg(B)$。令 $(R_0, R_1, \ldots, R_k, 0, \ldots)$ 是 $A$ 与 $B$ 的多项式剩余序列，其中 $R_k \neq 0$。那么对 $0 \leq i, j \leq k$，有 $gcd(R_i, R_{i+1})$ 相似于 $gcd(R_j, R_{j+1})$。特别地 ($i = 0, j = k$ 时)，$R_k$ 相似于 $gcd(A, B)$。

*定理1.5.1的证明.* 设 $0 \leq i < k$，$G = \gcd(R_i, R_{i+1})$，$H = \gcd(R_{i+1}, R_{i+2})$。因为 $i < k$ 且 $R_{i+1} \neq 0$，所以由多项式剩余序列和伪余数的定义知，存在 $\alpha, \beta \in D \setminus 0$ 和 $Q \in D[x]$ 使得

$$\alpha R_i = R_{i+1} Q + \beta R_{i+2}.$$

因此 $H \mid \alpha R_i$，又 $H \mid \alpha R_{i+1}$，那么 $\alpha G$ 是 $\alpha R_i$ 和 $\alpha R_{i+1}$ 的最大公因式，所以 $H \mid \alpha G$。从上面的等式我们得到 $G \mid \beta R_{i+2}$。又 $G \mid \beta R_{i+1}$ 所以 $G \mid \beta H$。所以存在 $Q_1, Q_2 \in D[x]$ 使得 $\alpha G = H Q_1$，$\beta H = G Q_2$。由此可得 $\alpha \beta G = G Q_1 Q_2$，所以 $Q_1, Q_2 \in D$，因此 $G$ 相似于 $H$。因此，当 $j = i + 1$ 时定理成立。由于相似性是可传递的，那么定理对 $0 \leq i < j \leq k$ 都成立。由于相似性是对称的，那么定理对 $0 \leq i \neq j \leq k$ 都成立。$i = j$ 时显然成立。所以定理对 $0 \leq i, j \leq k$ 成立。 $\square$

所以，$A$ 与 $B$ 的任何多项式剩余序列都包含 $\gcd(A, B)$。同时，所有 $A$ 与 $B$ 的非零子结式都相似于序列中的某一元素。下面的多项式剩余序列基本定理给出了相似系数的明确公式。

**定理 1.5.2.** 设 $A, B \in D[x]$，$B \neq 0$ 且 $deg(A) \geq deg(B)$。令 $(R_0, R_1, \ldots, R_k, 0, \ldots)$ 是 $A$ 与 $B$ 的多项式剩余序列，其中 $R_k \neq 0$。令 $i = 1, \ldots k$，设 $n_i = \deg(A) \geq \deg(B)$，$r_i$ 是 $R_i$ 的首项系数。那么，对于任意的 $j \in \{0, \ldots, \deg(B) - 1\}$，

$$S_j(A, B) = \begin{cases} \eta_i R_i & \text{如果 } j = n_{i-1} - 1 \\ \tau_i R_i & \text{如果 } j = n_i \\ 0 & \text{其他情况} \end{cases}$$

其中

$$\eta_i = (-1)^{\phi_i} r_{i-1}^{1 - n_{i-1} + n_i} \prod_{j=1}^{i-1} \left[ \left( \frac{\beta_j}{r_j^{1 + n_{j-1} - n_j}} \right)^{1 + n_j - n_{i-1}} r_j^{n_{j-1} - n_{j+1}} \right]$$

$$\tau_i = (-1)^{\sigma_i} r_i^{n_{i-1} - n_i - 1} \prod_{j=1}^{i-1} \left[ \left( \frac{\beta_j}{r_j^{1 + n_{j-1} - n_j}} \right)^{n_j - n_i} r_j^{n_{j-1} - n_{j+1}} \right] \quad (1.9)$$

$$\phi_i = \sum_{j=1}^{i-1} (n_j - n_{i-1} + 1)(n_{j-1} - n_{i-1} + 1), \sigma_i = \sum_{j=1}^{i-1} (n_{j-1} - n_i)(n_j - n_i) \quad (1.10)$$

$A$ 与 $B$ 的子结式多项式剩余序列是一个特殊的多项式剩余序列，它是由 Collins 和 Brown 提出的，即定理 1.5.2 中 $\eta_i = 1$ 的情况。它是通过如下对 $\beta_i$ 的递归得到的：

$$R_0 = A, R_1 = B, \gamma_1 = -1, \beta_1 = (-1)^{\delta_1 + 1}$$

且

$$\begin{cases} \gamma_{i+1} = (-\text{lc}(R_i))^{\delta_i} \gamma_i^{1-\delta_i} \\ \beta_{i+1} = -\text{lc}(R_i) \gamma_{i+1}^{\delta_{i+1}} \end{cases}$$

对于任意的 $i \geq 1$ 有 $\delta_i = deg(R_{i-1}) - deg(R_i)$。它的重要性质有下面的定理给出。

**定理 1.5.3.** 设 $A, B \in D[x]$，且 $deg(A) \geq deg(B)$，$(R_0, R_1, R_2, \ldots, R_k, 0, \ldots)$ 是 $A$ 与 $B$ 的 PRS 的子结式，其中 $R_k \neq 0$，且 $n_i = deg(R_i)$，$i = 1, \ldots, k$。那么，

$$\forall j \in \{0, \ldots, deg(B) - 1\},$$

$$S_j(A, B) = \begin{cases} R_i & \text{当} j = n_{i-1} - 1 \\ \tau_i R_i & \text{当} j = n_i \\ 0 & \text{其他} \end{cases}$$

其中 $\tau_i$ 由 1.9 给出。

这个定理产生了计算 $A$ 与 $B$ 结式的所谓的子结式算法：如果 $deg(A) \geq deg(B)$，那么由定义，$A$ 与 $B$ 的子结式为 $S_0(A, B)$，因此我们可以计算 $A$ 与 $B$ 的 PRS 子结式。如果 $deg(R_k) > 0$，那么 $A$ 与 $B$ 含有公因子，所以 $A$ 与 $B$ 的子结式为 0。否则，由定理 1.5.3 可知 $S_0(A, B)$ 要么等于 $R_k$，如果 $deg(R_{k-1}) = 1$；要么等于 $\tau_k R_k$，如果 $deg(R_{k-1}) > 1$。在后一种情况，由于 $n_k = 0$，$\tau_k$ 的计算可以简化：式 1.10 变为 $\sigma_k = \sum_{j=1}^{k-1} n_{j-1} n_j$，所以 $(-1)^{\sigma_k} = \Pi_{j=1}^{k-1}(-1)^{n_{j-1}n_j}$。当 $n_{j-1}$ 和 $n_j$ 都是奇数时，系数 $-1$ 会出现在结果中。进一步，由于 $\deg(R_k) = 0$，$r_k = R_k$ 和 1.9 变为

$$\tau_k = (-1)^{\sigma_k} R_k^{n_{k-1}-1} \prod_{j=1}^{k-1} \left[ \left( \frac{\beta_j}{r_j^{1+n_{j-1}-n_j}} \right)^{n_j} r_j^{n_{j-1}-n_j+1} \right]$$

如果 $\deg(A) < \deg(B)$，我们可以计算出 $B$ 和 $A$ 的子结式 PRS，以及由定理 1.4.1 可知 $resultant(A, B) = (-1)^{deg(A) deg(B)} resultant(B, A)$。

**SubResultant**$(A, B)$    (* 子结式算法 *)
(* 给定整环 $D$ 和 $A, B \in D[x]$ 且 $B \neq 0$, $deg(A) > deg(B)$, 返回 resultant$(A, B)$ 和 $A$ 和 $B$ 子结式 PRS $(R_0, R_1, \ldots, R_k, 0)$ *)
  $R_0 \leftarrow A, R_1 \leftarrow B$
  $i \leftarrow 1, \gamma_1 \leftarrow -1$
  $\delta_1 \leftarrow deg(A) - deg(B)$
  $\beta_1 \leftarrow (-1)^{\delta_1+1}$
  **while** $R_i \neq 0$ **do**
    $r_i \leftarrow lc(R_i)$
    $(Q, R) \leftarrow$ **PolyPseudoDivide**$(R_{i-1}, R_i)$
    $R_{i+1} \leftarrow R/\beta_i$ (* 这一除法总能整除 *)
    $i \leftarrow i+1$
    $\gamma_i \leftarrow (-r_{i-1})^{\delta_{i-1}} \gamma_{i-1}^{1-\delta_{i-1}}$
    $\delta_i \leftarrow deg(R_{i-1}) - deg(R_i)$
    $\beta_i \leftarrow -r_{i-1}\gamma_i^{\delta_i}$
  **end while**
  $k \leftarrow i-1$
  **if** $deg(R_k) > 0$ **then**
    **return**$(0, (R_0, R_1, \ldots, R_k, 0))$
  **end if**
  **if** $deg(R_{k-1}) = 1$ **then**
    **return**$(R_k, (R_0, R_1, \ldots, R_k, 0))$
  **end if**
  $s \leftarrow 1, c \leftarrow 1$
  **for** $j \leftarrow 1$ to $k-1$ **do**
    **if** $deg(R_{j-1})$ 是奇数且 $deg(R_j)$ 是奇数 **then**
      $s \leftarrow s$
    **end if**
    $c \leftarrow c \left(\beta_j/r_j^{1+\delta_j}\right)^{deg(R_j)} r_j^{deg(R_{j-1})-deg(R_{j+1})}$
  **end for**
  **return** $(scR_k^{deg(R_{k-1})}, (R_0, R_1, \ldots, R_k, 0))$

**例 1.5.1.** 下面是 $A = x^2 + 1$ 和 $B = x^2 - 1 \in \mathbb{Z}[x]$ 的子结式算法过程:

| $i$ | $R_i$ | $\gamma_i$ | $\delta_i$ | $\beta_i$ | $r_i$ | $r_i^{1+\delta_i}$ |
|---|---|---|---|---|---|---|
| 0 | $x^2+1$ | | | | 1 | |
| 1 | $x^2-1$ | $-1$ | 0 | $-1$ | 1 | 1 |
| 2 | $-2$ | $-1$ | 2 | $-1$ | $-2$ | $-8$ |
| 3 | 0 | | | | | |

我们得到 $k = 2$, $deg(R_2) = 0$ 和 $deg(R_1) = 2$, 所以我们计算 $s$ 和 $c$:

| $j$ | $deg(R_{j-1})$ | $deg(R_j)$ | $s$ | $c$ |
|---|---|---|---|---|
| 1 | 2 | 2 | 1 | 1 |
| 2 | 2 | 0 | 1 | 1 |

所以 $R = sc R_2^2 = 4 = resultant(x^2 + 1, x^2 - 1)$。

**例 1.5.2.** 下面是 $A = 3tx^2 - t^3 - 4$ 和 $B = x^2 + t^3 x - 9 \in D[x]$ 且 $D = \mathbb{Z}[t]$ 的子结式算法过程：

| $i$ | $R_i$ | $\gamma_i$ | $\delta_i$ | $\beta_i$ | $r_i$ | $r_i^{1+\delta_i}$ |
|---|---|---|---|---|---|---|
| 0 | $A$ | | | | $3t$ | |
| 1 | $B$ | $-1$ | $0$ | $-1$ | $1$ | $1$ |
| 2 | $3t^4 x + t^3 - 27t + 4$ | $-1$ | $1$ | $1$ | $3t^4$ | $9t^8$ |
| 3 | $R$ | $-3t^4$ | $1$ | $9t^8$ | $R$ | $R^2$ |
| 4 | $0$ | | | | | |

其中 $R = -3t^{10} - 12t^7 + t^6 - 54t^4 + 8t^3 + 729t^2 - 216t + 16 \in D$。我们得到 $k = 3$，$deg(R_3) = 0$ 和 $deg(R_2) = 1$，所以 $R = resultant_x(A, B)$，如例 1.4.3。

## 1.6 本原多项式

设 $D$ 为唯一分解整环，$x$ 为 $D$ 上的一个变量。那么由定理 1.1.3，最大公因式总是存在的。在这一部分，我们研究 $D[x]$ 中元素的系数的最大公因数的性质。

**定义 1.6.1.** 设 $A = \Sigma_{i=0}^n a_i x^i \in D[x] \setminus 0$。$A$ 的容量是

$$\text{content}(A) = \gcd(a_0, \ldots, a_n) \in D$$

我们也称 $A$ 是本原的，如果 $\text{content}(A) \in D*$。最终，$A$ 的本原部分为

$$\text{pp}(A) = \frac{A}{\text{content}(A)} \in D[x]$$

按照惯例，$\text{content}(0) = pp(0) = 0$ 且 $0$ 不是本原的。

注意，像最大公因子一样，容量和本原的部分都是通过乘以一个单位来定义的。在这本书中我们假设一个单位是前后一致的使得 $A = \text{content}(A) pp(A)$，对任意 $A \in D[x]$。同时，本原性依赖于环 $D$。当 $D$ 扩展为更大的唯一分解整环时，非本原多项式可以成为本原的：作为 $\mathbb{Z}[x]$ 的元素，$4x + 6$ 不是本原的，但作为 $\mathbb{Q}[x]$ 的元素是本原的。实际上，如果 $D$ 是一个域，那么每一个非零多项式都是本原的。设 $P \in D[x] \setminus D$ 为不可约的。由于 $P = content(P) pp(P)$ 和 $pp(P)$ 不是一个单位，因而 $content(P)$ 一定是一个单位，因此 $P$ 是本原的。

容量的基本性质是他们是可乘的。这一经典的结果来自于高斯，被人熟知为高斯引理：

**引理 1.6.1.**
$$\text{content}(AB) = \text{content}(A)\text{content}(B) \quad A, B \in D[x]$$

因此，本原多项式的乘积也是本原的。这对 $D[x]$ 中质因式分解的首项系数有所影响：设 $A \in D[x]$ 为非零元，$A = u \Pi_{j=1}^m p_j^{d_j} \Pi_{i=1}^n P_i^{e_i}$ 是它的质因式分解，其中 $u \in D^*$，每一个 $p_j$ 在 $D$ 中都不可约，每一个 $P_i$ 在 $D[X] \setminus D$ 上都不可约。每一

1 代数基础

$P_i$ 如上所述都是本原的，因此 $\Pi_i = 1^n P_i^{e_i}$ 是本原的，所以由引理 1.6.1，对于一些 $v \in D*$，$\text{content}(A) = uv\Pi_{j=1}^m p_j^{d_j}$。如果 $A$ 是本原的，由 $D$ 上质因式分解的唯一性可知 $m = 0$，所以我们可以为容量选择合适的单位使得 $\text{pp}(A)$ 的质因式分解有如下形式：$\text{pp}(A) = \Pi_{i=1}^n P_i e_i$，其中 $P_i$ 是互质的，且 $\deg(P_i) > 0$。在下面的定义中以及当我们在积分算法中使用本原部分的时候，我们利用这一事实。

**定义 1.6.2.** 设 $A \in D[x]$ 且 $\text{pp}(A) = \Pi_{i=1}^n P_i^{e_i}$ 是它本原部分的质因式分解，其中对任意 $i$，$e_i \geq 1$。我们定义 $A$ 的无平方部分为

$$A^* = \prod_{i=1}^n P_i$$

且对 $k \in \mathbb{Z}, k \geq 0$，$A$ 的 $k-$ 降阶为

$$A^{-k} = \prod_{i=1}^n P_i^{\max(0, e_i - k)} = \prod_{i|e_i > k} P_i^{e_i - k}$$

注意 $A^{-0} = \text{pp}(A)$。为了方便我们简称 $A^{-1}$ 为 $A$ 的降阶，且记为 $A^-$，即

$$A^- = A^{-1} = \prod_{i=1}^n P_i^{e_i - 1}$$

作为定义的结论我们得到如下有用的关系：

$$A^* A^- = \text{pp}(A) \tag{1.11}$$

$$A^{-k} = A^{-i-j} \ \ i, j \geq 0 \ i + j = k$$

上面的关系中一个特殊的情况是：

$$A^{-k+1} = A^{-k^-} \tag{1.12}$$

上式与 1.11 共同得到

$$A^{-k+1} = \frac{A^{-k}}{A^{-k*}} \tag{1.13}$$

尽管无平方部分与降阶是根据质因式分解定义的，它们可以通过在 $D[x]$ 最大公因子的计算来得到。最基本的想法是 $A$ 的一个质因式除以 $dA/dx$ 比 $A$ 少一次。

**定理 1.6.1.** 设 $A, P \in D[x] \setminus D$ 且 $n > 0$ 为整数。于是

(i) $P^{n+1}|A \Rightarrow P^n|\gcd(A, dA/dx)$，

(ii) 如果 $P$ 是不可分解的，且 $\text{char}(D) = 0$，那么 $P^n|\gcd(A, dA/dx) \Rightarrow P^{n+1}|A$。

*定理 1.6.1的证明.* (i) 假设 $P^{n+1}|A$，那么存在 $B \in D[x]$ 使得 $A = P^{n+1}B$。因此，

$$\frac{dA}{dx} = P^{n+1}\frac{dB}{dx} + (n+1)P^n B \frac{dP}{dx}$$

所以 $P^n|dA/dx$，由此可得 $P^n|\gcd(A, dA/dx)$。(ii) 假设 $D$ 有特征 $0$，$P$ 是不可分解的，且 $P^n|\gcd(A, dA/dx)$。设 $m > 0$ 是一个唯一的整数使得 $P^m|A$ 且 $P_{m+1} \nmid A$。那么存在 $B \in D[x]$ 使得 $A = P^m B$ 且 $P \nmid B$。正如 (i)，我们有

$$\frac{dA}{dx} = P^m \frac{dB}{dx} + mP^{m-1} B \frac{dP}{dx}$$

我们有 $m \geq n$ 由于 $P^n|A$。假设 $m = n$。那么，

$$\frac{dA}{dx} - P^n \frac{dB}{dx} = nP^{n-1} B \frac{dP}{dx}$$

我们有 $P^n|dA/dx$，所以 $P^n|nP^{n-1}B(dP/dx)$，因此 $P|nB(dP/dx)$。但是 $P$ 是不可分解的并且 $P \nmid B$，所以 $P|n(dP/dx)$。在特征 $0$，$n(dP/dx)$ 是非零的并且含有比 $P$ 更小的阶，所以 $P \nmid n(dP/dx)$。因此 $m \neq n$，所以 $m > n$，因此 $P^{n+1}|A$。 □

由定理 1.6.1得到的直接结论是当 $D$ 含有特征 $0$ 时，

$$A^{-} = \gcd\left(A, \frac{dA}{dx}\right) \tag{1.14}$$

对任意本原的 $A$，并且 $A^*$ 可以通过 1.11计算得到。$A$ 的进一步的降阶可以通过 1.12递归得到。无平方部分和降阶因此比质因式分解更容易计算。我们将在下一节介绍无平方分解的概念时运用这种方法。

## 1.7 无平方因子分解

令 $D$ 是唯一分解整环，且 $x$ 是 $D$ 上的不定元。从而 $D[x]$ 是一个唯一分解整环，即每个 $A \in D[x]$ 有分解到不可约元素的因子分解。总体上来说这样的因子分解是很难计算的，但是有另外的因子分解在很多场合可以作为替代使用，并且更易于计算。在这一节我们将介绍无平方因子分解，它主要被用在积分算法中。

**定义 1.7.1.** 称 $A \in D[x]$ 是无双的，如果不存在 $B \in D[x] \setminus D$ 使得在 $D[x]$ 中 $B^2|A$。

等价的，$A$ 是无双的若 A 在 D 上的任何素因子分解有 $e_i = 1$，$i = 1, \ldots, n$。

**定义 1.7.2.** 令 $A \in D[x]$。$A$ 的一个无双因子分解是形如 $A = \prod_{i=1}^{m} A_i^i$ 的一个因子分解，其中每个 $A_i$ 是无双的并且对 $i \neq j$ 有 $\gcd(A_i, A_j) \in D$。

注意由定义 $D$ 中元素很自然是无双的，从而没有必要像素因子分解那样要求一个 $D*$ 中分离的首项系数和 $D$ 中的素因子。更进一步，如果 $A$ 本原部分的无双因子分解形如 $pp(A) = \prod_{i=1}^{m} A_i^i$，那么

$$A = (content(A)A_1)\prod_{i=2}^{m} A_i^i$$

1 代数基础是 A 的一个无双因子分解，如此计算本原部分的无双因子分解足够了。下面，我们认为 $D$ 的特征为 0（对于正特征的无双因子分解请参见………）。在特征为 0 的情况下，因为 $A$ 的一零点必然为某一确定 $A_i$ 的一零点，并且其在 $A$ 中的阶数为 $i$，所以 $A$ 的无双因子分解将 $A$ 的零点按阶数分离。我们用这个事实的目的是依据 $A$ 的缩减（？）表达这些 $A_i$，反之亦然。

**引理 1.7.1.** 令 $A \in D[x] \setminus D$，$pp(A) = \prod_{i=1}^{n} P_i^{e_i}$ 是 $(ppA)$ 的的素因子分解，$m = max(e_1,\ldots,e_m)$ 并且对 $1 \leq i \leq m$ 有 $A_i = \prod_{j|e_j=i} P_j$。那么，

**(i)** 对任何整数 $k \geq 0$ 有 $A^{-k} = \prod_{i=k+1}^{m} A_i^{i-k} = A_{k+1} A_{k+2}^2 \cdots A_m^{m-k}$。

**(ii)**
$$\text{对} 1 \leq i \leq m \text{有} A_i = \frac{A^{-i-1*}}{A^{-i*}}。 \tag{1.15}$$

**(iii)** $pp(A) = \prod_{i=1}^{m} A_i^i$ 是 $pp(A)$ 的一个无双因子分解。

由于缩减和无平方部分都可通过前面小节所阐述的 gcd 来计算，我们得到了对本原的 $A$ 计算无平方因子分解的算法如下：由(1.14)，我们有 $A^{-1} = A^- = \gcd(A, dA/dx)$，从而 $A^{-0*} = A^* = pp(A)/A^-$。一旦对 $k \geq 0$ 我们得到了 $A^{-k*}$ 和 $A^{-k+1}$，则序列接下来是

$$\gcd(A^{-k*}, A^{-k+1}) = \gcd(A_{k+1} \cdots A_m, A_{k+2} A_{k+3}^2 \cdots A_m^{m-k-1}) = A^{-k+1*},$$

并且 $A_{k+1}$ 和 $A^{-k+2}$ 分别通过(1.15)和(1.13)得到。我们延续该序列直到 $A^{-k+1} \in D$，也就是说 $A^{-k}$ 是无平方的，此情形下有 $k = m-1$ 和 $A_m = A^{-k}$。这个无平方因子分解算法仅用到了有理数操作以及在 $D[x]$ 中的 gcd 计算。

---

**Squarefree**($A$)    (\* Musser 无平方因子分解 \*)
(\* 给出特征为 0 的唯一析因整环 $D$ 和 $A \in D[x]$，返回 $A_1,\ldots,A_m \in D$，使得 $A = \prod_{k=1}^{m} A_k^k$ 是 A 的无平方因子分解。\*)
  $c \leftarrow \text{content}(A), S \leftarrow A/c$                                                   (\* $S = pp(A)$ \*)
  $S^- \leftarrow \gcd(S, dS/dx)$
  $S^* \leftarrow S/S^-$
  $k \leftarrow 1$
  **while** $\deg(S^-) > 0$ **do**                   (\* $S^- = A^{-k}, S^* = A^{-k-1}$ \*)
    $Y \leftarrow \gcd(S^*, S^-)$                                       (\* $Y = A^{-k*}$ \*)
    $A_k \leftarrow S^*/Y$                                 (\* $A_k = A^{-k-1*}/A^{-k*}$ \*)
    $S^* \leftarrow Y$                                                  (\* $S^* = A^{-k*}$ \*)
    $S^- \leftarrow S^-/Y$                                          (\* $S^- = A^{-k+1}$ \*)
    $k \leftarrow k+1$
  **end while**
  $A_k \leftarrow S^*$
  **return**$((cS^-)A_1,\ldots,A_k)$

**例 1.7.1.** 对 $A = x^8 + 6x^6 + 12x^4 + 8x^2 \in \mathbb{Q}[x]$ 使用 **Squarefree** 算法，我们有 $c = 1, S = A, dS/dx = 8x^7 + 36x^5 + 48x^3 + 16x$,

$$S^- = \gcd(S, \frac{dS}{dx}) = x^5 + 4x^3 + 4x$$

并且 $S^* = S/S^- = x^3 + 2x$。接下来，

| $k$ | $S^*$ | $S^-$ | $Y$ | $A_j$ |
|---|---|---|---|---|
| 1 | $x^3 + 2x$ | $x^5 + 4x^3 + 4x$ | $x^3 + 2x$ | 1 |
| 2 | $x^3 + 2x$ | $x^2 + 2$ | $x^2 + 2$ | $x$ |
| 3 | $x^2 + 2$ | 1 | | $x^2 + 2$ |

从而有:
$$A = x^8 + 6x^6 + 12x^4 + 8x^2 = x^2 + (x^2 + 2)^3$$

在 Yun[95] 中提到，可以通过减少循环中 gcd 出现的多项式次数来提高效率。他的想法来源于下面的多项式序列

$$Y_k = \sum_{i=k}^m (i-k+1)\frac{dA_i}{dx}\frac{A^{-k-1*}}{A_i} = \sum_{i=k}^m (i-k+1) A_k \cdots A_{i-1}\frac{dA_i}{dx} A_{i+1} \cdots A_m \quad \text{对} k \geq 1 \quad (1.16)$$

下面的引理终结了序列的性质。

**引理 1.7.2.** 记号同前，

$$\gcd(A^{-i-1*}, Y_i) \in D, dA^{-i-1}/dx = A^{-i}Y_i, \quad (1.17)$$

并且对引理 2.2 中所定义的 $A_i$，有

$$Y_i - \frac{dA^{-i-1*}}{dx} = A_i Y_{i+1} \quad (1.18)$$

对 $1 \leq i \leq m$ 成立。

*引理 1.7.2.* 令 $1 \leq i \leq j \leq m$。那么，由于 $A_j$ 是无平方的，并且所有 $A_i$ 两两互素，有

$$\gcd(A_j, A_i \cdots A_{j-1}\frac{dA_j}{dx} A_{j+1 \cdots A_m}) \in D$$

其中 $A_i$ 两两互素是因为

$$j \neq k \ A_j \mid A_i \cdots A_{k-1}\frac{dA_k}{dx} A_{k+1} \cdots A_m$$

这推出 $\gcd(A_j, Y_i) \in D$，从而 $\gcd(A^{-i-1*}, Y_i) \in D$。
令 $1 \leq i \leq m$。用引理 2.2 和 (1.11) 我们有

$$\frac{dA^{-i-1}}{dx} = \frac{d}{dx}(\prod_{j=i}^m A_j^{j-i+1}) = \sum_{j=i}^m (j-i+1)\frac{dA_j}{dx}\frac{A^{-i-1}}{A_j}$$

$$\sum_{j=i}^{m}(j-i+1)\frac{dA_j}{dx}\frac{A^{-i-1^{-}}A^{-i-1^{*}}}{A_j}$$

$$=A^{-i-1^{-}}\sum_{j=i}^{m}(j-i+1)\frac{dA_j}{dx}\frac{A^{-i-1^{*}}}{A_j}=A^{-i-1^{-}}Y_i$$

从

$$\frac{dA^{-k-1^{*}}}{dx}=\frac{d}{dx}(\prod_{j=k}^{m}A_j)=\sum_{j=k}^{m}\frac{dA_j}{dx}\frac{A^{-k-1^{*}}}{A_j}$$

得

$$Y_i-\frac{dA^{-i-1^{*}}}{dx}=\sum_{j=i}^{m}(j-i+1)\frac{dA_j}{dx}\frac{A^{-i-1^{*}}}{A_j}-\sum_{j=k}^{m}\frac{dA_j}{dx}\frac{A^{-k-1^{*}}}{A_j}$$

$$=\sum_{j=i+1}^{m}(j-i)\frac{dA_j}{dx}\frac{A^{-i-1^{*}}}{A_j}$$

$$=A_i\sum_{j=i+1}^{m}(j-i)\frac{dA_j}{dx}\frac{A^{-i^{*}}}{A_j}=A_iY_{i+1}$$

$\square$

由于 $A^{-i-1^{*}}=A_iA^{-i^{*}}$ 并且 $\gcd(A^{-i^{*}},Y_{i+1})\in D$，我们从(1.18)得出

$$\gcd(A^{-i-1^{*}},Y_i-\frac{dA^{-i-1^{*}}}{dx})=A_i \qquad (1.19)$$

从而导出 Yun 的无平方因子分解算法：如前假设 $A$ 是本原的，我们有 $A^{-}=\gcd(A,dA/dx)$，进而由(1.17)得

$$A^{-0^{*}}=A^{*}=\text{pp}(A)/A^{-} \text{ 与} Y_1=\frac{dA/dx}{A^{-}}$$

一旦我们得到了 $A^{-k-1^{*}}$ 和 $Y_k$，$A_k$ 可以通过(1.19)来计算，并且 $Y_{k+1}$ 和 $A^{-k^{*}}$ 可以分别通过(1.18)和(1.15)得到。这样延续序列直到 $Y_k=dA^{-k-1^{*}}/dx$，也即 $A^{-k-1^{*}}$ 是无平方因子的，并且此时 $k=m$，$A_k=A^{-k-1}=A^{-k-1^{*}}$。这个无平方因子分解算法和前一个的区别在于，这里在主 gcd 计算中 $Y_k-dA^{-k-1^{*}}/dx$ 代替了 $A^{-k}$。

```
Squarefree(A)      (* Yun 无平方因子分解 *)
(* 给出特征为 0 的唯一析因整环 D 和 A ∈ D[x]，返回 A_1,…,A_m ∈ D，使得
A = ∏_{k=1}^m A_k^k 是 A 的无平方因子分解。*)
  c ← content(A), S ← A/c                              (* S = pp(A) *)
  S' ← dS/dx
  S⁻ ← gcd(S, S')
  S* ← S/S⁻
  Y ← S'/S⁻
  k ← 1
  while Z ← Y − dS*/dx ≠ 0 do            (* S⁻ = A^{-k-1}, Y ← Y_k *)
    A_k ← gcd(S*, Z)                                       (* (1.19) *)
    S* ← S*/A_k                                         (* S* = A^{-k*} *)
    Y ← Z/A_k                                            (* Y = Y_{k+1} *)
    k ← k + 1
  end while
  A_k ← S*
  return(cA_1,…,A_k)
```

**例 1.7.2.** 这里给出 Yun 算法对例 1.7.1中 $A$ 的逐步运行过程。我们首先得到 $c = 1, S = A, S' = dS/dx = 8x^7 + 36x^5 + 48x^3 + 16x$，以及 $S^- = \gcd(S, S') = x^5 + 4x^3 + 4x$。接下来，

| $k$ | $S^*$ | $Y$ | $Z$ | $A_j$ |
|---|---|---|---|---|
| 1 | $x^3 + 2x$ | $8x^2 + 4$ | $5x^2 + 2$ | 1 |
| 2 | $x^3 + 2x$ | $5x^2 + 2$ | $2x^2$ | $x$ |
| 3 | $x^2 + 2$ | $2x$ | 0 | $x^2 + 2$ |

因此有，
$$A = x^8 + 6x^6 + 12x^4 + 8x^2 = x^2(x^2 + 2)^3$$

循环中 gcd 计算得第二参数在表格中为 $Z$ 列，它们的度小于例 1.7.1中相应的 $S^-$ 列的度。

## 1.8 习题

**习题 1.8.1.** 用欧几里得辗转相除法计算 217 和 413 在整数中的最大公因子。

**习题 1.8.2.** 找出符合下列方程的 $x$ 和 $y$:

(a) $12x + 19y = 1$.

(b) $3x + 2y = 5$.

**习题 1.8.3.** 找出 14 在 $\mathbb{Z}_{37}$ 中的逆元。

**习题 1.8.4.** 找出 $2x^3 - \frac{19}{5}x^2 - x + \frac{6}{5}$ 和 $x^2 + \frac{1}{3} - \frac{14}{3}$ 在 $\mathbb{Q}[x]$ 中的最大公因子。

1 代数基础

**习题 1.8.5.** 计算在 $\mathbb{Z}[x]$ 中 $x^4 - 7x + 7$ 除以 $3x^2 - 7$ 得到的伪商和伪余数。

**习题 1.8.6.** 分别计算出 $7x^5 + 4x^3 + 2x + 1$ 除以 $2x^3 + 3$ 在 $\mathbb{Z}_5[x]$，$\mathbb{Z}_{11}$，$\mathbb{Z}[x]$ 和 $\mathbb{Q}$ 中的商和余数（或伪商和伪余数）。考虑在每种情况下分别在什么结构下进行。

**习题 1.8.7.** 计算 $x^4 + x^3 - t$ 和 $x^3 + 2x^2 + 3tx - t - 1$ 在 $\mathbb{Z}[t][x]$ 中的本原 PRS（多项式剩余子列）和 PRS 的子结式。

**习题 1.8.8.** 分别计算出在下列代数结构中 $4x^4 + 13x^3 + 15x^2 + 7x + 1$ 和 $2x^3 + x^2 - 4x - 3$ 的最大公因子: $a)\mathbb{Q}[x]$；$b)\mathbb{Z}[x]$。

**习题 1.8.9.** 找出 $x^8 - 5x^6 + 6x^4 + 4x^2 - 8$ 的一种无平方因子分解。

**习题 1.8.10.** 证明 2 是不可约的，但在 $\mathbb{Z}[\sqrt{-5}]$ 并不是素数。

**习题 1.8.11.** 证明定义 1.5.2 中的相似性是一个等价关系。

**习题 1.8.12.** 证明: $a$，$b$ 在一个欧几里得整环 $D$ 中，如果对于 $D$ 中 $q$，$r$，$a = qb + r$，那么 $gcd(a,b) = gcd(b,r)$。

**习题 1.8.13.** 使用拓展欧几里得算法和定理 1.4.1 证明定理 1.4.2。

**习题 1.8.14.** 使用循环不变量证明拓展欧几里得算法是正确的。

**习题 1.8.15.** 设 $D$ 是一个唯一分解整环，$F$ 是它的一个商域（见例 1.1.14），$x$ 是 $D$ 中一个不定元。证明引理 1.6.1 的一个结论: 对于 $D[x]$ 中的任意 $A$，$B$，其中 $A$ 是本原多项式，那么 $A$ 在 $D[x]$ 中整除 $B$ 当且仅当 $A$ 在 $F[x]$ 中整除 $B$。

**习题 1.8.16.** 设 $R$ 是一个整环，$x_1, \ldots, x_n \in R$。如果对于 $1 \leq i \leq n$，$x_i | z$，并且

$$\forall t \in R, x_i | t, 1 \leq i \leq n \longrightarrow z | t.$$

证明唯一分解整环 $R$ 中的任意两个元素 $x$ 和 $y$ 都有一个最小公倍数。

# 第 2 章 有理函数积分

在这一章中我们介绍有理函数积分的算法。由于有理函数几乎都有最基本的部分，因而这种情况是最简单的；并且它也是很重要的，因为更加复杂的函数积分的算法是有理函数算法中运用的技巧的延伸。由于这一章中的算法和定理是 Risch 算法的特殊情形，我们先不证这些算法的正确性，证明将放在后面讨论超越函数积分的章节中。在这一章中，我们假定 $K$ 是一个域，其特征为 $0$，$x$ 是其上的变元，并且 $'$ 为 $K(x)$ 上 $d/dx$ 的符号，所以 $x$ 是积分变量。我们称两个关于 $x$ 的多项式的商为一个关于 $x$ 的有理函数，同时允许含有其他不含 $x$ 的表达式在其中。例如，$\log(y)/(x-e-\pi)$ 是一个关于 $x$ 的有理函数，其中 $K = \mathbb{Q}(\log(y), e, \pi)$。从这个例子中我们可以看到，一般情况下，在 $K$ 的代数闭包 $\overline{K}$ 中运算效率很低或者不切实际。因此，只要有可能，现代算法试图避免在 $K$ 的延拓中进行运算。

## 介绍

有理函数积分的问题从表面上看与微分一样古老。根据 Ostrogradsky，牛顿 (Newton) 和莱布尼兹 (Leibniz) 都试图去计算有理函数的不定积分，然而没有一个人成功地设计出完整的算法。莱布尼兹的方法是先在实数域上计算分母的不可分解因式，然后将分式分解为分母只含 $x$ 一次或二次项的分式之和，最后分别对每一部分分式积分。然而，他不能完全解决分母为二次的分式的情况。在 18 世纪早期，Johan Bernoulli 完善了分式分解方法并使完全实现了莱布尼兹的方法，这应该是历史记载中最早的积分算法。令人惊奇地是，现今的微积分课本仍介绍这种方法，在分析课程的开始时它仍被教授给高中和大学学生。但很明显，这个方法的主要计算问题是计算一个多项式在实数域上的完全因式分解。这个问题是 19 世纪活跃的研究方向，并且早在 1845 年，俄罗斯数学家 M. W. Ostrogradsky 提出一个新算法，这个算法中直接计算有理部分的积分而不再分解。尽管他的方法传授给了俄罗斯学生，并在早些的俄罗斯分析课本中出现，但它并没有在世界的其他地方传授。这是因为积分算法或者类似的算法被独立地发现。因此，Hermite 在 1872 年发表了一个实现相同功能的不同算法，即不用因式分解计算有理部分积分。在近期，E. Horowitz 独立地发现了本质上的 Ostrogradsky 方法并附以详细的复杂度分析。不用因式分解计算超越部分积分的问题经过一个世纪仍未解决，但近期的几篇论文最终解决了这个问题。

## 2.1 伯努利算法

这种方法是最早也是最简单的。由于在 $\mathbb{R}[x]$ 上分解带来的花销，它在实际中并不经常被使用；但它是重要的，因为它为随后的所有算法提供了理论基础。设 $f \in \mathbb{R}(x)$ 是被积函数，并且 $f = P + A/D$，其中 $P, A, D \in \mathbb{R}[x]$，$\gcd(A, D) = 1$，同时 $\deg(A) < \deg(D)$。设

*2 有理函数积分*

$$D = c \prod_{i=1}^{n} (x - a_i)^{e_i} \prod_{j=1}^{m} \left(x^2 + b_j x + c_j\right)^{f_j}$$

为 $D$ 在 $\mathbb{R}$ 上的最简因式分解，其中 $c, a_i, b_j$ 和 $c_j \in \mathbb{R}$，$e_i, f_j$ 为正整数。计算 $f$ 的部分分式分解，我们得到

$$f = P + \sum_{i=1}^{n} \sum_{k=1}^{e_i} \frac{A_{ik}}{(x - a_i)^k} + \sum_{j=1}^{m} \sum_{k=1}^{f_j} \frac{B_{jk} x + C_{jk}}{(x^2 + b_j x + c_j)^k}$$

其中 $A_{ik}, B_{jk}, C_{jk} \in \mathbb{R}$。因此，

$$\int f = \int P + \sum_{i=1}^{n} \sum_{k=1}^{e_i} \int \frac{A_{ik}}{(x - a_i)^k} + \sum_{j=1}^{m} \sum_{k=1}^{f_j} \int \frac{B_{jk} x + C_{jk}}{(x^2 + b_j x + c_j)^k}$$

计算 $\int P$ 不会产生问题（对于其他函数类将会出现问题），对于其他项我们有

$$\int \frac{A_{ik}}{(x - a_i)^k} = \begin{cases} A_{ik} (x - a_i)^{1-k} / (1 - k) & \text{如果} k > 1 \\ A_{i1} \log (x - a_i) & \text{如果} k = 1 \end{cases}$$

同时，注意到 $b_j^2 - 4c_j < 0$ 因为 $x^2 + b_j x + c_j$ 在 $\mathbb{R}[x]$ 上是最简的，

$$\int \frac{B_{j1} x + C_{j1}}{x^2 + b_j x + c_j} = \frac{B_{j1}}{2} \log \left(x^2 + b_j x + c_j\right) + \frac{2C_{j1} - b_j B_{j1}}{\sqrt{4c_j - b_j^2}} \arctan \left( \frac{2x + b_j}{\sqrt{4c_j - b_j^2}} \right)$$

对于 $k > 1$，

$$\begin{aligned}
\int \frac{B_{jk} x + C_{jk}}{(x^2 + b_j x + c_j)^k} &= \frac{(2C_{jk} - b_j B_{jk}) x + b_j C_{jk} - 2c_j B_{jk}}{(k-1) \left(4c_j - b_j^2\right) (x^2 + b_j x + c_j)^{k-1}} \\
&+ \int \frac{(2k-3) (2C_{jk} - b_j B_{jk})}{(k-1) \left(4c_j - b_j^2\right) (x^2 + b_j x + c_j)^{k-1}}
\end{aligned}$$

最后一个等式可以递归调用直至 $k = 1$，因此产生完整的积分。

**例 2.1.1.** 考察 $f = 1/(x^3 + x) \in \mathbb{Q}(x)$。$f$ 的分母在 $\mathbb{R}$ 的因式分解为 $x^3 + x = x(x^2 + 1)$，则 $f$ 的部分分式分解为

$$\frac{1}{x^3 + x} = \frac{1}{x} - \frac{x}{x^2 + 1}$$

所以从上面的等式我们得到

$$\int \frac{dx}{x^3 + x} = \log(x) - \frac{1}{2} \log(x^2 + 1)$$

**例 2.1.2.** 考察 $f = 1/(x^2+1)^2 \in \mathbb{Q}(x)$。$f$ 的分母在 $\mathbb{R}$ 上为 $(x^2+1)^2$，$f$ 的部分分式分解为 $1/(x^2+1)^2$ 所以令上面的等式中 $j=1, k=2, b_1 = B_{12} = 0, c_1 = C_{12} = 1$，我们得到

$$\int \frac{dx}{(x^2+1)^2} = \frac{2x}{4(x^2+1)} + \int \frac{2dx}{4(x^2+1)} = \frac{x}{2(x^2+1)} + \frac{1}{2}\arctan(x)$$

适合于含特征 0 的自由域 $K$ 的伯努利算法的一个变体是通过在 $K$ 的代数闭包上线性分解 $D = \prod_{i=1}^{q}(x-\alpha_i)^{e_i}$，然后对下面的 $f$ 的部分分式分解的各项运用式 (2.1)：

## 2.2 埃尔米特约化

从前面对伯努利算法变体的讨论，我们知道对任何 $f \in K(x)$ 其有积分形如

$$\int f = v + \sum_{i=1}^{m} c_i log(u_i) \tag{2.1}$$

其中 $v, u_1, \ldots, u_m \in K(x)$ 并且 $c_1, \ldots, c_m \in K$。称 $v$ 为积分的有理部分，而对数的和称为积分的超越部分。埃尔米特 (Hermite[43]) 给出了如下计算 $v$ 的有理算法：将被积函数写为 $f = A/D$，其中 $A, D \in K[x]$ 并且 $\gcd(A, D) = 1$。令 $D = D_1 D_2^2 \cdots D_n^n$ 为 $D$ 的无平方因子分解。对 $f$ 进行遵从 $D_1, D_2^2, \ldots, D_n^n$ 的部分分式分解，得到

$$f = P + \sum_{k=1}^{n} \frac{A_k}{D_k^k}$$

其中 $P$ 和所有 $A_k$ 都属于 $K[x]$，并且对每个 $k$ 有或者 $A_k = 0$，或者 $\deg(A_k) < \deg(D_k^k)$ 进而

$$\int f = \int P + \sum_{k=1}^{n} \int \frac{A_k}{D_k^k}$$

从而问题转化为对形如 $Q/V^k$ 的分式积分，其中 $\deg(Q) < \deg(V^k)$ 并且 $V$ 是无平方的，也就是说 $\gcd(V, V') = 1$ 因此，若 $k > 1$ 则可以用拓展欧几里德算法来找到 $B, C \in K[x]$ 使得

$$\frac{Q}{1-k} = BV' + CV$$

并且 $\deg(B) < \deg(V)$。由此有 $\deg(BV') < \deg(V^2) \leq \deg(V^k)$，故而 $\deg(C) < \deg(V^{k-1})$。两边同乘 $(1-k)/V^k$ 我们得到

$$\frac{Q}{V^k} = -\frac{(k-1)BV'}{V^k} + \frac{(1-k)C}{V^{k-1}}$$

在右边加上然后减去 $B'/V^{k-1}$ 我们有

$$\frac{Q}{V^k} = \left(\frac{B'}{V^{k-1}} - \frac{(k-1)BV'}{V^k}\right) + \frac{(1-k)C - B'}{V^{k-1}}$$

*2 有理函数积分*

两边同时积分则有

$$\int \frac{Q}{V^k} = \frac{B}{V^{k-1}} + \int \frac{(1-k)C - B'}{V^{k-1}}$$

由于 $\deg((1-k)C-B') < \deg(V^{k-1})$，故而被积函数被化为了分母的 $V$ 的次数更小的类似分式。从而不断重复直到 $k=1$，我们得到满足 $\deg(E) < \deg(V)$ 和 $Q/V^k = y' + E/V$ 的 $y \in K(x)$ 与 $E \in K[x]$。对每一项 $A_i/D_i^i$ 进行这样的操作，我们得到 $g, h \in K(x)$ 使得 $f = g' + P + h$ 并且 $h$ 有一个无平方的分母并且没有多项式部分，从而 $\int h$ 是有常系数的对数的线性组合。那么(2.1)中的 $v$ 便仅是 $g + \int P$。埃尔米特并没有给出任何新的技术来计算对 $h$ 的积分，故在那时问题 Q2 仍然尚未解决。

---

**HermitReduce**$(A, D)$ （* Hermit 约化——原始版本 *）
(* 对给出的域 $K$ 以及 $A, D \in K[x]$ 使得 $D$ 非零并且与 $A$ 互素，返回 $g, h \in K(x)$ 使得 $\frac{A}{D} = \frac{dg}{dx} + h$ 并且 $h$ 有一个无平方的分母。*)
 $(D_1, \ldots, D_n) \leftarrow$ **SquareFree**$(D)$
 $(P, A_1, A_2, \ldots, A_n) \leftarrow$ **PartialFraction**$(A, D_1, D_2^2, \ldots, D_n^n)$
 $g \leftarrow 0$
 $h \leftarrow P + A_1/D_1$
 **for** $k \leftarrow 2$ **to** $n$ 使得 $\deg(D_k) > 0$ **do**
   $V \leftarrow D_k$
   **for** $j \leftarrow k-1$ **downto** 1 **do**
     $(B, C) \leftarrow$ **ExtendedEuclidean**$(\frac{dV}{dx}, V, -A_k/j)$
     $g \leftarrow g + B/V^j$
     $A_k \leftarrow -jC - \frac{dB}{dx}$
   **end for**
   $h \leftarrow h + A_k/V$
 **end for**
 **return**$(g, h)$

---

**例 2.2.1.** 这里我们对

$$f = \frac{x^7 - 24x^4 - 4x^2 + 8x - 8}{x^8 + 6x^6 + 12x^4 + 8x^2} \in \mathbb{Q}(x)$$

使用 **HermitReduce**。$f$ 分母的一个无平方因式分解为

$$D = x^8 + 6x^6 + 12x^4 + 8x^2 = x^2(x^2+2)^3 = D_2^2 D_3^3$$

$f$ 的部分分式分解为：$f = \frac{x-1}{x^2} + \frac{x^4 - 6x^3 - 13x^2 - 12x + 8}{(x^2+2)^3}$ 下面是对 $f$ 的 Hermite 约化的剩余部分：

| $i$ | $V$ | $j$ | $A_i$ | $B$ | $C$ |
|---|---|---|---|---|---|
| 2 | $x$ | 1 | $x-1$ | 1 | $-1$ |
| 3 | $x^2+2$ | 2 | $x^4 - 6x^3 - 18x^2 - 12x + 8$ | $6x$ | $-\frac{x^2}{2} + 3x - 2$ |
| 3 | $x^2+2$ | 1 | $x^2 - 6x - 2$ | $-x+3$ | 1 |

从而，
$$\int \frac{x^7 - 24x^4 - 4x^2 + 8x - 8}{x^8 + 6x^6 + 12x^4 + 8x^2} dx = \frac{1}{x} + \frac{6x}{(x^2+2)^2} - \frac{x-3}{x^2+2} + \int \frac{dx}{x}$$

值得一提的是，下面 Hermit 算法的变体不需要对 $f$ 进行部分分式分解：令 $D = D_1 D_2^2 \cdots D_m^m$ 为 $D$ 的无平方因式分解并假设 $m \geq 2$（否则 $D$ 已为无平方的了）。再令 $V = D_m$ 并且 $U = D/V^m$。由于 $\gcd(UV', V) = 1$，我们可以用拓展欧几里得算法来找到满足
$$\frac{A}{1-m} = BUV' + CV$$
及 $\deg(B) < \deg(V)$ 的 $B, C \in K[x]$。两边同乘以 $(1-m)/(UV^m)$ 可得
$$\frac{A}{UV^m} = \frac{(1-m)BV'}{V^m} + \frac{(1-m)C - UB'}{UV^{m-1}}$$
两边同时积分推出
$$\int \frac{A}{UV^m} = \frac{B}{V^{m-1}} + \int \frac{(1-m)C - UB'}{UV^{m-1}}$$
故被积函数被化为了分母中 $V$ 次数更小的一个。这个过程被重复执行直到分母是无平方的。由于每一次某一个无平方因式的指数减 1，最差的约化步数是 $1+2+\cdots+(m-1)$，也即 $O(m^2)$，从而我们称这个变体为二次埃尔米特约化。

---

**HermitReduce**$(A, D)$ (* Hermit 约化——二次版本 *)
(* 对给出的域 $K$ 以及 $A, D \in K[x]$ 使得 $D$ 非零并且与 $A$ 互素，返回 $g, h \in K(x)$ 使得 $\frac{A}{D} = \frac{dg}{dx} + h$ 并且 $h$ 有一个无平方的分母。*)
  $g \leftarrow 0, (D_1, \ldots, D_m) \leftarrow$ **SquareFree**$(D)$
  **for** $i \leftarrow 2$ **to** $m$ 使得 $\deg(D_i) > 0$ **do**
    $V \leftarrow D_i, U \leftarrow D/V^i$
    **for** $j \leftarrow i-1$ **downto** $1$ **do**
      $(B, C) \leftarrow$ **ExtendedEuclidean**$(U\frac{dV}{dx}, V, -A/j)$
      $g \leftarrow g + B/V^j, A \leftarrow -jC - U\frac{dB}{dx}$
    **end for**
    $D \leftarrow UV$
  **end for**
  **return**$(g, A/D)$

---

**例 2.2.2.** 同样考虑例 2.2.1 的被积函数，二次埃尔米特约化如下执行，其中 $D_3 = x^2+2$：

| $i$ | $V$ | $U$ | $j$ | $B$ | $C$ | $A$ |
|---|---|---|---|---|---|---|
| 2 | $x$ | $D_3^3$ | 1 | 1 | $-x^6 - x^5 + 18x^3 - 8x - 8$ | $x^6 + x^5 - 18x^3 + 8x + 8$ |
| 3 | $D_3$ | $x$ | 2 | $6x$ | $-\frac{x^4}{2} - \frac{x^3}{2} + x^2 - 2x - 2$ | $x^4 + x^3 - 2x^2 - 2x + 4$ |
| 3 | $D_3$ | $x$ | 1 | $-x+3$ | $-x^2 + x - 2$ | $x^2 + 2$ |

因而有
$$\int \frac{x^7 - 24x^4 - 4x^2 + 8x - 8}{x^8 + 6x^6 + 12x^4 + 8x^2} dx = \frac{1}{x} + \frac{6x}{(x^2+2)^2} + \frac{3-x}{x^2+2} + \int \frac{dx}{x}$$

## 2 有理函数积分

和例 2.2.1中的结果一样，但却不需要部分分式分解了。

假设被积函数的分母 $D$ 有形如 $D = D_1 D_2^2 \cdots D_m^m$ 的无平方因式分解，其中每个 $D_i$ 有正的度（这是埃尔米特约化的最坏情况）。在之前两个版本的算法中，所需约化步数都是 $m$ 的二次多项式。然而由 Mack[62]，有另一个只需要 $m-1$ 约化步骤的变体算法，故称之为线性埃尔米特约化。另外，Mack 的变体既不需要 $f$ 的部分分式分解，也不需要其分母的无平方因式分解（曾在约化中计算）。令 $D = D_1 D_2^2 \cdots D_m^m$ 为 $f$ 分母的一个无平方因式分解（实际上我们并不需要计算它），同时回想一下定义 1.6.2中的记号，也就是

$$P^* = \prod_{i=1}^n P_i \text{ 和} P^{-k} = \prod_{i=1}^n P_i^{max(0, e_i - k)}$$

对任何 $P \in K[x]$ $K$，其中 $pp(P) = \prod_{i=1}^n P_i^{e_i}$ 是 $pp(P)$ 的质因式分解。由于正在域 $K$ 中工作，我们可以认为 $D = pp(D)$。正如在无平方因子分解算法中，我们首先计算 $D^- = \gcd(D, D')$ 和 $D^* = D/D^-$。若 $\deg(D^-) = 0$，则 $D$ 是无平方的，否则由于来自(1.11)的 $D^- = D^{-*} D^{-2}$，来自引理 1.7.2的 $D^{-'} = D^{-2} Y_2$（其中 $Y_2$ 由(1.16)给出），来自引理的 $D_1 = D^*/D^{-*}$，我们有

$$\frac{D^* D^{-'}}{D^-} = \frac{D^* D^{-2} Y_2}{D^-} = \frac{D^* D^{-2} Y_2}{D^{-*} D^{-2}} = \frac{D^*}{D^{-*} Y_2} = D_1 Y_2 \in K[x] \tag{2.2}$$

更进一步地，由引理有 $\gcd(D_1, D^-) = 1$，由引理 1.7.2有 $gcd(Y_2, D^{-*}) = 1$，这些可推出

$$\gcd(\frac{D^* D^{-'}}{D^-}, D^{-*}) = gcd(D_1 Y_2, D^{-*}) = 1$$

因而，我们可以用拓展欧几里德算法来找 $B, C \in K[x]$ 以满足

$$A = B(-\frac{D^* D^{-'}}{D^-}) + C D^{-*}$$

和之前类似，在两边同除以 $D = D^* D^- = D_1 D^{-*} D^-$，得到

$$\frac{A}{D} = -\frac{B D^{-'}}{D^{-2}} + \frac{C}{D_1 D^-}$$

在右边加上然后减去 $B'/D^-$，我们有

$$\frac{A}{D} = (\frac{B'}{D^-} - \frac{B D^{-'}}{D^{-2}}) + \frac{C - D_1 B'}{D_1 D^-}$$

两边同时积分有

$$\int \frac{A}{D} = \frac{B}{D^-} + \int \frac{C - D_1 B'}{D_1 D^-}$$

鉴于 $D_1 D^- = (D_1 D_2) D_3^2 \cdots D_m^{m-1}$，被积函数被化为了一个分母有最多 $m-1$ 个不同指数的无平方因式分解（相对于初始被积函数的 $m$）。因此，该过程重复最多 $m-1$ 次可

以得到一个有无平方因式分解的分母。通过该轮参数计算下一轮迭代的参数可以做到更进一步的优化：新的被积函数是

$$\frac{C - D_1 B'}{D_1 D^-} = \frac{\overline{A}}{\overline{D}}$$

其中

$$\overline{A} = C - D_1 B' = C - \frac{D^*}{D^{-*}} B'$$

并且

$$\overline{D} = D_1 D^- = D_1 D_2 D_3^2 \cdots D_m^{m-1}$$

我们从而可得

$$\overline{D}^* = D_1 D_2 \cdots = D^*$$

这表示 $D^*$ 在整个约化过程中没有改变。更进一步的，

$$\overline{D}^- = D_3 D_4^2 \cdots D_m^{m-2} = D^{-2}$$

表示 $D^-$ 在整个约化过程中每一步被其降阶 (deflation) 所替换。

值得一提的是，相对一个域，在一个 UFD 上执行所有埃尔米特约化的变体是可能的，结果则在其商域中表达（?）。在这种情况下，Mack 的变体要求分母 $D$ 是本原的（之前的变体并不需要这一点）。

---

**HermitReduce**$(A, D)$    (* Hermit 约化——Mack 线性版本 *)
(* 对给出的域 $K$ 以及 $A, D \in K[x]$ 使得 $D$ 非零并且与 $A$ 互素，返回 $g, h \in K(x)$ 使得 $\frac{A}{D} = \frac{dg}{dx} + h$ 并且 $h$ 有一个无平方的分母。*)
   $g \leftarrow 0$
   $D^- \leftarrow \gcd(D, \frac{dD}{dx})$
   $D^* \leftarrow \frac{D}{D^-}$
   **while** $\deg(D^-) > 0$ **do**
     $D^{-2} \leftarrow \gcd(D^-, \frac{dD^-}{dx})$
     $D^{-*} \leftarrow \frac{D^-}{D^{-2}}$
     $(B, C) \leftarrow$ **ExtendedEuclidean**$(-D^* \frac{dD^-}{dx}/D^-, D^{-*}, A)$
     $A \leftarrow C - \frac{dB}{dx} D^*/D^{-*}$      (* 新的分子 *)
     $g \leftarrow g + B/D^-$
     $D^- \leftarrow D^{-2}$      (* $\overline{D}^- = D^{-2}$ *)
   **end while**
   **return**$(g, A/D^*)$

---

**例 2.2.3.** 考虑同例 2.1 中相同的被积函数。Mack 算法给出下列的步骤：

1. $g = 0$

2. $D^- = \gcd(D, dD/dx) = x^5 + 4x^3 + 4x$

3. $D^* = D/D^- = x^3 + 2x$

4. 第一次约化：
   $D^{-2} = \gcd(x^5 + 4x^3 + 4x, 5x^4 + 12x^2 + 4) = x^2 + 2$

5. $D^{-*} = D^-/D^{-2} = x^3 + 2x$

6. 
$$(B, C) = \textbf{ExtendedEuclidean}(-5x^2 - 2, x^3 + 2x, A)$$
$$= (8x^2 + 4, x^4 - 2x^2 + 16x + 4)$$

7. $A = x^4 - 2x^2 + 16x + 4 - 16x = x^4 - 2x^2 + 4$

8. 
$$g = g + \frac{B}{D^-} = \frac{8x^2 + 4}{x^5 + 4x^3 + 4x}$$

9. $D^- = D^{-2} = x^2 + 2$

10. 第二次约化：
    $D^{-2} = \gcd(x^2 + 2, 2x) = 1$

11. $D^{-*} = D^-/S_3 = x^2 + 2$

12. $(B, C) = \textbf{ExtendedEuclidean}(-2x^2, x^2 + 2, x^4 - 2x^2 + 4) = (3, x^2 + 2)$

13. $A = x^2 + 2$

14. 
$$g = g + \frac{B}{D^-} = \frac{8x^2 + 4}{x^5 + 4x^3 + 4x} + \frac{3}{x^2 + 2}$$

15. $D^- = D^{-2} = 1$

因此有

$$\int \frac{x^7 - 24x^4 - 4x^2 + 8x - 8}{x^8 + 6x^6 + 12x^4 + 8x^2} dx = \frac{8x^2 + 4}{x^5 + 4x^3 + 4x} + \frac{3}{x^2 + 2} + \int \frac{dx}{x}$$

和埃尔米特约化的结果相同，但相比原来的 3 步，现在只需要 2 步约化。

## 2.3 霍罗威茨-奥斯特洛格拉茨基算法

奥斯特洛格拉茨基算法也可以计算积分的实数部分，但是它只能计算 $K$ 上的线形代数方程而不是类似 (1.5) 的多项式丢番图方程系统。假设被积函数 $f$ 为 $A/D$ 的形式，并且 $deg(A) < deg(D)$。像之前一样，设 $D = D_1 D_2^2 \ldots D_m^m$ 是分母 $f$ 的无平方因子分解（事实上这种算法并没有真正去求解它）。使用定义 1.6.2 的符号 $P^*$ 和 $P^-$，我们有 $D^- = gcd(D, D')$，$D^* = D/D^-$。从埃尔米特约化的步骤来看，很明显如何 $f = g' + h$（$h$ 是一个无平方因子的分母），则 $g$ 的分母整除 $D^-$，$h$ 的分母整除 $D^*$，于是我们可以得到 $g = B/D^-$ 和 $h = C/D*$，其中 $B, C \in K[x]$ 都是未知的。此外，由 $deg(A) < deg(D)$，我们可以猜到 $deg(B) < deg(D^-)$ 和 $deg(C) < deg(D^*)$。写下 $f = g' + h$，我们可以看到

$$\frac{A}{D} = \frac{B'}{D^-} - \frac{BD^{-'}}{(D^-)^2} + \frac{C}{D^*}$$

然后两边乘以 $D = D^*D^-$，

$$A = B'D^* - B\left(\frac{D^*D^{-'}}{D^-}\right) + CD^- \tag{2.3}$$

由于 $D^- | D^*D^{-'}$，上面的等式对于带有多项式系数 $B$ 和 $C$ 来说是一个线性方程。此外，由埃尔米特约化总能得到这样的一个结果可知，该方程在 $K[x]$ 中总是有一个解。由于 $B$ 和 $C$ 的系数之间有关系，我们用

$$\sum_{i=0}^{deg(D^-)-1} b_i x^i \text{和} \sum_{j=0}^{deg(D^*)-1} c_j x^j$$

替换 $B$ 和 $C$，其中 $b_i$' 和 $c_j$' 是 $K$ 中未定的常量。(2.6) 等式两边相等产生了有关 $b_i$' 和 $c_j$' 的线形方程系统，并且这个系统的任何解均给出 $B$ 和 $C$，从而得到 $g$ 和 $h$。

---

**HorowitzOstrogradsky(A, D)**  (* 霍洛维茨-奥斯特罗格拉茨基算法 *)
(* 给定一个域 $K$，$A, D \in K$，$deg(A) < deg(D)$，$D$ 非零且与 $A$ 互质，返回 $g, h \in K(x)$ 使得 $\frac{A}{D} = \frac{dg}{dx}$，$h$ 有一个无平方因子的分母 *)
  $D^- \leftarrow gcd(D, \frac{dD}{dx})$
  $D^* \leftarrow D/D^-$
  $n \leftarrow deg(D^-) - 1$
  $m \leftarrow deg(D^*) - 1$
  $d \leftarrow deg(D)$
  $B \leftarrow \sum_{i=0}^{n} b_i x^i$
  $C \leftarrow \sum_{j=0}^{m} c_j x^j$
  $H \leftarrow A - B'D^* + BD^*D^{-'}/D^- - CD^-$ (* 正合除法 *)
  $(b_0, \ldots, b_n, c_0, \ldots, c_m) \leftarrow \mathbf{solve}(\mathbf{coefficient})(H, x^k) = 0, 0 \leq k \leq d)$
  **return**$(\sum_{i=0}^{n} b_i x^i / D^-, \sum_{j=0}^{m} c_j x^j / D^*)$

---

**例 2.3.1.** 给定与例 2.2.1 相同的被积函数，霍洛维茨-奥斯特罗格拉茨基算法进行步骤如下：

**1.** $D^- = gcd(D, dD/dx) = x^5 + 4x^3 + 4x$

**2.** $D^* = D/D^- = x^3 + 2x$

**3.** $n = deg(D^-) - 1 = 4, m = deg(D^*) - 1 = 2$

**4.**

$$H = A - D^* \left(\sum_{i=0}^{n} b_i x^i\right)' + \left(\sum_{i=0}^{n} b_i x^i\right) \frac{D^* D^{-'}}{D^-} - D^- \sum_{j=0}^{m} c_j x^j$$
$$= (1-c_2)x^7 + (b_4 - c_1)^6 + (2b_3 - c_0 - 4c_2)x^5 + (3b_2 - 6b_4 - 4c_1 - 24)x^4$$
$$+ 4(b_1 - b_3 - c_0 - c_2)x^3 + (5b_0 - 2b_2 - 4c_1 - 4)x^2 + 4(2-c_0)x + 2(b_0 - 4)$$

**5.** 这由令 $H$ 等于 0 得到的系统有特殊解

$$(b_0, b_1, b_2, b_3, b_4, c_0, c_1, c_2) = (4, 6, 8, 3, 0, 2, 0, 1)$$

因此,

$$\int \frac{x^7 - 24x^4 - 4x^2 + 8x - 8}{x^8 + 6x^6 + 12x^4 + 8x^2} dx = \frac{3x^3 + 8x^2 + 6x + 4}{x^5 + 4x^3 + 4x} + \int \frac{x^2 + 2}{x^3 + 2x} dx$$
$$= \frac{3x^3 + 8x^2 + 6x + 4}{x^5 + 4x^3 + 4x} + \int \frac{dx}{x}$$

这与埃尔米特约化得到的结果相一致。

尽管就复杂性来说,这种算法对解有理函数 [45] 是非常有利的,但就普适性来说,在解大型函数时并没有埃尔米特约化简单,故我们在以后的算法中通常采用线性埃尔米特约化的方法。

## 2.4 罗思坦-特拉杰算法

根据埃尔米特约化,我们认为现在的分数的形式 $f = A/D$,并且 $deg(A) < deg(D)$,$D$ 是非平方数。设 $\alpha_1, \ldots, \alpha_n \in \bar{K}$ 是 $D$ 在 $\bar{K}$ 中的零点,$f$ 的部分分式分解必须是如下形式

$$f = \sum_{i=1}^{n} \frac{a_i}{x - \alpha_i}$$

其中 $a_1, \ldots, a_n \in \bar{K}$. 通过与复值函数类比,$a_i$ 叫做 $f$ 在 $x = \alpha_i$ 时的剩余其中. 通过简单的算法,我们知道

$$\int f = \sum_{i=1}^{n} a_i \log(x - \alpha_i).$$

问题是这样计算的不含有理部分 $D$ 的 $f$ 的剩余的。

特拉米 [59] 和罗思坦 [83] 分别证明了如下定理。

**定理 2.4.1.** *([83,89]). 设 $t$ 是 $K(x)$ 上的一个未定元和 $A$, $D$ 在 $K[x]$ 中,$deg(D) > 0$, $D$ 是无双的,$gcd(A, D) = 1$. 设*

$$R = resultant_x(D, A - tD') \in K[t].$$

然后,

***(i)*** *$R$ 在 $\bar{K}$ 中的零点都是 $A/D$ 在 $D$ 的所有在 $\bar{K}$ 中零点上余式的值。*

***(ii)*** *设 $a \in \bar{K}$ 是 $R$ 的一个零点，$G_a = \gcd(D, A - aD') \in K(a)[x]$。那么 $deg(G_a) > 0$，且 $G_a$ 在 $K$ 中的零点都是 $D$ 的零点，且此时 $A/D$ 的余式等于 $a$*

***(iii)*** *任何一个具有 (2.4) 形式的包含 $A/D$ 的整式部分的域同样包含 $R$ 在 $K$ 中的所有零点。*

由于这个定理是一个特殊情况的结果将在第四章和第五章中的证明，我们将在那里对它进行证明。这个定理的一个直接结果是

$$\int \frac{A}{D} = \sum_{a|R(a)=0} a \log(\gcd(D, A - aD'))$$

其中总和被不同的根所影响。罗思坦-特拉米算法结合一平方分母的有理函数多项式部分并没有给出公式。经过适当的修改，罗思坦-载体的算法能像埃尔米特剩余，适用于在一个 $UFD$ 上应用有理函数，而不是字段。定理 2.4.1 的第三条表明 $R$ 的分裂域是最小的 $K$ 代数扩张要表达的 $A/D$ 使用对数积分，从而本质上回答问题 $Q2$. 当然它可能表示积分在一个较小的连续域，于其上定义了除了对数积分意外其它函数，例如 $\int dx/(x^2 + 1) = \arctan(x)$ 但由于一个函数的不定积分是可以正式毗邻的一个字段（第三章），问题 $Q2$ 只有在涉及到积分的具体形式才有意义。如果在反三角函数的积分，那么 Rioboo 的算法（第二章第八节）表明，积分可以在一个含有所有根中的实部和虚部的数域中表示，这一结果，连同定理 2.4.1 的第三部分给问题 $Q2$ 提供了一个完整的有理函数的初等积分（初等积分将被定义第五章）的答案。请注意，此算法需要一个最大公约数算法在 $K(a)[x]$ 其中 $a$ 实数中的零点，在 $K$ 上的代数常数。因此一个定义在 $K$ 上的素因子分解式 $R = uR_1^{e_1} \ldots R_m^{e_m}$，因此我们必须为每一个 $R_i$ 计算出一个相应的零点。由于答案可以被看作是每个 $R_i$ 零点的总和，就没有必要去计算 $R_i$ 的分裂域。

---

**IntRationalLogPart**$(A, D)$    (* 罗思坦-特拉米算法 *)
(* 给定一个特征为 0 的域 $K$，$A, D \in K[x]$，其中 $deg(A) < deg(D), D$ 是非零的，且与 $A$ 互质，无平方因子分解，返回 $\int A/Ddx$. *)
   $u \leftarrow$ a new indeterminate over $K$
   $R \leftarrow resultant_x(D, A - t\frac{dD}{dx})$
   $uR_1^{e_1} \ldots R_m^{e_m} \leftarrow \text{factor}(R)$      (* *factorization into irreducibles* *)
   $(C, G) \leftarrow$ **ExtendedEuclidean**$(F', F, 1)$      (* $CF' + GF = 1$ *)
   **for** $i \leftarrow 0$ **to** $m$ **do**
     $a \leftarrow a|R_i(a) = 0$
     $v_j \leftarrow F^{(j+1)}/(j+1)$
     $G_i \leftarrow \gcd(D, A - a\frac{dD}{dx})$      (* *algebraic gcd computation* *)
   **end for**
   **return**$(\sum_{i=1}^{m} \sum_{a|R_i(a)=0} a \log(G_i))$

---

**例 2.4.1.** 考虑
$$f = \frac{x^4 - 3x^2 + 6}{x^6 - 5x^4 + 5x^2 + 4} \in \mathbb{Q}(x)$$

$f$ 的分母，$D = x^6 - 5x^4 + 5x^2 + 4$ 是一个无平方因式分解（事实上受 Q 约束），然后罗思坦-特拉米的结果为

$$resultant_x(x^6 - 5x^4 + 5x^2 + 4, x^4 - 3x^2 + 6 - t(6x^5 - 20x^3 + 10x)) = 45796(4t^2 + 1)^3$$

使 $a$ 为一个代数数比如 $4a^2 + 1 = 0$, 我们发现

$$G_a = \gcd(x^6 - 5x^4 + 5x^2 + 4, x^4 - 3x^2 + 6 - a(6x^5 - 20x^3 + 10x)) = x^3 + 2ax^2 - 3x - 4a$$

所以，

$$\int \frac{x^4 - 3x^2 + 6}{x^6 - 5x^4 + 5x^2 + 4} = \sum_{a|4a^2+1=0} a\log(x^3 + 2ax^2 - 3x - 4a)$$
$$= \frac{\sqrt{-1}}{2}\log(x^3 + x^2\sqrt{-1} - 3x - 2\sqrt{-1})$$
$$- \frac{\sqrt{-1}}{2}\log(x^3 - x^2\sqrt{-1} - 3x + 2\sqrt{-1}).$$

## 2.5 拉扎德-雷欧布-特拉杰算法

罗思坦-特拉杰算法计算完整表示下的最小代数扩展，Trager 和 Rioboo 独立的发现罗思坦-特拉杰算法计算结果的质因数分解和在代数扩展中计算最大公约数是可以避免的，只要我们使用 PRS 子结式去计算 (2.7)。他们的算法通过下面的定理修正。

**定理 2.5.1.** ([56,66]) 设 $\overline{K}$ 为 $K$ 的代数闭包,$t$ 是 $K(x)$ 中的变量,$A, B, C \in K[x] \setminus \{0\}$ 满足 $\gcd(A, C) = \gcd(B, C) = 1$, $\deg(A) < \deg(C)$ 并且 $C$ 无平方因子分解. 设

$$R = resultant_x(C, A - tB) \in K[t]$$

当 $deg(B) < deg(C)$ 时,$(R_0, R_1, \ldots, R_k \neq 0, 0, \ldots)$ 表示 PRS 子结式关于 $x$ 对于 $C$ 和 $A$-$tB$, 当 $deg(B) \leq deg(C)$ 时,$(R_0, R_1, \ldots, R_k \neq 0, 0, \ldots)$ 表示 PRS 子结式关于 $x$ 对于 $C$ 和 $A$-$tB$. 设 $\alpha \in \overline{K}$. 那么,

**(i)** n=deg(C), 在这种情况下

$$\gcd(C, A - \alpha B) = C \in K(\alpha)[x].$$

**(ii)** n<deg(C), 在这种情况下存在唯一 $m \leq 1$ 满足 $deg_x(R_m) = n$, 且

$$\gcd(C, A - \alpha B) = pp_x(R_m)(\alpha, x) \in K(\alpha)[x].$$

在这里 $pp_x(R_m)$ 是 $R_m$ 关于 $x$ 的基元

*Proof.* 设 $R = resultant_x(C, A - tB) \in K[t]$ 和当 $deg(B) < deg(C)$ 时,$(R_0, R_1, \ldots, R_k \neq 0, 0, \ldots)$ 表示 PRS 子结式关于 x 对于 C 和 A-tB, 当 $deg(B) \leq deg(C)$ 时,$(R_0, R_1, \ldots, R_k \neq 0, 0, \ldots)$ 表示 PRS 子结式关于 x 对于 C 和 A-tB. 设 $\alpha \in \overline{K}$. 设 $q = deg(C), c \in K*$ 表

示 C 的首项系数, $C = c\prod_{i=1}^{q} x - \beta_i$ 为 C 在 $\overline{K}$ 中的线性因数, 在这里 $\beta_i$ 互不相同因为 C 无平方因子分解. 通过定理, 我们有

$$R = c^p \prod_{i=1}^{q}(A(\beta_i) - tB(\beta_i))$$

在这里, $p = deg_x(A - tB)$. 因此,R 的首项系数为 $\pm c^p \prod_{i=1}^{q} B(\beta_i)$, 由于 $gcd(B,C) = 1$ 我们可以知道 R 的首项系数非零. 因此 $R \neq 0$, 所以设 $\alpha \in \overline{K}$ 为对于 R 的 0 多样性. 我们注意到 trailing monomail of R 是 $c^p \prod_{i=1}^{q} A(\beta_i)$, 由于 $gcd(A,C) = 1$ 我们可以知道 trailing monomail of R 非 0, 所以 $\alpha \neq 0$. 由于 $\alpha$ 带有 n 的重数, 因此存在子集 $I_\alpha \subseteq \{1,\ldots,q\}$ 关于基数 n 满足 $A(\beta_i) - \alpha B(\beta_i) = 0$ 当且仅当 $i \in I_\alpha$. 因此, $G_\alpha = \prod_{i \in I_\alpha}(x - \beta_i)$ 在 $\overline{K}[x]$ 中对 $A - \alpha B$ 进行拆分. 但是 $x - \beta_i A - \alpha B$ 对任意的 $i \in I_\alpha$, 所以 $G_\alpha$ 是 C 和 $A - \alpha B$ 在 $K(\alpha)[x]$ 中的最大公约数. 因此, $deg_x(gcd(C, A - \alpha B)) = n$, $n \leq deg(C)$.

**(i)** 如果 $n = deg(C)$, 则 $gcd(C, A - \alpha B)$ 是 C 在 $deg(C)$ 下的最大公约数, 因此 $gcd(C, A - \alpha B) = c$.

**(ii)** 考虑 $n < deg(C)$. 则 $A - \alpha B \neq 0$, 否则我们有 $gcd(C, A - \alpha B) = gcd(C, 0) = C$, 度大于 n. 设 $S_n \in K[t][x]$ 是 C 和 $A - tB$ 关于 x 的子结式, $\sigma : K[t] \to K(\alpha)$ 是定义在 K 上的由 t 映射到 $\alpha$ 的环同态, $\overline{\sigma} : K[t][x] \to K(\alpha)[x]$ 为 $\overline{\sum a_j x^j} = \Sigma \sigma(a_j)x^j$. 因为 A,B,C 与 t 无关, $\overline{(\sigma)}(C) = C$ 且 $\overline{(\sigma)}(A - tB) = A - \alpha B$, 因此 $deg(\overline{\sigma(C)}) = q$, 并且定理说明 $\overline{\sigma}(S_n) = c^r \overline{S_n}$, 其中 r 是非负整数并且 $\overline{S_n}$ 是 C 和 $A - \alpha B$ 的第 n 个子结式. 当 $deg(B) < deg(C)$ 时,$(Q_0, Q_1, \ldots, Q_l \neq 0, 0, \ldots)$ 表示 PRS 子结式关于 x 对于 C 和 A-tB, 当 $deg(B) \leq deg(C)$ 时,$(Q_0, Q_1, \ldots, Q_l \neq 0, 0, \ldots)$ 表示 PRS 子结式关于 x 对于 C 和 A-tB. 通过定理,$Q_l$ 是 C 和 $A - \alpha B$ 的最大公约数, 所以 $deg_x(Q_l) = n$. 因此, 通过定理 $\overline{\sigma}(S_n)$ 相似于 $Q_l$, 从而 $\overline{\sigma}(S_n) \neq 0$ 并且 $\overline{\sigma}(S_n)$ 为 C 和 $A - \alpha B$ 的最大公约数, 实际上,$deg(\overline{\sigma}(S_n)) = n$. 通过定义 $deg_x(S_n) \leq n$ 并且 $deg(\overline{\sigma}(S_n)) \leq deg_x(S_n)$, 我们有 $deg_x(S_n) = n$. 通过定理,$S_n$ 相似于 $R_m$ 对一些 $m \geq 0$, 从而有 $deg_x(R_m) = n$. 因为 $deg(R_0) \geq deg(C) > n$, 我们有 $n \geq 1$, 从而我们推出 m 唯一, 因为 $deg(R_i) > deg(R_{i+1})$ 对于 $i \geq 1$ 在任何 PRS 中. 写 $\rho_1 S_n = \rho_2 PP_x(R_m)$ 对于 $\rho_1, \rho_2 \in K[t]$ 满足 $gcd(\rho_1, \rho_2) = 1$. 因此,$\sigma(\rho_1)\overline{\sigma}(S_n) = \sigma(\rho_2)\overline{\sigma}(pp_x(R_m))$. 注意 $\overline{\sigma}(S_n) \neq 0$ 且 $\overline{pp_x(R_m)} \neq 0$ 因为 $pp_x(R_m)$ 为基元. 进一步, 我们不能有 $\sigma(\rho_1) = \sigma(\rho_2) = 0$ 因为 $gcd(\rho_1, \rho_2) = 1$. 因此,$\sigma(\rho_1) \neq 0$ 且 $\sigma(\rho_2) \neq 0$, 所以 $\overline{\sigma}(pp_x(R_m)) = pp_x(R_m)(\alpha, x)$ 是 C 和 $A - \alpha B$ 的最大公约数

□

现在设 $A, D \in K[x]\setminus\{0\}$ 满足 $gcd(A, D) = 1$.D 无平方因子分解且 $deg(A) < deg(D)$. 应用定理和 $A = A, B = D', C = D$, 我们有 $R = resultant_x(D, A - tD')$ 并且对任何 R 的根 $\alpha$, 我们有 $i \leq deg(D)$ 和

**i** 如果 $i = deg(D)$, 那么 $gcd(D, A - \alpha D') = D$,

**ii** 如果 $i < deg(D)$, 那么 $gcd(D, A - \alpha D') = pp_x(R_m)(\alpha, x)$ 其中 $m \geq 1$ 是唯一的正整数 $deg_x(R_m) = i$.

因此, 我们并不需要去计算 (2.8) 中出现的 $gc$, 我们可以使用在 PRS 子结式中出现的余数替代, 所以我们用 D 和 $A - tD'$ 的 PRS 子结式. 只要返回的结果是一些多项式根的形式和, 所有的计算就可以在 K 中完成, 不要求任何的代数扩展, 并且由和引入的形式代数是最小可能的代数扩展.

实际上, 我们定义一个无平方因子分解 $R = \prod_{i=1}^{n} Q_i^i$, 所以 R 的根恰好是 $Q_i$ 的根. 计算 $pp_x(R_m)$ 对于 $t = \alpha$ ($\alpha$ 是 $Q_i$ 的根) 等价于减少系数当 x 为 $pp_x(R_m)$ modulo $Q_i$. 我们不需要计算 $pp_x(R_m)$, 这事实上足够保证 $Q_i$ 和 $R_m$ 的首系数没有非平凡最大公约数, 从而 $Q_i$ 的余数非零.

---

**IntRationalLogPart(A, D)**  (\* Lazard-Rioboo-Trager 算法 \*)
(\* 给定一个域 $K$, $A, D \in K[x]$ 满足 $deg(A) < deg(D)$, D 非零、无平方因子分解、与 A 互质, 返回 $\int A/D dx$.\*)

$t \leftarrow$ a new indetermiate over K
$(R, (R_0, R_1, \ldots, R_k, 0)) \leftarrow \mathbf{SubResultant}_x(D, A - t\frac{dD}{dx})$
$(Q_1, \ldots, Q_n) \leftarrow \mathbf{SquareFree}(R)$
**for** $i \leftarrow$ n such that $deg_t(Q_i) > 0$ **do**
**if** $i = deg(D)$ **then** $S_i \leftarrow D$
**else**
$S_i \leftarrow R_m$ where $deg_x(R_m) = i, 1 \leq m \leq k$
$(A_1, \ldots, A_q) \leftarrow \mathbf{SquareFree}(lc_x(S_i))$
**for** $j \leftarrow \mathbf{1 to q do} S_i \leftarrow S_i/gcd(A_j, Q_i)^j$ (\*exact quotient\*)
**return** $(\sum_{i=1}^{n} \sum_{a|Q_i(a)=0} a \log S_i(a, x))$

---

**例 2.5.1.** 考虑与 example2.4.1 相同的积分表. D 无平方分解, D 和 $A - tD'$ 的 PR 子结式是 Rothstein-Trager 结式是 $R = R_6$, 且他的无平方因子分解因式是

| $i$ | $R_i$ |
|---|---|
| 0 | $x^6 - 5x^4 + 5x^2 + 4$ |
| 1 | $-6tx^5 + x^4 + 20tx^3 - 3x^2 - 10tx + 6$ |
| 2 | $(-60t^2 + 1)x^4 + 2tx^3 + (120t^2 - 3)x^2 + 26tx + 144t^2 + 6$ |
| 3 | $(800t^3 - 14t)x^3 - 400t^2 - 7)x^2 - (2440t^3 - 32t)x + 792t^2 - 16$ |
| 4 | $(-11200t^4 - 2604t^2 + 49)x^2 + 25600t^4 + 5952t^2 - 112$ |
| 5 | $(-119840t^5 - 59920t^3 - 7490t)x - 23968t^4 - 11984t^2 - 1498$ |
| 6 | $2930944t^6 + 2198208t^4 + 549552t^2 + 45796$ |

$R = 2930944t^6 + 2198208t^4 + 549552t^2 + 45796 = 45796(4t^2 + 1)^3 = 45796Q_3^3$ 并且他的 PRS 三阶余数是

$$R_3 = (800t^3 - 14t)x^3 - 400t^2 - 7)x^2 - (2440t^3 - 32t)x + 792t^2 - 16$$

因为

$$gcd(lc_x(R_3, Q_3)) = gcd(800t^3 - 14t, 4t^2 + 1) = 1,$$

$S_3 = R_3$. 对于 t 计算 $Q_3(a) = 4a^2 + 1 = 0$ 的根, 我们有

$$S_3(a, x) = -214ax^3 + 107x^2 + 642ax - 214$$

所以,一个积分是

$$\int \frac{x^4 - 3x^2 + 6}{x^6 - 5x^4 + 5x^2 + 4} = \sum_{a|4a^2+1=0} a\log\left(-214ax^3 + 107x^2 + 642ax - 214\right)$$

使 $S_3(a,x)$monic 我们有 $S_3(a,x) = -214a(x^3 + 2ax^2 - 3x - 4a)$ 所以积分相同于例 2.4.1.

---

**IntegrateRationalFunction(A, D)**  (* Rational function integration *)
(* 给定一个域 $K$ 和 $f \in K(x)$, 返回 $\int f dx$.*)
  $(g, h) \leftarrow$ **HermiteReduce**($numerator(f), denominator(f)$)
  $(Q, R) \leftarrow$ **PolyDivide**($numerator(h), denominator(h)$)
  **if** $R = 0$ **then return**($g + \int Q dx$)
  **return**($g + \int Q dx +$ **IntRationalLogPart**($R, denominator(h)$))

---

**例 2.5.2.** 让我们计算积分

$$f = \frac{A}{D} = \frac{36}{x^5 - 2x^4 - 2x^3 + 4x^2 + x - 2} \in Q(x).$$

1.**HermiteReduce**$(A,D)$ 返回 $g = (12x+6)/(x^2-1)$ 和 $h = 12/E$ 其中 $E = x^2 - x - 2$.
2.**PolyDivide**$(12, E)$ 返回 $Q = 0$ 和 $R = 12$.
3.**IntRationalLogPart**$(12, E)$ 返回 $\sum_{a|a^2-16=0} a\log(x - 1/2 - 3a/8)$.

所以，我们有

$$\int \frac{36}{x^5 - 2x^4 - 2x^3 + 4x^2 + x - 2} dx = \frac{12x+6}{x^2 - 1} + \sum_{a|a^-16=0} a\log(x - \frac{1}{2} - \frac{3a}{8})$$

一个更简单的逻辑部分会在 2.7 节中给出.

## 2.6 次秋斯基算法

次秋斯基已经指出在 $K[x,z]$ 中, 罗思坦-特拉杰算法和拉扎德-雷欧布-特拉杰算法的结式和子结式运算可以被格罗布纳基运算取代。由于有关格罗布纳基的理论超出了本书的范围, 所以我们只是给出它和它的算法, 而没有进行证明。有兴趣的读者可以参考 [7,25] 格罗布纳基德导论和 [26] 有关下面定理的证明。

**定理 2.6.1.** *([26]).* 设 $t$ 是 $K(x)$ 上的一个未定元, $A$, $D$ 在 $K[x]$ 中, 且满足 $deg(D) > 0$, $D$ 是无双的, $gcd(A,D) = 1$, $\mathcal{B}$ 是 $K[t,x]$ 中由 $D$ 和 $A - tD'$ 生成的理想在 $x > t$ 下字典排序的即约的格罗布纳基. 记 $\mathcal{B} = P_1, \ldots, P_m$, 其中 $P_i$ 是在 $K[x,t]$ 中, 并且对于任一 $i$, 在 $x > t$ 下的字典排序中 $P_{i+1}$ 的最高项大于 $P_i$ 的最高项。则,

**(i)** $lc_x(pp_x(P_i)) = 1$, $1 \leq i \leq m$。

**(ii)** 在 $K[x,z]$ 中,$content_x(p_{i+1})$ 整除 $content_x(p_i)$, $1 \leq i \leq m$。

*(iii)*

$$\int \frac{A}{D}dx = \sum_{i=1}^{m-1} \sum_{a|Q_i(a)=0} (alog(pp_x(p_{i+1})(a,x))$$

这里 $Q_i = content_x(P_i)/content_x(P_{i+1}) \in K[t]$.

*注: 定理 2.6.1 的 (i) 即是说明次秋斯基的算法引出了算法中的首一多项式.*

---

**IntRationalLogPart**$(A, D)$   *(\* 次秋斯基算法 \*)*
*(\* 给定一个特征为 0 的域 K, 和 K[x] 中的 A,B 且满足:$det(A) < det(D)$, 且 D 非零, 无双, 与 A 互素. 返回 $\int A/Ddx$ \*)*
 *(\* 计算既约格罗布纳基 \*)*
 $(P_1,\ldots,P_m) < -ReducedGrobner(D, A - t\frac{dD}{dx}, pure\ lex, x > t)$
 *(\*$(P_1,\ldots p_m)$ 必须按最高项的增序排列 \*)*
 **for** $j \leftarrow 0$ **to** $m-1$ **do**
    $Q_i = content_x(P_i)/content_x(P_{i+1})$ *(\* 精确计算商 \*)*
    $S_i = pp_x(p_{i+1})$
 **end for**
 **return**$\sum_{i=1}^{m-1} \sum_{a|Q_i(a)=0}(alogS_i(a,x))$

---

**例 2.6.1.** 考虑与例 2.4.1 相同的积分, 则 $(D, A - tdD/dx)$ 的格罗布纳基为:
$(x^6 - 5x^4 + 5x^2 + 4, -6tx^5 + x^4 + 20tx^3 - 3x^2 - 10tx + 6)$ 关于:$x > t$ 下字典排序:
B= $(P_1, P_2)$ =$(4t^2 + 1, x^3 + 2tx^2 - 3x - 4t)$
$Q_1 = 4t^2 + 1/1 = 4t^2 + 1$ 且 $S_1 = P_2$, 这说明:

$$\int \frac{x^4 - 3x^2 + 6}{x^6 - 5x^4 + 5x^2 + 4} = \sum_{a|4a^2+1=0} alog(x^3 + 2ax^2 - 3x - 4a)$$

这显然是和 2.4.1 中的结果是相同的.

## 2.7 重访牛顿 -莱布尼兹 -伯努利

我们已经注意到, 运用公式 (2.1) 的困难是在被积函数极点处劳伦级数展开的计算, 这在牛顿莱布尼茨时代就已经被人们所知. 然而, Bronstein 和 Salvy 给出了一个合理的算法, 可以得到全微分有理分式, 在此我们有必要介绍这种算法. 最基本的结论是

$$f = \frac{A_{ie_i}}{(x - \alpha_i)^{e_i}} + \cdots + \frac{A_{i2}}{(x - \alpha_i)^2} + \frac{A_{i1}}{(x - \alpha_i)^1} + \cdots$$

的 $A'_{ij}s$ 可以经计算得到不可因式分解的 $\alpha'_i s$ 的函数。

**定理 2.7.1.** 设 $A$, $D \in K[x]$, 其中 $D$ 是首一的且非零, $gcd(A, D) = 1$, 并且设 $D$ 的一个无平方因式分解为 $D = D_1 D_2^2 \cdots D_n{}^n$。那么, 仅通过在 $K$ 上的有理运算, 我

们可以计算 $H_{ij} \in K[x]\ 1 \leq j \leq i \leq n$ 得到 $A/D$ 的部分分式分解为

$$\frac{A}{D} = P + \sum_{i=1}^{n} \sum_{\alpha | D_i(\alpha)=0} \left(\frac{H_{ii}(\alpha)}{(x-\alpha)^i} + \cdots + \frac{H_{i1}(\alpha)}{x-\alpha}\right)$$

其中，$P$ 是 $A$ 除以 $D$ 得到的商。

定理 2.7.1 的证明. 我们首先来描述所有 $H_{ij}$ 的结构：设 $i \in 1, \ldots, n$，$E_i = D/D_i^i$，

$$h_i = \frac{A}{u^i E_i} \in K(x)<u>$$

其中 $u$ 是 $K(x)$ 上的一个微分变量（例如，$u$ 和它的所有导数 $u', u'', \ldots$ 是 $K(x)$ 上的独立变元）。根据无平方因子分解的定义，每一个 $D_i$ 都是无双的且与其它所有 $D_k$ 互质，所以 $gcd(E_i, D_i) = gcd(D_i', D_i) = 1$。如此，使用扩张欧几里得算法来计算 $B_i, C_i \in K[x]$，可得

$$B_i E_i \equiv 1 (\mod D_i) \text{ and } C_i D_i' \equiv 1 (\mod D_i). \tag{2.4}$$

$j = 1, \ldots, i$，计算 $h_i^{i-j}/(i-j)!$ 并写成

$$\frac{h_i^{i-j}}{(i-j)!} = \frac{P_{ij}(x, u, u', u'', \ldots, u^{(i-j)})}{u^{2i-j} E_i^{i-j+1}} \tag{2.5}$$

其中 $P_{ij}$ 是一个系数在 $K$ 中的多项式。然后

$$Q_{ij} = P_{ij}(x, D_i', \frac{D_i''}{2}, \frac{D_i^3}{3}, \cdots, \frac{D_i^{(i-j+1)}}{i-j+1}) \in K[x]$$

最后

$$H_{ij} = Q_{ij} B_i^{i-j+1} C_i^{2i-j} (\mod D_i) \tag{2.6}$$

其中 $B_i$ 和 $C_i$ 由 (2.10) 给出。

我们现在已经证明了由 (2.12) 给出的所有 $H_{ij}$ 符合该定理。设 $\overline{K}$ 是 $K$ 的代数封闭，$\alpha \in \overline{K}$ 是 $D_i$ 的一个根，$D_{i,a} = D_i/(x-\alpha)$，

$$h_{i,a} = \frac{A}{D_{i,a}^i E_i} = \frac{A}{D}(x-\alpha)^i$$

既然 $h_{i,a}$ 仅是 $h_i$ 在 $u = D_{i,a}$ 计算得到的，我们有

$$\frac{h_{i,a}^{(i-j)}}{(i-j)!} = \frac{P_{ij}(x, D_{i,a}, D_{i,a}', D_{i,a}'', \ldots, D_{i,a}^{(i-j)})}{D_{i,a}^{2i-j} E_i^{i-j+1}}$$

其中 $P_{ij}$ 就像在 (2.11) 中是与之相同的多项式。我们有 $D_i = (x-\alpha)D_{i,a}$，所以对于 $k > 0$，

$$D_k^{(k)} = \sum_{j=0}^{k} \binom{k}{j}(x-\alpha)^{(j)} D_{i,a}^{(k-j)} = (x-\alpha)D_{i,a}^{(k)} + k D_{i,a}^{(k-1)}$$

*2 有理函数积分*

对于 $j > 1$ 有 $(x-\alpha)^{(j)} = 0$。因此，

$$\frac{D_i^{(k)}(\alpha)}{k} = D_{i,a}^{(k-1)}(\alpha)$$

其中 $k > 0$，这意味着

$$Q_{ij}(\alpha) = P_{ij}(\alpha, D_i'(\alpha), \frac{D_i''(\alpha)}{2}, \frac{D_i^{(3)}(\alpha)}{3}, \ldots, \frac{D_i^{(i-j+1)}(\alpha)}{i-j+1}) \quad (2.7)$$

$$= P_{ij}(\alpha, D_{i,a}(\alpha), D_{i,a}'(\alpha), D_{i,a}''(\alpha), \ldots, D_{i,a}^{(i-j)}(\alpha)) \quad (2.8)$$

另外，(2.10) 意味着 $B_i(\alpha) = 1/E_i(\alpha)$ 和 $C_i(\alpha) = 1/D_i'(\alpha)$，所有

$$H_{ij}(\alpha) = Q_{ij}(\alpha)B_i(\alpha)^{i-j+1}C_i\alpha^{2i-j} = \frac{h_{i,a}^{(i-j)}(\alpha)}{(i-j)!}$$

$x - \alpha$ 不能整除 $h_{i,a}$ 的分母，$h_{i,a}$ 在 $x = \alpha$ 出由泰勒级数，有泰勒公式可得，

$$h_{i,a} = \sum_{k \geq 0} \frac{h_{i,a}^{(i-j)}(\alpha)}{k!}(x-\alpha)^k$$

所以 $A/D$ 在 $x = \alpha$ 处的劳伦级数为

$$\frac{A}{D} = \frac{h_{i,a}}{(x-\alpha)^i} = \sum_{j=1}^{i} \frac{h_{i,a}^{(i-j)}(\alpha)}{(i-j)!} \frac{1}{(x-\alpha)^j} + \cdots = \sum_{j=1}^{i} \frac{H_{ij}(\alpha)}{(x-\alpha)^j} + \cdots$$

这样我们就证明了这个定理。$\qquad\square$

这个定理给我们提供了一种计算有理函数的劳伦级数展开式的算法。我们做进一步的推断：这是可能的，找到 $i$，$j$'s 使得 $G_{ij} = \gcd H_{ij}, D_i$ 是非平凡的。这意味着，对于 $G_{ij}$ 的一个根 $\alpha$，在 $A/D$ 的展开式中 $1/(x-\alpha)^j$ 的系数是零。这时，我们用 $D_{ij} = D_i/G_{ij}$ 代替 $D_i$，然后代回 $A/D$ 的部分分式展开式

$$\frac{A}{D} = P + \sum_{i=1}^{n} \sum_{j=1}^{i} \sum_{\alpha | D_i(\alpha)=0} \frac{H_{ij}(\alpha)}{(x-\alpha)^j}$$

中，其中所有被加数都是确保非零的。

---

**LaurentSeries**$(A, D, F, n)$  (* $F$ 关于 $A/D$ 的完全部分分式分解的基值 *)
(* 给定一个特征为 0 的域 $K$, $A, D\ F \in K[x]$, $D$ 是首一非零的, 且与 $A$ 互质, $F$ 是 $D$ 的无平方因子分解的 $n$ 阶因子, 返回 $A/D$ 在 $F$ 中任意非零处的劳伦级数的主要部分 *)

  **if** $deg(F) = 0$ **then return** $0$
  **end if**
  $u \leftarrow$ 微分未知元, $\sigma \leftarrow 0$
  $E \leftarrow D/F^n, h \leftarrow A/(u^n E)$
  $(B, G) \leftarrow$ **ExtendedEuclidean**$(E, F, 1)$     (* $BE + GF = 1$ *)
  $(C, G) \leftarrow$ **ExtendedEuclidean**$(F', F, 1)$     (* $CF' + GF = 1$ *)
  **for** $j \leftarrow 0$ **to** $n - 1$ **do**
    $P \leftarrow u^{n+j} E^{1+j} h$     (* $P \in K[x, u, u', u'', \ldots, u^{(j)}]$ *)
    $v_j \leftarrow F^{(j+1)}/(j+1)$
    $Q \leftarrow$ **eval**$(P, u \leftarrow v_0, \ldots, u^{(j)} \leftarrow v_j)$
    $h \leftarrow h'/(j+1)$
    **if** $deg(F^*) > 0$ **then**
      $H \leftarrow QB^{1+j}C^{n+j} \bmod F^*$
      $\sigma \leftarrow \sigma + \sum_{F^*(\alpha)=0} H(\alpha)/(x-\alpha)^{(n-j)}$
    **end if**
  **end for**
  **return**$\sigma$

---

**例 2.7.1.** 考虑
$$f = \frac{36}{(x-2)(x^2-1)^2} \in \mathbb{Q}(x)$$
运用劳伦级数算法, $A = 36$, $D = (x-2)(x^2-1)^2$, $F = x^2 - 1$, $n = 2$, 我们可以得到:

**1.** $\sigma = 0, E = x - 2, h = 36/(u^2(x-2))$,

**2.** $(x^2-1)/3 - (x/3 + 2/3)(x-2) = 1$, 所以 $B = -(x+2)/3$,

**3.** $(x/2)2x - (x^2-1) = 1$, 所以 $C = x/2$,

**4.** $P = u^2(x-2)h = 36$,

**5.** $v_0 = F' = 2x$,

**6.** $Q = eval(36, u \leftarrow v_0) = 36$, 所以 $gcd(x^2-1, Q) = 1$,

**7.** $F^* = x^2 - 1$,

**8.** $H = -36(x+2)/3(x/2)^2 \bmod x^2-1 = -3x-6$, 所以 $\sigma = \sum_{\alpha^2-1=0}(-3\alpha-6)(x-\alpha)^2$,

**9.** $h = h' = ((-72x+144)u' - 36u)/(u^3(x-2)^2)$,

**10.** $P = u^3(x-2)^2 h = (-72x + 144)u' - 36u$,

**11.** $v_1 = F''/2 = 1$,

**12.** $Q = eval(P, u \leftarrow v_0, u' \leftarrow v_1) = -144x + 144$, 所以 $gcd(x^2 - 1, Q) = x - 1$,

**13.** $F^* = (x^2 - 1)/(x - 1) = x + 1$,

**14.** $H = (-144x+144)((x+2)/3)^2(x/2)^3 \bmod (x+1) = -4$, 所以 $\sigma = \sigma + \sum_{\alpha+1=0} -4/(x-\alpha)$

因此，$f$ 在 $x^2 - 1 = 0$ 的根上的劳伦级数的和为

$$\frac{36}{x^5 - 2x^4 - 2x^3 + 4x^2 + x - 2} = \left(\sum_{\alpha^2-1=0} \frac{-3\alpha - 6}{(x-\alpha)^2}\right) - \frac{4}{x+1} + \cdots$$

---

**FullPartialFraction**($f$)  (\* $f$ 的完全部分分式分解 \*)
(\* 给定一个特征为 0 的域 $K$，$f \in K(x)$，返回 $f$ 的完全部分分式分解 \*)
  $D \leftarrow denominator(f)$
  $(Q, R) \leftarrow$ **PloyDivide**($numerator(f), D$)
  $(D_1, \ldots, D_m) \leftarrow$ **Squarefree**($D$)
  $\mathbf{return}(Q + \sum_{i=1}^{m} \mathbf{LaurentSeries}(R, D, D_i, i))$

---

**例 2.7.2.** 对下式运用完全部分分式分解

$$f = \frac{36}{x^5 - 2x^4 - 2x^3 + 4x^2 + x - 2} \in \mathbb{Q}(x)$$

得到：

**1.** $D = x^5 - 2x^4 - 2x^3 + 4x^2 + x - 2$,

**2.** $(Q, R) = $ **PloyDivide**$(36, D) = (0, 36)$,

**3.** $D_1 D_2^2 = $ **SquareFree**$(D) = (x-2)(x^2-1)^2$,

**4.** **LaurentSeries**$(36, D, x-2, 1)$ returns $4/(x-2)$,

**5.** **LaurentSeries**$(36, D, x^2-1, 2)$ returns

$$\left(\sum_{\alpha^2-1=0} \frac{-3\alpha - 6}{(x-\alpha)^2}\right) - \frac{4}{x+1}$$

这正如例 2.7.1 中所示。

因此，$f$ 的完全部分分式分解为

$$\frac{36}{x^5 - 2x^4 - 2x^3 + 4x^2 + x - 2} = \left(\sum_{\alpha^2-1=0} \frac{-3\alpha - 6}{(x-\alpha)^2}\right) - \frac{4}{x+1} + \frac{4}{x-2} \quad (2.9)$$

---

**IntegrateRationalFunction**($f$)　(* 完全部分分式积分 *)
(* 给定一个特征为 0 的域 $K$，$f \in K(x)$，返回 $\int f dx$ 的完全部分分式分解 *)

$$P + \sum_{i=1}^{n} \sum_{j=1}^{i} \sum_{\alpha|D_i(\alpha)=0} \frac{H_{ij}(\alpha)}{(x-\alpha)^j} \leftarrow \textbf{FullPartialFraction}(f)$$

$$\textbf{return} \int P \ + \ \sum_{i=1}^{n} \sum_{\alpha|D_i(\alpha)=0} H_{i1}(\alpha) log(x-\alpha)$$

$$+ \sum_{i=2}^{n} \sum_{j=2}^{i} \sum_{\alpha|D_i(\alpha)=0} \frac{H_{ij}(\alpha)}{(1-j)(x-\alpha)^{j-1}}$$

---

**例 2.7.3.** 对于例 2.7.2 的分数 $f$，完全部分分式分解返回 (2.13)，因此 $f$ 的积分是

$$\int \frac{36}{x^5 - 2x^4 - 2x^3 + 4x^2 + x - 2} dx \ = $$

$$4log(x-2) \ - \ 4log(x+1) + \sum_{\alpha^2-1=0} \frac{3\alpha+6}{x-\alpha}$$

大家可以将此算法对比一下之前小节中对于相同积分返回 (2.9) 的算法。

尽管积分结果被返回成含有分数 $v$ 的 (2.4) 的形式，当然也可以被展开成含有线形分母的部分分数，但是相比这章出现的其它算法并没有更多的优势。无论如何，此算法实现了部分分式算法的无因子化。如此，这些有理分数积分的算法只能在有理运算下被执行。

## 2.8 雷欧布的实有理函数算法

上述算法返回一个实基元，给出了一个 (2.19) 的表达式。但是积分算法返回了下列形式

$$\sum_{\alpha|R(\alpha)=0} \alpha \log S(\alpha, x)$$

在上式中 $R \in K[t]$ 无平方因子分解，且 $S \in K[t,x]$。为了完成 Rioboo 算法，我们需要将上式中的和转化成 (2.19) 中的复数对数，这是实数在子域中的代数衍生。任何一个实数域有一个实闭包。这个定理也在 [92] 中给出证明，11.6 节的区别在于可数实域。注意到 $K$ 的闭包不是唯一的，也不全纯，除非 $K$ 已经排好序了。

## 2 有理函数积分

**定理 2.8.1.** *([54],Chap,XI).* 设 $L$ 为实闭域，那么

**(i)** $L$ 存在唯一序，由下式给出:$x > 0 \Leftrightarrow x = y^2$ 对于某一个 $y \in L$

**(ii)** $L(\sqrt{-1})$ 是 $L$ 的代数闭包

设 $K$ 是为上述部分剩下的实域，设 $\mathbb{K}$ 为 $K$ 的代数闭包且 $\bar{K} = \mathbb{K}(i)$ 当 $i^2 = -1$. 我们说 $\alpha \in \bar{K}$ 为真当 $\alpha \in \bar{K}$. 设 $f$ 是 (2.21) 形式的和，$R = \Sigma_j r_j x^j \in K[x]$, $S = \Sigma_{j,k} s_{jk} t_j x_k \in K[t,x]$, 并且设 $u,v$ 在 $K(x)$ 上未确定. 我们首先拆分和式 (2.21) 到一个实根和一个复根上：

$$f = g + \Sigma_{\alpha \notin \mathbb{K}, R(\alpha)=0} \alpha \log\left(S(\alpha, x)\right)$$

其中

$$g = \Sigma_{\alpha \in \mathbb{K}, R(\alpha)=0} \alpha \log\left(S(\alpha, x)\right)$$

为实方程，我们接下来计算 $P, Q \in K[u,v]$, 满足

$$R(u+iv) = \Sigma_j r_j (u+iv)^j = P(u,v) + iQ(u,v)$$

并且

$$S(u+iv, x) = \Sigma_{j,k} s_{jk}(u+iv)^j x^k = A(u,v,x) + iB(u,v,x)$$

因为 $\bar{K} = \mathbb{K}(i)$，它是二维向量空间中关于基 $(1,i)$ 的向量空间。所以对于 $\alpha \in \bar{K}$, $R(\alpha) = 0$ 当且仅当 $P(a,b) = Q(a,b) = 0$ 其中 $\alpha = a + ib$。目前为止，$\alpha \notin \mathbb{K}$ 当且仅当 $b \neq 0$。因此我们可以化简 (2.22)

$$f = g + \Sigma_{a,b \in \mathbb{K}, b \neq 0, P(a,b)=Q(a,b)=0} (a+ib) \log\left(S(a+ib, x)\right)$$

设 $\sigma$ 属于 $\bar{K}$ 满足 $\sigma(i) = -i$ 和 $\sigma(z) = z$ 对于任意 $z \in \mathbb{K}$, 定义 $\bar{\sigma} : \bar{K}[x] \to \bar{K}[x]$ 通过 $\bar{\sigma}(\Sigma a_j x^j) = \Sigma \sigma(a_j) x^j$。设 $a, b \in \mathbb{K}$.

$$A(a,b,x) - iB(a,b,x) = \bar{\sigma}(A(a,b,x) + iB(a,b,x)) = \bar{\sigma}(S(a+ib, x))$$
$$= S(\sigma(a+ib), x) = S(a-ib, x)$$

对于 $\sigma$

$$P(a,b) - iQ(a,b) = \sigma(P(a,b) + iQ(a,b))$$
$$= \sigma(R(a+ib)) = R(\sigma(a+ib)) = R(a-ib)$$

从而我们导出了 $R(a+ib) = 0$ 当且仅当 $R(a-ib) = 0$。因此，对于任意 $(a,b)$, $(a,-b)$ 也必须在和 (2.25) 中出现，所以我们可以把 (2.25) 重写成

$f = g + \Sigma_{a,b \in \mathbb{K}, b > 0, P(a,b)=Q(a,b)=0} \{(a+ib) \log S(a+ib, x) + (a-ib \log\left(S(a-ib, x)\right))\}$
$= g+$
$\Sigma_{a,b \in \mathbb{K}, b>0, P(a,b)=Q(a,b)=0} \{a(\log\left(A(a,b,x) + iB(a,b,x)\right)) + \log\left(A(a,b,x) - iB(a,b,x)\right)\}$

因此
$$f = g + h + \Sigma_{a,b\in\mathbb{K},b>0,P(a,b)=Q(a,b)=0}ib\log\left(\frac{A(a,b,x)+iB(a,b,x)}{A(a,b,x)-iB(a,b,x)}\right)$$

其中
$$h = \Sigma_{a,b\in\mathbb{K},b>0,P(a,b)=Q(a,b)=0}a\log\left(A(a,b,x)^2+B(a,b,x)^2\right)$$

为实函数。因为剩余的非实数部分都符合形式 (2.19)。我们可以使用定理 2.8.1 和它的对应算法将它们转化成实函数。注意，因为转化 (2.19) 到实函数要求计算 A 和 B 的最大公约式，我们在理论上可以使用 LogToAtan 计算。

**定理 2.8.2.** *([72]).* 设 $K$ 为实域，$\mathbb{K}$ 为 $K$ 的实闭包，$C, D \in K[x]$ 满足 $\deg(D) > 0, \deg(D) > \deg(C)$，$D$ 无平方因子分解且 $gcd(C,D)=1$，假设 $R, S$..

证明.. 对于某一个序列，我们可以应用 Rioboo 的定理

$$i\log\left(\frac{A(u,v,x)+iB(u,v,x)}{A(u,v,x)-iB(u,v,x)}\right)$$

其中 $u, v$ 独立且非确定性，观察到实函数 $\phi(u,v,x)$。我们可以把 (2.26) 改写成

$$f = g + h + \Sigma_{a,b\in\mathbb{K},b>0,P(a,b)=Q(a,b)=0}b\phi(a,b,x)$$

其中定理 2.8.4 保证 $\phi(u,v,x)$ 的存在性。通过把答案表示成和形式，我们不需要实际上解决这个系统，或者去引入任何的代数数字。实际上，每当 $P(u,v) = Q(u,v) = 0$ 的实根可以被计算的时候，他可以被更有效地计算第一个实根，然后调用 LogToAtan，而不是应用规约到更一般的问题。

**LogToReal**$(R, S)$   (* Rioboo *)

change $R(u+iv)$ into $P(u,v) + iQ(u,v)$
change $S(u+iv,x)$ into $A(u,v,x) + iB(u,v,x)$
return   $\Sigma_{a,b\in\mathbb{K},b>0,P(a,b)=Q(a,b)=0}a\log(A(a,b,x)^2+B(a,b,x)^2)$   $+$
$b$**LogToAtan**$(A,B)(a,b,x) + \Sigma_{\alpha\in\mathbb{K},R(\alpha)=0}a\log(S(a,x))$

**例 2.8.1.** 应用 **LogToReal** 到积分 (2.15)，我们有

$$R(t) = 4t^2+1 \in Q[t], S(t,x) = x^3 + 2tx^2 - 3x - 4t \in Q[t,x]$$

和 1. $R(u+iv) = 4(u+iv)^2+1 = 4u^2-4v^2+1+8iuv$，所以 $P = 4u^2-4v^2+1$ 和 $Q = 8uv$.
2. $S(u+iv), x = x^3 + 2(u+iv)x^2 - 3x - 4(u+iv) = x^3 + 2ux^2 - 3x - 4u + i(2vx^2 - 4v)$，所以 $A = x^3 + 2ux^2 - 3x - 4u$ 和 $B = 2vx^2 - 4v$  3. $H = resultant_v(p,q) = 256u^4 + 64u^2$ 只有 0 实根。$P(0,v) = 1 - 4v^2$ 只有 $\frac{1}{2}$ 这一个正实根。 4. $A(0,1/2,x) = x^3 - 3x$，$B(0,1/2,x) = x^2 - 2$ 和 **LogToAtan**$(x^3-3x, x^2-2)$ 返回

$$2arctan\left(\frac{x^5-3x^3+x}{2}\right) + 2arctan(x^3) + 2arctan(x)$$

不去解系统中的 $P(u,v) = Q(u,v) = 0$，我们可以调用 **LogToAtan**($\mathbf{x^3 + 2ux^2 - 3x - 4u, 2vx^2}$
返回

$$\phi(u,v,x) = \arctan(\frac{x}{2v} + \frac{u}{v})$$
$$+ 2\arctan(\frac{x^3}{2v} + \frac{2u}{v}x^2 + \frac{4u^2+4v^2-1}{2v}x - \frac{u}{v})$$
$$+ 2\arctan(\frac{x^5}{4v} + \frac{u}{v}x^4 + \frac{u^2+v^2-1}{v}x^3 - \frac{3u}{v}x^2 - \frac{8u^2+8v^2-3}{4v}x + \frac{u}{v})$$

积分的返回值

$$\int \frac{x^4 - 3x^2 + 6}{x^6 - 5x^4 + 5x^2 + 4}dx = \Sigma_{a,b\in\mathbb{R},b>0,4a^2-4b^2+1=8ab=0} b\phi(a,b,x)$$

是实函数, 带入 $a = 0$ 和 $b = \frac{1}{2}$，我们得到了和上面一样的解

---

**IntegrateRealRationalFunction**($f$)    (* Real rational function integration *)

$v + \Sigma_{i=1}^m \Sigma_{\alpha|R_i(\alpha)=0} a\log S_i(a,x) \leftarrow$ **IntegrateRealRationalFunction**($f$)
return($v + \Sigma_{i=1}^m$**LogToReal**($R_i, S_i$))

---

## 2.9 内场积分

在这一节中, 我们概述一个积分算法的微变形式. 这个变形被用于判定一个是函数是否是下列情况的种:
. 一个有理函数的导数
. 一个有理函数的导数的对数
这些问题的重要性源自更普遍函数的积分. 更进一步, 判定一个实函数是否是一个对数微分对解一个实函数系的的常微分方程式非常有帮助的.

可识别的导数:
第一个问题是给定一个 $f \in K(x)$ 中的函数, 确定是否存在 $u \in K(x)$ 使得 $du/dx = f$. 为了计算 $u$, 我们只要使用霍洛维兹-奥斯特罗格拉茨算法或者一下艾尔米特约化的变形, 对于 $f$, 我们得到 $g \in K(x)$ 和 $A, D \in K(x)$ 其中 D 是无双的, 使得 $f = dg/dx + A/D$. 在这一点上,$u \in K(x)$ 当且仅当 $D|A$ 时满足. 这里 $u = g + \int (A/D)dx$

这里同时存在一组可以用来在不对 $f$ 积分的条件下来判断 $f$ 是否是一个有理函数得微分的方法:
. 计算 $f$ 的分母的无双因式分解 $D_1 D_2^2..D_n^n$, 且对每个 i，多项式 $H_{i1}$ 属于定理 2.7.1 中的 $K[x]$, 在这里我们用了劳伦级数算法。我们记 $D_i = G_i E_i$ 其中 $G_i = gcd(H_{i1}, D_i)$ 且 $gcd(E_i, H_{i1}) = 1$。众所周知：$f$ 在某个根 $G_i$ 处的余数为 0, 且在 $E_i$ 的一个根 $\alpha$ 处为 $H_{i1}(\alpha)$ 非 0, 故 $f$ 为一个实函数的衍生式当且仅当对任意 i 满足 $E_i = 1$, 这也就是说对每个 i 我们有 $D_i|H_{i1}$。
. 计算 $f$ 的分母的无双因式分解 $D_1 D_2^2..D_n^n$, 且将 $f$ 写为如下的和式:
$f = \sum_{i=1}^n \frac{A_i}{D_i}$

若 $f$ 是一个实函数的产生式 ($Derivatives$)，则 $D_i|A_i$，此外在 $f$ 在 $D_1$ 的根处的余数将非 0。若 $D_i|A_i$，则 $f$ 为一个实函数的衍生式当且仅当对所有 $i>1$ 的情形有 $A_i/D_i^i$ 也是一个实函数的衍生式。同时我们应用马利克判定，该判定说明 $A/D_i$ 在 $m>1$ 的情形下，是一个实函数的派生式当且仅当：

$D|Wronskian(\frac{dD}{dx}, \frac{dD^2}{dx}, \ldots, \frac{dD^{m-1}}{dx}, A)$

然而这些判定不是艾尔米特约化或者霍洛维兹-奥斯特洛格拉茨基算法的实用方法。我们只是非常理论的关注。现在还没有找到对大多数函数普遍适用的一般性判定方法。而正是如此在一般情况下判定衍生式的问题是非常难的。（见 5.12 小节）

判定对数的衍生式

第二个问题是，给定一个 $f \in K(x)$，确定这里是否存在 $u \in K(x)*$ 满足 $du/dx = uf$. 我们将会在后面（见练习 4.2）给出证明：

$f$ 是一个实函数的对数衍生式当且仅当 $f = A/D$ 其中 $D$ 是无双的，$det(A) < det(D), gcd(A,D) = 1$，且所有罗斯坦-杰拉夫算法产生的根为整数。如果这样的话，任何罗斯坦-杰拉夫算法，拉扎德-雷布欧-特拉杰算法和次秋斯基算法的产生式 $u \in K(x)$ 满足 $du/dx = uf$。

## 2.10 习题

**习题 2.10.1.** *用埃尔米特归约和罗斯坦-特拉杰算法计算:*

$$\int \frac{x^5 - x^4 + 4x^3 + x^2 - x + 5}{x^4 - 2x^3 + 5x^2 - 4x + 4} dx$$

**习题 2.10.2.** *用拉扎德-雷布欧-特拉杰算法计算:*

$$\int \frac{8x^9 + x^8 - 12x^7 - 4x^6 - 26x^5 - 6x^4 + 30x^3 + 23x^2 - 2x - 7}{x^10 - 2x^8 - 2x^7 - 4x^6 + 7x^4 + 10x^3 - 4x - 2} dx$$

**习题 2.10.3.** *a) 用罗斯坦-特拉杰算法或者拉扎德-雷布欧-特拉杰算法计算:*

$$\int \frac{72x^7 + 256x^6 - 192x^5 - 1280x^4 - 312x^3 + 1440x^2 + 576x - 96}{9x^8 + 36x^7 - 32x^6 - 252x^5 - 78x^4 + 486x^3 + 288x^2 - 108x+)} dx$$

并用这个结果计算 $-2 \leq x \leq -2/3$ 的象征性定积分 ($symbolic definite integral$)，同时将其与直接用数值方法计算的结果进行比较。

*b) 对上述结果应用雷布欧算法并在此计算其在 $-2 \leq x \leq -2/3$ 的定积分。*

**习题 2.10.4.** *a) 计算*

$$\int \frac{dx}{1+x^4}$$

*b) 为 $\int \frac{dx}{1+x^n}$ 寻找一个闭模式 ($closed form$) 其中 $n \in N$。*

**习题 2.10.5.** ([66]) *计算*

$$\int \frac{x^4 + x^3 + x^2 + x + 1}{x^5 + x^4 + 2x^3 + 2x^2 - 2 + 4\sqrt{-1+\sqrt{3}}} dx$$

## Chapter 2.1.3 微分域

在这一章节中我们研究积分算法可以实现并能够被证明其正确性的代数结构。主要的思想来源于 J.F.Ritt，在纯代数结构中定义微分的概念（换句话说，不使用函数，极限，正切线这些分析中的概念），并且研究这样定义的微分在任一对象上作用的性质。用这种方法，之后我们可以将积分问题转换为在某个代数结构上的解方程问题，解方程问题可以通过代数算法解决。因为任意的超越函数都可以看做是在具有微分结构的域上的一个单变量有理函数，首先我们需要研究环上和域上微分结构的一般性质。这使得我们可以将有理函数的积分算法推广到到非常广泛的超越函数类上。

**3.1 微分**

尽管之后章节中的积分算法仅仅在特征为0的微分域上能够使用，但是这一章节的前两小节的讨论的对象是具有任意特征的环和域。如果在任一环上的映射满足通常的规则，那么我们称其为一个微分

定义 3.1.1. $R$ 是一个环（域），$R$ 上的一个微分是一个映射 $D: R \to R$，满足下列两条性质，对于 $\forall a, b \in R$
$(i) D(a+b) = Da + Db$
$(ii) D(ab) = aDb + bDa$
则 $(R, D)$ 叫做一个微分环（域）。

集合 $\text{Const}_D(R) = \{a \in R \ such \ that \ Da = 0\}$ 叫做 $R$ 关于映射 $D$ 的常数子环（子域）。
如果 $S$ 是 $R$ 的子环（域）并且 $DS \subseteq S$，那么 $S \subseteq R$ 就叫做 $R$ 的一个微分子环（域）。
如果微分不具有多值性，我们常常称 $R$ 是微分环（域），而不是说 $(R, D)$ 是微分环（域）。基于我们的定义，我们可以发现通常在分析中的微分的代数性质，对于我们定义的微分也成立。

定理 3.1.1. $(R, D)$ 是一个微分环（域）
$(i) D(ca) = cDa \ \ \forall a \in R, c \in \text{Const}_D(R)$
$(ii)$ 如果 $R$ 是一个域，
$$D\frac{a}{b} = \frac{bDa - aDb}{b^2} \ \ \forall a, b \in R, b \neq 0$$
$(iii) \text{Const}_D(R)$ 是 $R$ 的微分半环（半域）
$(iv) Da^n = na^{n-1} \ \ \forall a \in R\backslash\{0\}, \forall n \in \mathbb{N}$
$(v)$ 对数的微分特征：如果 $R$ 是一个整环，那么
$$\frac{D(u_1^{e_1} \cdots u_n^{e_n})}{u_1^{e_1} \cdots u_n^{e_n}} = e_1 \frac{Du_1}{u_1} + \cdots + e_n \frac{Du_n}{u_n}$$
$\forall u_1, \ldots, u_n \in R^*, \forall e_1, \ldots, e_n \in \mathbb{N}$
$(vi)$
$$DP(u_1, \ldots, u_n) = \sum_{i=1}^{n} \frac{\partial P}{\partial X_i}(u_1, \ldots, u_n) Du_i$$
$\forall u_1, \ldots, u_n \in R^*, \forall e_1, \ldots, e_n \in \mathbb{N}$

总之，一个环可以在其上有不止一个的微分定义，例如$Q[X,Y]$至少有微分$0, d/dX$和$d/dY$。但是它有更多的微分定义，例如$D = d/dX + d/dY$。实际上，任意以$R$中元素为系数的微分的线性组合也是$R$上的一个微分。

引理 3.1.1. $R$上所有微分映射的集合$\Omega(R)$是一个左$R-$模。

定义 3.1.2. $(R,D)$为一个微分环，如果$DI \subseteq I$，那么$R$的理想$I$称作微分理想。

引理 3.1.2. $(R,D)$为微分环，$I$是$R$的一个微分理想，$\pi: R \to R/I$为典范投影，那么$D$诱导出$R/I$上的一个微分$D^*$，使得$D^* \circ \pi = \pi \circ D$。

例 3.1.1. $R$为任意一个环，$D$是$R$上的零映射，那么$R$的任意一个理想是一个微分理想，诱导映射$D^*$是$R/I$上的零映射。

例 3.1.2. 令$X$为不定元，$D$为$R = \mathbb{Q}[X]$上的$d/dX$。$R$仅有的微分理想为$(0)$和$(1)$，诱导映射分别为$D$和零映射。

例 3.1.3. $(R,D)$为微分环，$X$是不定元，$\Delta : R[X] \to R[X]$是由$\Delta(\sum_n a_n X^n) = \sum_n (Da_n + na_n)X^n$定义的映射。可以验证$\Delta$是$R[X]$上的一个微分映射，$\forall m \in \mathbb{N}$，理想$I_m = (X^m)$是一个微分理想。$m = 1$，映射$\pi : R[X] \to R[X]/(X) \simeq R$是$X \to 0$的替代，$R$上的诱导映射$\Delta^*$满足
$$\Delta^*\pi(p) = \pi(\Delta p) = D(p(0)) \quad \forall p \in R[X]$$
所以在$R$上$\Delta^* = D$。

## 3.2 微分扩张

在这一部分，我们考虑如何讲一个给定的微分扩展到一个更大的环或域上。像前一小节一样，本小节中的环和域也都是有任意的特征。在古典代数中，为了得到一个更大的环，会将方程的解或者新的不定元加入到已有的环中。一个自然的问题就出现了，如果原始的环上定义了一个微分$D$，那么$D$可以扩展为新的更大的环上的一个新的微分吗？如果答案是肯定的，并且新的微分和原有的微分$D$相容，我们就说新的微分环是要有微分环的一个微分扩张。下面的定义说明了"和$D$相容"的概念。

定义 3.2.1 $(R,D)$和$(S,\Delta)$是微分环，如果$R$是$S$的一个子环并且$\Delta a = Da, \forall a \in R$，那我们称$(S,\Delta)$是$(R,D)$的一个微分扩张。

首先，我们证明整环上的任意微分能够唯一的扩张到自身的分式域，这个扩张是由通常的分式微分的规则定义的。

定理 3.2.1. $R$是一个整环，$F$是$R$的分式域，$D$是$R$上的微分，那么存在$F$上唯一的微分$\Delta$使得$(F,\Delta)$是$(R,D)$的微分扩张。

定义 3.2.2. $R$是一个环，$X$是$R$上的不定元。$D$是$R$上任意的一个微分，$D$在系数上的作用由映射$\kappa_D : R[X] \to R[X]$定义
$$\kappa_D(\sum_{i=0}^n a_i X^i) = \sum_{i=0}^n (Da_i)X^i$$
映射$\kappa_D$将微分$D$作用于$R$上多项式的每一个系数，注意到在$\kappa_D$作用下，多项式的次数

可能变化。

引理 3.2.1. $\kappa_D$是$R[X]$上的一个微分。

如果$R$是一个整环，那么$R[X]$是一个整环，所以，根据定理 3.2.1, $\kappa_D$可以唯一的扩展为其分式域上$R(X)$的一个微分，我们也将这个到$R(X)$的扩展记为$\kappa_D$。

引理 3.2.2. $(R,D)$为一个微分环，$(S,\Delta)$是$(R,D)$的一个微分扩张，$X$是$R$上的一个不定元。那么，

$$\Delta(P(\alpha)) = \kappa_D(P)(\alpha) + (\Delta\alpha)\frac{dP}{dX}(\alpha) \quad \forall \alpha \in S, \forall P \in R[X]$$

现在我们可以证明关于微分扩张的主要结论：$F(t)$是微分域$(F,D)$的一个简单扩张，如果$t$是$F$上的代数元，那么$D$扩展到$F(t)$的方式是唯一的，否则$D$可以不同的方式扩展到$F(t)$，但是从中选择$Dt$的值后，就可以使扩张唯一。我们在两个定理中证明这个结论，一个是针对超越元的情况，一个是针对代数元的情况。

定理 3.2.2. $(F,D)$是一个微分域，$t$是$F$上的超越元，那么$\forall w \in F(t)$，存在$F(t)$上的唯一微分$\Delta$，使得$\Delta t = w$，并且$(F(t),\Delta)$是$(F,D)$的一个微分扩张。

例 3.2.1. $F$为任意的一个域，$0_F$是$F$上零映射，$x$是$F$上的超越元。令$D$为$0_F$到$F(x)$的一个扩张，满足$Dx = 1$。因为$(F(x), d/dx)$是$(F, 0_F)$的一个微分扩张，并且$dx/dx = 1$，定理 3.2.2. 证明了$D = d/dx$，换言之，$F(x)$上的唯一微分为$F$上的$0$映射，$d/dx$将$x$映为 $1$.

例 3.2.2. $(F,D)$是一个微分域，$t$是$F$上的超越元，令$\Delta$为$D$到$F(t)$使得一个扩张，并且满足$\Delta t = 0$。因为$(F(t), \kappa_D)$是$(F,D)$的一个微分扩张，并且$\kappa_D t = 0$，定理 3.2.2 证明了$\Delta = \kappa_D$，换言之，$\kappa_D$是$D$到$F(t)$的唯一扩张，且在$\kappa_D$意义下$t$是常数。

下面我们考虑微分域的代数扩张情形。假设$E$在$F$上是可分的对于下一个定理来说是必须的，因为可能出现$F$有非零特征的情况，同时$E$在$F$上是可分的意味着对于$E$上的任意元素在$F$上的最小不可约多项式，没有重根。在特征为$0$的情况下，代数扩张总是可分的，所以仅仅对于这种情况可以忽略可分性的假设。另外我们在证明中使用$Zorn$引理来证明非有限扩张的情形。如果只考虑有限生成的代数扩张，那么那一部分的证明可以跳过。

定理 3.2.3. $(F,D)$是一个微分域，$E$是$F$的一个可分代数扩张，那么存在$E$上的唯一微分$\Delta$使得$(E,\Delta)$是$(F,D)$的一个微分扩张。

例 3.2.3. $(F,D)$是一个特征为$0$的微分域，$C = Const_D(F)$。$\alpha$是$C$上的代数元，$P \in C[x]$是$\alpha$在$C$上的最小不可约多项式。那么$D$有到$F(\alpha)$的唯一扩张，并且一定有

$$0 = D(P(\alpha)) = \kappa_D(P)(\alpha) + (D\alpha)\frac{dP}{dX}(\alpha) = (D\alpha)\frac{dP}{dX}(\alpha)$$

所以$D\alpha = 0$，这意味着任意任意在常数域上的代数元自身在微分$D$的意义下也是一个常数。

例 3.2.4. 令$F = \mathbb{Q}(x)$，$\alpha$是$Y^2 - x \in F[Y]$的一个跟，换言之，$\alpha$代表了函数$\pm\sqrt{x}$。那么$d/dx$

有唯一到$\mathbb{Q}(X,\alpha)$的扩展，并且必须满足

$$0 = \tfrac{d}{dx}(\alpha^2 - x) = \kappa_{d/dx}(Y^2-x)(\alpha) + \tfrac{d\alpha}{dx}\tfrac{d(Y^2-x)}{dY}(\alpha) = -1 + 2\alpha\tfrac{d\alpha}{dx}$$

所以
$$\tfrac{d\alpha}{dx} = \tfrac{1}{2\alpha}$$

这是$\alpha = \pm\sqrt{x}$在通常意义下关于$x$的微分。

作为定理 3.2.3.的结论，我们可以将任意用微分扩张的迭代生成的域用一个可分的代数扩张替代，并且我们仍然有一个合理的微分扩张的迭代。

**推论 3.2.1.** $(K,D)$为一个微分域，$(F,\Delta)$为$(K,D)$的微分扩张，$\overline{F}$是$F$的代数闭包，并且$E \subseteq \overline{F}$是$K$的一个代数可分扩张。那么$D$可以唯一的扩张为$E$上的映射，$\Delta$可以唯一的扩张为$EF$上的映射，$(EF,\Delta)$是$(E,D)$的一个微分扩张。

在微分域的一个代数扩张中，微分和共轭映射可交换。这说明了迹映射和微分映射可交换，并且可以给出一个关于对数微分的迹的方程。

**定理 3.2.4.** $(K,D)$是一个微分域。
$(i)F$为$K$的可分代数扩张，那么$F$在$K$上的任意域自同构和$D$可交换。
$(ii)E$是$K$生成的有限可分代数扩张，$Tr : E \to K, N : E \to K$是$E$到$K$的迹映射和范数映射，那么$Tr$和$D$可交换，并且
$$Tr(\tfrac{Da}{a}) = \tfrac{DN(a)}{N(a)}, \forall a \in E^*$$

### 3.3 常数和扩张

在这一部分，我们研究在微分域的常数子域上的扩张的作用。首先，我们不难发现，常数在扩张之后仍为常数。

**引理 3.3.1.** $(F,D)$为一个微分域，$(E,\Delta)$是$(F,D)$的一个微分扩张。那么$\mathrm{Const}_D(F) \subseteq \mathrm{Const}_\Delta E$

在接下来的几个引理中，我们考虑能够在微分扩张中出现的新的代数常量。首先，我们先证明一个代数常量实际上必须为初始常数域上的代数元,相应地初始常数域上的可分代数元必须是一个常量。

**引理 3.3.2.** $(F,D)$是一个微分域，$(E,\Delta)$是$(F,D)$的一个微分扩张，那么
$(i)c \in \mathrm{Const}_\Delta(E)$是$F$上的代数元$\Longrightarrow c$在$\mathrm{Const}_D(F)$上是代数的
$(ii)c \in E$在$\mathrm{Const}_D(F)$上是代数的，并且是可分的$\Longrightarrow c \in \mathrm{Const}_\Delta(\mathrm{E})$

所以，在进行微分域的可分代数扩张时，新的常数都恰好是由在初始常数子域上的代数元扩张生成的。特别的是，特征为0的微分域的代数闭包的常数恰好组成了初始常数子域的一个代数闭包。

**推论 3.3.1.** $(F,D)$是一个微分域，$E$是$F$的一个可分代数扩张。令$C = \mathrm{Const}_D(F)$，同时令

$\overline{C}^E$为$E$中$C$的代数闭包，换句话说$E$中所有元素组成的子域在$C$上都是代数的。那么$D$可以唯一地扩展为$E$上的一个映射，并且$\mathrm{Const}_D(E) = \overline{C}^E$。，如果$E$是代数封闭的，那么$\mathrm{Const}_D(E)$是$C$的一个代数闭包。

作为推论 3.2.1.的一个结论，我们可以将任何以微分扩张迭代形式表示出来的域用它的代数闭包替代（如果该域的代数闭包是可分的）。如果常数域在初始迭代中保持不变，那么在替代之后，常数子域保持不变。在下一个引理中，$F$是完全域的这个假设仅仅是为了保证$F$的代数闭包在$F$上是可分的。所有特征为0的域是完全域，所以只对特征为0的域感兴趣的读者可以忽略这个假设。

引理 3.3.3. $(F,D)$是一个完全微分域，$(E,\Delta)$是$(F,D)$的一个微分扩张，$\overline{E}$是$E$的一个代数闭包，$\overline{F}$是$F$的一个代数闭包，并且满足$\overline{F} \subseteq \overline{E}$。那么$(E\overline{F},\Delta)$是$(\overline{F},D)$的一个微分扩张，并且
$$\mathrm{Const}_D(F) = \mathrm{Const}_\Delta(E) \Longrightarrow \mathrm{Const}_D(\overline{F}) = \mathrm{Const}_\Delta(E\overline{F})$$

正如预料的那样，将一个常数加入到微分域中，得到的新的常数域仅仅比初始的常数域多了这么唯一的一个元素。

引理 3.3.4. $(F,D)$是一个微分域，$(E,\Delta)$是$(F,D)$的一个微分扩张。那么$\mathrm{Const}_\Delta(F(t)) = \mathrm{Const}_D(F)(t), \forall t \in \mathrm{Const}_\Delta(E)$。

最终，我们需要一些微分代数中性质的结果包括哪些常数可以在微分扩张中保持不变。

定义 3.3.1. $(F,D)$是一个微分域，$y_1,\ldots,y_n \in F$。$y_1,\ldots,y_n$的朗斯基行列式$W(y_1,\ldots,y_n) = \det(\mathcal{M}(y_1,\ldots,y_n))$
$$M(y_1,\ldots,y_n) = \begin{pmatrix} y_1 & y_2 & \cdots & y_n \\ Dy_1 & Dy_2 & \cdots & Dy_n \\ D^{n-1}y_1 & D^{n-1}y_2 & \cdots & D^{n-2}y_n \end{pmatrix} \tag{3.1}$$

朗斯基行列式是否恒等于零（当然若恒等于零，也不一定意味着函数组线性相关）在分析中是众所周知的可以用来检验函数组的线性独立性的的方法,可以证明的是这个性质在任意的微分域上也成立。

引理 3.3.5. $(F,D)$是一个微分域，那么$y_1,\ldots,y_n \in F$，且$y_1,\ldots,y_n$在$\mathrm{Const}_D(F)$上线性相关当且仅当$W(y_1,\ldots,y_n) = 0$。

常数上的线性独立性是由常数域上的线性独立性体现出来的,因此这种线性独立的性质在微分扩张下保持。

推论 3.3.2. $(F,D)$是一个微分域，$(E,\Delta)$是$(F,D)$的微分扩张。如果$S \subset F$在$\mathrm{Const}_D(F)$上是线性独立的，那么$S$在$\mathrm{Const}_\Delta(E)$上是线性独立的。

下一个引理强调了如果一个代数系统的常数能够满足一些等式或不等式，那么代数常数也能满足这些等式或不等式。

引理 3.3.6. $(F,D)$是一个微分域，并且常数域代数封闭，$(E,\Delta)$是$(F,D)$的微分扩张，$X_1,\ldots,X_m$是$F$上的独立不定元，$g \in F[X_1,\ldots,X_m]$, $S$是$F[X_1,\ldots,X_m]$的任意一个子集。如果存在$c_1,\ldots,c_m \in \text{Const}_\Delta(E)$使得对于$\forall f \in S$，$g(c_1,\ldots,c_m) \neq 0$，并且$f(c_1,\ldots,c_m) = 0$，那么在$\text{Const}_D(F)$中也有这样的元素$c_1,\ldots,c_m$。

## 3.4 单项式扩张

我们要研究形式为$k(t)$的简单超越微分扩张，对于$k(t)$来说，微分$D$和$d/dt$有一定的相似性，这使得我们可以将有理函数的积分算法可以应用到这样的扩张上。回顾一下，如果$k$是一个微分域，$K$是$k$的一个微分扩张，$t$是$K$的一个元素，那么如果在$K$上微分$D$的作用下$k(t)$能过保持封闭，那么$k(t)$本事就是一个微分域。微分$D$和$d/dt$之间的相似性成立的条件之一是$D$将$t$的多项式映为$t$的多项式，换言之，$k[t]$在$D$的作用下封闭。（这个条件可能不是必须的，但是积分算法还没有在[71]的扩张中实现）。因此，在这里我们研究的微分扩张都是满足多项式的微分为多项式这种情况。另外，我们将研究限定在特征为0的域上，所以对于剩下的章节来讲，$k$是一个特征为0的微分域，$K$是$k$的一个微分扩张，$D$表示$K$上的微分。首先，我们来证明$Dt \in k[t]$和$k[t]$是$k(t)$的微分子环这两个条件是等价的。

引理 3.4.1. $t \in K$，那么$Dt \in k[t] \iff k[t]$在$D$作用下封闭。

我们注意到在上面的这个引理中，不需要限定$t$是$k$上的超越元。现在我们可以定义之后积分算法所适用的微分扩张类。这个微分扩张类对于刻画微积分中通常的初等超越函数来说足够一般。这个微分扩张类是由简单的超越扩张组成的，在这些超越扩张中$k[t]$在$D$的作用下是封闭的。

定义 3.4.1. $t \in K$，如果满足
$(i)t$在$k$上是超越的
$(ii)D[t] \in k[t]$
那么我们称$t$是$k$上的一个有关$D$的单项式。
另外我们定义$t$的$D-degree$（$D-$次数）为$\delta(t) = \deg_t(Dt)$，$t$的$D-leading\ coefficient$（$D-$首项系数）为$\lambda(t) = lc_t(Dt)$。如果$\delta(t) \leq 1$，我们称$t$是线性的，反之，则称$t$是非线性的。此外，我们令$\mathcal{H}_t \in k[X]$，$\mathcal{H}_t$为多项式满足$Dt = \mathcal{H}_t(t)$。
因为在单项式扩张中，多项式的微分为多项式，我们经常需要知道一个微分的次数和首项系数。

引理 3.4.2. $t$为$k$上的单项式，$p \in k[t]$
$(i)\deg(Dp) \leq \deg(p) + \max(0, \delta(t) - 1)$
$(ii)$如果$t$非线性，并且$\deg(p) > 0$，那么$(i)$中等号成立，$Dp$的首项系数为$\deg(p)\text{lc}(p)\lambda(t)$

另外我们引入下面的概念，定义$special$（特殊）多项式和首一不可约的$special$（特殊）多项式集合：
$$\mathcal{S}_{k[t]:k} = \{p \in k[t]\ |p\text{是}special\text{（特殊）的}\}$$

$$\mathcal{S}_{k[t]:k}^{irr} = \{p \in \mathcal{S}_{k[t]:k},\ p\text{是首一的且不可约的}\}$$

当可以清晰地从上下文中看出表达单项式扩张的概念时，我们忽略记号，简单地记为$\mathcal{S}$和$\mathcal{S}^{irr}$。一个多项式不一定必须 normal（普通）或 special（特殊），但是一个不可约多项式$p \in k[t]$必须或者 normal（普通）或 special（特殊），因为$\gcd(p,Dp)$是$p$的一个因子。注意到$k \subseteq \mathcal{S}$，$p \in k[t]$既 normal（普通）又 special（特殊）当且仅当$(p)=(1)$，这等价于说$p \in k^*$。Special（特殊）的多项式生成微分理想，所以在分式环上存在一个诱导出来的微分（引理 3.1.2.）。更重要的是，这个诱导出来的微分被证明是$D$的一个扩张。

**引理 3.4.3.** $p \in \mathcal{S} \setminus k$，那么$(p)$是$k[t]$的一个微分理想，并且$(k[t]/(p), D^*)$是$(k, D)$的一个微分扩张，$D^*$是诱导微分映射。

**引理 3.4.4.** $p_1, \ldots, p_m \in k[t]$满足$\gcd(p_i, p_j)=1, i \neq j$，令$p = \prod_{i=1}^m p_i^{e_i}$，其中$e_i$为正整数，那么

$$\gcd(p, Dp) = \left(\prod_{i=1}^m p_i^{e_i-1}\right) \prod_{i=1}^m \gcd(p_i, Dp_i)$$

我们可以看出，任意 normal（普通）的多项式必须是无平方因子。另外我们得到了 normal（普通）和 special（特殊）多项式的可乘性，特别$\mathcal{S}$是由$k$和$\mathcal{S}^{irr}$生成的乘法幺半群。

**定理 3.4.1.**
(i) 任意有限个 normal（普通）的多项式（两两互素）的乘积是 normal（普通）的。任意一个 normal（普通）多项式的任意因子是 normal（普通）的。
(ii) $p_1, \ldots, p_n \in \mathcal{S} \implies \prod_{i=1}^n p_i \in \mathcal{S}$
(iii) $p \in \mathcal{S} \setminus \{0\} \implies \forall q \in k[t],\ q\text{整除}p,\ q \in \mathcal{S}$

正如上面提到的那样，每一个 normal（普通）的多项式必须是无平方因子的。但是无平方因子的多项式并不总是 normal（普通）的，并且一个无平方因子多项式的正则性和它的根的微分性质之间有着重要的联系。这种联系用以下的两个定理说明。

**定理 3.4.2.** $\overline{k}$是$k$的代数闭包，$p \in k[t]$是无平方因子的，那么
$p$是 normal（普通）的 $\iff D\alpha \neq \mathcal{H}_t(\alpha)$，$\alpha$为$p$在$\overline{k}$中的任意一个根

**定理 3.4.3.** $\overline{k}$是$k$的代数闭包，$p \in k[t] \setminus \{0\}$，那么
$p \in \mathcal{S} \iff D\alpha = \mathcal{H}_t(\alpha)$，$\alpha$为$p$在$\overline{k}$中的任意一个根

在之后我们会经常用到这个平凡的结论，在进行$k$的代数扩张中，special（特殊）和 normal（普通）多项式保持自身特性不变。（保持 special 或者 normal）。

**推论 3.4.1.** $E$是$k$的一个代数闭包。那么$t$是$E$上的一个单项式。此外$\mathcal{S}_{k[t]:k} \subseteq \mathcal{S}_{E[t]:E}$，如果$p \in k[t]$是 normal（普通）的，那么将$p$视为$E[t]$的一个元素，$p$依然是 normal（普通）的。

作为推论的结论，当$k$的所有元素都是常数时，特殊的不可约多项式恰好是$\mathcal{H}_t$的因子，普通多项式恰好是和$\mathcal{H}_t$互质的无平方因子多项式。

推论 3.4.2. 假设$\forall a \in k$，$Da = 0$

$(i) p \in k[t]$，且$p$是首一且不可约，那么$p \in \mathcal{S}^{irr} \iff p | \mathcal{H}_t$

$(ii) p \in k[t]$，且$p$无平方因子。那么$p$ normal（普通）$\iff \gcd(p, \mathcal{H}_t) = 1$

特别的是，将上述推论应用到$D = d/dt$的情况中，我们可以得到$Dt = 1 = \mathcal{H}_t$，所以$k[t]$中的每一个无平方因子多项式在$d/dt$的意义下是normal（普通）的。

我们还没有对于单项式扩张中常数域可能发生的扩张作出假设，所以我们现在将注意力放在$k(t)$可能的新常量。结果证明是新的常量和special（特殊）多项式紧密联系。回顾一个单项式$t$，如果$\deg_t(Dt) \geq 2$，那么$t$是非线性的。

引理 3.4.5. 如果$c \in \mathrm{Const}_D(k(t))$，那么$c$的分子和分母都属于$\mathcal{S}$。此外，如果$c \notin 0$并且$t$是非线性的，那么$c$的分子和分母有同样的次数。

因此，$k(t) \backslash k$中新常量的存在说明$\mathcal{S}^{irr}$非空。反之，非平凡special（特殊）多项式的存在能否证明新常量的存在，是一个更加困难的问题：达布提出的一个定理强调了如果$k$是自身常数域的一个纯超越扩张，那么就能够充分说明$\mathcal{S}^{irr}$中元素的存在性等价于$k(t) \backslash k$中新常量的存在性。幸运的是，在积分算法中出现的关键性的单项式扩张的情况要简单的多，正如下一个引理说明的那样，这种情况下$\mathcal{S}^{irr}$中的任意元素产生一个新的常量。

引理 3.4.6. 假设$dt/t \in k$，$p \in k[t]$且非零。那么
$$p \in \mathcal{S} \iff D\left(\frac{p}{\mathrm{lc}(p) t^{\deg(p)}}\right) = 0$$

为了之后的需要，我们需要特别定义一类special（特殊）多项式。首先我们定义一些有用的术语。

定义 3.4.3. 如果存在$v \in k^*$，并且存在一个整数$n \neq 0$使得$nu = Dv/v$，那么我们称$u \in k$是一个具有$k-$根的对数微分。

例 3.4.1. $k = \mathbb{Q}(x)$，定义微分$D = d/dx$，$u = 1/(2x) \in k$。因为$2u = Dx/x$，$u$是具有一个$\mathbb{Q}(x)-$根的对数微分。事实上，$u$是$\sqrt{x}$的对数微分，$\sqrt{x}$是$\mathbb{Q}(x)$上的一个根式。另一方面，$Dv/v \notin \mathbb{Z}, \forall v \in k^*$，所以1不是一个具有$\mathbb{Q}(x)-$根的对数微分。

从定义中我们可以清晰地看到，如果我们将$k$扩张到某个扩张域$E$，那么具有$k-$根的对数微分就变成了具有$E-$根的对数微分。但是，当$E$是$k$的代数扩域时，$k$中不是具有$k-$根的对数微分的元素不能成为具有$E-$根的对数微分。

引理 3.4.8. $E$是$k$的代数扩域，$a \in k$。如果$a$不是具有$k-$根的对数微分，那么$a$不是具有$E-$根的对数微分。

定义 3.4.4. 我们称$q \in k[t]$是关于$D$第一类special（特殊）的，如果$q \in \mathcal{S}$，对于在$k$的代数闭包中$q$的任意根$\alpha$，$p_\alpha(\alpha)$不是具有$k(\alpha)-$根的对数微分，$p_\alpha$定义如下：
$$p_\alpha = \frac{Dt - D\alpha}{t - \alpha} \in k(\alpha)[t]$$

另外，我们引入以下概念：
$$\mathcal{S}_{1,k[t]:k} = \{p \in \mathcal{S}_{k[t]:k},\ p\text{是第一类}special（特殊）\text{的}\}$$
$$\mathcal{S}_{1,k[t]:k}^{irr} = \{p \in \mathcal{S}_{1,k[t]:k},\ p\text{是首一且不可约的}\}$$

当可以清晰地从上下文中看出表达单项式扩张的概念时，我们忽略扩张的下标，简单地记作$\mathcal{S}_1$和$\mathcal{S}_1^{irr}$。注意到既然$\alpha$是多项式$Dt - D\alpha$的一个根，在$k(\alpha)[t]$中$t-\alpha | Dt - D\alpha$，所以$p_\alpha(\alpha)$永远都是有定义的。另外我们注意到根据定义，$k^* \subseteq \mathcal{S}_1$，在上面的定义中我们可以用$k-$根代替$k(\alpha)-$根。引理 3.4.8., 定理 3.4.1., 推论 3.4.1.可以方便地推广到第一类$special$（特殊）的多项式。$\mathcal{S}_1$是由$k^*$和$\mathcal{S}_1^{irr}$生成的乘法半群。

**定理 3.4.4.**
$(i) p_1, \ldots, p_n \in \mathcal{S}_1 \Longrightarrow \prod_{i=1}^n p_i \in \mathcal{S}_1$
$(ii) p \in \mathcal{S}_1 \Longrightarrow q \in \mathcal{S}_1, \forall q \in k[t], q\text{整除}p$
$(iii)$如果$E$是$k$的代数扩域，那么$\mathcal{S}_{1,k[t]:k} \subseteq \mathcal{S}_{1,E[t]:E}$

## 3.5 典型表示

已知$p \in k[t]$，我们想要分离$p$的$special$（特殊）和$normal$（普通）部分。下面正式定义分离的概念。

**定义 3.5.1.** $p \in k[t]$。如果$p_n, p_s \in k[t], p_s \in \mathcal{S}$，$p_n$的每一个无平方因子是$normal$（普通）的，那么我们称$p = p_s p_n$是$p$的一个分裂分解。

定理 3.4.2.和定理 3.4.3 的结论之一就是$p$在$k$上的一个分裂分解同时也是$p$在$k$的任意一个代数扩张上的分裂分解，因为在$\overline{k}$中对于$p_s$的所有根来讲：$D\alpha = \mathcal{H}_t(\alpha)$，对于$p_n$的所有根来讲：$D\alpha \neq \mathcal{H}_t(\alpha)$。基于同一个原因，在$p$的一个分裂分解中我们可以得到结论$\gcd(p_n, p_s) = 1$，像素数分解一样，这样的分解之间仅仅相差$k$中的单位因子。显而易见，$p$的一个素数分解生成$p$的一个分裂分解，但是结果证明是分裂分解只能通过公约数进行计算，和无平方因子分解相近。

**定理 3.5.1.** $p \in k[t]$，那么
$(i)$
$$\frac{\gcd(p, Dp)}{\gcd(p, dp/dt)}$$
是由$p$的所有互素的$special$（特殊）不可约因子生成。
$(ii)$ 如果 $p$ 是无平方因子的，那么 $p = p_s p_n$ 是 $p$ 的一个分裂分解，其中$p_s = \gcd(p, Dp), p_n = p/p_s$

这个定理给了我们两个可以计算分裂分解的算法：第一个是计算$\mathcal{S} = \gcd(p, Dp)/\gcd(p, dp/dt)$和$q = p/\mathcal{S}$。如果$\mathcal{S} \in k$，那么$p$没有$special$（特殊）不可约因子，所以返回$p_n = p, p_s = 1$。否则$\deg(q) < \deg(p)$，所以递归地计算$q$的一个分裂分解$q = q_n q_s$，返回$p_n = q_n, p_s = \mathcal{S} q_s$

```
SplitFactor(p, D)        (* Splitting Factorization *)

(* Given a derivation D on k[t] and p ∈ k[t], return (p_n, p_s) ∈ k[t]^2 such
that p = p_n p_s, p_s is special, and each squarefree factor of p_n is normal. *)

S ← gcd(p, Dp)/ gcd(p, dp/dt)                    (* exact division *)
if deg(S) = 0 then return(p, 1)
(q_n, q_s) ← SplitFactor(p/S, D)                 (* exact division *)
return(q_n, Sq_s)
```

第二种算法首先计算$p$的一个无平方分解$p = p_1 p_2^2 \cdots p_m^m$，然后计算$S_i = \gcd(p_i, Dp_i)$和$N_i = p_i/S_i$。根据定理 3.5.1. $p = p_s p_n$ 是 $p$ 的一个分裂分解，其中 $p_s = S_1 S_2^2 \cdots S_m^m, p_n = N_1 N_2^2 \cdots N_m^m$。这种方法可以提供$p_s$和$p_n$的无平方分解。此外，Yun算法可以用来计算初始的无平方分解。

```
SplitSquarefreeFactor(p, D)     (* Splitting Squarefree Factorization *)

(* Given a derivation D on k[t] and p ∈ k[t], return (N_1,...,N_m) and
(S_1,...,S_m) in k[t]^m such that p = (N_1 N_2^2 ⋯ N_m^m)(S_1 S_2^2 ⋯ S_m^m) is a split-
ting factorization of p and the N_i and S_i are squarefree and coprime. *)

(p_1,...,p_m) ← Squarefree(p)
for i ← 1 to m do
    S_i ← gcd(p_i, Dp_i)
    N_i ← p_i/S_i                                (* exact division *)
return((N_1,...,N_m),(S_1,...,S_m))
```

现在我们可以定义$k(t)$中元素的一个分解，这个分解生成了$f$的有理典型表示 $f = p + a/d$，$f \in k(t)\backslash\{0\}$，将$f$写成$f = a/d$的形式，其中$a, d \in k[t], \gcd(a, d) = 1$，同时$d$是首一的（这样的一个表示是唯一的）。令$d = d_s d_n$为$d$的一个关于$D$的分裂分解，其中$d_s, d_n$是首一的，这使得分解是唯一的。这样的话，存在唯一的$p, b, c \in k[t]$，使得 $\deg(b) < \deg(d_s), \deg(c) < \deg(d_n)$，$f$的表示为

$$f = \frac{a}{d} = p + \frac{b}{d_s} + \frac{c}{d_n}$$

我们称这个关于$D$的，并且唯一的分解为$f$关于$D$的典型表示。我们引入概念$f_p = p$（$f$的多项式部分），$f_s = b/d_s$（$f$的$special$（特殊）部分），和$f_n = c/d_n$（$f$的$normal$（普通）部分）。

```
CanonicalRepresentation(f, D)      (* Canonical Representation *)

(* Given a derivation D on k[t] and f ∈ k(t), return (f_p, f_s, f_n) ∈ k[t] ×
k(t)^2 such that f = f_p + f_s + f_n is the canonical representation of f. *)

(a, d) ← (numerator(f), denominator(f))         (* d is monic *)
(q, r) ← PolyDivide(a, d)
(d_n, d_s) ← SplitFactor(d, D)
(b, c) ← ExtendedEuclidean(d_n, d_s, r)         (* deg(b) < deg(d_s) *)
return(q, b/d_s, c/d_n)
```

我们需要定义一些在以后经常使用的术语。$\mathbb{C}$上的有理函数是简单的如果它仅有可去的仿射极点，换句话说只有一阶极点。这和分母无平方因子是等价的。因为在单项式扩张中，*normal*（普通）多项式和无平方因子多项式是相似的，所以如果$k(t)$中的元素有一个*normal*（普通）的分母，那么称其*simple*（简单）就是很自然的。相似地，一个普通的多项式可以看做是没有可去极点的有理函数，或者是一个没有分母的有理函数。在单项式扩张中相似的情况是函数没有*normal*（普通）的可去极点，换句话说，有至多有限处的极点或者在*special*（特殊）多项式的情况下有极点。这意味着这是一个有着*special*（特殊）分母的函数。

**定义 3.5.2.** $f \in k(t)$，如果$f$的分母关于$D$是*normal*（普通）的，那么我们称$f$关于$D$ *simple*（平凡）。如果$f$的分母关于$D$ *special*（特殊），那么我们称$f$关于$D$既约。另外，我们将$k(t)$的所有既约元素记为$k\langle t \rangle$。

显然，$k[t] \subseteq k\langle t \rangle$，在下一章我们可以看到，$k[t], k\langle t \rangle$是$k(t)$的微分子环。

有一个在之后分裂分解有用的的地方：在一个微分域中，分裂分解可以将常数根从多项式的非常数根中分离出来。$K$是一个特征为$0$的微分域，$X$是$K$上的不定变元，$p \in K[X]$。假设我们想要将常数根从根中分离出来。为实现这一点，仅仅需要在$D$意义下，$K[X]$上的系数映射$\kappa_D$作用下的分裂分解。

**定理 3.5.2.** $(K, D)$是特征为$0$的微分域，$\overline{K}$是$K$的代数闭包，$X$是$K$上的不定变元。$\forall p \in K[X] \backslash \{0\}$，令$p = p_s p_n$为在$\kappa_D$作用下$p$的分裂分解。那么对于$p$在$\overline{K}$中的任意根$\alpha$

$$\begin{cases} D\alpha = 0 \iff p_s(\alpha) = 0, \\ D\alpha \neq 0 \iff p_n(\alpha) = 0. \end{cases}$$

*Chapter* 2.1.4 **阶函数**

在本章中，我们引入阶函数的概念，阶函数是之后我们证明积分算法正确性的主要工具。这个函数的有用之处在于它可以将任意唯一分解整环($UFD$)的元素映为整数，所以将阶函数作用于一个方程的左右两端，得到含整数的等式或不等式，这样的话就使得证明原始方程没有根，或计算估计原始方程解的阶数变为可能。因此阶函数在很多积分之外的情况下得到应用，例如在解微分方程的算法中。一边我们在积分算法中，将阶函数运用到多项式上，另外我们在具有任意特征的唯一分解整环($UFD$)中引入阶函数，在一个可能可约的元素的阶数被取出来的时候，研究它最一般的性质。

**4.1 基本性质**

这一小节和下一小节，$D$为具有任意特征的一个唯一分解整环($UFD$)，$D^*$是$D$的单位组成的集合，$F$是$D$的分式域，$a \in D$且$a \neq 0, a \notin D^*$

定义 4.1.1. $a$的阶数是映射$\nu_a : D \Rightarrow \mathbb{Z} \cup \{+\infty\}$，$\nu_a$定义如下
$(i) \nu_a(0) = +\infty$
$(ii) x \in D \backslash \{0\}$，$\nu_a(x) = \max\{n \in \mathbb{N}$，使得$a^n | x\}$

尽管映射$\nu_a$只取非负值，我们将其定义为映射至$\mathbb{Z} \cup \{+\infty\}$，是为了将其最终扩展到$D$的分式域上。首先我们证明集合$S_a(x) = \{n \in \mathbb{N}$，使得$a^n | x\}$对于$x \in D \backslash \{0\}$是有限且非空的。因为$a \neq 0$并且$a$不是$D$中的单位，令$p \in D$是$a$的不可约因子。那么存在$x$的一个不可约分解，其中$p$具有指数$e \geq 0$。令$n > e$，并且假设$p^n | x$。那么$x = p^n y, y \in D$。令$y = u \prod_{i=1}^{m} p_i^{e_i}$为$y$不可约分解，其中$p_i$是互质的，$u$是一个单位。之后我们有$x$的一个不可约分解$x = u p^n \prod_{i=1}^{m} p_i^{e_i}$，$p$的指数至少保证$n > e$，与$D$是唯一分解整环($UFD$)相矛盾。因此，$\forall q \in S_a(x)$满足$q \leq e$，所以$S_a(x)$是有限的。另外$0 \in S_a(x)$，所以$\nu_a$在$D$上是良定义的。

引理 4.1.1. $x, y \in D$，那么
$(i) \nu_a(xy) \geq \nu_a(x) + \nu_a(y)$，如果$a$是不可约元，等式成立
$(ii) \nu_a(x + y) \geq \min(\nu_a(x), \nu_a(y))$，如果$\nu_a(x) \neq \nu_a(y)$，等式成立
$(iii)$如果$x | y$，那么$\nu_a(x) \leq \nu_a(y)$
$(iv) \nu_a(\gcd(x, y)) = \min(\nu_a(x), \nu_a(y))$

例 4.1.1. $\mathbb{Z}$中，$\nu_6(12) = \nu_6(18) = 1$，$\nu_6(12 \times 18) = \nu_6(216) = 3$，这说明$(i)$中的等式并不总是成立，如果$a$是可约的。另一方面，$\nu_3(12) = 1, \nu_3(18) = 2, \nu_3(216) = 3 = 1 + 2$，同样$\nu_2(12) = 2, \nu_2(18) = 1, \nu_2(216) = 3 = 1 + 2$。

接下来的这个引理说明了单位乘以$a$或者单位乘以$\nu_a$作用的变量并不改变阶函数。这个性质在将$\nu_a$的定义扩展到$F$上是非常必要的。

引理 4.1.2. $u \in D^*, x \in D$，那么：
$(i) \nu_a(ux) = \nu_a(x) = \nu_{ua}(x)$
$(ii) \nu_a(u) = 0$

在接下来的定义和本章剩下的部分中，我们称两个$D$中元素$y$和$z$没有公约数，如果$\gcd(y,z)$是$D$中的单位。

定义 4.1.2. $x \in F^*$，将$x$记作$x = y/z$，满足$y,z \in D$且$y,z$没有公约数，$z \neq 0$。那么我们将$\nu_a(x) = \nu_a(y) - \nu_a(z)$。

$x \in F^*$，$y,z,t,w \in D$，其中$y$和$z$没有公约数，$t$和$w$没有公约数，$x = y/z = t/w$。那么$y/t = z/w = u \in D^*$。所以根据引理 4.1.2. $\nu_a(y) = \nu_a(ut) = \nu_a(t)$，$\nu_a(z) = \nu_a(uw) = \nu_a(w)$，所以$\nu_a(y) - \nu_a(z) = \nu_a(t) - \nu_a(w)$，这说明$\nu_a$在$F$上是良定义的。另外，根据引理 4.1.2.$\nu_a(1) = 0$，所以当$x \in D$时，令$y = x, z = 1$，我们可以看见以上的定义和$D$上$\nu_a$的定义是相容的。我们发现引理 4.1.1.的$(i),(ii)$两部分在$F$上并不成立：$\nu_6(5/3) = \nu_6(1/2) = 0$，但是$\nu_6(5/3 \times 1/2) = \nu_6(5/3 + 1/2) = -1 < 0$。但是当$a$是不可约时，这些陈述仍然是正确的。

定理 4.1.1. $x, y \in F$，假设$a$是$D$中的不可约元，那么
$(i)\nu_a(xy) = \nu_a(x) + \nu_a(y)$
$(ii)$如果$x \neq 0$，那么$\nu_a(x^m) = m\nu_a(x), \forall m \in \mathbb{Z}$
$(iii)\nu_a(x+y) \geq \min(\nu_a(x), \nu_a(y))$，如果$\nu_a(x) \neq \nu_a(y)$，则等式成立。

上述定理的$(i),(ii)$的两部分说明如果$a$是可约的，那么在定义 4.1.2.中$y,z$没有公约数的这条限制可以去掉。$\forall y, z \in F, x = y/z$，$\nu_a(x) = \nu_a(yz^{-1}) = \nu_a(y) - \nu_a(z)$。

在多项式环中，我们需要研究扩展常数域对于阶函数的影响。结果证明是，当一个不可约多项式在一个代数扩张中分解时，新的不可约因子处的阶数和之前只在基域上有定义的情况下是一样的。

定理 4.1.2. $F$是一个域，$E$是$F$的一个可分代数扩张，$x$是$E$上的一个不定变元。如果$p \in F[x]$是$F$上的不可约多项式，那么对于$p$在$E[x]$中任意不可约因子$q \in E[x]$，$\nu_p(f) = \nu_q(f), \forall f \in F(x)$。

## 4.2 局部化

定义 4.2.1. 我们定义$a$处的局部化为
$$\mathcal{O}_a = \cap_{p|a} = \{x \in F, \nu_p(x) \geq 0\}$$
其中交集取为$a$在$D$中的所有不可约因子。

显而易见，$a$处的局部化是$F$中所有可以表示为分母与$a$无公约数形式的的分式集合。如果$a$是不可约的，局部化就变成了一个局部环，也可以看成$F$中所有在$a$处有正阶数的分式的集合。

引理 4.2.1.
$(i)\mathcal{O}$是$F$中包含$D$的子环
$(ii)x \in \mathcal{O}_a \Longrightarrow \nu_a(x) \geq 0$
$(iii)x \in a\mathcal{O}_a \Longleftrightarrow \nu_a(x) \geq 1$,其中$a\mathcal{O}_a$是$a$在$\mathcal{O}_a$中生成的理想
$(iv)$如果$a$是不可约的，那么$x \in \mathcal{O}_a \Longleftrightarrow \nu_a(x) \geq 0$

$(v)$如果$a$是不可约的，那么$xa^{-\nu_a(x)} \in \mathcal{O}_a, \forall x \in F^*$

$(vi)$如果$\Delta$是$D$上的任意微分，那么$\Delta\mathcal{O}_a \subseteq \mathcal{O}_a$

例 4.2.1. $D = \mathbb{Z}$
$$\mathcal{O}_6 = \mathcal{O}_2 \cap \mathcal{O}_3 = \{x \in \mathbb{Q}, x = b/c \text{且} b, c \in Z, 2 \nmid c, 3 \nmid c\}$$

所以$1/3 \notin \mathcal{O}_6$，尽管$\nu_6(1/3) = 0$。这说明如果$a$是可约的，以上引理的部分$(iv)$和$(v)$并不总是成立。这使得我们注意到对于部分$(iii)$中两个方向对于$a$可约也是成立的。

当$D$是一个主理想整环$(PID)$，对于$D$的任意非零非平凡理想$I$，典范映射$\pi_I: D \to D/I$可以自然地扩张到局部化$\mathcal{O}_a$上，其中$a$为$I$的任意生成元。下一个定义构造了这个扩张。

定义 4.2.2. $D$是一个主理想整环$(PID)$，$I$是$D$的一个非零非平凡理想，换句话说，$I \neq D$，且$I \neq (0)$，$a$是$I$的一个生成元，换句话说$I = (a)$。我们定义$a$处的值由映射$\pi_a: \mathcal{O}_a \to D/I$给出：令$x \in \mathcal{O}_a$，将$x$写作$x = b/c$，$b, c \in D$，且$b, c$没有公约数。我们定义$\pi_a(x)$为$\pi_I(bd)$，其中$d, e \in D$，使得$cd + ae = 1$，$\pi_I$是从$D$到$D/I$的典范映射。

为了说明$\pi_a$是良定义的，我们需要证明这样的$d$和$e$总是存在的，$\pi_a(x)$的值和$b, c, d, e$的选择无关。首先因为$I \neq (0)$，$a \neq 0$，因为$I \neq D$，$a \notin D^*$，所以$\mathcal{O}_a$是有定义的。令$x \in \mathcal{O}_a$，记作$x = b/c$，其中$b, c \in D$，且$b, c$没有公约数。令$p$为$a$的任意不可约元。因为$x \in \mathcal{O}_a$，所以$\nu_p(x) = \nu_p(b) - \nu_p(c) \geq 0$。但是因为$b$和$c$没有公约数，所以$\nu_p(b)$和$\nu_p(c)$至少有一个为零，所以$\nu_p(c) = 0$，这说明$p \nmid c$。因为这个对于$a$的任意不可约元$p$成立，$\gcd(a, c) = 1$，所以存在$d, e \in D$，使得$cd + ae = 1$。假设对于$d, e, f, g \in D$，$cd + ae = cd + ag = 1$。那么$a(g - e) = c(d - f)$。令$p$为$a$任意不可约因子。我们得到
$$\nu_p(c) + \nu_p(d - f) = \nu_p(c(d - f)) = \nu_p(a(g - e)) = \nu_p(a) + \nu_p(g - e) \geq \nu_p(a)$$
但是像之前一样，$\nu_p(c) = 0$，所以$\nu_p(d - f) \geq \nu_q(a)$，这说明在$a$的分解中任意有一个正次数$n$的不可约$p \in D$必须在$d - f$的分解中出现，并且有一个指数$m \geq n$。因此$a | d - f$，换句话说$d - f \in I$，所以$\pi_I(d - f) = 0$。因为$\pi_I$是一个环同态，我们得到$\pi_I(bd) = \pi_I(bf)$，所以$\pi_a(x)$的值不依赖于$d$和$e$的选择。假设最终$x = b/c = b'/c', b, c, b', c' \in D$，并且$\gcd(b, c) = \gcd(b', c') = 1$。像之前一样，这说明$b' = ub, c' = uc, u \in D^*$。令$d, e \in D$，使得$cd + ae = 1$，那么$c'd' + ae = 1, d' = u^{-1}d \in D$，我们又$b'd' = ubu^{-1}d = bd$，所以$\pi_a(x)$的值不依赖于$b$和$c$的选择，所以$\pi_a$在$\mathcal{O}_a$上是良定义的。

下面我们证明$\pi_a$是$\pi_I$到$\mathcal{O}_a$的一个扩张，诱导出$\mathcal{O}_a/a\mathcal{O}_a$和$D/I$之间的一个同态，换句话说，我们可以得到一下的图像：

$$\begin{array}{ccc} \mathcal{O}_a & \longrightarrow & \mathcal{O}_a/a\mathcal{O}_a \\ \uparrow & \searrow_{\pi_a} & \parallel \\ D & \xrightarrow{\pi_I} & D/I \end{array}$$

定理 4.2.1. $D$是一个主理想整环$(PID)$，$I$是$D$的一个非平凡非零理想，$a \in D$是$I$的一个生成元，那么

$(i) \pi_a(b) = \pi_I(b), \forall b \in D$（换句话说，$\pi_a$扩张了$\pi_I$）

$(ii) \ker(\pi_a) = a\mathcal{O}_a$

$(iii)\pi_a$是从$\mathcal{O}_a$到$D/I$的一个满环同态，因此是$\mathcal{O}_a/a\mathcal{O}_a$到$D/I$的一个环同构（如果$I$是极大理想，那么$\pi_a$是一个域同构）

$(iv)$如果$\Delta$是$D$的一个微分，并且$\Delta I \subseteq I$，那么$\Delta^* \circ \pi_a = \pi_a \circ \Delta$，其中$\Delta^*$是$D/I$上的诱导微分。（引理 3.1.2.）

当$D$是一个欧几里得整环($ED$)时，我们称$\pi_a(x)$是$x$在$a$处的余数。可以通过下面的算法进行计算，这个算法和$D$中的扩展欧几里得算法复杂度相同。

```
Remainder(x, a)      (* Local remainder at a point *)

(* Given a Euclidean domain D, a ∈ D \ {0} with a ∉ D* and x ∈ O_a,
   return π_a(x) as an element of D. *)

(b, c) ← ExtendedEuclidean(a, denominator(x), 1)
(q, r) ← PolyDivide(numerator(x) c, a)
return r
```

### 4.3 无穷处的阶

在多项式环中，我们引入一个额外的阶函数，称为在无穷处的阶。这个阶函数和之前章节的阶函数性质类似。虽然可以使用通常的多项式函数的指数函数进行代替，但是之后无穷处的阶函数的性质可以在代数曲线上推广到无穷远点，这时次数就是没有定义的。在这一部分中，令$D$是一个具有任意特征的整环，$x$是$D$上的一个不定变元。对于$a \in D[x]$，我们用$\text{lc}(a)$表示$a$的首项系数，换句话说，如果$a = a_0 + a_1 x + \cdots + a_n x^n$，且$a_n \neq 0$，那么$\text{lc}(a) = a_n$。

**定义 4.3.1.** $\infty$处的阶是由映射$\nu_\infty : D(x) \to \mathbb{Z} \cup \{+\infty\}$给出，其中$\nu_\infty(0) = +\infty, \nu_\infty(b/c) = c - b, \forall b, c \in D[x]\backslash\{0\}$。

假设$f = b/c = d/e$，其中$b, c, d, e \in D[x]$。那么$be = cd$，所以$\deg(b) + \deg(e) = \deg(c) + \deg(d)$，所以$\deg(c) - \deg(b) = \deg(e) - \deg(d)$，这说明$\nu_\infty$在$D(x)$上是良定义的。接下来我们说明对于$D[x]$上的一个不可约元$a$，$\nu_\infty$和$\nu_a$满足同样的性质。

**定理 4.3.1.** $f, g \in D(x)$，那么
$(i)\nu_\infty(fg) = \nu_\infty(f) + \nu_\infty(g)$
$(ii)\nu_\infty(f+g) \geq \min(\nu_\infty(f), \nu_\infty(g))$，等式成立条件为$\nu_\infty(f) \neq \nu_\infty(g)$
$(iii)$如果$f \neq 0$，那么$\nu_\infty(f^m) = m\nu_\infty(f), \forall m \in \mathcal{Z}$

因为$\nu_\infty$满足和$\nu_a$相似的性质，所以可以自然地定义局部化的概念，并且用和前面小节定义在一点处值映射相似的方法定义在无穷处的值映射。

**定义 4.3.2.** 我们定义在无穷处的局部化为
$$\mathcal{O}_\infty = \{f \in D(x), \text{ 使得}\nu_\infty(f) \geq 0\}$$

直观上讲，$\mathcal{O}_\infty$作为一个局部环，是由$D(x)$中所有分母次数至少与分母次数相同的有理函数组成的。换句话说，没有在无穷处的极点。不出所料，对于$D[x]$中不可约元$a$，$\mathcal{O}_\infty$满

足和$\mathcal{O}_a$相似的性质。

**引理 4.3.1.**
$(i)$ $\mathcal{O}_\infty$是$D(x)$的子环
$(ii)$
$$f \in x^{-1}\mathcal{O}_\infty \iff \nu_\infty(f) \geq 1$$
其中$x^{-1}\mathcal{O}_\infty$是由$x^{-1}$在$\mathcal{O}_\infty$中生成的理想。
$(iii)$ $fx^{\nu_\infty(f)} \in \mathcal{O}_\infty, \forall f \in D(x)^*$

**定义 4.3.3.** $F$是$D$的分式域，我们定义无穷处的值由映射$\pi_\infty : \mathcal{O}_\infty \to F$决定：
$$\pi_\infty(f) = \begin{cases} \mathrm{lc}(b)/\mathrm{lc}(c), & \nu_\infty(f) = 0 \\ 0, & \nu_\infty(f) > 0 \end{cases}$$
其中$b, c \in D[x], f = b/c$

假设$f = b/c = d/e$，其中$b, c, d, e \in D[x]$，$\nu_\infty(f) = 0$。那么$be = cd$，所以$\mathrm{lc}(b)\mathrm{lc}(e) = \mathrm{lc}(c)\mathrm{lc}(d)$，所以$\mathrm{lc}(b)/\mathrm{lc}(c) = \mathrm{lc}(d)/\mathrm{lc}(e)$，这说明$\pi_\infty$在$\mathcal{O}_\infty$上是良定义的。

**定理 4.3.2.**
$(i)$ $\ker(\pi_\infty) = x^{-1}\mathcal{O}_\infty$
$(ii)$ $\pi_\infty$是从$\mathcal{O}_\infty$到$D$的分式域$F$的满环同态，因此是$\mathcal{O}_\infty/x^{-1}\mathcal{O}_\infty$和$F$之间的域同构。

```
ValueAtInfinity(f)     (* Value at infinity *)
   (* Given a Euclidean domain D, and f ∈ O_∞, return π_∞(x). *)
   if f = 0 then return 0
   a ← numerator(f), b ← denominator(f)
   if deg(b) > deg(a) then return 0
   return(lc(a)/lc(b))
```

## 4.4 留数和$Rothstein-Trager$结式（罗斯坦-特雷格结式）

我们在这一部分给出阶函数在积分中所用到的性质，换句话说就是一个函数在一点的阶和它在这一点的微分之间的关系。并且我们也给出在单项式扩张中留数的基本理论，相当于$Rothstein-Trager$结式的基本性质。这种联系和各种各样的留数方程使我们将函数的极点和在它积分中出现的函数的极点联系起来。这一部分自始至终，$K$是一个特征为0的微分域，在上面定义微分$D$，$t$是$K$上的一个单项式。首先我们定义一个$normal$（普通）多项式处留数的概念。

**定义 4.4.1.** $p \in K[t] \backslash K$，$p$是$normal$（普通）的，$\mathcal{R}_p$是一个集合，定义如下：
$$\mathcal{R}_p = \{f \in K(t), \text{使得} pf \in \mathcal{O}_p\}$$
我们定义$p$处的留数为映射$\mathrm{residue}_p : \mathcal{R}_p \to K[t]/(p)$，定义如下
$$\mathrm{residue}_p(f) = \pi_p(f\frac{p}{Dp})$$

$q \in K[t]$是$p$的任意不可约元。因为$p$是normal（普通）的，那么$q \nmid Dp$，所以$1/Dp \in \mathcal{O}_q$。因为这个对于$p$的任意不可约元成立，我们可以得到$1/Dp \in \mathcal{O}_p$。对于$f \in \mathcal{R}_p, pf \in \mathcal{O}_p$，那么$fp/Dp \in \mathcal{O}_p$，这以为着residue$_p$是良定义的。因为对于任意$a \in K$，$\pi_p(a) = a$，当设计留数概念时，我们将$K$和$\pi_p(K) \subseteq K[t]/(p)$视为一致。因此，在这一小节的剩余部分中我们称$f$有一个留数$\alpha \in K$，这意味着$f$的留数是$K$中的一个元素在$\pi_p$作用下的像。

定理 4.4.1.
$p \in K[t]\backslash K$, $p$是normal（普通）的。那么$\mathcal{R}_p$是$K$上的一个向量空间，$\ker(\text{residue}_p) = \mathcal{O}_p$，residue$_p$是$R_p/\mathcal{O}_p$和$K[t]/(p)$之间的一个$K-$向量空间同构。

例 4.1.1. $K = \mathbb{Q}$，$t$是$K$上的单项式，且$Dt = 1$（也就是说$D = d/dt$），$p = t \in K[t]$，且$p$是normal（普通）且不可约的。我们有$f = 1/t \in \mathcal{R}_p$，但是$f^2 = 1/t^2 \notin \mathcal{R}_p$。所以$\mathcal{R}_p$甚至当$p$是normal（普通）且不可约时也不是一个环。

接下来的公式给出了一个normal（普通）多项式的留数和它的非平凡因子之间的关系。

引理 4.4.1. $p \in K[t]\backslash K$是normal（普通）的，并且$q \in K[t]\backslash K$是$p$的一个因子。那么$\mathcal{R}_p \subseteq \mathcal{R}_q$，并且对于任意$f \in \mathcal{R}_p$, $\text{residue}_q(f) = \pi_q(\text{residue}_p(f))$。

例 4.4.2. 令$K = \mathbb{Q}$，$t$是$K$上的一个单项式，$Dt = 1$（也就是说$D = d/dt$），$f = (t-2)/(t^2 - 1) \in K[t]$，那么
当
$$\text{residue}_{t-1}(f) = \pi_{t-1}\left(\frac{t-2}{t+1}\right) = -\frac{1}{2} = \pi_{t-1}\left(\frac{1-2t}{2}\right)$$
且
$$\text{residue}_{t+1}(f) = \pi_{t+1}\left(\frac{t-2}{t-1}\right) = -\frac{3}{2} = \pi_{t+1}\left(\frac{1-2t}{2}\right)$$
时
$$\text{residue}_{t^2-1}(f) = \pi_{t^2-1}\left(\frac{t-2}{2t}\right) = \frac{1-2t}{2}$$

定理 4.4.2. $f \in K(t)\backslash\{0\}$，且$p \in K[t]$是不可约的。
$(i)$如果$p$是normal（普通）的，且$\nu_p(f) \neq 0$，那么$\nu_p(Df) - \nu_p(f) - 1$，如果$\nu_p(f) = 0$，那么$\nu_p(Df) \geq 0$。此外
$$\pi_p(p^{1-\nu_p(f)}Df) = \nu_p(f)\pi_p(p^{-\nu_p(f)}f)\pi_p(Dp)$$
$(ii) p \in \mathcal{S} \Longrightarrow \nu_p(Df) \geq \nu_p(f)$
$(ii) p \in \mathcal{S}_1$, 并且$\nu_p(f) \neq 0 \Longrightarrow \nu_p(Df) = \nu_p(f)$

例 4.4.3. 令$K = \mathbb{Q}$，$t$是$K$上的单项式，且$Dt = 1$（换句话说$D = d/dt$），$p = t \in K[t]$是normal（普通）的，并且不可约，对于任意的一个整数$m > 0$，$f = t^m + 1 \in K(t)$。我们有$\nu_t(f) = 0$，但是$Df = mt^{m-1}$，所以$\nu_t(Df) = m - 1$。这说明当$\nu_p(f) = 0$时，我们不等得到$\nu_p(Df)$的一个上界。

定理 4.4.2.能得到一些有用的推论：$K\langle t\rangle$必须是一个$K(t)$微分子环，我们得到了对数倒数的阶数和留数的公式，也得到了在一个给定的$p$处的留数的公式。

推论 4.4.1. 令$f \in K(t)$
$(i)f$关于$D$ $simple$（简单）$\Longrightarrow \nu_p(f) \geq -1, \forall p \in K[t]$，$p$是$normal$（普通）且不可约的。
$(ii)f \in K\langle t\rangle \Longleftrightarrow \nu_p(f) \geq 0, \forall p \in K[t]$，$p$是$normal$（普通）且不可约的。
$(iii)K\langle t\rangle$是$K(t)$的一个微分子环。

推论 4.4.2. 令$f \in K(t)\backslash\{0\}$，$p \in K[t]$是不可约的，那么
$(i)\nu_p(Df/f) \geq -1$
$(ii)\nu_p(Df/f) = -1 \Longleftrightarrow \nu_p(f) \neq 0$，$p$是$normal$（普通）的
$(iii)$如果$p$是$normal$（普通）的，那么$\nu_p(Df) \neq -1$且$\text{residue}_p(Df/f) = \nu_p(f)$

引理 4.4.2. 令$p \in K[t]$是$normal$（普通）且不可约的，$g \in \mathcal{O}_p$，且$d \in K[t]$，满足$\nu_p(d) = 1$。那么$\text{residue}_p(g/d) = \pi_p(g/Dd)$。

引理 4.4.3. 令$q \in K[t]$是$normal$（普通）且不可约的，$f \in K(t)$，满足$\nu_q(f) = -1$。将$f$改写为$f = p + a/d$，其中$p, a, d \in K[t], d \neq 0, \deg(a) < \deg(d)$，且$\gcd(a, d) = 1$。那么对于$\forall \alpha \in K$，
$$q|\gcd(a - \alpha Dd, d) \Longleftrightarrow \text{residue}_q(f) = \alpha$$

现在我们可以陈述$Rothstein-Trager$结式的基本形式，换句话说从任意$simple$（简单）函数出发，可以构造出$K$上的一个多项式，满足在$K$中的非零根正好就是$f$在$K$中的留数。值得注意的是，并不是$f$的所有留数都在$K$中，除非$K$是代数封闭的。

定理 4.4.3. 令$f \in K(t)$关于$D$是$simple$（简单）的，将$f$改写为$f = p + a/d$，其中$p, a, d \in K[t], d \neq 0, \deg(a) < \deg(d)$，且$\gcd(a, d) = 1$，令
$$r = \text{resultant}_t(a - zDd, d) \in K[z]$$
其中$z$是$K$上的一个不定变元。那么，对于$\forall \alpha \in K^*$

$r(\alpha) = 0 \Longleftrightarrow \text{residue}_q(f) = \alpha$，其中$q$是$K[t]$中的某个$normal$（普通），不可约多项式我们称由上式得到的多项式$r$为$f$的$Rothstein-Trager$结式。

令$F$是一个特征为$0$的域，$x$是$F$上的一个不定变元，$D$是$F(x)$上的微分$/dx$。因为$F[x]$中的每一个不可约多项式$q$都是关于$d/dx$是$normal$（普通）的，将上面证明的结果应用到$K = \overline{F}$，我们可以发现 定理 4.4.3. 和引理 4.4.3. 分别证明了定理 2.4.1. 的$(i)$部分和$(ii)$部分。

同样存在关于$K(t)$中的元素在无穷处的阶数和关于它的微分相似的结论。

定理 4.4.4. 令$f \in K(t)\backslash\{0\}$，那么
$(i)\nu_\infty(Df) \geq \nu_\infty(f) - \max(0, \delta(t) - 1)$
$(ii)$如果$t$是非线性的，且$\nu_\infty(f) \neq 0$，那么$(i)$中的等式成立，且
$$\pi_\infty\left(t^{1-\delta(t)}\frac{Df}{f}\right) = -\nu_\infty(f)\lambda(t)$$
$(iii)$如果$t$是非线性的，且$\nu_\infty(f) = 0$，那么$(i)$中的严格不等式成立，换句话说，

$\nu_\infty(Df) > 1 - \delta(t)$, 且
$$\pi_\infty\left(t^{1-\delta(t)}\frac{Df}{f}\right) = 0$$

# *Chapter* 2.1.5 超越函数积分

在之前章节我们已经完整地提出了所需要的代数结构,现在我们可以描述超越函数的积分算法。在这一章节中,我们从代数的角度正式描述积分问题,证明积分理论的主定理(Liouville's Theorem 刘维尔定理),描述积分算法的主体部分。

在之后的叙述中,如果没有特殊的说明,所有提到的域都是特征为 0 的域,同时我们约定 $deg(0) = -\infty$。

## 5.1 初等刘维尔扩张

在这一部分,我们给出初等函数的精确定义,和 integrating functions in finite terms(能够用有限形式表示出被积分式,即被积分式可积,之后会有准确的定义)的精确定义

定义 5.1.1 如果 $Dt \in K, t \in K^*$,那么称 $t \in K$ 在 $k$ 上是一个 *primitive*(基本式)。如果 $Dt/t \in k$,那么称 $t \in K^*$ 是一个 *hyperexponential*(超越指数式)。如果 $t$ 在 $k$ 上是代数的,或者 $t$ 是 $k$ 上的一个基本式或者超越指数式,那么称 $t \in K$ 在 $k$ 上是刘维尔的。如果存在 $t_1, \ldots, t_n \in K$,使得 $K = k(t_1, \ldots, t_n)$,且 $t_i$ 在 $k(t_1, \ldots, t_{i-1})$ 上是刘维尔的,对于 $i \in \{1, \ldots, n\}$,那么称 $K$ 是 $k$ 的一个刘维尔扩张。

我们记作 $t = \int a$,当 $t$ 是 $k$ 上的一个基本式时,使得 $Dt = a$;记作 $t = e^{\int a}$,当 $t$ 是 $k$ 上的一个超越指数式,使得 $Dt/d = a$。在已知 $t$ 在 $k$ 上是刘维尔的,我们需要知道 $t$ 在 $k$ 上是代数的还是超越的。我们将会说明有一些简单,必须,充分的条件保证一个基本式或者指数式实际上是 $k$ 上的一个单项式。

引理 5.1.1. 如果 $t$ 是 $k$ 上的一个基本式,$Dt$ 不是 $k$ 中某个元素的微分,那么 $Dt$ 不是 $k$ 的任意一个代数扩张中的元素的微分。

定理 5.1.1. 如果 $t$ 是 $k$ 上的一个基本式,$Dt$ 不是 $k$ 中某个元素的微分,那么 $t$ 是 $k$ 上的一个单项式, $\mathrm{Const}(k(t)) = \mathrm{Const}(k)$,且 $\mathcal{S} = k$(换句话说,$\mathcal{S}^{irr} = \mathcal{S}_1^{irr} = \emptyset$)。反之,如果 $t$ 在 $k$ 上是一个超越基本式,且 $\mathrm{Const}(k(t)) = \mathrm{Const}(k)$,那么 $Dt$ 不是 $k$ 中某个元素的微分。

定理 5.1.2. 如果 $t$ 是 $k$ 上的一个超越指数式,$Dt/t$ 不是一个 $k-$根的对数微分,那么 $t$ 是 $k$ 上的一个单项式, $\mathrm{Const}(k(t)) = \mathrm{Const}(k)$,且 $\mathcal{S}^{irr} = \mathcal{S}_1^{irr} = \{t\}$。反之,如果 $t$ 是 $k$ 的一个超越指数式,$\mathrm{Const}(k(t)) = \mathrm{Const}(k)$,那么 $Dt/t$ 不是一个 $k-$根的对数微分。

实际中我们只考虑满足定理 5.1.1. 或定理 5.1.2.假设的基本式和超越指数式。像已证明的一样,这样的满足额外条件 $\mathrm{Const}(k(t)) = \mathrm{Const}(k)$ 的基本式和超越指数式都是单项式。这些单项式在文献中习惯上被称为刘维尔单项式。

定义 5.1.2. 如果 $t \in K$ 是 $k$ 上的超越元,在 $k$ 上是刘维尔的,并且满足 $\mathrm{Const}(k(t)) = \mathrm{Const}(k)$,那么 $t$ 是一个刘维尔单项式。

我们应该注意到在第三章中单项式的定理并没有要求 $\mathrm{Const}(k(t)) = \mathrm{Const}(k)$,所以第

三章意义下的单项式在$k$上是刘维尔的，却不是定义 5.1.2.中的刘维尔单项式（例如$\log(2)$在$\mathbb{Q}$上）。定理 5.1.1.和定理 5.1.2.可以看作一个基本式或一个超越指数式为一个刘维尔单项式的充要条件。而且，这两个定理描述了在这样的扩张中所有的*special*（特殊）多项式，并且它们都是第一类的。我们也得到：

$$k\langle t\rangle = \begin{cases} k[t], & Dt \in k, \\ k[t, t^{-1}], & Dt/t \in k \end{cases}$$

$k$和$k(t)$有同样的常数域，这个事实使我们可以改善在刘维尔单项式扩张中多项式的次数和它的微分之间的关系，并且加强定理 4.4.4.

引理 5.1.2. 令$t$为$k$上的一个刘维尔单项式，$f \in k(t)$满足$Df \neq 0$，将$f$写作$f = p/q$，其中，$p, q \in k[t]$，并且$q$是首一的。如果$\nu_\infty(f) = 0$，那么$\nu_\infty(Df) \geq 0$。否则$\nu_\infty(f) \neq 0$并且

$$\nu_\infty(Df) = \begin{cases} \nu_\infty(f), & Dt/t \in k \ or \ (\mathrm{lc}(p)) \neq 0 \\ \nu_\infty(f) + 1, & Dt \in k \ and \ (\mathrm{lc}(p)) = 0 \end{cases}$$

我们注意到引理 5.1.2.应用到多项式$p \in k[t]$，其中$t$是$k$上的一个刘维尔单项式这种情况时，意味着

$$\deg(Dp) = \begin{cases} \deg(p), & Dt/t \in k \ or \ D(\mathrm{lc}(p)) \neq 0 \\ \deg(p) - 1, & Dt \in k \ and \ D(\mathrm{lc}(p)) = 0 \end{cases}$$

只要$Dp \neq 0$，在之后我们会经常用到这个结论。

现在我们介绍特殊情况下的刘维尔扩张,这种扩张明确了初等不定积分是否存在的问题，也就是说初等扩张。

定义 5.1.3. 如果$t \in K$，$\exists b \in k^*$使得$Dt = Db/b$，那么称$t$是$k$上的一个对数。如果$t \in K^*$，$\exists b \in k$使得$Dt/t = Db$，那么称$t$是$k$上的一个指数。如果$t \in k$,，且$t$是$k$上的代数元或对数，或指数，那么称$t$在$k$上是初等的。$t \in K$，如果$t$在$k$上是超越的且是初等的，并且$\mathrm{Const}(k(t)) = \mathrm{Const}(k)$，那么称$t$是$k$上的初等单项式。

当$t$是$k$上的对数，即$Dt = Db/b$时，我们将$t$记为$t = \log(b)$，当$t$是$k$上的指数，即$Dt/t = b$时我们将$t$记作$t = e^b$。因为对数为基本式，指数为超越指数式，初等单项式为刘维尔单项式，这一部分的所有结论都可以运用在它们上面。

定义 5.1.4. 如果存在$t_1, \ldots, t_n \in K$，使得$K = k(t_1, \ldots, t_n)$，且$t_i$在$k(t_1, \ldots, t_{i-1})$上是初等的，对于$i \in \{1, \ldots, n\}$，我们称$K$是$k$的一个初等扩张。如果存在一个$k$的一个初等扩张$E$，$g \in E$，使得$Dg = f$，我们称$f \in k$在$k$上有一个初等不定积分。$(\mathbb{C}(x), d/dx)$的任意初等扩张的任意元素称为一个初等函数。

我们现在可以准确地定义初等不定积分存在的问题：确定一个微分域$k$，一个被积函数$f \in k$，以有限步确定$f$在$k$上是否有一个初等的积分，如果有的话，将初等不定积分计算出来。注意在$k$上有一个初等不定积分和有一个初等不定积分的区别：考虑$k = \mathbb{C}(x, t_1, t_2)$，其中$x, t_1, t_2$是$\mathbb{C}$上的不定变元，用微分定义为$Dx = 1, Dt_1 = t_1$和$Dt_2 = t_1/x$（换句话说，即$t_1 = e^x, t_2 = Ei(x)$），那么

$$\int \frac{e^x Ei(x)}{x} dx = \frac{(Ei(x))^2}{2} \in k$$

所以在$k$上有一个初等不定积分，尽管它的积分并不是一个初等函数。这两个概念仅仅当$k$本身是一个初等函数域时是一致的。

注意到定义 5.1.4.中的初等函数包括分析中所有通常的初等函数，因为三角函数和它们的反函数可以根据欧拉方程$e^{f\sqrt{-1}} = cos(f) + sin(f)\sqrt{-1}$的衍生方程用复指数和对数的形式重写。这些变换将$\sqrt{-1}$引入计算，造成计算不便，结果证明是当对实三角函数积分时，这些变换是可以避免的。

## 5.2 积分算法的整体框架

在这一部分我们将给出积分算法的整体框架，使得之后章节的内容，结构便于理解。已知一个被积函数$f(x)dx$，首先我们需要构造一个包含$f$的微分域，我们之后描述的积分算法需要$f$被包含在一个形式为$K = C(t_1, t_2, \ldots, t_n)$的微分域中，其中$C = \text{Const}(K)$，$Dt_1 = 1$（换句话说$t_1 = x$为被积变量），并且每一个$t_i$是$C(t_1, \ldots, t_{i-1})$上的单项式。如果$f(x)$的方程中只包含刘维尔运算，这一条可以在将每一个基本式或超越指数式变量加入扩张之前，通过对它们的变量进行递归式的积分操作进行判断，使用定理 5.1.1.或定理 5.1.2.进行查证，是否为一个刘维尔单项式。另一个整体上更加有效的可替代方法是，在可适用的情况下，应用从各种各样结构理论中衍生出来的算法进行验证。

例 5.2.1. 考虑

$$\int \log(x)\log(x+1)\log(2x^2+2x)dx$$

我们通过

$$Dx = 1, Dt_1 = \frac{1}{x}, Dt_2 = \frac{1}{x+1}, Dt_3 = \frac{2x+1}{x^2+x}$$

构造微分域$K = \mathbb{Q}(x, t_1, t_2, t_3)$
我们一边构造$K$，每一步我们都进行积分，并进行以下的验证：
$\int dx \notin \mathbb{Q}$，所以$x$是$\mathbb{Q}$上的一个刘维尔单项式
$\int dx/x \notin \mathbb{Q}(x)$，所以$t_1$是$\mathbb{Q}(x)$上的一个刘维尔单项式
$\int dx/(x+1) \notin \mathbb{Q}(x, t_1)$，所以$t_2$是$\mathbb{Q}(x, t_1)$上的一个刘维尔单项式

$$\int \frac{2x+1}{x^2+x} dx = t_1 + t_2 \in \mathbb{Q}(x, t_1, t_2)$$

所以$t_3$不是$\mathbb{Q}(x, t_1, t_2)$上的一个刘维尔单项式，$K$作为一个微分域同构于$\mathbb{Q}(c)(x, t_1, t_2)$，其中$c = t_3 - t_1 - t_2 \in \text{Const}(K)$

或者，应用$Risch$结构定理（推论 9.3.1.），对于$a = 2x^2 + 2x$我们的方程为

$$\frac{r_1}{x} + \frac{r+2}{x+1} = \frac{2x+1}{x^2+x}$$

方程有有理解$r_1 = r_2 = 1$。这说明$Dt_3$是$K$中某个元素的微分，$c = t_3 - t_1 - t_2 \in \text{Const}(K)$。

注意到每一个$t_i$都是单项式的要求消除了算法结果表示中出现代数函数的情况。尽管初等函数的不定积分问题中包含代数函数的情况也是可判定的，但是代数函数的不定积分算法

的介绍已经超过了这本书的范畴。

一旦我们得到了单项式的塔状扩张 $K = C(t_1, \ldots, t_n)$，本章中出现的积分算法将 $K$ 中元素的积分问题简化为多种多样涉及 $C(t_1, \ldots, t_{n-1})$ 中元素并和积分相关的问题，从而消除单项式 $t_n$。因为简化问题包括在 $C$ 上的一个具有较小超越指数的塔状扩张中的被积函数问题，我们可以递归地使用算法对它们进行处理，算法能够在有限步结束，这一点是有保障的。为了避免在整本书的行文过程中以完整形式记录完整的塔状扩张，我们记 $K = k(t)$，其中 $k = C(t_1, \ldots, t_{n-1})$，$t = t_n$ 是 $k$ 上的一个单项式，本章中出现的算法的任务是将对 $k(t)$ 中一个给定元素的积分问题转化为 $k$ 上与积分相关的问题。如果 $t$ 在 $k$ 上是初等的，那么在 $k(t)$ 上有一个初等不定积分等价于在 $k$ 上有一个初等积分，所以本书中的算法提供了一个完整的决策过程，以解决判断 $(C(x), d/dx)$ 的一个纯超越初等扩张是否在 $C(x)$ 上有一个初等积分。对于更一般的函数，当 $t$ 在 $k$ 上并不是初等的情况下，可以证明如果 $t$ 是 $k$ 上的一个超越指数单项式或非线性单项式，同时满足 $\mathcal{S}_1^{irr} = \mathcal{S}^{irr}$，那么在 $k(t)$ 上有一个初等不定积分等价于在 $k$ 上有一个初等不定积分。所以算法对于由超越对数，反正切，超越指数，正切函数组成的被积函数来说是完备的。唯一对于刘维尔被积函数的完整算法来说的障碍是当 $t$ 是 $k$ 上的一个非初等基本式的情况，尽管我们可以将问题简化为 $k$ 上的一个被积函数，问题随之变为要决心 $f \in k$ 是否在 $k(t)$ 上有一个初等的不定积分，尽管有针对特殊类型的基本单项式的算法，这个问题对于一般的单项式来讲还没有被完全地解决。正像这本书中很多例子说明的那样，这个算法仍然可以针对很多被积函数包括非初等单项式，的情况下成功地使用。但是当 $t$ 是 $k$ 上的一个废除等基本式时，它不能总是提供一个 $k(t)$ 上初等不定积分不存在的证明。从 $k(t)$ 到 $t$ 的化简对于一般的非线性单项式来讲也是没有完成的，但是对于正切和双曲正切，这个工作已经完成。

不定积分算法的整体轮廓为进行成功的化简，全部的手段都是致力于将被积函数进行形式上更为简单的化解，知道剩余的被积函数是在 $k$ 中。

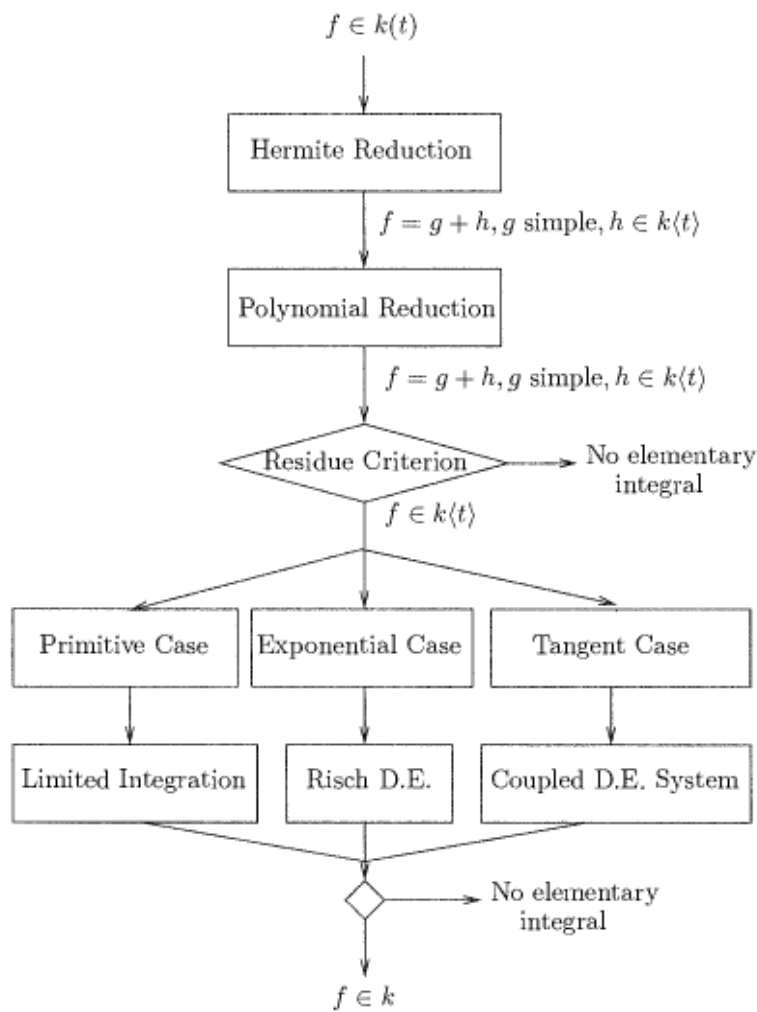

Fig. 5.1. General outline of the integration algorithm

除了最后一部分，多种多样的约化对于任意的单项式扩张都是适用的。

*Hermite reduction*（厄米特约化）可以应用于任一单项式，将整体的被积函数转化为一个*simple*（简单）函数和一个既约函数和的形式

*polynomial reduction*（多项式约化）可以应用于非线性单项式，降低被积函数中多项式部分的次数

*residue criterion*（留数准则）可以应用于任一单项式，或者证明被积函数没有一个$k(t)$上的初等积分，或者将被积函数转化为一个既约被积函数（换句话说，一个$k\langle t\rangle$中的被积函数）

既约被积函数针对刘维尔式或超切向量单项式的各种情况通过特定的算法进行积分。这些算法或者证明不存在$k(t)$上的初等不定积分，或者将问题简化为多种多样的$k$上的积分相关问题。解决这些相关问题的算法在第六章，第七章和第八章中进行描述。

### 5.3 厄米特约化

我们已经在 2.2 节中看到厄米特约化将任意有理函数重新写为一个微分和一个有理函数

和的形式，其中有理函数有一个无平方因子的分母。在这一部分，我们将会说明厄米特约化可以应用于任意单项式扩张中任意元素的 *normal*（正常）部分。对于以下两节，令$(k, D)$为一个微分域，$t$为$k$上的一个单项式。

定义 5.3.1. 对于$f(t)$，我们定义$f$的极点重数为
$$\mu(f) = -\min_{p \in k[t] \setminus k}(\nu_p(f))$$
注意到$\mu(0) = -\infty$，对于任意$f \neq 0$, $\mu(f) \geq 0$，因为在这种情况下，总是存在多项式$p \in k[t]$，使得$\nu_p(f) = 0$。同样上述定义中的最小值可以从$f$分母的不可约因子或无平方因子中取出。容易看出对于$f \neq 0$，$\mu(f)$恰好是$f$分母的任意无平方分解中出现的最高次幂。

定理 5.3.1. 令$f \in k(t)$。在$k[t]$仅适用扩展欧几里得算法，可以找到$g, h, r \in k(t)$，其中$h$是 *simple*（简单）的，$r$是既约的，$f = Dg + h + r$。此外$g, h$和$r$的分母是$f$分母的因子，或者$g = 0$，或者$\mu(g) < \mu(f)$。

尽管我们在以上的证明中已经使用了厄米特约化的二次形式，其它版本在单项式扩张中依然适用。取代将一个有理函数分解为一个微分和一个*simple*（简单）有理函数的处理方法，厄米特约化将$k(t)$中的任意一个元素分解为一个微分，一个*simple*（简单）和一个既约元素和的形式。因此，厄米特约化将任意积分问题简化为被积函数为一个*simple*（简单）和一个既约元素和的形式。

```
HermiteReduce(f, D)        (* Hermite Reduction – quadratic version *)
(* Given a derivation D on k(t) and f ∈ k(t), return g, h, r ∈ k(t) such
   that f = Dg + h + r, h is simple and r is reduced. *)
(f_p, f_s, f_n) ← CanonicalRepresentation(f, D)
(a, d) ← (numerator(f_n), denominator(f_n))        (* d is monic *)
(d_1, ..., d_m) ← SquareFree(d)
g ← 0
for i ← 2 to m such that deg(d_i) > 0 do
    v ← d_i
    u ← d/v^i
    for j ← i − 1 to 1 step −1 do
        (b, c) ← ExtendedEuclidean(u Dv, v, −a/j)
        g ← g + b/v^j
        a ← −jc − u Db
    d ← uv
(q, r) ← PolyDivide(a, uv)
return(g, r/(uv), q + f_p + f_s)
```

例 5.3.1. 令$k = \mathbb{Q}(x)$, $D = d/dx$，令$t$为$k$上的一个单项式，满足$Dt = 1 + t^2$，换句话说$t = \tan(x)$，考虑
$$f = \frac{x - \tan(x)}{\tan(x)^2} = \frac{x - t}{t^2} \in k(t)$$
因为$f$没有多项式部分，$t$在$k[t]$中是 *normal*（普通）的，$f$的典型表示为$(f_p, f_s, f_n) = (0, 0, f)$，所以我们得到$a = x - t$, $d = t^2 = d_2^2$，其中$d_2 = t$，接下来我们得到

| i | v | u | j | b | c | a |
|---|---|---|---|---|---|---|
| 2 | t | 1 | 1 | $-x$ | $xt+1$ | $-xt$ |

$a/uv = -xt/t = -x$，所以厄米特约化返回$(-x/t, 0, -x)$，这就意味着

$$\int \frac{x - tan(x)}{tan(x)^2} dx = -\frac{x}{tan(x)} - \int x dx$$

剩下的被积函数属于$k\langle t \rangle$。

厄米特约化可以被反复使用，将$f$分解为$k(t)$中既约，$simple$（简单）元素的更高阶微分和的形式。

### 5.4 多项式约化

在非线性单项式的情况下，另一种化简的手段使我们可以将$k[t]$中的任意多项式改写为一个微分和一个次数低于$\delta(t)$的多项式的和的形式。

定理 5.4.1. 如果$t$是一个非线性单项式，那么$\forall p \in k[t]$，我们可以找到$q, r \in k[t]$，使得$p = Dq + r$，满足$\deg(r) < \delta(t)$

```
PolynomialReduce(p, D)      (* Polynomial Reduction *)

(* Given a derivation D on k(t) and p ∈ k[t] where t is a nonlinear
monomial over k, return q, r ∈ k[t] such that p = Dq + r, and deg(r) <
δ(t). *)

if deg(p) < δ(t) then return(0, p)
m ← deg(p) − δ(t) + 1
q₀ ← (lc(p)/(mλ(t))) t^m
(q, r) ← PolynomialReduce(p − Dq₀, D)
return(q₀ + q, r)
```

例 5.4.1. 令$k = \mathbb{Q}(x)$，$D = d/dx$，令$t$为$k$上的一个单项式，满足$Dt = 1 + t^2$，换句话说$t = \tan(x)$，考虑

$$p = 1 + x tan(x) + tan(x)^2 = 1 + xt + t^2 \in k[t]$$

我们得到$\delta(t) = 2, \lambda(t) = 1$，应用多项式约化，我们得到
$m = \deg(p) - 1 = 1, q_0 = t, Dq_0 = 1 + t^2$，所以$p - Dq_0 = xt$，次数为1，因此

$$\int (1 + x tan(x) + tan(x)^2) dx = tan(x) + \int x tan(x) dx$$

之后我们会证明剩下的积分并不是一个初等函数。

如果$\int \neq k, \mathcal{S}^{irr} \neq \emptyset$，那么$\mathcal{S}$的任意非平凡元素可以用来消除一个多项式中次数为$\delta(t) - 1$的项。

定理 5.4.2. 假设$t$是一个非线性单项式，令$p \in k[t]$，且$\deg(p) < \delta(t)$，$a \in k$为$p$中$t^{\delta(t)-1}$的

系数，$c = a/\lambda(t)$，那么

$$\deg\left(p - \frac{c}{\deg(q)}\frac{Dq}{q}\right) < \delta(t) - 1$$

$\forall q \in \mathcal{S}\backslash k$

### 5.5 刘维尔定理

确定一个微分域$K$和$K$上的一个被积函数$f$，如果能够找到一个初等不定积分，可以通过微分的方法非常简单地进行验证。此外，当初等不定积分存在的情况下，有几种不同的方法找到初等不定积分。但是证明$f$没有初等不定积分是一个截然不同的问题，因为我们需要一个结果将初等不定积分的存在性和被积函数的特殊形式联系起来。第一个这样的结论是拉普拉斯原理，原理大致说明我们可以将积分问题通过积分中只出现新的线性表示的对数，其余函数必须已经出现在被积函数中进行积分形式上的简化。刘维尔是第一个从这个观察中陈述并且证明了一个严格的定理，首先是在被积函数为代数函数的情况下，接下来扩展到更一般的被积函数的情况下。可以通过阅读第九章了解在 19 世纪，刘维尔定理迷人的历史。这个定理成为了证明对于给定函数，初等不定积分不存在不存在的主要工具。而且，因为定理提供了一个明确的初等扩张类,可供寻找一个不定积分,刘维尔定理组成了积分算法的基础。虽然刘维尔使用的是分析的陈述，但是现在可以在微分域中用代数的方法进行证明。代数技巧首先被$Ostrowski$（奥斯特洛夫斯基）使用，奥斯特洛夫斯基提供了一个刘维尔定理的现代证明，同时当$t$是$k$上的一个基本单项式时，也给出了一个将$k(t)$中的积分问题简化为$k$中的积分问题的算法。刘维尔定理的第一个完整的代数证明由$Rosenlicht$在之后发表，刘维尔定理的强化形式的首次证明由$Risch$完成，$Risch$将证明发布在一起，同时发表了一个针对纯超越初等函数的完整算法。两个人的证明方法都会在这里得到体现，首先我们介绍$Rosenlicht$对若刘维尔定理的证明，接下来逐渐地接触对于常数域的限制，得到$Risch$对强刘维尔定理的证明。我们注意到刘维尔定理已经被扩展到不同的方向上，但是这些扩张已经超越了这本书的范围。

定理 5.5.1. （刘维尔定理） $K$为一个微分域，$f \in K$。如果存在$K$的一个初等扩张$E$，同时满足$\mathrm{Const}(E) = \mathrm{Const}(K)$，$g \in E$使得$Dg = f$，那么存在$v \in K, u_1, \ldots, u_n \in K^*$，$c_1, \ldots, c_n \in \mathrm{Const}(K)$，使得

$$f = Dv + \sum_{i=1}^{n} c_i \frac{Du_i}{u_i}$$

当然，在实际使用中，为了计算积分，正如我们在第二章中看到的那样，我们也许必须添加新的常量。首先，我们指出为了表示出一个初等不定积分，新的超越常量是不必要的。

定理 5.5.2. 令$K$为一个微分域，$K$的常数域代数封闭，$f \in K$。如果存在$K$的一个初等扩张$E$,$g \in E$，是的$Dg = f$，那么存在$v \in K, u_1, \ldots, u_n \in K^*$，$c_1, \ldots, c_n \in \mathrm{Const}(K)$，使得

$$f = Dv + \sum_{i=1}^{n} c_i \frac{Du_i}{u_i}$$

最终我们可以在刘维尔定理中对于常数域做出的所有限制，指出对于任意常数子域，刘

维尔定理表达式：$f = Dv + \sum_{i=1}^{n} c_i \frac{Du_i}{u_i}$中的$v$可以在$K$中取到，$u_i$可以在$K(c_1,\ldots,c_n)$中取到。

定理 5.5.3.（刘维尔定理—强形式）$K$为一个微分域，$C = \text{Const}(K)$，$f \in K$。如果存在$K$的一个初等扩张$E$，$g \in E$，使得$Dg = f$，那么存在$v \in K$，$c_1,\ldots,c_n \in \overline{C}$，$u_1,\ldots,u_n \in K(c_1,\ldots,c_n)^*$，使得

$$f = Dv + \sum_{i=1}^{n} c_i \frac{Du_i}{u_i}$$

### 5.6 留数准则

现在刘维尔定理为我们提供了一种证明一个函数在一个给定域上没有初等不定积分的方法，我们可以完成积分算法。在本章的剩余部分，令$(k, D)$为一个微分域，$t$为$k$上的一个单项式。根据厄米特约化，不失一般性我们可以假设被积函数可以写作$k(t)$中的一个$simple$（简单）和一个既约元素和的形式。

我们在2.4节中已经看到，$Rothstein - Trager$算法将一个$simple$（简单），无多项式部分的有理函数的积分写作对数和的形式。在这一小节，我们将指出这个算法可以推广到任一单项式扩张上，既可以证明一个函数没有初等不定积分，或者可以将积分问题简化到$k\langle t\rangle$中元素的积分问题。$Rothstein$在他的论文中将这个算法推广到初等超越扩张上。

引理 5.6.1. $f \in k(t)$，且$f$是$simple$（简单）的，如果存在$h \in k\langle t\rangle$，$E$为$\text{Const}(k)$的一个代数扩张，$v \in k(t)$，$c_1,\ldots,c_n \in E$，并且$u_1,\ldots,u_n \in Ek(t)$，使得

$$f + h = Dv + \sum_{i=1}^{n} c_i \frac{Du_i}{u_i}$$

那么
$$\text{residue}_p(f) = \sum_{i=1}^{n} c_i \nu_p(u_i)(0)$$
对于$\forall p \in Ek[t]$，$p$为$normal$（普通）不可约的。

引理 5.6.2. 假设$\text{Const}(k)$是代数封闭的，令$f \in k(t)$，且$f$是$simple$（简单）的。如果存在$h \in k\langle t\rangle$使得$f + h$在$k(t)$有一个初等不定积分，那么$\text{residue}_p(f) \in \text{Const}(k)$，对于$\forall p \in k[t]$，$p$是$normal$（普通）且不可约的。

例 5.6.1. 令$k = \mathbb{Q}$，$t$是$k$上的一个单项式，且$Dt = 1$（换句话说，$D = d/dt$），并且

$$f = \frac{2t - 2}{t^2 + 1} \in k(t)$$

那么，$f$在$k(t)$上有一个初等不定积分：
$$\int \frac{2t - 2}{t^2 + 1} dt = (1 + \sqrt{-1})\log(1 + t\sqrt{-1}) + (1 - \sqrt{-1})\log(1 - t\sqrt{-1})$$

另一方面，$t^2 + 1$在$\mathbb{Q}$上是不可约的，但是

$$\text{residue}_{t^2+1}(f) = \pi_{t^2+1}\left(\frac{2t-2}{2t}\right) = t+1$$

并不是一个常数。这说明引理 5.6.2 关于$k$的常数域是代数封闭的这一个假设是必需的。如果我们将$\mathbb{Q}$替换为$\mathbb{C}$，那么$t^2+1 = (t-\sqrt{-1})(t+\sqrt{-1})$

$$\text{residue}_{t-\sqrt{-1}}(f) = \pi_{t-\sqrt{-1}}\left(\frac{2t-2}{t+\sqrt{-1}}\right) = 1+\sqrt{-1}$$

并且

$$\text{residue}_{t+\sqrt{-1}}(f) = \pi_{t+\sqrt{-1}}\left(\frac{2t-2}{t-\sqrt{-1}}\right) = 1-\sqrt{-1}$$

为常数。这说明引理 5.6.1.和引理 5.6.2.中$p$是不可约的这一个假设也是必须满足的。

定理 5.6.1. $f \in k(t)$ 是 $simple$（简单）的，将 $f$ 写作 $f = p + a/d$，其中 $p, a, d \in k[t], d \neq 0, \deg(a) < \deg(d)$，并且$\gcd(a, b) = 1$。令$z$是$k$上的一个不定变元，
$$r = \text{resultant}_t(a - zDd, d) \in k[z]$$
$r = r_s r_n$为$r$的一个关于$D$，映射到$k[z]$的系数映射$\kappa_D$作用下的分裂分解。且

$$g = \sum_{r_s \alpha = 0} \alpha \frac{Dg_\alpha}{g_\alpha}$$

其中$g_\alpha = \gcd(a - \alpha Dd, d) \in k(\alpha)[t]$，和由$r_s$的所有不同的根组成，那么
$(i)$ $g \in k(t)$，$g$的分母整除$d$，$f - g$是$simple$（简单）的
$(ii)$如果存在$h \in k\langle t\rangle$，使得$f + h$在$k(t)$上有一个初等不定积分，那么$r_n \in k$，且$f - g \in k[t]$
$(iii)$ 如果存在 $h \in k\langle t\rangle$，$\text{Const}(k)$ 的一个代数扩张 $E$，$v \in k(t), c_1, \ldots, c_n \in E, u_1, \ldots, u_n \in Ek(t)$，使得

$$f + h = Dv + \sum_{i=1}^{n} c_i \frac{Du_i}{u_i}$$

那么$r_s$在$E$上可以线性分解。

注意到因为根据定理 3.5.2，$r_s$的所有根都是常数，由$g = \sum_{r_s \alpha = 0} \alpha \dfrac{Dg_\alpha}{g_\alpha}$得到的$g$总是有一个初等不定积分，换句话说

$$\int g = \sum_{r_s(\alpha)=0} \alpha \log(\gcd(d, a - \alpha Dd))$$

这就是在有理函数情况下的$Rothstein-Trager$公式。定理 5.6.1.的第三部分应用到有理函数情况下证明了定理 2.4.1.的第三部分，因此完成了定理 2.4.1. 的证明。在有理函数的情况下，一个素数分解$r_s = u s_1^{e_1} \cdots s_m^{e_m}$是必须的，同样也需要对于每一个$i$，在$k(\alpha_i)[t]$中可以进行最大公约数计算，其中$\alpha_i$是$s_i$的一个根。但是没有必要计算$r_s$的分裂域。而且，$r_s$的首一部分总是有常值系数。

```
ResidueReduce(f, D)      (* Rothstein–Trager resultant reduction *)
(* Given a derivation D on k(t) and f ∈ k(t) simple, return g elementary
   over k(t) and a Boolean b ∈ {0, 1} such that f − Dg ∈ k[t] if b = 1, or
   f + h and f + h − Dg do not have an elementary integral over k(t) for
   any h ∈ k⟨t⟩ if b = 0. *)
d ← denominator(f)
(p, a) ← PolyDivide(numerator(f), d)                    (* f = p + a/d *)
z ← a new indeterminate over k(t)
r ← resultant_t(d, a − zDd)
(r_n, r_s) ← SplitFactor(r, κ_D)
u s_1^{e_1} ··· s_m^{e_m} ← factor(r_s)                 (* factorization into irreducibles *)
for i ← 1 to m do
    α ← α | s_i(α) = 0
    g_i ← gcd(d, a − αDd)                               (* algebraic gcd computation *)
if r_n ∈ k then b ← 1 else b ← 0
return(∑_{i=1}^{m} ∑_{α|s_i(α)=0} α log(g_i), b)
```

例 5.6.2. 考虑
$$\int \frac{2\log(x)^2 - \log(x) - x^2}{\log(x)^3 - x^2\log(x)} dx$$
令$k = \mathbb{Q}(x)$，$D = d/dx$，令$t$为$k$上的一个单项式，满足$Dt = 1/x$，换句话说$t = \log(x)$，我们的被积函数变为
$$f = \frac{2t^2 - t - x^2}{t^3 - x^2 t} \in k(t)$$
为$simple$（简单）的，因为$t^3 - x^2 t$是无平方因子的。我们得到
$$d = t^3 - x^2 t, p = 0, a = 2t^2 - t - x^2$$
同时
$$r = \text{resultant}_t \left((t^3 - x^2 t, \tfrac{2x-3z}{x}t^2 + (2xz - 1)t + x(z - x)\right)$$
$$= 4x^3(1 - x^2)\left(z^3 - xz^2 - \tfrac{1}{4}z + \tfrac{x}{4}\right)$$
为无平方因子的，那么
$$\kappa_D r = -x^2(4(5x^2 + 3)z^3 + 8x(3x^2 - 2)z^2 + (5x^2 - 3)z - 2x(3x^2 - 2))$$
所以$r$关于$\kappa_D$的分裂分解为
$$r_s = \gcd(r, \kappa_D r) = x^2\left(z^2 - \frac{1}{4}\right)$$
并且
$$r_n = \frac{r}{r_s} = -4x(x^2 - 1)(z - x) \notin k$$
因此，$f$没有一个初等不定积分，更进一步地我们得到
$$g_1 = \gcd\left(t^3 + x^2 t, \frac{2x - 3\alpha}{x}t^2 + (2x\alpha - 1)t + x(\alpha - x)\right) = t + 2\alpha x$$
其中$\alpha^2 - 1/4 = 0$，所以

$$g = \sum_{\alpha|\alpha^2-1/4=0} \alpha\log(t+2\alpha x) = \frac{1}{2}\log(t+x) - \frac{1}{2}\log(t-x)$$

计算$f-Dg$，我们发现
$$\int \frac{2\log(x)^2 - \log(x) - x^2}{\log(x)^3 - x^2\log(x)}dx = \frac{1}{2}\log\left(\frac{\log(x)+x}{\log(x)-x}\right) + \int \frac{dx}{\log(x)}$$
$$= \tfrac{1}{2}\log\left(\tfrac{\log(x)+x}{\log(x)-x}\right) + \text{Li}(x)$$

其中$\text{Li}(x)$是对数积分，因为$r_n \notin k$，所以可以证明$\text{Li}(x)$是非初等函数。

使用定理 5.6.1. 中的符号，我们有$\gcd(r_s, r_n) = 1$，所以$r_s$的任意$n$重根$\alpha$也是$r$的一个$n$重根。因为$\gcd(a, d) = \gcd(d, Dd) = 1$，且$\deg(a) < \deg(d)$，我们可以应用定理 2.5.1. ，其中$A = a, B = Dd, C = d$，对于$r$的任意$i(i>0)$重根我们得到，
$$\gcd(d, a - \alpha Dd) = pp_t(R_m)(\alpha, t)$$

其中$\deg_t(R_m) = i$，如果$\deg(Dd) \leq \deg(d)$，那么$R_m$在$d$和$a - zDd$的$PRS$子结式中，如果$\deg(Dd) > \deg(d)$，那么$R_m$在$a - zDd$和$d$的$PRS$子结式中。因此，$Lazard - Rioboo - Trager$算法对于任意单项式扩张来说是适用的，并且没有必要计算$r_s$的素数分解，或者由式$g = \sum_{r_s\alpha=0} \alpha\frac{Dg_\alpha}{g_\alpha}$得到的$g_\alpha$，我们可以使用子结式中出现的多种多样的余数加以代替。如同有理函数的情况，我们使用$r_s$的一个无平方分解$r_s = \prod_{i=1}^n q_i^i$来分解$g = \sum_{r_s\alpha=0} \alpha\frac{Dg_\alpha}{g_\alpha}$这个和形式，得到一些被加数，每一个都可以用$q_i$的根的指数形式表示出来。

我们也可以避免计算$pp_t(R_m)$，只要它的首项系数和对应的$q_i$互质。因为$g = \sum_{r_s\alpha=0} \alpha\frac{Dg_\alpha}{g_\alpha}$式中的任意一个$g_\alpha$乘上$k(\alpha)$中的任意一个非零元素并不改变定理 5.6.1. 的结论。我们可以令$pp_t(R_m)$首一，以便简化答案。这最后一步需要求$k[\alpha]$中一个元素的逆，并且最后一步不是必须的。正如有理函数情况下，结果证明$pp_t(R_m)(\alpha, t)$的首项系数在$k[\alpha]$中总是可逆的。

```
ResidueReduce(f, D)
(* Lazard–Rioboo–Rothstein–Trager resultant reduction *)

  (* Given a derivation D on k(t) and f ∈ k(t) simple, return g elementary
  over k(t) and a Boolean b ∈ {0, 1} such that f − Dg ∈ k[t] if b = 1, or
  f + h and f + h − Dg do not have an elementary integral over k(t) for
  any h ∈ k⟨t⟩ if b = 0. *)

  d ← denominator(f)
  (p, a) ← PolyDivide(numerator(f), d)              (* f = p + a/d *)
  z ← a new indeterminate over k(t)
  if deg(Dd) ≤ deg(d)
    then (r, (R_0, R_1, ..., R_q, 0)) ← SubResultant_x(d, a − zDd)
    else (r, (R_0, R_1, ..., R_q, 0)) ← SubResultant_x(a − zDd, d)
  ((n_1, ..., n_n), (s_1, ..., s_n)) ← SplitSquarefreeFactor(r, κ_D)
  for i ← 1 to n such that deg(s_i) > 0 do
    if i = deg(d) then S_i ← d
    else
        S_i ← R_m where deg_t(R_m) = i,   1 ≤ m < q
        (A_1, ..., A_s) ← SquareFree(lc_t(S_i))
        for j ← 1 to s do S_i ← S_i / gcd_z(A_j, s_i)^j     (* exact quotient *)
  if ∏_{i=1}^n n_i ∈ k then b ← 1 else b ← 0
  return(∑_{i=1}^m ∑_{α|s_i(α)=0} α log(S_i(α, t)), b)
```

例 5.6.3. 考虑和例 5.6.2.中相同的被积函数

$$\int \frac{2\log(x)^2 - \log(x) - x^2}{\log(x)^3 - x^2\log(x)} dx$$

我们有$\mathrm{Dd} < \deg(d)$,$d$和$a - zDd$的$PRS$子结式为

| $i$ | $R_i$ |
|---|---|
| 0 | $t^3 - x^2 t$ |
| 1 | $(2 - 3z/x)t^2 + (2xz - 1)t + x(z - x)$ |
| 2 | $(4x^2 - 6)z^2 + 3xz - 2x^2 + 1)t + x(z - x)(2xz - 1)$ |
| 3 | $4x^3(1 - x^2)\left(z^3 - xz^2 - \frac{1}{4}z + \frac{1}{4}x\right)$ |

$Rothstein - Trager$结式为$r = R_3$,它关于$\kappa_D$的无平方因式分解为

$$s_1 = \gcd(r, \kappa_D r) = x^2 \left(z^2 - \frac{1}{4}\right), n_1 = \frac{r}{s_1} = -4x(x^2 - 1)(z - x) \notin k$$

因此,$f$没有初等不定积分。更进一步我们可以发现$s_1$是无平方因子的,在$PRS$中,$t$的次数为 1 的余数为

$$R_2 = ((4x^2 - 6)z^2 + 3xz - 2x^2 + 1)t + x(z - x)(2xz - 1)$$

因为

$$\gcd(\mathrm{lc}_t(R_2), s_1) = \gcd\left((4x^2 - 6)z^2 + 3xz - 2x^2 + 1, x^2\left(z^2 - \tfrac{1}{4}\right)\right) = 1$$

$S_1 = R_2$。在求在$z^2 - 1/4 = 0$的根$\alpha$处的值,我们有

$$S_1(\alpha, t) = -\frac{1}{2}((2x^2 - 6\alpha x + 1)t + 4\alpha x^3 - 3x^2 + 2\alpha x)$$

所以
$$g = \sum_{\alpha|\alpha^2-1/4=0} \alpha\log\left(-\tfrac{1}{2}((2x^2-6\alpha x+1)t+x(4\alpha x^2-3x+2\alpha))\right)$$
$$= \tfrac{1}{2}\log\left(-\tfrac{(2x^2-3x+1)(t+x)}{2}\right) - \tfrac{1}{2}\log\left(-\tfrac{(2x^2+3x+1)(t-x)}{2}\right)$$

计算 $f-Dg$ 我们发现

$$\int \frac{2\log(x)^2-\log(x)-x^2}{\log(x)^3-x^2\log(x)}dx = \tfrac{1}{2}\log\left(\frac{(2x^2-3x+1)(\log(x)+x)}{(2x^2+3x+1)(\log(x)-x)}\right) + \int\left(\frac{1}{\log(x)} - \frac{6x^2-3}{4x^4-5x^2+1}\right)dx$$

剩余的积分部分已经证明是非初等的。实际上，它是一个有理函数的积分加上一个对数积分。

如果我们已经决定令 $S_1(\alpha,t)$ 首一，我们将会得到

$$S_1(\alpha,x) = -\frac{1}{2}(2x^2-6\alpha x+1)(t+2\alpha x)$$

所以积分结果和例 5.6.2. 中的结果是一致的。

## 5.7 既约函数的积分

在之前小节结果的基础上，我们所要解决的剩余问题就是在一个单项式扩张中既约函数的积分问题。对于这样的元素，我们使用刘维尔定理的一个特殊的版本。

定理 5.7.1. 令 $k$ 为一个微分域，$t$ 是 $k$ 上的一个单项式，$C = \text{Const}(k(t))$，$f \in k\langle t\rangle$。如果存在 $k(t)$ 的一个初等扩张 $E$，$g \in E$，使得，那么存在 $v \in k\langle t\rangle$，$c_1,\ldots,c_n \in \overline{C}$ 和 $u_1,\ldots,u_n \in S_{k(c_1,\ldots,c_n)[t]:k(c_1,\ldots,c_n)}$ 使得

$$f = Dv + \sum_{i=1}^{n}\frac{Du_i}{u_i}$$

在非线性单项式的情况下，我们已经指出我们总可以将 $k[t]$ 中的一个多项式 $p$ 改写为一个微分和一个次数比 $\delta(t)$ 低的多项式的和的形式。我们就得到了与留数准则相似的一种情况，或者证明这样的一个既约函数没有一个初等的不定积分，或者从它的多项式部分中消除次数为 $\delta(t)-1$ 的项。

定理 5.7.2. 假设 $t$ 是一个非线性单项式。令 $f \in k\langle t\rangle$，将 $f$ 改写为 $f = p + a/d$ 的形式，其中 $p,a,d \in k[t], d \neq 0, \deg(p) < \delta(t)$，且 $\deg(a) < \deg(d)$。令 $b \in k$ 为 $p$ 中 $t^{\delta(t)-1}$ 的系数，$c = b/\lambda(t)$。如果 $f$ 在 $k(t)$ 上有一个初等的不定积分，那么 $Dc = 0$。

如果 $c$ 是一个常数，那么定理 5.4.2. 说明

$$f - D\left(\frac{c}{\deg(q)}\log(q)\right)$$

对于 $\forall q \in S\backslash k$，次数至多为 $\delta(t)-2$，所以在非线性单项式的情况下，假若我们至少知道一个非平凡的 *special*（特殊）多项式，我们所要解决的既约函数的多项式部分的次数至多为 $\delta(t)-2$。如果我们知道不存在非平凡 *special*（特殊）多项式，那么事实上，这样的非线性扩张的既约元素的积分问题就会更加简单，相应的算法将会在 5.11 中给出。

现在我们已经掌握多有完善积分算法的工具。在接下来的小节中，我们将会给出算法。

已知$k(t)$中的一个被积函数，$t$为一个单项式，我们或者证明$f$在$k(t)$上没有初等不定积分，或者计算$k(t)$的一个初等扩张和$E$中元素$g$，使得$f - Dg \in k$。这个方法将被积函数中的$t$消除，因此将问题简化为$k$上元素的积分问题，可以通过递归式的方法进行解决，换句话说，本章的算法可以不断地应用于$k$上的元素知道我们只需要解决常数积分的问题。注意到当$t$自身不是$k$上的初等元时，$k$中元素是否在$k$或$k(t)$上有初等的不定积分的问题就变得从本质上讲是不同的。所以只有当被积函数自身是一个初等函数时，我们的算法会给出不可积的证明。但是这些算法可以应用于更大的函数类上。

同时，假设一些积分相关问题对于$k$上元素是可解决的是必需的。这些问题依赖于我们处理的单项式的类型，所以在这一点上，我们需要分别处理各种各样的情况。所有积分相关问题的算法将会在之后的章节给出。

## 5.8 基本式情况

在微分域$k$上的 *primitive*（基本）单项式的情况下，我们需要解决的$k$上的相关问题为 *limited integration problem*（受限积分问题）：回忆初等不定积分是否存在的定义为：已知$f \in k$，判定是否存在$k$的一个初等扩张$E$和$g \in E$，是的$\mathrm{Const}(E)$在$\mathrm{Const}(k)$上是代数的，且$Dg = f$。令$w_1, \ldots, w_n \in k$固定。关于$w_1, \ldots, w_n$的受限积分问题是：已知$f \in k$，判定是否存在$g \in k$，$c_1, \ldots, c_n \in \mathrm{Const}(k)$，使得$Dg = f - c_1 w_1 - \ldots - c_n w_n$，如果存在的话计算$g$和$c_i$。这与初等不定积分是否存在这个问题很像，除非为积分进行特定的微分扩张$k(\int w_1, \ldots, \int w_n)$。在这一部分我们提供一个算法，对$k$做出恰当的假设，当$t$是$k$上的一个基本单项式时，对$k(t)$的元素进行积分。首先我们描述对$k[t]$元素进行积分的算法。

定理 5.8.1. 令$k$为一个微分域，$t$是$k$上的一个基本式。如果关于$Dt$的受限积分问题对于$k$中元素是可判定的，并且$Dt$不是$k$中元素的微分，那么对于$\forall p \in k[t]$，或者我们可以证明$p$在$k(t)$上没有初等不定积分，或者计算$q \in k[t]$，使得$p - Dq \in k$。

```
IntegratePrimitivePolynomial(p, D)
(* Integration of polynomials in a primitive extension *)

    (* Given a is a primitive monomial t over k, and p ∈ k[t], return q ∈ k[t]
    and a Boolean β ∈ {0, 1} such that p − Dq ∈ k if β = 1, or p − Dq does
    not have an elementary integral over k(t) if β = 0. *)

    if p ∈ k then return(0, 1)
    a ← lc(p)
    (* LimitedIntegrate will be given in Chap. 7 *)
    (b, c) ← LimitedIntegrate(a, Dt, D)          (* a = Db + cDt *)
    if (b, c) = "no solution" then return(0, 0)
    m ← deg(p)
    q_0 ← ct^{m+1}/(m + 1) + bt^m
    (q, β) ← IntegratePrimitivePolynomial(p − Dq_0, D)
    return(q + q_0, β)
```

例 5.8.1 考虑

$$\int \left( \left( \log(x) + \frac{1}{\log(x)} \right) \mathrm{Li}(x) - \frac{x}{\log(x)} \right) dx$$

其中$\text{Li}(x) = \int dx/\log(x)$是对数积分。令$k = \mathbb{Q}(x, t_0)$，$D = d/dx$，其中$t_0$是$\mathbb{Q}(x)$上的单项式，满足$Dt_0 = 1/x$，换句话说$t_0 = \log(x)$，令$t$是$k$上的一个单项式满足$Dt = 1/t_0$，换句话说$t = \text{Li}(x)$，我们被积函数变为

$$p = \left(t_0 + \frac{1}{t_0}\right)t - \frac{x}{t_0} \in k[t]$$

我们得到

1. $a = \text{lc}(p) = t_0 = 1/t_0$

2. $\left(t_0 = \frac{1}{t_0)}\right) - \frac{1}{t_0} = t_0 = \log(x) = \frac{d}{dx}(x\log(x) - x) = D(xt_0 - x)$

   所以$(b, c) = \textbf{LimitedIntegrate}(t_0 + 1/t_0, 1/t_0, D) = (xt_0 - x, 1)$

3. $q_0 = ct^2/2 + bt = t^2/2 + (xt_0 - x)t$

4. $p - Dq_0 = -x \in k$，所以调用$\textbf{IntegratePrimitivePolynomial}(-x, D)$，返回$(q, \beta) = (0, 1)$

因此，

$$\int\left(\left(\log(x) + \frac{1}{\log}\right)(x)\text{Li}(x) - \frac{x}{\log}(x)\right)dx$$

$$= \frac{\text{Li}(x)^2}{2} + (x\log(x) - x)\text{Li}(x) - \int x\, dx$$

$$= \frac{\text{Li}(x)^2}{2} + (x\log(x) - x)\text{Li}(x) - \frac{x^2}{2}$$

将所有部分组合在一起，我们就得到了对$k(t)$元素做积分的算法。

**定理 5.8.2.** 令$k$为一个微分域，$t$为$k$上的一个基本式。如果关于$Dt$的受限积分问题对于$k$中元素是可判定的，并且$Dt$不是$k$中元素的微分，那么对于$\forall f \in k(t)$，我们或者可以证明$f$在$k(t)$上没有初等不定积分，或者计算出$k(t)$的一个初等扩张$E$和$g \in E$，使得$f - Dg \in k$。

```
IntegratePrimitive(f, D)      (* Integration of primitive functions *)

(* Given a is a primitive monomial t over k, and f ∈ k(t), return g
   elementary over k(t) and β ∈ {0, 1} such that f − Dg ∈ k if β = 1, or
   f − Dg does not have an elementary integral over k(t) if β = 0. *)

(g₁, h, r) ← HermiteReduce(f, D)
(g₂, β) ← ResidueReduce(h, D)
if β = 0 then return(g₁ + g₂, 0)
(q, β) ← IntegratePrimitivePolynomial(h − Dg₂ + r, D)
return(g₁ + g₂ + q, β)
```

### 5.9 超越指数式情况

在微分域$k$的超越指数式的情况下，我们在$k$上需要解决的相关问题为$Risch\ differential\ equation\ problem$（$Risch$微分方程问题）：已知$f, g \in k$，判定是否存在$y \in k$使得

$$Dy + fy = g$$

如果存在的话，计算出$y$。通常$Dy + fy = g$在$k$中有多余一个的解，所以首先我们需要验证多解的情况什么时候能出现。

引理 5.9.1. 令 $(K,D)$ 为一个微分域。如果存在 $\alpha, y, z \in K$，使得 $y \neq z$ 并且 $Dy + \alpha y = Dz + \alpha z$，那么存在 $u \in K^*$，使得 $\alpha = Du/u$。

在这一部分，我们在 $k$ 上进行适当的假设，提出一个算法，当 $t$ 是 $k$ 上的一个超越指数单项式，进行 $k(t)$ 上元素的积分。首先我们描述一个对 $k\langle t\rangle$ 元素进行积分的算法。

定理 5.9.1. 令 $k$ 为一个微分域，$t$ 为 $k$ 上的一个超越指数式。如果在 $k$ 上我们可以解出 $Risch$ 微分方程，并且 $Dt/t$ 不是一个 $k-$根的对数积分，那么对于 $\forall p \in k\langle t\rangle$，或者我们可以证明 $p$ 在 $k(t)$ 上没有初等不定积分，或者计算出 $q \in k\langle t\rangle$，使得 $p - Dq \in k$。

```
IntegrateHyperexponentialPolynomial(p, D)
(* Integration of hyperexponential polynomials *)

(* Given an hyperexponential monomial t over k and p ∈ k[t, t⁻¹] return
   q ∈ k[t, t⁻¹] and a Boolean β ∈ {0,1} such that p − Dq ∈ k if β = 1, or
   p − Dq does not have an elementary integral over k(t) if β = 0. *)

q ← 0, β ← 1
for i ← ν_t(p) to −ν_∞(p) such that i ≠ 0 do
    a ← coefficient(p, tⁱ)
    (* RischDE will be given in Chap. 6 *)
    v ← RischDE(iDt/t, a)              (* a = Dv + ivDt/t *)
    if v = "no solution" then β ← 0 else q ← q + vtⁱ
return(q, β)
```

例 5.9.1. 考虑
$$\int \left( (tan(x)^3 + (x+1)tan(x)^2 + tan(x) + x + 2)e^{tan(x)} + \frac{1}{x^2+1} \right) dx$$
令 $k = \mathbb{Q}(x, t_0)$，$D = d/dx$，其中 $t_0$ 是 $\mathbb{Q}(x)$ 上的单项式，满足 $Dt_0 = 1 + t_0^2$，换句话说 $t_0 = tan(x)$，令 $t$ 为 $k$ 上单项式，满足 $Dt = (1+t_0^2)t$，换句话说，$t = e^{tan(x)}$，被积函数变为

$$p = (t_0^3 + (x+1)t_0^2 + t_0 + x + 2)t + \frac{1}{x^2+1} \in k[t]$$

我们得到
1. $q = 0, \beta = 1$
2. $\nu_t(p) = -\nu_\infty(p) = 1$
3. $i = 1$
4. $a = \text{lc}(p) = t_0^3 + (x+1)t_0^2 + t_0 + x + 2$
5. $D(t_0 + x) + (1+t_0^2)(t_0 + x) = a$，所以 $v = \textbf{RischDE}(1+t_0^2, a) = t_0 + x$
6. $q = vt = (t_0 + x)t$
7. $p - Dq = 1/(x^2+1)$

因此，
$$\int \left( (tan(x)^3 + (x+1)tan(x)^2 + tan(x) + x + 2)e^{tan(x) + \frac{1}{x^2+1}} \right) dx$$

$$= (tan(x) + x)e^{tan(x)} + \int \frac{dx}{x^2+1}$$
$$= (tan(x) + x)e^{tan(x)} + arctan(x)$$

将所有部分组合在一起，我们就得到了一个对$k(t)$元素进行积分的算法。

定理 5.9.2. 令$k$为一个微分域，$t$为$k$上的一个超越指数式。如果我们可以解出$k$上的$Risch$微分方程，并且$Dt/t$不是一个$k-$根的对数积分，那么对于$\forall f \in k(t)$，我们或者可以证明$f$在$k(t)$没有初等不定积分，或者可以计算出$k(t)$的一个初等扩张$E$和$g \in E$，满足$f - Dg \in k$。

**IntegrateHyperexponential**$(f, D)$
(* Integration of hyperexponential functions *)

(* Given an hyperexponential monomial $t$ over $k$ and $f \in k(t)$, return $g$ elementary over $k(t)$ and a Boolean $\beta \in \{0,1\}$ such that $f - Dg \in k$ if $\beta = 1$, or $f - Dg$ does not have an elementary integral over $k(t)$ if $\beta = 0$. *)

$(g_1, h, r) \leftarrow$ **HermiteReduce**$(f, D)$
$(g_2, \beta) \leftarrow$ **ResidueReduce**$(h, D)$
if $\beta = 0$ then return$(g_1 + g_2, 0)$
$(q, \beta) \leftarrow$ **IntegrateHyperexponentialPolynomial**$(h - Dg_2 + r, D)$
return$(g_1 + g_2 + q, \beta)$

### 5.10. 超切向量情况

正切和三角函数可以转化为负对数和复指数，再进行积分，但是单项式扩张的理论使我们可以对它们进行直接积分，并不需要引入代数元$\sqrt{-1}$。我们首先定义正切单项式，计算$special$（特殊）多项式。令$k$为一个微分域，$K$是$k$的一个微分扩张。

定义 5.10.1. 令$t \in K$，满足$t^2 + 1 \neq 0$。$t$为$k$上的一个超切向量，如果$Dt/(t^2+1) \in k$。如果$Dt/(t^2+1) = Db, \exists b \in k$，那么$t$是$k$上的一个正切。如果$t$是$k$上的一个超切向量（正切），并且$t$在$k$上是超越的，且$\text{Const}(k(t)) = \text{Const}(k)$，那么$t$是$k$上的一个正切向量（正切）单项式。

当$t$是$k$上的一个超切向量时，满足$Dt/(t^2+1) = a$，我们将$t$记作$t = tan(\int a)$。如果$t$是$k$上的一个正切，满足$Dt/(t^2+1) = Db$，我们将$t$记作$t = tan(b)$。

引理 5.10.1. 令$(F, D)$为一个包含$\sqrt{-1}$的微分域，$a \in F$满足$a^2 + 1 \neq 0$，并且$b = (\sqrt{-1} - a)/(\sqrt{-1} + a)$，那么$b \neq 0$并且
$$\frac{Db}{b} = 2\sqrt{-1}\frac{Da}{a^2+1}$$

定理 5.10.1. 如果$t$是$k$上的一个超切向量，并且$\sqrt{-1}Dt/(t^2+1)$不是一个$k(\sqrt{-1})-$根的对数微分，那么$t$是$k$上的一个单项式，$\text{Const}(k(t)) = \text{Const}(k)$，并且$\forall p \in \mathcal{S}^{irr}$在$k[t]$中整除$t^2 + 1$。此外，$\mathcal{S}_1^{irr} = \mathcal{S}^{irr}$。反之，如果$t$在$k$上是超越的，且是一个超切向量，并且$\text{Const}(k(t)) = \text{Const}(k)$，那么$\sqrt{-1}Dt/(t^2+1)$不是一个$k(\sqrt{-1})-$根的对数微分。

因此，我们得到
$$k\langle t\rangle = \{f \in k(t), 存在整数 n \geq 0, 使得 (t^2+1)^n \in k[t]\}$$
其中$t$是$k$上的一个超切向量单项式。现在我们可以在$k$上做出合适假设的情况下，当$t$是$k$上的一个超切向量单项式时，提出一个对$k(t)$中元素做积分的算法。首先注意到，多项式$X^2+1$在$k$上可分解的话，那么$\sqrt{-1} \in k$，所以$k(t) = k(\theta)$，其中$\theta = (\sqrt{-1} - t)/(\sqrt{-1} + t)$是$k$上的一个超越指数单项式。因此我们在这种情况下使用对超越指数扩张元素做积分的算法，所以我们可以在这一小节的剩余部分中假设$X^2+1$在$k$上不可约，换句话说即是$\sqrt{-1} \notin k$。因为超切向量是非线性单项式，对于$k[t]$中元素做积分是直截了当的。

定理 5.10.2. 令$k$为一个不包含$\sqrt{-1}$的微分域，$t$是$k$上的一个超切向量。如果$\sqrt{-1}Dt/(t^2+1)$不是一个$k(\sqrt{-1})-$根的对数微分，那么对于$\forall p \in k[t]$，我们可以计算$q \in k[t]$和$c \in k$满足
$$p - Dq - c\frac{D(t^2+1)}{t^2+1} \in k$$
而且，如果$Dc \neq 0$，那么$p$在$k$上没有初等不定积分。

```
IntegrateHypertangentPolynomial(p, D)
(* Integration of hypertangent polynomials *)

(* Given a differential field k such that √-1 ∉ k, a hypertangent mono-
mial t over k and p ∈ k[t], return q ∈ k[t] and c ∈ k such that
p − Dq − cD(t² + 1)/(t² + 1) ∈ k and p − Dq does not have an elementary
integral over k(t) if Dc ≠ 0. *)

(q, r) ← PolynomialReduce(p, D)              (* deg(r) ≤ 1 *)
α ← Dt/(t² + 1)
c ← coefficient(r, t)/(2α)
return(q, c)
```

例 5.10.1. 考虑
$$\int (tan(x)^2 + xtan(x) + 1)dx$$
令$k = \mathbb{Q}(x)$, 令$D = d/dx$, 令$t$为$k$上的一个单项式, 满足$Dt = 1+t^2$, 换句话说$t = tan(x)$, 我们的被积函数可以改写为
$$p = t^2 + xt + 1 \in k[t]$$
我们得到

1.$(q, r) = \textbf{PolynomialReduce}(t^2 + xt + 1) = (t, xt)$
2.$\alpha = Dt/(t^2+1) = 1$
3.$c = x/2$

因为$Dc = 1/2 \neq 0$，我们得到
$$\int (tan(x)^2 + xtan(x) + 1)dx = tan(x) + \int xtan(x)dx$$
后面的积分不是一个初等函数。

对于一个超切向量扩张中的既约元素来说，我们在$k$上需要解决的相关问题为

*coupled defferential system problem*（耦合微分系统问题）：已知$f_1, f_2, g_1, g_2 \in k$，判定是否存在$y_1, y_2 \in k$满足

$$\begin{pmatrix} Dy_1 \\ Dy_2 \end{pmatrix} + \begin{pmatrix} f_1 & -f_2 \\ f_2 & f_1 \end{pmatrix} \begin{pmatrix} y_1 \\ y_2 \end{pmatrix} = \begin{pmatrix} g_1 \\ g_2 \end{pmatrix}$$

如果$y_1, y_2$存在的话，将它们计算出来。

定理 5.10.3. 令$k$为一个不包含$\sqrt{-1}$的微分域，$t$是$k$上的一个超切向量。如果我们可以解决$k$上的耦合微分系统，并且$\sqrt{-1}Dt/(t^2+1)$不是一个$k(\sqrt{-1})-$根的对数微分，那么对于$\forall p \in k\langle t \rangle$，我们或者可以证明$p$在$k(t)$上没有初等不定积分，或者计算出$q \in k\langle t \rangle$，满足$p - Dq \in k[t]$。

```
IntegrateHypertangentReduced(p, D)
(* Integration of hypertangent reduced elements *)

(* Given a differential field k such that √-1 ∉ k, a hypertangent mono-
mial t over k and p ∈ k⟨t⟩, return q ∈ k⟨t⟩ and a Boolean β ∈ {0, 1} such
that p - Dq ∈ k[t] if β = 1, or p - Dq does not have an elementary integral
over k(t) if β = 0. *)

m ← -ν_{t²+1}(p)
if m ≤ 0 then return(0, 1)
h ← (t² + 1)^m p                                    (* h ∈ k[t] *)
(q, r) ← PolyDivide(h, t² + 1)        (* h = (t² + 1)q + r, deg(r) ≤ 1 *)
a ← coefficient(r, t), b ← r - at                   (* r = at + b *)
(* CoupledDESystem will be given in Chap. 8 *)
(c, d) ← CoupledDESystem(0, 2mDt/(t² + 1), a, b)
(* Dc - 2mDt/(t² + 1)d = a, Dd + 2mDt/(t² + 1)c = b *)
if (c, d) = "no solution" then return(0, 0)
q₀ ← (ct + d)/(t² + 1)^m
(q, β) ← IntegrateHypertangentReduced(p - Dq₀, D)
return(q + q₀, β)
```

例 5.10.2. 考虑

$$\int \frac{sin(x)}{x} dx$$

令$k = \mathbb{Q}(x)$，$D = d/dx$，令$t$为$k$上的一个单项式，满足$Dt = (1+t^2)/2$，换句话说$t = tan(x/2)$。使用经典的半角公式，被积函数变为

$$p = \frac{sin(x)}{x} = \frac{2tan(x/2)}{x(tan(x/2)^2 + 1)} = \frac{2t/x}{t^2 + 1} \in k\langle t \rangle$$

我们得到$Dt/(t^2+1) = 1/2$，并且
1. $m = -\nu_{t^2+1}(p) = 1$
2. $h = p(t^2 + 1) = 2t/x$
3. $(1, r) = \textbf{PolyDivide}(2t/x, t^2 + 1) = (0, 2t/x)$，所以$(a, b) = (2/x, 0)$
4. 因为

$$\begin{pmatrix} Dc \\ Dd \end{pmatrix} + \begin{pmatrix} 0 & -1 \\ 1 & 0 \end{pmatrix} \begin{pmatrix} c \\ d \end{pmatrix} = \begin{pmatrix} 2/x \\ 0 \end{pmatrix}$$

在$\mathbb{Q}(x)$中无解，**CoupledDESystem**$(0, 1, 2/x, 0)$返回无解。

因此

$$\int \frac{sin(x)}{x} dx$$

不是一个初等函数。

将所有的部分组合在一起，我们得到一个对$k(t)$中元素做积分的算法。

定理 5.10.4. 令$k$是一个不包含$\sqrt{-1}$的微分域，$t$是$k$上的一个超切向量。如果我们能够解决$k$上的耦合微分系统，并且$\sqrt{-1}Dt/(t^2+1)$不是一个$k(\sqrt{-1})-$根的对数微分，那么对于$\forall f \in k(t)$，我们或者可以证明$f$没有在$k(t)$上的初等不定积分，或者计算出$k(t)$的一个初等扩张$E$和$g \in E$，满足$f - Dg \in k$。

```
IntegrateHypertangent(f, D)      (* Integration of hypertangent functions *)
(* Given a differential field k such that √-1 ∉ k, a hypertangent mono-
mial t over k and f ∈ k(t), return g elementary over k(t) and a Boolean
β ∈ {0, 1} such that f − Dg ∈ k if β = 1, or f − Dg does not have an
elementary integral over k(t) if β = 0. *)
(g₁, h, r) ← HermiteReduce(f, D)
(g₂, β) ← ResidueReduce(h, D)
if β = 0 then return(g₁ + g₂, 0)
p ← h − Dg₂ + r
(q₁, β) ← IntegrateHypertangentReduced(p, D)
if β = 0 then return(g₁ + g₂ + q₁, 0)
(q₂, c) ← IntegrateHypertangentPolynomial(p − Dq₁, D)
if Dc = 0 then return(g₁ + g₂ + q₁ + q₂ + c log(t² + 1), 1)
else return(g₁ + g₂ + q₁ + q₂, 0)
```

**5.11 无$Specials$（特殊元素）的非线性情况**

在一个微分域$k$上的非线性单项式情况，我们可以将问题简化为对形式为$p + a/d$，其中$p, a \in k[t], d \in S\backslash\{0\}, \deg(p) < \delta(t)$，且$\deg(a) < \deg(d)$的既约元素的积分问题。此外，定理 5.7.2.提供了一个对于不可积情况的准则，如果$S\backslash k$的一个元素一直，使得我们可以将问题简化为$\deg(p) < \delta(t) - 1$。在这一部分，我们在这一部分涉及$S = k$这种情况，换句话说$S^{irr} = \emptyset$，符合一系列引人注意的函数类，之后会通过例子进行阐明。注意到如果$S^{irr} = \emptyset$，那么$k\langle t\rangle = k[t]$，所以作为多项式约化的结果，我们考虑形式为$p \in k[t]$的被积函数，满足$\deg(p) < \delta(t)$。结果证明是如果这样的元素可积，那么他们一定包含在$k$中。

推论 5.11.1. 假设$t$是非线性单项式，并且$S^{irr} = \emptyset$。令$p \in k[t]$满足$\deg(p) < \delta(t)$，如果$p$在$k(t)$上有初等不定积分，那么$p \in k$。

这提供了对于$k(t)$中元素进行积分的完整算法。

定理 5.11.1. 令$k$为一个微分域，$t$是$k$上的一个非线性单项式，满足$S^{irr} = \emptyset$。那么对于

$\forall f \in k(t)$，我们或者可以证明$f$在$k(t)$上没有初等的不定积分，或者计算出$k(t)$的一个初等扩张和$g \in E$，满足$f - Dg \in k$。

```
IntegrateNonLinearNoSpecial(f, D)
(* Integration of nonlinear monomials with no specials *)

(* Given a is a nonlinear monomial t over k with S^irr = ∅, and f ∈ k(t),
return g elementary over k(t) and a Boolean β ∈ {0, 1} such that f - Dg ∈
k if β = 1, or f - Dg does not have an elementary integral over k(t) if
β = 0. *)

(g_1, h, r) ← HermiteReduce(f, D)
(g_2, β) ← ResidueReduce(h, D)
if β = 0 then return(g_1 + g_2, 0)
(q_1, q_2) ← PolynomialReduce(h - Dg_2 + r, D)
if q_2 ∈ k then β ← 1 else β ← 0
return(g_1 + g_2 + q_1, β)
```

例 5.11.1. 令$\nu \in \mathbb{Z}$为任意整数，考虑

$$\int \frac{J_{\nu+1}(x)}{J_\nu(x)} dx$$

其中$J_\nu(x)$为阶数为$\nu$的第一类$Bessel$（贝塞尔）函数，从

$$\frac{dJ_\nu(x)}{dx} = -J_{\nu+1}(x) + \frac{\nu}{x} J_\nu(x)$$

我们得到

$$\int \frac{J_{\nu+1}(x)}{J_\nu(x)} dx = \int \nu \frac{dx}{x} - \int \frac{dJ_n u(x)/dx}{J_\nu(x)} dx = \nu \log(x) - \int \phi_\nu(x) dx$$

其中$\phi_\nu(x)$是$J_\nu(x)$的对数微分。因为$J_\nu(x)$是贝塞尔方程的一个解

$$y'(x) + y(x)^2 + \frac{1}{x} y(x) + \left(1 - \frac{v^2}{x^2}\right) = 0$$

令$k = \mathbb{Q}(x)$，$D = d/dx$，令$t$是$k$上的一个单项式，满足$Dt = -t^2 - t/x - (1 - \nu^2/x^2)$，换句话说$t = \phi_\nu(x)$。可以证明在这个扩张中$\mathcal{S}^{irr} = \emptyset$，所以根据推论 5.11.1. 说明$t$在$k$上没有初等不定积分，因此

$$\int \frac{J(\nu+1)(x)}{J_\nu(x)} dx = \nu_{\log}(x) - \int \phi_\nu(x) dx$$

其中剩余积分不是$\mathbb{Q}(x, \phi_\nu(x))$上的初等积分。

以上的例子使用了贝塞尔函数，但实际中本章的算法无论被积函数是否可以表示一个函数的对数微分项的形式都可以应用，其中该函数由一个二阶线性常微分方程定义。如果典型方程在求积过程中无解（例如$Airy$（亚里）函数），如本章中注解 4 所说明的一样，$\mathcal{S}^{irr} = \emptyset$。

## 5.12 内场积分

在这一部分中，我们介绍用以判定$k(t)$中元素性质的积分算法所延伸出的一些子变形算法，判定选项为：

$k(t)$中元素是否为$k(t)$中元素的微分

$k(t)$中元素时候为$k(t)$中元素的对数微分

$k(t)$中元素是否为$k(t)$—根的对数微分

正如我们从 5.2 节了解到的一样，当构造包含被积函数的塔状域扩张时，这样的过程是需要的。此外，剩下章节的积分算法中，在不同的地方这样的过程也是需要的，尤其在限定结束和次数时。

注意到第九章的结构定理提供了改良积分算法的有效的替代方案，在一些情况下，这个方案是唯一确定对数微分的完整算法。

确定微分

第一个问题是，已知$f \in k(t)$，判定是否存在$u \in k(t)$满足$Du = f$，如果存在的话计算这样的$u$。首先我们对$f$使用厄米特约化，得到$g \in k(t)$，一个$simple$（简单）元素$h \in k(t)$和$r \in k\langle t \rangle$，满足$f = Dg + h + r$。在这一点上，我们可以证明如果存在$u \in k(t)$，使得$f = Du$，那么$h \in k[t]$，所以我们要对既约的$h + r$进行积分运算。5.7 至 5.11 的算法可以应用（在非线性情况下有一个附加的修正，防止引入新的对数），或者证明不存在这样的$u$，或者将问题简化为判定$k$中元素$a$是否在$k(t)$中有不定积分存在。

如果$t$是$k$上的基本式，那么它符合定理 4.4.2.和引理 5.1.2.的条件，如果$a$在$k(t)$中有一个不定积分，那么$a = Dv + cDt$，其中$v \in k, c \in \text{Const}(k)$，我们将问题简化为一个$k$中的受限积分问题。否则，$\delta(t) \geq 1$，它符合定理 4.4.2.和引理 3.4.2.，引理 5.1.2.的条件，如果$a$在$k(t)$中有一个不定积分，那么$a = Dv$，其中$v \in k$，我们将问题$k$上的一个相似的问题。

当存在$a \in k(t)^*$，使得$f = Da/a$，那么推论 9.3.1.，推论 9.3.2.或推论 9.4.1.提供了替代算法: 存在$u \in k(t)$, $f = Du$，当且仅当在$\mathbb{Q}$中线性方程(9.8), (9.12)或(9.21)有解。推论 9.3.2.也提供了一个替代性算法：如果存在$b \in k(t)$, $f = Db/(b^2 + 1)$，换句话说$f = arctan(b)$。

显然解$u$不是唯一的，但是如果对于$u, v \in k(t)$, $f = Du = Dv$，那么$u - v \in \text{Const}(k(t))$。

确认对数微分

第二个问题是，已知$f \in k(t)$，判定是否存在一个$k(t)$中的非零元$u$，满足$Du/u = f$，如果$u$存在，将$u$计算出来。我们可以证明如果存在$k(t)$中非零元$u$, $f = Du/u$，那么$f$是$simple$（简单）的，$Rothstein - Trager$结式的所有根都是整数。在这种情况下，留数化简说明

$$g = \sum_{r_s(\alpha)=0} \alpha \frac{Dg_\alpha}{g_\alpha} = \frac{D(\prod_{r_s(\alpha)=0} g_\alpha^\alpha)}{\prod_{r_s(\alpha)=0} g_\alpha^\alpha} = \frac{Dv}{v}$$

因为$\alpha$都是整数，$v \in k(t)$。此外定理 5.6.1. 说明如果存在$u \in k(t)$, $f = Du/u$，那么$f - g \in k[t]$，所以我们要判定$k[t]$中元素$p$是否$k(t)$中某个元素的对数微分。如果存在$u \in k(t)$, $p = Du/u$，那么从练习 4.2 结论可知$\deg(p) < \max(1, \delta(t))$，从推论 4.4.2.结论可知$u = p_1^{e_1} \cdots p_n^{e_n}$，其中$p_i \in \mathcal{S}, e_i \in \mathbb{Z}$。

如果$t$是$k$上的基本式，那么因为$\mathcal{S} = k$，所以$p$和$u$都必须在$k$中，所以我们将问题简化为$k$上相类似的问题。

如果$t$是$k$上的一个超越指数式，那么$p \in k, u = vt^e$，其中$v \in k^*$，并且$e \in \mathbb{Z}$，因为$\mathcal{S}^{irr} = \{t\}$，因此我们将问题简化为判定$p \in k$是否可以写成下列形式

$$p = \frac{Dv}{v} + e\frac{Dt}{t}$$

$v \in k^*, e \in \mathbb{Z}$。这是含参数对数微分问题的一个特殊情况，是受限积分问题的一个变形，将会在第 7 章中进行讨论。

如果$t$是$k$的一个超切向量，$\sqrt{-1} \notin k$，那么$p = a + bt, a, b \in k$，且对于$v \in k^*$，$u = v(t^2 + 1)$, $e \in \mathbb{Z}$，因为$\mathcal{S}^{irr} = (t^2 + 1)$。因此，我们将问题简化为判定$a + bt$是否能表示为下列形式：

$$a + bt = \frac{Dv}{v} + e\frac{D(t^2+1)}{t^2+1} = \frac{Dv}{v} + 2e\frac{Dt}{t^2+1}t$$

这等价于

$$a = \frac{Dv}{v} \quad \frac{b}{2}\frac{t^2+1}{Dt} \in \mathbb{Z}$$

第二个条件可以立刻得到验证，而第一个条件相当于判定$k$中的一个元素是否是$k$中一个元素的对数微分。

当存在$b \in k(t)$，使得$f = Db$，那么推论 9.3.1.，推论 9.3.2.或推论 9.4.2. 提供了替代算法：$f$是一个$k(t)-$根的对数微分当且仅当线性方程(9.9), (9.13)或(9.22)在$\mathbb{Q}$中有解。

解$u$不是唯一的，但是如果$f = Du/u = Dv/v$，对于$u, v \in k(t)\backslash\{0\}$，那么$u/v \in \text{Const}(k(t))$（这是接下来的引理 5.12.1.中$n = m = 1$这种情况）。

确定一个$k(t)-$根的对数微分

第三个问题是，已知$f \in k(t)$，判定是否存在一个非零$n \in \mathbb{Z}$和一个非零$u \in k(t)$，使得$Du/u = nf$，如果这样的$n$和$u$存在，那么将它们计算出来。我们可以证明如果存在非零$n \in \mathbb{Z}$和$u \in k(t)$，使得$nf = Du/u$，那么$f$是$simple$（简单）的，并且$Rothstein-Trager$结式的所有跟为有理数。在这种情况下，令$m$为$Rothstein-Trager$结式的根的公分母。那么，留数化简使我们得到

$$g = \sum_{r_s(\alpha)=0} \alpha \frac{Dg_\alpha}{g_\alpha} = \frac{1}{m} \frac{D\left(\prod_{r_s(\alpha)=0} g_\alpha^{m_\alpha}\right)}{\prod_{r_s(\alpha)=0} g_\alpha^{m_\alpha}} = \frac{1}{m}\frac{Dv}{v}$$

因为$m_\alpha$对于每一个$\alpha$来说，都是一个整数，那么$v \in k(t)$。此外定理 5.6.1.说明如果对于$n \in \mathbb{Z}, u \in k(t)$, $f = Du/(nu)$，那么$f - Dg \in k[t]$。所以我们要解决的剩下的问题为判定$k[t]$中的一个元素$p$是否是一个$k(t)-$根的对数微分。如果对于$n \in \mathbb{Z}, u \in k(t)$, $p = Du/(nu)$，那么从练习 4.2.的结论中我们可以知道$\deg(p) < \max(1, \delta(t))$，从推论 4.4.2.的结论中我们可以得出$u = p_1^{e_1} \cdots p_s^{e_s}$，其中$p_i \in \mathcal{S}, e_i \in \mathbb{Z}$。

如果$t$是$k$上的一个初等式，那么因为$\mathcal{S} = k$，那么$p$和$u$都必须在$k$中，所以我们将问题简化为$k$上的一个相似的问题。

如果$t$是$k$的一个超越指数式，那么因为$\mathcal{S}^{irr} = t$，那么$p \in k, u = vt^e$，其中$v \in k^*, e \in \mathbb{Z}$。因此我们将问题简化为判定$p \in k$是否可以记为：

$$p = \frac{1}{n}\frac{Dv}{v} + \frac{e}{n}\frac{Dt}{t}$$

其中$v \in k^*, n, e \in \mathbb{Z}$。这是含参数的对数导数问题，受限积分问题的一个变形，将会在第七章中进行讨论。

如果$t$是$k$上的一个超切向量，并且$\sqrt{-1} \notin k$，那么对于$a, b \in k$, $p = a + bt$，且因为$\mathcal{S}^{irr} = \{t^2 + 1\}$，所以$u = v(t^2 + 1), v \in k^*, e \in \mathbb{Z}$。因此我们将问题简化为判定$a + bt$是否可以写作：

$$a + bt = \frac{1}{n}\frac{Dv}{v} + \frac{e}{n}\frac{D(t^2+1)}{t^2+1} = \frac{1}{n}\frac{Dv}{v} + \frac{2e}{n}\frac{Dt}{t^2+1}t$$

这等价于
$$na = \frac{Dv}{v} \quad \frac{b}{2}\frac{t^2+1}{Dt} \in \mathbb{Q}$$

第二个条件可以立刻得到验证，第一个条件相当于判定$k$中一个元素是否是一个$k-$根的对数微分。

当存在$b \in k(t)$，使得$f = Db$，那么推论 9.3.1.，推论 9.3.2. 或推论 9.4.1.提供了替代性算法：$f$是一个$k(t)-$根的对数微分当且仅当线性方程(9.9), (9.13)或(9.22)在$\mathbb{Q}$中有解。

解$(n, u)$不是唯一的，但是任意两个解可以由以下的引理联系起来。

引理 5.12.1. 令$(K, D)$为一个微分域，$u, v \in K^*$，如果
$$\frac{1}{n}\frac{Du}{u} = \frac{1}{m}\frac{Dv}{v}$$
其中$n, m$为非零整数，那么
$$\frac{u^{\text{lcm}(n,m)/n}}{v^{\text{lcm}(n,m)/m}} \in \text{Const}(K)$$

# *Chapter* 2.1.6 *Risch*微分方程

在这一部分我们讨论Risch微分方程解的问题，换句话说，已知一个特征为0的微分域$K$，$f,g \in K$，判定方程
$$Dy + fy = g \tag{6.1}$$
在$K$中是否有解，如果存在的话，找到一个解。我们只在超越情况下研究方程$Dy + fy = g$，换句话说，当$K$是一个微分子域$k$的$simple$（简单）单项式扩张的情况下。所以在本章剩余部分中，令$k$为一个特征为0的微分域，$t$是$k$上的一个单项式。我们假设在方程中$f, g$的系数是$k(t)$中的元素，并且寻找一个在$k(t)$中的解$y$。我们要介绍的算法如下：

1.计算任意一个解的$normal$（普通）部分。这将问题简化为在$k\langle t\rangle$中寻找一个解。
2.计算任意一个解的$special$（特殊）部分。这将问题简化为在$k[t]$中寻找一个解。
3.限定$k[t]$中任意解的次数。
4.将方程(6.1)转化为相似的形式，但是$f, g \in k[t]$
5.在$k[t]$中找到次数受限的约化方程的解。

**6.1 分母的$normal$（普通）部分**
  在这一部分我们将指出如果方程经过充分的处理，那么在一个单项式扩张中的一个$Risch$微分方程的任意解的分母的$normal$（普通）部分，是由一个用方程系数组成的显式方程决定的。首先我们介绍需要的预处理。

定义 6.1.1. 如果$\text{residue}_p(f)$对于任意$normal$（普通）不可约的$p \in k[t]$，满足$f \in \mathcal{R}_p$的情况不是一个正整数，那么我们称$f(t)$关于$t$弱正规化。

  这个定义的动机是接下来的这个引理，这个引理给出了$Dy + fy$在一个$normal$（普通）多项式处的阶数方程，只要$f$是弱正规化的。

引理 6.1.1. 令$f \in k(t)\setminus\{0\}$关于$t$是弱正规化的，$y \in k(t)\setminus\{0\}$，$p \in k[t]$是$normal$（普通）可约的，那么
$$\nu_p(y) < 0 \Longrightarrow \nu_p(Dy + fy) + \min(\nu_p(f), -1)$$

  当然，下一步是，已知$f \in k(t)$，检验$f$是否关于$t$是弱正规化的，否则找一个一个充分的变换使得$f$弱正规化。接下来的定义说明将一个恰当的对数微分加到任意一个$f \in k(t)$上可以使$f$弱正规化，并且给出了一个明确的变量代换，将方程(6.1)转化为一个具有若正规化系数的相似方程。

定理 6.1.1. 对于任意$f \in k(t)$，我们可以计算出$q \in k[t]$，使得$\overline{f} = f - Dq/q$关于$t$弱正规化。而且，对于$\forall g, y \in k(t)$，
$$Dy + fy = g \Longleftrightarrow Dz + \overline{f}z = qg$$
其中$z = gy$。

  我们注意到在实际中，$d_n$的一个完全无平方分解是不必须的，因为对于计算$q$来说只有$d_1$是需要的。上面的证明给出了一个将$k(t)$中任意元素进行弱正规化的算法。

```
WeakNormalizer(f, D)      (* Weak normalization *)
(* Given a derivation D on k[t] and f ∈ k(t), return q ∈ k[t] such that
f − Dq/q is weakly normalized with respect to t. *)
(d_n, d_s) ← SplitFactor(denominator(f), D)
g ← gcd(d_n, dd_n/dt)
d* ← d_n/g
d_1 ← d*/ gcd(d*, g)
(a, b) ← ExtendedEuclidean(denominator(f)/d_1, d_1, numerator(f))
r ← resultant_t(a − zDd_1, d_1)
(n_1, ..., n_s) ← positive integer roots of r
return(∏_{i=1}^{s} gcd(a − n_i Dd_1, d_1)^{n_i})
```

现在我们可以假设在方程(6.1)中，$f$是关于$t$弱正规化的。那么接下来的定理给出了一个解的分母的$normal$（普通）部分的明确方程。

定理 6.1.2. 令$f \in k(t)$是关于$t$弱正规化的，$g \in k(t)$，令$y \in k(t)$满足于$Dy + fy = g$。令$d = d_s d_n$为$f$分母的一个分裂分解，$e = e_s e_n$为$g$分母的一个分裂分解。令$c = \gcd(d_n, e_n)$，并且
$$h = \frac{\gcd(e_n, de_n/dt)}{\gcd(c, dc/dt)} \in k[t]$$
那么
$(i) yh \in k\langle t \rangle$

$(ii)$对于$\forall q \in k[t]$，满足$q | h$，$\frac{yh}{q} \notin k\langle t \rangle$

推论 6.1.1. 令$f \in k(t)$关于$t$为弱正规化的，$g \in k(t)$，$d_n, e_n$和$h$由定理 6.1.2.定义，那么
$(i)$对于$Dy + fy = g$的任意解$y \in k(t)$，$q = yh \in \langle t \rangle$，$q$是如下方程的一个解：
$$d_n h Dq + (d_n hf − d_n Dh) = d_n h^2 g \tag{6.2}$$
反之，对于方程(6.2)的任意解$q \in k\langle t \rangle$，$y = q/h$是$Dy + fy = g$的一个解。

$(ii)$如果$Dy + fy = g$在$k(t)$中有一个解，那么$e_n | d_n h^2$

上面的定理与推论为我们提供一个算法，或者证明一个给定的$Risch$微分方程在一个给定的单项式扩张中无解，或者将问题简化为在$k\langle t \rangle$上方程求解的问题。

```
RdeNormalDenominator(f, g, D)
(* Normal part of the denominator *)

(* Given a derivation D on k[t] and f, g ∈ k(t) with f weakly normalized
with respect to t, return either "no solution", in which case the equation
Dy + fy = g has no solution in k(t), or the quadruplet (a, b, c, h) such
that a, h ∈ k[t], b, c ∈ k⟨t⟩, and for any solution y ∈ k(t) of Dy + fy = g,
q = yh ∈ k⟨t⟩ satisfies aDq + bq = c. *)

(d_n, d_s) ← SplitFactor(denominator(f), D)
(e_n, e_s) ← SplitFactor(denominator(g), D)
p ← gcd(d_n, e_n)
h ← gcd(e_n, de_n/dt) / gcd(p, dp/dt)
if e_n ∤ d_n h^2 then return "no solution"
return(d_n h, d_n h f − d_n Dh, d_n h^2 g, h)
```

例 6.1.1. 令$k = \mathbb{Q}$，$t$为$k$上的一个单项式满足$Dt = 1$，换言之$D = d/dt$，考虑方程

$$Dy + y = \frac{1}{t}$$

这个方程由$e^t/t$的积分生成，我们有$f = 1, g = 1/t$，所以：

1.$(d_n, d_s) =$ **SplitFactor**$(1, d/dt) = (1, 1)$

2.$(e_n, e_s) =$ **SplitFactor**$(t, d/dt) = (t, 1)$

3.$p = \gcd(1, t) = 1$

4.$h = \gcd(t, 1)/\gcd(1, 1) = 1$

因为$t \nmid 1$，所以我们得出结论方程在$\mathbb{Q}(t)$中无解，因此$\int e^t/t \, dt$不是一个初等函数。

## 6.2 分母的 special（特殊）部分

由推论 6.1.1.，我们可以将问题简化为在$k\langle t \rangle$中找到方程(6.2)的一个解，我们将方程重新记为

$$aDq + bq = c \quad (6.6)$$

其中$a \in k[t]$没有 special（特殊）因子，$b, c \in k\langle t \rangle, a \neq 0$，$t$是$k$上的一个单项式。在这一小节中我们给出对于特定类型单项式，在$k\langle t \rangle$中方程(6.6)的解的分母的计算算法。首先我们先给出一个对任意单项式扩张都正确的结果。

引理 6.2.1. 令$t$为$k$上的一个单项式，$p \in \mathcal{S}^{irr}$，$a, b, y \in k(t)$，其中$a \neq 0, \nu_p(y) \neq 0$，那么

$(i)$如果$\nu_p(b) < \nu_p(a)$，那么$\nu_p(aDy + by) = \nu_p(b) + \nu_p(y)$

$(ii)$如果$p \in \mathcal{S}^{irr}$，且$\nu_p(b) > \nu_p(a)$，那么$\nu_p(aDy + by) = \nu_p(a) + \nu_p(y)$

$(iii)$如果$\nu_p(b) = \nu_p(a)$，那么或者$\nu_p(aDy + by) = \nu_p(a) + \nu_p(y)$，或者

$$\pi_p\left(-\frac{b}{a}\right) = \nu_p(y)\pi_p\left(\frac{Dp}{p}\right) + \frac{D^*u}{u}$$

其中非零元$u \in k[t]/(p)$，且$D^*$是诱导微分。

因为$a \in k[t]$，并且在方程(6.6)中没有 special（特殊）因子，这意味着对于$\forall p \in \mathcal{S}$，$\nu_p(a) = 0$，所以引理 6.2.1.对于$\nu_p(q)$给出了一个更低的限制，其中$q \in k\langle t \rangle$在以下情况中是

方程(6.6)的一个解：
$(i)$如果$\nu_p(b) < 0$，那么$\nu_p(q) \in \{0, \nu_p(c) - \nu_p(b)\}$
$(ii)$如果$\nu_p(b) > 0$，且$p \in \mathcal{S}^{irr}$，那么$\nu_p(q) \in \{0, \nu_p(c)\}$

对于$p \in \mathcal{S}^{irr}$，一旦我们得到了一个更低限制，存在$n \leq 0, \nu_p(q) \geq n$，在方程(6.6)中用$hp^n$替换$q$，得到
$$a(p^n Dh + np^{n-1}hDp) + bhp^n = c$$
因此
$$aDh + \left(b + na\frac{Dp}{p}\right)h = cp^{-n} \tag{6.7}$$

此外，因为$q \in k\langle t \rangle$，所以$h \in k\langle t \rangle$，因为$\nu_p(q) \geq n$，所以$h \in \mathcal{O}_p$。因此我们将问题简化为找到方程(6.7)的一个根$h$, $h \in k\langle t \rangle \cap \mathcal{O}_p$。注意到因为$c \in k\langle t \rangle$，所以$cp^{-n} \in k\langle t \rangle$，因为$b \in k\langle t \rangle, a \in k[t], p \in \mathcal{S}$，所以$b + naDp/p \in k\langle t \rangle$。$p$在$b + naDp/p$分母和$cp^{-n}$中的幂指数可以通过将方程(6.7)左右两边乘上$p^N$进行消除，其中$N = \max(0, -\nu_p(b), -\nu_p(c))$，保证了方程(6.7)的系数也在$k\langle t \rangle \cap \mathcal{O}_p$中。

因为在这一部分我们所考虑的单项式扩张中，所有的 *special*（特殊）多项式都是第一类的，我们只需要在可能的消去情况，换言之$\nu_p(b) = 0$的情况下，找到对于$\nu_p(q)$一个更低的限制。我们对于不同的单项式扩张的情况分别进行讨论。

基本式情况

如果$Dt \in k$，那么每一个无平方因子多项式都是 *normal*（普通）的，所以$k\langle t \rangle = k[t]$，这意味着$a, b, c \in k[t]$，方程(6.6)在$k\langle t \rangle$中解一定在$k[t]$中。

超越指数式情况

如果$Dt/t \in k$，那么$k\langle t \rangle = k[t, t^{-1}]$，所以我们需要计算$\nu_t(q)$的一个较低的限制，其中$q \in k\langle t \rangle$是方程(6.6)的一个解。为了计算这样的一个限制边界，我们需要能够判定$k$中的任意一个元素$f$是否能写成下列形式
$$f = m\eta + \frac{Du}{u} \tag{6.8}$$
其中$m$为非负整数，$u \in k^*$，并且$\eta = Dt/t \in k$。正如 5.12 节说明的一样，这是一个含参数的对数微分问题，一个受限积分问题的变形，将会在第七章中进行讨论。因为方程(6.8)的一个解中非负整数$m$可以在一个较低的限制边界计算中出现，首先我们需要确定在方程(6.8)的所有解中$m$都是一致的。

引理 6.2.2. $K$是一个特征为$0$的微分域，假设$\eta \in k^*$不是一个$K-$根的对数微分，那么对于$f \in K$和方程(6.8)在$\mathbb{Z} \times K^*$中的任意解$(m, u)$和$(n, v)$，我们有$n = m$和$v/u \in \text{Const}(K)$。

引理 6.2.3. 假设$t$是$k$上的一个超越指数式，满足$\eta = Dt/t$不是一个$k-$根的对数微分。令$a \in k[t]$，$b, q \in \langle t \rangle$，满足$\gcd(a, t) = 1, \nu_t(b) = 0$, 且$\nu_t(q) \neq 0$，那么，或者
$$\nu_t(aDq + bq) = \nu_t(q)$$
或者
$$-\frac{b(0)}{a(0)} = \nu_t(q)\eta + \frac{Du}{u}, u \in k^*$$

因为根据定理 5.1.2.，我们有 $t \in \mathcal{S}^{irr}$，引理 6.2.1.和引理 6.2.3.总是能够使我们求出 $\nu_t(q)$ 的一个较低受限边界，其中 $q \in k\langle t \rangle$ 是方程(6.6)的一个解：如果 $\nu_t(b) \neq 0$、那么引理 6.2.1. 像之前说明的一样给出了一个边界。否则，当 $\nu_t(b) = 0$，对于 $m \in \mathbb{Z}$ 和 $u \in k^*$，或者在 $\nu_t(q) \in \{0, m, \nu_t(c)\}$ 的情况下，有 $-b(0)/a(0) = m\eta + Du/u$，或者 $\nu_t(q) \in \{0, \nu_t(c)\}$。注意到根据引理 6.2.2.，这样的 $m$ 在 $k$ 上是唯一的。因为 $\mathcal{S}^{irr} = \{t\}, k\langle t \rangle \cap \mathcal{O}_t = k[t]$，所以在决定对于 $\nu_t(q)$ 的更小限定边界的情况下，我们需要找到方程(6.7)的解 $h \in k[t]$。

```
RdeSpecialDenomExp(a, b, c, D)
(* Special part of the denominator – hyperexponential case *)

(* Given a derivation D on k[t] and a ∈ k[t], b, c ∈ k⟨t⟩ with Dt/t ∈
   k, a ≠ 0 and gcd(a, t) = 1, return the quadruplet (ā, b̄, c̄, h) such that
   ā, b̄, c̄, h ∈ k[t] and for any solution q ∈ k⟨t⟩ of aDq + bq = c, r = qh ∈ k[t]
   satisfies āDr + b̄r = c̄. *)

p ← t                              (* the monic irreducible special polynomial *)
n_b ← ν_p(b), n_c ← ν_p(c)
n ← min(0, n_c − min(0, n_b))                                 (* n ≤ 0 *)
if n_b = 0 then                    (* possible cancellation *)
    α ← Remainder(−b/a, p)         (* α = −b(0)/a(0) ∈ k *)
    if α = mDt/t + Dz/z for z ∈ k* and m ∈ ℤ then n ← min(n, m)
N ← max(0, −n_b, n − n_c)          (* N ≥ 0, for clearing denominators *)
return(ap^N, (b + naDp/p)p^N, cp^{N−n}, p^{−n})
```

例 6.2.1. 令 $k = \mathbb{Q}(x)$，$D = d/dx$，令 $t$ 是 $k$ 上的一个单项式，满足 $Dt = t$，换言之 $t = e^x$，考虑方程

$$(t^2 + 2xt + x^2)Dq + \left(\left(1 + \frac{1}{x^2}\right)t^2 + \left(2x - 1 + \frac{2}{x}\right)t + x^2\right)q = \frac{t}{x^2} - 1 + \frac{2}{x} \quad (6.9)$$

方程(6.9)由如下函数的积分生成

$$\frac{e^x - x^2 + 2x}{(e^x + x)^2 x^2} e^f$$

其中

$$f = \frac{x^2 - 1}{x} + \frac{1}{e^x + x}$$

我们有 $a = t^2 + 2xt + x^2, b = (1 + 1/x^2)t^2 + (2x - 1 + 2/x)t + x^2$，并且 $c = t/x^2 - 1 + 2/x$，因此

1. $n_b = \nu_t(b) = 0, n_c = \nu_t(c) = 0$
2. $n = \min(0, n_c - \min(0, n_b)) = 0$
3. $n_b = 0$，所以 $\alpha = -b(0)/a(0) = -x^2/x^2 = -1$
4. $-1 = -Dt/t$，所以 $m = -1, n = \min(n, m) = -1$
5. $N = \max(0, -n_b, n - n_c) = 0$

因此，方程(6.9)的任意一个解 $q \in k\langle t \rangle$ 一定具有 $q = p/t$ 的形式，其中 $p \in k[t]$ 满足

$$(t^2 + 2xt + x^2)Dp + \left(\frac{t^2}{x^2} + \left(\frac{2}{x} - 1\right)t\right)p = \frac{t^2}{x^2} + \left(\frac{2}{x} - 1\right)t \quad (6.10)$$

超切向量情况

如果$Dt/(t^2+1) \in k$，且$\sqrt{-1} \notin k$，那么唯一的首一$special$（特殊）不可约元是$t^2+1$，所以我们需要计算一个$\nu_{t^2+1}(q)$上更小的限制，其中$q \in k\langle t\rangle$是方程(6.6)的一个解。

引理 6.2.4. 假设$\sqrt{-1} \notin k$，$t$是$k$上的一个超切向量，满足$\eta\sqrt{-1}$不是一个$k(\sqrt{-1})-$根的对数微分，其中$\eta = Dt/(t^2+1) \in k$。令$a \in k[t], b, q \in k\langle t\rangle$满足$\gcd(a, t^2+1) = 1, \nu_{t^2+1}(b) = 0$，且$\nu_{t^2+1}(q) \neq 0$，那么，或者
$$\nu_{t^2+1}(aDq + bq) = \nu_{t^2+1}(q)$$
或者将$D$扩张到$k(\sqrt{-1})$，通过$D\sqrt{-1} = 0$，并且将$-b(\sqrt{-1})/a(\sqrt{-1})$写作$\alpha\sqrt{-1} + \beta$，其中$\alpha, \beta \in k$，我们有
$$-\frac{b(\sqrt{-1})}{a(\sqrt{-1})} = 2\nu_{t^2+1}(q)\eta\sqrt{-1} + \frac{Du}{u}, 2\beta = \frac{Dv}{v} \tag{6.11}$$
其中$u \in k(\sqrt{-1})^*, v \in k^*$

因为根据定理 5.10.1.，$t^2+1 \in \mathcal{S}_1^{irr}$，引理 6.2.1，引理 6.2.4.总是能使我们得到$\nu_{t^2+1}(q)$的一个较小限制，其中$q \in k\langle t\rangle$是方程(6.6)的一个解：如果$\nu_{t^2+1}(b) \neq 0$，那么引理 6.2.1.像之前说明的一样给出限制。否则，在$\nu_{t^2+1}(b) = 0$的情况下，所以或者在$\nu_{t^2+1}(q) \in \{0, m, \nu_{t^2+1}(c)\}$的情况下，存在$m \in \mathbb{Z}$和$u \in k(\sqrt{-1})^*$，使得$-b(\sqrt{-1})/a(\sqrt{-1}) = m\eta\sqrt{-1} + Du/u$，或者$\nu_{t^2+1}(q) \in \{0, \nu_{t^2+1}(c)\}$。注意到根据引理 6.2.2.这样的$m$关于$k(\sqrt{-1})$唯一。我们也注意到(6.11)中的验证意味着在$k(\sqrt{-1})$上解决一个含参数的对数微分问题。因为只有非负整数$\nu_{t^2+1}(q)$在结果中得到了使用，所以$\sqrt{-1}$的伴随是一时的，所以一旦边界确定下来，算法在$k$上是适用的。因为必要条件$2\beta = Dv/v$是在$k$上定义的，所以可以首先验证这个条件，只要这个条件满足，$\sqrt{-1}$就必须被引入。因为$\mathcal{S}^{irr} = \{t^2+1\}, k\langle t\rangle \cap \mathcal{O}_{t^2+1} = k[t]$，所以在已经找到$\nu_{t^2+1}(q)$的一个更小限制条件下，问题简化为在$k[t]$中找到(6.7)的一个解。

存在引理 6.2.4.的相似情况，相应的针对含$\sqrt{-1}$的域的算法为

---

**RdeSpecialDenomTan**$(a, b, c, D)$
(* Special part of the denominator – hypertangent case *)

(* Given a derivation $D$ on $k[t]$ and $a \in k[t], b, c \in k\langle t\rangle$ with $Dt/(t^2+1) \in k$, $\sqrt{-1} \notin k$, $a \neq 0$ and $\gcd(a, t^2+1) = 1$, return the quadruplet $(\bar{a}, \bar{b}, \bar{c}, h)$ such that $\bar{a}, \bar{b}, \bar{c}, h \in k[t]$ and for any solution $q \in k\langle t\rangle$ of $aDq + bq = c$, $r = qh \in k[t]$ satisfies $\bar{a}Dr + \bar{b}r = \bar{c}$. *)

$p \leftarrow t^2 + 1$     (* the monic irreducible special polynomial *)
$n_b \leftarrow \nu_p(b), n_c \leftarrow \nu_p(c)$
$n \leftarrow \min(0, n_c - \min(0, n_b))$     (* $n \leq 0$ *)
**if** $n_b = 0$ **then**     (* possible cancellation *)
    $\alpha\sqrt{-1} + \beta \leftarrow$ **Remainder**$(-b/a, p)$     (* $\alpha, \beta \in k$ *)
    $\eta \leftarrow Dt/(t^2+1)$     (* $\eta \in k$ *)
    **if** $2\beta = Dv/v$ for $v \in k^*$
        and $\alpha\sqrt{-1} + \beta = 2m\eta\sqrt{-1} + Dz/z$ for $z \in k(\sqrt{-1})^*$ and $m \in \mathbb{Z}$
        **then** $n \leftarrow \min(n, m)$
$N \leftarrow \max(0, -n_b, n - n_c)$     (* $N \geq 0$, for clearing denominators *)
**return**$(ap^N, (b + naDp/p)p^N, cp^{N-n}, p^{-n})$

例 6.3.2. 继续对例 6.1.2.进行处理，令$k = \mathbb{Q}(x)$，$D = d/dx$，$t$是$k$上的一个单项式，满足$Dt = 1 + t^2$，换句话说$t = tan(x)$，考虑(6.5)在$k\langle t\rangle$中的解$q$，由$e^{tan(x)}/tan(x)^2$的积分生成，我们有$a = t, b = (t-1)(t^2+1)$和$c = 1$，因此

1.$n_b = \nu_{t^2+1}(b) = 1, n_c = \nu_{t^2+1}(b) = 0$

2.$n = \min(0, n_c - \min(0, n_b)) = 0$

3.$n_b = 0$，所以$N = \max(0, -n_b, n - n_c) = 0$

因此(6.5)在$k\langle t\rangle$中的任意解$q$一定属于$k\langle t\rangle \cap \mathcal{O}_{t^2+1} = k[t]$。

## 6.3 次数限制

由之前小节结论可知，我们已经将问题简化为在$k[t]$中找到(6.7)的解$q$，方程(6.7)改写为
$$aDq + bq = c \tag{6.12}$$

其中$a, b, c \in k[t], a \neq 0$，$t$是$k$上的一个单项式。在这一小节中我们给出对于特定单项式类型，计算方程(6.12)在$k[t]$中的任意解中$t$的次数上界的算法，首先我们给出对于任意单项式扩张都正确的一个结果。

引理 6.3.1. 令$t$为$k$上的一个单项式，$a, b, q \in k[t]$，其中$a \neq 0$，且$\deg(q) > 0$，那么

(i)如果$\deg(b) > \deg(a) + \max(0, \delta(t) - 1)$，那么
$$\deg(aDq + bq) = \deg(b) + \deg(q)$$

(ii)如果$t$是非线性的，并且$\deg(b) < \deg(a) + \delta(t) - 1$，那么
$$\deg(aDq + bq) = \deg(a) + \deg(q) + \delta(t) - 1$$

(iii)如果$\delta(a) \geq 1$，并且$\deg(b) = \deg(a) + \delta(t) - 1$，那么或者
$$\deg(aDq + bq) = \deg(b) + \deg(q)$$
或者
$$-\frac{\text{lc}(b)}{\text{lc}(a)} = \pi_\infty\left(\frac{Dq}{qt^{\delta(t)-1}}\right)$$

引理 6.3.1.在如下情况，给出了方程(6.12)在$k[t]$中的一个解$q$的次数$\deg(q)$的一个上界：

(i)如果$\deg(b) > \deg(a) + \max(0, \delta(t) - 1)$，那么$\deg(q) \in \{0, \deg(c) - \deg(b)\}$

(ii)如果$\deg(b) < \deg(a0 + \delta(t) - 1$，并且$\delta(t) \geq 2$，那么
$$\deg(q) \in \{0, \deg(c) - \deg(a) + 1 - \delta(t)\}$$

结果说明，对于刘维尔单项式，我们只需要考虑$\deg(b) \leq \deg(a)$这种情况，对于非线性单项式，$\deg(b) = \deg(a) + \delta(t) - 1$。对于不同的单项式扩张的情况，我们分别进行讨论。

基本式情况

如果$Dt \in k$，为了计算出$\deg(q)$的一个上界，我们需要判定$k$中任意元素$f$是否可以写成下面形式：
$$f = m\eta + Du \tag{6.13}$$

其中$m$为非负整数，$u \in k$，$\eta = Dt \in k$。注意到(6.13)是$k$中的一个受限积分问题，所以将第七章的算法作用于$f$和$\eta$上便可以解决问题。因为(6.13)的解中的非负整数$m$可以在上界计算中出现，我们首先需要确定在(6.13)的所有解中，$m$是一致的。

引理 6.3.2. 假设$t$是$k$上的一个基本式，满足$\eta = Dt$不是$k$中一个元素的微分。那么，对于$f \in k(t)$，(6.13)在$\mathbb{Z} \times k$中的任意解$(m, u)$和$(n, v)$，我们可以得到$n = m$，且

$v - u \in \text{Const}(k)$。

引理 6.3.3. 假设$t$是$k$上的一个基本式,满足$\eta = Dt$不是$k$中一个元素的微分。令$a, b, q \in k[t]$,满足$a \neq 0, \deg(b) \leq \deg(a)$,且$\deg(q) > 0$,那么

$(i)$如果$\deg(b) < \deg(a) - 1$,那么
$$\deg(aDq + bq) \in \{\deg(a) + \deg(q), \deg(a) + \deg(q) - 1\}$$

$(ii)$如果$\deg(b) = \deg(a) - 1$,那么或者
$$\deg(aDq + bq) \in \{\deg(a) + \deg(q), \deg(a) + \deg(q) - 1\}$$
或者
$$-\frac{\text{lc}(b)}{\text{lc}(a)} = \deg(q)\eta + Du, \exists u \in k$$

$(iii)$如果$\deg(b) = \deg(a)$,那么或者
$$\deg(aDq + bq) \in \{\deg(a) + \deg(q), \deg(a) + \deg(q) - 1\}$$
或者
$$-\frac{\text{lc}(b)}{\text{lc}(a)} = \frac{D((lc)(q))}{\text{lc}(q)}, \text{ 且} -\frac{\text{lc}(aD(\text{lc}(q)) + b\text{lc}(q))}{\text{lc}(a)\text{lc}(b)} = \deg(q)\eta + Du, \exists u \in k$$

引理 6.3.1.和引理 6.3.3.总是能够给出$\deg(q)$的一个上界,其中$q$为(6.12)在$k[t]$中的一个解:如果$\deg(b) > \deg(a)$,那么引理 6.3.1.说明$\deg(q) \in \{0, \deg(c) - \deg(b)\}$。如果$\deg(b) < \deg(a) - 1$,那么引理 6.3.3.说明$\deg(q) \in \{0, \deg(c) - \deg(a), \deg(c) - \deg(a) + 1\}$。如果$\deg(b) = \deg(a) - 1$,那么或者在$\deg(q) \in \{0, m, \deg(c) - \deg(a), \deg(c) - \deg(a) + 1\}$的情况下,存在$m \in \mathbb{Z}, u \in k$,使得$-\text{lc}(b)/\text{lc}(a) = m\eta + Du$,或者$\deg(q) \in \{0, \deg(c) - \deg(a), \deg(c) - \deg(a) + 1\}$。注意由引理 6.3.2.可知这样的$m$是唯一的。

最终,如果$\deg(b) = \deg(a)$,那么或者在$\deg(q) \in \{0, m, \deg(c) - \deg(a), \deg(c) - \deg(a) + 1\}$的情况下,存在$u \in k^*$,使得$-\text{lc}(b)/\text{lc}(a) = Du/u$,存在$m \in \mathbb{Z}, v \in k$,使得$-\text{lc}(aDu + bu)/(u\text{lc}(a)) = m\eta + Dv$。或者$\deg(q) \in \{0, \deg(c) - \deg(a), \deg(c) - \deg(a) + 1\}$。我们可以用一个积分算法的变形计算这样的$u$。尽管不是唯一的,但是如果$-\text{lc}(b)/\text{lc}(a) = Du/u = Dv/v, \exists u, v \in k^*$,那么根据引理存在$c \in \text{Const}(k)$,使得$u = cv$,这说明
$$\frac{\text{lc}(aDu + bu)}{\text{lc}(a)u} = \frac{\text{lc}(acDv + bcv)}{\text{lc}(a)cv} = \frac{c(\text{lc}(aDv + bv))}{c\text{lc}(a)v} = \frac{\text{lc}(aDv + bv)}{\text{lc}(a)v}$$

所以说我们使用的结果不影响边界$m$,根据引理 6.3.2.,$m$是唯一的。

```
RdeBoundDegreePrim(a, b, c, D)
(* Bound on polynomial solutions – primitive case *)

   (* Given a derivation D on k[t] and a, b, c ∈ k[t] with Dt ∈ k and a ≠ 0,
   return n ∈ ℤ such that deg(q) ≤ n for any solution q ∈ k[t] of aDq+bq = c.
   *)

   d_a ← deg(a), d_b ← deg(b), d_c ← deg(c)
   if d_b > d_a then n ← max(0, d_c − d_b) else n ← max(0, d_c − d_a + 1)
   if d_b = d_a − 1 then                           (* possible cancellation *)
      α ← −lc(b)/lc(a)
      if α = mDt + Dz for z ∈ k and m ∈ ℤ then n ← max(n, m)
   if d_b = d_a then                               (* possible cancellation *)
      α ← −lc(b)/lc(a)
      if α = Dz/z for z ∈ k* then
         β ← −lc(aDz + bz)/(z lc(a))
         if β = mDt + Dw for w ∈ k and m ∈ ℤ then n ← max(n, m)
   return n
```

在特定的情况下，$D = d/dt$，那么对于$\forall u \in k$，$Du = 0$。所以特别的是对于$u \in k$ $-\mathrm{lc}(b)/\mathrm{lc}(a)$不存在$Du/u$这种形式。这产生了引理 6.3.3.在这种情况下的一个简单形式，同时也产生了一个相比简单的算法。

推论 6.3.1. 假设$t$是$k$上超越元，$D = d/dt$，令$a, b, q \in k[t]$满足$a \neq 0, \deg(q) > 0$，那么
$(i)$如果$\deg(b) > \deg(a) - 1$，那么$\deg(aDq + bq) = \deg(b) + \deg(q)$
$(ii)$如果$\deg(b) < \deg(a) - 1$，那么$\deg(aDq + bq) = \deg(a) + \deg(q) - 1$
$(iii)$如果$\deg(b) = \deg(a) - 1$，那么或者$\deg(aDq + bq) = \deg(b) + \deg(q)$，或者
$$-\frac{\mathrm{lc}(b)}{\mathrm{lc}(a)} = \deg(q)$$

```
RdeBoundDegreeBase(a, b, c)
(* Bound on polynomial solutions – base case *)

   (* Given a, b, c ∈ k[t] with a ≠ 0, return n ∈ ℤ such that deg(q) ≤ n for
   any solution q ∈ k[t] of adq/dt + bq = c. *)

   d_a ← deg(a), d_b ← deg(b), d_c ← deg(c)
   n ← max(0, d_c − max(d_b, d_a − 1))
   if d_b = d_a − 1 then                           (* possible cancellation *)
      m ← −lc(b)/lc(a)
      if m ∈ ℤ then n ← max(0, m, d_c − d_b)
   return n
```

超越指数式情况
引理 6.3.4. 假设$t$是$k$上的一个超越指数式，满足$\eta = Dt/t$不是一个$k-$根的对数微分。令$a, b, q \in k[t]$，满足$a \neq 0, \deg(b) \leq \deg(a), \deg(q) > 0$，那么
$(i)$如果$\deg(b) < \deg(a)$，那么$\deg(aDq + bq) = \deg(a) + \deg(q)$
$(ii)$如果$\deg(b) = \deg(a)$，那么或者$\deg(aDq + bq) = \deg(a) + \deg(q)$，或者

$$-\frac{\mathrm{lc}(b)}{\mathrm{lc}(a)} = \deg(q)\eta + \frac{D(\mathrm{lc}(q))}{\mathrm{lc}(q)}$$

引理 6.3.1.和引理 6.3.4.总是能够给出$\deg(q)$的一个上界，其中$q$为方程(6.12)在$k[t]$中的一个解：如果$\deg(b) > \deg(a)$，那么引理 6.3.1.意味着$\deg(q) \in \{0, \deg(c) - \deg(b)\}$。如果$\deg(b) < \deg(a)$，那么引理 6.3.4.意味着$\deg(q) \in \{0, \deg(c) - \deg(a)\}$。最终，如果$\deg(b) = \deg(a)$，或者在$\deg(q) \in \{0, m, \deg(c) - \deg(b)\}$的情况下，$\exists m \in \mathbb{Z}, u \in k^*$，使得$-\mathrm{lc}(b)/\mathrm{lc}(a) = m\eta + Du/u$，或者$\deg(q) \in \{0, \deg(c) - \deg(b)\}$。注意到根据引理 6.2.2.，这样的$m$是唯一的。

```
RdeBoundDegreeExp(a, b, c, D)
(* Bound on polynomial solutions – hyperexponential case *)

    (* Given a derivation D on k[t] and a, b, c ∈ k[t] with Dt/t ∈ k and
    a ≠ 0, return n ∈ ℤ such that deg(q) ≤ n for any solution q ∈ k[t] of
    aDq + bq = c. *)
    d_a ← deg(a), d_b ← deg(b), d_c ← deg(c), n ← max(0, d_c − max(d_b, d_a))
    if d_a = d_b then                                (* possible cancellation *)
        α ← −lc(b)/lc(a)
        if α = mDt/t + Dz/z for z ∈ k* and m ∈ ℤ then n ← max(n, m)
    return n
```

非线性式情况

引理 6.3.5. 假设$t$是$k$上的一个非线性单项式，令$a, b, q \in k[t]$，满足$a \neq 0, \deg(b) = \deg(a) + \delta(t) - 1, \deg(q) > 0$，那么或者$\deg(aDq + bq) = \deg(b) + \deg(q)$，或者

$$-\frac{\mathrm{lc}(b)}{\mathrm{lc}(a)} = \deg(q)\lambda(t)$$

引理 6.3.1.和引理 6.3.5 总是能够给出$\deg(q)$的一个上界，其中$q$是方程(6.12)在$k[t]$中的一个解 L 如果$\deg(b) \neq \deg(a) + \delta(t) - 1$，那么引理 6.3.1.像之前说明的一样，给出上界。否则，或者在$\deg(1) \in \{0, m, \deg(c) - \deg(b)\}$的情况下，存在$m \in \mathbb{Z}$，使得$-\mathrm{lc}(b)/\mathrm{lc}(a) = m\lambda(t)$，或者$\deg(q) \in \{0, \deg(c) - \deg(b)\}$。

```
RdeBoundDegreeNonLinear(a, b, c, D)
(* Bound on polynomial solutions – nonlinear case *)

    (* Given a derivation D on k[t] and a, b, c ∈ k[t] with deg(Dt) ≥ 2 and
    a ≠ 0, return n ∈ ℤ such that deg(q) ≤ n for any solution q ∈ k[t] of
    aDq + bq = c. *)
    d_a ← deg(a), d_b ← deg(b), d_c ← deg(c), δ ← deg(Dt), λ ← lc(Dt)
    n ← max(0, d_c − max(d_a + δ − 1, d_b))
    if d_b = d_a + δ − 1 then                        (* possible cancellation *)
        m ← −lc(b)/(λ lc(a))
        if m ∈ ℤ then n ← max(0, m, d_c − d_b)
    return n
```

## 6.4 The SPDE Algorithm（随机偏微分方程算法）

现在我们将问题简化为在$k[t]$中找到(6.12)的解$q$，并且我们已经得到$\deg(q)$的一个上界$n$。我们介绍$Rothstein$的一个算法，或者将方程(6.12)简化为$a = 1$这种情况，或者证明方程在$k[t]$中没有次数至多为$n$的解。这个算法以下的定理为基础。

定理 6.4.1. 令$a, b, c \in k[t]$，其中$a \neq 0$，并且$\gcd(a,b) = 1$。令$z, r \in k[t]$，满足$c = az + br$。同时或者$r = 0$，或者$\deg(r) < \deg(a)$。那么对于$aDq + bq = c$在$k[t]$的任意解$q$，$h = (q-r)/a \in k[t]$，$h$为下述方程的解：
$$aDh + (b + Da)h = z - Dr \tag{6.16}$$
反之，对于方程(6.16)的任意解$h \in k[t]$，$q = ah + r$为$aDq + bq = c$的一个解。

定理 6.4.1. 将方程(6.12)简化为(6.16)，(6.16)也是和(6.12)同样类型的一个方程。但是，如果(6.12)有一个次数$n$的解$q$，那么因为$q = ah + r, \deg(r) < \deg(a)$，与$q$相对应的(6.16)的解$h$的次数之多为$n - \deg(a)$。因此，如果$\deg(a) > 0, \gcd(a,b) = 1$，那么我们可以用定理6.4.1.降低未知多项式的次数。$\gcd(a,b) = 1$这个假设不是一个限制：如果(6.12)在$k[t]$中有一个解，那么$c \in (a,b)$，多以$g = \gcd(a,b)$一定整除$c$，在这种情况下我们可以用$g$去除$a, b$和$c$，这样我们可以得到一个$\gcd(a,b) = 1$的等价方程。注意到这一步降低了$a$的次数。如果$\gcd(a,b) \nmid c$，我们可以断定(6.12)在$k[t]$中无解。我们可以不断地重复上述步骤知道或者我们证明(6.12)在$k[t]$中没有次数至多为$n$的解，或者直到$\deg(a) = 0$，换言之$a \in k^*$，此时我们用$a$除以方程两端，得到与(6.12)相同类型的方程，且$a = 1$。这就是$Rothstein$的$SPDE$算法。

```
SPDE(a, b, c, D, n)        (* Rothstein's SPDE algorithm *)

(* Given a derivation D on k[t], an integer n and a, b, c ∈ k[t] with a ≠ 0,
return either "no solution", in which case the equation aDq + bq = c has
no solution of degree at most n in k[t], or the tuple (b̄, c̄, m, α, β) such that
b̄, c̄, α, β ∈ k[t], m ∈ ℤ, and any solution q ∈ k[t] of degree at most n of
aDq + bq = c must be of the form q = αh + β, where h ∈ k[t], deg(h) ≤ m
and Dh + b̄h = c̄. *)

if n < 0 then
    if c = 0 then return(0,0,0,0,0) else return "no solution"
g ← gcd(a, b)
if g ∤ c then return "no solution"
a ← a/g, b ← b/g, c ← c/g
if deg(a) = 0 then return(b/a, c/a, n, 1, 0)
(r, z) ← ExtendedEuclidean(b, a, c) (* br + az = c, deg(r) < deg(a) *)
u ← SPDE(a, b + Da, z − Dr, D, n − deg(a))
if u = "no solution" then return "no solution"
(b̄, c̄, m, α, β) ← u
(* The solutions of (6.16) are h = αs + β where Ds + b̄s = c̄ *)
return(b̄, c̄, m, aα, aβ + r)        (* ah + r = aαs + aβ + r *)
```

## 6.5 *The Non-Cancellation Cases*（不可消去情况）

我们现在将问题简化为在$k[t]$中找到下述方程的解：
$$Dq + bq = c \tag{6.19}$$

其中$b, c \in k[t]$，$t$是$k$上的一个单项式。此外，我们已经得到了$\deg(q)$的一个上界$n$。在这一部分。我们介绍可以在任意单项式扩张中使用的算法，只要$Dq$的第一项和$bq$的和不为0。这种情况的充分条件或者为$D = d/dt$，或者$\deg(b) > \max(0, \delta(t) - 1)$，或者$t$是非线性的，且$\deg(b) \neq \delta(t) - 1$或者$\deg(b)/\lambda(t)$不是一个负整数。因为在上述这些情况中，$Dq$的首项和$bq$不会相消，我们称之为不可消去情况。

引理 6.5.1. 令$b, q \in k[t]$，其中$q \neq 0$
($i$) 假设$b \neq 0$。如果$D = d/dt$或者$\deg(b) > \max(0, \delta(t) - 1)$，那么$Dq + bq$的第一单项式为
$$\mathrm{lc}(b)\mathrm{lc}(q)t^{\deg(q)+\deg(b)}$$
($ii$) 如果$\deg(q) > 0$，$\deg(q) < \delta(t) - 1$，并且或者$\delta(t) \geq 2$，或者$D = d/dt$，那么$Dq + bq$的第一单项式为
$$\deg(q)\mathrm{lc}(q)\lambda(t)t^{\deg(q)+\delta(t)-1}$$
($iii$) 如果$\delta(t) \geq 2$，$\deg(b) = \delta(t) - 1$，$\deg(q) > 0$，并且$\deg(q) \neq -\mathrm{lc}(b)/\lambda(t)$，那么$Dq + bq$的第一单项式为
$$(\deg(q)\lambda(t) + \mathrm{lc}(b))\mathrm{lc}(q)t^{\deg(q)+\delta(t)-1}$$

引理 6.5.1.产生了下面寻找方程(6.19)解的算法，只要满足引理 6.5.1.的假设之一即可。

当$\deg(b)$足够大时

假设$b \neq 0$，$D = d/dt$或者$\deg(b) > \max(0, \delta(t) - 1)$。那么对于$Dq + bq = c$在$k[t]\setminus\{0\}$中任意解$q$，我们一定得到$\deg(q) + \deg(b) = \deg(c)$，所以$\deg(q) = \deg(c) - \deg(b)$，并且$\mathrm{lc}(b)\mathrm{lc}(q) = \mathrm{lc}(c)$。这给出了任意这样的$q$的第一单项式$ut^n$，并且在方程(6.19)中用$ut^n + h$代替$q$，我们得到
$$D(ut^n) + Dh + but^n + bh = c$$
所以
$$Dh + bh = c - D(ut^n) - but^n$$
这个方程与(6.19)为相同类型的方程，且有之前一样的$b$。因此引理 6.5.1. 的第一部分假设再次满足，所以我们可以继续重复这个过彻骨，但是$\deg(h)$的限制边界降为$n - 1$。这个限制边界在这个过程中，每进行一次都会减小，保证会在有限步结束。

```
PolyRischDENoCancel1(b, c, D, n)        (* Poly Risch d.e. – no cancellation *)
(* Given a derivation D on k[t], n either an integer or +∞, and b, c ∈ k[t]
with b ≠ 0 and either D = d/dt or deg(b) > max(0, δ(t)−1), return either
"no solution", in which case the equation Dq + bq = c has no solution
of degree at most n in k[t], or a solution q ∈ k[t] of this equation with
deg(q) ≤ n. *)
q ← 0
while c ≠ 0 do
    m ← deg(c) − deg(b)
    if n < 0 or m < 0 or m > n then return "no solution"
    p ← (lc(c)/lc(b)) t^m
    q ← q + p
    n ← m − 1
    c ← c − Dp − bp
return q
```

当$\deg(b)$足够小时

假设$\deg(b) < \delta(t) - 1$，或者$D = d/dt$。这说明$b = 0$，或者$\delta(t) \geq 2$。令$q$为$Dq + bq = c$在$k[t]$中的一个解。

如果$\deg(q) > 0$，那么$\deg(q) + \delta(t) - 1 = \deg(c)$，所以$\deg(q) = \deg(c) + 1 - \delta(t)$，并且$\deg(q)\mathrm{lc}(q)\lambda(t) = \mathrm{lc}(c)$。这产生了$q$的第一单项式$ut^n$，并且在方程中用$ut^n + h$代替$q$产生了一个相似的方程，新方程的解的次数限制的更低。

如果$q \in k$，那么：如果$b \in k^*$，那么$Dq + bq \in k$，所以当我们将问题简化为在$k$上解一个类型为(6.1)的$Risch$微分方程时，$c \in k$，或者$\deg(c) > 0$，(6.19)在$k$中无解，因此在$k[t]$中无解。如果$\deg(b) > 0$，那么$Dq + bq$的第一单项式为$q\mathrm{lc}(b)t^{\deg(b)}$，所以在$q = \mathrm{lc}(c)/\mathrm{lc}(b)$为唯一可能解的情况下，$\deg(c) = \deg(b)$，或者$\deg(c) \neq \deg(b)$，(6.19)在$k$中没有解，进而在$k[t]$中没有解。

```
PolyRischDENoCancel2(b, c, D, n)        (* Poly Risch d.e. – no cancellation *)

(* Given a derivation D on k[t], n either an integer or +∞, and b, c ∈ k[t]
with deg(b) < δ(t) − 1 and either D = d/dt or δ(t) ≥ 2, return either
"no solution", in which case the equation Dq + bq = c has no solution
of degree at most n in k[t], or a solution q ∈ k[t] of this equation with
deg(q) ≤ n, or the tuple (h, b_0, c_0) such that h ∈ k[t], b_0, c_0 ∈ k, and for
any solution q ∈ k[t] of degree at most n of Dq + bq = c, y = q − h is a
solution in k of Dy + b_0 y = c_0. *)

q ← 0
while c ≠ 0 do
    if n = 0 then m ← 0 else m ← deg(c) − δ(t) + 1
    if n < 0 or m < 0 or m > n then return "no solution"
    if m > 0 then p ← (lc(c)/(m λ(t))) t^m
    else                                                            (* m = 0 *)
        if deg(b) ≠ deg(c) then return "no solution"
        if deg(b) = 0 then return(q, b, c)
        p ← lc(c)/lc(b)
    q ← q + p
    n ← m − 1
    c ← c − Dp − bp
return q
```

当 $\delta(t) \geq 2$ 且 $\deg(b) = \delta t - 1$ 时

在这种情况下，我们只有当 $\deg(q) = -\mathrm{lc}(b)/\lambda(t)$ 时，进行消去。这特别说明了 $-\mathrm{lc}(b)/\lambda(t)$ 为正整数。令 $q$ 为 $Dq + bq = c$ 在 $k[t]$ 中的一个解。

如果 $\deg(q) > 0$，并且 $\deg \neq -\mathrm{lc}(b)/\lambda(t)$，那么 $\deg(q) + \delta(t) - 1 = \deg(c)$，所以 $\deg(q) = \deg(c) + 1 - \delta(t)$，且 $(\deg(q)\lambda(t) + \mathrm{lc}(b))\mathrm{lc}(q) = \mathrm{lc}(c)$。这产生了 $q$ 的第一单项式 $ut^n$，在方程中用 $ut^n + h$ 代替 $q$ 得到一个相似的方程，且方程的解的次数限制更低。只要新的次数比 $-\mathrm{lc}(b)/\lambda(t)$ 的次数大，我们就可以重复这个过程，或者直到 $-\mathrm{lc}(b)/\lambda(t)$ 不是一个正整数时，我们得到了一个完整的解。

如果 $q \in k$，那么 $Dq + bq$ 的首项为 $q\mathrm{lc}(b)t^{\delta(t)-1}$，所以或者在 $q = \mathrm{lc}(c)/\mathrm{lc}(b)$ 为唯一可能解的这种情况下，$\deg(c) = \delta(t) - 1$，或者 $\deg(c) \neq \delta(t) - 1$，且 (6.19) 在 $k$ 中无解，进而在 $k[t]$ 中无解。

```
PolyRischDENoCancel3(b, c, D, n)          (* Poly Risch d.e. – no cancellation *)
(* Given a derivation D on k[t] with δ(t) ≥ 2, n either an integer or
+∞, and b, c ∈ k[t] with deg(b) = δ(t) − 1, return either "no solution",
in which case the equation Dq + bq = c has no solution of degree at most
n in k[t], or a solution q ∈ k[t] of this equation with deg(q) ≤ n, or the
tuple (h, m, c̄) such that h ∈ k[t], m ∈ ℤ, c̄ ∈ k[t], and for any solution
q ∈ k[t] of degree at most n of Dq + bq = c, y = q − h is a solution in k[t]
of degree at most m of Dy + by = c̄. *)

q ← 0
if −lc(b)/λ(t) ∈ ℕ then M ← −lc(b)/λ(t) else M ← −1
while c ≠ 0 do
    m ← max(M, deg(c) − δ(t) + 1)
    if n < 0 or m < 0 or m > n then return "no solution"
    u ← mλ(t) + lc(b)
    if u = 0 then return(q, m, c)
    if m > 0 then p ← (lc(c)/u) t^m
    else                                                    (* m = 0 *)
        if deg(c) ≠ δ(t) − 1 then return "no solution"
        p ← lc(c)/lc(b)
    q ← q + p
    n ← m − 1
    c ← c − Dp − bp
return q
```

## 6.6 The Cancellation Cases（可消去情况）

最终我们研究方程(6.19)在不可消去情况没有出现的这种情况，换言之，在下列情况之一：

1. $\delta(t) \leq 1, b \in k$，且 $D \neq d/dt$

2. $\delta(t) \leq 2, \deg(b) = \delta(t) - 1$，且 $\deg(q) = -lc(b)/\lambda(t)$

我们在这一部分给对于特定类型的单项式，在以上情形下的算法。

基本式情况

如果 $Dt \in k$，那么 $\delta(t) = 0$，所以引理 6.5.1.唯一没有处理的一种情况是 $b = 0$ 或 $b \in k^*$。如果 $b = 0$，那么(6.19)变为 $Dq = c, c \in k[t]$，这是一个在 $k[t]$ 中的积分问题，判定它在 $k[t]$ 中是否有一个解可以通过内场积分算法解决，所以现在假设 $b \in k^*$。

如果存在 $u \in k^*$，$b = Du/u$，这也可以通过一系列积分算法进行验证，那么(6.19)变为 $Dq + qDu/d = c$，换言之，$D(uq) = uc$，如前面说的那样，这是一个 $k[t]$ 中的积分问题。

如果 $b$ 不是 $Du/u$ 这种形式，其中 $u \in k^*$，那么 $D(lc(q)) + blc(q) \neq 0$，所以 $Dq + fq$ 的第一单项式为 $(D(lc(q)) + blc(q))t^{\deg(q)}$。这说明 $\deg(b) = \deg(c)$，$lc(q)$ 是下列方程在 $k^*$ 中的一个解。

$$Dy + by = lc(c) \tag{6.23}$$

这是一个在 $k$ 中的Risch微分方程。如果在 $k$ 中无解，那么(6.19)在 $k[t]$ 中无解。否则，引理 5.9.1.说明方程有唯一解，必须是 $lc(q)$。这给出了任意解 $q$ 的第一单项式 $yt^{\deg(c)}$，像之前一样，在(6.19)中用 $yt^{\deg(c)} + h$ 代替 $q$ 产生一个相同类型的方程，新的方程的解的次数限制

更低，方程右侧次数更低。

```
PolyRischDECancelPrim(b, c, D, n)
(* Poly Risch d.e., cancellation – primitive case *)

(* Given a derivation D on k[t], n either an integer or +∞, b ∈ k and
   c ∈ k[t] with Dt ∈ k and b ≠ 0, return either "no solution", in which case
   the equation Dq + bq = c has no solution of degree at most n in k[t], or
   a solution q ∈ k[t] of this equation with deg(q) ≤ n. *)

if b = Dz/z for z ∈ k* then
    if zc = Dp for p ∈ k[t] and deg(p) ≤ n then return(p/z)
    else return "no solution"
if c = 0 then return 0
if n < deg(c) then return "no solution"
q ← 0
while c ≠ 0 do
    m ← deg(c)
    if n < m then return "no solution"
    s ← RischDE(b, lc(c))              (* Ds + bs = lc(c) *)
    if s = "no solution" then return "no solution"
    q ← q + st^m
    n ← m − 1
    c ← c − bst^m − D(st^m)            (* deg(c) becomes smaller *)
return q
```

超越指数式情况

如果$Dt/t = \eta \in k$，那么$\delta(t) = 1$，所以引理 6.5.1.唯一没有处理的情况是$b = 0$或$b \in k^*$。如果$b = 0$，那么(6.19)变为$Dq = c$，$c \in k[t]$，这是一个$k[t]$上的积分问题，可以通过一系列的积分算法判定它是否在$k[t]$中有一个解，所以假设$b \in k^*$。

如果$b = Du/u + m\eta$，存在$u \in k^*, m \in \mathbb{Z}$，那么(6.19)变为$Dq + (Du/u + m\eta)q = c$，换句话说$D(uqt^m) = uct^m$，这是一个在$k\langle t\rangle$中的积分问题，可以通过一系列的积分算法判定它是否在$k\langle t\rangle$中有一个解。

最后假设$b$没有$Du/u + m\eta$这种形式，$u \in k^*, m \in \mathbb{Z}$。那么$D(\mathrm{lc}(q) + +\deg(q)\eta\mathrm{lc}(q) + b\mathrm{lc}(q)) \neq 0$，所以$Dq + bq$的第一单项式是$D(\mathrm{lc}(q) + +\deg(q)\eta\mathrm{lc}(q) + b\mathrm{lc}(q))t^{\deg(q)}$。这说明$\deg(q) = \deg(c)$，$\mathrm{lc}(q)$是以下方程在$k^*$中的一个解：

$$Dy + (b + \deg(q)\eta)y = \mathrm{lc}(c) \qquad (6.24)$$

这是在$k$中的一个Risch微分方程。如果它在$k$中无解，那么(6.19)在$k[t]$中无解。否则，引理 5.9.1.说明它有唯一的一个解，必须为$\mathrm{lc}(q)$。这给出了任意解$q$的第一单项式为$yt^{\deg(c)}$，正如前面讨论的一样，在(6.19)中用$yt^{\deg(c)} + h$代替$q$产生了一个相同类型的方程，解的次数限制更低，方程右侧次数更低。

```
PolyRischDECancelExp(b, c, D, n)
(* Poly Risch d.e., cancellation – hyperexponential case *)

(* Given a derivation D on k[t], n either an integer or +∞, b ∈ k and
c ∈ k[t] with Dt/t ∈ k and b ≠ 0, return either "no solution", in which
case the equation Dq + bq = c has no solution of degree at most n in k[t],
or a solution q ∈ k[t] of this equation with deg(q) ≤ n. *)
if b = Dz/z + mDt/t for z ∈ k* and m ∈ ℤ then
    if czt^m = Dp for p ∈ k⟨t⟩ and q = p/(zt^m) ∈ k[t] and deg(q) ≤ n then
        return(q)
    else return "no solution"
if c = 0 then return 0
if n < deg(c) then return "no solution"
q ← 0
while c ≠ 0 do
    m ← deg(c)
    if n < m then return "no solution"
    s ← RischDE(b + mDt/t, lc(c))      (* Ds + (b + mDt/t)s = lc(c) *)
    if s = "no solution" then return "no solution"
    q ← q + st^m
    n ← m - 1
    c ← c - bst^m - D(st^m)             (* deg(c) becomes smaller *)
return q
```

非线性情况

如果$\delta(t) \geq 2$，那么我们得到$\deg(b) = \delta(t) - 1$，且$\text{lc}(b) = -n\lambda(t)$，其中$n > 0$为$\deg(q)$上的限制。没有在这种情况下解决方程(6.19)的一般性算法。但是如果$\mathcal{S}^{irr} \neq \emptyset$，那么我们可以：对于$p \in \mathcal{S}^{irr}$，将$\pi_p$作用于方程(6.19)上，利用等式$D^* \circ \pi_p = \pi_p \circ D$，其中$D^*$是$k[t]/(p)$上的诱导微分，我们得到

$$D^*q* + \pi_p(b)q^* = \pi_p(c) \quad (6.25)$$

其中$q^* = \pi_p(q)$。假设我们有在$k[t]/p$中解方程(6.25)的算法，我们可以按如下步骤解方程(6.19)：如果(6.25)在$k[t]/(p)$中没有解，那么(6.19)在$k[t]$中没有解。否则，令$q^* \in k[t]/(p)$为(6.25)的一个解，令$r \in k[t]$满足$\deg(r) < \deg(p)$，且$\pi_p(r) = q^*$。注意到$\pi_p(Dr + br) = \pi_p(c)$，所以$p | c - Dr - br$。另外，$\pi_p(q) = \pi_p(r)$，所以$h = (q - r)/p \in k[t]$，并且我们有

$$c = Dq + bq = p\left(Dh + \left(b + \frac{Dp}{p}\right)h\right) + Dr + br$$

所以$h$是如下方程在$k[t]$中一个次数至多为$\deg(q) - \deg(p)$的解

$$Dh + \left(b + \frac{Dp}{p}\right)h = \frac{c - Dr - br}{p} \quad (6.26)$$

这和方程(6.19)的类型相同，但是解的次数的限制更低。

存在方程(6.25)可以解出的情况，例如存在$p \in \mathcal{S}^{irr}, \deg(p) = 1$。那么$k[t]/(p) \simeq k$，所以(6.25)是一个$k$中的$Risch$微分方程。另一个可能性是如果$\mathcal{S}^{irr} \cap \text{Const}(k)[t] \neq \emptyset$，在这种情况下取$p = t - \alpha$，其中$\alpha$是一个不可约$special$（特殊）多项式的常数根，我们得到$k[t]/(p) \simeq k(\alpha)$，所以(6.25)是$k(\alpha)$中的一个$Risch$微分方程。这是当$t$是一个超切向量单项

式的情况，其中$\alpha = \pm\sqrt{-1}$。取$p = t - \alpha$，也可以得到这样的结论，只是$\alpha$不是一个常数，但是(6.25)会是$k(t)$的一个非常熟代数扩张中的一个$Risch$微分方程，当$t$是一个非线性单项式时，还没有适用于这种情况的算法。尽管某些技巧可以推广到这种情况下，但是它们不会产生一个具有它们目前形式的实用性算法。

超切向量式情况

如果$Dt/(t^2+1) = \eta \in k$，那么$\delta(t) = 2$，所以引理 6.5.1.没有处理的情况为 $b = b_0 - n\eta t$，其中$b_0 \in k, n > 0$为$\deg(q)$的边界。在这样的扩张中，上面列出的方法的轮廓给出了一个完整的算法：如果$\sqrt{-1} \in k$，那么$\mathcal{S}^{irr} = \{t - \sqrt{-1}, t + \sqrt{-1}\}$，(6.25)是$k$上的一个简单的$Risch$微分方程。

如果$\sqrt{-1} \notin k$，那么取$p = t^2 + 1 \in \mathcal{S}^{irr}$，(6.25)变为
$$Dq^* + (b_0 - n\eta\sqrt{-1})q^* = c(\sqrt{-1}) \tag{6.27}$$
其中$D$通过$D\sqrt{-1} = 0$扩张到$k[t]/(p) \simeq k(\sqrt{-1})$上。一种可能性是将(6.27)看做$k(\sqrt{-1})$中的$Risch$微分方程，用递归式的方法解方程。如果它在$k(\sqrt{-1})$中没有解，那么(6.19)在$k[t]$中没有解。否则，如果$u + v\sqrt{-1}$是(6.27)的一个解，其中$u, v \in k$，那么令$r = u + vt, h = (q - r)/p$是方程(6.26)在$k[t]$中次数至多为$n - 2$的一个解。也是有可能避免引入$\sqrt{-1}$，方法是考虑(6.27)的实部和虚部：将$q^*$写作$q^* = u + v\sqrt{-1}$，我们得到
$$\begin{pmatrix} Du \\ Dv \end{pmatrix} + \begin{pmatrix} b_0 & n\eta \\ -n\eta & b_0 \end{pmatrix} \begin{pmatrix} u \\ v \end{pmatrix} = \begin{pmatrix} c_0 \\ c_1 \end{pmatrix} \tag{6.28}$$
其中$c_0 + c_1 t$是$t^2 + 1$除以$c$的余数。这是在5.10节中提到的耦合微分系统问题。如果在$k$中无解，那么(6.19)在$k[t]$中无解。否则，如果$(u, v) \in k^2$是(6.28)的一个解，那么令
$r = u + vt, h = (q - r)/p$是(6.26)在$k[t]$中次数至多为$n - 2$的一个解。

**PolyRischDECancelTan**($b_0, c, D, n$)
(* Poly Risch d.e., degenerate cancellation – tangent case *)

(* Given a derivation $D$ on $k[t]$, $n \in \mathbb{Z}$, $b_0 \in k$ and $c \in k[t]$ with $Dt/(t^2 + 1) \in k$, $\sqrt{-1} \notin k$ and $n \geq 0$, return either "no solution", in which case the equation $Dq + (b_0 - ntDt/(t^2 + 1))q = c$ has no solution of degree at most $n$ in $k[t]$, or a solution $q \in k[t]$ of this equation with $\deg(q) \leq n$. *)

if $n = 0$ then
  if $c \in k$ then
    if $b_0 \neq 0$ then return **RischDE**($b_0, c$)
    else if $\int c = q \in k$ then return($q$) else return "no solution"
  else return "no solution"
$p \leftarrow t^2 + 1$     (* the monic irreducible special polynomial *)
$\eta \leftarrow Dt/p$     (* $t = \tan(\int \eta)$ *)
$(\bar{c}, c_1 t + c_0) \leftarrow$ **PolyDivide**($c, p$)   (* $c(\sqrt{-1}) = c_1 \sqrt{-1} + c_0$ *)
(* CoupledDESystem will be given in Chap. 8 *)
$(u, v) \leftarrow$ **CoupledDESystem**($b_0, -n\eta, c_0, c_1$)
(* $Du + b_0 u + n\eta v = c_0$, $Dv - n\eta u + b_0 v = c_1$ *)
if $(u, v) =$ "no solution" then return "no solution"
if $n = 1$ then return($u + vt$)
$r = u + vt$
$c \leftarrow (c - Dr - (b_0 - n\eta t)r)/p$   (* this division is always exact *)
$h \leftarrow$ **PolyRischDECancelTan**($b_0, c, D, n - 2$)
if $h =$ "no solution" then return "no solution"
return($ph + r$)

# *Chapter* 2.1.7 **参数问题**

在这一章节中，我们说明含参数积分相关问题的解。这些问题是作为积分算法的子问题提出的：受限积分问题，这是在一个基本扩张中对多项式做积分的背景下提出的；含参数的对数微分问题，这是在确定对数微分和研究 $Risch$ 微分方程的解的受限边界的阶数和解的次的背景下提出的。这些问题之间的普遍联系是它们的本质都是确定是否存在常数，使得对于一个给定的含参数的微分方程在一个给定的微分域中有解存在。

**7.1.含参数的 $Risch$ 微分方程**

首先我们提出经典的含参数问题，换言之含参数的 $Risch$ 微分方程，是指我们将一个 $Risch$ 微分方程的右端的 $g \in K$ 用线性组合 $\sum_{i=1}^m c_i g_i, g_i \in K$ 代替。问题就是确定所有 $\mathrm{Const}(K)$ 中的常数 $c_i$，使得方程

$$Dy + fy = \sum_{i=1}^{m} c_i g_i \tag{7.1}$$

在 $K$ 中有解，当然计算出这个解。这个问题，当我们仅仅是在对初等超越函数做积分没有出现，但是在对非初等函数或者在对一非初等函数组成的函数做积分时出现了。另外，受限积分问题可以看做这个问题的一种特殊情况。注意到使得(7.1)在 $K$ 中有解常数集合 $(c_1, \ldots, c_m)$ 组成了 $\mathrm{Const}(K)^m$ 的一个线性子空间。这就促使产生了下面含参数的 $Risch$ 微分方程问题的正式定义：给定一个特征为0的微分域 $K, f, g_1, \ldots, g_m \in K$，计算 $h_1, \ldots, h_r \in K$，一个属于 $\mathrm{Const}(K)$ 有着 $m+r$ 列和行的矩阵 $A$ 使得 (7.1) 有一个解 $c_1, \ldots, c_m \in \mathrm{Const}(K), y \in K$ 当且仅当 $y = \sum_{j=1}^{r} d_j h_j$，其中 $d_1, \ldots, y_m \in \mathrm{Const}(K), A(c_1, \ldots, c_m, d_z, \ldots, d_r)^T = 0$

和第六章一样，我们只在超越情况下研究方程(7.1)，换言之，当 $K$ 是一个微分子域 $k$ 的 $simple$（简单）单项式扩张这一种情况，所以在剩下这一部分中，令 $k$ 为一个特征为0的微分域，$t$ 是 $k$ 上的一个单项式。此外我们假设 $\mathrm{Const}(k(t)) = \mathrm{Const}(k)$。我们假设方程的系数 $f$ 和 $g_1, \ldots, g_m$ 属于 $k(t)$，寻找解 $c_1, \ldots, c_n \in \mathrm{Const}(k)$ 和 $y \in k(t)$。结果证明第六章的算法可以简单地扩展到含参数问题上。

分母的 $normal$（普通）部分

因为对于任意不可约元 $p \in k[t]$，$\nu_p(\sum_{i=1}^m c_i g_i) \geq \min_{1 \leq i \leq m}(\nu_p(g_i))$，定理 6.1.2.的第一部分可以扩展到含参数问题上。当然因为上面的不等式为严格不等式，所以定理 6.1.2.的第二部分不能扩展到含参数问题上。

定理 7.1.1. 令 $f \in k(t)$ 关于 $t$ 弱正规化，$g_1, \ldots, g_m \in k(t)$。令 $c_1, \ldots, c_m \in \mathrm{Const}(k), y \in k(t)$ 满足 $Dy + fy = \sum_{i=1}^{m} c_i g_i$。令 $d = d_s d_n$ 为 $f$ 分母的一个分裂分解，$e$ 是 $g_i$ 分母的一个最小公倍数，$e = e_s e_n$ 为 $e$ 的一个分裂分解。令 $c = \gcd(d_n, e_n)$，且

$$h \frac{\gcd(e_n, de_n/dt)}{\gcd(c, dc/dt)} \in k[t]$$

那么，$yh \in k\langle t \rangle$。

推论 7.1.1. 令 $f \in k(t)$ 关于 $t$ 弱正规化，$g_1, \ldots, g_m \in k(t)$，$d_n, e_n$ 和 $h$ 定义和定理 7.1.1.一致，那么，对于 $Dy + fy = \sum_{i=1}^{m} c_i g_i$ 的任意解 $c_1, \ldots, c_m \in \mathrm{Const}(k)$ 和 $y \in k(t)$，$q = yh \in k\langle t \rangle$，

并且$q$是以下方程的一个解：

$$d_n h Dq + (d_n hf - d_n Dh)q = \sum_{i=1}^{m} c_i(d_n h^2 g_i) \qquad (7.2)$$

反之，对于(7.2)的任意解$c_1,\ldots,c_m \in \text{Const}(k)$和$q \in k\langle t\rangle$，$y = q/h$是$Dy + fy = \sum_{i=1}^{m} c_i g_i$的一个解。

上面的定理和推论为我们提供了一个可以将一个给定的$Risch$微分方程问题简化为$k\langle t\rangle$上的$Risch$微分方程问题的算法。

**ParamRdeNormalDenominator**$(f, g_1, \ldots, g_m, D)$
(* Normal part of the denominator *)

(* Given a derivation $D$ on $k[t]$ and $f, g_1, \ldots, g_m \in k(t)$ with $f$ weakly normalized with respect to $t$, return the tuple $(a, b, G_1, \ldots, G_m, h)$ such that $a, h \in k[t]$, $b \in k\langle t\rangle$, $G_1, \ldots, G_m \in k(t)$, and for any solution $c_1, \ldots, c_m \in \text{Const}(k)$ and $y \in k(t)$ of $Dy + fy = \sum_{i=1}^{m} c_i g_i$, $q = yh \in k\langle t\rangle$ satisfies $aDq + bq = \sum_{i=1}^{m} c_i G_i$. *)

$(d_n, d_s) \leftarrow$ **SplitFactor**$(\text{denominator}(f), D)$
$(e_n, e_s) \leftarrow$ **SplitFactor**$(\text{lcm}(\text{denominator}(g_1), \ldots, \text{denominator}(g_m)), D)$
$p \leftarrow \gcd(d_n, e_n)$
$h \leftarrow \gcd(e_n, de_n/dt)/\gcd(p, dp/dt)$
**return**$(d_n h, d_n hf - d_n Dh, d_n h^2 g_1, \ldots, d_n h^2 g_m, h)$

*分母的special（特殊）部分*

作为推论 7.1.1.的一个结论，现在我们可以将问题简化为找到方程(7.2)的解$c_1, \ldots, c_m \in \text{Const}(k)$和$q \in k\langle t\rangle$，我们将问题重写为

$$aDq + bq = \sum_{i=1}^{m} c_i g_i \qquad (7.3)$$

其中$a \in k[t]$没有$special$（特殊）因子，$b \in k\langle t\rangle$，$g_1, \ldots, g_m \in k(t), a \neq 0$，并且$t$是$k$上的一个单项式。因为$\nu_p(\sum_{i=1}^{m}) \geq \min_{1 \leq i \leq m}(\nu_p(g_i))$，其中$p$为$k[t]$中的任意不可约元，因为$a \in k[t]$并且在(7.3)中没有$special$（特殊）因子，在不含参数的情况下，引理 6.2.1.给出了$\nu_p(q)$的一个更低的边界限制。
(1)如果$\nu_p(b) < 0$，那么$\nu_p(q) \geq \min_{1 \leq i \leq m}(\nu_p(g_i)) - \nu_p(b))$
(ii)如果$\nu_p(b) > 0$，并且$p \in \mathcal{S}_1^{irr}$，那么$\nu_p(q) \geq \min(0, \min_{1 \leq i \leq m}(\nu_p(g_i)))$

对于$p \in \mathcal{S}^{irr}$，一旦我们有一个更小的边界限制$\nu_p(q) \geq n, n \leq 0$，用$hp^n$代替$q$，带入(7.3)得到

$$a(p^n Dh + np^{n-1}hDp) + bhp^n = \sum_{i=1}^{m} c_i g_i$$

因此

$$aDh + \left(b + na\frac{Dp}{p}\right)h = \sum_{i=1}^{m} c_i(g_i p^{-n}) \qquad (7.4)$$

此外，因为$q \in k\langle t \rangle$，$h \in \langle t \rangle$，因为$\nu_p(q) \geq n$，所以$h \in \mathcal{O}_p$。因此我们将问题简化为找到(7.4)的解$c_1, \ldots, c_m \in \mathrm{Const}(k)$和$h \in k\langle t \rangle \cap \mathcal{O}_p$。注意到因为$b \in k\langle t \rangle, a \in k[t], p \in \mathcal{S}$，所以$b + naDp/p \in k\langle t \rangle$。$p$在$b + naDp/p$的分母中的最终的幂指数可以通过将$p^N$乘到(6.7)的两端进行消除，其中$N = \max(0, -\nu_p(b))$，确保(7.4)左端的系数也属于$k\langle t \rangle \cap \mathcal{O}_p$。

因为在我们考虑的单项式扩张的情形中，所有的*special*（特殊）多项式都是第一类的，我们只需要在可能的消去情形下对于$\nu_p(q)$更小的边界，换言之$\nu_p(b) = 0$。对于不同的单项式扩张，我们分别进行讨论。

基本式情况

如果$Dt \in k$，那么每一个无平方因子的多项式都是*normal*（普通）的，所以$k\langle t \rangle = k[t]$，这意味着$a, b \in k[t]$，(7.3)在$k\langle t \rangle$中的任意解$q$一定在$k[t]$中。

超越指数式情况

如果$Dt/t = \eta \in k$，那么$k\langle t \rangle = k[t, t^{-1}]$，所以我们需要计算$\nu_t(q)$的一个较小的边界限制，其中$c_1, \ldots, c_m \in \mathrm{Const}(k)$, $q \in k\langle t \rangle$是(7.3)的一个解。根据定理 5.1.2.，我们得到$t \in \mathcal{S}_1^{irr}$，引理 5.1.2.和引理 6.2.3. 总是能够给出$\nu_t(q)$的一个较小的边界限制：如果$\nu_t(b) \neq 0$，那么引理 6.2.1. 像之前说明的一样给出了一个限制边界。否则的话，$\nu_t(b) = 0$，所以或者在如下情况中$-b(0)/a(0) = s\eta + Du/u$，其中$s \in \mathbb{Z}, u \in k^*$

$$\nu_t(q) \geq \min\left(0, s, \min_{1 \leq i \leq m}(\nu_t(g_i))\right)$$

或者

$$\nu_t(q) \geq \min\left(0, \min_{1 \leq i \leq m}(\nu_t(g_i))\right)$$

注意到根据将引理 6.2.2.作用于$k$上得到的结果，这样的$s$是唯一的。因为$\mathcal{S}^{irr} = \{t\}$，$k\langle t \rangle \cap \mathcal{O}_t = k[t]$，所以在确定$\nu_t(q)$的一个更小的边界的前提下，我们需要(7.4)的解$c_1, \ldots, c_m \in \mathrm{Const}(k), h \in k[t]$。

**ParamRdeSpecialDenomExp**$(a, b, g_1, \ldots, g_m, D)$
(\* Special part of the denominator – hyperexponential case \*)

(\* Given a derivation $D$ on $k[t]$ and $a \in k[t]$, $b \in k\langle t \rangle$ and $g_1, \ldots, g_m \in k\langle t \rangle$ with $Dt/t \in k$, $a \neq 0$ and $\gcd(a, t) = 1$, return the tuple $(\bar{a}, \bar{b}, \overline{g_1}, \ldots, \overline{g_m}, h)$ such that $\bar{a}, \bar{b}, h \in k[t]$, $\overline{g_1}, \ldots, \overline{g_m} \in k\langle t \rangle$, and for any solution $c_1, \ldots, c_m \in \mathrm{Const}(k)$ and $q \in k\langle t \rangle$ of $aDq + bq = \sum_{i=1}^m c_i g_i$, $r = qh \in k[t]$ satisfies $\bar{a}Dr + \bar{b}r = \sum_{i=1}^m c_i \overline{g_i}$. \*)

$p \leftarrow t$                  (\* the monic irreducible special polynomial \*)
$n_b \leftarrow \nu_p(b)$, $n_c \leftarrow \min_{1 \leq i \leq m}(\nu_p(g_i))$
$n \leftarrow \min(0, n_c - \min(0, n_b))$              (\* $n \leq 0$ \*)
**if** $n_b = 0$ **then**           (\* possible cancellation \*)
    $\alpha \leftarrow$ **Remainder**$(-b/a, p)$        (\* $\alpha = -b(0)/a(0) \in k$ \*)
       **if** $\alpha = sDt/t + Dz/z$ for $z \in k^*$ and $s \in \mathbb{Z}$ **then** $n \leftarrow \min(n, s)$
$N \leftarrow \max(0, -n_b)$              (\* $N \geq 0$, for clearing denominators \*)
**return**$(ap^N, (b + naDp/p)p^N, g_1 p^{N-n}, \ldots, g_m p^{N-n}, p^{-n})$

超切向量式情况

如果$Dt/(t^2+1) = \eta \in k$，且$\sqrt{-1} \notin k$，那么唯一的首一*special*（特殊）不可约元为$t^2+1$，所以我们需要计算出$\nu_{t^2+1}(q)$的一个更小的边界，其中$c_1,\ldots,c_m \in \text{Const}(k), q \in k\langle t\rangle$为(7.3)的解。因为根据定理 5.10.1.，$t^2+1 \in \mathcal{S}_1^{irr}$，引理 6.2.1.和引理 6.2.4.总是能给出$\nu_{t^2+1}(q)$的一个更小边界：如果$\nu_{t^2+1}(b) \neq 0$，那么引理 6.2.1.像之前说明的一样给出了边界。否则，$\nu_{t^2+1}(b) = 0$，所以或者在以下情况中，存在$s \in \mathbb{Z}, u \in k(\sqrt{-1})^*$，使得$-b(\sqrt{-1})/a(\sqrt{-1}) = s\eta\sqrt{-1} + Du/u$

$$\nu_{t^2+1}(q) \geq \min\left(0, s, \min_{1\leq i \leq m}(\nu_{t^2+1}(g_i))\right)$$

或者

$$\nu_{t^2+1}(q) \geq \min\left(0, \min_{1\leq i \leq m}(\nu_{t^2+1}(g_i))\right)$$

注意到将引理 6.2.2.作用于$k(\sqrt{-1})$可知这样的$s$是唯一的。在无参数情况下，关于临时添加$\sqrt{-1}$的注记在这里依然是成立的。因为$\mathcal{S}^{irr} = \{t^2+1\}, k\langle t\rangle \cap \mathcal{O}_{t^2+1} = k[t]$，所以在确定$\nu_{t^2+1}(q)$的一个较小边界的情况下，我们需要找到(7.4)的解$c_1,\ldots,c_m \in \text{Const}(k), h \in k[t]$。

**ParamRdeSpecialDenomTan**$(a, b, g_1, \ldots, g_m, D)$
(* Special part of the denominator – hypertangent case *)

(* Given a derivation $D$ on $k[t]$ and $a \in k[t], b \in k\langle t\rangle$ and $g_1, \ldots, g_m \in k(t)$ with $Dt/(t^2+1) \in k$, $\sqrt{-1} \notin k$, $a \neq 0$ and $\gcd(a, t^2+1) = 1$, return the tuple $(\bar{a}, \bar{b}, \overline{g_1}, \ldots, \overline{g_m}, h)$ such that $\bar{a}, \bar{b}, h \in k[t], \overline{g_1}, \ldots, \overline{g_m} \in k(t)$, and for any solution $c_1, \ldots, c_m \in \text{Const}(k)$ and $q \in k\langle t\rangle$ of $aDq + bq = \sum_{i=1}^m c_i g_i$, $r = qh \in k[t]$ satisfies $\bar{a}Dr + \bar{b}r = \sum_{i=1}^m c_i \overline{g_i}$. *)

$p \leftarrow t^2 + 1$    (* the monic irreducible special polynomial *)
$n_b \leftarrow \nu_p(b), n_c \leftarrow \min_{1\leq i \leq m}(\nu_p(g_i))$
$n \leftarrow \min(0, n_c - \min(0, n_b))$    (* $n \leq 0$ *)
**if** $n_b = 0$ **then**    (* possible cancellation *)
  $\alpha\sqrt{-1} + \beta \leftarrow$ **Remainder**$(-b/a, p)$    (* $\alpha, \beta \in k$ *)
  $\eta \leftarrow Dt/(t^2+1)$    (* $\eta \in k$ *)
  **if** $2\beta = Dv/v$ for $v \in k^*$
    and $\alpha\sqrt{-1} + \beta = 2s\eta\sqrt{-1} + Dz/z$ for $z \in k(\sqrt{-1})^*$ and $s \in \mathbb{Z}$
  **then** $n \leftarrow \min(n, s)$
$N \leftarrow \max(0, -n_b)$    (* $N \geq 0$, for clearing denominators *)
**return**$(ap^N, (b + naDp/p)p^N, g_1 p^{N-n}, \ldots, g_m p^{N-n}, p^{-n})$

**常数上的线性约束**

作为之前部分的结论之一，我们将问题简化为找到(7.4)的解$c_1, \ldots, c_m \in \text{Const}(k)$和$q \in k[t]$。我们将方程改写为：

$$aDq + bq = c_1 g_1 + \cdots + c_m g_m \qquad (7.5)$$

其中$a, b \in k[t], g_1, \ldots, g_m \in k(t), a \neq 0$，且$t$是$k$上的一个单项式。另外，如果需要的话用$\gcd(a, b) = 1$除以(7.5)两端，不失一般性，我们可以假设$\gcd(a, b) = 1$。我们说明如果存在$g_i \notin k[t]$，那么我们得到在$c_i$上的线性约束，将(7.5)简化为一个相似的方程，方程右侧属于$k[t]$。

**引理 7.1.1.** 令$a, b, q \in k[t]$，$g_1, \ldots, g_m \in k(t)$，且$c_1, \ldots, c_m \in \text{Const}(k)$满足

$aDq + bq = c_1g_1 + \cdots + c_mg_m$。令$d_i$为$g_i$的分母，$1 \leq i \leq m, d = \text{lcm}(d_1, \ldots, d_m)$，且 $q_1, \ldots, q_m, r_1, \ldots, r_m$满足$dg_i = dq_i + r_i$，对于每一个$i$或者$r_i = 0$，或者$\deg(r_i) < \deg(d)$，那么

$$\sum_{i=1}^{m} c_i r_i = 0 \tag{7.6}$$

且
$$aDq + bq = c_1q_1 + \cdots + c_mq_m \tag{7.7}$$

将(7.6)左右两侧的$t$的幂指数的系数用方程列出，产生了一个$c_i$组成的齐次线性方程组，换言之，一个系数在$k$中的矩阵$M$满足

$$M \begin{pmatrix} c_1 \\ c_2 \\ \vdots \\ c_m \end{pmatrix} = 0 \tag{7.8}$$

**LinearConstraints**$(a, b, g_1, \ldots, g_m, D)$
(\* Generate linear constraints on the constants \*)

    (\* Given a derivation $D$ on $k(t)$, $a, b \in k[t]$ and $g_1, \ldots, g_m \in k(t)$, return $q_1, \ldots, q_m \in k[t]$ and a matrix $M$ with entries in $k(t)$ such that for any solution $c_1, \ldots, c_m \in \text{Const}(k)$ and $p \in k[t]$ of $aDp + bp = c_1g_1 + \ldots + c_mg_m$, $(c_1, \ldots, c_m)$ is a solution of $Mx = 0$, and $p$ and the $c_i$ satisfy $aDp + bp = c_1q_1 + \ldots + c_mq_m$. \*)

    $d \leftarrow \text{lcm}(\text{denominator}(g_1), \ldots, \text{denominator}(g_m))$
    **for** $i \leftarrow 1$ **to** $m$ **do**
        $(q_i, r_i) \leftarrow \textbf{PolyDivide}(dg_i, d)$         (\* $dg_i = q_i d + r_i$ \*)
    **if** $r_1 = \ldots = r_m = 0$ **then** $N = -1$ **else** $N \leftarrow \max(\deg(r_1), \ldots, \deg(r_m))$
    **for** $i \leftarrow 0$ **to** $N$ **do**
        **for** $j \leftarrow 1$ **to** $m$ **do** $M_{ij} \leftarrow \textbf{coefficient}(r_j, t^i)$
    **return**$(q_1, \ldots, q_m, M)$

因为我们只关心(7.8)的常数解，我们需将其简化为一个等价的方程组，其中系数属于$\text{Const}(k)$。化简的算法由以下引理给出。

引理 7.1.2.令$(K, D)$为一个微分域，$A$是一个系数在$K$中的矩阵，$u$是一个系数在$K$中的向量。那么对$A$和$u$只使用初等的行变换，我们或者可以证明$Ax = u$没有常数解，或者我们能够计算出一个矩阵$B$和一个向量$v$，系数都在$\text{Const}(K)$中，满足$Ax = u$的常数解正好是$Bx = v$的所有常数解。此外，如果$u = 0$，那么$v = 0$。

```
ConstantSystem(M, u, D)
(* Generate a system for the constant solutions *)

  (* Given a differential field (K, D) with constant field C, a matrix A and
  a vector u with coefficients in K, returns a matrix B with coefficients in
  C and a vector v such that either v has coefficients in C, in which case
  the solutions in C of Ax = u are exactly all the solutions of Bx = v, or
  v has a nonconstant coefficient, in which case Ax = u has no constant
  solution. *)

  (A, u) ← RowEchelon(A, u)
  m ← number of rows of A
  while A is not constant do
      j ← minimal index such that the j^th column of A is not constant
      i ← any index such that a_{ij} ∉ C,
      R_i ← i^th row of A
      R_{m+1} = D(R_i)/D(a_{ij}), u_{m+1} ← D(u_i)/D(a_{ij})
      for s ← 1 to m do
          R_s ← R_s − a_{sj}R_{m+1}
          u_s ← u_s − a_{sj}u_{m+1}
      A ← A ∪ R_{m+1}, u ← u ∪ u_{m+1}          (* vertical concatenation *)
  return(A, u)
```

使用引理 7.1.1. 和引理 7.1.2. ，我们能够得到一个$c_i$的齐次线性方程组。如果矩阵的核为0维，那么(7.5)的唯一解为$q = c_1 = \cdots = c_m = 0$。否则，核的一组基可以让我们用一些$c_i$表示出其它的$c_i$，因此减小$m$，将问题简化为解方程(7.7)。

次数限制

作为引理 7.1.1.的结论，我们将问题简化为找到(7.7)的解$c_1, \ldots, c_m \in \mathrm{Const}(k)$和$q \in k[t]$，其中$a, b, q_1, \ldots, q_m \in k[t], a \neq 0$，$t$是$k$上的一个单项式。因为$\deg(\sum_{i=1}^{m} c_i q_i) \leq \max_{1 \leq i \leq m}(\deg(q_i))$，和无参数情况下一样，引理 6.3.1.给出了$\deg(q)$的一个上界。

$(i)$如果$\deg(b) > \deg(a) + \max(0, \delta(t) - 1)$，那么
$$\deg(q) \leq \max(0, \max_{1 \leq i \leq m}(\deg(q_i) - \deg(b))$$
$(ii)$如果$\deg(b) < \deg(a) + \delta(t) - 1$，且$\delta(t) \geq 2$，那么
$$\deg(q) \leq \max(0, \max_{1 \leq i \leq m}(\deg(q_i) - \deg(a) + 1 - \delta(t))$$
由结果可知，我们只需要考虑刘维尔单项式，$\deg(b) \leq \deg(a)$这种情况，和非线性单项式，$\deg(b) = \deg(a) + \delta(t) - 1$这种情况。我们对于不同的单项式扩张的情况，分别进行讨论。

基本式情况

如果$Dt = \eta \in k$，那么和无参数情况相同，引理 6.3.1.和引理 6.3.3.总是给出$\deg(q)$的一个上界：

如果$\deg(b) > \deg(a)$，那么引理 6.3.1. 说明
$$\deg(q) \geq \max(0, \max_{1 \leq i \leq m}(\deg(q_i) - \deg(b))$$
如果$\deg(b) < \deg(a) - 1$，那么引理 6.3.3.说明
$$\deg(q) \geq \max(0, \max_{1 \leq i \leq m}(\deg(q_i) - \deg(a) + 1)$$

如果 $\deg(b) = \deg(a) - 1$，那么或者在以下情况，存在 $s \in \mathbb{Z}, u \in k$，使得 $-\mathrm{lc}(b)/\mathrm{lc}(a) = s\eta + Du$
$$\deg(q) \geq \max(0, s, \max_{1 \leq i \leq m}(\deg(q_i) - \deg(a) + 1)$$
或者
$$\deg(q) \geq \max(0, \max_{1 \leq i \leq m}(\deg(q_i) - \deg(a) + 1)$$
注意到根据引理 6.3.2.，这样的 $s$ 是唯一的。

最后，如果 $\deg(b) = \deg(a)$，那么在下述情况中
$$-\frac{\mathrm{lc}(b)}{\mathrm{lc}(b)} = \frac{Du}{u} \text{且} -\frac{\mathrm{lc}(aDu + bu)}{u\mathrm{lc}(a)} = s\eta + Dv$$
其中 $u \in k^*, v \in k$，且 $s \in \mathbb{Z}$
$$\deg(q) \geq \max(0, s, \max_{1 \leq i \leq m}(\deg(q_i) - \deg(a) + 1)$$
或者
$$\deg(q) \geq \max(0, \max_{1 \leq i \leq m}(\deg(q_i) - \deg(a) + 1)$$

我们可以通过一系列的积分算法计算出这样的 $u$。像前面在无参数情况下说明的一样，尽管 $u$ 不是唯一的，引理 5.12.1. 说明 $u$ 的选择不会影响 $\mathrm{lc}(aDu + bu)/(u\mathrm{lc}(a))$，所以根据引理 6.3.2.，$s$ 是唯一的。

---

**ParamRdeBoundDegreePrim**$(a, b, q_1, \ldots, q_m, D)$
(* Bound on polynomial solutions – primitive case *)

(* Given a derivation $D$ on $k[t]$ and $a, b, q_1, \ldots, q_m \in k[t]$ with $Dt \in k$ and $a \neq 0$, return $n \in \mathbb{Z}$ such that $\deg(q) \leq n$ for any solution $c_1, \ldots, c_m \in \mathrm{Const}(k)$ and $q \in k[t]$ of $aDq + bq = \sum_{i=1}^{m} c_i q_i$. *)

$d_a \leftarrow \deg(a), d_b \leftarrow \deg(b), d_c \leftarrow \max_{1 \leq i \leq m}(\deg(q_i))$
**if** $d_b > d_a$ **then** $n \leftarrow \max(0, d_c - d_b)$ **else** $n \leftarrow \max(0, d_c - d_a + 1)$
**if** $d_b = d_a - 1$ **then** (* possible cancellation *)
   $\alpha \leftarrow -\mathrm{lc}(b)/\mathrm{lc}(a)$
   **if** $\alpha = sDt + Dz$ for $z \in k$ and $s \in \mathbb{Z}$ **then** $n \leftarrow \max(n, s)$
**if** $d_b = d_a$ **then** (* possible cancellation *)
   $\alpha \leftarrow -\mathrm{lc}(b)/\mathrm{lc}(a)$
   **if** $\alpha = Dz/z$ for $z \in k^*$ **then**
     $\beta \leftarrow -\mathrm{lc}(aDz + bz)/(z\,\mathrm{lc}(a))$
     **if** $\beta = sDt + Dw$ for $w \in k$ and $s \in \mathbb{Z}$ **then** $n \leftarrow \max(n, s)$
**return** $n$

---

在特定情况 $D = d/dt$ 时，和无参数情况一样，推论 6.3.1. 产生了一个更为简单的算法。

```
ParamRdeBoundDegreeBase(a, b, q_1, ..., q_m)
(* Bound on polynomial solutions – base case *)

    (* Given a, b, q_1, ..., q_m ∈ k[t] with a ≠ 0, return n ∈ ℤ such that deg(q) ≤
    n for any solution c_1, ..., c_m ∈ k and q ∈ k[t] of a dq/dt + bq = ∑_{i=1}^m c_i q_i. *)
    d_a ← deg(a), d_b ← deg(b), d_c ← max_{1≤i≤m}(deg(q_i))
    n ← max(0, d_c − max(d_b, d_a − 1))
    if d_b = d_a − 1 then                              (* possible cancellation *)
        s ← −lc(b)/lc(a)
        if s ∈ ℤ then n ← max(0, s, d_c − d_b)
    return n
```

超越指数式情况

如果$Dt/t = \eta \in k$，那么引理 6.3.1.和引理 6.3.4.和无参数情况下一样，总是给出$\deg(q)$的一个上界：

如果$\deg(b) > \deg(a)$，那么引理 6.3.1. 说明
$$\deg(q) \geq \max(0, \max_{1 \leq i \leq m}(\deg(q_i) - \deg(b))$$

如果$\deg(b) < \deg(a)$，那么引理 6.3.4.说明
$$\deg(q) \geq \max(0, \max_{1 \leq i \leq m}(\deg(q_i) - \deg(a))$$

最后，如果$\deg(b) = \deg(a)$，那么或者在下述情况中$-lc(b)/lc(a) = s\eta + Du/u$其中$u \in k^*$，且$s \in \mathbb{Z}$
$$\deg(q) \geq \max(0, s, \max_{1 \leq i \leq m}(\deg(q_i) - \deg(b))$$

或者
$$\deg(q) \geq \max(0, \max_{1 \leq i \leq m}(\deg(q_i) - \deg(b))$$

注意到根据引理 6.2.2.，这样的$s$是唯一的。

```
ParamRdeBoundDegreeExp(a, b, q_1, ..., q_m, D)
(* Bound on polynomial solutions – hyperexponential case *)

    (* Given a derivation D on k[t] and a, b, q_1, ..., q_m ∈ k[t] with Dt/t ∈ k
    and a ≠ 0, return n ∈ ℤ such that deg(q) ≤ n for any solution c_1, ..., c_m ∈
    Const(k) and q ∈ k[t] of aDq + bq = ∑_{i=1}^m c_i q_i. *)
    d_a ← deg(a), d_b ← deg(b), d_c ← max_{1≤i≤m}(deg(q_i))
    n ← max(0, d_c − max(d_b, d_a))                    (* n ≥ 0 *)
    if d_a = d_b then                                  (* possible cancellation *)
        α ← −lc(b)/lc(a)
        if α = sDt/t + Dz/z for z ∈ k^* and s ∈ ℤ then n ← max(n, s)
    return n
```

非线性式情况

如果$\delta(t) \geq 2$，那么引理 6.3.1.和引理 6.3.5. 和无参数情况下一样总是给出$\deg(q)$的一个上界：

如果$\deg(b) \neq \deg(a) + \delta(t) - 1$，那么引理 6.3.1.像之前说明的一样给出了一个边界限制，否则，或者在下述情况中$-lc(b)/lc(a) = s\lambda(t)$，其中$s \in \mathbb{Z}$
$$\deg(q) \geq \max(0, s, \max_{1 \leq i \leq m}(\deg(q_i) - \deg(b))$$

或者
$$\deg(q) \geq \max(0, \max_{1 \leq i \leq m}(\deg(q_i) - \deg(b))$$

```
ParamRdeBoundDegreeNonLinear(a, b, q_1, ..., q_m, D)
(* Bound on polynomial solutions - nonlinear case *)

    (* Given a derivation D on k[t] and a, b, q_1, ..., q_m ∈ k[t] with deg(Dt) ≥ 2
    and a ≠ 0, return n ∈ ℤ such that deg(q) ≤ n for any solution c_1, ..., c_m ∈
    Const(k) and q ∈ k[t] of aDq + bq = Σ_{i=1}^m c_i q_i. *)

    d_a ← deg(a), d_b ← deg(b), d_c ← max_{1≤i≤m}(deg(q_i))
    δ ← deg(Dt), λ ← lc(Dt), n ← max(0, d_c - max(d_a + δ - 1, d_b))
    if d_b = d_a + δ - 1 then                    (* possible cancellation *)
        s ← -lc(b)/(λ lc(a))
        if s ∈ ℤ then n ← max(0, s, d_c - d_b)
    return n
```

**含参数的随机偏微分方程算法**

现在我们将问题简化为求(7.7)的解$c_1, ..., c_m \in \text{Const}(k)$和$q \in k[t]$,并且我们有$\deg(q)$的一个上界$n$。定理 6.4.1.和 6.4 节的随机偏微分方程算法可以推广到含参数的情况下。

**定理 7.1.2.** 令$a, b, q_1, ..., q_m \in k[t]$,其中$a \neq 0, \gcd(a, b) = 1$。令$z_1, ..., z_m, r_1, ..., r_m \in [t]$满足对于每一个$i$,$q_i = az_i + br_i$,或者$r_i = 0$,或者$\deg(r_i) < \deg(a)$,令$r = \sum_{i=1}^m c_i r_i$。那么,对于$aDq + bq = \sum_{i=1}^m c_i q_i$的任意解$x_1, ..., c_m \in \text{Const}(k)$和$q \in k[t]$,$p = (q - r)/a \in k[t]$,$p$是下述方程的解

$$aDq + (b + Da)p = c_1(z_1 - Dr_1) + \cdots + c_m(z_m - Dr_m) \quad (7.14)$$

反之,对于(7.14)的任意解$c_1, ..., c_m \in \text{Const}(k)$和$p \in k[t]$,$q = ap + r$是(7.7)的一个解。

定理 7.1.2 将方程(7.7)简化为(7.14),(7.14)和(7.7)类型一致。如果新方程的系数$a$和$b$有一个非平凡的最大公约数,那么我们用这个最大公约数去除新方程,得到一个和(7.5)类型一样的方程,再次应用线性约束算法,得到一个和(7.7)类型一致的新方程。但是,在所有情况下,如果(7.7)有一个次数为$n$的解$q$,那么因为$q = ap + r, \deg(r) < \deg(a)$,所以新方程相应的解的次数一定至多为$n - \deg(a)$。因此,如果$\deg(a) > 0$,我们可以使用定理 7.1.2.和引理 7.1.1.降低未知多项式的次数。我们可以重复知道$\deg(a) = 0$,换言之$a \in k^*$,在这种情况下我们用$a$去除方程两端,得到一个和(7.7)类型一致的方程,其中$a = 1$。

```
ParSPDE(a, b, q_1, ..., q_m, D, n)      (* Parametric SPDE algorithm *)

    (* Given a derivation D on k[t], an integer n and a, b, q_1, ..., q_m ∈ k[t]
    with deg(a) > 0 and gcd(a, b) = 1, return (ā, b̄, q̄_1, ..., q̄_m, r_1, ..., r_m, n̄)
    such that for any solution c_1, ..., c_m ∈ Const(k) and q ∈ k[t] of degree at
    most n of aDq + bq = c_1 q_1 + ... + c_m q_m, p = (q - c_1 r_1 - ... - c_m r_m)/a
    has degree at most n̄ and satisfies āDp + b̄p = c_1 q̄_1 + ... + c_m q̄_m. *)

    for i ← 1 to m do               (* br_i + az_i = q_i, deg(r_i) < deg(a) *)
        (r_i, z_i) ← ExtendedEuclidean(b, a, q_i)
    return(a, b + Da, z_1 - Dr_1, ..., z_m - Dr_m, r_1, ..., r_m, n - deg(a))
```

不可消去情况

现在我们将问题简化为找到下列方程的解 $c_1, \ldots, c_m \in \mathrm{Const}(k), q \in k[t]$

$$Dq + bq = \sum_{i=1}^{m} c_i q_i \qquad (7.16)$$

其中 $b, q_1, \ldots, q_m \in k[t]$,$t$ 是 $k$ 上的一个单项式。此外，我们有 $\deg(q)$ 的一个上界 $n$。像无参数情况下一样，引理 6.5.1. 给出了一个对于所有不可消去情况都适用的算法。

当 $\deg(b)$ 足够大时

假设 $b \neq 0$，而且或者 $D = d/dt$，或者 $\deg(b) > \max(0, \delta(t) - 1)$。那么，对于 (7.16) 的任意解 $q = y_n t^n + \cdots + y_0 \in k[t]$，引理 6.5.1. 说明 $\deg(Dq + bq) \leq n + \deg(b)$，且由方程两侧 $t^{n+\deg(b)}$ 的系数列出方程，得到

$$\mathrm{lc}(b) y_n = \sum_{i=1}^{m} \mathrm{coefficient}(q_i, t^{n+\deg(b)})$$

在 (7.16) 中用 $h + \sum_{i=1}^{m} c_i s_{in} t^n$ 替换 $q$，其中

$$s_{in} = \frac{\mathrm{coefficient}(q_i, t^{n+\deg(b)})}{\mathrm{lc}(b)} \in k \qquad (7.17)$$

我们得到

$$Dh + \sum_{i=1}^{m} c_i D(s_{in} t^n) + \sum_{i=1}^{m} c_i b s_{in} t^n + bh = \sum_{i=1}^{m} c_i q_i$$

这等价于

$$Dh + bh = \sum_{i=1}^{m} c_i (q_i - D(s_{in} t^n) - b s_{in} t^n)$$

这是和 (7.16) 类型一致的方程，其中 $b$ 保持一致。因此引理 6.5.1. 第一部分的假设再次满足，所以我们可以重复这个过程，但是 $\deg(h)$ 的边界变为 $n-1$。注意到尽管 $b$ 保持不变，(7.16) 的右端在每一步操作中都发生变化，所以我们必须再次计算在 (7.17) 中出现的 $q_i$。在整个过程中，每一次运算，$\deg(q)$ 的边界限制都会减小，保证了可以在有限步结束。在结束 $n = 0$ 的情况后，我们得到的初始方程在 $k[t]$ 中的任意解 $q$，满足 $\deg(q) \leq n$，其形式必须为 $q = \sum_{i=1}^{m} c_i h_i$，其中 $h_i = \sum_{j=0}^{n} s_{ij} t^j \in k[t]$。在 (7.16) 中用这种形式替换 $q$，附加原始的 $q_i$ 产生

$$\sum_{i=1}^{m} c_i (q_i - Dh_i - bh_i) = 0$$

左侧部分是 $k[t]$ 中的元素，所以将它的所有系数设为 0 产生了一个齐次线性方程组，形式为 $M(c_1, \ldots, c_m)^T = 0$，其中 $M$ 含有 $k$ 中的元素。同样的方程组也可以通过 $n = 0$ 时，$q_i$ 和方程 $\sum_{i=1}^{m} c_i (q_i - D s_{i0} - b s_{i0}) = 0$ 得到，并且这就是在下述算法中得到该方程组的方法。根据引理 7.1.2.，我们可以计算出一个等价的方程组，形式为 $A(c_1, \ldots, c_m)^T = 0$，其中 $A$ 含有 $\mathrm{Const}(k)$ 中的元素。初始问题的解随之变为 $q = \sum_{i=1}^{m} d_i h_i$，其中附加方程 $d_i = c_i, 1 \leq i \leq m$ 被添加到 $A$ 中，换言之，一个 $m \times 2m$，形式如下的矩阵块

$$\begin{pmatrix} 1 & 0 & \cdots & \cdots & 0 & -1 & 0 & 0 & \cdots & 0 \\ 0 & 1 & 0 & \cdots & \cdots & 0 & -1 & 0 & \cdots & 0 \\ \vdots & \ddots & \ddots & \ddots & & & & \ddots & \ddots & \vdots \\ \vdots & & \ddots & \ddots & \ddots & & & & \ddots & \vdots \\ 0 & \cdots & \cdots & 0 & 1 & 0 & \cdots & \cdots & 0 & -1 \end{pmatrix} \tag{7.18}$$

被附加到 $A$ 的下部，同时也在其右侧附加零矩阵块。最终的线性约束方程组的形式为 $A(c_1,\ldots,c_m,d_1,\ldots,d_m)^T=0$。

```
ParamPolyRischDENoCancel1(b, q_1, ..., q_m, D, n)
(* Parametric Poly Risch d.e. – no cancellation *)

(* Given a derivation D on k[t], n ∈ ℤ and b, q_1, ..., q_m ∈ k[t] with b ≠ 0
and either D = d/dt or deg(b) > max(0, δ(t) − 1), returns h_1, ..., h_r ∈ k[t]
and a matrix A with coefficients in Const(k) such that if c_1, ..., c_m ∈
Const(k) and q ∈ k[t] satisfy deg(q) ≤ n and Dq + bq = Σ_{i=1}^m c_i q_i then q =
Σ_{j=1}^r d_j h_j where d_1, ..., d_r ∈ Const(k) and A(c_1, ..., c_m, d_1, ..., d_r)^T =
0. *)

d_b ← deg(b), b_d ← lc(b)
for i ← 1 to m do h_i ← 0
while n ≥ 0 do
    for i ← 1 to m do
        s_i ← coefficient(q_i, t^{n+d_b})/b_d
        h_i ← h_i + s_i t^n
        q_i ← q_i − D(s_i t^n) − b s_i t^n
    n ← n − 1
(* The remaining linear constraints are Σ_{i=1}^m c_i q_i = 0 *)
if q_1 = ... = q_m = 0 then d_c ← −1 else d_c ← max_{1 ≤ i ≤ m}(deg(q_i))
for i ← 0 to d_c do for j ← 1 to m do M_{i+1,j} ← coefficient(q_j, t^i)
(A, u) ← ConstantSystem(M, 0)                     (* u = 0 *)
(* Add the constraints c_i − d_i = 0 for 1 ≤ i ≤ m *)
n_{eq} ← number of rows(A)
for i ← 1 to m do A_{i+n_{eq}, i} ← 1, A_{i+n_{eq}, m+i} ← −1
return(h_1, ..., h_m, A)
```

当 $\deg(b)$ 足够小时

假设 $\deg(b) < \delta(t) - 1$，或者 $D = d/dt$，这说明 $b = 0$，或者 $\delta(t) \geq 2$。令 $q = y_n t^n + \cdots + y_0 \in k[t]$ 为(7.16)的一个解。

如果 $n > 0$，那么引理 6.5.1. 说明 $\deg(Dq+bq) \leq n+\delta(t)-1$，以左右两端 $t^{n+\delta(t)-1}$ 的系数列方程，得到

$$n\lambda(t) y_n = \sum_{i=1}^m c_i \text{coefficient}(q_i, t^{n+\delta(t)-1})$$

在(7.16)中用 $h + \sum_{i=1}^m c_i s_{in} t^n$ 替换 $q$，其中

$$s_{in} = \frac{\text{coefficient}(q_i, t^{n+\delta(t)-1})}{n\lambda(t)} \in k \tag{7.19}$$

我们得到

$$Dh + bh = \sum_{i=1}^{m} c_i(q_i - D(s_{in}t^n) - bs_{in}t^n)$$

这是和(7.16)相同类型的一个方程，且$b$相同。因此引理 6.5.1.的第二部分假设再次满足，所以我们可以重复这个过程，但是$\deg(h)$的边界变为$n-1$。注意到尽管$b$保持不变，(7.16)的右端在每一步操作中都发生变化，所以我们必须再次计算在(7.19)中出现的$q_i$。在整个过程中，每一次运算，$\deg(q)$的边界限制都会减小，直到$n=0$。换言之，我们寻找在$k$中的解$q=y_0$。在这一点上，在$\deg(b)>0$和$b\in k$两种情况下算法运行不同。

如果$\deg(b)>0$，那么以方程两端$t^{\deg(b)}$的系数列方程，产生

$$\text{lc}(b)y_0 = \sum_{i=1}^{m} c_i\text{coefficient}(q_i, t^{\deg(b)})$$

所以任意解$y_0 \in k$一定具有$y_0 = \sum_{i=1}^{m} c_i s_{i0}$这种形式，其中

$$s_{i0} = \frac{\text{coefficient}(q_i, t^{\deg(b)})}{\text{lc}(b)} \in k$$

这说明原始方程的任意解$q \in k[t]$，满足$\deg(q) \leq n$，一定有$q = \sum_{i=1}^{m} c_i h_i$这种形式，其中

$$h_i = \sum_{j=0}^{n} s_{ij} t^j \in k[t]$$

在(7.16)中用这种形式替换$q$，$q_i$保持不变，我们得到

$$\sum_{i=1}^{m} c_i(q_i - Dh_i - bh_i) = 0$$

正像之前我们所看到的一样，这个可以转为一个齐次线性方程组，形式为$A(c_1,\ldots,c_m)^T = 0$，其中$A$含有$\text{Const}(k)$中的元素。原始方程的解为$q = \sum_{i=1}^{m} d_i h_i$，其中像之前一样附加方程$d_i = c_i, 1 \leq i \leq m$添加到$A$中。最终的线性约束方程组为$A(c_1,\ldots,c_m,d_1,\ldots,d_m)^T = 0$。

如果$b \in k$，那么(7.16)的任意解$y_0 \in k$满足

$$Dy_0 + by_0 = \sum_{i=1}^{m} c_i q_i(0) \qquad (7.20)$$

这是一个$k$上的类型与(7.1)一致的含参数的$Risch$微分方程。假设我们可以在$k$上解决这样的问题，我们得到$f_1,\ldots,f_r \in k$和系数属于$\text{Const}(k)$的矩阵$B$，满足(7.20)的任意解$y_0 \in k$形式为

$$y_0 = \sum_{j=1}^{r} d_j f_j$$

其中$d_1,\ldots,d_j \in \text{Const}(k), B(c_1,\ldots,c_m,d_1,\ldots,d_r)^T = 0$。这说明原始方程满足$\deg(q) \leq n$的解$q \in k[t]$的形式一定为$q = \sum_{j=1}^{r} d_j f_j + \sum_{i=1}^{m} c_i h_i$，其中

$$h_i = \sum_{j=1}^{n} s_{ij} t^j \in k[t]$$

在(7.16)中用这种形式替换$q$,和原始的$q_i$产生

$$\sum_{i=1}^{m} c_i(q_i - Dh_i - bh_i) - \sum_{j=1}^{r}(Df_j + bf_j) = 0$$

用在前面情况中使用的相似的方法,这个可以转换为一个齐次线性方程组,形式为$A(c_1, \ldots, c_m, d_1, \ldots, d_r)^T = 0$,其中$A$含有$\text{Const}(k)$中的元素。初始问题的解变为

$$q = \sum_{j=1}^{r} d_j f_j + \sum_{i=1}^{m} e_i h_i$$

其中附加方程$B(c_1, \ldots, c_m, d_1, \ldots, d_r)^T = 0$添加到$A$中,同时也将方程$e_i = c_i, q \leq i \leq m$附加上去。最终的线性约束方程组为$A(c_1, \ldots, c_m, d_1, \ldots, d_r, e_1, \ldots, e_m)^T = 0$。

---

**ParamPolyRischDENoCancel2**$(b, q_1, \ldots, q_m, D, n)$
(* Parametric Poly Risch d.e. – no cancellation *)

(* Given a derivation $D$ on $k[t]$, $n \in \mathbb{Z}$ and $b, q_1, \ldots, q_m \in k[t]$ with $\deg(b) < \delta(t) - 1$ and either $D = d/dt$ or $\delta(t) \geq 2$, returns $h_1, \ldots, h_r \in k[t]$ and a matrix $A$ with coefficients in $\text{Const}(k)$ such that if $c_1, \ldots, c_m \in \text{Const}(k)$ and $q \in k[t]$ satisfy $\deg(q) \leq n$ and $Dq + bq = \sum_{i=1}^{m} c_i q_i$ then $q = \sum_{j=1}^{r} d_j h_j$ where $d_1, \ldots, d_r \in \text{Const}(k)$ and $A(c_1, \ldots, c_m, d_1, \ldots, d_r)^T = 0$. *)

$\delta \leftarrow \delta(t), \lambda \leftarrow \lambda(t)$
**for** $i \leftarrow 1$ **to** $m$ **do** $h_i \leftarrow 0$
**while** $n > 0$ **do**
　　**for** $i \leftarrow 1$ **to** $m$ **do**
　　　　$s_i \leftarrow \textbf{coefficient}(q_i, t^{n+\delta-1})/(n\lambda)$
　　　　$h_i \leftarrow h_i + s_i t^n, q_i \leftarrow q_i - D(s_i t^n) - bs_i t^n$
　　$n \leftarrow n - 1$
**if** $\deg(b) > 0$ **then**
　　**for** $i \leftarrow 1$ **to** $m$ **do**
　　　　$s_i \leftarrow \textbf{coefficient}(q_i, t^{\deg(b)})/\text{lc}(b)$
　　　　$h_i \leftarrow h_i + s_i, q_i \leftarrow q_i - Ds_i - bs_i$
　　**if** $q_1 = \ldots = q_m = 0$ **then** $d_c \leftarrow -1$ **else** $d_c \leftarrow \max_{1 \leq i \leq m}(\deg(q_i))$
　　**for** $i \leftarrow 0$ **to** $d_c$ **do for** $j \leftarrow 1$ **to** $m$ **do** $M_{i+1,j} \leftarrow \textbf{coefficient}(q_j, t^i)$
　　$(A, u) \leftarrow \textbf{ConstantSystem}(M, 0)$　　　　　　　　　(* $u = 0$ *)
　　$n_{eq} \leftarrow$ number of rows$(A)$
　　**for** $i \leftarrow 1$ **to** $m$ **do** $A_{i+n_{eq},i} \leftarrow 1, A_{i+n_{eq},m+i} \leftarrow -1$
　　**return**$(h_1, \ldots, h_m, A)$
**else**　　　　　　　　　　　　　　　　　　　　　　　　(* $b \in k$ *)
　　$(f_1, \ldots, f_r, B) \leftarrow \textbf{ParamRischDE}(b, q_1(0), \ldots, q_m(0))$
　　**if** $q_1 = \ldots = q_m = 0$ **then**
　　　　**if** $Df_1 + bf_1 = \ldots = Df_r + bf_r = 0$ **then** $d_c \leftarrow -1$ **else** $d_c \leftarrow 0$

```
else d_c ← max_{1≤i≤m}(deg(q_i))
for i ← 0 to d_c do for j ← 1 to m do M_{i+1,j} ← coefficient(q_j, t^i)
for j ← 1 to r do M_{1,j+m} ← -Df_j - bf_j
(A, u) ← ConstantSystem(M,0)                    (* u = 0 *)
A ← A ∪ B                                        (* vertical concatenation *)
(* Add the constraints c_i - e_i = 0 for 1 ≤ i ≤ m *)
n_eq ← number of rows(A)
for i ← 1 to m do A_{i+n_eq,i} ← 1, A_{i+n_eq,m+r+i} ← -1
return(f_1, ..., f_r, h_1, ..., h_m, A)
```

当 $\delta(t) \geq 2$, 且 $\deg(b) = \delta(t) - 1$

在这种情况下，我们只有当 $\deg(q) = -\mathrm{lc}(b)/\lambda(t)$ 时可以消去，这特别说明了 $-\mathrm{lc}(b)/\lambda(t)$ 是1和我们的次数限制 $n$ 之间的一个整数。令 $q = y_n t^n + \cdots + y_0 \in k[t]$ 为(7.16)的一个解。如果 $n \neq -\mathrm{lc}(b)/\lambda(t)$，那么引理 6.5.1. 说明 $\deg(Dq + bq) \leq n + \delta(t) - 1$，方程两端 $t^{n+\deg(b)}$ 的系数产生方程

$$(n\lambda(t) + \mathrm{lc}(b))y_n = \sum_{i=1}^{m} c_i \mathrm{coefficient}(q_i, t^{n+\delta(t)-1})$$

在(7.16)中用 $h + \sum_{i=1}^{m} c_i s_{in} t^{n+\delta(t)-1}$ 替换 $q$，其中

$$s_{in} = \frac{\mathrm{coefficient}(q_i, t^{n+\delta(t)-1})}{n\lambda(t) + \mathrm{lc}(b)} \in k \tag{7.21}$$

我们得到

$$Dh + bh = \sum_{i=1}^{m} c_i(q_i - D(s_{in}t^n) - bs_{in}t^n)$$

这是和(7.16)类型相同的等价方程，$b$ 不变，但 $\deg(h)$ 的边界变为 $n-1$。只要次数边界不合 $-\mathrm{lc}(b)/\lambda(t)$ 相等，引理 6.5.1.的第三部分假设再次满足，所以我们可以重复这个过程直到或者我们完成 $n = 0$ 这种情况，或者我们打到 $n = -\mathrm{lc}/\lambda(t)$ 这种情况。

如果我们完成 $n = 0$ 这种情况，那么原始方程的任意解 $q \in k[t]$，且满足 $\deg(q) \leq n$，一定有 $q = \sum_{i=1}^{m} c_i h_i$ 这种形式，其中

$$h_i = \sum_{j=0}^{n} s_{ij} t^j \in k[t]$$

在(7.16)中用这种形式替换 $q$，$q_i$ 保持不变，我们得到

$$\sum_{i=1}^{m} c_i(q_i - Dh_i - bh_i) = 0$$

正像之前我们所看到的一样，这个可以转为一个齐次线性方程组，形式为 $A(c_1, \ldots, c_m)^T = 0$，其中 $A$ 含有 $\mathrm{Const}(k)$ 中的元素。原始方程的解为 $q = \sum_{i=1}^{m} d_i h_i$，其中像之前一样附加方程 $d_i = c_i, 1 \leq i \leq m$ 添加到 $A$ 中。

如果我们要处理 $n = -\mathrm{lc}(b)/\lambda(t) > 0$ 这种情况，那么下一小节针对消去情况的算法生成

$f_1, \ldots, f_r \in k[t]$和一个矩阵$B$，矩阵$B$的系数在$\text{Const}(k)$中，使得任意次数至多为$n$的解$q \in k[t]$，一定具有下述形式

$$q = \sum_{j=1}^{r} d_j f_j$$

其中$d_1, \ldots, d_j \in \text{Const}(k)$, $B(c_1, \ldots, c_m, d_1, \ldots, d_r)^T = 0$。这说明原始方程的解一定具有下述形式

$$q = \sum_{j=1}^{r} d_j f_j + \sum_{i=1}^{m} c_i h_i$$

其中

$$h_i = \sum_{j=1-\text{lc}(b)/\lambda(t)}^{n} s_{ij} t^j \in k[t]$$

在(7.16)中用这种形式替换$q$，和原始的$q_i$生成

$$\sum_{i=1}^{m} c_i(q_i - Dh_i - bh_i) - \sum_{j=1}^{r} d_j(Df_j + bf_j) = 0$$

真如我们之前看到的一样，这等价于一个齐次线性方程组，形式为$A(c_1, \ldots, c_m, d_1, \ldots, d_r)^T = 0$，其中$A$含有$\text{Const}(k)$中的项。原始方程的解变为

$$q = \sum_{j=1}^{r} d_j f_j + \sum_{i=1}^{m} e_i h_i$$

其中附加方程$B(c_1, \ldots, c_m, d_1, \ldots, d_r)^T = 0$被添加到$A$上，同时添加的还有方程$e_i = c_i, 1 \leq i \leq m$。

如果$\deg(q) > 0$，且$\deg(q) \neq -\text{lc}(b)/\lambda(t)$，那么$\deg(q) + \delta(t) - 1 = \deg(c)$，所以$\deg(q) = \deg(c) + 1 - \delta(t)$，且$(\deg(q)\lambda(t) + \text{lc}(b))\text{lc}(q) = \text{lc}(c)$。这生成了$q$的第一单项式$ut^n$，在方程中用$ut^n + h$替换$q$，得到一个相似方程，解的次数限制更小。只要新的次数限制$\deg(q) \neq -\text{lc}(b)/\lambda(t)$，我们就可以重复这一过程。

如果$q \in k$，那么$Dq + bq$的首项为$q\text{lc}(b)t^{\delta(t)-1}$，所以或者在$q = \text{lc}(c)/\text{lc}(b)$为唯一可能解的情况下，$\deg(c) = \delta(t) - 1$，或者$\deg(c) \neq \delta(t) - 1$，且(6.19)在$k$中无解，因此在$k[t]$中无解。

可消去情况
最终我们研究方程(7.16)，只要不可消去情况不成立，换言之，以下情况之一成立：
1. $\delta(t) \leq 1, b \in k$且$D \neq d/dt$
2. $\delta(t) \geq 2, \deg(b) = \delta(t) - 1$，且$\deg(q) = -\text{lc}(b)/\lambda(t)$

刘维尔情况
如果$D \neq d/dt$，且$Dt \in k$或$Dt/t \in k$，那么$\delta(t) \leq 1$，所以唯一一种引理 6.5.1.不能处理的情况为$b \in k$。那么对于任意(7.16)的解$q = y_n t^n + \cdots + y_0 \in k[t]$，引理 5.1.2.说明$\deg(Dq + bq) \leq n$，用方程两端$t^n$的系数列方程得到：

$$Dy_n + by_n = \sum_{i=1}^{m} c_i \text{coefficient}(q_i, t^n) \tag{7.22}$$

如果$Dt \in k$，且

$$Dy_n + \left(b + n\frac{Dt}{t}\right)y_n = \sum_{i=1}^{m} c_i \text{coefficient}(q_i, t^n) \quad （7.23）$$

如果$Dt/t \in k$，(7.22)和(7.23)都是在$k$上和(7.1)类型相同的含参数的$Risch$微分方程问题。假设我们可以在$k$上解决这样的问题，我们得到$f_{1n}, \ldots, f_{2n} \in k$和矩阵$A_n$，其系数在$\text{Const}(k)$中，$y_n$具有下述形式

$$y_n = \sum_{j=1}^{r_n} d_{jn} f_{jn}$$

其中$d_{1n}, \ldots, d_{r_n,n} \in \text{Const}(k)$，且$A_n(c_1, \ldots, c_m, d_{1n}, \ldots, d_{r_n,n})^T = 0$。在(7.16)中用$h + \sum_{j=1}^{r_n} d_{jn} f_{jn} t^n$替换$q$，我们得到

$$Dh + bh = \sum_{i=1}^{m} c_i q_i - \sum_{j=1}^{r} d_{jn}(D(f_{jn}t^n) - bf_{jn}t^n)$$

这是一个和(7.16)相同类型的方程，且$b$保持不变。因此，我们可以重复这个过程，但是$\deg(h)$的边界变为$n-1$。注意到尽管$b$保持不变，但是(7.16)的右端在每一次变换下都会改变，所以必须再次计算(7.22)或(7.23)中出现的$q_i$。也注意到在右端不确定的常数的数目在每一步结束后都会增多。整个过程中$\deg$的边界在每一步中都会减小，这保证了能够在有限步后运行结束。在结束$n = 0$这种情况后，我们得到原始方程在$k[t]$中的任意解$q$，且满足$\deg(q) \leq n$一定具有下述形式

$$q = \sum_{i=0}^{n} \sum_{j=1}^{r_i} d_{ji} h_{ji} \quad （7.24）$$

其中$h_{ji} = f_{ji} t^i$
在(7.16)中用这种形式替换$q$，和原始的$q_i$产生下列方程

$$\sum_{i=1}^{m} c_i q_i - \sum_{i=0}^{n} \sum_{j=1}^{r_i} d_{ji}(Dh_{ji} - bh_{ji}) = 0.$$

正如我们在不可消去情况下看到的一样，这个方程可以转化为一个齐次线性方程组，形式为$A(c_1, \ldots, c_m, d_{11}, \ldots, d_{r_n,n})^T = 0$，其中$A$有在$\text{Const}(k)$中的元素。初始问题的解由(7.24)提供，其中附加方程$A_i(c_1, \ldots, c_m, d_{1i}, \ldots, d_{r_n,i})^T = 0, 0 \leq i \leq n$添加到$A$上。

非线性情况

如果$\delta(t) \geq 2$，那么我们一定有$\deg(b) = \delta(t) - 1, \text{lc}(b) = -n\lambda(t)$，其中$n > 0$为$\deg(q)$的边界。和无参数情况一致，在这种情况下，不存在解方程(7.16)的一般性算法。但是如果$\mathcal{S}^{irr} \neq \emptyset$，那么和无参数情况一致，将(7.16)映射到$k[t]/(p), p \in \mathcal{S}^{irr}$是可以实现的。因为$k[t]/(p)$是$k$的一个有限代数扩张，根据推论 3.3.1. $\text{Const}(k[t]/(p))$是$\text{Const}(k)$的一个有限代数扩张，所以令$b_1, \ldots, b_s$是$\text{Const}(k[t]/(p))$在$\text{Const}(k)$上的一组向量基。现在，$D^*$是$k[t]/(p)$上的诱导微分，我们得到

$$D^* q^* + \pi_p(b) q^* = \sum_{i=1}^{m} c_i \pi_p(q_i) \quad （7.25）$$

其中$q^* = \pi_p(q)$。假设我们有在$k[t]/(p)$中解(7.25)的算法，我们得到$h_1, \ldots, h_r \in k[t]/(p)$和

一个矩阵$B$，其中矩阵的系数在$\text{Const}(k[t]/(p))$中满足(7.25)在$k[t]/(p)$中的任意解一定具有$q^* = \sum_{j=1}^{r} d_j h_j$这样的形式，其中$d_1, \ldots, d_r \in \text{Const}(k[t]/(p))$，同时$B(c_1, \ldots, c_m, d_1, \ldots, d_r)^T = 0$。将常数$d_1, \ldots, d_r$和$B$的元素关于基$b_1, \ldots, b_s$展开，我们得到矩阵$A$，$A$的系数在$\text{Const}(k)$中，满足方程组$B(c_1, \ldots, c_m, d_1, \ldots, d_r)^T = 0$和$A(c_1, \ldots, c_m, d_{11}, \ldots, d_{rs})^T = 0$是等价的，其中

$$d_j = \sum_{l=1}^{s} d_{jl} b_l, 1 \leq j \leq r$$

(7.25)在$k[t]/(p)$中任意解一定具有下述形式

$$q^* = \sum_{j=1}^{r} \left( \sum_{l=1}^{s} d_{jl} b_l \right) h_j = \sum_{j=1}^{r} \sum_{l=1}^{s} d_{jl} h_{jl} \tag{7.26}$$

其中$h_{jl} = b_l h_j \in k[t]/(p)$。对于每一个$j$和$l$，令$r_{jl} \in k[t]$满足$\deg(r_{jl}) < \deg(p)$和$\pi_p(r_{jl}) = h_{jl}$，令

$$u = \sum_{j=1}^{r} \sum_{l=1}^{s} d_{jl} r_{jl} \in k[t]$$

我们得到$\deg(u) < \deg(p)$，(7.26)说明$\pi_p(q) = \pi_p(u)$，因此$h = (q-u)/p \in k[t]$，在(7.16)中用$ph+u$代替$q$，我们得到

$$\sum_{i=1}^{m} c_i q_i = Dq + bq = p\left( Dh + \left( b + \frac{Dp}{p} \right) h \right) + Du + bu$$

所以$h$是下述方程在$k[t]$中的解，次数至多为$\deg(q) - \deg(p)$

$$Dh + \left( b + \frac{Dp}{p} \right) h = \frac{\sum_{i=1}^{m} c_i q_i - \sum_{j=1}^{r} \sum_{l=1}^{s} d_{jl}(Dr_{jl} + br_{jl})}{p}. \tag{7.27}$$

现在将$q_i$写作$q_i = p\overline{q_i} + \hat{q_i}$，将$Dr_{jl} + br_{jl}$写作$Dr_{jl} + br_{jl} = p\overline{r_{jl}} + \hat{r_{jl}}$其中$\hat{q_i}, \hat{r_{jl}} \in k[t], \deg(\hat{q_i}) < \deg(p)$，且$\deg(\hat{r_{jl}}) < \deg(p)$，(7.27)的右端变为

$$\frac{\sum_{i=1}^{m} c_i q_i - \sum_{j=1}^{r} \sum_{l=1}^{s} d_{jl}(Dr_{jl} + br_{jl})}{p} =$$

$$\sum_{i=1}^{m} c_i \overline{q_i} - \sum_{j=1}^{r} \sum_{l=1}^{s} d_{jl} \overline{r_{jl}} + \frac{\sum_{i=1}^{m} c_i \hat{q_i} - \sum_{j=1}^{r} \sum_{l=1}^{s} d_{jl} \hat{r_{jl}}}{p}.$$

因为$\pi_p(u) = q^*$是(7.25)的一个解，我们有

$$0 = \pi_p\left( \sum_{i=1}^{m} c_i q_i - Du - bu \right) = \pi_p\left( \sum_{i=1}^{m} c_i \hat{q_i} - \sum_{j=1}^{r} \sum_{l=1}^{s} d_{jl} \hat{r_{jl}} \right)$$

因为$\deg(\hat{q_i}) < \deg(p)$，且$\deg(\hat{r_{jl}}) < \deg(p)$，且满足下列等式

$$\sum_{i=1}^{m} c_i \hat{q_i} - \sum_{j=1}^{r} \sum_{l=1}^{s} d_{jl} \hat{r_{jl}} = 0$$

所以(7.27)变为

$$Dh + \left(b + \frac{Dp}{p}\right)h = \sum_{i=1}^{m} c_i \overline{q_i} - \sum_{j=1}^{r}\sum_{l=1}^{s} d_{jl}\overline{r_{jl}}$$

超切向量式情况

如果$Dt/(t^2+1) = \eta \in k$，那么$\delta(t) = 2$，所以引理 6.5.1.唯一不能处理的情况为$b = b_0 - n\eta t$，其中$b_0 \in k, n > 0$为$\deg(q)$的边界。在这样的扩张中，上面列出的方法给出了一个完整的算法：令$p = t^2 + 1 \in \mathcal{S}^{irr}$，(7.25)变为

$$Dq^* + (b_0 - n\eta\sqrt{-1})q^* = \sum_{i=1}^{m} c_i q_i(\sqrt{-1}) \qquad (7.28)$$

其中$D$通过$D\sqrt{-1} = 0$扩张到$k[t]/(p) \simeq k(\sqrt{-1})$，一种可能性是将(7.28)看做$k(\sqrt{-1})$的一个含参数的$Risch$微分方程，并用递归式的方法解它。在将结果关于基$\{1, \sqrt{-1}\}$扩张后（只要$\sqrt{-1} \notin k$），我们得到$h_1, \ldots, h_r \in k(\sqrt{-1})$和一个矩阵$A$，$A$的元素在$\text{Const}(k)$中，使得(7.28)在$k(\sqrt{-1})$中的所有解的形式都是$q^* = \sum_{j=1}^{r} d_j h_j$，其中$d_j \in \text{Const}(k)$，$A(c_1, \ldots, c_m, d_1, \ldots, d_r)^T = 0$。对于每一个$j$, $h_{j0}, h_{j1} \in k$，将$h_j$记作$h_j = h_{j0} + h_{j1}\sqrt{-1}$。接下来，令$r_j = h_{j0} + h_{j1}t \in k[t]$，$\overline{q_i}$为$q_i$除以$p$的商，$\overline{r_j}$为$Dr_j + br_j$除以$p$的商，$h = (q - \sum_{j=1}^{m} d_j r_j)/p$为下列方程在$k[t]$中次数至多为$n - 2$的一个解。

$$Dh + (b_0 - (n-2)\eta t)h = \sum_{i=1}^{m} c_i \overline{q_i} - \sum_{j=1}^{r} d_j \overline{r_j}.$$

如果$\sqrt{-1} \notin k$，那么通过考虑(7.28)的实部和虚部可以避免引入$\sqrt{-1}$：将$q^*$记为$q^* = u + v\sqrt{-1}$，我们得到

$$\begin{pmatrix} Du \\ Dv \end{pmatrix} + \begin{pmatrix} b_0 & n\eta \\ -n\eta & b_0 \end{pmatrix} \begin{pmatrix} u \\ v \end{pmatrix} = \sum_{i=1}^{m} c_i \begin{pmatrix} q_{i0} \\ q_{i1} \end{pmatrix} \qquad (7.29)$$

其中$q_{i0} + q_{i1}t$是$q_i$除以$t^2 + 1$的余数。这是在5.10节提到的耦合微分系统问题的含参数版本，第八章的算法可以以在这章中处理$Risch$微分方程相似的方法将算法扩展到含参数的情况下。

### 7.2 受限积分问题

在这一部分我们描述首先积分问题的一个解，换言之，给定一个特征为0的微分域$K$，$f, w_1, \ldots, w_m \in K$，判定是否存在常数$c_1, \ldots, c_m \in \text{Const}(K)$，满足
$$f = Dv + c_1 w_1 + \cdots + c_m w_m \qquad (7.30)$$
有一个解$v \in K$，如果有解的话找到这样的一个解。正如我们在第五章中看到的一样，这个问题的背景是在基本扩张中对多项式做积分。有以下几条可能的途径解决这个问题：

1.如果所有的$w_i$都是$K$中元素的对数微分，那么(7.30)的一个解的存在说明$f$在$K$的一个初等扩张中存在不定积分，所以方程(7.30)可以看做是一个初等不定积分问题，第五章的算法可以使用，并进一步地用$w_i$尝试重写积分结果。

2.方程(7.30)可以看做是一个关于$v$的含参数的$Risch$微分方程问题，可以通过7.1小节的算法解决。

只有当对初等函数做积分时，第一种方法是可行的，因为在被积函数中唯一的基本单项式为对数，实际上这是$Risch$最初使用的方法，并且出现在在大多数计算机代数系统和文献中。但是怎样将一个初等不定积分以$w_i$的形式重写从来没有一个明确的算法，并且不断地有新的困难和复杂度出现。第二种方法对于任意的$w_i$都是适用的，所以它允许被积函数中出现任意的基本式。此外，对于一些非初等函数，例如$Erf, Ei, Li$和二重对数的算法，首先产生候选的特殊函数，接下来对于这些特殊函数解决受限积分问题。因为这些优势，我们在这里使用这一种方法。但是含参数的$Risch$微分方程算法可以依据以下事实进行简化：在$v$的极点处（包括无穷处）的消去不能发生，因为只有$Dv$出现在方程中，没有$v$的倍数出现。所以限制阶数和次数的难度显著变小。在这一部分我们提供一种7.1小节算法的简化版本，这个版本充分利用了这个事实。

我们只在超越情况下研究方程(7.30)，换言之，当$K$是一个微分子域$k$的一个$simple$（简单）单项式扩张，所以在这一小节的剩余部分中，令$k$是一个特征为0的微分域，$t$是$k$上的一个单项式，另外我们可以假设$\mathrm{Const}(k(t)) = \mathrm{Const}(k)$。我们假设我们方程的系数$f$和$w_1, \ldots, w_m$属于$k(t)$，寻找解$c_1, \ldots, c_m \in \mathrm{Const}(k)$和$v \in k(t)$。

因为(7.30)的特殊形式，定义 7.1.1.可以加强到不仅仅是分母的$normal$（普通）部分，只要$\mathcal{S}_1^{irr} = \mathcal{S}^{irr}$，对于$special$（特殊）部分也适用，并且次数有限制，只要$t$是一个刘维尔单项式或是一个非线性单项式。

定理 7.2.1. 令 $v, f, w_1, \ldots, w_m \in k(t), c_1, \ldots, c_m \in \mathrm{Const}(k)$ 满足 $f = Dv + c_1 w_1 + \cdots + c_m w_m$。令$d = d_s d_n$为$f$分母的一个分裂分解，$e_i = e_{s,i} e_{n,i}$为$w_i$的分母的一个分裂分解。令$c = \mathrm{lcm}(d_n, e_{n,1}, \ldots, e_{n,m})$，$h_s = \mathrm{lcm}(d_s, e_{s,1}, \ldots, e_{s,m})$，同时

$$h_n \gcd\left(c, \frac{dc}{dt}\right)$$

那么
$(i) vh_n \in k\langle t \rangle$
$(ii)$如果$\mathcal{S}_1^{irr} = \mathcal{S}^{irr}$，那么$vh_n h_s \in k[t]$
$(iii)$如果$t$在$k$上是非线性的或是刘维尔的，那么或者$\nu_\infty(v) = 0$或者
$$\nu_\infty(v) \geq \min(\nu_\infty(f), \nu_\infty(w_1), \ldots, \nu_\infty(w_m)) + \delta(t) - 1$$

推论 7.2.1. 令$f, w_1, \ldots, w_m, c, h_n$和$h_s$定义同定理 7.2.1.，那么
$(i)$对于(7.30)的任意解$c_1, \ldots, c_m \in \mathrm{Const}(k)$和$v \in k(t)$，$q = vh_n \in k\langle t \rangle$，并且$q$实现数方程的一个解

$$h_n Dq - qDh_n = h_n^2 f - \sum_{i=1}^{m} c_i h_n^2 w_i. \qquad (7.31)$$

反之，对于(7.31)的任意解，满足$q \in k\langle t \rangle$，$v = q/h_n$产生(7.30)的一个解。
$(ii)$ 如果 $\mathcal{S}_1^{irr} = \mathcal{S}^{irr}$，那么对于(7.30)的任意解 $c_1, \ldots, c_m \in \mathrm{Const}(k), v \in k(t)$，$p = vh_n h_s \in k[t]$，$p$是下述方程的一个解

$$h_n h_s Dp - \left(Dh_n + h_n \frac{Dh_s}{h_s}\right) p = h_n^2 h_s f - \sum_{i=1}^{m} c_i h_n^2 h_s w_i. \qquad (7.32)$$

另外，如果$t$在$k$上是非线性的或者是刘维尔的，那么或者
$$\deg(p) = \deg(h_n) + \deg(h_s)$$
或者

$$\deg(p) \leq \deg(h_n) + \deg(h_s) + 1 - \delta(t) - \min\left(\nu_\infty(f), \nu_\infty(w_1), \ldots, \nu_\infty(w_m)\right)$$

反之，对于(7.32)的任意解满足$p \in k[t]$，$v = p/(h_n h_s)$产生(7.30)的一个解。

这给出了一个将受限积分问题简化为$k\langle t\rangle$上的问题，或者$k[t]$上的问题，如果$\mathcal{S}_1^{irr} = \mathcal{S}^{irr}$的算法。

```
LimitedIntegrateReduce(f, w_1, ..., w_m, D)
(* Reduction to a polynomial problem *)

(* Given a derivation D on k(t) and f, w_1, ..., w_m ∈ k(t), return
   (a, b, h, N, g, v_1, ..., v_m) such that a, b, h ∈ k[t], N ∈ ℕ, g, v_1, ..., v_m ∈
   k(t), and for any solution v ∈ k(t), c_1, ..., c_m ∈ C of f = Dv +
   c_1 w_1 + ... c_m w_m, p = vh ∈ k⟨t⟩, and p and the c_i satisfy aDp + bp =
   g + c_1 v_1 + ... + c_m v_m. Furthermore, if S_1^irr = S^irr, then p ∈ k[t], and if t
   is nonlinear or Liouvillian over k, then deg(p) ≤ N. *)

(d_n, d_s) ← SplitFactor(denominator(f), D)
for i ← 1 to m do (e_{n,i}, e_{s,i}) ← SplitFactor(denominator(w_i), D)
c ← lcm(d_n, e_{n,1}, ..., e_{n,m})
h_n ← gcd(c, dc/dt)
a ← h_n, b ← -Dh_n, N ← 0
if S_1^irr = S^irr then
    h_s ← lcm(d_s, e_{s,1}, ..., e_{s,m})
    a ← h_n h_s, b ← -Dh_n - h_n Dh_s / h_s           (* exact division *)
    μ ← min(ν_∞(f), ν_∞(w_1), ..., ν_∞(w_m))
    N ← deg(h_n) + deg(h_s) + max(0, 1 - δ(t) - μ)
return(a, b, a, N, ah_n f, -ah_n w_1, ..., -ah_n w_m)
```

针对受限积分问题所产生的含参数的$Risch$微分方程算法可以进行一些修正：如果线性约束算法生成一个零维零空间，那么不存在$c_0 = 1$的解，我们可以结束算法。如果生成一个一维零空间，那么存在唯一满足$c_0 = 1$的解，所以在(7.35)中用唯一解代替$c_1, \ldots, c_m$，产生一个无参数问题，对于该问题，6.4节的随机偏微分方程算法是适用的。同样也可以在(7.30)中用这个唯一解代替$c_1, \ldots, c_m$，应用5.12节的内场积分算法，但是这意味着再次计算$v$的分母。最终我们可以使用由推论 7.2.1.给出的$\deg(p)$的上界，而不是再次进行上界计算。

### 7.3 含参数的对数微分问题

在这一部分我们给出对于含参数的对数微分问题的一种解法，换言之，给定一个特征为0的微分域，$K$上的一个超越指数单项式$\theta$和$f \in K$，判定是否存在整数$n, m \in \mathbb{Z}, n \neq 0$，满足

$$nf = \frac{Dv}{v} + m\frac{D\theta}{\theta} \tag{7.37}$$

在$K$中有解$v$，如果存在的话，找到这样的一个解。正如我们在第五章看到的一样，这个问题的背景是判定$K(\theta)$中元素是否是$K(\theta)$中元素的对数微分或者$K(\theta)-$根的对数微分。因此我们能够假设我们能够递归地判定$K$中的元素是够是$K-$根的对数微分。尽管方程(7.37)和(7.30)非常类似，但是7.2节的受限积分算法不能直接应用在这个问题上。但是，因为未知的

常数被限定是整数，所以第九章的结构定理提供了针对这个问题的一个完整的解，只要它们是可应用的。实际上他们提供了唯一已知的完整解，但是首先我们给出一系列的线性约束算法，这些算法在大多数情况下可以给出唯一可能解，因此可以将问题解决掉。这种方法不是一个完整的算法，因为也许会在判定$m/n$时失败，在这种情况下我们必须回到结构定理和结构定理的相关算法这个出发点上。

和之前一样，我们只在超越情况下研究方程(7.37)，换言之当$K$是一个微分子域$k$的一个$simple$（简单）单项式扩张，所以在本小节的剩余部分中，令$k$为一个特征为0的微分域，$t$是$k$上的一个单项式。另外我们假设$\text{Const}(k(t)) = \text{Const}(k)$。我们假设我们的方程中$f$和$D\theta/\theta$的系数都在$k(t)$中，我们要找到解$n, m \in \mathbb{Z}$和$v \in k(t)$。

引理 7.3.1. 令$u, v, w \in k(t)$，$c, \bar{c} \in \text{Const}(k)$满足$v \neq 0, \bar{c} \neq 0$，并且满足

$$u = \bar{c}\frac{Dv}{v} + cw \quad (7.38)$$

将$u$写作$u = p + a/d$，将$w$写作$w = q + b/e$，其中$p, q, a, b, d, e \in k[t]$，$d \neq 0, e \neq 0, \gcd(a, b) = \gcd(b, e) = 1$，且$\deg(a) < \deg(d), \deg(b) < \gcd(e)$，那么

$$\deg(p - cq) \leq \max(0, \delta(t) - 1) \quad (7.39)$$

此外，令$l = l_n l_s$为$l = \text{lcm}(d, e)$的一个分裂分解，$l_n^-$为$l_n$的紧缩映射（定义 1.6.2.），那么

$$lu - clw \equiv 0 \pmod{l_s l_n^-}$$

给定$u, w \in k(t)$，引理 7.3.1.或者证明(7.38)没有解$v \in k(t), c, \bar{c} \in \text{Const}(k)$，或者在以下情况中生成唯一的一个$c \in \text{Const}(k)$作为后备解。

1.如果$\deg(q) > \max(0, \delta(t) - 1)$：将$p - cq$中所有次数比$\max(0, \delta(t) - 1)$大的项当做0，产生一个关于$c$的超定线性代数方程组。如果这个方程组在$\text{Const}(k)$中无解，那么(7.38)无解，否则我们得到$c$的唯一可能解。

2.如果$\deg(p) > \max(0, \delta(t) - 1) \geq \deg(q)$：那么(7.39)永远不满足，所以(7.38)无解。

3.如果$\deg(l_s l_n^-) > 0$，接着令$r \in k[t]$为$lu - clw$模$l_s l_n^-$的余数。如果$r$与0等价，那么$lu \equiv lw \pmod{l_s l_n^-}$，这说明$l_n^* u \in k[t], l_n^* w \in k[t]$，其中$l_n^*$是$l_n$的无平方因子部分，因此$d$和$e$都是$normal$（普通）的，与$\deg(l_s l_n^-) > 0$是矛盾的。因此$r$不是与0等价，所以将它的所有系数当做0产生一个关于$c$的超定线性代数方程组。如果这个方程组在$\text{Const}(k)$中无解，那么(7.38)无解，否则我们得到$c$的唯一可能解。

这些情况轮流产生解(7.37)的一种方法：给定$f$和$\theta$，将引理 7.3.1.应用到$u = f$和$w = D\theta/\theta$上，我们或者能证明(7.38)没有$c \in \mathbb{Q}$的解，在这种情况下(7.37)无解，或者得到一个对于$c = m/n$唯一的可能解，或者如果以上任何一种情况都没有满足的话，就无法得到$c$的任何信息。如果我们得到唯一可能解$c \in \mathbb{Q}$，将$c$写作$c = M/N$，其中$M, N \in \mathbb{Z}, N > 0$，且$\gcd(M, N) = 1$。那么对于(7.37)的任意解，我们一定有$n = QN, m = QM$，$Q$为非零整数。这说明$QNf = Dv/v + QMD\theta/\theta$，因此$Nf - MD\theta/\theta$为一个$k$-根的对数微分，这个我们可以通过递归式的方法检验出来。

注意到如果$\theta$是$k(t)$上的一个超越指数式，那么$D\theta/\theta = D\eta, \eta \in k(t)$。如果$\nu_\infty(\eta) < 0$，根据定理 4.4.4.，$\nu_\infty(D\eta) < -\max(0, \delta(t) - 1)$，所以$\deg(q) > \max(0, \delta(t) - 1)$，且上述方法成立。如果对于$k[t]$中任意$normal$（普通）不可约多项式$p$，$\nu_p(\eta) < 0$，那么根据定理 4.4.2.，$\nu_p(D\eta) < -1$，所以$p | l_n^-$，且上面的方法成立。如果对于任意的$special$（特殊）的$p \in \mathcal{S}^{irr}$，$\nu_p(\eta) < 0$，那么根据定理 4.4.2.，$\nu_p(D\eta) < 0$，所以$p | l_s$，且上面的方法成立。因此上面方法失败的唯一情况是当$\theta$是$k(t)$上的一个超越指数式的情况，且如果

$\eta \in k, \mathcal{S}_1^{irr} = \mathcal{S}^{irr}$。换言之，$\eta$是$k$上的一个超越指数式。

在相同情况下，我们可以看到如果$f = Dg, g \in k(t)$，那么上面的方法当$\mathcal{S}_1^{irr} = \mathcal{S}^{irr}$时时成立，除非$g \in k$。因此如果$\mathcal{S}_1^{irr} = \mathcal{S}^{irr}$，$f = Dg$，且$\theta$是$k(t)$上的一个超越指数式，在这种情况下一个和5.12小节相似的分析说明对于(7.37)的任意解，$v$一定在$k^*$中。如果$Dt \in k$，在这种情况下，我们将问题简化为在$k$上解决相似的问题，或者$v$的形式一定为$v = wt^q, q \in k^*$，或一个整数$q$，如果$Dt/t \in k$。在后一种情况中，我们将问题简化为解以下形式的方程

$$nDg = \frac{Dw}{w} + m\frac{D\theta}{\theta} + q\frac{Dt}{t} \tag{7.40}$$

其中$\theta$和$t$都是$k$上的超越指数式。引理 7.3.1. 可以扩展到任意数目的$w_i$上，并可以作用于(7.40)。当我们到达常数域时，这个过程就停止了，因为在这一点上$Dw = 0$，(7.40)变为一个含未知整数的线性代数方程组。但是在实际中，如果引理 7.3.1.在第一轮没有给出$c$，我们倾向使用结构定理。

```
ParametricLogarithmicDerivative(f, θ, D)
(* Parametric Logarithmic Derivative Heuristic *)

(* Given a derivation D on k[t], f ∈ k(t) and a hyperexponential mono-
mial θ over k(t), returns either "failed", or "no solution", in which case
nf = Dv/v + mDθ/θ has no solution v ∈ k(t)* and n, m ∈ ℤ with n ≠ 0,
or a solution (n, m, v) of that equation. *)

w ← Dθ/θ                                              (* w ∈ k(t) *)
d ← denominator(f), e ← denominator(w)
(p, a) ← PolyDivide(numerator(f), d)                  (* f = p + a/d *)
(q, b) ← PolyDivide(numerator(w), e)                  (* w = q + b/e *)
B ← max(0, deg(Dt) - 1)
C ← max(deg(p), deg(q))
if deg(q) > B then
    s ← solve( coefficient(p, t^i) = c coefficient(q, t^i), B + 1 ≤ i ≤ C)
    if s = ∅ or s ∉ ℚ then return "no solution"
    N ← numerator(s), M ← denominator(s)              (* s ∈ ℚ *)
    if Q(Nf - Mw) = Dv/v for some Q ∈ ℤ and v ∈ k(t) with Q ≠ 0 and
        v ≠ 0 then return(QN, QM, v) else return "no solution"
if deg(p) > B then return "no solution"               (* deg(q) ≤ B *)
l ← lcm(d, e)
(l_n, l_s) ← SplitFactor(l, D)
z ← l_s gcd(l_n, dl_n/dt)                             (* z = l_s l_n^- *)
if z ∈ k then return "failed"
(u_1, r_1) ← PolyDivide(lf, z)                        (* r_1 ≡ lf (mod l_s l_n^-) *)
(u_2, r_2) ← PolyDivide(lw, z)                        (* r_2 ≡ lw (mod l_s l_n^-) *)
s ← solve( coefficient(r_1, t^i) = c coefficient(r_2, t^i), 0 ≤ i < deg(z))
if s = ∅ or s ∉ ℚ then return "no solution"
M ← numerator(s), N ← denominator(s)                  (* s ∈ ℚ *)
if Q(Nf - Mw) = Dv/v for some Q ∈ ℤ and v ∈ k(t) with Q ≠ 0 and
    v ≠ 0 then return(QN, QM, v) else return "no solution"
```

第九章的结构定理给出了一个解(7.37)的有效的替代方法：首先假设$f$在$K$上有一个初等不定积分，这种出现在含参数对数微分问题的情况的背景就是对一个初等函数求不定积分。令$F$为$K(\theta)$的一个初等扩张，$g \in F$满足$f = Dg$。那么如果(7.37)有一个$n \neq 0$的解，我们得到

$$nf = \frac{Dv}{v} + m\frac{D\theta}{\theta} = \frac{D(c\theta^m)}{v\theta^m}$$

这说明$f = Dg$是一个$F-$根的对数微分。如果$F = C(x)(t_1, \ldots, t_n)$，其中$C = \text{Const}(K), Dx = 1$，并且每一个$t_i$或者是袋鼠的，或者是一个$C(x)(t_1, \ldots, t_{i-1})$上的初等或实初等，或者一个非初等基本单项式，那么可以证明$f$是一个$F-$根的对数微分当且仅当存在$r_i \in \mathbb{Q}$，满足

$$\sum_{i \in L} r_i Dt_i + \sum_{i \in E} r_i \frac{Dt_i}{t_i} = f \tag{7.44}$$

其中

$E = \{i \in \{1, \ldots, n\}$，使得$t_i$是$C(x)(t_1, \ldots, t_{i-1})$上的超越指数单项式$\}$

同时

$L = \{i \in \{1, \ldots, n\}$，使得$t_i$是$C(x)(t_1, \ldots, t_{i-1})$上的对数单项式$\}$

可以通过将其看做一个系数在$F$中的，关于$r_i$的线性方程，从而找到(7.44)的有理解，那么应用引理 7.1.2.，可以得到一个方程组，方程组的系数都在$C$中，并且常数解与原方程组相同。假设我们有一在$\mathbb{Q}$上以$C$为基础的向量基，其中包含1，将方程组映射到1上得到一个系数在$\mathbb{Q}$中的线性方程组，并且和(7.44)有相同的有理解。这个方法也可以应用到与(7.40)类型相同的方程，在方程右端有任意数目的项，因为解的存在说明$f$是一个$F-$根的对数微分。

# *Chapter* 2.1.8 **平行积分**

在这一章中我们描述一个可替代实现积分的方法，该方法同样基于刘维尔定理，但可以尝试避免递归式方法的损失和在之前章节中介绍算法的关联的计算时间损失。这个方法由 $Risch$ 和 $Norman$ 在 $SYMSAC'76$ 的会议上提出，尝试以一种平行的方式处理包含被积函数的微分域的所有生成元，换言之，同时处理所有的生成元，而是不是只考虑目前处于最顶端的一个生成元。同样，这种方法在文献中被称为"新 $Risch$ 算法"，"$Risch - Normna$ 算法"或者"平行 $Risch$ 算法" 。实际上，它的启发性远远强于它的可编程性，因为它可能在计算初等不定积分的理论和实践中同时失败。但是，它实现的便捷性，速度上的优势以及成功率令人满意的特性使得它成为对于计算机代数系统的设计者而言及其具有吸引力的一种代替方案。结果证明是，在几个代数系统中，这套方法都得到了实现，或者作为一个替代品，或者作为完成积分算法的预处理程序。

平行算法背后的整体思想是通过将含有被积函数的微分域看做是一个在其常数域上多元的有理函数域，从而避免计算算法递归的本性。回顾 5.2 节，我们被积函数 $f$ 属于一个形式为 $K = C(t_1, \ldots, t_n)$ 的微分域，其中 $C = \mathrm{Const}(K)$，其中每一个 $t_i$ 在 $C(t_1, \ldots, t_{i-1})$ 上是超越的。根据强刘维尔定理 5.5.3.，如果 $f$ 的不定积分在 $K$ 是初等的，那么存在 $v \in K$，$C$ 上的代数元 $c_1, \ldots, c_m$ 和 $u_1, \ldots, u_m \in K(c_1, \ldots, c_m)^*$ 使得

$$f = Dv + \sum_{i=1}^{m} c_i \frac{Du_i}{u_i} \tag{10.1}$$

因为 $v$ 是一个含 $t_1, \ldots, t_n$ 的多元多项式的商的形式，可以不是一般性的假设 $u_i$ 为 $t_1, \ldots, t_n$ 的多项式（对数微分恒等式），平行的方法包括对 $u_i$ 和 $v$ 的分母做出有根据的假设，同时对它们的分子也做出同样的假设。这个将解方程 (10.1) 的问题简化为找到 $c_i$ 和 $v$ 的分子的常值系数，这个可以通过初等线性代数实现。如果对未知系数的线性方程有一个解，那么就找到了 $f$ 的一个不定积分。另一方面，解的不存在性并不总是说明 $f$ 在 $K$ 上没有一个初等的不定积分，可能只是说明猜测是错误的。现在已经发布了平行算法的几个变形，区别在于假设或者解未知常值系数的细节不同，但是所有的方法都有共同的性质，那就是存在有初等不定积分的函数通过这种方法不能找到其不定积分的现象。总而言之，平行积分是一个方便的递归式，与实现完整积分算法相比是明显简单的一种方法。将其变为对于一类被积函数通行的算法需要证明，给定的猜测方法能够覆盖该函数类中所有有初等不定积分的函数。这依然是一个开放性的问题，虽然对于对数性的被积函数已经给出了部分的结果，在这里我们将其扩展到更一般的微分域上。

## 10.1 多项式环的微分

首先我们描述单项式扩张在多元多项式环上的微分更一般定义下保持不变的几个性质。令 $F$ 是一个域，$t_1, \ldots, t_n$ 是 $F$ 上的独立的不定变元，$D$ 是多项式环 $R = F[t_1, \ldots, t_n]$ 上的一个微分。因为 $F$ 是一个域，$R$ 是一个唯一分解整环（$UFD$）。在计算机代数书中我们可以找到计算 $\gcd$，$\mathrm{lcm}$ 和无平方因子分解，多元分解的算法，我们只使用这些算法，而不进行更进一步的陈述。容度和本源部分的概念能够关于每一个变元 $t_i$ 扩展为相似的概念，通过将 $R$ 记为 $R = R_i[t_i]$，其中 $R_i = F[t_1, \ldots, t_{i-1}, t_{i+1}, \ldots, t_n]$。我们分别用 $\mathrm{content}(p, t_i)$ 和 $\mathrm{pp}(p, t_i)$ 表示它们，如果 $\mathrm{content}(p, t_i) \in R_i^* = F^*$，那么我们称 $p \in R$ 关于 $t_i$ 是 *pmimitive*（本原）

的。首先我们注意到引理 3.4.4.在这种定义下依然是成立的。

引理 10.1.1. 令 $p_1, \ldots, p_m \in F[t_1, \ldots, t_n]$ 满足 $\gcd(p_i, p_j) = 1, i \neq j$, 令 $p = \prod_{i=1}^m p_i^{e_i}$, 其中 $e_i$ 为正整数，那么

$$\gcd(p, Dp) = \left(\prod_{i=1}^m p_i^{e_i-1}\right) \prod_{i=1}^m \gcd(p_i, Dp_i)$$

类比与单项式情形，如果 $\gcd(p, Dp) = 1$，我们称 $p \in R$ 关于 $D$ 是 *normal*（普通）的，如果 $p | Dp$，那么称 $p$ 关于 $D$ 是 *special*（特殊）的。因为定理 3.4.1.是引理 3.4.4.的一个结论，所以在这里定理还是成立的。

定理 10.1.1.
($i$)任意有限个 *normal*（普通）多项式（两两互素）的乘积是 *normal*（普通）的。任意一个 *normal*（普通）多项式的任意因子是 *normal*（普通）的。
($ii$)任意有限个 *special*（特殊）多项式的乘积是 *special*（特殊）的。任意一个 *special*（特殊）多项式的任意因子是 *special*（特殊）的。

我们称 $p = p_s p_n$ 是 $p$ 的一个分裂分解，如果 $p_2, p_n \in R$，且 $p_s$ 是 *special*（特殊）的，$p_n$ 的每一个无平方因子是 *normal*（普通）的。我们继续说明，给定容度求根的附加步骤，3.5 节的分裂分解，可以扩展到 $R$ 上。

定理 10.1.2. 令 $p \in F[t_1, \ldots, t_n]$，且存在 $j$ 使得 $p$ 关于 $j$ 本原，那么
($i$)

$$\frac{\gcd(p, Dp)}{\gcd(p, dp/dt_j)}$$

是所有 $p$ 的所有互质 *special*（特殊）不可约因子的乘积
($ii$)另外如果 $p$ 无平方因子，那么 $p = p_s p_n$ 是 $p$ 的一个分裂分解，其中 $p_s = \gcd(p, Dp)$ 且 $p_n = p/p_s$。
其中 $p$ 关于 $t_j$ 是本原的假设是必需的。

对于 $\forall p \in F[t_1, \ldots, t_n] \backslash F$，定义 $p$ 的主变量，定义 $\mathrm{mainVar}(p)$ 为 $t_j$，其中 $j$ 为满足 $\deg_{t_j}(p) > 0$ 的最高指数。将 $p$ 写为 $p = \mathrm{content}(p, t_j) \mathrm{pp}(p, t_j)$，其中 $t_j = \mathrm{mainVar}(p)$，定理 10.1.2.说明 $SplitFactor$（分解因式）算法可以应用到 $\mathrm{pp}(p, t_j)$ 上，生成 $\mathrm{pp}(p, t_j)$ 的分裂分解 $\mathrm{pp}(p, t_j) = q_s q_n$。递归地计算 $\mathrm{content}(p, t_j)$ 的分裂分解，其中 $\mathrm{content}(p, t_j)$ 比 $p$ 少一个变量，生成 $\mathrm{content}(p, t_j) = h_s h_n$，并且 $p = p_s p_n$ 是 $p$ 的一个分裂分解，其中 $p_s = q_s h_s$，且 $p_n = q_n h_n$。

```
SplitFactor(p, D)        (* Splitting Factorization *)

(* Given a derivation D on F[t_1,...,t_n] and p ∈ F[t_1,...,t_n], return
(p_n, p_s) ∈ F[t_1,...,t_n]^2 such that p = p_n p_s, p_s is special, and each square-
free factor of p_n is normal. *)

if p ∈ F then return(1, p)
t ← mainVar(p)
c ← content(p, t), q ← p/c                          (* q = pp(p, t) *)
(h_n, h_s) ← SplitFactor(c, D)
S ← gcd(q, Dq)/ gcd(q, dq/dt)                       (* exact division *)
if deg_t(S) = 0 then return(h_n p, h_s)
(q_n, q_s) ← SplitFactor(q/S, D)                    (* exact division *)
return(h_n q_n, S h_s q_s)
```

$SplitSquarefreeFactor$（无平方因子分解）算法可以按照下面的方法，求出每一个无平方因子的根，进而推广为 $F[t_1,\dots,t_n]$ 上的一种相似形式。另外一种可替代的方法为计算 $p$ 本原部分的无平方因式分解，接着分别计算它的容度的一个分裂分解并重组两个因式分解。

```
SplitSquarefreeFactor(p, D)        (* Splitting Squarefree Factorization *)

(* Given a derivation D on F[t_1,...,t_n] and p ∈ F[t_1,...,t_n], re-
turn (N_1,...,N_m) and (S_1,...,S_m) in F[t_1,...,t_n]^m such that p =
(N_1 N_2^2 ··· N_m^m)(S_1 S_2^2 ··· S_m^m) is a splitting factorization of p and the N_i
and S_i are squarefree and coprime. *)

(p_1,...,p_m) ← Squarefree(p)
for i ← 1 to m do
    t ← mainVar(p_i)
    c_i ← content(p_i, t)
    q_i ← p_i/c_i                                   (* q_i = pp(p_i, t) *)
    (M_i, T_i) ← SplitFactor(c_i, D)                (* c_i is already squarefree *)
    S_i ← gcd(q_i, Dq_i)
    N_i ← q_i/S_i                                   (* exact division *)
return((M_1 N_1,...,M_m N_m), (T_1 S_1,...,T_m S_m))
```

现在令 $K = F(t_1,\dots,t_n)$ 为 $F[t_1,\dots,t_n]$ 的商域，令 $D$ 为 $K$ 上的一个微分，满足 $DF \subseteq F$。因为 $F[t_1,\dots,t_n]$ 在 $D$ 的作用下封闭不是必需的，我们定义关于 $D$, $K$ 的分母，定义 $\text{den}_D(K)$ 为 $Dt_1,\dots,Dt_n$ 分母的最小公倍数。因为 $D = \kappa_D + \sum_{i=1}^{n}(Dt_i)/dt_i$，其中 $\kappa_D$ 为 $F[t_1,\dots,t_n]$ 上微分，定义由下式给出

$$\kappa_D\left(\sum_e a_e t_1^{e_1}\cdots t_n^{e_n}\right) = \sum_e (Da_e) t_1^{e_1}\cdots t_n^{e_n}$$

$\overline{D} = \text{den}_D(K)D$ 为 $F[t_1,\dots,t_n]$ 的一个微分，所以我们可以使用关于 $\overline{D}$ 的分裂分解。

## 10.2 初等不定积分的结构

现在我们回到形式为$K = F(t_1, \ldots, t_n)$形式的微分域上的积分问题，其中$DF \subseteq F$，且每一个$t_i$在$C(t_1, \ldots, t_{i-1})$上都是超越的。正如在前一小节的结尾看到的一样，$\overline{D} = \text{den}_D(K)D$是$F[t_1, \ldots, t_n]$的一个微分。这说明 *normal*（普通）多项式的一个关键性质依然在多元情况下成立。

**引理 10.2.1.** 令$p \in F[t_1, \ldots, t_n]$为关于$\overline{D}$不可约且 *normal*（普通）的。如果$p^m, \exists m > 0$整除$f \in K^*$的分母，那么$p^{m+1}$整除$\overline{D}f$和$Df$的分母。

现在我们陈述一个初等不定积分的结构。我们用$C$代表$\text{Const}_D(K)$，$\overline{C}$代表它的代数闭包。

**定理 10.2.1.** 假设$\text{Const}_D(K) \subseteq F$，令$f = a/d \in K^*$，其中$a, d \in F[t_1, \ldots, t_n]$为互质的。令$d = d_s d_n$为$d$关于$\overline{D}$的一个分裂分解，$\prod_{j=1}^{e} d_j^j$和$\prod_{k=1}^{t} p_i^{e_i}$分别为$d_n$在$\overline{C}F[t_1, \ldots, t_n]$中的无平方因子分解和它的不可约因子分解。如果$f$在$K$中有一个初等不定积分，那么存在$b, s \in F[t_1, \ldots, t_n], w_1, \ldots, w_r \in \overline{C}F^*$，满足$s, s_1, \ldots, s_m$关于$\overline{D}$是 *special*（特殊）的，且

$$f = D\left(\frac{b}{s \prod_{j=2}^{e} d_j^{j-1}}\right) + \sum_{i=1}^{m} \alpha_i \frac{Ds_i}{s_i} + \sum_{i=1}^{t} \beta_i \frac{Dp_i}{p_i} + \sum_{i=1}^{r} \gamma_i \frac{Dw_i}{w_i}. \quad (10.2)$$

定理 10.2.1 给出了$f$的初等不定积分的一部分。当$F = \text{Const}_D(K)$时，我们用这个来计算$f$的不定积分，此时$\sum_i \gamma_i Dw_i/w_i = 0$，所以这一项可以从(10.2)中移除，剩下的未知部分为$v$的分子和一些 *special*（特殊）多项式。对于一般微分而言，*special*（特殊）多项式是未知的，尽管要求$F = \text{Const}_D(K)$说明只可能存在有限多互质不可约 *special*（特殊）多项式。当$K = F(t_1, \ldots, t_n)$是$F$上一个嵌套式的塔状单项式扩张时，那么情况变得更加简单，我们能够描述所有的 *special*（特殊）多项式。首先注意到如果$F$在$D$的作用下是封闭的，每一个$t_i$是$F(t_1, \ldots, t_{i-1})$上的单项式，那么$F(t_1, \ldots, t_{i-1})$和$F(t_1, \ldots, t_{i-1})[t_i]$在$D$的作用下封闭，所以对于$F(t_1, \ldots, t_{i-1})[t_i]$中元素关于$D$是 *special*（特殊）这个概念是良定义的。对于任意非零$p \in F(t_1, \ldots, t_{i-1})[t_i]$，记$p^+ \in F[t_1, \ldots, t_i]$为$p$的系数关于$t_i$的分母的最小公倍数。我们引入概念

$$\mathcal{S}_{K:F}^{\text{irr}} = \{p^+ \text{ for } p \in \bigcup_{1 \leq i \leq n} \mathcal{S}_{F(t_1, \ldots, t_{i-1})[t_i]:F(t_1, \ldots, t_{i-1})}^{\text{irr}}\} \subset F[t_1, \ldots, t_n].$$

当$t_i$都是本原式，超越指数式或超切向量时，那么$\mathcal{S}_{K:F}^{irr}$是有限的，正好由为超越指数式的$t_i$和为超切向量的$t_j$，$1 + t_j^2$的因子组成。

**定理 10.2.2.** 令$K = F(t_1, \ldots, t_n)$，其中$DF \subseteq F$，每一个$t_i$为$F(t_1, \ldots, t_{i-1})$上的单项式。那么，关于$\overline{D}$的不可约 *special*（特殊）多项式正好是$\mathcal{S}_{K:F}^{irr}$的所有元素，和$\text{den}_D(K)$的所有不可约元。

因此，甚至当$\mathcal{S}_{K:F}^{irr}$是空集时，可能存在关于$\overline{D}$的 *special*（特殊）多项式，如在嵌套本原扩张中，但是它们的不可约因子都限制为$\text{den}(K)$的因子。例如考虑$K = \mathbb{Q}(x, t), Dx = 1,$

对$q \in \mathbb{Q}[x]$，$Dt = Dq/q$。因为$\text{den}_D(K)$是$q$的无平方因子部分，$q$的每一个不可约因子都是 *special*（特殊）的，可以在积分中作为一个 *logand* 出现。这解释了$\log(x+1)$在下列积分中出现的原因。

$$\int \log(x^2 - 1)dx = (x-1)\log(x^2-1) + 2\log(x+1) - 2x$$

这同时也解释了异常项$\log(1+tan^2\theta)$当对含$tan\theta$的表达式做积分时出现的原因，因为$1+tan^2\theta$是 *special*（特殊）的。

作为定理 10.2.2.的结论之一，当$K$是$F$的一个塔状嵌套单项式扩张时，方程(10.2)可以更加精细。$s_i$可以取为$\overline{C}F$上关于$\text{den}_D(K)$，$\mathcal{S}_{K:F}^{irr}$中元素的不可约因子，其中$s$的形式如下

$$s = \prod_{p|\text{den}_D(K)} p^{e_p} \prod_{q \in \mathcal{S}_{K:F}^{irr}} q^{e_q} \qquad (10.3)$$

其中$e_p \geq 0, e_q \geq 0$，其中的第一个乘积取遍$\text{den}_D(K)$在$F$上所有首一不可约因子，而不是在$\overline{C}F$上。这些指数可以在微分域中被限制，包括嵌套对数扩张，这将在 10.4 小节中进行说明。

## 10.3 积分方法

现在，平行方法由之前小节的结论产生，并可以应用到多元有理函数域上，只要保证$\text{Const}_D(K) = F$，微分可以是任意的。平行方法包括对(10.2)中的$s, s_i$和$p_i$做出有根据的猜测，限定$v$的分子$b$的次数，最终解决关于$b$系数和未知常数$\alpha_i, \beta_i$的线性方程。

因为$s$关于$\overline{D}$是 *special*（特殊）的，$s$的形式一定为$s = \prod_p p^{e_p}$，其中$p$取遍所有首一不可约 *special*（特殊）多项式，其中只有有限个$e_p$为0。因为$e_p$的边界是未知的，文献中通常进行的猜测是去最大的$w$，满足$p^e|d$。这种猜测是有附加优势的，使得$s = d_s$，且不需要对$d_s$进行分解操作。并且该猜测也是有一定道理的，因为任一更大的指数，或者$s$的一个 *special*（特殊）不可约因子不整除$d$将会需要$Dv$中的消去，使得该因子消失，所以$s = d_s$这个猜测将这个过程变为一个递归式的过程，可能会导致计算初等不定积分的失败。

例 10.3.1. 考虑

$$\int \frac{e^x - x^2 + 2x}{(e^x + x^2)x^2} e^{(x^2-1)/x + 1/(e^x+x)} dx$$

我们有$K = \mathbb{Q}(x, t_1, t_2)$，$Dx = 1, Dt_1 = t_1$（换言之，$t_1 = e^x$）和

$$Dt_2 = \frac{(x^2+1)t_1^2 + (2x^3 - x^2 + 2x)t_1 + x^4}{x^2(x+t_1)^2} t_2$$

换言之，$t_2 = e^{(x^2-1)/x + 1/(e^x+x)}$。因此，$\text{den}_D(K) = x^2(x+t_1)^2$，多项式$x$和$x+t_1$关于$\text{den}_D(K)D$是 *special*（特殊）的。我们的被积函数变为

$$f = \frac{(t_1 - x^2 + 2x)t_2}{x^2(x+t_1)^2}$$

分母为$d = x^2(x+t_1)^2$，$d$是 *special*（特殊）的。结果证明是 *special*（特殊）分母$s = t_1$不能整除$d$，但是

$$f = D\frac{t_2}{t_1}$$

因此，$f$的初等不定积分不能$s = d_s$的猜测下，通过平行方法进行计算。

接下来要处理首一 *special*（特殊）不可约元 $s_i$。当 $K$ 是 $F$ 上的一个本原，超越指数或超切向量的嵌套塔状扩张时，那么定理 10.2.2 给出了所有这样的扩张的一个详细且有限的列表，换言之 $\text{den}_D(K)$ 的所有因子和超越指数式 $t_i$，$1 + t_j^2$ 的因子，其中 $t_j$ 为超切向量。根据定理 10.2.1，确实应该取它们在 $\overline{F}[t_1, \ldots, t_n]$ 中的不可约因子（回顾 $F = \text{Const}_D(K)$），而不是在 $F[t_1, \ldots, t_n]$，下面的例子可以说明这个问题

例 10.3.2. 令 $K = \mathbb{Q}(x, t)$，其中 $Dx = 1. Dt = 2x/(x^2 - 2)$（换言之，$t = \log(x^2 - 2)$），那么，$\text{den}_D(K) = x^2 - 2$ 关于 $\text{den}_D(K)D$ 是 *special*（特殊）的，并在 $\mathbb{Q}(x, t_1)$ 中是不可约的，但是

$$t = D((t-2)x) + \sqrt{2}\frac{D(x+\sqrt{2})}{x+\sqrt{2}} - \sqrt{2}\frac{D(x-\sqrt{2})}{x-\sqrt{2}}$$

这说明 $\text{den}_D(K)$ 在 $\overline{\mathbb{Q}}[x, t]$ 中的不可约因子一定是考虑在内的。

相似地，$p_i$ 应该为 $d_n$ 在 $\overline{F}[t_1, \ldots, t_n]$ 中的不可约因子，而不是 $F[t_1, \ldots, t_n]$ 中，由下式说明

$$\int \frac{dx}{x^2 - 2} = \frac{1}{2\sqrt{2}} \log(\sqrt{2} - x) - \frac{1}{2\sqrt{2}} \log(\sqrt{2} + x)$$

但是，当 $K$ 是一个塔状嵌套的本原，超越指数和超切向量的单项式扩张时，对只在 $F$ 上分解 $d_n$ 和 $\text{den}_D(K)$ 而不在 $\overline{F}$ 进行分解存在争论。因为无能我们怎样选择，平行方法都可能失败，如果平行方法没有找到一个初等不定积分，那么我们必须转换到完整的积分算法进行求解。所以只有当它的速度比完整算法明显快时，平行方法才是有用的。虽然存在在 $\overline{F}[t_1, \ldots, t_n]$ 中进行不可约因子因式分解计算的算法（称为绝对分解），但是它们的时间花费比完整的积分算法都要高。另外，这个分解可以计算 $F$ 的代数扩张，这个对于计算积分来说不是必需的，正如在下列情况中

$$\int \frac{2x}{x^2 - 2} = \log(x^2 - 2)$$

基于这些原因，两个方法都出现在文献中：关于实际实现的文献选择在 $F$ 上分解，分析方法的文献选择在 $\overline{F}$ 上分解。

最终我们需要确定 $v$ 的分子 $b$ 的次数边界。当 $K$ 是一个塔状嵌套单项式扩张时，引理 3.4.2. 提出了以下的自然猜测，这个猜测也是文献中所使用的猜测：

$$\deg_{t_i}(b) \leq \begin{cases} 1 + \max(\deg_{t_i}(a), \deg_{t_i}(d)) & \text{if } \delta(t_i) = 0, \\ \max(\deg_{t_i}(a), \deg_{t_i}(d)) & \text{if } \delta(t_i) > 0. \end{cases} \quad (10.4)$$

其中 $f = a/d$，一旦我们得到了 $\deg_{t_i}(b)$ 的边界 $b_i$，我们记为

$$b = \sum_{i_1=0}^{b_1} \cdots \sum_{i_n=0}^{b_n} u_{i_1 \cdots i_n} t_1^{i_1} \cdots t_n^{i_n} \quad (10.5)$$

其中 $u_{i_1 \cdots, i_n}$ 为 $F$ 中而不是 $\overline{F}$ 中的不定常量。在 (10.2) 的右端，代入我们所有的猜测，使之与 $f$ 等同，消除分母，产生下列形式的方程

$$q = \sum_{i_1=0}^{b_1} \cdots \sum_{i_n=0}^{b_n} u_{i_1 \cdots i_n} q_{i_1 \cdots i_n} + \sum_{i=1}^{m} \alpha_i r_i + \sum_{i=1}^{t} \beta_i w_i \quad (10.6)$$

其中$q$和$q_{i_1\cdots i_n}$都是在$F[t_1,\ldots,t_n]$中，$r_i$和$w_i$或者在$F[t_1,\ldots,t_n]$中，或者在$\overline{F}[t_1,\ldots,t_n]$中，要依据之前作出的选择。以(10.6)两端的相同单项式列方程，产生了一个关于未知常量$u_{i_1\cdots i_n}, \alpha_i$和$\beta_i$的非齐次线性方程组。有几个关于怎样解这个线性方程组的变形方法：原始的方法首先解总的最高次数的单项式，然后在得到的方程中解具有最低权值的$u_{i_1,\cdots,i_n}$。另外一种方法则是直接解整个线性方程组，在实践中至少和上一个方法效果相同。

平行方法可以很容易适用于不是嵌套单项式扩张的微分域，因为定理 10.2.1 在这样的域上也是成立的。在这种情况下，因为如果平行方法失败了，是没有完整的积分算法的，所以应该在$\overline{F}$上对(10.2)潜在的$s_i$和$p_i$进行不可约的因式分解，而不是在$F$上进行因式分解。另外，(10.4)的边界限制应该替换为

$$\deg(b) \leq 1 + \deg(a) + \max(0, \deg(\operatorname{den}_D(K)) - \max_{1\leq i\leq n}(\deg(\overline{D}t_i)))$$

这是基于次数总和没有出现相消的这个假设得到的（deg代表次数总和）。最终，(10.5)中$b$的形式应该替换为

$$b = \sum_{i_1+\cdots+i_n\leq \deg(b)} u_{i_1\cdots i_n}t_1^{i_1}\cdots t_n^{i_n}$$

在这样的扩张中出现的另外一个问题是关于$\overline{D}$的$special$（特殊）多项式是未知的。开始的时候，应该取遍$\operatorname{den}_D(K)$的$special$（特殊）部分在$\overline{F}$上的所有不可约因子（并不总是$special$（特殊）的，见例 10.3.4.）。另外，可能通过将$p$取为一个含未知常值系数，次数固定的多项式来寻找总次数更低的$special$（特殊）多项式。条件$p|\overline{D}p$可以转化为一个求解$p$的系数的非线性代数方程组，这个方程组可以通过代数的技巧求解。这个计算只依赖于域$K$，所以$special$（特殊）部分可以再次计算并为了一些非单项式扩张的情况记录下来。一个有限的，由首一不可约$special$（特殊）元素组成，可能的空集$\mathcal{S}$是可以通过这种方法找到的，并且在非单项式情形代替集合$\mathcal{S}_{K:F}^{irr}$。最终，$s = d_s$的猜测，应该被替换为

$$s = d_s \prod_{p\in\mathcal{S}, p|d_s} p$$

这样可以体现出在$d_s$因子之外可能的消去情况。

```
ParallelIntegrate(f, D)        (* Parallel Integration *)
(* Given a derivation D on K = F(t_1,...,t_n) such that Const_D(K) = F,
return either an elementary integral of f, or "failed", in which case it is
unknown whether f has an elementary integral over K. *)
                               (* The first 4 steps need to be done once per field K *)
h ← lcm(denominator(Dt1),...,denominator(Dt_n))
if each t_i is a monomial then h_s ← h else
    (h_n, h_s) ← SplitFactor(h, D)
    Δ ← 1 + max(0, deg(h) − max_{1≤i≤n}(deg(hDt_i)))    (* total degree *)
if each t_i is a primitive, hyperexponential or hypertangent monomial then
    S ← S^{irr}_{K:F}
else S ← FindSpecials(hD)      (* exhaustive search for low degree *)
S ← S ∪ IrreducibleFactors(h_s)                  (* over F or F̄ *)

                               (* The remaining steps are done for each integrand *)
a ← numerator(f), d ← denominator(f)
(d_n, d_s) ← SplitFactor(d, D)
(d_1,...,d_e) ← SquareFree(d_n)
{p_1,...,p_t} ← IrreducibleFactors(d_1···d_e)    (* over F or F̄ *)
if each t_i is a primitive, hyperexponential or hypertangent monomial then
    v_s ← d_s
    for i ← 1 to n do
        b_i ← max(deg_{t_i}(a), deg_{t_i}(d))
        if deg_{t_i}(Dt_i) = 0 then b_i ← b_i + 1
    b ← Σ_{i_1=0}^{b_1} ··· Σ_{i_n=0}^{b_n} u_{i_1···i_n} t_1^{i_1} ··· t_n^{i_n}
else                                             (* for general fields *)
    v_s ← d_s ∏_{p∈S, p∤d_s} p
    δ ← deg(a) + Δ                               (* total degree *)
    b ← Σ_{i_1+...+i_n≤δ} u_{i_1···i_n} t_1^{i_1} ··· t_n^{i_n}
v ← b / (v_s ∏_{j=2}^{e} d_j^{j−1})
Solve the inhomogeneous linear system for the u_{i_0···i_n}, α_s and β_i obtained
by clearing denominators and equating coefficients of equal monomials in
f = Dv + Σ_{s∈S} α_s Ds/s + Σ_{i=1}^{t} β_i Dp_i/p_i. if it has no solution then
return "failed" else return  v + Σ_{s∈S} α_s log(s) + Σ_{i=1}^{t} β_i log(p_i).
```

### 10.4 简单的微分域

在这一小节中我们介绍定理 10.2.1 怎样在一类包含嵌套对数扩张的微分域上得到改进，同样也包含更一般的扩张情况。对于一般的微分，*special*（特殊）多项式不总是满足引理 10.2.1，例如 $p = e^x \in \mathbb{Q}(x, e^x)$。我们对满足每一个不可约特殊元都能整除 $\text{den}_D$，且满足引理 10.2.1 的微分域感兴趣。在这一小节中，$(K, D)$ 是一个微分域，形式为 $K = F(t_1, \ldots, t_n)$，其中 $DF \subseteq F$，每一个 $t_i$ 都在 $F(t_1, \ldots, t_{i-1})$ 上为超越元。

定义 10.4.1. 如果存在 $m > 0$，$p^m$ 整除 $f \in K^*$ 的分母说明 $p^{m+1}$ 整除 $Df$ 的分母，那么我们

称一个不可约$p \in F[t_1,\ldots,t_n]$关于$D$是$simple$（简单）的。如果$F[t_1,\ldots,t_n]$的每一个不可约$special$（特殊）多项式能够整除$\mathrm{den}_D(K)$，并且是$simple$（简单）的，那么我们称$(K,D)$是$simple$（简单）的。

一个没有非平凡$special$（特殊）多项式的微分域显然是$simple$（简单）的。这种情况是针对$(C(x), d/dx)$，同时也是针对当$\mathcal{S}_{K:F}^{irr}$是空集，$\mathrm{den}_D(K) = 1$时的嵌套单项式扩张。

简易性的一个关键性质是，它能够在本原扩张中保持，只要没有引入新的常量，$\mathrm{den}(D)$是无平方因子的，这等价于需要$Dt_i$的分母都是无平方因子的。

引理 10.4.1. 令$K = F(t_1,\ldots,t_n)$，其中$DF \subseteq F$，每一个$t_i$都是在$F(t_1,\ldots,t_{i-1})$上是超越的。假设$\mathrm{Const}(K) \subseteq F$，那么$E = F(t_1,\ldots,t_{n-1})$在$D$下是封闭的，且$Dt_n \in E$。如果$E$是$simple$（简单）的且$\mathrm{den}_D(K)$是无平方因子的，那么$K$是$simple$（简单）的。

现在我们得到了一大类的$simple$（简单）微分域，换言之为嵌套本原扩张，且$\mathrm{den}_D$是无平方因子的。

定理 10.4.1. 令$K = F(t_1,\ldots,t_n)$，其中$DF \subseteq F$，$\mathrm{Const}_D(K) \subseteq F$且每一个$t_i$都是在$F(t_1,\ldots,t_{i-1})$上是超越的。如果$\mathrm{den}_D(K)$是无平方因子的，那么$(K,D)$是$simple$（简单）的，而且$Df/f$的分母对于$\forall f \in K^*$是无平方因子的。

$\mathrm{den}_D(K)$无平方因子这个条件尤其在本原扩张中的嵌套对数扩张的情况下是满足的，因此这种扩张情况是$simple$（简单）的。

推论 10.4.1. 令$K = F(t_1,\ldots,t_n)$，其中$DF \subseteq F$，$\mathrm{Const}_D(K) \subseteq F$，$t_1$是$F$上的一个本原（基本）单项式，每一个$t_i$都是$F(t_1,\ldots,t_{i-1})$上的对数单项式，其中$2 \leq i \leq n$。那么$\mathrm{den}_D(K)$是无平方因子的，而且$(K,D)$是$simple$（简单）的。

最终我们可以使定理 10.2.1.在满足$\mathrm{den}_D(K)$是无平方因子的$simple$（简单）微分域上变得更加精细。像之前一样，我们将$\mathrm{Const}_D(K)$的代数闭包记为$\overline{C}$。

定理 10.4.2. 假设$\mathrm{Const}_D(K) \subseteq F$，$(K,D)$是$simple$（简单）的，而且$\mathrm{den}_D(K)$是无平方因子的。令$f = a/d \in K^*$，其中$a, d \in F[t_1,\ldots,t_n]$是互质的。令$d = d_s d_n$为$d$关于$\overline{D}$的一个分裂分解，$d = \prod_{j=1}^{e} d_j^j$为$d$的无平方因式分解，$d_n = \prod_{k=1}^{t} p_i^{e_i}$为$d_n$在$\overline{C}F[t_1,\ldots,t_n]$中的不可约因式分解，$h_s = \prod_{l=1}^{m} s_i^{f_i}$为$h_s$在$\overline{C}F[t_1,\ldots,t_n]$中的不可约因式分解。如果$f$在$K$上有一个初等不定积分，那么存在$w_1,\ldots,w_r \in \overline{C}F^*, b \in F[t_1,\ldots,t_n]$和$\alpha_1,\ldots,\alpha_m, \beta_1,\ldots,\beta_t, \gamma_1,\ldots,\gamma_r \in \overline{C}$，使得

$$f = D\left(\frac{b}{\prod_{j=2}^{e} d_j^{j-1}}\right) + \sum_{i=1}^{m} \alpha_i \frac{Ds_i}{s_i} + \sum_{i=1}^{t} \beta_i \frac{Dp_i}{p_i} + \sum_{i=1}^{r} \gamma_i \frac{Dw_i}{w_i} \quad （10.11）$$

像之前提到的一样，当$F = \mathrm{Const}_D(K)$时，$\sum_i \gamma_i Dw_i/w_i = 0$，所以这一项可以从(10.11)中

去掉。$ParallelIntegrate$ 算法对于满足 $\text{den}_D$ 无平方因子的 $simple$（简单）微分域来说是容易改进的：集合 $\mathcal{S}$ 只包含 $h_s$ 的不可约因子，我们计算 $d$ 的无平方因式分解和 $d_n$ 的不可约因式分解而不是 $d_n$ 的无平方因式分解。最终 $v_s$ 这一项可以在可能的不定积分中忽略掉。

## 2.1.9 Albi 后记

以上便是 Albi 系统针对超越函数的理论及算法的系统总结，针对代数函数的理论介绍会在之后的章节给出。下面这一部分内容则是主要是有关 Albi 系统实现的这部分内容。下面先介绍一下我实现的思路。

由于 maTHmU 内核还没有完全成熟，如果直接在 maTHmU 内核上进行代码的编写，那么很可能会因为内核中底层函数的缺失影响代码的编写和调试。所以我决定先在成熟的 Mathematica 内核上进行代码编写和调试工作，选择 Mathematica 内核主要是因为 Mathematica 和 MaTHmU 语法一致，内核环境基本一致，所以可以将 Mathematica 内核上实现的代码移植到 maTHmU 内核上进行调试。为此我也进行了系统框架，所需底层函数的整理与总结。保证了 Albi 系统的可移植性。

在 Mathematica 上完成代码后，我在 maTHmU 内核上进行了调试。因为 maTHmU 内核中底层函数的缺失，导致代码不能正常运行。为了解决这个问题，我通过一些调试技巧将所有的内核 Bug 都找了出来，将之后的工作重点转向对 maTHmU 内核的修正，使 Albi 系统在 maTHmU 内核上能够尽快实现。

为实现 Albi 系统，我选择 pmint 框架作为我的实现思路，我所实现的代码在实际运行的过程中主要出现两个问题：一是递归层数难于把握，二是结果表达比较复杂，三是只在被积函数为超越函数的情况下，运行效果较好，对于其它情况则效果欠佳。为了解决这些问题，我采取了下列措施：一，首先人工设置递归层数，进行效果比较，选择一个比较好的边界值，避免无限扩张的情况的出现。二是单独对结果表达进行处理，尤其针对三角函数这种情况。三是在 Albi 系统的实现理论短时间内难于突破的情况下，考虑引入模式匹配型算法进行辅助计算，弥补 Albi 系统的实现缺陷。

下面给出 Albi 系统的 pmint 实现框架所需的底层函数及标识符列表，pmint 实现分析以及在 maTHmU 上的 Bug 列表，作为下一步工作的基础。

**1. pmint 实现框架所需的底层函数及标识符列表**

| | |
|---|---|
| Alternatives | |
| All | 是用于某些选项的设置，在 part 以及相关的函数中，All 指定特定层中的所有部分 |
| ArrayRules | |
| Array | [f,n]生成长度为 n，元素为 f[i]的列表 [f,(n1,n2,…)]生成嵌套列表 |
| ArcTan | |
| Apply | |
| Block | |
| Clear | Clear[symbol1,symbol2,…] clears values and definitions for the symbol |
| Collect | [expr,x] 把匹配 x 的对象的相同幂的项组合到一起 有扩展形式 |
| Csc | |
| Cot | |
| Cos | |
| ConstantArray | |
| ComplexExpand | |
| cofficient | |
| CoefficientList | |
| CoefficientArrays | |
| Catch | |
| Cases | |
| Derivative | 微分函数 |
| D() | gives the partial derivative |
| Denominator | 取分母 |
| DeleteCases | [expr,pattern] 移除 expr 中任意满足形式的元素，可以设置等级，数量 |
| Drop | [list,n] 去掉 list 的前 n 个元素 [list,-n]去掉 list 的后 n 个元素 [list,(n)] 去掉 list 的第 n 个元素还有其它的复杂结构 |
| Delete | |
| Dimensions | |
| Exp | |
| Extension | 是各种多项式和代数函数的一个可选项，它指定所使用的代数数域的生成器， 是一个非常重要的函数 |
| Exponent | [expr,form]用来给出 expr 展开式中出现的 form 的最大幂，[expr,form,h]把 h 应用到 expr 中 form 所出现的指数集上 |
| Evaluate | |
| Expand | |
| Except | |
| Extract | |
| Flattern | flattens out nested lists |
| Fold | 给出 Foldlist 的最后一个元素 |
| First | [expr] 返回表达式中的第一个元素 |
| FactorList | [list]给出一个多项式的因子及它们的指数组成的列表 |
| FactorSquareFreeList | [poly]给出了一个多项式的无平方因子以及它们的指数组成的列表 |
| Function | |
| FactorTermList | |
| FreeQ | |
| GCD | |
| HoldPattern | [expr]HoldPattern[expr] is equivalent to expr for pattern marching, but maintain expr in an unevaluated form |
| Head | [expr] 给出 expr 的头部 |
| Internal`SubresultantPRS | |

| 函数 | 说明 |
| --- | --- |
| Intersection | |
| Integer | |
| Im | |
| IntegerQ | |
| Join | 合并集合，可以设置等级 |
| Log | 自然对数函数 |
| Length | [expr]计数功能 gives the number of elements in expr |
| List | 列表 |
| Last | 返回 expr 的最后一个元素 |
| LinearSolve | [m,b]求解矩阵方程 mx==b 的 x |
| Listable | |
| Module | 指定 expr 中符号 x、y、… 出现的位置应被当作局部值. |
| Map | 将 f 作用于每一个元素上 |
| Max | 取最大值 |
| Most | [expr]给出去掉最后一个元素的 expr |
| Numerator | 返回分子 |
| Negative | [x]判断 x 是否是负数，若是则返回 True |
| NumberQ | |
| Outer | [f,list1,list2,…] 给出 listi 的广义外积，形成列表最底层元素的所有可能组合，并把它们作为 f 的自变量，有复杂结构 |
| OddQ | [expr]如果 expr 是一个奇数，则给出 True，否则给出 False |
| PolynomialQ | [expr,var] 如果 expr 是 var 的多项式，返回 true，否则返回 false |
| Power | 指数函数 |
| PolynomialLCM | 最小公倍数 |
| PolynomialQuotient | [p,q,x]求关于 x 的多项式 p 除以 q 的商，去掉余项 |
| PolynomialGCD | [poly1,poly2,…]给出多项式 polyi 的最大公约式<br>[poly1,poly2,…, Module->p]计算按素数 p 求模的最大公约数 |
| PolynomialQuotientRemainder | [p,q,x]作为 x 的多项式，给出 p 和 q 的商和余式的列表 |
| Print | [expr]输出 expr |
| PolynomialMod | |
| Position | |
| PadRight | |
| Quiet | [expr]在后台处理 expr，不输出多产生的任何信息，有复杂结构 |
| Range | 在一个范围内列出数表 |
| Return | 返回函数值 |
| Rest | [expr]移除第一个元素，返回剩余元素 |
| Reap | [expr] 给出表达式 expr 的值 |

| | | |
|---|---|---|
| Replace | | [expr,rules]应用一个规则或规则列表来转换整个表达式 expr |
| RootSum | | |
| RandomInteger | | |
| Reverse | | |
| Rational | | |
| RandomChoice | | |
| ReplacePart | | |
| Re | | |
| Resultant | | |
| Select | | [list,crit] 选择在 crit 函数作用下正确的变量 |
| Sequence | | [expr1,expr2,…] 将参数序列自动拼接到函数 |
| Scan | | [f,expr] 将 f 应用到 expr 的每个元素上并对其进行计算 |
| Sow | | |
| Sort | | [list]按标准次序对 list 元素进行排序，[list,p]用排序函数 p 对元素排序 |
| Split | | [list]将 list 分割为有相同元素构成的子列表 |
| SetAttributes | | |
| Sec | | |
| Sin | | |
| SparseArray | | |
| Thread | | |
| Times | | 连乘 |
| Together | | 计算和式并以公分母分式形式表示，并消去公因子 |
| Tan | | |
| Total | | 计算 list 中元素之和 |
| Table | | 列表运算 |
| Throw | | |
| Through | | |
| Union | | give a sorted list of all the distinct elements that appear in any of the listi |
| Variables | | [poly]给出在一个多项式中所有独立变量的列表 |
| With | | 用 x 等替代 expr 中出现的符号 |
| _ | | 可替代部分 |
| :> | | |
| $Failed | | aspecial symbol returned by certain functions when they cannot do what they were asked |
| lhs->rhs | | |

| | |
|---|---|
| @@ | 指数函数 |
| @@ | apply 作用的符号 |
| Null | 用来指明一个表达式或结果不存在，在普通输出中它不显示 |
| I | 虚数单位 |
| Slot # | |

在具体实现过程中可能略有偏差，需要针对不同的计算机代数系统进行调整。底层函数及标识符的证明说明了 Albi 实现的可移植性，减小具体实现过程中将要遇到的困难。使我们的工作具有一般性。同时底层函数及标识符的统计可能略有偏差，其中没有注释的函数可以通过 Mathematica 中的帮助菜单进行查找。

**2.pmint 实现的分析**

分析主要针对实现过程中的四个主要文件进行分析，分别是 AlgebraicPrelim.m，Misc.m，RationalIntegration.m 和 pmint.m 进行分析。分析的目的是为了帮助开发者更好的理解代码，并明确地说明和系统底层函数及标识符的依赖关系，以及内置函数的具体工鞡呢。

2.1 AlgebraicPrelim.m 分析

Limitations: Missing The Partial Fraction Decomposition's Algorithm

Comments:速度问题存在，现在版本虽然速度慢但是有效，为了避免高速版本出现错误，所以保留了现在的版本

【说明】Integration of Elementary Function，algorithms from book Symbolic Integration
只是 Chapter1 里的一些基本算法

【Kernel】

PolynomialQuotientRemainder [p,q,x]作为 x 的多项式，给出 p 和 q 的商和余式的列表

Variables [poly]给出在一个多项式中所有独立变量的列表,例

求出多项式变量的列表：

In[1]:= Variables[(x + y)^2 + 3 z^2 - y z + 7]

Out[1]= {x, y, z}

Print [expr]输出 expr

Exponent [expr,form] 用来给出 expr 展开式中出现的 form 的最大幂

PolynomialExtendedGCD [poly1,poly2,x]将 poly1 和 poly2 看成是关于 x 的单变量都像是，给出扩展的最大公因式

计算扩展的最大公约式：

In[1]:= {f, g} = {2 x^5 - 2 x, (x^2 - 1)^2};

In[2]:= {d, {a, b}} = PolynomialExtendedGCD[f, g, x]

Out[2]= $\{-1 + x^2, \{\frac{x}{4}, \frac{1}{4}(-4 - 2 x^2)\}\}$

第二部分给多项式一个线性组合的系数，使得该线性组合生成最大公约式：

In[3]:= a f + b g == d // Expand

Out[3]= True

Sort   [list]按标准次序对 list 元素进行排序，[list,p]用排序函数 p 对元素排序
    很活的例子：

In[1]:= Sort[{{a, 2}, {c, 1}, {d, 3}}, #1[[2]] < #2[[2]] &]

Out[1]= {{c, 1}, {a, 2}, {d, 3}}

All   是用于某些选项的设置，在 part 以及相关的函数中，All 指定特定层中的所有部分

利用 Part 提取第一列向量：

In[1]:= {{1, 2}, {3, 4}}[[All, 1]]

Out[1]= {1, 3}

利用 Take 提取直列矩阵：

In[2]:= Take[{{1, 2}, {3, 4}}, All, {1}]

Out[2]= {{1}, {3}}

例：

FactorSquareFreeList   [poly]给出了一个多项式的无平方因子以及它们的指数组成的列表

Split   [list]将 list 分割为有相同元素构成的子列表
    例 :

### 例 (1)

```
In[1]:= Split[{a, a, a, b, b, a, a, c}]
Out[1]= {{a, a, a}, {b, b}, {a, a}, {c}}
```

### 推广和延伸 (5)

分割为递增的子列表：

```
In[1]:= Split[{1, 2, 3, 4, 3, 2, 1, 5, 6, 7, 4, 3}, Less]
Out[1]= {{1, 2, 3, 4}, {3}, {2}, {1, 5, 6, 7}, {4}, {3}}
```

分割为递减的子列表：

```
In[2]:= Split[{1, 2, 3, 4, 3, 2, 1, 5, 6, 7, 4, 3}, Greater]
Out[2]= {{1}, {2}, {3}, {4, 3, 2, 1}, {5}, {6}, {7, 4, 3}}
```

基于第一个元素的分割：

```
In[1]:= Split[{1 → a, 1 → b, 2 → a, 2 → c, 3 → a}, First[#1] === First[#2] &]
Out[1]= {{1 → a, 1 → b}, {2 → a, 2 → c}, {3 → a}}
```

跳跃分割：

```
In[1]:= Split[Array[Prime, 20], #2 - #1 < 4 &]
Out[1]= {{2, 3, 5, 7}, {11, 13}, {17, 19}, {23},
         {29, 31}, {37}, {41, 43}, {47}, {53}, {59, 61}, {67}, {71}}
```

分割列表，使得每个子列表的连续元素不相同：

```
In[1]:= Split[{a, a, a, b, a, b, b, a, a, a, c, a}, UnsameQ]
Out[1]= {{a}, {a}, {a, b, a, b}, {b, a}, {a}, {a, c, a}}
```

Replace [expr,rules]应用一个规则或规则列表来转换整个表达式 expr

缺省下 Replace 将规则仅应用到整个表达式上：

```
In[1]:= Replace[x^2, x^2 -> a + b]
Out[1]= a + b
```

它并不作用到子集中：

```
In[2]:= Replace[1 + x^2, x^2 -> a + b]
Out[2]= 1 + x^2
```

Negative [x]判断 x 是否是负数，若是则返回 True
Array    [f,n]生成长度为 n，元素为 f[i]的列表  [f,(n1,n2,…)]生成嵌套列表

▼ 例 (3)

```
In[1]:= Array[f, 10]
Out[1]= {f[1], f[2], f[3], f[4], f[5], f[6], f[7], f[8], f[9], f[10]}

In[2]:= Array[1 + #^2 &, 10]
Out[2]= {2, 5, 10, 17, 26, 37, 50, 65, 82, 101}
```

生成一个 3×2 数组：

```
In[1]:= Array[f, {3, 2}]
Out[1]= {{f[1, 1], f[1, 2]}, {f[2, 1], f[2, 2]}, {f[3, 1], f[3, 2]}}
```

生成一个 3×4 数组：

```
In[2]:= Array[10 #1 + #2 &, {3, 4}]
Out[2]= {{11, 12, 13, 14}, {21, 22, 23, 24}, {31, 32, 33, 34}}
```

例：

Range  {$i_{max}$} 生成列表{1,2,…, $i_{max}$}，有复杂形式

OddQ  [expr]如果 expr 是一个奇数，则给出 True，否则给出 False

Internal`SubresultantPRS

【Definite】

PolyDivide

PolyPseudoDivide

newPolyPseudoDivide

HalfExtendedEuclidean

ExtendedEuclidean

HalfExtendEculidean

fastExtendedEculidean

SquareFree

SquareFreePositiveDegrees

squarefree

YunSquarefree

subResultant

fastSubREsultantPRS

【From other places】

lc

deg

NonZeroPolynomialQ

PolyContent

PolyPP

content

pp

【说明】

定义 PolyDivide 函数

定义 PolyPseudoDivide 函数：Euclidean Polynomial Pseudo-Division

定义 newPolyPseudoDivide 函数

定义 HalfExtendeEuclidean 函数

定义 ExtendedEuclidean 函数

定义 HalfExtendedEculidean 函数

定义 fastExtendedEculidean 函数

定义 SquareFree 函数

定义 SquareFreePositiveDegrees 函数

其中注释掉了一段函数 newSquareFree，应该是 SquareFree 函数的新版，但是应该有一些 bug

定义 Squarefree 函数　Musser's squarefree

定义 YunSquarefree 函数　Yun's squarefree

定义 subResultant 函数　返回有理系数多项式 A，B 的结式和子结式

定义 fastRubResultantPRS 函数

2.2 Misc.m 分析

【Kernel】

SetAttributes

Listable

Print

RandomInteger

cofficient

GCD

CoefficientList

FactorTermList

Times

Most

Through

Numerator

Denominator

Reverse

NumberQ

Expand

Intersection

Csc

Sec

Tan

Cot

Sin

Cos

Slot #

HoldPattern

Apply

SetAttributes

Block

CoefficientArrays

ArrayRules

Rational

Integer

Except

FreeQ

Delete

Dimensions

ConstantArray

PadRight

RandomChoice

Range

SparseArray

ArrayRules

ReplacePart

Log

_

ComplexExpand

Re

Im

Extract

Position

PolynomialQ

IntegerQ

Resultant

【Definite】

NoSolution

DRPrint

RandomPoly

RandomRF

【From other places】

deg

DebugRisch

lc

deg

content

Algebra`Polynomial`NestedTermsList

Subset

CLog

NonZeroPolynomialQ
NoSolution

【说明】
定义符号 NNoSolution=$Failed
定义 DRPrint
定义 RandomPoly
　　输入指数和系数的边界值
　　输出一个随机的关于 x 的多项式，指数为输入指数，系数为整数，系数的绝对值小于等于输入的系数的边界值

进行操作:返回一个随机的有理多项式，deg(num)=deg1,deg(den)=deg2,系数的绝对值小于等于输入的系数的边界值
实现：定义 RandomRF

返回多项式的积分
实现：？PolyInt[f_,x_]这种形式

返回 x 的多项式的首项系数

定义变量
content,pp

Returns the content and the primitive part of A as a poly in t
定义 PolyContentPP
定义 PolyContent
定义 PolyPP

定义 NumeratorDenominator
定义　NumeratorPrimitiveDenominator

定义 Derivation
定义 mainVar

定义 ZeroPolynomialQ

定义 SubSet[A_, B_] := Length[Intersection[A,B]] == Length[Union[A]];
目的是需要 Union 这个函数来应对应对一些情形：返回重复列表

定义 TrigToTan
【注释】这个对于多项式的操作，在目前为止没有任何限制，因为只使用于多项式情况，此函数并没有打算去掉，所以有必要进行改进

检测有理数

实现：定义 RationalNumberQ

检测有理函数
实现：定义 RationalFunctionQ

To delete colum of index 'index' of a matrix M
实现：定义 DeleteColum
　　　定义 Blockjoin
　　　定义 RandomMatrix
　　　定义 ZeroColumnQ
　　　定义 PaddedSparseJoin

注释掉的部分
*PaddedSparseJoin[a_,b_,level_:1] := (
  If[!FreeQ[{a,b,level},Join,Heads->True],
      Print["Faulty arguments ", {a,b,level}];
      1/0;
  ];
  1 /; False)
*)
(* To put together the logs.    This will be used in recognition of (Radical) Log Derivatives. *)
定义 Clog 函数
定义 CollectLog 函数

流程：To do Re[x+I y] = x instead of -Im[y]+Re[x]
定义 myRe 函数
定义 myIm 函数

流程：To extract the first vector with the first entry non-zero
定义 FirstWithNonZeroFirstEntry

流程：  To extract the first vector with the first two entries non-zero

流程：To extract the first vector with the first entry 0 and second entry non-zero
定义  FirstWithZeroFirstEntryAndNonZeroSecond 函数
定义  RationalMultipleCoefficients 函数

流程：
Return a non-zero rational number r such that the coefficients of t^i, i = n,..,m in q are the same as those of
　　p times r, otherwise returns $Failed.
　　Example: p = 1 + 7 x + 3 x^2 + 4 x^3 and
　　q = 2/3 + (14/3)x + 2 x^2 + (8/3) x^3

  then RationalMultipleCoefficients[p,q,x,1,3] returns 3/2 i.e. {3, 4} = (3/2){2, 8/3} *)

流程：Returns the linear factors of 'p' as a polynomial in 'z'
定义 PolynomialLinearFactors 函数
引用 Algebra`Polynomial`NestedTermsList

流程：Returns the integer roots of a polynomial 'p' in 'z'.
定义 IntegerRootsPolynomialQ 函数

流程：Returns the rational roots of a polynomial 'p' in 'z'
定义 RationalRootsPolynomialQ 函数

*********************Recognizing The Monomials Extensions *********************
定义 PrimitiveMonomialQ 函数
定义 HyperExponentialMonomialQ 函数
定义 HyperTangentMonomialQ 函数

*********************Recognizing Normal And Special Elements*********************
定义 SpecialElementQ 函数
定义 NormalElementQ 函数
定义 ReduceElementQ 函数
定义 SimpleElementQ 函数

*************Recognizing Log Derivatives And Log Derivatives Of k (t)-Radicals*************
**** Auxilary Functions for recognition of Log Derivatives and of Radical Log Derivatives*******
定义 ToCollectLog 函数

流程：Primitive Case for Log Derivatives
定义 PolynomialLogDerivativeTest 函数
定义 PrimitiveMonomialQ 函数
注释：This function has very specific entries and scope. It is intended to be used by LogarithmicDerivativeQ and NOT to be visible to the user. For that reason it doesn't have conditionals that check that the entries are valid.
定义 LogarithmicDerivative 函数

流程：HyperTangent Case for Log Derivatives
定义 PolynomialLogDerivativeTest 函数
定义 HyperTangentMonomialQ 函数
注释：This function has very specific entries and scope. It is intended to be used by LogarithmicDerivativeQ and NOT to be visible to the user. For that reason it doesn't have conditionals that check that the entries are valid.

流程：HyperExponential Case for Log Derivatives
定义 PolynomialLogDerivativeTest 函数

定义 HyperExponentialMonomialQ 函数

注释：This function has very specific entries and scope. It is intended to be used by LogarithmicDerivativeQ and NOT to be visible to the user. For that reason it doesn't have conditionals that check that the entries are valid.

定义 ParametricLogarithmicDerivative 函数

流程：Primitive Case for Radical Log Derivatives

定义 PolynomialRadicalLogDerivativeTest 函数

定义 PrimitiveMonomialQ 函数

注释：This function has very specific entries and scope. It is intended to be used by LogarithmicDerivativeQ and NOT to be visible to the user. For that reason it doesn't have conditionals that check that the entries are valid.

定义（或引用）RadicalLogarithmicDerivativeQ 函数

流程：HyperTangent Case for Radical Log Derivatives

定义 PolynomialRadicalLogDerivativeTest 函数

定义 HyperTangentMonomialQ 函数

注释：This function has very specific entries and scope. It is intended to be used by LogarithmicDerivativeQ and NOT to be visible to the user. For that reason it doesn't have conditionals that check that the entries are valid.

定义 RadicalLogarithmicDerivativeQ 函数

流程：HyperExponential Case for Log Derivatives

定义 PolynomialRadicalLogDerivativeTest 函数

定义 HyperExponentialMonomialQ 函数

注释：This function has very specific entries and scope. It is intended to be used by LogarithmicDerivativeQ and NOT to be visible to the user. For that reason it doesn't have conditionals that check that the entries are valid.

定义 ParametricLogarithmicDerivative 函数

*** End of Auxilary Functions for recognition of Log Derivatives and of Radical Log Derivatives***

Recognition of Logarithmic Derivatives

【注释】If the last two entries are empty, then we are in the constant field (call it C). So, D (c) = 0 for every element and therefore if f !=0 then it cannot be a logarithimic derivative of an element in C. Also, since we know that D (1) = 0 always (doesn't matter what type of derivation we have) then it is safe to assume the following.

定义 LogarithmicDerivativeQ 函数
定义 LogarithmicDerivativeQ 函数
定义 IntegerRootsPolynomialQ 函数
引用 ResidueReduce 函数
引用 PolynomialLogDerivativeTest 函数

【注释】If the last two entries are empty, then we are in the constant field (call it C). So, D (c) =

0 for every element and therefore if f !=0 then it cannot be a logarithimic derivative of C - radical. Also, since we know that D (1) = 0 always (doesn't matter what type of derivation we have) then it is safe to assume the following。

定义 RadicalLogarithmicDerivativeQ 函数

流程：Recognition of Log Derivatives for k (t)-radicals
定义 RadicalLogarithmicDerivativeQ 函数
引用 RationalRootsPolynomialQ 函数
引用 ResidueReduce 函数
*********End of Recognizing Log Derivatives and Log Derivatives of k (t)-radicals*************

*************************Recognizing    Derivatives*****************************
******************* Auxilary Functions for recognition of derivatives*****************
流程：primitve and hyperexponential case
定义 FirstDerivativeTest 函数
定义 PrimitiveMonomialQ 函数
定义 HyperExponentialMonomialQ 函数
定义 IntegratePolynomial 函数

流程：hypertangent case
定义 FirstDerivativeTest 函数
定义 HyperTangentMonomialQ 函数
引用 IntegrateHypertangentReduced 函数

流程：primitive case
定义  SecondDerivativeTest 函数
定义  PrimitiveMonomialQ 函数
注释：NOTE: This function has very specific entries and scope.  It is intended to be used by DerivativeQ and not by the users.
定义 LimitedIntegrate 函数

when deg (Dt) >= 1
定义 SecondDerivativeTest 函数

************** End of Auxilary Functions for recognition of derivatives ******************
流程：If the last two entries are empty, then we are in the constant field (call it C).   So, D (c) = 0 for every element    and therefore if f !=0 then it cannot be a derivative of an element in C.   Also, since we know that D (1) = 0 always (doesn't matter what type of derivation we have) then it is safe to assume the following.

定义 DerivativeQ 函数
定义 PrimitiveMonomialQ 函数

【注释，可以改进 的地方】

This function uses IntegratePolynomial which, as of today, is only implemented for Primitive monomials.

引用 RationalFunctionQ 函数

引用 Derivation 函数

引用 MonomialHermiteReduce 函数

引用 FirstDerivativeTest 函数

************************End of Recognizing of Derivatives ************************

*************** Recognizing    If Irreducible Specials are of the first kind******************

流程：Primitive Case

引用 IrrSpecialsAreOfFirstKindQ 函数

引用 定义 PrimitiveMonomialQ 函数

流程：HyperExponential Case

引用 IrrSpecialsAreOfFirstKindQ 函数

引用 定义 HyperExponentialMonomialQ 函数

【注释】In the HyperExponential case i.e. D[t] = a*t for a in k, we have Sirr = {t}.   Since the only root of t is alpha=0, then p_alpha[t] = (D[t]-D[alpha])/(t - alpha) = a*t/t = a.   So we only need to check that D[t]/t is not a logarithimic derivative of k-radicals. NOTE:   We need to implement RadicalLogarithmicDerivativeQ for Hyper Exponential Monomials

NOTE:    We need to implement RadicalLogarithmicDerivativeQ for Hyper Exponential Monomials

引用 定义 RadicalLogarithmicDerivativeQ 函数

【注释】In the HyperTangent case i.e. D[t] = a (t^2+1) for a in k, we have (assuming I = Sqrt[-1] not in k) Sirr = {t^2+1}. Since the only roots of t^2+1 are alpha (+-) = (+-)I, then p_alpha (+-)[t] = (D[t]-D[alpha])/(t - alpha) = a (t^2+1)/(t -+ I) = +- 2aI. So we only need to check that 2I (D[t]/(t^2+1)) = 2aI is not a logarithimic derivative    of a k (I)-radicals.

2.3 RationalIntegration.m 分析

Comments:速度问题存在，现在版本虽然速度慢但是有效，为了避免高速版本出现错误，所以保留了现在的版本

【说明】Chapter 2 的一些算法

********************************Chapter 2**********************************

【Kernel】

Function

Cases

RootSum

Evaluate

Catch

Throw

PolynomialMod

2ArcTan

【Definite】

HermiteReduce

IntRationalLogPart

LazardRiobooTrager

IntegrateRationalFunction

LaurentSeries

FullPartialFraction

LogToAtan

Horowitz-Ostrongadsky （被注释掉了）

slowIntRationalLogPart（被注释掉了）

IntRationalLogPart（被注释掉了）

【From other places】

PolyDivide

PolyContentPP

deg

fastExtendedEuclidean

PolyInt

NumeratorDenominator

IntRationalLogPart

fastSubResultantPRS

Square

NumeratorDenominator

ZeroPolynomialQ

PolyInt

DRPrint

****很重要的一个函数: Calculus`Albi`Rational`Log2ArcTan****************

【说明】

定义 HermiteReduce 函数

定义 IntRationalLogPart 函数

  注：

  ？奇怪的地方： First /@ DeleteCases[FactorList[Q],u_/;FreeQ[u,t]]

    假设：num 的次数小于 den 的次数，den 不等于 0，并且 den is squarefree and coprime with num

定义 LazardRiobooTrager 函数（Lazard—Rioboo—Trager algorithm）

定义 IntegrateRationalFunction 函数（有理函数积分，输入 f（属于 Q（x），返回它的积分值）

  中间有奇怪的注释： Get[NotebookDirectory[]<>"\\Albi\\Rational\\intSubRat/m"]

      ans(not simplified)

        通过 Calculus`Albi`Rational`Log2ArcTan[ans,x]输出答案

定义 LaurentSeries 函数
定义 FullPartialFraction 函数
定义 LogToAtan 函数
定义 Horwitz—Ostrongradsky Algorithm 函数（被注释掉了）
　　Given A,D in Q[x] with deg (A)< deg (D), D nonzero and
　coprime with A, return g,h in Q (x) such that A/D-dg/dx+h
　and h has a squarefree denominator. *)
定义 slowIntRationalLogPart 函数 Lazard-Rioboo-Trager algorithm（被注释掉了）
　　　Given A,d in Q[x] with deg (A)<deg (d), d nonzero, squarefree and coprime
　with A, return Integrate[A/d,x] *)
定义 slowIntRationalLogPart 函数（被注释掉了）
　　Lazard-Rioboo-Trager algorithm

## 2.4 pmint.m 分析
【准备，整体说明】
引入库
<<Misc.m
<<AlgebraicPrelim.m
<<RationalIntegration.m

This package was written by Luis A. Medina and Sasha Pavlyk.

Functions included in this package are:
pmint[f,x]

******************* Auxiliary Functions 辅助函数****************************
【程序流程】Auxiliary Functions 辅助函数
【Kernel】
Derivative 微分函数
Log 自然对数函数
_ 可替代部分

【definite】
exp 函数及规则

【说明】 定义 exp：1.exp 的微分 is exp　2.exp^(a+b)=exp^a*exp^b 3.exp^(log x)=x

[definite]candidateTower
candidateTower[f_,x_,t_]:=Module[{vars,terms,deriv,newF,limitList,list,F,dterms,count=0}

[Kernel] Module
Module [{x, y, ...}, expr]
指定 expr 中符号 x、y、... 出现的位置应被当作局部值.

[Kernel]Exp

[Kernel]:>

[Kernel]Select[list,crit] 选择在 crit 函数作用下正确的变量

[Kernel]PolynomialQ[expr,var] 如果 expr 是 var 的多项式，返回 true，否则返回 false

[Kernel]Union  **Usage**: give a sorted list of all the distinct elements that appear in any of the list$_i$

[Kernel]Flattern  **Usage**:flattens out nested lists

【特殊符号 Kernel】$Failed is aspecial symbol returned by certain functions when they cannot do what they were asked

[Kernel]D()  **Usage**:gives the partial derivative $\partial f/\partial x$

[Kernel]Map  **Usage**:Map[f,expr] of f/@expr applies f to each element on the first level in expr 将 f 作用于每一个元素上

Map[f,expr,levelspec]  **Usage**:applies f to parts of expr specified by levelspec

[Kernel] Range  在一个范围内列出数表

[Kernel] Length[expr]  **Usage**:计数功能 gives the number of elements in expr

[Kernel] Thread[f [args]]  **Usage**:"threads" f over any lists that appear in args 与 Map 相比，threads 更注重于单纯的匹配而不是实际的作用

【Rule】[Kernel] ->  lhs->rhs or lhs->rhs represents a rule that transform lhs to rhs

[Kernel] HoldPattern[expr]  **Usage**:HoldPattern[expr] is equivalent to expr for pattern marching, but maintain expr in an unevaluated form 模式的定义和模式相同性的判断

[Kernel] Power(^)  指数函数

[Kernel] Times  连乘

【符号】[Kernel] @@  apply 作用的符号

[Kernel] List {}  列表

[Kernel] Fold **Usage**：给出 Foldlist 的最后一个元素，Foldlist 是一个很奇怪的展开操作

[Kernel] Join: 合并集合，可以设置等级

**[From other places]**

myVariables

Subset

RationalFunctionQ

【说明】candidateTower： 1.F is the integrand ($\int f(x)\,dx$, $f(x)$ is the integrand) 2.x 是积分中的自变量 3.t 用于定义新的自变量为 t[1],t[2]…用于进行域扩张。

输出结果为 1.将 F 用新的变量表示出来 2.加入的新变量是为了积分的计算，在最终结果中还要换成最初的变量 3.最后的输出结果的最后两个元素的形式为{x,t[1],t[2],..} and {1,D[t[1]],D[t[2]],..}，目的是 produce the Total Derivation

计数器设置 <=15（这个应该是可以通过大量统计进行优化的），如果高于 15 次就结束，return Failed

[Kernel] Clear **Usage**:Clear[symbol1,symbol2,…] clears values and definitions for the symbol

**[From other places or definite]**
ToIndertsAndDerivation，在这里重新定义了一下 ToIndertsAndDerivation

【说明】先 clear ToIndertsAndDerivation，进行重新定义

[definite] ToTerms
[definite]denD
[Kernel]PolynomialLCM 最小公倍数
[Kernel]Denominator 取分母
[Kernel]Together 计算和式并以公分母分式形式表示，并消去公因子

【说明】定义 derivations_List derivations 是 t[i]微分的列表
************************ End of Auxiliary Functions ****************************

*************************Parallel RIsch*****************************************
【Kernel】
Return    Usage:返回函数值
DeleteCases   [expr,pattern]  移除 expr 中任意满足形式的元素，可以设置等级，数量
Rest [expr]移除第一个元素，返回剩余元素
Tan

【Definite】
pmint[f,x]

【From other places】
RationalFunctionQ
IntegrateRationalFunction
TrigToTan
ToIndetsAndDerivation
INT
denD 在上文的辅助函数部分定义了这个函数
getSpecial
ToTerms
pmIntegrate

【说明】This is a version of Parallel Integration.这是平行算法的一个版本
首先定义 pmint 函数

*************************得到 Darboux 多项式******************************
【Kernel】
With [{x=$x_0$,y=$y_0$, … …},expr]   用$x_0$等替代 expr 中出现的符号
Sequence[expr1,expr2,…] 将参数序列自动拼接到函数
Null 是一个符号，用来指明一个表达式或结果不存在，在普通输出中它不显示
First [expr] 返回表达式中的第一个元素

Scan [f,expr]  将 f 应用到 expr 的每个元素上并对其进行计算
【辨析】Map [f,expr]或 f/@expr  将 f 应用到 expr 中第一层的每个元素
　　　　Thread[f[args]]  将 f 线性作用于任意出现在 args 中的列表，线性这个说明很关键，下面以这个例子进行说明

```
In[3]:= Thread[f[{a, b, c}, {x, y, z}]]

Out[3]= {f[a, x], f[b, y], f[c, z]}
```

Last　　返回 expr 的最后一个元素
Max　　取最大值
Numerator　　返回分子
Total　　计算 list 中元素之和
Table　　列表运算
I　　虚数单位
【再次说明】Module[{x,y,…},expr]
　　　　指定 expr 中符号 x,y,…出现的位置应被当做局部值

Reap [expr]  给出表达式 expr 的值，
Sow

Collect [expr,x]  把匹配 x 的对象的相同幂的项组合到一起  有扩展形式  Collect[expr,{x1,x2,…}]
　　　　Collect[expr,var,h]
　　　　例: In[1]:=Collect[ax+by+cx,x]
　　　　　　Out[1]:=(a+c)x+by

Alternatives　　一个很神奇的东西  p1|p2|…是一个模式对象，用于代表任意模式 pi，例子
```
In[1]:= {a, b, c, d, a, b, b, b} /. a | b → x
Out[1]= {x, x, c, d, x, x, x, x}
```
Flatten　　[list]压平嵌套列表，有复杂结构[list,n],[list,n,h],[list{{s11,ss12,…},{s21,s22,…}}]
　　　　例 Flatten[{{a, b}, {c, {d}, e}, {f, {g, h}}}]={a,b,c,d,e,f,g,h}
DeleteCases　　[expr,pattern] 删除 expr 中与 pattern 匹配的所有元素，有复杂结构
　　　　[expr,pattern,levelspec],[expr,levelspec,n]
Outer　　[f,list1,list2,…] 给出 listi 的广义外积，形成列表最底层元素的所有可能组合，并把它们作为 f 的自变量，有复杂结构 Outer[f,list1,list2,…,n],Outer[f,list1,list2,..,n1,n2,…]
Quiet　　[expr]在后台处理 expr，不输出多产生的任何信息，有复杂结构[expr,{s1::t1,s2::t2,…}],[expr,"name"]例：

求值但不输出信息：

In[1]:= Quiet[1/0]

Out[1]= ComplexInfinity

求值且输出信息：

In[2]:= 1/0

Power::infy: Infinite expression $\frac{1}{0}$ encountered. »

Out[2]= ComplexInfinity

LinearSolve  [m,b]求解矩阵方程 mx==b 的 x

Head  [expr] 给出 expr 的头部，例

In[1]:= Head[f[a, b]]

Out[1]= f

In[2]:= Head[a + b + c]

Out[2]= Plus

In[3]:= Head[{a, b, c}]

Out[3]= List

---

In[1]:= Head[45]

Out[1]= Integer

In[2]:= Head[x]

Out[2]= Symbol

Drop  [list,n]  去掉 list 的前 n 个元素  [list,-n]去掉 list 的后 n 个元素  [list,(n)]  去掉 list 的第 n 个元素还有其它的复杂结构

FactorList  [list]给出一个多项式的因子及它们的指数组成的列表，例：

In[1]:= FactorList[x^2 - 1]

Out[1]= {{1, 1}, {-1 + x, 1}, {1 + x, 1}}

In[2]:= FactorList[2 x^3 + 2 x^2 - 2 x - 2]

Out[2]= {{2, 1}, {-1 + x, 1}, {1 + x, 2}}

Extension  是各种多项式和代数函数的一个可选项，它指定所使用的代数数域的生成器是一个非常重要的函数，例

**例 (2)**

Factor 是 $\mathbb{Q}[\sqrt{2}]$ 中的一个多项式：

In[1]:= `Factor[x^2 - 2, Extension -> Sqrt[2]]`

Out[1]= $-(\sqrt{2} - x)(\sqrt{2} + x)$

---

PolynomialGCD 在代数数产生域中表示当前系数：

In[1]:= `PolynomialGCD[x^2 - 2 Sqrt[3] x + 3, x^2 - 3, Extension -> Automatic]`

Out[1]= $\sqrt{3} - x$

Most   [expr]给出去掉最后一个元素的 expr
Squence   [expr1,expr2,…] 表示将参数序列自动拼接到函数，例

Sequence 自动拼接：

In[1]:= `f[a, Sequence[b, c], d]`

Out[1]= `f[a, b, c, d]`

【Definite】
tryInteger[f_,lv_List,ld_List,q_,cand_,lunk_,l1_,l2_,ls_,k_]
　　　k 是要加入的新变量

【From other places】
getSpecial
pmInteger
DRPrint
splitFactor
deflation
totalDeg
enumerateMonoms
INT
tryIntegral
Derivation
myFactors
PolyCoeffs

【这部分潜在的问题】
(* can we estimate the expected number of monomials ?    *)
(* we should not attempt solving the linear system if the number of such monomials exceeds 500.
-- Sasha *)

【说明】

根据 DRPrint 输出的内容可以大致推断程序的目的，这里输出的就是如果单项式的数目大于 800，就输出积分失败，当然积分本身是存在潜在的可行性的，只是可能需要太长的时间，800 这个限制是可调的，和前面的 15 是一样的

lu=Union[lunk, Table[BB[i].{I,Length[candlog]}]] (funky way of avoiding doing big Together…一种避免过大的有效的方法)

【截止标识】
enumerateMonoms[ {}, deg_ ]:= {1};
enumerateMonoms[lv_List,deg_] := Module[{i, v = Most[lv]},
(* Calculate all the monomials in variables "lv" such that the TOTAL deg
    is less than or equal to "deg" *)
     Union[enumerateMonoms[v,deg],
         Sequence @@ Table[Last[lv]^i*enumerateMonoms[v,deg-i], {i,deg} ]
     ]
]

*****************************************************************************
【Kernel】
PolynomialQuotient    [p,q,x]求关于 x 的多项式 p 除以 q 的商，去掉余项
PolynomialGCD     [poly1,poly2,…]给出多项式 polyi 的最大公约式 [poly1,poly2,…, Module->p]
            计算按素数 p 求模的最大公约数
Exponent    [expr,form]用来给出 expr 展开式中出现的 form 的最大幂，[expr,form,h]把 h 应用
            到 expr 中 form 所出现的指数集上，例

求 $x$ 的最高指数：

In[1]:= Exponent[1 + x^2 + a x^3, x]
Out[1]= 3

【Definite】
splitFactor  和之前的 splitFactor 不一样
mainVar

【From Other Places】
Derivation
PolyContentPP

【说明】
最开始定义的 splitFactor 和之前的 splitFactor 非常像，但是不是同一个

【Kernel】

【Definite】
deflation
Derivation
PolyContentPP

【From other places】
minVar

【说明】该部分的目的是 Calculate the deflation of p with respect to the "new derivation"

根据以上的分析及说明，整理，为之后的改进工作打下了一个良好的基础。需要注意的是上述实现只是 AIbi 实现的一种实现过程，具体的流程会依照实现的不同而略有差异。

### 3.在 maTHmU 内核上调试出现的问题
注：行号为代码行的大致位置
由于 maTHmU 内核的底层函数及标识符出现问题，所以并没有完成在 maTHmU 内核上的实现工作。主要出现的问题为识别，匹配函数出现了问题，例如一个典型问题为三个下划线___匹配模式的未实现所产生的 Bug 在整个程序中出现了多次。下面所列举的 Bug 为调试中出现的全部 Bug，为下一步开展内核的完善工作打下了基础。
***************************************************************
【Riboo.m】
在 Calculus 目录下注明了一处 Bug

***************************************************************
【Misc.m】
112-118
TanToTrig[tan[y_]/(1 + tan[y_]^2)] := (1/2) Sin[2y];
TanToTrig[2 tan[y_]/(1 + tan[y_]^2)] := Sin[2y];
TanToTrig[(1 - tan[y_]^2)/(1 + tan[y_]^2)]:= Cos[2y];
TanToTrig[1+tan[x_]^2]:= Sec[x]^2;
TanToTrig[(1 + tan[y_]^2)/(1 - tan[y_]^2)]:=Sec[2y];
TanToTrig[(1 + tan[y_]^2)/(2 tan[y_])]:=Csc[2y];
TanToTrig[(1 - tan[y_]^2)/(2 tan[y_])]:= Cot[2y];
TanToTrig[e_]:=e

经过调试，发现问题在于/的使用，像 TanToTrig[Tan[y_]] := (1/2) Sin[2y]是可行的

152 RationalNumberQ[___] := False
  修正为 RationalNumberQ[_] := False
  准确的说是三个下划线的情况没有实现

214 CLog[a_ Log[b_]] := Log[b^a];
  修正为 CLog[a_Log[b_]] := Log[b^a];去掉一个空格

```
*****************************************************************
```
【RischDE.m】

592 RischDE[___]:= NoSolution;
  修正为 RischDE[_]:= NoSolution;
  准确的说是三个下划线的情况没有实现

```
*****************************************************************
```
【ParametricProblems.m】

90  ParamRdeSpecialDenominator[___]:= NoSolution
  修正为 ParamRdeSpecialDenominator[_]:= NoSolution
  准确的说是三个下划线的情况没有实现

344 行  {h, DeleteCases[A, {0..}]}
384 行  {h, DeleteCases[A, {0..}]}
407 行  {h, DeleteCases[A, {0..}]}
462 行  {H, DeleteCases[AA, {0..}]}
550 行  indx = Intersection[Position[h, 0, {1}], Position[Transpose[AA], {0..}, {1}] ];
另外的 bugs 参见前一文档

```
*****************************************************************
```
【RischDE.m】

562 ParamRischDE[___]:=NoSolution;
  修正为 ParamRischDE[_]:=NoSolution;
  准确的说是三个下划线的情况没有实现

588 LimitedIntegrateReduce[___]:= NoSolution
  修正为 LimitedIntegrateReduce[_]:= NoSolution
  准确的说是三个下划线的情况没有实现

667 LimitedIntegrate[___] := NoSolution
  修正为 LimitedIntegrate[_] := NoSolution
  准确的说是三个下划线的情况没有实现

```
*****************************************************************
```
【SymbolicIntBook.m】
65 OrderNu[___] := $Failed
  修正为 OrderNu[_] := $Failed
  准确的说是三个下划线的情况没有实现

```
*****************************************************************
```
【pmint】
178 getSpecial[___] := Null;

修正为 getSpecial[_] := Null;
准确的说是三个下划线的情况没有实现

*************************************************************
【Integrate.m】
IntegrateList 出问题了

*************************************************************
<< Path["../Polynomial/mUPolynomial.m"]出问题了
在执行$SolutionFormula 这一行的时候出现问题了

## 2.3 Rubi 系统

如之前所说 Albi 系统在代数函数方面现在有一定的理论和实现的不完备性，例如在 pmint 实现的框架下，连最简单的 $\sqrt{x}$ 也是无法处理的，另外加之 pmint 实现只是对完整算法的一个近似实现，所以我们考虑在 Albi 系统的基础上，引入规则匹配式算法进行补充，最终达到建立起一个尽可能完善的符号积分系统的目的。

### 2.3.1 Rubi 系统综述

由于 Rubi 系统不是我们研究工作的重点，所以对于 Rubi 系统，我们不进行详细地介绍，只从总体出发，针对关键性地方进行阐述。

Rubi 系统的关键部分为：

   Ⅰ.Look-up tables     查表
   Ⅱ.Rule-based rewriting   规则设定
   Ⅲ.Algorithmic methods   算法

Ⅰ. Look-up tables

Rubi 系统的一个突出的优势地方就是采取了新的模式匹配的方法，采取一种辨识网的机制，对新型进行储存和检索，从而达到最快模式匹配的目的。例如，所有在数学表述上适用于 sin(u)这种形式的表达式都会被当做树状结构的一个分支，所有的微分规则在另一个分支，所有的积分规则在其它分支。

Ⅱ. Rule-based rewriting

常规的规则匹配的符号积分系统，由于实现起来较为粗糙简单，导致大量被积函数无法通过运算得到积分，同时可能因为规则设置不当的问题，导致出现死循环的情况，严重影响了系统的运行效果。Rubi 系统则是通过建立明确的规则，使用更为有效的模式匹配的方法大大优化了模式匹配理念的具体实施过程。

In RUBI (Rule-Based Integrator):

Rubi 系统中包含约 1400 约化规则 (A->B),，大小约为 554Kb。具体结构包括以下三个部分：

- Conditions: 1. 转换有效性条件  2. 约化限制条件
- Transformation
- Comments

另外规则还要满足：使用函数式定义、限制有效域、限于化简、使用局部变量、保持规则之间的互斥性。

以具体的规则为例说明 Rubi 设置规则的合理性：

$$N: m+1 \neq 0$$
$$T: \int (a+bx)^m dx \to \frac{(a+bx)^{m+1}}{(m+1)b}.$$

1.每一条规则形式简单，左右各有明确的数学表达，避免循环，条件分支结构的出现而导致规则的复杂化，表达清楚，简单，便于计算机代数系统进行匹配理解。

2.规则中将变量限制在特定区域，保证了变量定义的合法性，比如 $\sqrt{z^2} \to z$ 说这条规则使用的条件必须是"Z 是纯虚数或者位于复平面的右半平面"，避免了例如 Maple 中出现'square-root bug'的问题。

3.限制化简和对局部变量的引用。

4.规则之间相互独立，保证规则的添加，移除或修改不影响其它规则。一个规则作用的结果不会是另一条规则生效的条件。

以上的这些措施保证了变换规则定义的实用性，在规则定义方面起到了极大的优化效果。

下面使用规则的具体例子以及一部分规则：

例：

$$\int \frac{x^m \mathrm{d}x}{(a+bx)^{12}} \quad (m \in \mathbb{Z})$$

$$I(a,b,c,d,m,n) = \int (a+bx)^m (c+dx)^n \mathrm{d}x \quad (m,n \in \mathbb{Z}, a,b,c,d \in \mathbb{C})$$

规则：

1. **T**: $\int \dfrac{dx}{a+bx} \to \dfrac{\ln(a+bx)}{b}$ .
2. **N**: $m+1 \neq 0$
   **T**: $\int (a+bx)^m dx \to \dfrac{(a+bx)^{m+1}}{(m+1)b}$ .
3. **N**: $bc-ad = 0, m+n+1 = 0$
   **T**: $\int (a+bx)^m (c+dx)^n \, \mathrm{d}x \to (a+bx)^{m+1}(c+dx)^n \ln(a+bx)/b$ .
4. **N**: $bc-ad = 0, m+n+1 \neq 0$
   **T**: $\int (a+bx)^m (c+dx)^n \, \mathrm{d}x \to \dfrac{(a+bx)^{m+1}(c+dx)^n}{b(m+n+1)}$ .
5. **N**: $bc-ad \neq 0$
   **T**: $\int (a+bx)^{-1} (c+dx)^{-1} \, \mathrm{d}x \to \dfrac{\ln(a+bx) - \ln(c+dx)}{bc-ad}$ .
6. **N**: $bc-ad \neq 0, m+n+2 = 0, n \neq -1$
   **T**: $\int (a+bx)^m (c+dx)^n \, \mathrm{d}x \to -\dfrac{(a+bx)^{m+1}(c+dx)^{n+1}}{(n+1)(bc-ad)}$ .
7. **N**: $m+n+1 = 0, m > 0, bc-ad \neq 0$
   **T**: $\int (a+bx)^m (c+dx)^n \, \mathrm{d}x \to -\dfrac{(a+bx)^m}{dm(c+dx)^m}$
   $\qquad + \dfrac{b}{d}\int (a+bx)^{m-1}(c+dx)^{-m} \, \mathrm{d}x$ .
8. **N**: $bc-ad \neq 0, m+n+1 \neq 0, n > 0$
   **T**: $\int (a+bx)^m (c+dx)^n \, \mathrm{d}x \to \dfrac{(a+bx)^{m+1}(c+dx)^n}{b(m+n+1)}$
   $\qquad + \dfrac{n(bc-ad)}{b(m+n+1)}\int (a+bx)^m (c+dx)^{n-1} \, \mathrm{d}x$ .
   **S**: $(2n+m+1 < 0 \lor m+n+1 > 0) \land (m < 0 \lor n \leq m)$
9. **N**: $bc-ad \neq 0, n+1 \neq 0$
   **T**: $\int (a+bx)^m (c+dx)^n \, \mathrm{d}x \to -\dfrac{(a+bx)^{m+1}(c+dx)^{n+1}}{(n+1)(bc-ad)}$
   $\qquad + \dfrac{(m+n+2)b}{(bc-ad)(n+1)}\int (a+bx)^m (c+dx)^{n+1} \, \mathrm{d}x$ .
   **S**: $n < -1, m < 0 \lor 2m+n+1 \geq 0$.

Ⅲ. Algorithmic methods

在实际运行过程中，每一个分支将会根据其具体形式进行递归式的细分操作，采取这种机制，时间复杂度会保持在 log(n)的水平，n 取决于知识库中规则的数量。

当然不能否认的是，*Rubi* 的问题也是客观存在的，以下是一个被虚拟机发现的 *Rubi/Rubi*2（最新的 *Rubi* 版本为 *Rubi*3）中 Bug 的类型总结：

1.崩溃（Mathematica 内核崩溃，关闭）

2.几种类型结果糟糕的输出

3.（无限运行）Loops 在 10 个小时之后依然没有结果的输出

4.数学上错误的输出

5.基本数学性质的错误

6.$Rubi$产生自己的警告信息

7.高达几百兆的输出结果

8.不必要的大系数 10^500

9.独立的操作影响$Rubi$的输出结果

10.不能计算一些新的被积函数

Rubi 系统的流程非常清晰，被积分式—>知识库，并在知识库中反复进行递归匹配操作，虽然理念简单，但是规则设立的科学性和新模式匹配结构的使用，使 Rubi 系统的效率非常高，同时可积函数范围随着知识库的扩充而不断变大。在实际测试中，Rubi 系统以超过 99%的成功率明显优于 Maple，Mathematica 等主流计算机代数系统所采用的积分规则系统（2011年 9 月，Maple13，Mathematica 7）。当然这个测试有其片面性存在，但从这引出了一个符号积分系统甚至是计算机代数系统的一个新的分支—测试系统的建立，这将在 2.3.3 小节中进行详细的说明。

### 2.3.2 未来改进的方向

Ⅰ.动态规则加载

现阶段 Rubi 在运行的时候要将所有的规则（52 个脚本文件）加载进来，非常占用启动时间。规则的编写是以函数类型进行分类，所以在使用过程中大部分规则没有使用，只有和被积函数类型相关的一些规则在具体的操作过程中能够起到作用。

所以为优化 Rubi 系统的启动效果，考虑引入动态启动模式，虽然 Rubi 系统在实际运行中可能会调用多条规则，但是调用的规则都是与被积函数的形式相关的，或者是与被积函数的类型相同，或者是被积函数在进行规则转换过程中可能出现的函数形式，所以根据输入的被积分式类型就可以过滤掉相当一部分的无用规则，比如输入三角函数作为原函数时，就可以将特殊函数部分的规则过滤掉。

Ⅱ.Rubi 规则的添加

Rubi 系统能够解决的积分问题的范围严重依赖于规则中是否有针对该种情况的规则的存在，所以知识库的范围决定了 Rubi 系统解决积分问题的能力。这也是 Rubi 系统的局限性所在，Rubi 系统的理念就是针对可能出现的问题预备好处理方案，而不是完整的 Albi 系统在理论上可以针对出现的问题预备好处理的方法。当然，通常出现的实际问题的范围毕竟是有限的，所以 Rubi 系统这种思想也是有很强的使用价值和实现的便捷性的。所以增强 Rubi 系统的功能的关键就是在与进一步完善知识库。

由于 Rubi 系统的被动性，Rubi 系统的改良也相对被动。最好的改良方案就是通过不断的测试寻找具体的，不能解决的实际问题，再结合实际问题完成规则的添加，从而增强 Rubi 系统的功能。

这种优化并不是没有意义的，如果 Rubi 系统的实现效果能够进行进一步的改进，那么就可以在完整的积分系统的构建过程中，作为预处理和第一步处理采取的方案，而避免 Albi 系统在时间上损耗较大这个问题。这种改良对于积分系统的整体来讲，是非常有价值的。

### 2.3.3 新概念：建立测试系统

在刚刚的介绍中也提到了一个测试的问题。这个不定积分的测试包括超过 1 万个积分问题，包括有理函数，代数函数，初等函数和特殊函数积分。测试样例的生成是通过 $Rubi$ 实现的。随着新的规则被添加到 $Rubi$ 中，每一个方面关于新规则的测试样例也会加入到测试之中。

在这些文件中对比了三个系统的运行结果。根据 ADR 的反馈，$Rubi$ 产生了 0.2%的麻烦结果和 0.0%的错误结果。与之相对的，根据 ADR，在 Maple 13 的运行结果中，28.6%的结果形式是糟糕的，2.6%的结果是错误的。在 Mathematica 7 的运行结果中，这两项数值分别为 24.4%和 0.9%。以我们的观点看，0.2%和 0.0%这两个数字说明了甚至世界顶尖的程序员也没有能力识别自己代码中的错误。

因为测试样例是由 $Rubi$ 生成的，所以 $Rubi$ 在测试中的良好变现是自然的。但是毫无疑问的是仍然有很多 $Rubi$ 不能解决的积分问题。当然，在这里提供的和主流商业软件的对比结果依然是有一定参考价值的。

构造任意多的特定计算机代数系统可以解决而其它计算机代数系统不能解决的问题是简单的，所以当对比不同计算机代数系统的性能的关键是测试样例的中和性和在测试文件中的最佳结果的质量。

因此，一个不定积分测试的文件的建立是有必要的，每一个人可以贡献无冗余的问题和最佳的不定积分结果。那么当一个计算机代数系统的新版本发布的时候，公众可以通过这个文件对其进行测试。一个典型的计算机代数系统的实现者也会通过测试文件提高产品的质量，所以对于每一个这个提议都是有好处的。最终，这样的公共测试网站应该对于数学的所有领域都建立起来，从而优化计算机数学的功能和实现。

从这个例子中，我们可以认识到计算机代数系统的一个新的重要分支的萌芽：建立一个客观，丰富，质量较高的测试系统也是计算机代数系统的测试过程所要达到的一个新的层次。从例子中看出，现在的主流测试存在一个主要问题就是测试的客观性不能保证。构造任意多的特定计算机代数系统可以解决而其它计算机代数系统不能解决的问题是简单的。这样的测试的价值是有限的。其它的测试手段，比如将被积函数进行积分操作之后，再进行求导操作，并与被积函数进行比较。都具有一定的局限性，因为根据计算数学的基本原理，不存在比较任意两个数学表达式是否相同的方法。不能通过这种方法绕过一个完备测试系统建立这一环节。主流的计算机代数系统，如 Maple 都有自己的一套测试系统。并且在这种测试系统的环境下可以人为地改变积分结果的形式，从而以一个最优，最人性化的结果来进一步衡量积分系统的优劣性，从而进一步改良积分系统。

同时这种思想，即计算机代数系统和测试系统的对称建立也是一条可行的开发一个计算机代数系统的途径，可以扩展到计算机代数系统的各个功能上去，这说明了这种思想的普遍性。

## 2.4 如何建立一个完整的符号积分系统

由于现阶段 Albi 系统和 Rubi 系统各自的局限性，单靠其中一个系统实现一个较为完整的符号积分系统是不现实的。所以在 SAINT 和 SIN 的系统框架及刘维尔定理思想的启发下，自然产生一种结合两者之长建立完整符号积分系统的想法，这个想法的实现的关键就是合理调用两套系统，合理分配子任务。

为解决这个关键性问题，引入总控制台的理念，实现任务的合理划分。初始设计流程如下：

1.分别使用两套系统对问题进行独立求解

进行这一步骤的时候自然有一个调用顺序的问题，调用的顺序要依靠两套系统的实现情况的比较，需要比较的因素主要有，可积分的函数范围，结果表达形式的优劣以及时间花费这三个方面。这提供了设计具体步骤的一个自由度，同时这也为下一步的拆分提供了一个依据。就现阶段而言，Rubi 系统的使用应该优于 Albi 系统，另外如果被积函数是超越函数，那么自然优先调用 Albi 系统，反之是代数函数，自然优先调用 Rubi 系统。由此可见，比较的原则是很丰富的，既有客观的实现情况，更有对于具体问题的分析。所以这一步的实现类似于 Rubi 系统的规则比较，选择最优的分支进行操作

那么如果第一次尝试就得出了一个结果，为什么还要再调用第二套系统并进行之后的流程呢？这就涉及到一个实用性的问题。主要因为至少要进行一次积分结果优劣性的比较。在保证时间和结果的正确性的前提下，需要进行结果形式的比较。因为通过不同途径得到的结果，其形式上的差别是相当大的，当然其中会有一个结果是最接近与使用者心中的结果的。例如，下列情况：

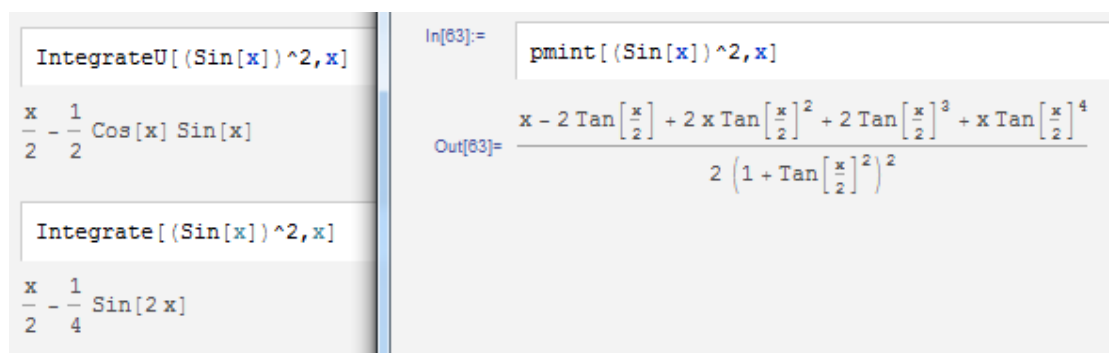

通过 Mathematica 内置的 Integrate 函数，Rubi 系统，Albi 系统结果的比较有明显的差别。当然如果时间因素是主导因素，自然越快输出越好，结果形式自然不那么重要了。

那么怎样比较结果的优劣性呢？衡量的标准自然多种多样：1.根据结果的长度进行判断，当函数的复杂程度相差不大的情况下，自然答案越短越好。2.根据一定规则进行判断，将使用者的标准转化为具体的评价函数，比如在上例中$\frac{x}{2} - \frac{1}{2}\text{Cos}[x]\text{Sin}[x]$和$\frac{x}{2} - \frac{1}{4}\text{Sin}[2x]$这两个结果，绝大多数情况下，我们认为第二个结果好一些，可以描述为：在使用函数复杂程度相差不多的情况下，使用的函数数目越少越好。在现阶段，实现了第一条衡量标准。

另外设计这一流程还有另外的一个原因，如果将被积函数进行分拆，可能会产生数学上的错误，例如$\int x^x dx$，$\int x^x \ln x dx$均非初等函数，但是$\int x^x(1+\ln(x))dx = x^x$为初等函数。

2.主控制台利用刘维尔定理对于积分结构的估计，将被积分式的结构进行分拆，根据一定的规则将子问题进行分配，利用两个子模块对不同的子问题分别进行处理。

分拆的主要对象还是和形式的被积函数，分拆的主要根据还是函数的类型，即函数是代数函数还是超越函数。

当然流程 2 是否调用，主要是根据流程 1 是否成功地得到一个答案和得到答案的形式是否足够令人满意来决定是否调用流程 2，毕竟还是要有一定的时间损失的。同样如果时间因素是主导因素，自然越快输出越好，结果形式自然不那么重要了。如果调用了流程 2，且得到一个结果，注意要和流程 1 得到的结果进行比较，判断结果形式的优劣性，最后输出最优的结果。

可见，总控制台的实际运行结果依赖于两套子系统实现的效果，如果 Albi，Rubi 两套系统能够产生一个自己所能产生的最好的结果，那么总控制台的运行效果将达到最优，时间的损失也会降低到最低。

当然,在最终答案输出之前,可以对结果在进行一次尝试化简,力争最优化的结果输出。

由此可见，符号积分模块在计算机代数系统中一定程度上不是独立存在的，算法效率，化简函数功能是否强大等各个因素都会影响符号积分系统的整体效果。

# 第三章

这一部分主要介绍的是对于 Albi 实现的一个改进方向及所需的数学知识，以及在前面提到的代数函数积分的一些基本理论知识和符号积分系统下一步研究的一些方向。

## 3.1. 平行$Risch$算法的一个扩展版本

由于完整的 Albi 理论还有不完善的地方，实现的过程也相对困难，在实现过程中常用$Risch$平行算法进行近似的实现。在我的实现过程中，就选择了 pmint 的思路进行实现，当然由于是近似的实现，衍生出了很多扩展版本即功能上的完善版本，换言之即是对于完整 Albi 系统的不同方向的逼近实现，其目的都是为了尽可能地发挥 Albi 系统的优势。在这里我们给出其中的一个扩展版本。

这个版本可以处理一大类递归版本无法处理的特殊函数。它们适用于一类特殊函数，该类函数满足一个常微分方程组，并且允许函数之间存在一定的代数依赖性。这个结果就产生了这种算法，例如，可以直接处理$sin$和$cos$，而不需要将他们重写为不同的函数。

**1.1.介绍**

令$f$为一个$x$的初等函数，换言之，只使用指数函数，对数函数，根式和标准的算术操作。用初等项表示不定积分问题是指判定$\int fdx$是否是一个初等函数，如果有的话，将$\int fdx$计算出来。这个问题首先由$Risch$解决，$Risch$提供的算法用$F(x)$的一系列塔状微分域扩张中的元素表示被积函数$f$，其中$F$是一个常数域

$$f \in K = F(x)(\theta_1)\cdots(\theta_n)$$

每一个$\theta_i$活着是一个超越指数式（换言之，$\theta_i'/theta_i = u', u \in F(x)(\theta_1)\cdots(\theta_{i-1})$)，或者是一个超越对数（换言之，$\theta_i' = u'/u, u \in F(x)(\theta_1)\cdots(\theta_{i-1})$）或者在$F(x)(\theta_1)\cdots(\theta_n)$上为代数的。

现在令$f = p/q$，其中$p$和$q$为$\theta_n$中的多项式，$p$和$q$的系数在$F(x)(\theta_1)\cdots(\theta_n)$中。算法通过将求$f$的不定积分问题简化为找到$F(x)(\theta_1)\cdots(\theta_n)$中函数的积分问题，但是算法解决问题的方式使高度非平凡的。这意味着算法必须用一种递归式的方法运行。

平行$Risch$算法有时被称为$Risch-Norman$算法，不是将$K$作为一个塔状扩张，而是作为一个同时用$x, \theta_1, \ldots, \theta_n$对$F$进行的扩张：

$$K = F(x, \theta_1, \ldots, \theta_n)$$

算法的思想是通过检查被积函数的分母来找到积分的分母和任意可能在积分中可能出现的对数项。接下来对数项的分子和常数因子可以通过解一个线性方程组得到，线性方程组由系数比较生成。

$Davenport$已经提出过允许$tan$直接在$Risch$算法的平行版本中使用这样的建议，目的是避免用复指数表达三角函数，这意味着引入一个$K$上的生成元$\theta$，满足$\theta' = 1 + \theta^2$。因为这既不是一个超越指数式或一个超越对数式，这个域扩张不适用于$Risch$算法的的经典版本。但是这只是一个将平行$Risch$算法的一种可能的向一个更大范围的函数类上扩张的例子。

1.1.1 函数满足一个常微分方程组

因为算法的平行版本不以递归式实现为基础，所以没有必要将$\theta_i'$限制到只依靠$x, \theta_1, \ldots, \theta_i$，我们允许它依赖于所有的$\theta_j$：

$$\theta_1' = R_1(x, \theta_1, \ldots, \theta_n)$$

$$\vdots$$
$$\theta'_n = R_n(x, \theta_1, \ldots, \theta_n)$$

$R_i$是$x$和$\theta_1, \ldots, \theta_n$的有理函数。这允许任意满足以上形式的一个非线性微分方程组的函数类的使用。已经注意到在平行$Risch$算法的情况下，$\theta_i$不仅限于单项式扩张，作者没有看到这一个在文献中提到的可能性，并且相信这个是一个新的观点。现在可能处理的函数包括满足一个线性微分方程（具有任意阶数）的任意方程，例如，这使得可以将$sin$和$cos$直接引入，通过

$$\theta'_1 = \theta_2$$
$$\theta'_2 = -\theta_1$$

而不是用$e^{ix}$或$tan\ x/2$将他们表示出来。在这种情况下，这是以引入域的生成元之间的代数相关性为代价做到的。注意到在这个例子中$\mathbb{Q}(x) \subset \mathbb{Q}(x,\theta_1) \subset \mathbb{Q}(x,\theta_1,\theta_2)$不是一个塔状的微分域扩张，因为$\theta'_1 \notin \mathbb{Q}(x,\theta_1)$。

其它能够以这种方式支持的函数为$elliptic$（椭圆积分），$Jacobian\ elliptic$（雅克比积分函数），$Bessel$（贝塞尔函数）和相关函数，还有$Lambert\ W$（朗伯函数）。用之前的算法处理的话，这些函数将会用它们的微分方程和他们之间的代数关系描述出来。通过以下的例子可以看到具体的细节。

### 1.1.2 代数相关性

我们方法的另一个创新点是有一步额外的消去过程，将微分域的生成元之间的代数关系生成的理想模掉，允许这样的代数扩张并且兼顾到函数之间的代数相关性，例如$\sin x$和$\cos x$。

例如，如果我们在微分域$\mathbb{Q}(x,y)$中进行函数处理，其中$y = \sqrt{1+x}$，那么$y^2$和$x+1$的表达式应该是视为一致的。因此我们在比较系数时，要考虑$y^2 = x+1$这个关系的存在。

我们通过计算代数关系的理想的$Gröbner$基来完成这一过程。在算法中，在系数比较之前，将多项式通过用$Gröbner$基简化为一个正常形式。

附注：
$Fitch$提到了他的实现可以处理更复杂的超越函数，如$Dilogarithms$函数，但是$\theta_i$只依赖于$\theta_j, j \leq i$。$Bronstein$的"$Poor\ Man's\ Integrator$"和推荐算法一样也是可以处理相当大的一类积分。

### 1.2.算法的轮廓

和$Risch$算法的递归版本一样，平行算法以刘维尔定理为基础

定理（强刘维尔定理）令$K$为一个微分域，$D$为微分，常数域为$F$，$\overline{F}$为$F$的代数闭包。令$f \in K$，假设存在在$K$上的初等函数$g$，$Dg = f$。那么存在$v_0 \in K, \lambda_1, \ldots, \lambda_n \in \overline{F}$，和$v_1, \ldots, v_n \in \overline{F}K$，满足

$$f = Dv_0 + \sum_{i=1}^{n} \lambda_i \frac{Dv_i}{v_i}$$

可以通过例1我们可以得到一个证明。从这个定理中我们可以得到，如果$f$有一个初等不定积分，那么

$$\int f dx = v_0 + \sum_{i=1}^{n} \lambda_i \log v_i$$

这给出了一个出发点，因为它限制了一个积分可能的形式。这个算法的轮廓概率地遵循[6]（附注：Keith O. Geddes and L. Yohanes Stefanus. On the Risch-Norman integration method and its implementation in MAPLE. In ISSAC '89 : Proceedings of the ACM-SIGSAM 1989 international symposium on Symbolic and algebraic computation, pages 212-217, New York, NY, USA, 1989. ACM.）

(1)被积函数改写为$f = p/q$，其中$p$和$q$为生成元$x, \theta_1, \ldots, \theta_n$组成的多项式

(2)分母$q$通过不可约因式分解进行检验

$$q = \prod_{i=1}^{m} q_i^{v_i}$$

(3)如果$D(1/q_i)$的分母整除$q_i$，那么令$v_i^* = v_i$，否则的话令$v_i^* = v_i - 1$

(4)$v_0$的分母变为

$$q_0 = \prod_{i=1}^{m} q_i^{v_i^*}$$

(5)每一个$q_i$产生相应的$v_i$作为一个对数项出现，但是附加的元素$v_{m+1}, \ldots, v_{m^*}$可能必须要考虑到。

(6)基于刘维尔定理，我们作出假设

$$\int f dx = \frac{u_0(x, \theta_1, \ldots, \theta_n)}{v_0} + \sum_{i=1}^{m^*} \lambda_i \log v_i(x, \theta_1, \ldots, \theta_n)$$

我们将$D$作用于等式两端，得到

$$\frac{p}{q} = D\frac{u_0(x, \theta_1, \ldots, \theta_n)}{v_0} + \sum_{i=1}^{m^*} \lambda_i \frac{Dv_i(x, \theta_1, \ldots, \theta_n)}{v_i(x, \theta_1, \ldots, \theta_n)}$$

(7)通过将方程乘以元素的公分母消去分母

(8)如果$x, \theta_1, \ldots, \theta_n$之间存在代数相关性，通过在等式两端模掉由代数相关性生成的理想的$Gröbner$基进行两端的化简。

(9)通过检验$u$，确定$u_0$次数的边界限制

(10)得到一个$\lambda_1, \ldots, \lambda_{m^*}$的线性方程组，$u_0$的系数通过方程两端的单项式匹配得到

(11)解出$\lambda_1, \ldots, \lambda_{m^*}$和$u_0$的系数

1.2.1 一个例子，假设我们希望计算出

$$\int sin^2 x dx$$

我们需要在$\mathbb{Q}(x, \theta_1, \theta_2)$域中进行计算，其中$\theta_1 = sinx, \theta_x = cosx$。描述代数关系的理想$I$为$I = (\theta_1^2 + \theta_2^2 - 1)$。为了达到多项式模掉$I$进行化简的目的，我们指定单项式的序为$\theta_1 < \theta_2$，所以$\theta_2^2$是$I$的生成元的首项。没有需要处理的分母，也不会出现对数项。为了使这个例子尽可能地短，我们只考虑恰好出现在积分中的单项式$x$和$\theta_1\theta_2$，并在我们的假设中省略其它的情况：

$$\int \theta_1^2 dx = \lambda x + \mu \theta_1 \theta_2$$

将微分$D$作用于等式两端，得到

$$\theta_1^2 = \lambda + \mu(\theta_2^2 - \theta_1^2)$$

用$I$对等式右端进行化简，得到
$$\theta_1^2 = \lambda + \mu(1 - 2\theta_1^2)$$
我们得到方程组
$$0 = \lambda + \mu$$
$$1 = -2\mu$$
有解$\lambda = -1/2$和$\mu = -1/2$，因此
$$\int sin^2 x dx = \frac{x}{2} - \frac{sinx cosx}{2}$$

### 1.2.2 杂散的对数项

*Davenport*在他的包含$tanx$的扩张中，必须要处理一种特殊情况，即在被积函数的分母中没有对数项相应的因子。
$$\int tanx dx = \tfrac{1}{2}\log(1 + tan^2 x)$$
*Davenport*处理这个问题的方式是假设$\log(1 + tan^2 x)$的存在，只要在被积函数含有因子$tanx$。这个过程完全是递归式的，并且只对这种特殊情况起作用。通过观察我们发现，在推荐扩张中出现的函数充当被积函数因子时，这种现象越来越频繁地出现。特别的是当代数相关性也包含在内时，例如
$$\int \frac{1}{sinx} = \frac{1}{2}\log(cosx - 1) - \frac{1}{2}\log(cosx + 1)$$

**1.3.现在版本的局限性**

对杂散的对数项的合适处理仍然需要研究，现在的实现在计算被积函数，如果被积函数需要表示相应的对数项，那么就会失败。

如果系数域包含一个参数，就会在分解多项多项式时失败。因此对于积分中参数的支持在目前还非常受限。在以下的例子中任意任意参数都会被一个特定的值代替。这不是算法本身限制，而是计算机代数系统提供的潜在的函数功能的限制。

同样的，目前，微分域的构造必须通过人工构造。一个用以对给定的被积表达式构造正确的微分域，并将表达式由域中元素重新表达的分析程序还没有完成。此外，任意必需的代数元必须一定包含在域中，现在如果需要添加新的代数元，域不会进行自动扩张。

**1.4.例子**

### 1.4.1 三角函数

当标准的$Risch$算法，用复指数$e^{ix}$表示三角函数，在平行算法的版本中已经以用$\theta = tan\frac{x}{2}$在进行三角函数表示，避免了引入代数元$i$，并使得结果与用户期待的形式更加贴近。其中
$$sinx = \frac{2\theta}{1+\theta^2}, cosx = \frac{1-\theta^2}{1+\theta^2}$$
现在$sin$和$cos$的积分会以$x$和$\theta$的有理函数的形式表示出来，这个结果并不是最想要得到的一个结果。新的推荐的扩张允许直接引入$sinx$和$cosx$，但是必须要引入代数相关性。

**1.4.1.1.** $x, sinx$和$cosx$的有理函数，我们定义域
$$F = \mathbb{Q}(x, sinx, cosx)/I$$
理想$I$是由关系$sin^2 x + cos^2 x - 1 = 0$定义。

**1.4.1.2.** 理论上可还原为椭圆积分，这一小节考虑的积分形式为

$$\int R(sinx, cosx, \Delta(x))dx$$

$\Delta(x) = \sqrt{1-k^2sin^2x}$，$R$是一个有理函数。接下来的讨论只考虑特殊情况$k^2 = 1/2$。为了处理尽可能多的情况，我们考虑以下定义的域

$$K = \mathbb{Q}(sinx, cosx, \Delta(x), F(x,k), E(x,k))/I$$

其中理想$I$由代数关系$sin^2x + cos^2x - 1 = 0$的左端和$\Delta(x)^2 - (1-k^2sin^2x) = 0$生成，$F$和$E$是第一型和第二型的不完备的椭圆积分：

$$F(x,k) = \int_0^x \frac{1}{\sqrt{1-k^w sin^2 t}}dt$$

和

$$E(x,k) = \int_0^x \sqrt{1-k^2sin^2t}dt$$

1.4.1.3.反三角函数。这一小节考虑下述形式的积分

$$\int R(x, \sqrt{a^2-x^2}, arcsin\frac{x}{a})$$

其中$R$是一个有理函数。因为在下列的例子中$a$的幂指数只在分子中引入，我们可以通过将其引入我们的域$\mathbb{Q}(a, x, \sqrt{1-x^2}, arcsinx)$进行参数的处理。

1.4.2 特殊函数的积分

上述的扩张特别适合于特殊函数$f_1, \ldots, f_n$的集合，$f_1, \ldots, f_n$满足下述形式的微分方程

$$\frac{d}{dx}f_i = R_i(x, f_1, \ldots, f_n)$$

这包括完整的椭圆积分，雅可比椭圆积分，贝塞尔和相关函数和马提厄函数。

1.4.2.1.完整的椭圆积分。第一型和第二型的完整的椭圆积分

$$K(k) = \int_0^{\frac{\pi}{2}} \frac{1}{\sqrt{1-k^2sin^2\theta}}d\theta$$

$$E(k) = \int_0^{\frac{\pi}{2}} \sqrt{1-k^2sin^2\theta}d\theta$$

满足

$$\frac{d}{dk}K(k) = -\frac{K(k)}{k} + \frac{E(k)}{k(1-k^2)}$$

$$\frac{d}{dk}E(k) = -\frac{K(k)}{k} + \frac{E(k)}{k}$$

1.4.2.2 雅可比椭圆积分

现在我们关注满足下列的微分方程组的雅可比椭圆积分$snu$, $cnu$和$dnu$

$$\frac{d}{du}snu = cnudnu$$

$$\frac{d}{du}cnu = -snudnu$$

$$\frac{d}{du}dnu = -k^2snucnu$$

代数相关性为
$$\text{sn}^2 u + \text{cn}^2 u = 1, \text{dn}^2 u = 1 - k^2 \text{sn}^2 u$$

**1.5.总结和未来的研究**

虽然平行的$Risch$算法版本和它的递归版本相比，认为是效率较差的，因为它不能判定一个给定的积分是否是初等的，但我们相信这种新的扩张形式使得它非常具有吸引力，因为在实际运用中，处理一大类积分的能力比证明一个初等表示形式不存在要重要的多。

因为这个方法和朴素的平行$Risch$算法都具有递归特性，所以要使其能够处理尽可能多的积分情况。从这一点上讲，失败产生的一个共同的原因是在假设中缺少对数项。因此之后主要的研究重点是如何自动判断在积分中可能出现的任意对数项。

现阶段只是在理论上提出了这种想法，并进行了一定的计算验证。虽然这个方法只是针对 pmint 框架给出了一个局部的优化，但是它的想法可以继续进行扩展。如果在之后构造出自动添加域的生成元的方法，对于提高这种方法的可行性，以及在解决相关问题方面都会提供很大的帮助。在这一节中设计到的主要数学知识$Gröbner$理论，在下一小节中进行介绍。

## 3.2. $Gröbner$理论

这一部分主要补充介绍上一小节中提到的$Gröbner$理论，并给出相应的算法。因为，在计算机代数系统中$Gröbner$理论应用广泛，了解这一部分内容也是有必要的。

$Gröbner$基是什么？

$Gröbner$基是一个有理想的算法特性的多元多项式的集合。每一个多项式集合可以转化为一个$Gröbner$基。这个过程形成了三个相似的技巧：解线性方程组的高斯消元法，计算两个单变量多项式的最大公约数的欧几里得算法，线性规划的单纯型算法。例如进行高斯消元法，输入一些线性方程，形式为
$$\mathcal{F} = \{2x + 3y + 4z - 5, 3x + 4y + 5z - 2\}$$
算法将$\mathcal{F}$转化为$Gröbner$基
$$\mathcal{G} = \{\underline{x} - z + 14, \underline{y} + 2z - 11\}\}$$
令$K$为任意一个域，例如实数域$K = \mathbb{R}$，复数域$K = \mathbb{C}$，有理数域$K = \mathbb{Q}$或者一个有限域$K = \mathbb{F}_p$。我们将$n$元多项式环记为$K[x_1, \ldots, x_n]$，其中多项式系数属于$K$。如果$\mathcal{F}$是任意一个多项式集合，那么由$\mathcal{F}$生成的理想为集合$\langle \mathcal{F} \rangle$，包括所有多项式的线性组合：
$$\langle \mathcal{F} \rangle = \{p_1 f_1 + \cdots + p_r f_r : f_1, \ldots, f_r \in \mathcal{F}, p_1, \ldots, p_n \in K[x_1, \ldots, x_n]\}$$
在我们的例子中集合$\mathcal{F}$和它的$Gröbner$基$\mathcal{G}$生成相同的理想：$\langle \mathcal{G} \rangle = \langle \mathcal{F} \rangle$。根据希尔伯特基定理，每一个$K[x_1, \ldots, x_n]$中的理想的形式都为$I = \langle \mathcal{F} \rangle$，也就是说理想是由多项式的一些有限集合$\mathcal{F}$生成的。

$k[x_1, \ldots, x_n]$上的一个序为所有单项式$x^a = x_1^{a_1} \cdots x_n^{a_n}$集合上的全序$\prec$，具有两个性质：
(1)可乘性：换言之，$x^a \prec x^b$说明对于所有的$a, b, c \in \mathbb{N}^n$，$x^{a+c} \prec x^{b+c}$。
(2)常数单项式是最小的：换言之，$1 \prec x^a, \forall a \in \mathbb{N}^n \setminus \{0\}$

一个次序的例子$(n = 2)$为次数的字典序
$$1 \prec x_1 \prec x_2 \prec x_1^2 \prec x_1 x_2 \prec x_2^2 \prec x_1^3 \prec x_1^2 x_2 \prec \cdots$$

如果我们确定了一个次序，那么每一个多项式$f$有唯一的一个初项$\prec(f) = x^a$。这是$\prec -$最高次单项式$x^a$，在$f$中系数非零。我们将$f$的项以$\prec -$递减的顺序写出，我们经常用下划

线标出首项。例如，一个二次多项式写为

$$f = \underline{3x_2^2} + 5x_1x_2 + 7x_1^2 + 11x_1 + 13x_2 + 17$$

现在假设$I$是$K[x_1,\ldots,x_n]$中的一个理想。那么在$(I)_\prec$中的初始理想由所有$I$中多项式的首项生成

$$in(I)_\prec = \langle in(f)_\prec : f \in I \rangle$$

$I$的一个有限子集$\mathcal{G}$是关于次序$\prec$的一个$Gröbner$基，如果$\mathcal{G}$中元素的首项能够生成初始理想：

$$in(I)_\prec = \langle in(g)_\prec : g \in \mathcal{G} \rangle$$

不存在成为一个$Gröbner$基的最低要求。如果$\mathcal{G}$是一个$I$的$Gröbner$基，那么$I$中包含$\mathcal{G}$的任意有限子集也是一个$Gröbner$基。为了补救无极小性，我们称满足以下条件的$\mathcal{G}$是一个约化$Gröbner$基

(1)对于$\forall g \in \mathcal{G}$，$in(g)_\prec$在$g$中的系数为1
(2)集合$\{in(g)_\prec : g \in \mathcal{G}\}$最小生成$in(I)_\prec$
(3)$\forall g \in \mathcal{G}g$，$g$的尾项不属于$in(I)_\prec$

在上述定义下，我们得到下列定理：如果序$\prec$是固定的，那么$K[x_1,\ldots,x_n]$中的任意理想$I$有唯一的一个$Gröbner$基。

约化$Gröbner$基$\mathcal{G}$可以通过集合$I$的任意生成集合，通过$Bruno\ Buchberger$在1965年论文中介绍的方法计算出来。$Buchberger$以他导师的名字$Wolfgang\ Gröbner$命名了这种方法。事后看来，$Gröbner$基的思想可以追溯到更早些时候，包括由永恒的理论家$Paul\ Gordan$在1900年写的一篇论文中就提到了$Gröbner$基的类似想法。但是Buchberger是第一个给出了计算$Gröbner$基的算法。

$Gröbner$基对于解多项式方程组是非常有效的。假设$K \subseteq \mathbb{C}$，令$\mathcal{F}$为一个$K[x_1,\ldots,x_n]$中多项式集合的有限集。$\mathcal{F}$的簇为所有所有零元素组成的集合，定义如下

$$\mathcal{V}(\mathcal{F}) = \{(z_1,\ldots,z_n) \in \mathbb{C}^n : f(z_1,\ldots,z_n) = 0, \forall f \in \mathcal{F}\}$$

如果我们另一个生成$K[x_1,\ldots,x_n]$中相同理想的多项式集合替换$\mathcal{F}$，簇不发生改变。尤其，理想$\langle \mathcal{F} \rangle$的约化$Gröbner$基有同样的性质。

$$\mathcal{V}(\mathcal{F}) = \mathcal{V}(\langle \mathcal{F} \rangle) = \mathcal{V}(\langle \mathcal{G} \rangle) = \mathcal{V}(\mathcal{G})$$

$\mathcal{G}$的优势在于它揭示了$\mathcal{F}$的簇的几何特性。第一个关于簇$\mathcal{V}(\mathcal{F})$的问题可能是簇是否非空。希尔伯特零点定理说明了

簇$\mathcal{V}(\mathcal{F})$是空的当且仅当$\mathcal{G}$和$\{1\}$等价

怎样计数一个给定方程组的零点个数？为了回答这个问题，我们需要更多的定义。给定$K[x_1,\ldots,x_n]$中一个固定理想$I$和一个序$\prec$，一个单项式$x^a = x_1^{a_1} \cdots x_n^{a_n}$称为是标准的，当且仅当它不属于初始理想$in(I)_\prec$。标准单项式的个数是有限的当且仅当每一个变量$x_i$在初始理想中以幂的形式单独出现。例如，如果$in(I)_\prec = \langle x_1^3, x_2^4, x_3^5 \rangle$，那么有六十个标准单项式，但是如果$in(I)_\prec = \langle x_1^3, x_2^4, x_1x_3^4 \rangle$，那么标准单项式的集合是无限集。

簇$\mathcal{V}(I)$是有限的当且仅当标准单项式的集合是有限的，且标准单项式的数目和$\mathcal{V}(\mathcal{I})$的势是等价的，其中零点是计数重数的。对于$n=1$，这就是代数基本定理，代数基本定理说明了$K[x]$中单变量次数为$d$的多项式$f$的簇$\mathcal{V}(f)$，包含$d$个复数。在这里单元素集合$\{f\}$是一个$Gröbner$基，标准单项式为$1, x, x^2, \ldots, x^{d-1}$。

我们判定一个簇是否是有限的准则生成了下列关于簇的维数的方程。考虑变量$\{x_1,\ldots,x_n\}$的子集$S$，使得$S$中变量的单项式部分不会出现在$in(I)_\prec$中，假设$S$是所有具有该性质的子集中势为最大的。这个最大的势$|S|$等价于$\mathcal{V}(I)$的维数。

标准多项式的集合是剩余环$K[x_1,\ldots,x_n]/I$的一个$K$维的向量空间基。多项式$p$模$I$的项可以唯一的表示为一个$K$维标准单项式的线性组合。这个表示形式是$p$的正规形式。计算

正规形式的算法就是辗转相除法。在只有一个变量$x$的相似情况下，其中$I = \langle f \rangle$，且$f$的次数为$d$，辗转相除法将任意多项式$p \in K[x]$写为$1, x, x^2, \ldots, x^{d-1}$的$K$维线性组合的形式。但是辗转相除法在涉及任意数目变元的$Gröbner$基的计算时都是有效的。

我们怎样检验给定的多项式集合$\mathcal{G}$是否是一个$Gröbner$基呢？考虑$\mathcal{G}$中的任意两个多项式$g$和$g'$，计算它们的$S-$多项式$m'g - mg'$，这里$m$和$m'$是可能存在的最小次数的单项式，满足$m' \cdot in(g)_\prec = m \cdot in(g')_\prec$。$S-$多项式$m'g - mg'$属于理想$\langle \mathcal{G} \rangle$。我们使用关于$m'g - mg'$可能的$Gröbner$基的辗转相除法。得到的结果的正规形式为一个$K$维单项式的线性组合，其中任意一个单项式不能被$\mathcal{G}$中的初始单项式整除。$\mathcal{G}$为一个$Gröbner$基的必要条件是

关于$\mathcal{G}$的正常形式$(m'g - ng;) = 0, \forall g, g' \in \mathcal{G}$

$Buchberger$规则强调了这个必要条件也是充分条件：一个多项式集合$\mathcal{G}$是一个$Gröbner$基当且仅当所有的$S-$多项式的正规形式都为0。从这个规则出发，我们可以得到计算任意输入集合$\mathcal{F}$的$Gröbner$基。

总之，$Gröbner$基和计算$Gröbner$基的$Buchberger$算法是代数中的基本概念。他们提供了在代数几何中更为有效的计算手段，例如消元理论，同调计算，奇点分析等等。给定多项式形式进行计算在科学和工程领域是广泛存在的。$Gröbner$基已经被各个领域的研究者使用，例如优化，编程，机器人学，控制理论，统计，分子生物学等等。

因为$Gröbner$基不是本书所要介绍的主要数学理论，所以下面引用相关材料中对于$Gröbner$基理论及相关算法的详细介绍，感兴趣的读者可以通过阅读进行进一步的了解。

## 13.3 Gröbner 基

Gröbner 基是 Buchberger 于 1965 年在其博士毕业论文中提出, 最初是用来解决多项式方程组的问题. 其后经过发展, 它在多元多项式环的理想等问题上也有重要的应用. [17] 一书对此有详细介绍, 另外 [174], [13] 对此也有介绍.

### 13.3.1 一些概念与介绍

我们主要是为了处理多元多项式而引入 Gröbner 基, 为了方便起见, 我们先给出一些记号上的说明.

设 $F$ 为一域, 以 $X$ 表示 $n$ 个不定元 $x_1, x_2, \ldots, x_n$, 则记多项式环 $R = F[X] = F[x_1, x_2, \ldots, x_n]$, 设有 $s$ 个多项式 $f_1, \ldots, f_s \in R$, 由它们生成的理想记作

$$I = \langle f_1, f_2, \ldots, f_s \rangle = \left\{ \sum_{1 \leq i \leq s} q_i f_i | q_i \in R \right\}.$$

**定义13.14.** 对于上面的理想 $I$, 定义其仿射簇为

$$V(I) = \{a = (a_1, a_2, \ldots, a_n) \in F^n | f_i(a) = 0, i = 1, 2, \ldots, s\}.$$

显然我们有 $V(f_1, f_2, \ldots, f_s) = \bigcap_{i=1}^{s} V(f_i)$, 且 $\forall f \in I(f(V(I)) = 0)$.

采用上面的记号, $f_1, \ldots, f_s$ 显然是 $I$ 的基, 我们知道, 在一元多项式环中, 由于其是主理想环, 我们有

$$\langle f_1, \ldots, f_s \rangle = \langle \gcd(f_1, \ldots, f_s) \rangle = \langle g \rangle,$$

且对于任何一个多项式 $f$, 将其对 $g$ 作 Euclid 除法得到 $f = qg + r$, 则 $f \in \langle g \rangle \Leftrightarrow r = 0$. 但对于多元情形, 这些良好的性质未必成立, 比如

$$\langle x, y \rangle \neq F[x, y] = \langle 1 \rangle = \langle \gcd(x, y) \rangle.$$

对于指标 $\alpha = (\alpha_1, \alpha_2, \ldots, \alpha_n) \in \mathbb{N}^n$, 定义 $X^\alpha = x_1^{\alpha_1} x_2^{\alpha_2} \cdots x_n^{\alpha_n}$, 称为单项式(Monomial), 将全体单项式集合记作 $M \subset R$.

我们可以定义 $M$ 中的一种良序 $<$, 使其满足与加法的和谐性, 即对任何三个指标 $\alpha, \beta, \gamma \in \mathbb{N}^n$, 有 $\alpha < \beta \Rightarrow \alpha + \gamma < \beta + \gamma$. 下面给出几个序的例子:

**例13.2.** $M$ 上的字典序(Lexicographic order) $<_{lex}$ : $\alpha <_{lex} \beta \Leftrightarrow \alpha - \beta$ 左边第一非零分量为负.

**例13.3.** $M$ 上的分级字典序(Graded lexicographic order)$<_{lex1}$:$\alpha <_{lex1} \beta \Leftrightarrow \|\alpha\| < \|\beta\| \vee (\|\alpha\| = \|\beta\| \wedge \alpha <_{lex} \beta)$, 其中 $\|\cdot\|$ 是 1-范数.

**例13.4.** $M$ 上的分级逆字典序(Graded reverse lexicographic order)$<_{lex2}$:$\alpha <_{lex2} \beta \Leftrightarrow \|\alpha\| < \|\beta\| \vee (\|\alpha\| = \|\beta\| \wedge \alpha - \beta$ 最右非零分量为零$)$.

我们一般取字典序即可, 就以 $<$ 来表示. 在该序下, 我们可以定义多项式 $f$ 的领项 $\text{lt}(f)$ 为 $f$ 最大的单项式, 类似地可定义领项系数 $\text{lc}(f)$ 和领项单项式 $\text{lm}(f)$. 一个多项式次数的定义为 $\deg f = \deg \text{lt}(f) \in \mathbb{N}^n$. 有了这些概念, 我们可以像在一元多项式环中那样做带余除法, 下面给出 $R$ 中带余除法的算法:

**算法13.2** (带余除法).

输入:$f, f_1, f_2, \ldots, f_s \in R$,

输出:$q_1, q_2, \ldots, q_s, r \in R$ 使得 $f = q_1 f_1 + \cdots + q_s f_s + r$ 且 $r$ 中任何单项不被 $\text{lt}(f_1), \ldots, \text{lt}(f_s)$ 中任何一个整除, 即不可再约化.

1. $r = 0, p = f, q_i = 0 (i = 1, \ldots, s)$,

2. 当 $p \neq 0$ 时, 循环做第 3 步,

3. 若存在某个 $i$ 使 $\text{lt}(f_i) | \text{lt}(p)$ 则

$$q_i = q_i + \frac{\text{lt}(p)}{\text{lt}(f_i)}, \quad p = p - \frac{\text{lt}(p)}{\text{lt}(f_i)} f_i,$$

否则 $r = r + \text{lt}(p), p = p - \text{lt}(p)$,

4. 输出 $q_1, q_2, \ldots, q_s, r$.

**例13.5.** 考虑 $f = x^2 y + xy^2 + y^2$, $f_1 = xy - 1$, $f_2 = y^2 - 1$.

**解:** 用上面的算法计算除法, 我们发现, 第一步只可以用 $f_1$ 来约化, 得到 $f = xf_1 + (xy^2 + x + y^2)$, 第二步我们可以用 $f_1$ 或 $f_2$ 来约化, 简单计算我们发现若这

一步用 $f_1$ 来约化, 得到结果

$$f = (x+y)f_1 + f_2 + (x+y+1),$$

反之则得到

$$f = xf_1 + (x+1)f_2 + (2x+1),$$

我们看到, 约化的顺序不同会导致结果不同, 这也是多元多项式环区别于一元多项式环的性质之一. ◇

**定义13.15.** 在算法 13.2 第 3 步中若选取满足领项能整除 $\mathrm{lt}(p)$ 的最小的指标 $i$ 对应的多项式进行约化, 则定义此时得到的 $r$ 为余式 $f\,\mathrm{rem}(f_1, f_2, \ldots, f_s) = r$.

由前面约化结果的不唯一性我们知道, 并不能用余式是否为零来判断一个多项式是否在所考察的理想中, 为了解决种种在多元多项式环中出现的问题, 我们需要引入 Gröbner 基.

### 13.3.2 单项式理想及一些准备定理

单项理想即是指由一些单项式生成的理想, 若 $A$ 是 $\mathbb{N}^n$ 的一个子集, 定义 $\langle x^A \rangle = \langle \{x^\alpha | \alpha \in A\} \rangle$.

**引理13.1.** $x^\beta \in I \Leftrightarrow \exists \alpha \in A(x^\alpha | x^\beta)$.

**引理13.2.** 设 $I$ 是一个单项理想, 以下三个命题是等价的:
   (1) $f \in I$,
   (2) $f$ 中每个单项都属于理想 $I$,
   (3) $f$ 是 $I$ 中某些多项式的 $F$-线性组合.

证明. (1)⇒(2) 设 $I = \langle x^A \rangle$, 则必有 $f = \sum_{\alpha \in A} q_\alpha x^\alpha$, 其中 $q_\alpha$ 是多项式. 由此可知 $f$ 中每个单项必可被某个 $x^\alpha$ 整除.

(2)⇒(3)和(3)⇒(1)均显然. □

由引理第二个等价条件得到:

**推论13.2.** 两个单项理想 $I_1, I_2$ 相等的充要条件是它们含有相同的单项式.

**定理13.7** (Dickson 引理). $\forall A \subset \mathbb{N}^n, \exists$ 有限集 $B \subset A$ 使得 $\langle x^A \rangle = \langle x^B \rangle$.

证明. 为了便于证明, 我们引入 $\mathbb{N}^n$ 上的偏序 $\preccurlyeq$ 满足 $\alpha \preccurlyeq \beta \Leftrightarrow \forall i \in \{1,2,\ldots,n\}(\alpha_i \leq \beta_i)$. 由此我们知道 $\alpha \preccurlyeq \beta \Leftrightarrow x^\alpha | x^\beta$.

设 $B$ 为 $A$ 的极小元集合, 即 $B = \{\beta \in A | \forall \alpha \in A(\alpha \not\prec \beta)\}$.

于是 $\forall \alpha \in A, \exists \beta \in B(\beta \preccurlyeq \alpha)$, 如若不然, 首先 $\beta \neq \alpha \Rightarrow \alpha \notin B$, 即 $\alpha$ 非极小元, 必存在 $\beta' \in B(\beta' \prec \alpha)$, 矛盾. 因此 $\forall \alpha \in A, \exists \beta \in B(x^\beta | x^\alpha)$, 即 $x^\alpha \in \langle x^B \rangle \Rightarrow \langle x^A \rangle \subset \langle x^B \rangle$. 又 $B \subset A \Rightarrow \langle x^B \rangle \subset \langle x^A \rangle$, 因而 $\langle x^A \rangle = \langle x^B \rangle$.

下面我们证明 $B$ 是有限集. 对于 $n = 1$ 的情况, 由于 $\preccurlyeq$ 是全序, 则 $B$ 是单元集, 命题显然成立. 下面假设命题对于 $n-1$ 的情况也是成立的, 命

$$A^* = \{(\alpha_1, \alpha_2, \ldots, \alpha_{n-1}) \in \mathbb{N}^{n-1} | \exists \alpha_n \in \mathbb{N}, (\alpha_1, \ldots, \alpha_n) \in A\},$$

则 $A^*$ 的极小元集 $B^*$ 是有限集. $\forall \beta^* = (\beta_1, \ldots, \beta_{n-1}) \in B^*$, 我们可取 $b_{\beta^*} \in \mathbb{N}$ 使得 $(\beta^*, b_{\beta^*}) \in A$, 由于 $B^*$ 的有限性, 我们可取最大值

$$b = \max\{b_{\beta^*} | \beta^* \in B^*\}.$$

于是 $\forall \alpha \in A, \exists \beta^* \in B^*$ 使得 $\beta^* \preccurlyeq (\alpha_1, \ldots, \alpha_{n-1})$, 假设 $\alpha_n > b$, 则

$$(\beta^*, b_{\beta^*}) \preccurlyeq (\beta^*, b) < \alpha,$$

即 $\alpha$ 不是极小元. 因此 $A$ 中任一极小元 $\alpha$ 必满足 $\alpha_n \leq b$, 那么 $\#B \leq (\#B^*) \times (b+1)$ 是有限的. 由归纳法, 本定理得证. $\square$

**引理13.3.** *$I$ 是 $R$ 中任一理想, 若 $G \subset I$ 是有限集且 $\langle \mathrm{lt}(G) \rangle = \langle \mathrm{lt}(I) \rangle$, 则 $\langle G \rangle = I$.*

证明. 设 $G = \{g_1, \ldots, g_t\}$, 则 $\forall f \in I$, 由带余除法可得到

$$f = q_1 g_1 + \cdots + q_t g_t + r,$$

其中 $r$ 不可再被 $G$ 约化. 而由于 $r = f - q_1 g_1 - \cdots - q_t g_t \in I$, 于是 $\mathrm{lt}(r) \in \mathrm{lt}(I) \Rightarrow \mathrm{lt}(r) \in \langle \mathrm{lt}(G) \rangle$, 即 $r$ 中每个单项都在 $\langle \mathrm{lt}(G) \rangle$ 中, 因此 $r = 0$, 则 $f \in \langle G \rangle \Rightarrow \langle G \rangle = I$. $\square$

由于任何单项理想均可有限生成, 我们有下面的:

**定理13.8** (Hilbert 基定理). *$R$ 中任何理想 $I$ 均可有限生成.*

**推论13.3** (理想升链定理). *设 $R$ 中有一理想升链 $I_1 \subset I_2 \subset \cdots \subset I_n \subset \cdots$, 则存在 $n \in \mathbb{N}(\forall m > n, I_m = I_n)$.*

这可由 $I = \bigcup_{i=1}^\infty I_i$ 是有限生成的得到. 满足这样条件的环也叫 Noether 环(Noetherian Domain).

### 13.3.3 Gröbner 基及其性质

现在引入 Gröbner 基的定义:

**定义13.16.** 设有多项式环 $R$ 中的理想 $I$ 和某一单项序 $<$, $I$ 的有限子集 $G$ 当满足 $\langle \text{lt}(G) \rangle = \langle \text{lt}(I) \rangle$ 时, 称为 $I$ 的 Grönber 基.

我们记理想 $I$ 的全体 Gröbner 基为 $GB(I)$, 即

$$GB(I) = \{G \in 2^I | G \text{是} I \text{的 Grönber 基}\}.$$

Gröbner 基的存在性是由 Hilbert 基定理和引理 13.3 保证的, 它有如下的性质:

**定理13.9.** 设 $G \in GB(I)$, $\forall f \in R$, 存在唯一的 $r \in R$ 使得 $f - r \in I$ 且 $r$ 中无单项可被 $\text{lt}(G)$ 中元素整除, 即不可被 $G$ 约化.

当考察的 $G$ 是 Gröbner 基时, 我们也记 $f \operatorname{rem} G = \overline{f}^G$ 或 $\overline{f}$.

证明. 存在性由带余除法算法得到, 对于唯一性, 可令 $f = h_1 + r_1 = h_2 + r_2$, 其中 $r_1, r_2$ 分别是两种不同途径约化的结果. 则 $r_1 - r_2 = h_2 - h_1 \in I \Rightarrow \text{lt}(r_1 - r_2)$ 可被 $\text{lt}(G)$ 中元素整除. 由 $r_1, r_2$ 的定义可知 $r_1 - r_2 = 0$. □

**推论13.4.** $f \in I \Leftrightarrow r = \overline{f} = f \operatorname{rem} G = 0$.

至此, 我们得到了 Gröbner 基的一个优美的性质.

构造理想的 Gröbner 基需要所谓的 S-多项式, 下面简要讨论之.

**定义13.17.** 对于非零多项式 $g, h$, 设 $\alpha = \deg g$, $\beta = \deg h$, $x^\gamma = \text{lcm}(x^\alpha, x^\beta)$, 即 $\gamma = (\max(\alpha_1, \beta_1), \ldots, \max(\alpha_n, \beta_n))$, 则定义 $g, h$ 的 S-多项式为

$$S(g, h) = \left( \frac{x^\gamma}{\text{lt}(g)} g - \frac{x^\gamma}{\text{lt}(h)} h \right) \in R.$$

**引理13.4.** 设 $g_1, \ldots, g_s \in R$, $\alpha_1, \ldots, \alpha_s \in \mathbb{N}^n$, $c_1, \ldots, c_s \in F \setminus \{0\}$,

$$f = \sum_{1 \leq i \leq s} c_i x^{\alpha_i} g_i \in R,$$

且有 $\delta \in \mathbb{N}^n$ 使 $\alpha_i + \deg g_i = \delta (1 \leq i \leq s)$, $\deg f < \delta$, 即这些多项式求和后领项消去(Leading term cancellation).

令 $x^{\gamma_{ij}} = \text{lcm}(\text{lm}(g_i), \text{lm}(g_j))$, 则存在 $c_{ij} \in F$ 使得 $x^{\gamma_{ij}} | x^\delta$ 且

$$f = \sum_{1 \leq i < j \leq s} c_{ij} x^{\delta - \gamma_{ij}} S(g_i, g_j),$$

且 $\deg x^{\delta - \gamma_{ij}} S(g_i, g_j) < \delta, (1 \leq i < j \leq s)$.

证明. 可假定 $\mathrm{lc}(g_i) = 1$, 否则可将其归并入 $c_i$ 中而使定理条件形式仍不变, 于是 $\mathrm{lt}(g_i) = \mathrm{lm}(g_i) = x^{\deg g_i}$. 由于 $x^\delta = x^{\alpha_i}\mathrm{lm}(g_i) = x^{\alpha_j}\mathrm{lm}(g_j)$, 则有 $x^{\gamma_{ij}}|x^\delta$.

由于
$$S(g_i, g_j) = \frac{x^{\gamma_{ij}}}{\mathrm{lt}(g_i)}g_i - \frac{x^{\gamma_{ij}}}{\mathrm{lt}(g_j)}g_j,$$
首项消去告诉我们 $\deg S(g_i, g_j) < \gamma_{ij}$, 因此 $\deg x^{\delta - \gamma_{ij}} S(g_i, g_j) \prec \delta$.

不妨设 $s \geq 2$, 则
$$\begin{aligned} g &= f - c_1 x^{\delta - \gamma_{ij}} S(g_1, g_2) \\ &= c_1 x^{\alpha_1} g_1 + c_2 x^{\alpha_2} g_2 + \sum_{3 \leq i \leq s} c_i x^{\alpha_i} g_i - c_1 x^{\delta - \gamma_{12}} \left( \frac{x^{\gamma_{12}}}{\mathrm{lt}(g_1)} g_1 - \frac{x^{\gamma_{12}}}{\mathrm{lt}(g_2)} g_2 \right) \\ &= c_1 (x^{\alpha_1} - x^{\delta - \deg g_1}) g_1 + (c_2 x^{\alpha_2} + c_1 x^{\delta - \deg g_2}) g_2 + \sum_{3 \leq i \leq s} c_i x^{\alpha_i} g_i \\ &= (c_1 + c_2) x^{\alpha_2} g_2 + \sum_{3 \leq i \leq s} c_i x^{\alpha_i} g_i, \end{aligned}$$
显然 $\deg g < \delta$, 由此时 $g$ 的形式和归纳法, 我们可以证明 $f$ 可以表成定理中的形式. $\square$

下面的定理给出了判别 Gröbner 基的一个充要条件:

**定理13.10.** $G = \{g_1, \ldots, g_s\} \in GB(I) \Leftrightarrow \forall (1 \leq i < j \leq s), S(g_i, g_j) \operatorname{rem} G = 0$.

证明. $\Rightarrow$. $S(g_i, g_j) \in I = \langle G \rangle \Rightarrow \overline{S(g_i, g_j)} = 0$.

$\Leftarrow$. 令 $f \in I \setminus \{0\}$, 由定义只需证明 $\mathrm{lt}(f) \in \langle \mathrm{lt}(G) \rangle$ 即可.

不妨设 $f = \sum_{1 \leq i \leq s} q_i g_i$, $\delta = \max\{\deg(q_i g_i) | 1 \leq i \leq s\}$, 则显然 $\deg f \leq \delta$. 倘若等号不成立, 即 $\deg f < \delta$, 定义
$$f^* = \sum_{1 \leq i \leq s, \deg(q_i g_i) = \delta} \mathrm{lt}(q_i) g_i,$$
它显然满足引理 13.4 的条件, 可写为 $S(g_i, g_j)$ 的线性组合, 因而其在 $G$ 下约化为零. 由带余除法, 存在 $q_i^*$ 使得 $f^* = \sum_{1 \leq i \leq s} q_i^* g_i$ 且 $\max\{\deg(q_i^* g_i) | 1 \leq i \leq s\} \leq \deg f^* < \delta$. 由 $f^*$ 的定义可知 $f - f^* = \sum_{1 \leq i \leq s} q_i^{**} g_i$, $\max\{\deg(q_i^{**} g_i) | 1 \leq i \leq s\} < \delta$. 于是 $f$ 也可以表示成
$$f = \sum_{1 \leq i \leq s} q_i g_i, \quad \max\{\deg(q_i g_i) | 1 \leq i \leq s\} < \delta,$$
这与 $\delta$ 的定义矛盾, 故 $\deg f = \delta$, 即 $\exists i$ 使 $\deg f = \deg(q_i g_i)$, 因此
$$\mathrm{lt}(f) = \sum_{1 \leq i \leq s, \deg(q_i g_i) = \delta} \mathrm{lt}(q_i) \mathrm{lt}(g_i) \in \langle \mathrm{lt}(G) \rangle.$$

证毕. $\square$

### 13.3.4 Buchberger 算法及约化 Gröbner 基

Buchberger 提出了 Gröbner 基的概念并给出了计算它的方法, 即下面的 Buchberger 算法:

**算法13.3** (Buchberger 算法).
输入:$f_1, \ldots, f_s \in R$,
输出:$I = \langle f_1, \ldots, f_s \rangle$ 的一个 Grönber 基 $G$.

1. $G = \{f_1, \ldots, f_s\}$,
2. 循环作后面所有步骤,
3. $S = \varnothing$, 将 $G$ 中元素编号为 $G = \{g_1, \ldots, g_t\}$,
4. 对于所有的序对 $(i, j), 1 \leq i < j \leq t$ 计算 $r = S(g_i, g_j) \operatorname{rem} G$, 若 $r \neq 0$ 则 $S = S \bigcup \{r\}$,
5. 若 $S = \varnothing$ 则输出 $G$, 否则 $G = G \bigcup S$.

**算法有效性.** 我们只需证明该算法循环是可以终止的, 因为终止时由定理 13.10 可知 $G$ 即是 Gröbner 基. 因为每步循环之后我们都可以得到一个新的集合 $G$, 我们给它们编上号, 记为 $G_1 \subset G_2 \subset \cdots$, 显然

$$\langle \operatorname{lt}(G_1) \rangle \subset \langle \operatorname{lt}(G_2) \rangle \subset \cdots,$$

由理想升链定理, $\exists n \in \mathbb{N}$ 使得 $\forall m > n$ 有 $\langle \operatorname{lt}(G_n) \rangle = \langle \operatorname{lt}(G_m) \rangle$, 此时 $\forall g, h \in G_n$, 令 $r = S(g, h) \operatorname{rem} G_n$, 则显然 $r = 0 \vee r \in G_{n+1}$, 于是 $\operatorname{lt}(r) \in \langle \operatorname{lt}(G_{n+1}) \rangle = \langle \operatorname{lt}(G_n) \rangle$, 由 $r$ 是多项式对 $G_n$ 的约化结果知道 $r = 0$, 于是算法终止. □

多项式理想的 Gröbner 基并不是唯一的, 而且在上面的算法中 $G$ 的规模的增长是十分迅速的, 求出的结果中可能含有大量的多项式, 因此我们要对 Gröbner 基做一定的优化, 下面我们一步一步对其进行约化化简.

**引理13.5.** 设 $G \in GB(I)$, $g \in G$ 且 $\operatorname{lt}(g) \in \langle \operatorname{lt}(G \setminus \{g\}) \rangle$, 则 $G \setminus \{g\} \in GB(I)$.

**证明.** $\operatorname{lt}(g) \in \langle \operatorname{lt}(G \setminus \{g\}) \rangle \Rightarrow \langle \operatorname{lt}(G \setminus \{g\}) \rangle = \langle \operatorname{lt}(G) \rangle = \langle \operatorname{lt}(I) \rangle \Rightarrow G \setminus \{g\} \in GB(I)$. □

**定义13.18.** 设 $G \in GB(I)$, 且 $\forall g \in G$, 有 $\text{lc}(g) = 1 \wedge \text{lt}(g) \notin \langle \text{lt}(G \setminus \{g\})\rangle$, 则称 $G$ 是 $I$ 的极小 Gröbner 基(Minimal Gröbner basis), 并简记为 $G \in MGB(I)$.

**定义13.19.** 设 $G \in MGB(I)$, 若对于 $g \in G$, $g$ 不可再被 $G \setminus \{g\}$ 约化, 即 $g$ 中任何一单项均不在理想 $\langle \text{lt}(G \setminus \{g\})\rangle$ 中, 则称 $g$ 关于 $G$ 是约化的. 若 $\forall g \in G$, $g$ 关于 $G$ 都是约化的, 则称 $G$ 是约化 Gröbner 基(Reduced Gröbner basis), 并简记为 $G \in RGB(I)$(或由下面的唯一性可记为 $G = RGB(I)$).

**定理13.11.** *每个多项式理想 $I$ 都有唯一的约化 Gröbner 基.*

*证明.* 存在性. 由引理 13.5 可将 $G$ 化为 $I$ 的极小 Gröbner 基, 设此时 $G = \{g_1, g_2, \ldots, g_s\}$, 然后对 $1 \leq i \leq s$ 归纳地做 $h_i = g_i \, \text{rem}\{h_1, \ldots, h_{i-1}, g_{i+1}, \ldots, g_s\}$. 由极小 Gröbner 基的条件知道 $\text{lt}(h_i) = \text{lt}(g_i), (1 \leq i \leq s)$. 于是由 $h_i$ 相对于 $\{h_1, \ldots, h_{i-1}, g_{i+1}, \ldots, g_s\}$ 约化可知其相对于 $G_s = \{h_1, \ldots, h_s\}$ 也是约化的.

唯一性. 设 $G, G^*$ 均是 $I$ 的约化 Gröbner 基, 则 $\forall g \in \text{lt}(G) \subset \langle \text{lt}(G)\rangle = \langle \text{lt}(G^*)\rangle$, $\exists g^* \in G^*$ 使 $\text{lt}(g^*)|\text{lt}(g)$, 同样地, $\exists g^{**} \in G$ 使 $\text{lt}(g^{**})|\text{lt}(g^*)$, 因此 $\text{lt}(g^{**})|\text{lt}(g)$, 由于约化 Gröbner 基也是极小 Gröbner 基, 我们知道 $\text{lt}(g) = \text{lt}(g^{**}) = \text{lt}(g^*) \in \text{lt}(G^*) \Rightarrow \text{lt}(G) \subset \text{lt}(G^*)$, 再由对称可证 $\text{lt}(G) = \text{lt}(G^*)$.

$\forall g \in G$, 取 $g^* \in G^*$ 使得 $\text{lt}(g) = \text{lt}(g^*)$. 由于 $G, G^*$ 约化, 则 $g - g^* \in I$ 中任一单项式均不能被 $\text{lt}(G \setminus \{g\}) = \text{lt}(G^* \setminus \{g^*\})$ 中元素约化. 于是 $g - g^* = g - g^* \, \text{rem} \, G = 0 \Rightarrow g = g^* \in G^* \Rightarrow G \subset G^* \Rightarrow G = G^*$. □

引理 13.5 给出了由 Gröbner 基求极小 Gröbner 基的方法, 定理 13.11 的存在性证明中也给出了极小 Gröbner 基构造约化 Gröbner 基的方法.

### 13.3.5 Buchberger 算法的两个改进

Buchberger 算法计算过程中集合 $G$ 中的多项式会越来越多, 呈指数规模增长, 因而需要对其作一定的优化. [17]3.3 节提出了两种改进的方法. 首先我们将算法 13.3 重新描述如下, 以便于我们后面叙述改进算法.

**算法13.4** (Buchberger 算法).

输入输出同算法 13.3,

1. $G = \{f_1, \ldots, f_s\}, H = \{\{i, j\} | i \neq j \wedge 1 \leq i, j \leq s\}$,

2. 当 $H \neq \varnothing$ 时循环做后面两步,

3. 任取 $h = \{i, j\} \in H, H = H \setminus \{h\}, r = S(f_i, f_j) \, \text{rem} \, G$,

4. 若 $r \neq 0$ 则 $f_{s+1} = r$, $H = H \bigcup \{\{i, s+1\}|1 \leq i \leq s\}$, $G = G\bigcup\{f_{s+1}\}$, $s = s+1$,

5. 输出 $G$.

注175. 设 $E = \{f_1,\ldots,f_s\}$, 如果我们要得到矩阵 $M$ 使得 $G = EM$, 那么首先令 $M_s = I_{s\times s}$, 然后在上面算法每次第 4 步判断成功后, 设带余除法给出 $r = S(f_i, f_j) \operatorname{rem} G = S(f_i, f_j) - q_1 f_1 - \cdots - q_s f_s$, 设 $S(f_i, f_j) = \sum_{1\leq k\leq s} S_k f_k$, 其中 $S_i = x^{\gamma_{ij}}/\operatorname{lt}(g_i)$, $S_j = -x^{\gamma_{ij}}/\operatorname{lt}(g_j)$, 其余的 $S_k = 0$. 则由

$$(f_1,\ldots,f_s,r) = (f_1,\ldots,f_s)\begin{pmatrix} 1 & & & S_1 - q_1 \\ & 1 & & S_2 - q_2 \\ & & \ddots & \vdots \\ & & & 1 & S_s - q_s \end{pmatrix} =: (f_1,\ldots,f_s)Q_s,$$

可知有迭代 $M_{s+1} = M_s Q_s$, 输出最后的 $M = M_s$ 即可.

注176. 产生极小 Gröbner 基时只要从 $M$ 中去掉相应的列, 而对于约化过程, 同样由带余除法得到 $q_1,\ldots,q_s$, 乘以相应的迭代矩阵.

## 第一个改进

**引理13.6.** 设有多项式 $f_1,\ldots,f_s$ 和 $d$, $I = \langle f_1,\ldots,f_s\rangle$, $J = \langle f_1 d,\ldots,f_s d\rangle$, 则 $\{f_1,\ldots,f_s\} \in GB(I) \Leftrightarrow \{f_1 d,\ldots,f_s d\} \in GB(J)$.

此引理由 Gröbner 基的定义 $\langle \operatorname{lt}(G)\rangle = \langle \operatorname{lt}(I)\rangle$ 可证.

**引理13.7.** 设有非零多项式 $f, g$, $I = \langle f, g\rangle$, $d = \gcd(f, g)$, 则下面两个条件等价:
(1) $\operatorname{lm}(f/d)$ 与 $\operatorname{lm}(g/d)$ 互素,
(2) $S(f, g) \operatorname{rem}\{f, g\} = 0$, 即 $\{f, g\} \in GB(I)$.

证明. (1)$\Rightarrow$(2). 先设 $d = \gcd(f, g) = 1$, 且 $f = aX + f'$, $g = bY + g'$, 其中 $\operatorname{lc}(f) = a, \operatorname{lm}(f) = X, \operatorname{lc} g = b, \operatorname{lm}(g) = Y$, $f', g'$ 是余下的部分, 于是

$$X = \frac{f - f'}{a}, \quad Y = \frac{g - g'}{b}.$$

(I)若 $f' = g' = 0$, 则 $S(f, g) = 0$,
(II)若 $f' = 0, g' \neq 0$, 由 $\gcd(\operatorname{lm}(f), \operatorname{lm}(g)) = 1$ 得

$$S(f, g) = \frac{1}{a}Yf - \frac{1}{b}Xg = \frac{1}{ab}(g - g')f - \frac{1}{ab}fg = -\frac{1}{ab}g'f,$$

若它能被 $g$ 约化, 则由 $\operatorname{lm}(g)|\operatorname{lm}(g'f) \wedge \gcd(\operatorname{lm}(g),\operatorname{lm}(f)) = 1$ 知 $\operatorname{lm}(g)|\operatorname{lm}(g')$, 这与 $\deg g' < \deg g \wedge g' \neq 0$ 矛盾.

(III)若 $f' \neq 0, g' = 0$, 这与情形(II)类似.

(IV)若 $f' \neq 0 \wedge g' \neq 0$, 此时经过计算可知
$$S(f,g) = \frac{1}{ab}(f'g - g'f),$$

我们断言 $\operatorname{lm}(f'g) \neq \operatorname{lm}(g'f)$. 假设二者相等, 则 $\operatorname{lm}(f')\operatorname{lm}(g) = \operatorname{lm}(g')\operatorname{lm}(f)$, 由于 $\operatorname{lm}(f)$ 和 $\operatorname{lm}(g)$ 互素, 我们得到 $\operatorname{lm}(f)|\operatorname{lm}(f')$, 矛盾. 不妨设此时 $\operatorname{lm}(f'g) > \operatorname{lm}(g'f)$, 则第一步仅可通过 $g$ 约化, 其结果为
$$S(f,g) \to \frac{1}{ab}[(f' - \operatorname{lt}(f'))g - g'f],$$

对于相反的情形如法炮制, 得到第一步的约化结果仍然可进行同样的讨论, 只需将其中的 $f' - \operatorname{lt}(f')$ 看作 $f'$ 即可. 于是这样的约化过程可一直继续, 直至约化为 $0$.

综上, 我们在 $d = 1$ 的情况下证明了 $(1)\Rightarrow(2)$. 当 $d \neq 1$ 时, 我们有 $\gcd(f/d, g/d) = 1$, 于是 $S(f/d, g/d) \operatorname{rem}\{f/d, g/d\} = 0$, 即 $\{f/d, g/d\}$ 是 Gröbner 基, 由引理 13.6 可知 $\{f, g\}$ 也是 Gröbner 基, 命题得证.

$(2)\Rightarrow(1)$. (I)首先我们仍然设 $d = \gcd(f, g) = 1$, 取单项式 $D, X, Y \in M$ 满足
$$\operatorname{lm}(f) = DX, \quad \operatorname{lm}(g) = DY, \quad \gcd(X, Y) = 1,$$

于是 $S(f, g) = \frac{Y}{\operatorname{lc}(f)} f - \frac{X}{\operatorname{lc}(g)} g$, 由假设 $\{f, g\}$ 是 Gröbner 基, 我们进行带余除法可以得到多项式 $u, v$ 使得 $S(f, g) = uf + vg$, 且 $\operatorname{lm}(uf), \operatorname{lm}(vg) \leq \operatorname{lm}(S(f, g))$. 整理可得
$$\left(\frac{X}{\operatorname{lc}(g)} + v\right) g = \left(\frac{Y}{\operatorname{lc}(f)} - u\right) f,$$

因此 $g|(Y/\operatorname{lc}(f) - u)$, 又
$$\operatorname{lm}(u)DX = \operatorname{lm}(uf) \leq \operatorname{lm}(S(f,g)) < \operatorname{lm}(Yf) = \operatorname{lm}(Xg) = DXY \Rightarrow \operatorname{lm}(u) < Y,$$

则 $g|(Y/\operatorname{lc}(f) - u) \Rightarrow DY|Y \Rightarrow D = 1$, 亦即 $\operatorname{lm}(f), \operatorname{lm}(g)$ 互素.

(II)当 $d \neq 1$ 时, 则 $\gcd(f/d, g/d) = 1$, 于是由 $\{f, g\}$ 是 Gröbner 基可知 $\{f/d, g/d\}$ 是 Gröbner 基, 因而 $\operatorname{lm}(f/d), \operatorname{lm}(g/d)$ 互素. □

**定理13.12.** $S(f, g) \operatorname{rem}\{f, g\} = 0$ 的一个充分条件是 $\gcd(\operatorname{lm}(f), \operatorname{lm}(g)) = 1$.

**注177.** 只需注意到 $\gcd(\operatorname{lm}(f), \operatorname{lm}(g)) = 1 \Rightarrow \gcd(f, g) = 1$.

我们可以由此得到 Buchberger 算法的第一个修正, 在计算 S-多项式并约化之前由 $\operatorname{lm}(f), \operatorname{lm}(g)$ 是否互素来决定是否要计算.

**第二个改进**

**定义13.20.** 对于多项式 $f_1,\ldots,f_s$, 设 $f=(f_1,\ldots,f_s)\in R^s$, 定义 $\operatorname{Syz}(E)=\{h=(h_1,\ldots,h_s)\in R^s|h\cdot f=0\}$.

**定理13.13.** 设 $c_1,\ldots,c_s\in F\setminus\{0\}$, $X_1,\ldots,X_s\in M$, $X_{ij}=\operatorname{lcm}(X_i,X_j)$, 则 $\operatorname{Syz}(c_1X_1,\ldots,c_sX_s)$ 由

$$\left\{\tau_{ij}=\frac{X_{ij}}{c_iX_i}e_i-\frac{X_{ij}}{c_jX_j}e_j\in R^s|1\le i<j\le s\right\}$$

生成, 其中 $e_i,e_j$ 为 $R^s$ 中的自然基矢.

证明. 首先易验证 $\langle\tau_{ij}\rangle\subset\operatorname{Syz}(c_1X_1,\ldots,c_sX_s)$. 现在假设 $h=(h_1,\ldots,h_s)\in\operatorname{Syz}(c_1X_1,\ldots,c_sX_s)$, 于是有

$$h_1c_1X_1+\cdots+h_sc_sX_s=0.$$

考虑任一单项式 $X\in M$, 则在上式中含 $X$ 的同类项合并之后为 0, 因此我们可只考虑这些项, 将其分离出来. 可设 $h_i=c_i'X_i'$, 其中 $c_i'=0$ 或 $X_i'X_i=X$, 设 $c_{i_1}',\ldots,c_{i_t}'$ 是 $c_i'$ 中不为零的系数重新编号, 于是有 $c_1'c_1+\cdots+c_s'c_s=c_{i_1}'c_{i_1}+\cdots+c_{i_t}'c_{i_t}=0$, 则

$$\begin{aligned}h=&(h_1,\ldots,h_s)=c_{i_1}'X_{i_1}'e_{i_1}+\cdots+c_{i_t}'X_{i_t}'e_{i_t}\\=&c_{i_1}'c_{i_1}\frac{X}{c_{i_1}X_{i_1}}e_{i_1}+\cdots+c_{i_t}'c_{i_t}\frac{X}{c_{i_t}X_{i_t}}e_{i_t}\\=&c_{i_1}'c_{i_1}\frac{X}{X_{i_1i_2}}(\frac{X_{i_1i_2}}{c_{i_1}X_{i_1}}e_{i_1}-\frac{X_{i_1i_2}}{c_{i_2}X_{i_2}}e_{i_2})\\&+(c_{i_1}'c_{i_1}+c_{i_2}'c_{i_2})\frac{X}{X_{i_2i_3}}(\frac{X_{i_2i_3}}{c_{i_2}X_{i_2}}e_{i_2}-\frac{X_{i_2i_3}}{C_{i_3}X_{i_3}}e_{i_3})+\cdots\\&+(c_{i_1}'c_{i_1}+\cdots+c_{i_{t-1}}'c_{i_{t-1}})\frac{X}{X_{i_{t-1}i_t}}(\frac{X_{i_{t-1}i_t}}{c_{i_{t-1}}X_{i_{t-1}}}e_{i_{t-1}}-\frac{X_{i_{t-1}i_t}}{c_{i_t}X_{i_t}}e_{i_t})\\&+(c_{i_1}'c_{i_1}+\cdots+c_{i_t}'c_{i_t})\frac{X}{c_{i_t}X_{i_t}}e_{i_t}.\end{aligned}$$

注意到上式最后一项为零, 得证. $\square$

**定理13.14.** 设 $G=\{g_1,\ldots,g_s\}$, $B\{\tau_{ij}|1\le i<j\le s\}$ 是 $\operatorname{Syz}(\operatorname{lt}(g_1),\ldots,\operatorname{lt}(g_s))$ 的生成元集, 则 $G\in GB(\langle G\rangle)\Leftrightarrow\forall h=(h_1,\ldots,h_s)\in B(h_1g_1+\cdots+h_sg_s\operatorname{rem}G=0)$.

证明. 注意到 $\tau_{ij}\cdot(g_1,\ldots,g_s)=S(g_i,g_j)$, 则命题显然. $\square$

**引理13.8.** 记 $X_{ij} = \mathrm{lcm}(X_i, X_j)$, $X_{ijl} = \mathrm{lcm}(X_i, X_j, X_l)$, 则

$$\frac{X_{ijl}}{X_{ij}}\tau_{ij} + \frac{X_{ijl}}{X_{jl}}\tau_{jl} + \frac{X_{ijl}}{X_{li}}\tau_{li} = 0,$$

若 $X_l | X_{ij}$, 则 $\tau_{ij}$ 在 $\tau_{jl}$ 和 $t_{li}$ 生成的 $R^s$ 的子模中.

注178. 等式可以由 $\tau_{ij}$ 的定义式直接验证, 对于第二个断言利用 $X_l | X_{ij}$ 时 $X_{ijl} = X_{ij}$ 代入等式直接得.

**推论13.5.** 设 $B \subset \{\tau_{ij} | 1 \le i < j \le s\}$ 是 $\mathrm{Syz}(c_1 X_1, \ldots, x_s X_s)$ 的生成元集, 若 $\exists i, j, l$ 使 $\tau_{ij}, \tau_{jl}, \tau_{li} \in B$, 且 $X_l | X_{ij}$, 则 $B \setminus \{\tau_{ij}\}$ 也是 $\mathrm{Syz}(c_1 X_1, \ldots, c_s X_s)$ 的生成元集.

由此推论我们可以得到 Buchberger 算法的第二个改进, 即某些 S-多项式可能是其它两个 S-多项式的线性组合, 可以不予计算. 这一步, 能够显著提高 Buchberger 算法的效率.

**改进后的算法**

鉴于前文的分析, 我们现在可以给出 Buchberger 算法的改进算法.

**算法13.5** (改进 Buchberger 算法).
输入: 多项式 $f_1, \ldots, f_s$,
输出: 理想 $\langle f_1, \ldots, f_s \rangle$ 的 Gröbner 基 $G$.

1. $G = \{f_1, \ldots, f_s\}$, $C = \varnothing$, $NC = \{\{1, 2\}\}$, $i = 2$,

2. 当 $i < s$ 时, 循环做: $NC = NC \bigcup \{\{j, i+1\} | 1 \le j \le i\}$, $NC = \mathrm{crit}(NC, C, i+1)$, $i = i + 1$,

3. 当 $NC \ne \varnothing$ 时, 循环做后面所有步骤, 否则输出 $G$ 退出,

4. 任选 $\{i, j\} \in NC$, 令 $NC = NC \setminus \{\{i, j\}\}$, $C = C \bigcup \{\{i, j\}\}$,

5. 若 $\gcd(\mathrm{lm}(f_i), \mathrm{lm}(f_j)) \ne 1$, 则做下面 6, 7 步,

6. $r = S(f_i, f_j) \, \mathrm{rem} \, G$,

7. 若 $r \ne 0$ 则 $f_{s+1} = r$, $G = G \bigcup \{f_{s+1}\}$, $s = s + 1$, $NC = NC \bigcup \{\{i, s\} | 1 \le i \le s - 1\}$, $NC = \mathrm{crit}(NC, C, s)$.

上面算法中 crit 函数是第二个改进判别法，由下面算法给出，其中的 $X_i = \text{lm}(f_i)$:

**算法13.6** (crit 函数).

crit$(NC, C, s)$, 输出简化后的 $NC$.

1. $l = 1$,

2. 当 $l < s$ 时循环做下面 3–7 步, 否则转 8 步,

3. 若 $\{l, s\} \in NC$ 则令 $i = 1$ 并做下面 4–6 步, 否则转 7 步,

4. 当 $i < s$ 时循环做下面 5, 6 步, 否则转 7 步,

5. 若 $\{i, l\} \in NC \bigcup \wedge \{i, s\} \in NC$ 则看 $X_l | \text{lcm}(X_i, X_s)$ 是否成立, 是则令 $NC = NC \setminus \{\{i, s\}\}$,

6. $i = i + 1$,

7. $l = l + 1$,

8. $i = 1$,

9. 若 $i < s$ 则做下面 10–14 步, 否则转 15 步,

10. 若 $\{i, s\} \in NC$ 则令 $j = i + 1$ 并做下面 11–13 步, 否则转 14 步,

11. 当 $j < s$ 时做下面 12, 13 步, 否则转 14 步,

12. 若 $\{j, s\} \in NC \wedge \{i, j\} \in NC$ 则看 $X_s | \text{lcm}(X_i, X_j)$ 是否成立, 是则令 $NC = NC \setminus \{\{i, j\}\}$,

13. $j = j + 1$,

14. $i = i + 1$,

15. 输出 $NC$.

### 13.3.6 Gröbner 基的应用

**一些简单的应用**

现在我们可以回到本章开头所提出的一些关于多元多项式理想的问题上来了. 我们要解决的第一个问题是对于任何一个多项式 $f$, 如何判断它在不在一个已知的理想 $I = \langle f_1, \ldots, f_s \rangle$ 中, 以及找出多项式 $v_1, \ldots, v_s$ 使得 $f = v_1 f_1 + \cdots + v_s f_s$.

首先我们根据前面生成 Gröbner 基的算法, 不仅得到了约化的 Gröbner 基 $G = \{g_1, \ldots, g_t\}$, 而且可以得到变换矩阵 $M_{t \times s}$ 使得 $(g_1, \ldots, g_t) = (f_1, \ldots, f_s) M$. 由推论 13.4 利用带余除法可判定 $f$ 是不是在理想 $I$ 中. 若 $f \in I$, 设利用除法算法得到的多项式为 $u_1, \ldots, u_t$, 满足 $f = u_1 g_1 + \cdots + u_t g_t$, 于是由

$$f = (g_1, \ldots, g_t) \begin{pmatrix} u_1 \\ \vdots \\ u_t \end{pmatrix} = (f_1, \ldots, f_s) M \begin{pmatrix} u_1 \\ \vdots \\ u_t \end{pmatrix}$$

知 $(v_1, \ldots, v_s)^T = M(u_1, \ldots, u_t)^T$.

对于两个多项式理想是否相等的判定, 也可由它们唯一的约化 Gröbner 基来进行. 而在商环 $R/I$ 中的算术需要用到代表元, 我们也可取为 $\overline{f} = f \operatorname{rem} G$. 下面定理给出了商环 $R/I$ 的一组基.

**定理13.15.** $\mathcal{B} = \{\overline{g} | g \in M \wedge g \notin \langle \operatorname{lt}(G) \rangle\}$ 是 $R/I$ 在 $F$ 上的一组基.

$R/I$ 中的求逆问题. 设 $f \in R/I$, 我们既已知道了该环的基, 则可设 $f^{-1}$ 为 $\mathcal{B}$ 的元素的线性组合来求解, 这比较复杂. 我们设 $g$ 是 $f$ 的逆, 注意到

$$fg - 1 \in I \Leftrightarrow 1 \in \langle I, f \rangle \Leftrightarrow \langle I, f \rangle = R,$$

则我们可以求 $\langle I, f \rangle$ 的 Gröbner 基, 看 1 是否在其中来确定是否存在逆. 再求出 1 在 $\langle I, f \rangle$ 中的线性表示, 即 $1 = v_1 f_1 + \cdots + v_s f_s + gf$, $g$ 即是 $f$ 在 $R/I$ 中的逆.

**Hilbert 零点定理及 Gröbner 基在解方程中的应用**

现在考虑 $F$ 的某个扩域 $k \supset F$(或$k = F$).

**定义13.21.** $V_k(I) = \{a \in k^n | \forall f \in I (f(a) = 0)\}$.

对于 $F^n$ 中子集 $V$, 定义 $R$ 的理想 $I(V) = \{f \in R | f(V) = 0\}$.

**定理13.16** (弱 Hilbert 零点定理). $I \subset R$, $k$ 是 $F$ 的代数闭域, 则 $V_k(I) = \varnothing \Leftrightarrow I = R = F[x]$.

**定义13.22.** 定义多项式理想 $I$ 的根理想(radical)为 $\sqrt{I} = \{f \in R | \exists e \in \mathbb{N}(f^e \in I)\}$.

显然有 $\forall k \supset F, V_k(I) = V_k(\sqrt{I})$.

**定理13.17** (强 Hilbert 零点定理). $I(V_k(I)) = \sqrt{I}$.

**推论13.6.** $V_k(I) = V_k(J) \Leftrightarrow \sqrt{I} = \sqrt{J}$.

注179. 强, 弱 Hilbert 零点定理的证明见有关代数几何的书, 如 [89] 及其中译本 [4].

**定理13.18.** 设 $G = \{g_1, \ldots, g_t\} \in GB(I)$, 下面三个命题等价:

(1) $V_k(I)$ 是有限集,

(2) $\forall i \in \{1, \ldots, n\}, \exists v_i \in \mathbb{N}, j \in \{1, \ldots, t\}$ 使得 $\text{lm}(g_j) = x_i^{v_i}$,

(3) $R/I$ 有限维.

当上面任何一个条件满足时, 我们也称理想 $I$ 是零维理想(zero-dimensional).

证明. (1)⇒(2). 若 $V_k(I) = \varnothing$ 则 $1 \in G$, 取 $v_i = 0$ 即可. 下面假设 $V_k(I) \neq \varnothing$, 取某个 $i \in \{1, \ldots, n\}$, 对于 $l = \#V_k(I)$, 取 $a_{im}(m = 1, 2, \ldots, l)$ 为 $V_k(I)$ 中点的第 $i$ 个分量, 取非零多项式 $f_m \in F[x_i] \subset F[x]$ 使得 $f_m(a_{im}) = 0$, 令 $f = f_1 \cdots f_l \in F[x_i] \subset F[x]$, 则 $f \in I(V_k(I)) = \sqrt{I}$, 于是 $\exists e \in \mathbb{N}$ 使得 $f^e \in I$, 则 $\text{lm}(f^e)$ 是 $x_i$ 的幂, $G$ 中有元素 $g_j$ 满足 $g_j = x_i^{v_i}$ 为 $x_i$ 的幂.

(2)⇒(3). 对于 $R/I$ 的基 $\prod_{1 \leq i \leq n} x_i^{\alpha_i}$, 一定满足 $\alpha_i < v_i$, 维数不超过 $\prod_{1 \leq i \leq n} v_i$, 因此它是有限维的.

(3)⇒(1). 考虑多项式集合 $\{\overline{x_i^j} | j \in \mathbb{N}\}$, 由于在 $R/I$ 中维数有限, 则它们一定线性相关, 即存在 $m \in \mathbb{N}$ 使得 $\sum_{0 \leq j \leq m} c_j \overline{x_i^j} = 0$, 亦即 $\sum_{0 \leq j \leq m} c_j x_i^j \in I$, 该多项式在 $F[x_i]$ 中零点有限, 至多为 $m$ 个, 不妨设其零点集为 $V_i$, 于是 $V_k(I) \subset \prod_{1 \leq i \leq n} V_i$ 是有限集. □

**推论13.7.** 设 $I$ 是零维理想, $G = \{g_1, \ldots, g_t\} = RGB(I)$, 规定的字典序为 $x_1 < x_2 < \cdots < x_n$, 则 $t \geq n$, 并可将 $g_1, \ldots, g_t \in G$ 重新编号使得 $g_i(1 \leq i \leq n)$ 只含 $x_1, x_2, \ldots, x_i$ 且 $\text{lm}(g_i)$ 为 $x_i$ 的幂.

例13.6. 考虑由 $f_1 = x^2 + y^2 + z^2 - 1, f_2 = x + y + z, f_3 = x^2 - 2x + y^2 - 2y + z^2 + 2z$ 生成的理想的 Gröbner 基.

解: 由 Buchberger 算法可得其 Gröbner 基为

$$G = \{g_1, g_2, g_3\} = \{16x^2 - 4x - 7, 4x + 4y - 1, 4z + 1\}.$$

显然, $V(I)$ 是一有限集, 此基的形式正如推论所说, 由此我们可以由第一个一元多项式方程解出其根, 代入第二个方程解出第二个变元, 依次迭代下去即可解出所有的根. 这里消元的结果和例 13.1 利用结式消元得到的结果是一样的. ◇

**方程组解的结构的一些讨论**

现在为了方便起见,将所讨论的域限定为复数域 $\mathbb{C}$, 由上一小节所讨论的内容, 我们很容易推广到如下结论:

**定理13.19.** $I$ 是零维理想当且仅当 $\forall i \in \{1,\ldots,n\}$, 消元理想 $I \bigcap \mathbb{C}[x] \neq \{0\}$.

证明. ($\Rightarrow$) 显然. 我们只要选取相应的字典序 $x_i < x_1 < \cdots < x_{i-1} < x_{i+1} < \cdots < x_n$ 即可.

($\Leftarrow$). 当消元理想是非平凡理想时, 其是一元多项式环 $\mathbb{C}[x_i]$ 上的理想, 因而是一个主理想, 可设其由 $f_i$ 生成, $f_i$ 首一且 $\deg f_i = m_i$, 则对任何单项序有 $x_i^{m_i} \in \langle \mathrm{lt}(I) \rangle = \langle \mathrm{lt}(G) \rangle$. 由此易得 $R/I$ 是 $\mathbb{C}$ 上有限维线性空间. □

注180. $f_i$ 可由满足 $\sum_{j=0}^{m_i} c_j \overline{x_i}^j = \overline{0}$ 的极小多项式得到.

**定义13.23.** 记 $p_{red}$ 为多项式 $p$ 的无平方部分, 即 $p_{red} := p/\gcd(p, p')$.

**定理13.20.** 当 $p$ 是一元多项式时, 有 $\sqrt{\langle p \rangle} = \langle p_{red} \rangle$.

证明. 首先由 $V(p) = V(p_{red}) \Rightarrow \sqrt{\langle p \rangle} = I(V(p_{red})) = \sqrt{\langle p_{red} \rangle}$. 而显然有 $\sqrt{\langle p_{red} \rangle} = \langle p_{red} \rangle$. □

[9]46 页给出了如下定理:

**定理13.21.** 设 $I \subset \mathbb{C}[x_1, x_2, \ldots, x_n]$ 是一多项式理想, 则有

$$\sqrt{I} = I + \langle p_{1,red}, p_{2,red}, \ldots, p_{n,red} \rangle,$$

其中 $p_i$ 是消元理想 $I \bigcap \mathbb{C}[x_i]$ 的唯一首一生成元, 且 $p_{i,red}$ 是 $p_i$ 的无平方部分.

证明. 设上面等式右边的理想为 $J$, 先证其为一根理想. 我们先将诸 $p_i$(记 $n_i = \deg p_i$)进行 $\mathbb{C}$ 上的因子分解, 得到

$$p_{i,red} = \prod_{1 \leq j \leq n_i} (x_i - a_{ij}),$$

其中由无平方部分的性质可知 $a_{i1}, a_{i2}, \ldots, a_{in_i}$ 互不相同. 由 $p_{1,red} \in J$ 可得

$$J = J + \langle p_{1,red} \rangle = \bigcap_{1 \leq j \leq n_1} (J + \langle x_1 - a_{1j} \rangle),$$

同样地对于其它 $p_{i,red}$ 可得

$$J = \bigcap_{1 \leq j_i \leq n_i, 1 \leq i \leq n} (J + \langle x_1 - a_{1j_1}, x_2 - a_{2j_2}, \ldots, x_n - a_{nj_n} \rangle).$$

对于每一项, $\langle x_1 - a_{1j_1}, \ldots, x_n - a_{nj_n} \rangle$ 都是极大理想, 因而 $J$ 是有限个极大理想的交集, 即有限个根理想的交集, 仍然是根理想.

由 $J$ 的定义显然有 $I \subset J$, 又由于 $J$ 的零点全在 $V(I)$ 中, 于是 $I \subset J \subset \sqrt{I}$, 取根理想有 $\sqrt{I} = \sqrt{J} = J$. □

下面我们给出一个有用的引理:

**引理13.9.** 设 $S = \{p_1, p_2, \ldots, p_m\} \subset \mathbb{C}^n$ 是 $m$ 个互不相同的点的集合, 则存在多项式组 $g_i \in \mathbb{C}[x_1, \ldots, x_n](1 \leq i \leq m)$, 使得 $g_i(p_j) = \delta_{ij}$.

证明. 回忆一元多项式情形下的 Lagrange 插值的构造方法, 我们可以类似地构造. 首先考虑 $p_1, p_2$ 两个点, 两个点肯定有某个分量不同, 不妨设 $p_{1,k} \neq p_{2,k}$, 则定义

$$g_{1,2} = \frac{x_k - p_{2,k}}{p_{1,k} - p_{2,k}},$$

其满足 $g_{1,2}(p_1) = 1$, $g_{1,2}(p_2) = 0$, 同样构造出 $g_{1,j}(2 \leq j \leq m)$ 后, 取 $g_1 = g_{1,2}g_{1,3} \cdots g_{1,m}$ 即可.

对于 $g_i(2 \leq i \leq m)$ 可类似构造. □

下面的定理给出了方程组根的个数的估计([9]47 页).

**定理13.22.** $I$ 是 $\mathbb{C}[X]$ 中的零维理想, $A = R/I$, 则 $\dim A \geq \#V(I)$, 取等号当且仅当 $I = \sqrt{I}$.

证明. 设 $V(I) = \{p_1, \ldots, p_m\}$, 其中 $m = \#V(I)$, 考虑线性映射:

$$\delta \subset \mathbb{C}[X]/I \times \mathbb{C}^m : \overline{f} \mapsto (\overline{f}(p_1), \ldots, \overline{f}(p_m)).$$

根据引理 13.9 设 $g_1, \ldots, g_m$ 满足 $g_i(p_j) = \delta_{ij}$, 则 $\overline{g_i}(p_j) = \delta_{ij}$. 由于 $\forall (\lambda_1, \ldots, \lambda_m) \in \mathbb{C}^m$, $\exists f = \sum_{1 \leq i \leq m} \lambda_i g_i$, 使得 $\delta(\overline{f}) = (\lambda_1, \ldots, \lambda_m)$, 因此 $\delta$ 是满射. 于是 $\dim A \geq \dim \mathbb{C}^m = m = \#V(I)$.

对于第二个等价条件, 先设 $I = \sqrt{I}$, 令 $\overline{f} \in \ker \delta$, 则 $f(V(I)) = 0 \Rightarrow f \in I(V(I)) = \sqrt{I} = I \Rightarrow \overline{f} = 0$, 从而 $\delta$ 是单射, 因而是同构, $\dim A = m$.

若 $\dim A = m$, 首先 $I \subset \sqrt{I}$, $\forall f \in \sqrt{I} = I(V(I))$, 有 $f(V(I)) = 0$, 则 $\delta(\overline{f}) = (0, \ldots, 0) \Rightarrow \overline{f} \in \ker \delta \Rightarrow f \in I \Rightarrow I = \sqrt{I}$. □

### 13.3.7 Gröbner 基和特征值法解方程组

在用 Gröbner 基理论将方程组化为三角形列后, 采用一步步代入来算会引入累积误差, [9]2.4 节给出了一种方法, 用于求解的任何一个分量, 同时也给出了求消元理想 $I \bigcap \mathbb{C}[x_i]$ 的首一生成元的快速方法. 设 $A = R/I$, 首先我们定义

**定义13.24.** 线性映射 $m_f \subset A \times A$ 使得 $m_f(\overline{g}) = \overline{fg}$, 显然 $m_f = 0 \Leftrightarrow \overline{f} = 0$. 以后 $m_f$ 既指线性映射, 也可指其在 $A$ 的某组基上的矩阵表示.

很显然可以验证这样定义的线性映射具有一定的和谐性, 即 $\forall h(t) \in \mathbb{C}[t]$, 有 $m_{h(f)} = h(m_f)$. 只需逐一验证 $m_{f+g} = m_f + m_g$, $m_{fg} = m_f m_g$ 即可.

下面的定理给出了求解的方法:

**定理13.23.** 设 $h_f$ 是 $m_f$ 的极小多项式, $\forall \lambda \in \mathbb{C}$, $I$ 是一零维理想, 则下面三个命题是等价的:

1. $h_f(\lambda) = 0$,

2. $\lambda$ 是 $m_f$ 的特征值,

3. $f$ 在 $V(I)$ 上取得值 $\lambda$, 即 $\lambda \in f(V(I))$.

*证明.* 由高等代数知识知道 1 和 2 等价是显然的.

$2 \Rightarrow 3$. 不妨设 $\lambda \notin f(V(I))$, 由特征值的定义知 $\exists \overline{z} \in A \setminus \{0\}$, 使得特征方程 $\overline{f - \lambda}\overline{z} = 0$ 成立. 令 $V(I) = \{p_1, \ldots, p_m\}$, 则 $f(p_i) \neq \lambda (\forall p_i \in V(I))$. 令 $g = f - \lambda$, 有 $g(p_i) \neq 0$, 设 $g_i(1 \leq i \leq m)$ 满足 $g_i(p_j) = \delta_{ij}$, 定义
$$g' = \sum_{i=1}^m \frac{1}{g(p_i)} g_i,$$
则 $g'(p_i)g(p_i) = 1 \Rightarrow 1 - g'g \in I(V(I)) = \sqrt{I} \Rightarrow \exists l \in \mathbb{N}((1 - g'g)^l \in I)$. 将其做二项式展开, 可得 $g''$ 满足 $1 - g''g \in I$, 即 $\overline{g''}\overline{g} = 1$, 故
$$\overline{f - \lambda}\overline{z} = 0 \Rightarrow \overline{g''}\overline{g}\overline{z} = 0 = \overline{z},$$
矛盾.

$3 \Rightarrow 1$. 设 $\lambda = f(p)(p \in V(I))$, 由 $h_f(m_f) = 0 \Rightarrow h_f(\overline{f}) = 0 \Rightarrow h_f(f) \in I$, 故 $h_f(\lambda) = h_f(f(p)) = 0$. □

**推论13.8.** 设 $I$ 是零维理想, 则 $m_{x_i}$ 的特征值集合 $V_i$ 即为 $V(I)$ 中点的 $x_i$ 分量, 且 $h_{x_i}(x_i)$ 是消元理想 $I \bigcap \mathbb{C}[x_i]$ 的唯一首一生成元.

注181. 使用上面方法的好处在于, 当我们只用 Gröbner 基方法时, 由于要求指定的字典序下的 Gröbner 基, 其计算是较为复杂的. 而我们的定理仅涉及 $A$ 的代数结构, 我们可以用计算量较小的分级字典序求出 Gröbner 基, 得到 $A$ 上的基, 以此来进行特征值的计算 [9].

注182. 特征值法另一优点是符号求解时不不多次计算多项式组的 Gröbner 基即可得到各个消元理想 $I \bigcap \mathbb{C}[x_i]$ 的首一生成元, 从而精确求解出 $x_i$.

## 3.3 代数函数的不定积分问题

  Bronstein 在出版的专著中，对于超越函数的不定积分问题进行了系统的阐述。与之相对，代数函数的不定积分的研究工作在今年来缺乏这一种系统的总结，加之理论研究的难度较大，所以现阶段还是停留在理论研究的层面上。这也导致了在实际的实现过程中 Albi 系统自身的缺陷和不完备性。毫无疑问，代数函数的不定积分的研究以及系统的梳理总结，进而产生实用算法是之后符号积分研究的重点。如果能够对于代数函数的不定积分问题进行系统的阐述，那么 Albi 才是真正意义上的完备系统。

  由于时间的关系，这一部分的学习只是刚刚开始，所以在这里不进行具体的介绍，只是提供一些理论上的学习，总结，以供参考。

# Chapter 4

# 符号积分II

## 4.1 关于对数函数的积分

### 4.1.1 介绍

在积分学中,我们涉及到一些不存在初等形式的积分,例如,$\mathrm{d}u/\log(u)$的积分。在这里我们利用:对数积分:$\mathrm{li}(u) = \sum_{i=1}^{\infty} \int \mathrm{d}u/\log(u)$,来研究一些关于对数生产的一些函数的积分。

**例4.1.** $\int (x/\log(x)^2)\mathrm{d}x$不存在初等函数的积分,而在本节中我们将得到

$$\int \frac{x}{\log(x)^2}\mathrm{d}x = 2\mathrm{li}(x^2) - \frac{x^2}{\log(x)}$$

### 4.1.2 li-初等扩张

**定义4.1.** $F$是一个微分域,即域$F$上定义了一个导数$'$,满足$\forall u, v \in F$ $(uv)' = uv' + u'v; (u+v)' = u' + v'$,特征为0,$E$是$F$上的一个微分扩张。称$E$为一个$F$的刘维尔初等扩张,若$\exists \theta_1, \ldots, \theta_n \in E$,使得$E = F(\theta_1, \ldots, \theta_n)$,且$\forall 1 \leq i \leq n$满足一下中的某一个

1. $\theta_i$是$F(\theta_1, \ldots, \theta_{i-1})$上的代数

2. 对某个$a \in F(\theta_1, \ldots, \theta_{i-1})$,有$\theta_i = \exp(a)$

3. 对某个$a \in F(\theta_1, \ldots, \theta_{i-1})$,有$\theta_i = \log(a)$

4. 对某个$a \in F(\theta_1, \ldots, \theta_{i-1})$,有$\theta_i = \mathrm{li}(a)$

注:在1, 2, 3中描述了初等函数,在4中我们引入了非初等元$\int \mathrm{d}u/\log(u)$

**例4.2.** 例 令$R$是实数域，$F = R(X)$是$R$上的有理函数集合，$g(x) = \int_0^x (e^t/t) \mathrm{d}t$，即$g(x) = \mathrm{li}(e^t)$。则$R(x, e^x, g(x))$，为一个li-初等函数域。

类似于刘维尔定理，我们有

**定理4.1.** $F$是$C(x)$的刘维尔扩张，$C$是$F$上常数的代数闭域，若$g \in F$在$F$的刘维尔初等扩张中有不定积分，则存在常数$c_i, d_i \in C$；$w_i, u_i, v_i \in F$，使得

$$g = w_0' + \sum c_i \frac{w_i'}{w_i} + \sum d_i \frac{u_i'}{v_i} \tag{4.1}$$

其中$v_i' = u_i'/u_i$，即$v_i = \log(u_i)$

很明显我们可以看出在(1)式中表达形式不唯一，实际上，若存在$c_i, d_i \in C$；$w_i, u_i, v_i \in F$，则$\lambda \in C$；$c_i, d_i, w_i, \bar{u}_i = \lambda u_i, v_i$同样可使(1)式成立。则

$$g = w_0' + \sum c_i \frac{w_i'}{w_i} + \sum \frac{d_i}{\lambda_i} \frac{u_i'}{v_i}$$

因此，我们调整$u_i$和$v_i$，使得它们满足$v_i = \ln(u_i)$。那么就可以保证上述刘维尔定理中分解的唯一性了。以下我们均以$u_i$和$v_i$已满足$v_i = \ln(u_i)$考虑。

### 4.1.3　初等函数塔

**定义4.2.** 初等函数塔 让$F = C(x, \theta_1, \ldots, \theta_n)$是$C$上的超越初等扩张，这里$C$是常数域，$x$是$x' = 1$的解。调整$\theta$使得$C(x) = F_0 \subset F_1 \subset \ldots \subset F_r = F$，这里$F_i = F_{i-1}(\theta_{i1}, \ldots, \theta_{im_i})$。每个$\theta_{ij}$满足下面其中一条：

1. $\theta_{ij}' = a_{ij}'/a_{ij}$，$a_{ij} \in F_{i-1}$但是$a_{ij} \notin F_{i-2}$

2. $\theta_{ij}' = \theta_{ij} a_{ij}'$，$a_{ij} \in F_{i-1}$但是$a_{ij} \notin F_{i-2}$

接着我们定义超越初等函数塔$F = C(x, \theta_1, \ldots, \theta_n)$的阶
$rank(F) = (m_r, \ldots, m_1, 1)$。

**注8.** $rank(F)$取决于$\theta_1, \ldots, \theta_n$的选取，阶的大小按字典排序法比较

**例4.3.** 域$E_1 = C(x, \exp(x), \exp(\exp(x) + x), \exp(\exp(x) + x^2))$；
$E_2 = C(x, \exp(x), \exp(x^2), \exp(\exp(x)))$，则$rank(E_1) = (2, 1, 1), rank(E_2) = (1, 2, 1)$

我们也可以定义$F$中的元素$a$的阶为$k$，若$a \in F_k$，但$a \notin F_{k-1}$.

令$F = C(x, \theta_1, \ldots, \theta_n)$是$C$上的初等扩张，我们称$F$是可分解的，如果每个对数单项式$\theta_i = \log(a_i)$，这里$a_i$是在$C[x, \theta_1, \ldots, \theta_{n-1}]$不可约的多项式。容易看出给定的任意一个$F$上的$C(x)$上的超越元，都可以做一个可分解的域$\tilde{F} = C(x, \tilde{\theta}_1, \ldots, \tilde{\theta}_n)$使得$F$同构于$\tilde{F}$的某个微分子域。

让$F$如上述定义，$\theta_i = \exp(a_i)$是一个k阶指数单项式，设$a_i = \sum(p_j/q_j)\theta_j + r$，其中$p_j$，$q_j$是整数，$\theta_j$是$k-1$阶对数多项式，$rank(r) < k-1$。称$F$是规范的，如果$0 < p_j/q_j < 1$，对于每个指数单项式成立。

**例4.4.** $F = C(x, \log(x), \exp(\frac{5}{2}\log(x) + x))$不是规范的，但它同构于
$F = C(x, \log(x), \exp(\frac{1}{2}\log(x) + x))$是规范的

如此可以看出，我们可以将$\theta_1, \ldots, \theta_n$换成$\tilde{\theta_1}, \ldots, \tilde{\theta_n}$使得$F = C(x, \tilde{\theta_1}, \ldots, \tilde{\theta_n})$是规范的。不难看出$rank(\tilde{F}) < rank(F)$

我们总假设以C中元素为系数的多项式总可以在有限步分解成不可约因子的乘积。

### 4.1.4 $\sum$−分解

**定义4.3.** $K$是一个特征为0的域，让$\sum = (f_1, \ldots, f_m)$是$K(x)$一列互不相同的不可约元，且$f_i \notin K$。$\Phi \in K(x)$，称$\Phi$在$K$上有一个$\sum$−分解，若存在$b_i \in K$，$a_{ij}$为整数，和自然数$n$，使得

$$\Phi = \sum_{i=1}^{n} b_i \prod_{i=1}^{m} f_j^{a_{ij}}$$

我们想要考虑这种$\sum$−分解的存在性，唯一性，和它的构造。
但事实并非如此

**例4.5.** 如果$\sum = (x^2+1)$，那么x就没有$\sum$−分解

**例4.6.** 令$K$是有理数域，$\sum = (x, x+1)$，则0有无数多组表示

$$0 = (x+1)^n - \sum_{k=1}^{n} \binom{n}{k} x^{n-k}, n = 1, 2, \ldots$$

令$T \subset Z$，$g : T \to Z^m$。我们说$\Phi$在$K(x)$中有一个$\sum$分解是被$g$限制，如果

$$\Phi = \sum_{i=1}^{n} b_i \prod_{j=1}^{m} f_j^{\alpha_{ij}}$$

对于任意的$i, \alpha_{i1} \in T$，$g(\alpha_{i1}) = (\alpha_{i1}, \alpha_{i2}, \ldots, \alpha_{im})$。除了平凡的情形外，$b_i \neq 0$。且假设$\alpha_{i1} < \alpha_{i2} < \ldots < \alpha_{im}$.

**定理4.2.** 令$\sum$和$g$如上所示。令$\Phi$是$K[x]$中的非零元，令$\sum_{i=1}^{n} b_i f_j^{a_{ij}}$是$\Phi$在$g(x)$上限制的一个$\sum$分解，且$\alpha_{11} < \alpha_{21} < \ldots < \alpha_{n1}$。那么$\alpha$是$f_1$在$\Phi$中的重数

证明. 当$n = 1$时显然。设$n > 1$，令$\Phi = f_1^\alpha p/q$，这里$f_1, p, q \in K[x]$是互素的，且q是首一的。我们有

$$f_1^\alpha \frac{p}{q} = b_1 f_1^{\alpha_{11}} \prod_{j=2}^{m} f_j^{\alpha_{1j}} + \sum_{i=2}^{n} b_i f_1^{\alpha_{i1}} \prod_{j=2}^{m} f_j^{\alpha_{ij}}$$

则

$$f_1^\alpha \frac{p}{q} = f_1^{\alpha_{11}} \left( b_1 \prod_{j=2}^{m} f_j^{\alpha_{1j}} + \sum_{i=2}^{n} b_i f_1^{\alpha_{i1} - \alpha_{11}} \prod_{j=2}^{m} f_j^{\alpha_{ij}} \right) \tag{4.2}$$

由于当$j > 1$时，$f_j$和$f_1$是互素的，我们有$b_1 \prod_{j=2}^{m} f_j^{\alpha_{1j}} = p_1/q_1$，这里$p_1, q_1 \in K[x]$是互素的。并且由$\forall i \geq 2 \quad \alpha_{i1} - \alpha_{11} > 0$，那么式4.2能写成

$$f_1^\alpha \frac{p}{q} = f_1^{\alpha_{11}} \left( \frac{p_1}{q_1} + f_1^\alpha \frac{p_2}{q_2} \right) = f_1^{\alpha_{11}} \left( \frac{p_1}{q_1} + f_1^\alpha \frac{p_2}{q_2} \right) = f_1^{\alpha_{11}} \frac{p_3}{q_3}$$

这里$f_1, p_3, q_3 \in K[x]$互素，且$q_3$是首一的，那么$\alpha = \alpha_{11}$ □

**定理4.3.** $\sum$和$g$如上所示，假设$\Phi$有一个被$g$限制的$\sum$分解。这是分解在不计次序的情况下是唯一的。

**定理4.4.** $\sum$和$g$如上所示，让$\Phi \in K(x)$且假设$\Phi$有一个被$g$限制的$\sum$分解。近一步假设$\forall \alpha \in T$，我们能够有限步的计算出$g(x)$，则我们也能够在有限步计算出$\Phi$的$\sum$分解。

证明. 令$\Phi = \sum_{i=1}^{n} b_i \prod f_j^{a_{ij}}$是$\Phi$在$g(x)$限制下的$\sum$分解。用递归的方式，只需证明$b_1, \alpha_{11}, \ldots, \alpha_{1m}$是可计算的。由定理4.2,$\alpha_{11}$是$f_1$在$\Phi$中的重数，又由于$g(x)$是可计算的，那么$\alpha_{12}, \ldots, \alpha_{1m}$可计算，那么下面我们只需计算$b_1$。

我们计算互素的$p, q$且，q是首一的，$f_1$和q是互素的。且

$$\frac{p}{q} = \frac{\Phi}{\prod f_j^{\alpha_{ij}}} = b_1 + \sum_{i=2}^{n} b_i \prod f_j^{\alpha_{ij} - \alpha_j}$$

由此得

$$p = b_1 q + q \sum_{i=2}^{n} b_i \prod f_j^{\alpha_{ij} - \alpha_j} \tag{4.3}$$

由于$p, b_1 q \in K[x]$，我们有$q \sum_{i=2}^{n} b_i \prod f_j^{\alpha_{ij} - \alpha_j} \in K[x]$。由于$\alpha_{ij} - \alpha_{11} > 0, \forall i > 1$我们能把式4.3写成

$$p = b_1 q + f_1 Q \tag{4.4}$$

这里$Q \in K[x]$。也就是说$b_1 = (p \mod f_1)(q \mod f_1)$。 □

4.1. 关于对数函数的积分101**定理4.5.** $\sum$和$g$如上所示，$d$是一个正整数。假设对于任意的整数$\alpha$我们能够在有限步决定，计算$g(\alpha)$。那么给定的$\Phi \in K(x)$如果$\Phi$有被$g$限制的$\sum$分解，则我们能在有限步决定。

证明. 我们计算$\alpha_{11}$是$f_1$在$\Phi$中的重数。如果$\alpha_{11} \notin T$停止，也就是$\Phi$没有这样的$\sum$分解。否则，我们计算$\alpha_{12}, \ldots, \alpha_{1m}$，如果$b_1 \notin K$，停止。如果$b \in K$，我们令$\Phi_2 = \Phi - b_1 \prod f_j^{\alpha_{ij}}$，并计算$f_1$在$\Phi_2$中的重数$\alpha_{21}$。如果$\alpha_{21} \in K$且$\alpha_{21} > \alpha_{11}$，我们计算$(b_1, \alpha_{21}, \ldots, \alpha_{2m})$。重复上述方式，我们能够计算$(b_3, \alpha_{31}, \ldots, \alpha_{3m})$，$\ldots, (b_d, \alpha_{d1}, \ldots, \alpha_{dm})$，和$p_1(\alpha), \ldots, p_m(\alpha)$。一旦知道知道这些多项式，我们能够找到一个整数$\alpha^*$，使得每个$p_j(\alpha)$在$(\alpha^*, \infty)$是单调的。如果有$p_j$是0阶的，我们用$\Phi_{d+1}/f_j^{\alpha_{ij}}$代替$\Phi_{d+1}$，保证$p_j$在这个区间中是严格单调的。继续计算，直到$\alpha_{k1} > \alpha^*$对于某个$k \geq d$。如果$\alpha_{ij} \neq p_j(\alpha_{i1})$，则$\Phi$没有这样的$\sum$分解。否则我们有一个元素$\Phi_{k+1}$我们期望写成$\sum_{k+1}^n b_i \prod f_j^{\alpha_{ij}}$，这里$(\alpha_{k+1,j}, \ldots, \alpha_{nj})$对于每个j都是严格单调的。

分为以下两种情形：

1. 假设第j列是单调下降的，令r是$f_j$在$\Phi_{k+1}$的重数。通过检验$\Phi$的部分分解我们能够看到$\alpha_{ij} = r$。因此如果对于某个$i > k$，有$\alpha_{ij} < k$，我们结束计算。

2. 假设所有的列都是单调增的，在这种情况在我们重复计算，直到$\alpha_{r1}, \ldots, \alpha_{rm}$对于某个r是正的。如果$\Phi_r \notin K[x]$，则没有$\sum$分解。否则，我们有$\deg(\prod f_j^{\alpha_{nj}}) = \deg(\Phi_r)$对于某个$i > r$我们有$\deg(\prod f_j^{\alpha_{ij}}) > \deg(\Phi_r)$我们能够终止计算。

$\square$

这些结果总结了多种情形。也就是说，让$\Sigma = (f_1, \ldots, f_m)$是一系列的两两互素的$K[x_1, \ldots, x_r]$，这里$f_i$均不在$K$中。

为了决定$\Phi$在$K(x_1, \ldots, x_r)$有一个严格被$d$阶$g$限制的$\Sigma$分解，我们首先单独考虑$x_k$，使得$f_1$不在$K[x_1, \ldots, x_{k-1}, x_{k+1}, \ldots, x_r]$。那么形式$\sum' = (f_1, f_{k1}, \ldots, f_{ks})$，这里$f_{kj}$不是在$K[x_1, \ldots, x_{k-1}, x_{k+1}, \ldots, x_r]$中。现在用定理4.5计算$K(x_1, \ldots, x_{k-1}, x_{k+1}, \ldots, x_r)$中的元素$B_i$和整数$\alpha_{ij}$使得

$$\Phi = \sum B_i f_1^{\alpha_{i1}} \prod f_{kj}^{\alpha_{ij}} \tag{4.5}$$

如果$\Phi$有一个被$d$阶$g$严格约束的$\sum$分解，那么4.5是严格被$g$在$K(x_1, \ldots, x_{k-1}, x_{k+1}, \ldots, x_r)$上的投影约束的$\sum'$分解。这样，我们只需在$K(x_1, \ldots, x_{k-1}, x_{k+1}, \ldots, x_r)$上分解$B_i$，然后检验这些因子和多项式来决定$\Phi$在$K$上是否有分解。并且这是分解是唯一的，且$B_i$也是唯一的。

### 4.1.5 对数函数积分

**引理4.1.** 让$F$如定理4.1所述，让$\theta$是$F$上的指数单项式，让$g$是$F(\theta)$中的元素，应用部分分解得到$g = A_m\theta^m + \ldots + A_{\tilde{m}}\theta^{\tilde{m}} + A_0 + P(\theta)/Q(\theta)$ 这里$A_k$在$F$中，$p,q$在$F(\theta)$中，$\deg(P(\theta)) < \deg(Q(\theta))$且$\gcd(Q(\theta),\theta) = 1$。那么$g$在$F(\theta)$的某个li–初等扩张中有不定积分，当且仅当每一项$A_m\theta^m,\ldots,A_{\tilde{m}}\theta^{\tilde{m}},\ldots,A_0 + P(\theta)/Q(\theta)$ 都在其中有不定积分。在这种情形下，对于每个$k$，和$A_k\theta^k$的积分相关的项$u_i'/v_i$是每个具有形式$f_i\theta^k$（这里$f_i$在$F$中），和$A_0 + P(\theta)/Q(\theta)$的积分相关的项$u_i'/v_i$在$F$中。

**引理4.2.** 让$K$是一个可计算的域，让$p$和$q$是$K[x]$中的两个互素的元素。给定$K[x]$中的$f$ 且$\deg(f) = n > 0$，我们在有限不决定是否存在一个阶小于$n$的$K[x]$中的多项式$d$，使得$f$整除$p + dq$。更进一步，如果存在，这个多项式是唯一的且是可计算的。

**定理4.6.** 让$E = C(x, \theta_1, \ldots, \theta_n)$是一个特征为0的微分域，$C$为代数闭域。假设$x$是$C$上的一个超越元，且是$x' = 1$的解。$\theta_i$是$C(x, \theta_1, \ldots, \theta_n)$的单项式，$E = C(x, \theta_1, \ldots, \theta_n)$是可分解且正则的。进一步假设$\theta_n$是$C(x, \theta_1, \ldots, \theta_{n-1})$的指数的，$\theta_n$的阶为$r$，$E$的阶为$(1, m_{r-1}, \ldots, 1)$。那么给定$A \in C(x, \theta_1, \ldots, \theta_n)$，如果其在$E$的某个刘维尔扩张中有不定积分我们能够在有限步决定$A\theta_n$。

*证明.* 令$F = C(x, \theta_1, \ldots, \theta_{n-1})$，让$e$和$l$是$F$的指数和对数的指标集

e={i—$\theta_i$是指数单项式},l={i—$\theta_i$是对数单项式}

并且令$\theta_i' = \theta_i\alpha_i'$对于所有的$i \in e$，令$\theta_i = a_i'/a_i$对于所有的$i \in l$，且令$\theta_n = \exp(a_n)$

现在假设$A\theta_i$在$E$的某个li初等扩张中有不定积分。从定理4.1和引理4.1的证明中我们有

$$A\theta_n = (B\theta_n)' + \sum d_i \frac{u_i'}{v_i} \quad (4.6)$$

这里$B \in F$，且$d_i$是常数，$v_i' = u_i'/u_i$，这里$u_i, v_i \in E$。由于$u_i$是$E$上为代数指数型的，我们由[Roca79,Thm3.1]有，存在有理数$r_i$和$r_{ij}$和一个常数$v_i$使得

$$v_i = r_i a_n + \sum_e r_{ij} a_j + \sum_l r_{ij}\theta_j + v_i$$

由于$u_i$是$v_i$的指数，则有

$$u_i = \eta_i \prod_l a_j^{r_{ij}} \prod_e \theta_j^{r_{ij}} \theta_n^{r_i} \quad (4.7)$$

这里$\eta_i \in C$。因此我们把式4.6重新写成

$$A\theta_n = (B\theta_n)' + \sum d_i \frac{(r_i a_n + \sum_e r_{ij} a_j + \sum_l r_{ij}\theta_j + v_i)'}{(r_i a_n + \sum_e r_{ij} a_j + \sum_l r_{ij}\theta_j + v_i)} \eta_i \prod_l a_j^{r_{ij}} \prod_e \theta_j^{r_{ij}} \theta_n^{r_i}$$

通过比较系数我们假设$r_i = 1$，由于$d_i$和$\eta_i$都是常数，我们可以令$\eta_i = 1$。又因为在4.7中的因子在$C[x, \theta_1, \ldots, \theta_n]$，那么可知$r_{ij}$都是整数。从条件，存在一个单项式$\theta$，阶为$r-1$使得$a_n$与$\theta$有关，其他的$a_j$都和$\theta$无关。让$F = D(\theta)$令$a = p(\theta)/q(\theta)$，这里$p(\theta)$和$q(\theta)$在$D[\theta]$中是互素的，且$q(\theta)$是首一的。我们有

$$A\theta_n = (B\theta_n)' + \sum d_i \frac{[p(\theta)/q(\theta) + \sum_e r_{ij}a_j + \sum_l r_{ij}\theta_j + v_i]'}{p(\theta)/q(\theta) + \sum_e r_{ij}a_j + \sum_l r_{ij}\theta_j + v_i} \prod_l a_j^{r_{ij}} \prod_e \theta_j^{r_{ij}} \theta_n^{r_i}$$

$$-------  \quad (4.8)$$

剩下的证明分成以下两个部分

**Case a** $\theta = \exp(a)$或者$\theta = x$，我们有

$$v_i = \frac{p(\theta)}{q(\theta) + \beta_i} = \frac{p(\theta) + \beta_i q(\theta)}{q(\theta)}$$

这里$\beta = + \sum_e r_{ij}a_j + \sum_l r_{ij}\theta_j + v_i$。由于$p(\theta)$和$q(\theta)$是互素的，则表达式的右边是已知的，则$v_i$是可以知道的。为了决定分子，首先，如果对于某个$\beta_0 \in D$使得$p(\theta) + \beta_i q(\theta) \in D$，则这样的$\beta_0$是唯一可计算的。因此，我们假设$p(\theta) + (\theta)$不在$D$中。它的分解为$\xi_i f_{i1}^{m_{i1}} \ldots f_{ik_i}^{m_{ik_i}}$，这里$\xi \in D$，每个$f_{ij}$是首一的、不可约的。那么等式4.7化为

$$A\theta_n = (B\theta_n') + \sum d_i \frac{(\xi_i f_{i1}^{m_{i1}} \ldots f_{ik_i}^{m_{ik_i}} q(\theta)^{-1})'}{(\xi_i f_{i1}^{m_{i1}} \ldots f_{ik_i}^{m_{ik_i}} q(\theta)^{-1})} \prod_l a_j^{r_{ij}} \prod_e \theta_j^{r_{ij}} \theta_n$$

或者

$$A\theta_n = (B\theta_n') + \sum d_i \left( \frac{\xi_i'}{\xi} - \frac{q(\theta)'}{q(\theta) + \sum_{j=1}^{k_i} m_{ij} \frac{f_{ij}'}{f_{ij}}} \right) \prod_l a_j^{r_{ij}} \prod_e \theta_j^{r_{ij}} \theta_n$$

我们指出，这里$f_{ij}$一定会整除$A$的分母，除非$f_{ij} = \theta$。首先指出，如果$f_{ij} \neq \theta$，那么$f_{ij}$不能整除$f_{ij}'\theta^k$，对于所有的$k \geq 0$。由引理4.2每个$f_{ij}$唯一的决定$\beta_i$，由$[Roca79, Thm.3.1]$，每个$\beta_i$唯一的决定$r_{ij}$和$v_i$。因此，每个$f_{ij} \neq \theta$一定是$\sum d_i u_i'/v_i$的分母的一个因子。现在考虑$(B\theta_n)'$，由部分分式可以看出$f_{ij}$或者是多重因子，或者不是$(B\theta_n)'$的分母的因子。在任一种情形下，$f_{ij}$都能够整除$A$的分子。

现在，对于任意一个不可约多项式$p_i$，在$A$的分母中，我们应用引理4.2，决定是否存在一个$\beta_i$，使得$p_i$整除$p(\theta) + \beta_i q(\theta)$。在这时，我们也让$p_i = \theta$。接下来我们决定是否存在整数$r_{ij}$和常数$v_i$，使得$\beta_i = \sum_e r_{ij}a_j + \sum_l r_{ij}\theta_j + v_i$

这样，对于给定的$A$，和所有的项$u_i'/v_i$，接下来，我们应用$[Risch69, Main\ Theorem, part(b)]$决定是否存在常数$d_i$和元素$B$满足4.6

注9. Risch69 Main Theorem 令$F$如上所示，每个$\theta_i$是$C(x,\theta_1,\ldots,\theta_{i-1})$上的单项式，我们有

**(a)** 令$f \in F$，那么能够在有限有限步决定是否存在$v_0 \in F$，$v_i \in \bar{C}(x,\theta_1,\ldots,\theta_n), i=1,\ldots,m$（其中$\bar{C}$是$C$的闭包），和$c_1,\ldots,c_m \in \bar{C}$，使得$f = [v_0 + \sum_{i=1}^{m} c_i \log v_i]'$。且如果存在，我们能够找到它们。

**(b)** 令$f, g_i, i=1,\ldots m$是$F$中的元素，那么能够在有限步内找到$h_1,\ldots,h_r \in F$，和一个以$C$中元素为系数、$m+r$个变量的线性代数方程组$\Psi$，使得对$y \in F$满足$y' + fy = \sum_{i=1}^{m} c_i g_i$，其中$c_i \in C$；当且仅当$y = \sum_{i=1}^{r} y_i h_i$这里$y_i \in K$，其中$c_1,\ldots,c_m, y_1,\ldots,y_r$满足方程组$\Psi$.

（具体参见[Risch69]）

**Case b**　$\theta = \log(a)$
让l'=$\theta_i|\theta_i$是$D$对数单项式的，我们有

$$v_i = \frac{p(\theta)}{q(\theta)} + r_i\theta + \beta_i = \frac{p(\theta) + r_i\theta q(\theta) + \beta_i q(\theta)}{q(\theta)}$$

这里$\beta_i = \sum_e r_{ij} a_j + \sum_{l'} r_{ij}\theta_j + c_i$。像前面一样，搜吗来决定$v_i$的分母。

**Case (b1)**　$q(\theta) \neq 0$或$\deg(p(\theta)) > 0$。

首先决定是否存在值$r_0$和$\beta_0$使得$p(\theta) + r_0\theta q(\theta) + \beta_0 q(\theta)$是在$D$中。这样$r_0 = -lc(p(\theta))$，（这里$lc$代表首项系数），$\beta_0 = -lc(p(\theta) + r_0\theta\theta q(\theta))$ 我们指出，当$q(\theta) = 1, p(\theta) + r_0\theta$在$D$中时，但这种情况在这里不会存在。因此，假设$p(\theta) + r_i\theta q(\theta) + \beta_i q(\theta)$不在$D$中，则设$\xi_i f_{i1}^{m_{i1}} \ldots f_{ik_i}^{m_{ik_i}}$是它的因式分解。这里$\xi \in D$，每个$f_{ij}$是首一的、不可约的。那么等式4.8化为

$$A\theta_n = (B\theta_n') + \sum d_i \left( \frac{\xi_i'}{\xi} - \frac{q(\theta)'}{q(\theta) + \sum_{j=1}^{k_i} m_{ij}\frac{f_{ij}'}{f_{ij}}} \right) \prod_l a_j^{r_{ij}} \prod_e \theta_j^{r_{ij}} \theta_n$$

注意，在这种情形下$\prod_l a_j^{r_{ij}} \prod_e \theta_j^{r_{ij}}$在$D$中，且$\deg_\theta(f_{ij}') < \deg_\theta(f_{ij})$。考虑$(B\theta_n)'$的部分分式分解，则$f_{ij}$整除$A$的分母，除非在项$\sum d_i u_i'/v_i$有约掉。如果有约化，则有下式之一成立

$$m_1\frac{f'}{f}\prod_l a_j^{r_{1j}}\prod_e \theta_j^{r_{1j}} + \ldots + m_l\frac{f'}{f}\prod_l a_j^{r_{1j}}\prod_e \theta_j^{r_{1j}} = 0$$

或者

$$m_1 a^{r_1}\prod_{l'} a_j^{r_{1j}}\prod_e \theta_j^{r_{1j}} + \ldots + m_l a^{r_1}\prod_{l'} a_j^{r_{1j}}\prod_e \theta_j^{r_{1j}} = 0$$

这里$m_i \in C$且不为0,且$F$整除$p(\theta) + r_i\theta q(\theta) + \beta_i q(\theta)$。$\{a\}\bigcup\{a_j\}_{l'}\bigcup\{\theta_j\}_e$是一个互素的不可约的集合,对于0是一个$\sum$分解,(这里$\sum = (a, \ldots, a_i, \ldots, \theta_j, \ldots)$, $i \in l', j \in e$)。根据引理4.2和[Roca79,Thm.3.1],给定$r_i$,我们能够唯一的决定$r_{ij}$。通过定理4.3这样的分解是唯一的,因此$t = 1, m_1 = 0$,则矛盾。因此,我们证明每个$v_i$的分子的因子都是$A$分母的因子。

我们能够重新地描述式4.6中的所有的$v_i$:这些有线性分子的$v_i$一定有$r_i = 0$或$r_i = -lc(p(\theta))$。用这些对于$r_i$的值用引理4.2 能及时所有的$\beta_i$使得$p(\theta) + r_i\theta q(\theta) + \beta_i q(\theta)$被$A$的分母的一个线性因子整除。接下来,计算所有分子阶数大于1的$v_i$,用引理4.2 我们能够直接从$A$的分母中得出。最后,由$\beta_i$的值,我们决定是否存在整数$r_{ij}$和常数$v_i$ 使得$\beta_i = \sum_e r_{ij}a_j + \sum_{l'} r_{ij}\theta_j + v_i$。这样所有的在4.6中的$u'_i/v_i$ 我们都能够决定,我们应用$[Risch69, MainTheorem, part(b)]$

**Case (b2)** $p(\theta) = r\theta + s$和$q(\theta) = 1$
首先,把式4.6分解成部分分式,得到

$$\left[ A_{m+1}\theta^{m+1} + \ldots + A_0 + \sum_{i=1}^{K}\sum_{j=1}^{k_i+1} \frac{A_{ij}}{p_i^j} \right] \theta_n$$
$$= \left[ \left( \sum_{i=0}^{m} B_i\theta^i + \sum_{i=1}^{K}\sum_{j=1}^{k_i} \frac{B_{ij}}{p_i^j} \right)' + (r\theta+s)'\left( \sum_{i=0}^{m} B_i\theta_i + \sum_{i=1}^{K}\sum_{j=1}^{k_i} \frac{B_{ij}}{p_i^j} \right) \right] \theta_n$$
$$+ \sum d_i \left[ \frac{(r+r_i)'}{(r+r_i)} + \frac{(\theta+(s+\beta_i)/(r+r_i))'}{(\theta+(s+\beta_i)/(r+r_i))} \right] \prod_l a_j^{r_{ij}} \prod_e \theta_j^{r_{ij}} \theta_n$$

这里$p_i$是首一不可约的。指出$r_i + r \neq 0$,由于$r_i$是不为0的整数,则$E$是正规的。通过比较等式两端我们有$A_{1k_1+1} = -k_i B_{ik_1} p'_1 \pmod{p_1}$。我们用$[Risch69, p.180]$,计算多项式$R$和$S$使得$Rp_1 + SP'_1 = A_{1k_1+1}$。那么$B_{1k_1} = -S/k_1$。用同样的方法,我们也能计算$B_{2k_2}, \ldots, B_{Kk_K}$。现在用$A\theta_n - [(B_1k_1/p_1^{k_1} + \ldots + B_{Kk_K}/p_K^{k_K})\theta_n]'$
代替$A\theta_n$重复上述过程。最终$B$化为一个多项式,也就是

$$A_{m+1}\theta^{m+1} + \ldots + A_0 + \frac{\bar{A}_{11}}{p_1} + \ldots + \frac{\bar{A}_{K1}}{p_K}$$
$$= (B_m\theta^m + \ldots + B_0)' + (r\theta+s)'(B_m\theta^m + \ldots + B_0)$$
$$+ \sum d_i \left[ \frac{(r+r_i)'}{(r+r_i)} + \frac{(\theta+(s+\beta_i)/(r+r_i))'}{(\theta+(s+\beta_i)/(r+r_i))} \right] \prod_l a_j^{r_{ij}} \prod_e \theta_j^{r_{ij}}$$

这里$\bar{A}_{k1}$在$D(\theta)$中且有$\deg(\bar{A}_{k1}) < \deg_\theta(p_k)$ 那么对于任意的$k$,我们有

$$\frac{\bar{A}_{k1}}{p_k} = \sum_{p_k} d_i \frac{(\theta+(s+\beta_i)/(r+r_i))'}{(\theta+(s+\beta_i)/(r+r_i))} \prod_l a_j^{r_{ij}} \prod_e \theta_j^{r_{ij}} \quad (4.9)$$

这里$\sum_{p_k}$表示和是遍历所有的指标有$p_k = \theta + (s+\beta_i)/(r+r_i)$。从式4.9有

$$\frac{\bar{A}_{k1}}{p'_k} = \sum_{p_k} d_i a^{r_i} \prod_l a_j^{r_{ij}} \prod_e \theta_j^{r_{ij}} \tag{4.10}$$

常数$r_{ij}, r_i, v_i$能在第4节计算。首先指出式4.14是$\bar{A}_{k1}/p'_k$的一个$\sum$分解.
（这里$\sum = (a_1, \ldots, a_i, \ldots, \theta_j, \ldots), i \in l', j \in e$）。进一步，我们有

$$\beta_i = (r+r_i)\delta_k - s \tag{4.11}$$

这里$p_k = \theta + \delta_k$，所以给定$r_i$能够决定$\beta_i$进而得出$r_{ij}$和$v_i$。因此式4.14是一个严格被一个可计算的函数限制的$\sum$分解。所以我们只有证明这个$\sum$分解是一阶的。

很明显有一个指标的$\sum$分解是一阶的。因此我们假设式4.11是被俩个不同的$r_1$和$r_2$满足的。现在定义集合$S$是$D$中所有能够被写成$\sum_e r_j a_j + \sum_{l'} r_j \theta_j + v$的元素集合。指出$S$在加法减法和数乘是封闭的。由式4.11我们有

$$(r+r_1)\delta_k - s \tag{4.12}$$

和

$$(r+r_2)\delta_k - s$$

都在$S$中。通过减法和乘以$1/(r_1 - r_2)$，我们有$\delta_k$在$S$。由4.12知，$r\delta_k - s$在$S$中。因此存在有理数$w_{1j}, w_{0j}$和常数$w_1, w_0$使得

$$\delta_k = \sum_e w_{1j} a_j + \sum_{l'} w_{1j} \theta_j + w_1$$

和

$$r\delta_k - s = \sum_e w_0 j a_j + \sum_{l'} w_{0j} \theta + w_0$$

把这些代入4.11，我们有

$$\sum_e r_{ij} a_j + \sum_{l'} r_{ij} \theta_j + v_i = \sum_e (w_{1j} r_i + w_{0j}) a_j + \sum_{l'} (w_{1j} r_i + w_{0j}) \theta_j + (w_1 r_i + w_0)$$

由于这个表示是在$S$中是唯一的，我们有$r_{ij} = w_{1j} r_i + w_{0j}$，证明在4.14中的$\sum$分解时一阶的。

我们能够决定式4.14中的$d_i$和$r_{ij}$是否存在。如果这些值存在，那么我们决定在式4.2中的所有的$u_i$和$v_i$，应用$[Risch69, Main Theorem, part(b)]$计算$B$和常数$d_i$，因此推出$Case(b2)$，从而证明了此定理。

□

这里$A_j \in D$，$\deg_\psi(P_j) < \deg_\psi(Q_j)$且$gcd(\psi, Q_j) = 1$。则有

$$A_j\theta^j = A_{m_j}\theta^j\psi^{m_j} + \ldots + A_{\bar{m}_j}\theta^j\psi^{m_j} + \theta^j A_{0_j} + \frac{\theta^j P_j(\psi)}{Q_j(\psi)}$$

又一次应用定理4.1我们有$A_j\theta^j$在$E = D(\theta)\psi$的某个刘维尔初等扩张中有不定积分，当且仅当$A_{m_j}\theta^j\psi^{m_j},\ldots,A_{\bar{m}_j}\theta^j\psi^{m_j}, \theta^j A_{0_j} + \dfrac{\theta^j P_j(\psi)}{Q_j(\psi)}$ 有不定积分。$\theta^j A_{0_j} + \dfrac{\theta^j P_j(\psi)}{Q_j(\psi)}$可用上述同样的方法处理。其他部分，考虑一般情形$A_{i_j}\theta^j\psi^{i_j}$，它在$E$的某个刘维尔初等函数扩张中有不定积分，当且仅当它$A_{i_j}\Theta_{i_j}$（$\Theta_{i_j} = \theta^j\psi^{i_j}$）在$D(\Theta_{i_j})$的某个刘维尔初等扩张中有不定积分。注意到在域$D(\Theta_{i_j})$中，$rank(\Theta_{i_j}) \leq r$则$rank(D(\Theta_{i_j})) \leq (m_r - 1, \ldots, m_1, 1)$。根据第二节的讨论，我们用$\tilde{D}_{i_j}$代替$D(\Theta_{i_j})$ 这里$\tilde{D}_{i_j}$和$D(\Theta_{i_j})$同构，且$\tilde{D}_{i_j}$是正则的，那么我们有

$$rank(\tilde{D}_{i_j}) \leq rank(D(\Theta_{i_j})) \leq (m_r - 1, \ldots, m_1, 1) < rank(E)$$

现在假设$\tilde{D}_{i_j}, A_{i_j}\Theta_{i_j}$有相同的代表元$\tilde{A}_{i_j}$。由于$rank(\tilde{D}_{i_j}) < rank(E)$，我们应用归纳假设证明$\tilde{A}_{i_j}$在$D_{i_j}$的某个刘维尔初等扩张中有不定积分。这种证明适用于$A_{m_j}\theta^j\psi^{m_j},\ldots,A_{\bar{m}_j}\theta^j\psi^{m_j}$，那么就证明了$m_r > 1$的情形。

**2** 假设$m_r = 1$，在$E$的某个刘维尔初等扩张中找到$A_j\theta^j$的不定积分，等价于在$F(\theta^{(j)})$中找到$A_j\theta^{(j)}$的不定积分，这里$\theta^j = \theta^{(j)}$。用$\tilde{\theta^{(j)}}$代替$\theta^{(j)}$使得$F(\theta^{(j)})$是正则的且同构于$F(\tilde{\theta^j})$。现在在$F(\tilde{\theta^{(j)}})$中，$A_j\theta^j$有一个代表元$\tilde{A}_j\tilde{\theta}^{(j)}$，这里$\tilde{A}_j \in F$，且有$rank(F(\tilde{\theta}^{(j)})) \leq rank(E)$。当$rank(F(\tilde{\theta}^{(j)})) < rank(E)$时，由归纳假设，我们可以决定$\tilde{A}_j\tilde{\theta}^{(j)}$是否有不定积分；当$rank(F(\tilde{\theta}^{(j)})) = rank(E)$时，由定理4.6可得结论。

$\square$

**例4.7.** $\int (x^3)/\log(x^2 - 1) dx$

证明. 首先$(x^3)/\log(x^2 - 1)dx = x^3/(\log(x+1) + \log(x-1))$，有$C(x, \theta_1 = \log(x+1) + \log(x-1))$。那么它的都是对数阶情形，由Case a 的步骤计算。设$\theta = \theta_2$，$K = 1$，$\bar{A}_{11} = x^3$和$p_1 = \log(x+1) + \log(x-1)$。那么此时4.15变为

$$\frac{(\bar{A}_{11}/p_1')'}{p_1'} = \frac{2x^4 - 3x^2 + 1}{2} = \Sigma d_i r_i (x+1)^{r_i}(x-1)^{r_{i1}}$$

由于$\beta_i/r_i = R_1 \log(x-1) + v = \log(x-1)$，我们有以上的$\Sigma$分解时严格地被函数$g(r_i) = (r_i, R_1 r_i) = (r_i, r_i)$。下面由定理4.5。我们令$\Phi_1 = (2x^4 - 3x^2 + 1)/2$，所以$r_1 = 1, r_{11} = 1, d_1 r_1 = 1/2$。接着$\Phi_3 = \Phi_2 - (x+1)^2(x-1)^2 = 0$。则我们计算得

$$\int \frac{x^3}{\log(x^2 - 1)} dx = \frac{1}{2}\text{li}(x^4 - 2x^2 + 1) + \frac{1}{2}\text{li}(x^2 - 1)$$

$\square$

**定理4.7.** *E*如上所述，给定$\gamma \in E$，如果其在*E*的某个刘维尔扩张中有不定积分我们能够在有限步决定，且能够找到常数$c_i, d_i$和$w_i, u_i, v_i \in E$使得

$$\gamma = w_0' + \sum c_i \frac{w_i'}{w_i} + \sum d_i \frac{v_i'}{v_i} \tag{4.13}$$

这里$v_i' = u_i'/u_i$。

*证明.* 用归纳法，对*E*的阶进行归纳。设定理对于阶小于$(m_r, \ldots, m_1, 1)$都成立，考虑$\gamma \in E = C(x, \theta_1, \ldots, \theta_n)$，这里*E*是可分解的、正则的，且阶为$(m_r, \ldots, m_1, 1)$。分成以下两种情况讨论

**Case a** 存在一个*r*阶对数单项式，记做$\theta$，且$\theta' = a'/a$，让$E = F(\theta)$ 由于$v_i$是$u_i$的一个对数，我们有

$$v_i = r_i \theta + \sum_e r_{ij} a_j + \sum_{l'} r_{ij} \theta_j + v_i$$

这里$l' = \{j|\theta_j = \log(a_j)$是一个*F*中的对数单项式$\}$，$e = \{j|\theta_j = \exp(a_j)$是一个*F*中的指数单项式$\}$，$r_{ij}$是一个有理数，$v_i \in C$，我们还有

$$u_i = \eta_i a^{r_i} \prod_{l'} a_j^{r_{ij}} \prod_e \theta^{r_{ij}}$$

这里$\eta \in C$，不是一般性，我们假设$\eta_i = 1$，对于任意的*i*成立。

现在将式4.1分解成关于$\theta$的部分分式。

$$A_m \theta^m + \ldots + A_0 + \sum_{i=1}^{K} \sum_{j=1}^{k_i+1} \frac{A_{ij}}{p_i^j}$$

$$= \left[ B_{m+1} \theta^{m+1} + \ldots + B_0 + \sum_{i=1}^{K} \sum_{j=1}^{k_i} \frac{B_{ij}}{p_i^j} \sum_{w_i \in F} c_i \int \frac{w_i'}{w_i} + \sum_{w_i \notin F} c_i \int \frac{w_i'}{w_i} \right.$$

$$\left. + \sum_{r_i=0} d_i \int \frac{u_i'}{v_i} + \sum_{r_i \neq 0} d_i \int \frac{(\theta + \beta_i/r_i)'}{(\theta + \beta_i/r_i)} a^{r_i} \prod_{l'} a_j^{r_{ij}} \prod_e \theta_j^{r_{ij}} \right]'$$

这里$\beta_i = \sum_e r_{ij} a_j + \sum_{l'} r_{ij} \theta_j + v_i$，且$p_i \in F(\theta)$是首一不可约的。这里$w_i \notin F$等于某个$p_i$，且每个$\theta + \beta_i/r_i$等于某个$p_i$，对于$1 \leq i \leq K$

首先计算$B_{m+1}, \ldots, B_1$。

接着去掉多项式部分，计算$B_{1k_1}, \ldots, B_{11}, \ldots, B_{Kk_K}, \ldots, B_{K1}$，使用Hermite约化方法。有等式

$$(\frac{\bar{A}_{11}}{P_1} + \ldots + \frac{\bar{A}_{K1}}{P_K}) = \sum c_i \frac{p_i'}{p_i} + \sum d_i \frac{(\theta + \beta_i/r_i)'}{(\theta + \beta_i/r_i)} a^{r_i} \prod_{l'} a_j^{r_{ij}} \prod_e \theta_j r_{ij}$$

这里$\bar{A}_{11}, \ldots, \bar{A}_{K1} \in F[\theta]$，且对于$\forall i$, $deg_\theta(\bar{A}_{i1}) < deg_\theta(p_i)$。由有部分分式分解的唯一性知，对于每个$p_k$, $1 \leq k \leq K$。如果$deg_\theta(p_k) > 1$那么对于某个$c_k$有$\bar{A}_{k1} = c_k p'_k$，因此考虑$p_k$关于$\theta$是线性的。则有

$$\frac{\bar{A}_{k1}}{p_k} = c_k \frac{p'_k}{p_k} + \sum_{p_k} d_i \frac{p'_k}{p_k} a^{r_i} \prod_{l'} a_j^{r_{ij}} \prod_e \theta_j^{r_{ij}}$$

这里$p_k = \theta + \beta_i/r_i$对于$\forall i$成立。因此

$$\frac{\bar{A}_{k1}}{p'_k} = c_k + \sum_{p_k} d_i a^{r_i} \prod_{l'} a_j^{r_{ij}} \prod_e \theta_j^{r_{ij}} \tag{4.14}$$

则

$$\frac{(\bar{A}_{k1}/p'_k)'}{p'_k} = \sum_{p_k} d_i r_i a^{r_i} \prod_{l'} a_j^{r_{ij}} \prod_e \theta_j^{r_{ij}} \tag{4.15}$$

我们现在用第三节的结果计算$r_i, r_{ij}, d_i$。我们仅仅需要证明式4.15是一个被限制的$\Sigma$分解（$\Sigma = (a, \ldots, a_i, \ldots, \theta_j, \ldots), i \in l', j \in e$）。然而$\beta_i/r_i$不依赖于$i$，能被写成$\sum_e R_j a_j + \sum_{l'} R_j \theta_j + v$，这里$R_j = r_{ij}/r_i$且$v = v_i/r_i$，对于所有的$i$和$j$成立。所以$r_{ij} = R_j r_i$证明了4.15是一个一阶的被限制的$\Sigma$分解，我们用4.14决定$c_k$，我们就证明了Case a

**Case b**　所有的$r$阶单项式都是指数的，选择其中的一个$\theta$，令$E = F(\theta)$，把$\gamma$分解成部分分式，得到

$$\gamma = A_m \theta^m + \ldots + A_{\bar{m}} \theta^{\bar{m}} + A_0 + \frac{P(\theta)}{Q(\theta)}$$

这里$\bar{m} \leq m$, $deg_\theta(P) < deg_\theta(Q)$且$gcd(\theta, Q(\theta)) = 1$。如果$\gamma$在$E$的某个刘维尔初等扩张中有一个不定积分，那么由定理4.1知每一个$A_m \theta^m, \ldots, A_{\bar{m}} \theta^{\bar{m}}$都有不定积分。

考虑第一部分$A_0 + P(\theta)/Q(\theta)$，从定理4.1我们有

$$A_0 + \frac{P(\theta)}{Q(\theta)} = w' + \sum c_i \frac{w'_i}{w_i} + \sum d_i \frac{u'_i}{v_i}$$

这里$c_i, d_i \in C$, $u_i, v_i \in E$。通过定理4.1，一定有$u_i, v_i \in F$，因此我们能够将其归结为指数情形的计算。

接下来考虑$A_j \theta_j$。我们将讨论分成两种不同的情形：$m_r > 1$或者$m_r = 1$。

**1**　首先设$m_r > 1$。这意味着有另一个$r$阶的指数单项式，设其为$\psi$，让$F = D(\psi)$，那么$E = F(\theta) = D(\psi)(\theta)$。对$A_j$做关于$\psi$的部分分式

$$A_j = A_{m_j} \psi^{m_j} + \ldots + A_{\bar{m}_j} \psi^{m_j} + A_{0_j} + \frac{P_j(\psi)}{Q_j(\psi)}$$

**例4.8.** $\int (x^2)/\log(x^2-1)dx$

证明. 根据上面有

$$\int \frac{x^2}{\log(x_2-1)}\mathrm{d}x = \int \frac{x^2}{\log(x+1)+\log(x-1)}\mathrm{d}x$$

且

$$\frac{(\bar{A}_{11}/p_1')'}{p_1'} = \frac{3x^4-4x^2+1}{4x} = \sum d_i r_i (x+1)^{r_i}(x-1)^{r_{i1}}$$

这里的$\sum$分解时严格被函数$g(r_i)=(r_i,r_i)$限制。应用定理4.5，我们有$\alpha_{11}=r_1=1$。由于$g(r_i)$对于它的每一个部分都是严格的单项式，$r_1=1>0$，且$\Phi$不是x的多项式，我们在上述定理中$\sum$分解不能终止。因此$\int(x^2)/\log(x^2-1)dx$不能写成对数和初等函数形式的积分。　　　　　　　　　　　　　　□

**例4.9.** 考虑$\int\left[\dfrac{2x+3}{3\log(x)+2x}e^{\log(x)/2+x}+\dfrac{1}{x+1}(e^{\log(x)/2+x})^2\right]$

证明. 我们有$C(x,\log(x),e^{\log(x)/2+x})$。则考虑定理4.6中Case b的情形。我们把问题分成两个部分：

**1** $\int((2x+3)/(3\log(x)+2x)e^{\log(x)/2+x})\mathrm{d}x$　　我们应用定理4.6中的Case b2的情形。这里$r=1/2,s=x,K=1,\bar{A}_{11}=(2x+3)/3p_1=\log(x)+2x/3$。这个$\sum$分解在等式4.14中变成

$$\frac{\bar{A}_{11}}{p_1'} = x = \sum d_i x^{r_i}$$

我们有$d_1=r_1=1$和

$$\int \frac{2x+3}{3\log(x)+2x}e^{\log(x)/2+x}\mathrm{d}x = d_1\mathrm{li}(x^{r_1}e^{\log(x)/2+x}) = \mathrm{li}(xe^{\log(x)/2+x})$$

**2** $\int 1/(x+1)(e^{\log(x)/2+x})^2\mathrm{d}x$　　这里，将这个问题化简得到$\int x/(x+1)e^{2x}\mathrm{d}x$，则在域$C(x,e^{2x})$中。这是定理4.6中的情形。这里$\theta=x,p(\theta)/q(\theta)=2x,A=x/(x+1)$且$\beta$是常数。$A$的唯一的因子是$x+1$。由引理4.2，我们令$\beta=2$且那样$x+1$能够整除$2x+\beta$。所以$u_1=e^{2x},v_1=2x+2$(指出这里$v_1\neq\ln(u_1)$)。应用$[Risch 69 Main Theorem, part(b)]$得到

$$\frac{x}{x+1}e^{2x} = (Be^{2x})' + d_1\frac{(2x)'e^{2x}}{2x+2}$$

这里得到$B=1/2,d_1=-1$。最终我们将$u_1$换为$\bar{u}_1=e^{2x+2}$，$d_1$换为$\bar{d}_1=-1/e^2$，使得$v_1=\ln(\bar{u}_!)$则

$$\int \frac{x}{x+1}e^{2x} = \frac{e^{2x}}{2} - e^{-2}\mathrm{li}(e^{2x+2})$$

　　　　　　　　　　　　　　　　　　　　　　　　　　　　　　　　　　　□

**例4.10.** $\int \left( \dfrac{2x^3 - x^2 - 6x}{x^2 + 3x + 2} + \dfrac{2x - 3}{\log(x) + 1} \right) e^{x \log(x) + x} \mathrm{d}x$

证明. 这里也是$Case b2$的情形，在定理4.6中，我们有$r = x, s = x, \bar{A}_{11} = 2x - 3, p_1 = \log(x) + 1$ 那么在4.14中的$\sum$分解变为

$$\frac{\bar{A}_{11}}{p'_1} = 2x^2 - 3x = \sum d_i x^{r_i}$$

这里$d_1 = -3, r_1 = 1, d_2 = 2, r_2 = 2$，所以

$$u_1 = xe^{x\log(x)+x}, \quad v_1 = x\log(x) + x + \log(x) + 1$$

且有

$$u_2 = x^2 e^{x\log(x)+x}, \quad v_2 = x\log(x) + x + 2\log(x) + 2$$

我们接下来应用$[Risch 69, part(b)]$，这里有

$$\left( \frac{2x^3 - x^2 - 6x}{x^2 + 3x + 2} + \frac{2x - 3}{\log(x) + 1} \right) e^{x\log(x)+x} = -3\frac{u'_1}{v_1} + 2\frac{u'_2}{v_2}$$

最后，我们令$\bar{u}_1 = eu_1, \bar{u}_2 = e^2 u_2$那么

$$\int \left( \frac{2x^3 - x^2 - 6x}{x^2 + 3x + 2} + \frac{2x - 3}{\log(x) + 1} \right) e^{x\log(x)+x} \mathrm{d}x$$
$$= -\frac{3}{e} \mathrm{li}(e^{x\log(x)+x+\log(x)+1}) + \frac{2}{e^2} \mathrm{li}(e^{x\log(x)+x+2\log(x)+2})$$

$\square$

### 4.1.6 总结

定理4.6也可应用于以下的一些特殊情形

- 指数形式的积分

$$ei(u) = \int \frac{u' \exp(u)}{u} \mathrm{d}(x) = \mathrm{li}(\exp(u))$$

- 正弦、余弦的积分

$$si = \int \frac{u' \sin(u)}{u} \mathrm{d}(x) = \frac{1}{2i}[ei(iu) - ei(-iu)]$$

$$ci = \int \frac{u' \cos(u)}{u} \mathrm{d}(x) = \frac{1}{2}[ei(iu) + ei(-iu)]$$

例如

$$\begin{aligned}
\int \frac{\cos(x)^2}{x^3} \mathrm{d}x &= -\frac{2x^2 ei(2ix) + 2x^2 ei(-2ix) - 2x \sin(2x) + \cos(2x) + 1}{4x^2} \\
&= -ci(2x) + \frac{\sin(2x)}{2x} - \frac{\cos(2x)}{4x^2} - \frac{1}{4x^2}
\end{aligned}$$

### 4.1.7　推广

**一些记号**

1. ：$\mathrm{erf}(u) = \int \exp(-u^2)\mathrm{d}u$

2. ：$\mathrm{li}_2(u) = -\int \log(u)\mathrm{d}u/(u-1)$

我们这研究一种算法能够求出一下的积分：

1. $\int \dfrac{\mathrm{d}x}{\log(x)\log(\mathrm{li}(x))} = \mathrm{li}(\mathrm{li}(x))$

2. $\int e^{-x^2-\mathrm{erf}^2(x)}\mathrm{d}x = \mathrm{erf}(\mathrm{erf}(x))$

3. $\int \dfrac{\exp(1/2)\log(\log(x)) - 1/\log(x)}{x\log^2(x)}\mathrm{d}x = 2\mathrm{erf}\left[\dfrac{1}{\log(x)}\right]$

**对数积分**

对于定义在li初等函数扩张中的函数我们同样有

**定理4.8.** $F$是$C(x)$的刘维尔扩张，$C$是$F$上常数的代数闭域，若$g \in F$在$F$的刘维尔初等扩张中有不定积分，则存在常数$c_i, d_i \in C$；　$w_i, u_i, v_i \in F$，使得

$$g = w_0^{'} + \Sigma c_i \dfrac{w_i^{'}}{w_i} + \Sigma d_i \dfrac{u_i^{'}}{v_i}$$

其中$v_i^{'} = u_i^{'}/u_i$，即$v_i = \log(u_i)$

**定理4.9.** $C(x)$是一个特征为零的微分域，且$C$是代数闭域，让$E = C(x, \theta_1, \ldots, \theta_n)$是$C(x)$一个的超越的刘维尔扩张。给定的$g \in E$如果有不定积分，能在有限步中找到$c_i, d_i \in C; u_i, v_i, w_i \in E$，使定理1中的等式成立。

**例4.11.**
$$\int \dfrac{\mathrm{d}x}{\log(x)\log(\mathrm{li}(x))}$$

我们先计算$\theta_i$，则有$\theta_1 = \log(x), \theta_2 = \mathrm{li}(x), \theta_3 = \log(\mathrm{li}(x))$ 则有$E = \mathbf{C}(x, \theta_1, \theta_2, \theta_3)$，其中$(C)$是复数域。

由定理1中的等式知，令$c_i = 0, d_1 = 0, u_1 = \mathrm{li}(x)$,则$g = u_1^{'}/\log(u_1)$（已经证明有且只有这一种表达），则上述积分为$\mathrm{li}(\mathrm{li}(x))$

对于

**误差函数**

我们再定义一个初等扩张域

**定义4.4.** $F$是一个微分域，满足$\forall u,v \in F \quad (uv)' = uv' + u'v; (u+v)' = u' + v'$，特征为0，$E$是$F$上的一个微分扩张。称$E$为一个$F$的误差初等扩张，若$\exists \theta_1,\ldots,\theta_n \in E$，使得$E = F(\theta_1,\ldots,\theta_n)$，且$\forall 1 \leq i \leq n$满足以下条件中的某一个

1. $\theta_i$是$F(\theta_1,\ldots,\theta_{i-1})$上的代数

2. 对某个$a \in F(\theta_1,\ldots,\theta_{i-1})$，有$\theta_i = \exp(a)$

3. 对某个$a \in F(\theta_1,\ldots,\theta_{i-1})$，有$\theta_i = \log(a)$

4. 对某个$a \in F(\theta_1,\ldots,\theta_{i-1})$，有$\theta_i = \mathrm{erf}(a)$

**定义4.5.** $E$是一个$C(x)$的正则对数显式刘维尔扩张，$E = C(x,\theta_1,\ldots,\theta_n)$，对于任意的$i$，满足$1,2,3$中的一个，或者

$\theta_i' = a$，对于某个$a \in F(\theta_1,\ldots,\theta_{i-1})$，且$\theta_i$在$F(\theta_1,\ldots,\theta_{i-1})$不是初等的

如果每个对数单项式$\theta_i = \log(a)$，使得$a$是一个$F(\theta_1,\ldots,\theta_{i-1})$上的不可约多项式，称$E$是可分解的。

**定义4.6.** $E$是一个$C(x)$的对数显式刘维尔扩张，且严格满足$C(x) = F_0 \subset F_1 \subset \ldots \subset F_r = E$，使得任意$i > 0$，$F_i = F_{i-1}(\theta_{i1},\ldots,\theta_{ik_i})$，满足任意$\theta_{ij} = \theta_k$对于某个$k$，且对于任意$j$满足以下条件中的一个

1. $\theta_{ij}$是$F_{i-1}$上的代数元，是$F_{i-2}$上的超越元

2. $\theta_{ij}'/\theta_{ij} = a_{ij}'$对于某个$a_{ij} \in F_{i-1}$，但$a_{ij} \notin F_{i-2}$

3. $\theta_{ij}' = a_{ij}'/a_{ij}$对于某个$a_{ij} \in F_{i-1}$，但$a_{ij} \notin F_{i-2}$

4. $\theta_{ij}' = a_{ij}$对于某个$a_{ij} \in F_{i-1}$，且$\theta_{ij}$不是$F_{i-2}$上的初等元

我们定义$E$中元素的阶

**定义4.7.** $a \in E$为$k$阶的，若$a \in F_k$，但$a \notin F_{k-1}$

我们定义$E$的阶

**定义4.8.** $E$的阶定义为$(m_r,\ldots,m_1,1)$

我们下面的定理

**定理4.10.** $F$是$C(x)$的刘维尔扩张，$C$是$F$上常数的代数闭域，若$g \in F$在$F$的误差初等扩张中有不定积分，则存在常数$c_i, d_i \in C$；　$w_i \in F$，$u_i, v_i$是$F$上的代数元，使得

$$g = w_0' + \Sigma c_i \frac{w_i'}{w_i} + \Sigma d_i u_i' v_i \tag{4.16}$$

其中$v_i'/v_i = -(u_i^2)'$，且$(u_i)^2, (v_i)^2, u_i' v_i$

为了得到我们想要的结果，我们定义某一种形式的刘维尔扩张

**定义4.9.** $E$是$C(x)$的一个超越的刘维尔扩张，其中$C$为$E$的常数域。让$\theta_m = \exp(a_m)$是一个$k$阶的指数单项式。我们说$\theta_m$是拟二次的，存在$k-1$阶对数单项式$\theta^* = \log(a^*)$和一个整数$l$，使得

$$a_m = (l/2)\theta^* + \frac{a\theta^2 + b\theta + c}{\theta + f}$$

这里

1. $\theta$是一个$k-2$阶的对数单项式

2. $a^* \in C[x, \theta_1, \ldots, \theta_{m-1}]$

3. $a \in C$

4. $b, c \in F_{k-3}[\theta^{1*}, \theta^{2*}, \ldots]$（这里$\theta_1{}^*$，$\theta_2{}^*$，…是除了$\theta$外的$k-2$阶对数），

5. 则对于任意$i$，$b$和$C$在$F_{k-3}[\theta_1{}^*, \ldots, \widehat{\theta_2}{}^*, \ldots]$（$\widehat{\theta_2}{}^*$表示以$\theta_2{}^*$为未知元）上至多是二次的

6. $F \in F_{k-3}[\theta_1{}^*, \theta_2{}^*, \ldots]$，使得于对任意$i$，$F$在$F_{k-3}[\theta_1{}^*, \ldots, \widehat{\theta_2}{}^*, \ldots]$上至多是线性的

7. $c - bf + af^2$是$F_{k-3}[\theta_1{}^*, \ldots] \setminus C$上的平方元

8. $\sqrt{c - bf + af^2}$是那些在$C$上至多$k$阶的指数单项式和至多$k-2$阶的对数单项式的变元（除$\theta_m$外）的线性组合

9. $a^*/(\theta + f) \in F_{k-3}$

**定义4.10.** 对于那些$k$或者$k-1$阶的指数单项式$\theta_i$，且对于任意整数列$\{n_i\}$，$(\Sigma_i \theta_i{}^{n_i})\theta_m{}^{n_m}$不是拟二次的，我们称$\theta_m$是$*-$可约的。如果$E$中的每个指数单项式都是$*-$可约的，我们称$E$是$*-$可约的。

关于误差函数积分我们有以下定理

**定理4.11.** 令$C(x)$如定理2中叙述，令$E = C(x, \theta_1, \ldots, \theta_n), n \geq 0$，是$C(x)$的一个刘维尔扩张。假设$E$是$*-$可约的，给定$g \in E$，如果$g$在$E$的某个误差初等函数扩张有不定积分，那么可以在有限步内找到常数$c_i, d_i$和$w_i \in E$和$E$上的代数元$u_i, v_i$，使得定理中的等式成立。

**例4.12.** $\int \exp(-x^2 - \mathrm{erf}^2(x))\mathrm{d}x$

证明. 我们令$g = \exp(-x^2 - \mathrm{erf}^2(x))$，决定$E = C(x, \theta_1 = \exp(-x^2), \theta_2 = \mathrm{erf}(x), \theta_3 = \exp(-x^2 - \mathrm{erf}^2(x))$ 通过决定步骤，每个满足定理3中(1)式的$u_i$，满足

$$-u_i^2 = -x^2 - \mathrm{erf}^2(x) + r_i(-x^2) + v_i$$

其中$r_i, v_i \in C$

我们还有$\exp(-tx^2)[-x^2-\mathrm{erf}^2(x)+r_i(-x^2)+v_i] = R_i^2$，其中$R_i \in F(x, \theta_1, \theta_2)$，$t = 0$或$1$

因此，我们得到$t = 0, r_i = -1, v_i = 0$。则$-u_i^2 = -\mathrm{erf}^2(x)$，所以$u_i = \pm\mathrm{erf}(x^2)$，至此我们得到一种可能的误差积分$\mathrm{erf}(\mathrm{erf}(x))$。

接着我们需要决定$b \in F, e_1 \in C$，使得

$$\begin{aligned}(b\theta_3)' &= g - e_1[\mathrm{erf}(\mathrm{erf}(x))]' \\ &= g = e_1\exp(-x^2)\exp(-\mathrm{erf}^2(x)) = \theta_3 - e_1\theta_3\end{aligned}$$

则我们得到$e_1 = 1, b = 0$，则$\int g = \mathrm{erf}(\mathrm{erf}(x))$ □

**例4.13.** $\int \dfrac{\exp[(1/2)\log(\log(x)) - 1/\log(x)]}{x\log^2(x)}$

证明. 设$g^* = 1/x\log^2(x)$，令$E = C(x, \theta_1 = \log(x), \theta_2 = \log(\log(x)), \theta_3 = \exp[(1/2)\log(\log(x)) - 1/\log(x)])$，我们要找到$\int g^*\theta_3$。

这里$E$是$C(x)$的不可约的初等扩张，所以Cherry的步骤就不起作用了，但是$E$是$*-$可约的，所以我们运用定理4

此时$u_i$满足 $-u_i^2 = 1/\log(x) + r_i\log(x) + v_i$（其中$r_i, v_i \in C$）且$x^i \cdot \log_2^t(x) \cdot (-u_i^2)$是一个平方元，对于任意$t_i = 0$或$1$。

接着我们令$g^*\log^2(x)$是一个$C(x)$上的分式，我们有$g^*\log^2(x) = 1/x$，下一步，有

$$\frac{1/x}{-1/x} = \Sigma e_i x^{r_i} \cdot \log^{r_{ij}}(x)$$

其中$r_i, r_{ij} \in C$

我们令$e_1 = -1, r_1 = r_{1j} = 0$，那么有$-u_i^2 = -1/\log(x) + v_i$，对于某个$v_i \in C$

且$(-u_i^2)\log(x) = -1 + v_i\log(x)$是完全平方，则$v_i = 0$。则可能的积分结果只能是$\mathrm{erf}(1/\sqrt{\log(x)})$。

接着，我们需要验证是否存在$b_i \in C(x, \log(x)), e_1 \in C$使得

$$\begin{aligned}
g * \theta_3 &= (b\theta_3)' + e_1 \cdot [\text{erf}(1/\sqrt{\log(x)})] \\
&= (b\theta_3)' + e_1 \cdot \frac{1}{2x\sqrt{\log^3(x)}} \exp(\frac{1}{\log(x)}) \\
&= (b\theta_3)' + e_1 \cdot \left[\frac{1}{2x\log^2(x)}\right] \cdot \theta_3
\end{aligned}$$

我们可以令$b = 0, e_1 = 2$，则

$$\int \frac{\exp[(1/2)\log(\log(x)) - 1/\log(x)]}{x\log^2(x)} = 2\text{erf}\left[\frac{1}{\sqrt{\log(x)}}\right]$$

$\square$

接下来我们给一个函数，$E$不是$*-$可约的，但是有初等误差的不定积分

**例4.14.** $\log g^* = \dfrac{x(x+1)}{\log(x)} - \dfrac{(x+1)^2 \log(x+1)}{2\log^2(x)} + \dfrac{(x+1)^2}{z\log(x)} + \dfrac{x}{(x+1)^3 \log(x)}$
$- \dfrac{\log(x+1)}{2(x+1)^2)\log^2(x)} + \dfrac{1}{2(x+1)^2 \log(x)}$

证明. 我们求$\int g^* \theta_3$，令$E = C(x, \theta_1 = \log(x), \theta_2 = \log(x+1), \theta_3 = \log(\log(x)), \theta_4 = \exp[(1/2)\log(\log(x)) + \log^2(x+1)/\log(x)])$，此时$E$是$*-$不可约的。我们需要找到

$g^*\theta_1 = x(x+1)\log(x) - 0.5(x+1)^2 \log(x+1) + 0.5(x+1)^2 \log(x) + x\log(x)/(x+1)^3 - 0.5\log(x+1/(x+1)^2) + 0.5\log(x)/(x+1)^2$ 我们需要得到

$$-0.5x(x+1)^2 - 0.5x/(x+1)^2 = \Sigma e_i x^{r_i}(x+1)^{r_{1i}}$$

这里$r_i, r_{1i} \in Z$，且$r_{1i}^2 = 4r_i, e_i \in C$。

我们得到满足这些条件的$r_1 = r_2 = 1, e_1 = -0.5, e_2 = -0.5, r_{1i} = 2, r_{2i} = -2$

而$g^*\theta_3$有积分为

$$\text{erf}\left[i\frac{\log(x+1) + \log(x)}{\log(x)}\right] + \text{erf}\left[i\frac{\log(x+1) - \log(x)}{\log(x)}\right]$$

$\square$

## 4.2 代数函数的积分

### 4.2.1 历史回顾

Liouville(1833)除给出重要的Liouville定理外，还对代数函数积分的一种特殊情形（可表为不含对数部分的形式）给出了积分算法。之后，在代数函数积

（$QF(V)$指的是$V$的商域）。令$\overline{v} = [v_1, \ldots, v_n]$是$V$在$R$上的一组基。$\overline{v}$的判别式生成$R$的理想称为$V$在$R$上的判别式。$V$在$R$上的基底的判别式仅相关一个$R$中单位的平方，因此生成同样的理想。

**定义4.12** (理想化子和逆). 设$m$为环$S$中的一个理想，我们定义$m$的理想化子$Id(m)$和逆$m^{-1}$分别为

$$Id(m) = \{u \in QF(S)|\ um \subseteq m\}$$

$$m^{-1} = \{u \in QF(S)|\ um \subseteq S\}$$

显然我们有包含关系$S \subseteq Id(m) \subseteq m^{-1}$。

**定理4.12.** $V$是整闭的当且仅当判别式的根式理想的理想化子等于$V$。

定理的证明过程暂时略去。

由此得到了我们计算$V$整闭包的算法：

---

**算法4.1** (计算整闭包).

输入：$V$的定义多项式$f(x, y)$。

输出：$V$的整闭包。

1. 令$d = \text{Res}_y(f, f'_y)$，$k = d$

2. 设$q = \prod p_i$，其中$p_i$是素的且$p_i \mid k$，$p_i^2 \mid d$，如果$q$是单位则输出$V$

3. 找到$(q)$在$V$中的根式$J_q(V)$

4. 找到$J_q(V)$的理想化子$\widehat{V}$，和将$\widehat{V}$变到$V$的变换基矩阵$M$

5. 令$k = \det(M)$，如果$k$是单位则返回$V$

6. 置$d = d/k^2$，$V = \widehat{V}$，返回2

---

**计算判别式的根式理想**

判别式是由$R$中某元素$d$在$V$中生成的主理想$(d)$。我们想要计算$(d)$的根式理想。由于$R$是主理想整环，从而是唯一分解整环。令$(p_1, \ldots, p_k)$为$d$在$R$中的不同的素因子。由于$(d)$的根式理想为所有包含$d$的素理想之交，从而也就是$p_i$的根式理想之交。因此我们可以先只考虑如何计算由$R$中的素元$p$在$V$中生成的主理想的根式理想。跟随Ford的记号，我们称这种理想为$V$的$p-radical$。

分领域做出重要贡献的是Risch(1968)，他概略地叙述了将代数函数积分问题转化成为决定代数函数域的除子群的挠子群这样一个代数几何问题的步骤，最终在引用一些最近的代数几何方面的结果后，他在理论上解决了这一问题。

### 4.2.2 整基

**概念引入与算法得出**

由于在之后的步骤中，我们会遇到转化函数极点的问题，因此我们需要判断并产生那些仅在无穷远点有极点的函数，这样的函数称为整性代数函数。下面给出整元的一般定义：

**定义4.11** (整元). 设$R$为一交换环，$F$为一域且包含$R$，$x \in K$称为在$R$上是整的，若$x$满足以下条件之一：

1. $x$满足方程
$$x^m + a_{m-1}x^{m-1} + \cdots + a_0 = 0$$
其中，$a_i \in R \quad \forall 0 \leq i < m.$

2. 存在有限生成的非零$R$-模$M \subseteq F$使得$xM \subseteq M$.

特别地，当$R$为一域时，整元的定义与代数元的定义相同。

设$K$为一域，则一个函数在$K(x)$上是代数的当且仅当它满足唯一的系数属于$K(x)$中的首一不可约多项式

$$Z^m + a_{m-1}Z^{m-1} + \cdots + a_0 \tag{4.17}$$

应用前面的定义，这样的一个函数在$K[x]$上是整的当且仅当$a_i \in K[x] \quad \forall 0 \leq i < n$，即系数均为$x$的多项式。

设$K(x,y)/K(x)$为$n$次代数扩张，那么$K(x,y)$中的整性函数全体构成一个阶为$n$的自由$K[x]$-模，即任一整性函数可以写为$n$个基元的系数为$x$的多项式的线性组合。

这样的基被称为整基。

容易看出，如果我们允许系数为$x$的有理函数，那么这组整基也构成了$K(x)$-线性空间$K(x,y)$的基底。

Trager的算法主要基于Zassenhaus和Ford的工作。

考虑$K(x,y)$，其中$K$为一数域，$x$为其上的超越元，$f(x,y)$为$K[x]$上的一个$n$次不可约的可分多项式。不失一般性，我们可以假定$f$为首一的，否则，可以令$\hat{y} = ay$，其中$a$为首项系数，则$\hat{y}$满足一个首一多项式，且生成同样的函数域。$K(x,y)$中的$K[x]$上整元构成了一个环，称为$K[x]$在$K(x,y)$中的整性闭

包。同前所述，这个环同样是一个阶为$n$的自由模。由于$y$在$K[x]$上是整的，整元的和与积仍是整元，因此$[1, y, \ldots, y^{n-1}]$构成整性$K[x]$-模的一组基。这是我们对于整基的初步"逼近"。下面算法的每一步都会生成一组严格更大的整性$K[x]$-模直到得到最终的整性闭包。

一个重要的对整性闭包的全子模（即阶为$n$的子模）的相对大小的度量可由判别式给出。令$[w_1, \ldots, w_n]$为$K(x,y)$中的元素。由于$K(x,y)$是$K(x)$的次数为$n$的可分代数扩张，因此有$n$种不同的方式嵌入映射$\sigma_i$映入的代数闭包。一个元素在这些映射下的像被称为它的共轭元。$\overline{w} = [w_1, \ldots, w_n]$可以定义为

$$M_{\overline{w}} = \begin{bmatrix} \sigma_1(w_1) & \cdots & \sigma_n(w_1) \\ \vdots & & \vdots \\ \sigma_1(w_n) & \cdots & \sigma_n(w_n) \end{bmatrix} \tag{4.18}$$

$\overline{w}$的判别式是如上述共轭矩阵行列式的平方。判别式非零当且仅当$\{w_i\}$生成一个完全模（即$w_1, \ldots, w_n$线性无关）。我们定义元素$w \in K(x,y)$的迹($sp$)为$sp(w) = \sum \sigma_i(w)$。由于这是共轭元的对称多项式，所以由简单的根与系数的关系知，迹总是$K(x)$中的元素。我们也可以定义如下的矩阵，换如下的观点来看判别式的含义：

$$\begin{aligned} SP_{\overline{w}} &= \begin{bmatrix} \sigma_1(w_1) & \cdots & \sigma_n(w_1) \\ \vdots & & \vdots \\ \sigma_1(w_n) & \cdots & \sigma_n(w_n) \end{bmatrix} \begin{bmatrix} \sigma_1(w_1) & \cdots & \sigma_1(w_n) \\ \vdots & & \vdots \\ \sigma_n(w_1) & \cdots & \sigma_n(w_n) \end{bmatrix} \\ &= \begin{bmatrix} sp(w_1^2) & \cdots & sp(w_1 w_n) \\ \vdots & & \vdots \\ sp(w_n w_1) & \cdots & sp(w_n^2) \end{bmatrix} \end{aligned} \tag{4.19}$$

这样$\overline{w}$的判别式就可以视为$SP_{\overline{w}}$行列式的平方。如果$w_i$'s都是整函数，则他们的迹就是$x$的多项式，因此判别式也是$x$的多项式。

如果$\overline{v} = [v_1, \ldots, v_n]$是包含$\overline{w}$的完全模的一组基，则每个$w_i$都可以写为$v_j$的以$x$的多项式为系数的线性组合，即有$\overline{w} = A\overline{v}$，其中变换矩阵$A$是$n \times n$的$x$的多项式的矩阵。因此，对于共轭矩阵，我们有$M_{\overline{w}} = A M_{\overline{v}}$以及$\operatorname{Disc}(\overline{w}) = \det(A)^2 \operatorname{Disc}(\overline{v})$。$\overline{w}$和$\overline{v}$生成相同的模当且仅当$A$为$K[x]$上的可逆阵，即$\det(A) \in K$。如果$\overline{v}$严格包含$\overline{w}$，那么$\det(A)$就是一次非零次的多项式$p(x)$，并且$\operatorname{Disc}(\overline{v}) = \operatorname{Disc}(\overline{w})/p(x)^2$。因此，每当我们可以产生一个严格大的$K[x]$-模，我们可以从判别式中消去一个平方因子，所以这一过程会在有限步后结束。

下面我们指出算法所基于的主要的代数结果。

令$R$为一主理想整环（PID），在我们的情形下即$K[x]$。令$V$为一个整环，且为$R$的有限整性扩张。则$V$也是阶等于$[QF(V) : QF(R)]$的自由$R$-模

$V$中的元素$u$属于$p-radical$当且仅当$u$所满足的首一极小多项式——式(4.17)中的系数$a_i$都可以被$p$整除。这给了我们判断$p-radical$中元素的方法，但我们关于的是理想的生成元。Zassenhaus和Ford发现了，在一定的条件下，从$V$到$R$的迹映射提供了$p-radical$中元素的线性约束。对于任意$V$中的元素$u$，$u$在$R$上的次数必须整除$V$在$R$上的阶数即$[QF(V):QF(R)]$。若$u$在$R$上的次数为$m$，则$sp(u) = -(n/m)a_{m-1}$。因此若$u \in p-radical$，则$p \mid sp(u)$。再次跟随Ford的记号，我们定义$p-trace-radical$为集合$\{u \in V \mid p \mid sp(uw) \quad \forall w \in V\}$我们有如下简单的引理：

**引理4.3.** $p-radical \subseteq p-trace-radical$

下面我们要找到引理4.3中两集合相等的条件。这样又导出了这样的定理：

**定理4.13.** 如果$R/(p)$的特征大于$V$在$R$上的阶数，则$p-radical = p-trace-radical$

而在我们关心的代数函数域的情形下，条件自动满足。

**计算p-trace-radical**

令$[w_1, \ldots, w_n]$为$V$在$R$上的一组基。$p-trace-radical\ J_p(V)$定义为集合$\{u \in V \mid p \mid sp(uw) \quad \forall w \in V\}$ 而

$$\begin{aligned} sp(uw) &\equiv 0 \mod p \quad \forall w \in V \quad \Longleftrightarrow \\ sp(uw_i) &\equiv 0 \mod p \quad \forall 1 \leq i \leq n \Longleftrightarrow \\ \sum_{j=1}^n u_j sp(w_j w_i) &\equiv 0 \mod p \quad \forall 1 \leq i \leq n \end{aligned}$$

利用由式(4.19)定义的迹矩阵$SP_{\overline{w}}$，我们可以将上式写为

$$SP_{\overline{w}} \cdot \overline{u} = \begin{bmatrix} sp(w_1^2) & \cdots & sp(w_1 w_n) \\ \vdots & & \vdots \\ sp(w_n w_1) & \cdots & sp(w_n^2) \end{bmatrix} \begin{bmatrix} u_1 \\ \vdots \\ u_n \end{bmatrix} \in pR^n \quad (4.20)$$

其中，$pR^n$表示由长度为$n$的每个分量均可被$p$整除的向量构成的集合。式(4.20)在$R^n$中的解$R$-模即为$J_p(V)$。而我们实际上要求的是$J_q(V)$，其中$q$是判别式的不同的素因子的乘积。

$$J_q(V) = \bigcap_{p_i \mid q} J_{p_i}(V)$$

因此，$J_q(V)$中的元素即为所有满足右端实际为$qR^n$的式(4.20)的所有$\overline{u} \in R^n$。为保证解向量的各分量确实落在$R$中，而不是$QF(R)$中，我们需要向式(4.20)增

加一些方程。令$I_n$为$n \times n$的单位矩阵，则$\overline{u} \in R^n$当且仅当$qI_n \cdot \overline{u} \in qR^n$。因此，令

$$M_q = \begin{pmatrix} SP_{\overline{w}} \\ qI_n \end{pmatrix}$$

则我们有

$$J_q(V) = \{u \in QF(V) \mid M_q \cdot \overline{u} \in qR^{2n}\}$$

如果我们左乘一个可逆$R$-矩阵，则解$R$-模是不变的。可逆$R$-矩阵被称为幺模阵，以行列式为$R$中单位元为特征。由于$R$是主理想整环，因此存在幺模阵将$M_q$转化为上三角矩阵。这一过程允许我们将$2n$个方程压缩为等价的$n$个相互独立的方程。上三角化完毕后，我们只需对约化后的$M_q$的前$n$行组成的方阵取其逆矩阵$M_q^{-1}$，则$qM_q^{-1}$的列向量即为式(4.20)的解$R$-模的一组基。

在我们关心的情形上，我们有$R$实际为欧几里德整环，这允许我们只经过行变换便可将$M_q$上三角化。这一过程称为Hermitian行消去法，与对域上矩阵的Gauss消去法类似。只不过，在Gauss消去法中，对于任意非零元，可利用第三种初等行变换将它所在列的其它元素均消为0，而在Hermitian行消去法中，对于任意非零元，只能利用第三种初等行变换将它所在列的其它元素消得比它小。而由于$R$为欧几里德整环，带余除法成立，因此经有限步可以找到一个元素能够整除它所在列的其它元素。

令$d$为与$R$相关的尺寸函数，对于$R = K[x]$我们取用多项式的次数，对于$R = \mathbb{Z}$我们取用整数的绝对值。为简化算法，我们规定$d(0) = \infty$。并用$nrows$和$ncols$分别代表给定矩阵$M$的行数和列数。根据实际问题，我们假定$nrows > ncols$且$M$的秩恰为$n$。则如下算法给出了Hermitian行消去法的算法。

---

**算法4.2** (Hermitian行消去法).
　　输入：矩阵$M$
　　输出：上三角化后的矩阵$M$

1. 对$j = 1$到$ncols$循环

    (a) 选择$k$使$d(M_{kj})$最小，对于$j \leq k \leq nrows$

    (b) 交换$j$行和$k$行

    (c) 对于$i = j + 1$到$nrows$循环

    　i. 设$q$为$M_{ij}/M_{jj}$的多项式部分

    　ii. 用$M_i - qM_j$代替$M_i$

> (d) 如果对于某个$M_{ij} \neq 0$，对于$j < i \leq m$返回a
>
> 2. 输出$M$

**计算理想化子**

给定$V$中理想$m$的一组$R$-基$overline m = (m_1, \ldots, m_n)$，我们希望能够计算出$m$的理想化子的一组$R$-基。回顾定义4.12。尽管在计算整基的过程中我们不需要计算$m^{-1}$，但我们会在后续的章节中用到它，而且两个算法是类似的，且计算逆的算法略微简单些，所以我们先给出求逆的算法。

我们假定$\overline v = (v_1, \ldots, v_n)$构成$V$在$R$上的一组基。$u \in m^{-1}$当且仅当$um_i = \sum r_{ij}v_j$，其中$r_{ij} \in R \quad \forall 1 \leq i \leq n$。乘以$m_i$为$V$的线性变换，令$M_i$为在基$\overline v$下对应的矩阵。由于$\overline v$也构成$QF(V)$在$QF(R)$上的一组基，$u$可表为$\sum u_i v_i$的形式，其中$u_i \in QF(R)$。则$M_i[u_1, \ldots, u_n]^T$生成$um_i$在基$\overline v$下的坐标表示。因此$u \in m^{-1}$当且仅当$M_i[u_1, \ldots, u_n]^T$的各分量均落在$R$中$\forall i$。令

$$M = \begin{pmatrix} M_1 \\ \vdots \\ M_n \end{pmatrix}$$

则$M\overline u \in R^{n^2}$的解空间即为$m^{-1}$。与求$p-trace-radical$时完全相同的道理，我们可以利用Hermitian行消去法，将$M$约化为$n \times n$的方阵$\widehat M$，从而$M\overline u \in R^{n^2}$当且仅当$\widehat M \overline u \in R^n$，则$\widehat M^{-1}$构成了$m^{-1}$的一组基。

完全相同的方式我们可以求得$Id(m)$。$u \in Id(m)$当且仅当$um_i = \sum r_{ij}m_j$，其中$r_{ij} \in R \quad \forall 1 \leq i \leq n$。$M_i$仍代表乘以$m_i$对应的矩阵，不过注意此时基底发生了变化。输入时是以$\overline v$为基底，输出时则以$m$的基底$\overline m$为基底。除此之外，两个算法完全相同。下面给出两个算法的摘要：

> **算法4.3** (求理想化子和逆).
>
> 输入：矩阵$M_i$，代表乘以$m_i$的矩阵，且取输入基底为$\overline v$，输出基底，对于计算逆，取为$\overline v$，对于计算理想化子，取为$\overline m$
>
> 输出：理想化子或逆的一组基
>
> 1. 取经算法4.2得到约化后的矩阵$M$，取其前$n$行构成的方阵为$\widehat M$
>
> 2. 返回$\widehat M^{-1}$的列向量组

注意到$\widehat M^{-1}$的转置矩阵就是算法4.1中第4步中所需要的变换矩阵。

**在∞处正则**

在之前的章节我们已经计算出了整基，因而可以识别被积函数的有限极点，但我们同样也需要处理在∞处的奇点。任给$K[x]$-模的一组基（整基为一特例），我们希望最小化所有基底元素在∞处的阶之和。一组整基$[w_1, \ldots, w_n]$的显著特征就是$\sum a_i(x) w_i$在$K[x]$整当且仅当每一项$a_i(x) w_i$是整的。换句话说，来自不同项之间的奇点是不能相消的。我们希望我们的基底能有$K(x)$在∞在的局部环拥有相同的性质。我们称拥有此种性质的基底在∞处正则。

**定义4.13** (赋值环). 域$F/K$的赋值环是指一个环$\mathcal{O} \subseteq F$并有如下性质：

1. $K \subsetneq \mathcal{O} \subsetneq F$
2. $\forall z \in F$有$z \in \mathcal{O}$或$z^{-1} \in \mathcal{O}$

特别地，在$K(x)$的情形下，在$p$处的赋值环被定义为$K(x)$中的在$p$处没有极点的函数构成的集合。回顾整性的定义，$K(x, y)$中的一个函数被称为在$p$的赋值环上是整的（或简称为在$p$处是整的）如果它满足首一多项式且系数均为$p$的赋值环中的元。

与整体的整基类似，$K(x)$的每一个$place$都存在一组局部整基，使得在$p$处整的元均写为它们的以$p$处赋值环中元为系数的线性组合。为方便起见，我们引入一个稍弱的概念——正则基。

**定义4.14** (正则基). 一组基$[w_1, \ldots, w_n]$被称为在$p$处正则，如果存在$r_i(x) \in K(x)$使得$r_i w_i$构成在$p$处的局部整基。换言之，正则基就是局部整基上允许有理分式因子的差异。

有了以上的术语，我们便可以知道一组基在∞处正则当且仅当基底元素经过乘以某个有理分式因子可以称为在∞处的局部整基。

令$[w_1, \ldots, w_n]$为$K[x]$-模的一组基，我们假定我们已经有了在∞处的一组局部整基$[v_1, \ldots, v_n]$。我们希望在不影响这组基在其它$place$处的性质的基础上，经过适当的修改，使之成为在∞处的正则基。

我们首先将$w_i$展为$v_j$的线性组合：

$$w_i = \sum_{j=1}^{n} M_{ij} v_j$$

其中$M_{ij} \in K(x)$ 按照正则基的定义，如果$w_i$是正则基，就一定存在$r_i \in K(x)$使得变换矩阵$(r_i M_{ij})$可逆，这等价于变换矩阵$(r_i M_{ij})$的行列式是在∞处赋值环的可逆元，即分子分母次数相同的有理分式。

令$k(w_i) = \min_{1 \leq j \leq n} ord_\infty M_{ij}$。我们先选定$r_i(x) = x^{-k(w_i)}$。这保证了$r_i w_i$在∞的整性。注意到$M$的行列式在∞处的阶总是$\geq \sum k(w_i)$。我们可以证明基底会变

为正则当有等号成立时。令$\widehat{M}=(r_iM_{ij})$为$r_iw_i$的变换矩阵。由于行列式在$\infty$处是整的等价于在$\infty$处有非零的值。令$N=(N_{ij})$，其中$N_{ij}$为$M_{ij}$在$\infty$处的值。由于对矩阵取行列式与赋值两种操作可以交换，$N$的行列式即为$\widehat{M}$的行列式在$\infty$处的值。因此$r_iw_i$是$\infty$处的局部整基当且仅当$N$有非零的行列式。如果$N$的行列式为零，则存在常数$c_i \in K$使得$\sum\limits_{i=1}^{n}c_iN_{ij}=0 \quad \forall 1 \leq j \leq n$。令$i0$为满足$c_i \neq 0$且$k(w_i)$最小的$i$，并定义

$$\widehat{w}_{i0} = \sum_{i=1}^{n} c_i x^{k(w_{i0})-k(w_i)} w_i$$

然后将$w_{i0}$替换为$\widehat{w}_{i0}$仍构成整体整基。类似地，将$\widehat{M}_{i0,j}$替换为$\sum c_i \widehat{M}_{ij}$会使得这一列各元素的阶严格正。因此$k(\widehat{w}_{i0}) > k(w_{i0})$，变换矩阵行列式的阶在我们的新基下保持不变。如此下去，经过有限步，变换矩阵行列式的阶将会等于$\sum k(w_i)$。

我们已经展示了在给定在$\infty$处局部整基的前提下，如何将任意一组基底改造为在$\infty$处正则。接下来，问题自然就变为如何计算$\infty$处局部整基。如果我们令$x=1/z$，则$\infty$就变为了$z$-空间的原点$0$。为了计算$0$处的局部整基，我们可以在经过一个最优化后使用前一节的算法。在$0$处的赋值环的可逆元即为不能被$z$整除的多项式。因此，我们可以用判别式所含的最高的$z$的幂次代替算法4.1中的判别式。最后再将$z$用$1/x$替换回即得到了$\infty$处的局部整基。

**对简单根式扩张情形的特别讨论**

如果$F$是一个域，且$y$在$F$上为$n$次代数元，并满足$y^n \in F$，我们就称$F(y)$为$F$的简单根式扩张。如果$F$的特征与$n$互素，则我们可以假设$F$包含一个$n$次单位根$\omega$，否则适当地扩张$F$即可。则存在唯一的微分自同构$\sigma: F(y) \to F$使得$\sigma(y)=\omega y$。引入算子：

$$T_i = \frac{1}{n}\sum_{j=0}^{n-1} \frac{\sigma^j}{\omega^{ij}}$$

注意到$T_i(y^j)=\delta_{ij}y^i$，其中$\delta_{ij}$为Dirac符号，在$i=j$时取1，在其它情形下取0。因此，$g=\sum g_iy^i$，其中$g_i \in F$，我们有$T_i(g)=g_iy^i$。由于$\sigma$总是将整函数映为整函数，整函数的和与积仍为整函数，因此我们可以知道$T_i$也将整函数映为整函数。如果$g$是一个整函数，则上面的讨论揭示了其展开式的每一项$g_iy^i$都必须也是整的，而这意味着基底$[1,y,\ldots,y^{n-1}]$处处正则，我们将其写为一个命题：

**命题4.1.** 如果$K(x,y)$是$K(x)$的$n$次简单根式扩张，且$n$与$K$的特征互素，则自然基$[1,y,\ldots,y^{n-1}]$处处正则。

不失一般性，我们可以假定$y$满足如下无平方因子分解：

$$y^n = \prod_{i=0}^{n-1} p_i^i$$

其中，$p_i \in K[x]$且不含平方因子。因此为使我们的自然基成为整基，我们只需找到次数尽可能大的多项式$d_i(x)$使得$y^i/d_i(x)$是整的。将这一表达式$n$次幂便得到$\prod p_j^{ij}/d_i^n \in K[x]$。这样便容易发现我们要求的次数最高的$d_i(x)$即为

$$d_i = \prod_{j=0}^{n-1} p_j^{[\frac{ij}{n}]}$$

因此，我们便统一得到了简单根式扩张下的一组整基$[1, \dfrac{y}{d_1(x)}, \ldots, \dfrac{y^{n-1}}{d_{n-1}(x)}]$

### 4.2.3 绝对不可约性

我们已经假定被积函数$y$的极小多项式$f(x,y)$在$K(x)$上不可约，但是在积分过程中，我们可能需要对系数域$K$进行域扩张，在扩张后的域上$f$可能不再是不可约的了。事实上，我们需要保证$f$在任何$K$的代数扩张域上仍是不可约的，被称为绝对不可约多项式。另一个困难在于确定用于表达的精确的系数域。起初，我们定义系数域$K$为$\mathbb{Q}$的用于表达$y$的极小多项式的最小扩张。任何函数域中的$k$-代数元也可以视为系数域的一部分。例如，如果$f(x,y) = y^4 - 2x^2$，则$y^2/x = \sqrt{2}$在$\mathbb{Q}$上代数。注意到一旦我们将$\sqrt{2}$加入$K$中，$f$就不再是不可约的，$y$将满足一个扩域上次数为2的多项式。尽可能地扩张系数域可以降低$y$的极小多项式的次数，由于我们的算法强烈依赖这个次数，所以这将明显地提升算法效率。我们现在定义$K(x,y)$的真系数域为$K^o$为$K$在$K(x,y)$的整闭包。由前面的例子我们可以看出$K(x,y)$的元素在$K$上是否代数（整）与$f(x,y)$的绝对不可约性有关。我们下面将证明事实确实如此，寻找$y$的绝对不可约极小多项式的过程将引领我们发现$K(x,y)$的真系数域。

鉴于上面的例子有明显人造的迹象，我们有理由相信在实际应用中，定义多项式是不可约却不是绝对不可约的情形很少见。事实确实如此，而上一章中计算所得的整基正好可以用来快速检验定义多项式的绝对不可约性。正如Duval在[18]中发现的那样，定义多项式的绝对不可约因子个数恰好等于除子(1)的重数，又由于我们已经计算得到在$\infty$处正则的事项，所以我们只需要检验其中有几个在$\infty$处没有极点。如果只有一个，我们就可以跳过本章的算法，因为我们已经保证了定义多项式的绝对不可约性。

在本章我们始终假定原系数域$K$是完全的，即$K$上的不可约多项式均没有重根，或$K$上的多项式都是可分的。在此假定下，$K$的任何有限代数扩张实际均为单扩张。

**真系数域和正则扩张**

我们首先介绍一些纯代数的结果。

**引理4.4.** 设$x$是域$K$上的超越元，则$K$在$K(x)$中整闭。

**引理4.5.** $F/K$为域扩张，$K$在$F$整闭，$K(\alpha)$为$K$的单代数扩张，则$[F(\alpha):F] = [K(\alpha):K]$。

**推论4.1.** 设$x$是域$K$上的超越元，$F/K$为单代数扩张，则$[F:K] = [F(x):K(x)]$。

**定理4.14.** 设$f(x,y)$是完全域$K$上的不可约多项式，则它是绝对不可约的当且仅当$K$在$K(x,y)$中整闭，即$K = K^o$。

**推论4.2.** 设$f(x,y)$在$K^o$上不可约，则$f$绝对不可约。

## 4.2.4　积分的有理部分

**Liouville定理和有理函数积分算法回顾**

**代数函数积分**

为避免使用繁琐的Puiseux展开式和不必要的代数数计算，我们将基于有理函数积分的Hermite方法得到计算代数函数积分有理部分的算法。在此，我们不使用Horowitz-Ostrogradsky方法的原因在于，即使最终积分没有初等函数表达，我们也可以将给出部分结果，将不可积的部分化的尽量简单，而并不是仅仅返回"不可积"。

为简单起见，我们将对被积函数进行变量替换，使之在$\infty$处没有极点或分支点。我们假定积分有形式$\int \sum \dfrac{R_i(x)}{Q(x)} y^i \mathrm{d}x$，其中$R_i$和$Q$都是$x$的多项式。设$a$为既不是$Q$的根也不是$f$的判别式的根的一个整数，并令$z = 1/(x-a)$，或等价地，$x = a + 1/z$。则积分则变为

$$\int \sum \frac{R_i(a + \frac{1}{z})}{Q(a + \frac{1}{z})} (-z^{-2}) y^i \mathrm{d}z$$

最后为得到最终结果，我们只需再经过一次相反的变量替换即可得到原本以$x$为项的答案。如果我们设$y$所满足的极小多项式$f$中$x$的最高幂次为$m$，则经变量替换后$y$所满足的极小多项式为$g(z,y) = z^m f(a+1/z, y)$。利用上一节的算法，我们可以找到$K[z]$在$K(z,y)$的整闭包的一组基（即整基）$[w_1, \ldots, w_n]$。因此，被积函数可表示为$\sum A_i(z) w_i / D(z)$，其中$A_i$和$D$都是$z$的多项式。由于经过变量替换后的被积函数在$\infty$处没有极点，因此$\deg(A_i) < \deg(D) \quad \forall i$

我们现在尝试模仿Hermite算法来求积分的有理部分。同样地，先对分母$D(z)$进行无平方因子分解

$$D = \prod_i D_i^i$$

并对某$k > 0$，令$V = D_{k+1}, U = D/V^{k+1}$。仿照Hermite算法，接下来我们需要寻找多项式$B_i$和$C_i$使得

$$\int \sum \frac{A_i}{UV^{k+1}} w_i \mathrm{d}z = \sum \frac{B_i}{V^k} w_i + \int \sum \frac{C_i}{UV^k} w_i \mathrm{d}z \qquad (4.21)$$

### 4.2.5 积分的对数部分

### 4.2.6 例子

**例4.15.** 计算$I = \int \sqrt{x + \sqrt{x}}\, \mathrm{d}x$

（法一：利用变量代换，将问题化为一次代数扩张的情形）

证明. 令$w = \sqrt{x}$，则积分化为$I = 2\int w\sqrt{w^2 + w}\, \mathrm{d}w$

令$y = \sqrt{w^2 + w}$，则被积函数变为$wy$，且$y$满足的极小多项式为$f(y, w) = y^2 - w(w + 1)$。

由算法，我们知$f(y, w)$为绝对不可约多项式。

根据简单根式扩张计算整基的简便算法，容易求得

$$p_1(w) = w(w + 1)$$

进而

$$d_1(w) = p_1(w)^{[\frac{1}{2}]} = 1$$

故整基为$1, y$。

为消去$\infty$处的极点，做变量代换$z = \dfrac{1}{w-1}$。此时$I = -2\int \dfrac{z+1}{z^3} y\, \mathrm{d}z$，且$y$满足的极小多项式变为$f(y, z) = z^2 y^2 - (z+1)(2z+1)$。由于

$$\frac{\mathrm{d}y}{\mathrm{d}z} = -\frac{\partial f/\partial z}{\partial f/\partial y} = -\frac{3z+2}{2z(z+1)(2z+1)} y$$

故知算法中的$E = z(z+1)(2z+1)$。

为使用Hermite方法，需考虑如下积分

$$I = -2\int \frac{(z+1)^2(2z+1)}{z^3(z+1)(2z+1)} y\, \mathrm{d}z$$

设

$$I = -2\int \frac{(z+1)^2(2z+1)}{z^3(z+1)(2z+1)} y\, \mathrm{d}z = \frac{B_0}{z^2} + \frac{B_1 y}{z^2} + \int \frac{C_0 + C_1 y}{z^2(z+1)(2z+1)}\, \mathrm{d}z$$

两边同时对$z$微分，乘以公母$z^3(z+1)(2z+1)$，并$mod\ z$，得

$$-2y \equiv -2B_0 - 2B_1 y + B_1 y' z(z+1)(2z+1) \quad (mod\ z)$$

而

$$y'z(z+1)(2z+1) \equiv -\frac{3z+2}{2}y \equiv -y \quad (mod\ z)$$

故可得

$$\begin{cases} 0 = B_0 \\ -2 = -2B_1 - B_1 \end{cases} \Rightarrow \begin{cases} B_0 = 0 \\ B_1 = \frac{2}{3} \end{cases}$$

带回原积分，得

$$I - \frac{2y}{3z^2} = -\frac{1}{3}\int \frac{12z^2 + 22z + 9}{z^2(z+1)(2z+1)} y\,\mathrm{d}z = \frac{B_0}{z} + \frac{B_1 y}{z} + \int \frac{C_0 + C_1 y}{z(z+1)(2z+1)}\,\mathrm{d}z$$

两边同时对$z$微分，乘以公母$z^2(z+1)(2z+1)$，并$mod\ z$，得

$$-3y \equiv -B_0 - B_1 y + B_1 y' z(z+1)(2z+1) \quad (mod\ z)$$

故可得

$$\begin{cases} B_0 = 0 \\ B_1 = \frac{3}{2} \end{cases}$$

带回原积分，得

$$I - \frac{y}{3z^2} - \frac{3y}{2z} = -\frac{1}{12}\int \frac{12z+7}{z(z+1)(2z+1)} y\,\mathrm{d}z = B_0 + B_1 y + \int \frac{C_0 + C_1 y}{(z+1)(2z+1)}\,\mathrm{d}z$$

两边同时对$z$微分，乘以公母$z(z+1)(2z+1)$，并$mod\ z$，得

$$-\frac{7}{12}y \equiv B_1 y' z(z+1)(2z+1) \quad (mod\ z)$$

故可得

$$\begin{cases} B_0 = 0 \\ B_1 = \frac{7}{12} \end{cases}$$

带回原积分，分别得到原积分的有理部分$I_1$和对数部分$I_2$

$$\begin{aligned}
I_1 &\triangleq \frac{2y}{3z^2} + \frac{3y}{2z} + \frac{7y}{12} \\
&= \frac{1}{12}(8w^2 + 2w - 3)y = \frac{1}{12}(8x + 2\sqrt{x} - 3)\sqrt{x + \sqrt{x}} \\
I_2 &\triangleq -\frac{1}{8}\int \frac{y}{(z+1)(2z+1)}\,\mathrm{d}z \\
&= -\frac{1}{8}\int \frac{y}{\frac{w}{w-1} \cdot \frac{w+1}{w-1}}(-\frac{1}{(w-1)^2})\,\mathrm{d}w \\
&= \frac{1}{8}\int \frac{1}{y}\,\mathrm{d}w = \frac{1}{8}\int \frac{y}{w(w+1)}\,\mathrm{d}w
\end{aligned}$$

下求积分的对数部分$-8I_2$：

$$D(w) = w(w+1), \quad g(y,w) = y, \quad f(y,w) = y^2 - w(w+1)$$

有

$$\begin{aligned}
\operatorname{Res}_y(ZD'(w) - g(y,w), f(y,w)) &= \operatorname{Res}_y((2w+1)Z - y, y^2 - w(w+1)) \\
&= \begin{vmatrix} -1 & 0 & 1 \\ (2w+1)Z & -1 & 0 \\ 0 & (2w+1)Z & -w(w+1) \end{vmatrix} \\
&= (2w+1)^2 Z^2 - w(w+1)
\end{aligned}$$

$$\begin{aligned}
R(Z) &= \operatorname{Res}_w(\operatorname{Res}_y(ZD'(w) - g(y,w), f(y,w)), D(w)) \\
&= \operatorname{Res}_w((2w+1)^2 Z^2 - w(w+1), w(w+1)) \\
&= \begin{vmatrix} 4Z^2 - 1 & 0 & 1 & 0 \\ 4Z^2 - 1 & 4Z^2 - 1 & 1 & 1 \\ Z^2 & 4Z^2 - 1 & 0 & 1 \\ 0 & Z^2 & 0 & 0 \end{vmatrix} \\
&=
\end{aligned}$$

□

（法二：直接考虑一次代数扩张）

证明. 令$y = \sqrt{x + \sqrt{x}}$，则积分化为$I = \displaystyle\int y\,\mathrm{d}x$，且$y$满足的极小多项式为$f(y,x) = y^4 - 2xy^2 + x^2 - x, 4y^3 - 4xy$。

由算法，我们知$f(y,x)$为绝对不可约多项式。

$$\begin{aligned}
d &= \operatorname{Res}_y(f, f'_y) = \operatorname{Res}_y(y^4 - 2xy^2 + x^2 - x, 4y^3 - 4xy) \\
&= \begin{vmatrix}
1 & 0 & 0 & 4 & 0 & 0 & 0 \\
0 & 1 & 0 & 0 & 4 & 0 & 0 \\
-2x & 0 & 1 & -4x & 0 & 4 & 0 \\
0 & -2x & 0 & 0 & -4x & 0 & 4 \\
x^2 - x & 0 & -2x & 0 & 0 & -4x & 0 \\
0 & x^2 - x & 0 & 0 & 0 & 0 & -4x \\
0 & 0 & x^2 - x & 0 & 0 & 0 & 0
\end{vmatrix} \\
&= 256x^3(x-1)
\end{aligned}$$

$$\overline{w} = [1, y, y^2, y^3]$$

$$M_{\overline{w}} = \begin{bmatrix} 1 & 1 & 1 & 1 \\ (x+\sqrt{x})^{\frac{1}{2}} & (x-\sqrt{x})^{\frac{1}{2}} & -(x+\sqrt{x})^{\frac{1}{2}} & -(x-\sqrt{x})^{\frac{1}{2}} \\ x+\sqrt{x} & x-\sqrt{x} & x+\sqrt{x} & x-\sqrt{x} \\ (x+\sqrt{x})^{\frac{3}{2}} & (x-\sqrt{x})^{\frac{3}{2}} & -(x+\sqrt{x})^{\frac{3}{2}} & -(x-\sqrt{x})^{\frac{3}{2}} \end{bmatrix}$$

$$SP_{\overline{w}} = \begin{bmatrix} 4 & 0 & 4x & 0 \\ 0 & 4x & 0 & 4(x^2+x) \\ 4x & 0 & 4(x^2+x) & 0 \\ 0 & 4(x^2+x) & 0 & 4(x^3+3x^2) \end{bmatrix}$$

$$\widehat{M_q} = \mathrm{diag}\{1, x, x, x\}$$

故$J_x(V)$的一组基为$\overline{m} = [x, y, y^2, y^3]$

$$M_1 = \mathrm{diag}\{1, x, x, x\} \quad, \quad M_2 = \begin{pmatrix} 0 & 0 & 0 & 1-x \\ 1 & 0 & 0 & 0 \\ 0 & 1 & 0 & 2x \\ 0 & 0 & 1 & 0 \end{pmatrix},$$

$$M_3 = \begin{pmatrix} 0 & 0 & 1-x & 0 \\ 0 & 0 & 0 & x(1-x) \\ 1 & 0 & 2x & 0 \\ 0 & 1 & 0 & 2x \end{pmatrix} \quad, \quad M_4 = \begin{pmatrix} 0 & 1-x & 0 & 2x(1-x) \\ 0 & 0 & x(1-x) & 0 \\ 0 & 2x & 0 & x(1+3x) \\ 1 & 0 & 2x & 0 \end{pmatrix}$$

$$\widehat{M} = I_4$$

故$V$是整闭的，一组整基为$\overline{v} = [1, y, y^2, y^3]$

$$\frac{\mathrm{d}y}{\mathrm{d}x} = -\frac{\partial f/\partial x}{\partial f/\partial y} = \frac{4xy^2 - 2x + 1}{4y^3 - 4xy} = \frac{y^3 + (1-3x)y}{4x(x-1)}$$

$$I_2 \triangleq \frac{1}{16} \int \frac{y^3 - (x+1)y}{x(x-1)} \, \mathrm{d}x$$

为消去$\infty$处的极点，做变量代换$z = \dfrac{1}{x+1}$。此时$I = -\int \dfrac{1}{z^2} y \, \mathrm{d}z$，且$y$满足的极小多项式变为$f(y, z) = z^2 y^4 + 2z(z-1)y^2 + 2z^2 - 3z + 1$。由于

$$\frac{\mathrm{d}y}{\mathrm{d}z} = \frac{\mathrm{d}y}{\mathrm{d}x} \cdot \frac{\mathrm{d}x}{\mathrm{d}z} = -\frac{zy^3 + (3z-2)y}{4z(z-1)(2z-1)}$$

故知算法中的$E = z(z-1)(2z-1)$。

下求积分的对数部分$16I_2$：

$$D(x) = x(x-1), g(x,y) = y^3 - (x+1)y, f(x,y) = y^4 - 2xy^2 + x^2 - x$$

有

$$\begin{aligned}
&\text{Res}_y(ZD'(x) - g(x,y), f(x,y)) \\
=\ &\text{Res}_y((2x-1)Z - y^3 + (x+1)y, y^4 - 2xy^2 + x^2 - x) \\
=\ &(-1+2x)^2 Z^2(2x - 2x^2 + Z^2 - 4xZ^2 + 4x^2Z^2) - x(x-1)^3
\end{aligned}$$

$$\begin{aligned}
R(Z) &= \text{Res}_x(\text{Res}_y(ZD'(x) - g(x,y), f(x,y)), D(x)) \\
&= \text{Res}_x((-1+2x)^2 Z^2(2x - 2x^2 + Z^2 - 4xZ^2 + 4x^2Z^2) - x(x-1)^3, x(x-1)) \\
&= Z^8
\end{aligned}$$

$\square$

## 3.4 潜在的研究方向

像之前提到的一样，不定积分是符号积分系统研究的核心功能，也是其它符号积分系统功能实现的基础。主流的计算机代数系统，如 Maple，Mathematica 等不仅实现了不定积分功能，定积分，路径积分，瑕积分等功能也都进行了相应的实现。毫无疑问，这些方向也是未来我们要关注并大力研究的方向。

例如定积分的实现过程中采用的方法之一——牛顿—莱布尼兹算法就是建立在不定积分基础上的一个算法，当然这不是唯一一种能够解决定积分问题的算法。我们可以认为不定积分为符号积分系统提供了一个基础，符号积分系统在不定积分的基础上进行发展，使得符号积分系统的功能越来越多，越来越强大，进而反作用于不定积分领域，提供不定积分理论研究与实现的动力，二者相辅相成，不可分割。

# 参考文献